\documentclass[aps,superscriptaddress,floatfix,nofootinbib,notitlepage]{revtex4-1}
\usepackage{amsmath,amssymb,amsfonts,mathtools} 
\usepackage[normalem]{ulem}
\usepackage[utf8]{inputenc} 
\usepackage[T1]{fontenc}
\usepackage{hyperref}

\usepackage{graphicx}
\usepackage{latexsym}
\usepackage{bbm}
\usepackage{epstopdf}
\usepackage{color}
\usepackage{braket}
\usepackage{hyperref}
\usepackage{lineno}

\usepackage{slashed}
\usepackage{placeins} 
\usepackage{diagbox}

\usepackage[dvipsnames]{xcolor}

\newcommand{\be}{\begin{equation}} 
\newcommand{\ee}{\end{equation}}
\newcommand{\bea}{\begin{eqnarray}} 
\newcommand{\eea}{\end{eqnarray}}
\def\lsim{\mathrel{\raise.3ex\hbox{$<$\kern-.75em\lower1ex\hbox{$\sim$}}}}
\def\gsim{\mathrel{\raise.3ex\hbox{$>$\kern-.75em\lower1ex\hbox{$\sim$}}}}

\newcommand{\bnl}{Physics Department, Brookhaven National Laboratory, Upton, NY 11973, USA} 
\newcommand{\catania}{INFN Sezione di Catania, I-95123 Catania, Italy}
\newcommand{\ceem}{Center for  Exploration  of  Energy  and  Matter,
Indiana  University,
Bloomington,  IN  47408,  USA}

\newcommand{\genoa}{INFN Sezione di Genova, I-16146 Genova, Italy} 
\newcommand{\gwu}{The George Washington University, Washington, DC 20052, USA}
\newcommand{\icn}{Instituto de Ciencias Nucleares, 
Universidad Nacional Aut\'onoma de M\'exico, Ciudad de M\'exico 04510, Mexico}

\newcommand{\indiana}{Department of Physics, Indiana  University, Bloomington,  IN  47405,  USA}
\newcommand{\jlabth}{Theory Center, Thomas  Jefferson  National  Accelerator  Facility, Newport  News,  VA  23606,  USA}

\newcommand{\lbnl}{Nuclear Science Division, Lawrence Berkeley National Laboratory, Berkeley, CA 94720, USA}
\newcommand{\mainz}{Institut f\"ur Kernphysik \& PRISMA$^+$  Cluster of Excellence, Johannes Gutenberg Universit\"at,  D-55099 Mainz, Germany}

\newcommand{\nmsu}{Department of Physics, New Mexico State University, Las Cruces, NM 88003, USA}

\newcommand{\uned}{Departamento de F\'isica Interdisciplinar, Universidad Nacional de Educaci\'on a Distancia (UNED), Madrid E-28040, Spain}

\newcommand{\lanl}{Theoretical Division, Los Alamos National Laboratory, Los Alamos, NM 87545, USA}

\newcommand{\llnl}{Nuclear and Chemical Sciences Division, Lawrence Livermore National Laboratory, Livermore, CA 94551, USA}

\newcommand{\messina}{Dipartimento di Scienze Matematiche e Informatiche, Scienze Fisiche e Scienze della Terra, 
Universit\`a degli Studi di Messina, I-98122 Messina, Italy}

\newcommand{\nikhef}{Nikhef Theory Group, Science Park 105, 1098 XG Amsterdam, The Netherlands}

\newcommand{\ub}{Departament de F\'isica Qu\`antica i Astrof\'isica and Institut de Ci\`encies del Cosmos, Universitat de Barcelona, Barcelona E-08028, Spain}

\newcommand{\uva}{University of Virginia, Charlottesville, VA 22904, USA}
\newcommand{\ucla}{Department of Physics and Astronomy, University of California, Los Angeles, CA 90095, USA}

\newcommand{\pittsburg}{Department of Physics and Astronomy, University of Pittsburgh, Pittsburgh PA 15260, USA}
\newcommand{\arizona}{Department of Physics, Arizona State University, Tempe, AZ 85287-1504, USA}

\newcommand{\ohio}{Department of Physics, The Ohio State University, Columbus, OH 43210, USA}
\newcommand{\graz}{Institut f\"ur Physik, Universit\"at Graz, 8010 Graz, Austria}
\newcommand{\giessen}{Institut f\"ur Theoretische Physik, Justus-Liebig-Universit\"at Giessen, 35392
Giessen, Germany}
\newcommand{\fairgiessen}{Helmholtz Forschungsakademie Hessen f\"ur FAIR (HFHF), GSI Helmholtzzentrum f\"ur Schwerionenforschung, Campus Giessen, 35392 Giessen, Germany}
\newcommand{\itp}{CAS Key Laboratory of Theoretical Physics, Institute of Theoretical Physics,
Chinese Academy of Sciences, Beijing 100190, China}
\newcommand{\ucas}{School of Physical Sciences, University of Chinese Academy of Sciences, Beijing 100049, China}

\newcommand{\ucd}{Department of Physics and Astronomy, University of California, Davis, CA 95616, USA}
\newcommand{\ut}{Institute for Theoretical Physics, T\"ubingen University, 72076 T\"ubingen, Germany}
\newcommand{\ur}{Institute for Theoretical Physics, Regensburg University, D-93040 Regensburg, Germany}
\newcommand{\uw}{Department of Physics, University of Washington, Seattle, WA 98195, USA}
\newcommand{\kielce}{Institute of Physics, Jan Kochanowski University, 25-406 Kielce, Poland}
\newcommand{\frankfurt}{Institute for Theoretical Physics, J. W. Goethe University, 60438 Frankfurt am Main, Germany}

\newcommand{\TUM}{Technical University of Munich, TUM School of Natural Sciences, Department of Physics T30f, James-Franck-Stra{\ss}e 1, 85748 Garching, Germany
}

\newcommand{\glas}{SUPA, School of Physics and Astronomy, University of Glasgow, Glasgow, G12 8QQ, United Kingdom}
\newcommand{\jag}{Institute of Theoretical Physics, Jagiellonian University, S. Lojasiewicza 11, 30-348 Krak{\'o}w, Poland}
\newcommand{\FIU}{Florida International University, Miami FL, 33199, USA}
\newcommand{\psuberks}{Physics, Division of Science, Penn State Berks, Reading, PA 19610, USA}
\newcommand{\CPHT}{CPHT, CNRS, \'Ecole polytechnique, Institut Polytechnique de Paris, 91120 Palaiseau, France}

\newcommand{\jyv}{Department of Physics, P.O. Box 35, 40014 University of Jyv{\"a}skyl{\"a}, Finland}

\begin{document}

\title{The case for an EIC Theory Alliance: Theoretical Challenges of the EIC}

\author{Raktim Abir}
\affiliation{Department of Physics, Aligarh Muslim University, Aligarh (U.P.)-202002, India}

\author{Igor Akushevich}
\affiliation{Department of Physics, Duke University, Durham, NC 27708, USA}

\author{Tolga Altinoluk}
\affiliation{Theoretical Physics Division, National Centre for Nuclear Research, Pasteura 7, Warsaw 02-093, Poland}

\author{Daniele Paolo Anderle }
\affiliation{South China Normal University, /no.55 Zhongshan West Avenue, Tianhe District, Guangzhou City, Guangdong Province, China}

\author{Fatma P. Aslan}
\affiliation{Department of Physics, University of Connecticut, Storrs, CT 06269-3046, USA}
\affiliation{\jlabth}

\author{Alessandro Bacchetta}
\affiliation{Dipartimento di Fisica, Universit `a degli Studi di Pavia, I-27100 Pavia, Italy}
 \affiliation{Istituto Nazionale di Fisica Nucleare, Sezione di Pavia, I-27100 Pavia, Italy}

\author{Baha Balantekin}
\affiliation{Department of Physics, University of Wisconsin–Madison, Madison, Wisconsin 53706, USA}

\author{Joao Barata}
\affiliation{\bnl}

\author{Marco Battaglieri}
\affiliation{\genoa} 

\author{Carlos A. Bertulani}
\affiliation{Department of Physics and Astronomy, Texas A\&M University-Commerce, TX 75429-3011, USA}

% ...

\date{\today}

\begin{abstract}
We outline the physics opportunities provided by the Electron Ion Collider (EIC). These include the study of the parton structure of the nucleon and nuclei, the onset of gluon saturation, the production of jets and heavy flavor,
hadron spectroscopy and tests of fundamental symmetries. We review the present status and future challenges in EIC theory that have to be addressed in order to realize this ambitious and impactful
physics program, 
including how to engage a diverse and inclusive workforce.
In order to address these many-fold challenges, we propose a coordinated effort involving theory groups with differing expertise is needed. We discuss the scientific goals and scope of such an EIC Theory Alliance.
\end{abstract}

\maketitle

\tableofcontents
\newpage

\pagenumbering{arabic}                
\pagestyle{plain} 

\section{Introduction}
\label{sec:intro}

The Electron-Ion Collider (EIC) was named the highest priority new construction project in the 2015 Long Range Plan for Nuclear Science \cite{NSAC}. The main goal of the EIC is to study the structure of hadrons and nuclei, including their 3-dimensional partonic structure and gluon saturation. The EIC will be the ultimate QCD machine, of far broader reach than other related facilities such as HERA, CEBAF, and RHIC. The EIC can make important contributions to other areas as well, including hadron spectroscopy, nuclear structure, and tests of fundamental symmetries. There are also synergies with high energy physics, such as the precision determination of parton densities. 

The key questions at the heart of EIC physics are deeply theoretical in nature and cannot be addressed by EIC experiments alone: it is crucial to drive a robust and synergistic theory effort alongside the experimental program.
The need for strong theory support to realize the full discovery potential of the EIC was pointed out in the National Academy of Sciences, Engineering, and Medicine report, ``An Assessment of U.S.-Based Electron-Ion Collider Science'' \cite{NAP25171}. 
Due to the long time frame for EIC construction, the prospect of decades of operation, and the broad interdisciplinary nature of EIC theory, it is critical to develop a strategic plan to advance the theory thrusts and meet the associated need for a strong, diverse theory workforce.

Over the last 5 years, interest in the EIC has resulted in seven faculty hires in related theory areas. This is a strong trend that must be sustained as the workforce develops. There is an acute need for growth at all levels, especially as the nature of EIC theory research differs from nuclear theory of the previous decades: the EIC will produce unprecedented amounts of data at unparalleled precision that needs to be understood, requiring researchers equipped with new theoretical and computational tools, such as AI/ML.

To assess these needs and challenges, a CFNS workshop ``Theory for the EIC in the next decade'' was organized at MIT September 20-22, 2022. The workshop was held in hybrid format, attracted 65 participants, and included 15 talks and 7 discussion sessions, see {\tt https://indico.bnl.gov/event/16740/}. The workshop concluded with a resolution session where it was determined that the optimal path forward is to create a Theory Alliance alongside the EIC facility. The Alliance was also discussed at the Hot and Cold QCD Town Hall meeting, held at MIT September 23-25, 2022, see {\tt https://indico.mit.edu/event/538/}, and gained strong support among the meeting participants.
Very recently the theory challanges of EIC have been discussed at the POETIC 2023 (Physics Opportunities at an Electron-Ion Collider 2023) conference, May 2-6, S\~ao Paulo,  Brazil, see {\tt https://www.ictp-saifr.org/poetic2023/}.

The central goal of the EIC Theory Alliance (EIC-TA) is to support the mission of the EIC by maximizing the scientific output from the experiment and deepening our understanding of QCD. 
To achieve these objectives, the EIC-TA will steward the EIC theory program; develop and mould the excellent and inclusive theory workforce necessary for the tasks at hand; and ensure that this workforce has the tools, resources, visibility, and support necessary to advance the EIC science program.

\subsection{Alliance structure}
The formation of a Theory Alliance funded throughout the lifetime of the EIC facility -- in addition to the base theory funding -- is the optimal way to advance EIC science from the theory front.
This is evidenced by the successes of the Theory Alliance formed alongside FRIB (FRIB-TA), which provides a model for the EIC-TA to build from. 
Topical collaborations in nuclear theory have been and will remain valuable resources for building collaborative efforts in various areas, including aspects of EIC-related theory; however, these collaborations address a targeted issue within a five-year period. 
EIC theory is broad and interdisciplinary, and is expected to require a sustained long-term effort given the timescales of the experiment. 

Moreover, the formation of an EIC-TA opens up a unique window to establish new norms in QCD theory and build a community that actively contributes to leadership, outreach, mentorship, and service efforts -- as we strive to grow nuclear theory and increasingly bring in talent that better reflects the demographic composition of the U.S.\ as a whole.
Topical collaborations are too narrowly focused and short in duration to make a system-wide impact in nuclear theory, which substantially lags behind other subfields with severe and longstanding issues of inclusion in physics. 
Organizations like the DNP are too broad in scope to address issues of climate and demographics in every local sector of the community. 
The EIC-TA would encompass an appropriately-sized  community to make a systemic impact on QCD theory during an important and exciting period of growth for the field.

The EIC-TA will be a membership organization run by a member-elected executive board. EIC-TA membership will be free and open to all students, postdocs, research staff, and faculty, from both within the US and internationally, who wish to join. 
The EIC-TA will have an open and easily accessible website, advertising itself and its activities, and explaining how new members can join. 

The executive board will determine the major scientific thrusts of the theory alliance and make decisions regarding EIC-TA sponsored fellowships and faculty bridge positions. Furthermore, the executive board will coordinate the organization of workshops and schools related to the research activities of the alliance. 
The EIC-TA will seek to recruit executive board candidates that are diverse along axes including but not limited to gender, race/ethnicity, topic of expertise, geographic, and type of institution. 
Executive members will have rotating multi-year terms to ensure continuity in governance. The executive board will also include short-term member-at-large positions for graduate, postdoc, and faculty representatives.

\subsection{Scientific targets}

The theoretical landscape for EIC physics is broad and compelling.
The theoretical understanding of the 3-dimensional partonic structure of hadrons, one key physics topic, has changed significantly over the last few years. In the past, global analyses based on perturbative QCD, as well as model calculations, have been the only tools available to obtain information about the parton distribution functions (PDFs), the generalized parton distribution functions (GPDs), and the transverse momentum distributions (TMDs). It has become clear in the last few years that lattice QCD (LQCD) calculations will play an important role in constraining
the PDFs, GPDs and TMDs. Next-to-leading order calculations in the color glass condensate framework also appeared recently, paving the way to quantitative studies of gluon saturation at the EIC. With all this progress, one emerging question to be addressed is that of the connection between TMD factorization and gluon saturation. While exploring the origin of nucleon mass and spin, the study of 3-dimensional partonic structure of hadrons, and the study of onset of gluon saturation are at core of the EIC physics program, there are many other exciting opportunities for studying QCD with the help of EIC. These include the precision determination of the parton distributions of hadrons, precision studies of jets, studies of hadronization and production of heavy flavor hadrons, resolving puzzles in hadron spectroscopy and a better understanding of the structure of nuclei, such as the role of short range correlations. In addition, tests of fundamental symmetries can be carried out with the help of the EIC.

In section \ref{WF+DEI}, we discuss in greater depth EIC-TA plans for developing an strong and diverse theory workforce. 
In the remaining sections we outline the science case for the EIC-TA based on broad range of physics topics, starting from 3-dimensional hadron structure (GPDs and TMDs) and gluon saturation 
at small $x$, which are the core topics in QCD to be explored
at EIC as explained in the NAS report. This discussion is
followed by sections on 
precision QCD and global analysis of hadron structure, 
which are important topics at intersctions of high energy and
nuclear physics. Furthermore, we discuss other theory 
opportunities related to EIC:
the study of jets in $e + p$ collisions, the study of hadronization and heavy flavor production, exploring aspects of nuclear structure, and tests of fundamental symmetries. We also elaborate on new directions such as intersections with Quantum Information Science (QIS) and AI/ML. 
Connections to lattice QCD calculations is highlighted in
all these sections.
We conclude with a brief summary in section \ref{NewSummary}.

\section{Workforce development and DEI}
\label{WF+DEI}

The EIC presents a powerful opportunity to advance fundamental scientific knowledge and strengthen the US
nuclear theory workforce. To accomplish these goals, it is critical to ensure that EIC scientists are reaching out to and
developing scientific talent from all backgrounds. The EIC-TA plans to do its part by focusing directly on the parts the
pipeline that a research-centered alliance is best poised to address: the career stages of graduate students, postdoctoral
scholars, university faculty, and national laboratory staff scientists. Our unusual combination of a shared community
research objective, coupled with a broad nationwide and international reach across a diverse swath of institutions,
presents a unique opportunity to effect systemic improvements in how we recruit and train nuclear scientists.
The EIC-TA plans to implement both general workforce development initiatives, as well as targeted efforts to
increase participation of physicists from historically underrepresented populations, including but not limited to
race/ethnicity, gender, LGBT+, disability status, veteran, age, nontraditional and socioeconomic background.\footnote{\normalsize For example, underrepresented racial minorities (URMs) and women earned 7\% and 21\% of US PhDs in physics in 2020, despite making up 39\% and about half of the 25-29 year-old population, respectively \cite{APS-stats}. Furthermore, physics faculty are more than 12 times likely to have a parent with a Ph.D. than the general population, and about twice as likely as other individuals who hold a Ph.D. \cite{Morgan2022}.}

 The EIC-TA will organize workforce development and DEI needs and activities into three categories: recruitment and
outreach; workforce development, support, and retention; as well as workplace and community climate and culture.
The EIC-TA will appoint a Workforce Development and DEI Committee to serve as an independent advisory council
on these efforts. STEM professional societies regularly convene task forces to intensively study the vast existing peer-reviewed literature about workforce development and DEI, including workforce statistics and research into
best practices, with the end goal of generating detailed reports about implementing their findings within
a STEM context. The Workforce Development and DEI committee will draw from such reports and literature to improve EIC-TA policies, practices, procedures, and initiatives.  

\subsection{Recruitment and outreach}
The cornerstone of recruitment efforts in the EIC-TA will be fellowships and bridge programs for outstanding
faculty, postdoctoral scholars, and graduate students that will provide recipients with both funding and mentoring
to build careers advancing EIC science. The bridge program will follow the successful programs developed by the
RIKEN-BNL Research Center, the JLab Theory Center, and the FRIB Theory Alliance (FRIB-TA), which have played a key role in 
research fields related to the CEBAF, RHIC and FRIB. 
The postdoctoral fellowships will likewise emulate the highly efficacious FRIB-TA Fellow program, and will help to
promote incoming talent and make EIC theory an attractive field for hiring junior faculty at universities.
The EIC-TA will seek to fund positions
at a diverse range of institutions, and will require
partner institutions to commit to hiring practices that actively account for the significant biases present in the field.
Further, from its outset, the EIC-TA will seek to develop relationships with minority-serving institutions with the
goal of placing faculty on these campuses.

Each bridge position will support new tenure-track faculty and staff through a cost-sharing arrangement between
the EIC-TA and the university. 
At the end of the bridge period, the position will be fully supported by the university. Recognizing
possible constraints on early career faculty with families, bridge position recipients will not be expected to spend
significant time at the EIC site, but will receive additional support to make such visits as they deem appropriate to
further their research and career goals.
The EIC postdoctoral fellow program will allow exceptionally qualified candidates to spend time
as a fellow at a specified partner institution. At the end of the fellowship, the
fellow should demonstrate the capability to direct their own independent research program. The EIC-TA will also
periodically offer fellowships to advanced graduate students that both faculty and students can apply for. Providing
multiple mentors is considered current best practice and partner institutions should identify two or more senior
researchers willing to provide scientific and/or career guidance to the fellow during their appointment; at least one of
the faculty mentors must work in EIC theory.

\subsection{Workforce development, support, and retention}
Members of the EIC-TA will be expected to actively help build a strong and diverse workforce through mentoring,
training, and supporting junior physicists. All senior members of the EIC-TA will also be expected to participate
in professional development as needed to become more effective and inclusive research mentors, leaders within their
community, and active contributors to EIC-TA initiatives.

The EIC-TA will host regular alliance meetings, generally in conjunction with EIC User Group meetings, as well as
topical workshops to collaborate on theory-specific topics. The annual meetings and focused workshops will provide
opportunities for theorists to share and learn about new scientific developments, build extensive networks, and better
advance EIC science. A regular summer school will also be held to train junior alliance members on techniques and
open questions in EIC theory. These schools should also include lectures introducing students to the intersections
of nuclear theory, experiment, and computational methods. These schools will be structured to help junior EIC-TA
members form extended networks of mentors who will further their growth within the field.

At each conference, workshop, and summer school, the EIC-TA will invite at least one outside speaker to run
professional development sessions aimed at the career stages of participants. These sessions will rotate through topics
of interest to the community, including but not limited to: effective leadership and management; inclusive teaching
and mentoring; conflict management and mediation; ethical conduct of research; facilitating an inclusive workplace
climate; bystander intervention; best practices in hiring and admissions; and career and application guidance for
junior members.

The EIC-TA website, along with a community-wide mailing list with open sign-up, will ensure equitable access
to important information about upcoming events, open funding opportunities, relevant job postings, and leadership
positions coming up for election. To supplement in-person scientific training and professional development events, the
EIC-TA will also dedicate part of its website to career guidance and resources for its junior members. These pages will
include information developed by the EIC-TA about practices within the field as well as helpful links to more general
websites and resources in the physics and academic communities. The EIC-TA will also create dedicated pages listing
resources and opportunities for physicists from underrepresented segments of the community.

\subsection{Workplace and community climate and culture}
Widespread community involvement is necessary to build a strong workforce and make substantive headway on
significant, persistent, and pervasive issues of inclusion in nuclear physics. The EIC-TA aims to build a culture that
supports excellence in science and values the effective mentoring, strong leadership, and culture of service to the
community necessary to train and maintain an excellent scientific workforce. 
Active participation in impactful service, leadership, DEI, and mentorship activities will be an important component in decision-making.
Nominees for and holders
of EIC-TA leadership positions and funding recipients are expected to meet standards of professional conduct and
integrity as described in the EIC-TA Code of Conduct. Violations of these standards may disqualify members from
holding office or may result in removal from office or removal of funding. Such requirements mirror those already
initiated by agencies such as NSF Broader Impacts statements, DOE PIER plans, service essays in many university
faculty applications, APS policy, and more.

The EIC-TA workforce development and DEI committee will facilitate engagement in such activities, through the
creation of guidance documents for community members. In particular, the committee will advise EIC-TA members
about impactful efforts they can participate in to fulfill a DOE PIER or NSF Broader Impacts plan. Such documents
will include ideas for what type of activities are effective in helping build a stronger and more diverse workforce as
well as what types of resources and administrative offices are commonly available at universities and national labs to
assist. The EIC-TA recognizes that the most conspicuous problems and the resources available for their resolution vary
significantly by the size, type, financial resources, and communities served by a given institution. Furthermore, there
are many ways for faculty to get involved, from being an enthusiastic participant in an existing effort to overseeing
the administrative details of starting a new initiative. Due to the wide range of useful activities, service and DEI
records may look very different for different members of the alliance.

The EIC-TA is dedicated to creating and maintaining an environment where all members can feel safe and are treated
with respect and dignity. Members of the alliance are expected to behave ethically and respectfully at all times while
participating in all professional activities. The EIC-TA will develop a code of conduct that its members must agree to
follow and will appoint a conduct panel whose tasks will include investigating claims of harassment, discrimination,
microaggressions, and bullying. The panel will actively monitor the climate of the EIC-TA by maintaining a log of
claims and communicating with the Workforce Development and DEI committee about observed trends as needed.
The EIC-TA will enforce consequences for verified code of conduct violations according to severity of the current
violation and any prior history of verified violations, up to removal from alliance membership and exclusion from
alliance-supported events.

\section{Generalized Parton Distributions and Nucleon Spin}
\label{sec:GPDs}

Generalized parton distributions (GPDs) were introduced as a tool to characterize the structure of hadrons in terms of their constituent partons~\cite{Ji:1996ek,Radyushkin:1997ki,Muller:1994ses}. 
They are generalizations of the 1-dimensional PDFs, with the initial and final hadron states 
carrying different momenta 
(off-forward kinematic). 
Thus, GPDs depend on the square of the invariant momentum transfer $t$ and the longitudinal momentum transfer $\xi$, in addition to the dependence on the fraction of the hadron momentum carried by the parton, $x$. 
Their multi-dimensionality relates them to different facets of hadron structure (tomography), describing the distribution of partons in position and momentum space, as well as the correlation between the spatial and momentum distributions~\cite{Burkardt:2000za, Ralston:2001xs, Diehl:2002he, Burkardt:2002hr}.
There is a wealth of information that can be accessed through GPDs~\cite{Goeke:2001tz, Diehl:2003ny, Belitsky:2005qn, Goeke:2007fp, Boffi:2007yc, Guidal:2013rya, Mueller:2014hsa, Kumericki:2016ehc} such as the spin~\cite{Ji:1996ek,Lorce:2021gxs} and the electromagnetic and gravitational form factors. The latter have been interpreted as a measure of the pressure and shear forces inside hadrons~\cite{Polyakov:2002wz, Polyakov:2002yz,Polyakov:2018zvc,Lorce:2018egm} as well as the momentum-current gravitational multipoles~\cite{Ji:2021mfb}; see~\cite{Burkert:2023wzr} for a recent review.

The appropriate high-energy 
processes to access GPDs are deeply virtual Compton scattering (DVCS)~\cite{Muller:1994ses, Ji:1996ek, Radyushkin:1996nd, Ji:1996nm, Collins:1998be} and hard exclusive meson production (DVMP)~\cite{Radyushkin:1996ru, Collins:1996fb, Mankiewicz:1997uy}.   Currently, limited information on GPDs is accessible from experiments. Fixed-target DVCS gives some information in the intermediate to high-$x$ region. Low-$x$ measurements exist from HERA. Upcoming data from the JLab 12~GeV program will offer more information on GPDs. GPDs are a core EIC physics topic and future measurements will provide a wealth of information.  However, obtaining data on GPDs and successfully disentangling them is very challenging experimentally. For example, there are strict requirements in luminosity, center-of-mass energy, and hadron beam parameters ~\cite{AbdulKhalek:2021gbh}. Thus, synergy with theory is essential to addressing the challenges of extracting GPDs. In particular, lattice QCD plays a complementary role to the EIC and can provide crucial information on GPDs in different kinematic regions than the experimental data sets. 

Theoretical studies of GPDs include dedicated LQCD calculations. The majority of these calculations are of Mellin moments that describe the electromagnetic and weak probes of hadrons. 
The first Mellin moments, the form factors, are the most reliably accessible. 
Information also exists on selected generalized form factors (GFFs) from the second Mellin moments which can serve as reliable constraints on the large- and small-$x$ extrapolations of the experimental data.
See Ref.~\cite{Constantinou:2020hdm} for a recent review of the status of the field. 
First principles lattice QCD calculations now take into account all  systematic uncertainties, including continuum and infinite volume extrapolations at the physical pion mass.  
Progress is ongoing on two fronts: calculations using ensembles at physical quark masses and decomposition by quark flavor that requires the computationally-intensive evaluation of disconnected-diagram contributions. Despite this progress, there are theoretical and computational limitations on calculating higher Mellin moments of GPDs. Therefore, reconstruction of the GPDs is very challenging at best. Instead of accessing GPDs through their Mellin moments, alternative methods to calculate the $x$-dependence of various distribution functions have been proposed over the years~\cite{Liu:1993cv, Ji:2013dva, Ji:2014gla, Detmold:2005gg, Braun:2007wv, Radyushkin:2017cyf, Orginos:2017kos, Chambers:2017dov, Ma:2017pxb}. In the last decade, the field has advanced significantly and is now being extended to calculations of $x$-dependent GPDs~\cite{Alexandrou:2020zbe, Lin:2020rxa, Alexandrou:2021bbo, CSSMQCDSFUKQCD:2021lkf, Bhattacharya:2021oyr, Lin:2021brq, Bhattacharya:2022aob}.
See Refs.~\cite{Cichy:2018mum, Ji:2020ect, Constantinou:2020pek, Cichy:2021lih} for reviews of recent results and novel developments in the field.

The current theoretical investigations have demonstrated the strength of the field.  The proposed theory alliance has the potential to coordinate reliable extraction of the GPDs and advance our knowledge of hadron tomography. Progress in GPD theory of will be essential for future experiments, in particular for EIC science. The theory alliance will facilitate and enhance progress through synergy between lattice QCD theorists, phenomenologists and experimentalists. Below we outline some of the topics that require synergy of the theory community to advance the field.

\begin{itemize} 
\item First principle calculations of lattice QCD have advanced significantly and, for certain quantities such as form factors, take into account all  systematic uncertainties (e.g., continuum and infinite volume extrapolations at the physical pion mass). Collaborative work with the global analysis community can lead to reliable constraints on the GPDs in the large- and small-$x$ regions. For more information, see Sec.~\ref{sec:glob_ana}. Coordination of lattice QCD results with a phenomenological approach based on Continuum Schwinger Functional Methods can deliver accurate predictions of the proton and neutron elastic form factors at high-$Q^2$~\cite{Rodriguez-Quintero:2018wma,Cui:2020rmu,Jang:2019jkn,Park:2021ypf,Salg:2022poa}. The synergy between discrete and continuum theoretical analyses can more easily be achieved through the EIC Theory Alliance.

\item Coordinated activities are required to effectively relate LQCD matrix elements to light-cone GPDs. This necessitates the development of new approaches to tackle theoretical and computational challenges. For example, the frame dependence of GPDs inherited in some approaches leads to computationally-costly calculations. Alternative definitions are imperative to overcome this problem and potentially provide fast convergence on physical quantities. A initial effort began this year~\cite{Bhattacharya:2022aob}, but a multi-component program is required.

\item Joint activities 
with lattice practitioners are necessary to develop model-independent properties of GPDs that can be verified by lattice QCD calculations (e.g., sum rules) and optimize the use of computational resources (e.g., exploitation of symmetries). 

\item A complete description of hadron structure must include multi-parton correlations. These are encoded in higher-twist distributions, where twist-3 is the most important. The field of twist-3 GPDs is largely unexplored experimentally, limiting our ability to properly map hadronic structure. In addition, theoretical investigations are restricted to the interpretation of cross- section data to the two-parton scattering approximation. First principles information on the twist-3 contribution is vital and requires a systematic program that combines expertise from theory and lattice QCD. Studies of twist-3 GPDs are also important for elucidating the orbital angular momentum of quarks and gluons inside the proton~\cite{Penttinen:2000dg,Kiptily:2002nx,Hatta:2012cs,Ji:2012ba,Leader:2013jra,Liu:2015xha,Rajan:2016tlg,Rajan:2017cpx}. The concept of transition distribution amplitudes (TDAs) \cite{Pire:2021hbl}, first introduced for pion to gamma
transition, extends the concept of GPDs to the near backward scattering kinematics \cite{Gayoso:2021rzj}. The TDAs are defined as twist-3 matrix elements of three-quark operators on the light cone and are expected to factorize in the scattering amplitude of backward DVCS or backward DVMP in proton-to-gamma transitions. Lattice studies of these objects are needed, as well as phenomenological estimates of observables that can be studied at the EIC, based on controllable theoretical input.  Proof of TDA factorization along the lines of Ref.~\cite{Qiu:2022bpq} is clearly needed.

\item To execute the GPD global analysis program at the EIC, dedicated efforts are needed to develop a reliable community-driven \emph{theory library} that allows us to match  experimental observables to the QCD-factorization framework. It is therefore necessary that flexible parametrizations, including valence, sea quark and gluon components which can be perturbatively evolved to the scale of the data, are made available.  
The parametric forms build on the previously determined valence distributions, modeled at a low initial scale, $Q_0^2 \approx 0.1$ GeV$^2$.
At this scale, only valence quarks are present. Gluons and sea quarks (quark-antiquark pairs) are resolved as independent degrees of freedom at larger scales, $Q_0^2 \approx 0.58$~GeV$^2$. These components subsequently undergo perturbative evolution and generate additional gluon and sea quarks dynamically through gluon Bremsstrahlung.
The GPD dynamical framework employs an initial parametrization based on the Reggeized-spectator model \cite{Ahmad:2006gn,Ahmad:2009fvg,Goldstein:2010gu,Gonzalez-Hernandez:2012xap,Kriesten:2019jep,Kriesten:2020wcx,Kriesten:2021sqc}.
The coordinated efforts of the EIC theory community have the potential to deliver important improvements such as higher-order corrections to fixed-order calculations, resummation, and includingpower corrections. For example, the sensitivity of the gluon GPD, $E_g$, needed to build the Ji sum rule begins as an $O(\alpha_s)$ correction to DVCS and DVMP at fixed-order perturbation theory, requiring a global analysis beyond the existing leading order analyses.           


\item Alternative processes to extract the $x$-dependence of GPDs need to be developed.  The DVCS and DVMP cross sections are known to not be particularly sensitive to the $x$-distribution of the relevant GPDs because $x$ is integrated over in the scattering amplitude~\cite{Bertone:2021yyz}. This problem can be avoided by measuring DDVCS processes. In addition, a new class of observables involving a pair of high-transverse momentum particles (jets) in the final state has been proposed to overcome this difficulty ~\cite{Beiyad:2010cxa, Pedrak:2017cpp, Boussarie:2016qop, Pedrak:2020mfm, Grocholski:2021man, Grocholski:2022rqj,Qiu:2022bpq,Qiu:2022pla}.  A framework needs to be developed for the phenomenological analysis of all these processes and incorporate them into global analysis. 

\item  
Coherent exclusive reactions with light ion beams available at the EIC will study light nuclei (Deuterium and Helium) GPDs, enabling a precise determination of the interior structure of these nuclei, in particular the non-nucleonic part of their wavefunctions. The Theory Alliance is a perfect venue to gather efforts to understand the quark and gluon content of light nuclei, to confront various models predictions, and to propose dedicated measurements at the EIC.

\item 
Exclusive reactions with ion beams at the EIC will reveal new aspects of  bound nucleons. The Theory Alliance will enable discussions between groups which have studied these effects, emphasizing the consequences of short range correlations of bound nucleons on the nuclear transparency ratios, as well as different ways to introduce and precisely model color transparency effects on various observables.

\item 
Investigation of pion GPDs through model calculations dates more than two decades ago~\cite{Polyakov:1999gs,Theussl:2002xp,Broniowski:2003rp}. More recent theoretical computations of pion and kaon GPDs have produced interesting preliminary results \cite{Polyakov:1999gs,Aguilar:2019teb, Arrington:2021biu, Burkert:2022hjz}. Recent studies \cite{Chavez:2021llq, Chavez:2021koz} have shown that the EIC will be able to provide significant information on the Sullivan process. The Theory Alliance is an ideal environment to promote constructive discussions between theorists and phenomenologists.

\item 
Non-perturbative computations of GPDs have generally proceeded by using lattice calculations~\cite{Alexandrou:2020zbe, Lin:2020rxa, Alexandrou:2021bbo, CSSMQCDSFUKQCD:2021lkf, Bhattacharya:2021oyr, Lin:2021brq, Bhattacharya:2022aob}, models~\cite{Alharazin:2020yjv,Pasquini:2014vua} including  the chiral quark soliton model where a complete tower of higher components of the wavefunctions of the partonic constituents was added~\cite{Petrov:1998kf,Penttinen:1999th,Ossmann:2004bp},  and Schwinger functions~\cite{Cui:2020rmu}. Although the last two are continuum approaches, they do not include ab initio studies of a complete tower of higher components of the wavefunctions of the partonic constituents. While lattice does take these into account, a continuum method would be desirable. Taking the infinite tower of wavefunctions into account involves studying the functional evolution equations~\cite{Kanatchikov:2018uoy,Ivanov:2020zkm,Kiefer:1991xy} which generalize and systematize the Schwinger-Dyson and Bethe-Salpeter hierarchies. This approach presents interesting challenges and opportunities. First, it may provide an ab initio method for non-perturbative physics. Second, it connects to deep mathematical problems. These problems include a number of nontrivial steps, such as analysis of singularities of amplitudes in a complex domain, construction of analytic forms of the multi-parton wavefunctions, number theory, and quantum chaos. Although this research is mostly mathematical at the moment (see, however, Ref.~\cite{de2009light}), the EIC theory alliance can bring the mathematical and phenomenological communities together because they are studying the same object from different angles. Research opportunities can be created for mathematically-intensive analysis that may provide physically-founded parameterizations of GPDs.

\item
At the precision frontier, 
the  NNLO coefficient functions for DVCS have been 
calculated~\cite{Braun:2022bpn} and  
work on the three-loop evolution equations for GPDs is in progress~\cite{Braun:2017cih,Braun:2022byg}.
Similar precision must be achieved for other processes to carry out a global analysis. Resummation of threshold logarithms~\cite{Schoenleber:2022myb} must also be pursued. The structure of kinematic higher-twist corrections to DVCS that restores Lorentz and electromagnetic gauge invariance of DVCS amplitudes is well understood~\cite{Braun:2012hq,Braun:2014sta,Braun:2022qly}. Synergistic activities are required to extend this analysis to other reactions and applied to lattice calculations of GPDs using the pseudo- or quasi-GPD approach, where 
large translation invariance-breaking effects have been found~\cite{Bhattacharya:2022aob}.

\item
The development of phenomenological methods, tools, and GPD models fulfilling all theory-driven constraints is necessary. The new models must be sufficiently flexible to describe data on various exclusive processes and accommodate LQCD information. The problem of model dependence and the deconvolution of GPDs from measured amplitudes must be carefully addressed, e.g., with the help of machine learning techniques~\cite{Dutrieux:2021wll}. Model parameters need to be constrained by data through a robust and precise (in terms of perturbative and twist expansions) description of exclusive processes. This task also requires further development of aggregate tools for theory-related developments such PARTONS~\cite{Berthou:2015oaw} and GeParD~\cite{Kumericki:2007sa}. These efforts require contributions from multiple groups to optimize the use of resources and make significant progress well before the EIC is operational.
\end{itemize}


\section{Transverse momentum distributions}

At the frontier of hadron structure studies is the three-dimensional (3D) structure of the nucleon.
Both the confined motion and the spatial distribution (see Section~\ref{sec:GPDs}) of quarks and gluons inside a bound nucleon characterize its 3D internal structure, which is  an immediate consequence of QCD dynamics.  To probe such 3D internal structure one utilizes  physical observables with  {\it two-scales}; 
a large momentum transfer $Q$ that ensures localization of the probe and manifestation of the particle nature of quarks and gluons, plus an additional well-measured soft momentum scale $q_T$ associated, for instance, with the transverse motion of quarks and gluons. Such two scale measurements provide much more sensitivity to the details of hadron's internal structure and to details of the inner mechanism of confinement in QCD. The distributions that encode both the longitudinal momentum fraction carried by the parton, $x$, and the transverse motion, ${\bf k}_T$ are called Transverse Momentum Dependent distribution (TMD PDFs) and fragmentation functions (TMD FFs), or collectively TMDs~\cite{Kotzinian:1994dv,Mulders:1995dh,Bacchetta:2006tn}.
 
Recently a great deal of progress was made in understanding the properties of TMDs from both the theoretical advances~\cite{Bacchetta:2019qkv,Ebert:2020dfc,Ebert:2021jhy,Vladimirov:2021hdn,Rodini:2022wki,Ebert:2022cku,Gao:2022bzi,Gamberg:2022lju} and phenomenological studies from global fits~\cite{Bacchetta:2017gcc,Scimemi:2019cmh,Echevarria:2020hpy,Bury:2020vhj,Bury:2021sue,Alrashed:2021csd,Bacchetta:2020gko,Bacchetta:2022awv,Barry:2023qqh}.  A crucial ingredient in  our exploration of hadron structure are experimental measurements provided by various facilities around the world~\cite{Amoroso:2022eow}, such as Tevatron at Fermilab~\cite{CDF:1988lbl}, HERMES at DESY~\cite{HERMES:2004vsf}, the LHC at CERN with its collider and fixed target~\cite{Jaeckel:2020dxj,Hadjidakis:2018ifr}, COMPASS experiments~\cite{COMPASS:2010shj}, RHIC at BNL~\cite{Aschenauer:2015eha,Aschenauer:2016our}, Jefferson Lab~\cite{Dudek:2012vr}, BELLE at KEK~\cite{Belle-II:2010dht}, Electron-ion collider in China~\cite{Anderle:2021wcy},etc. The EIC will provide essential information, with the promise to dramatically improve the precision of various measurements, and to enable the exploration of the role of the sea quarks and the gluons in a polarized nucleon~\cite{Boer:2011fh,Accardi:2012qut,Borsa:2020lsz,AbdulKhalek:2021gbh}.
 
Guiding and understanding the future experimental measurements will require a laborious and meticulous analysis of the data, new approaches and new methods in the theoretical treatment and in the phenomenological extraction of TMDs. The EIC Theory Alliance will provide an essential framework for guiding and organizing the broad theoretical and phenomenological efforts needed to tackle the challenges and opportunities provided by the future EIC. Research directions supported by the EIC Theory Alliance will also ensure that US remains at the forefront in studies of the inner 3D structure of matter.

Important theoretical topics for studies relevant to enabling the full potential of the EIC to be reached include:
\begin{itemize}
\item Rigorous theoretical exploration of bench mark TMD observables as well as new experimental observables related to TMD physics. This exploration includes studies of leading and sub-leading contributions to Semi-Inclusive Deep Inelastic Scattering process, individuation of the set of observables that allow precise extraction of the 3D structure for quarks and gluons.
\item Theoretical and phenomenological exploration of QCD factorization theorems and expanding the region of their applicability, for instance by inclusion of power corrections in $q_T/Q$. A crucial ingredient will be matching collinear factorization ($\Lambda_\mathrm{QCD} \ll q_T \sim Q$) and TMD factorization ($\Lambda_\mathrm{QCD} \lesssim q_T \ll Q$) in the overlap region $\Lambda_\mathrm{QCD} \ll q_T \ll Q$ in a stable and efficient way. Such a matching is needed for our ability to describe the measured quantities, differential in transverse momentum, in the widest possible region of phase space. In turn, this will lead to a much more reliable understanding of both collinear and TMD related functions and uncertainties in their determinations.
\item Exploring the QCD factorization theorem and phenomenology for distributions related to TMD-like Generalized TMDs (GTMDs).  
These distributions extend our understanding of multidimensional hadronization and can 
arise in exclusive processes like double Drell-Yan~\cite{Kanazawa:2014nha,Echevarria:2016mrc,Bertone:2022awq,Echevarria:2022ztg}, as well as being probed by exclusive diffractive processes that are sensitive to small-$x$ gluon GTMDs and gluon saturation~\cite{Hatta:2016dxp,Mantysaari:2019csc,Boer:2021upt}.
\item Development of theoretical methods to address various open issues is crucially needed, including: advancing new methods for perturbative calculations, developing formalism and calculations for TMD power corrections, the need to design new observables that can improve the comparison between theory and experiment, and a full exploration of the best way to parameterize nonperturbative TMDs. Methods used to tackle these problems include  effective theories, nonperturbative and computational methods in QCD, and feedback from carrying out fits to experimental data.
\item Creation of extraction frameworks that include modern techniques and methods from statistics (such as Bayesian statistical methods) and computer science (such as Artificial Intelligence and Machine Learning). Extraction frameworks are critical for phenomenological studies of TMDs. There exist already several frameworks such as NangaParbat\footnote{\url{https://github.com/MapCollaboration/NangaParbat}} of the MAP Collaboration, JAM Collaboration\footnote{\url{https://github.com/JeffersonLab/jam3d/}}, and arTeMiDe\footnote{\url{https://github.com/VladimirovAlexey/artemide-public}}. 
These publicly available frameworks will facilitate engagements of new groups in EIC related studies.
\item Supporting long terms commitments in the analysis of large data sets from the existing experiments and facilities. Global QCD analyses of the experimental data are usually multi year efforts of relatively large collaborations. Encouraging theory participation in such efforts is very important for the future development of new QCD analyses, both to preserve knowledge and to utilize advanced methods.
\item Encouraging calculations of higher order perturbative quantities, such as anomalous dimensions, spin-dependent cross-sections etc, needed for an accurate and precise extraction of 3D structure and for reliable predictions of future measurements.
\item 
Supporting the experimental community with the development of Monte-Carlo event generators is an essential task \cite{AbdulKhalek:2021gbh} that requires a multi-year commitment. The alliance will support studies based on the conventional techniques that take into account radiative corrections, such as Ref.~\cite{Byer:2022bqf}; and new frameworks that incorporate TMD and QED physics, such as Ref.~\cite{Liu:2021jfp}.


%
\item Comprehensive analysis of the nonperturbative behaviour of TMDs. As TMDs encode the consequence of confinement, it is very important to understand better the nonperturbative structure of the nucleon as prototype of baryons, and of the pion as prototype of mesons~\cite{Vladimirov:2019bfa,Cerutti:2022lmb}. It can be done in model or ab-initio calculations, such as lattice QCD, and in the global QCD analyses. The combination of model, lattice QCD, and phenomenological results will allow for a better understanding of the nature of the extracted quantities and for a better precision of extractions in case the experimental measurements are scarce for some observables.   For example, at present gluon TMDs are phenomenologically unknown and model calculations represent a useful tool to explore their features, confirm or falsify generally accepted assumptions, make reasonable predictions for experimental observables, and or guide the choice of functional forms in future gluon TMD fits~\cite{Bacchetta:2020vty,Bacchetta:2022esb}. Explore in more detail the impact of the nonperturbative behavior of TMDs (in particular, their flavor dependence) on the determination of some crucial Standard Model parameters like the $W$ boson mass~\cite{Bacchetta:2018lna}.
%
%
\item Understanding the flavor dependence of quark TMDs and the gluon sector of TMDs. Both quark and gluon TMDs present a vast field of exploration in terms of the TMDs that encode aspects of the internal structure, such as spin correlations, flavor dependence, etc. We know that there exist highly universal functions, such as Collins-Soper kernel~\cite{Collins:1981uk,Collins:1981va,Collins:1984kg,Collins:2014jpa} related to the properties of the vacuum of QCD, less universal non-perturbative functions that encode flavor or hadron dependence but not less interesting as they carry the footprint of the non-perturbative QCD interactions. Careful examination of the whole spectrum of TMDs is important for our final goal of understanding of the underlying 3D structure of hadrons.
\item Understanding nuclear TMDs with the methods developed for the nucleon. Following the methodology of well-established nuclear collinear PDFs, Ref.~\cite{Alrashed:2021csd} performed the first extraction of nuclear modified TMDs from the world set of data in semi-inclusive electron-nucleus deep inelastic scattering and Drell-Yan production in proton-nucleus collisions. It is important to advance and improve the methodology along this direction. The modification of TMDs in nuclei in comparison with those in the nucleon has important connections with the conventional transverse momentum broadening in nuclei, see e.g.~\cite{Liang:2008vz,Schafer:2013mza,Boer:2015kxa} and \cite{Guo:1998rd,Kang:2008us,Kang:2011bp,Xing:2012ii,Kang:2013raa} within different formalisms, and is highly relevant to the jet transport coefficient $\hat q$ in nuclei~\cite{Ru:2019qvz,Ru:2023ars}. In addition, the (non)universality of gluon TMDs~\cite{Bomhof:2007xt,Dominguez:2011wm,Scheihing-Hitschfeld:2022xqx} will also be studied within the context of small-$x$ physics and nuclei structure in Sec.~\ref{sec:gluon_sat} and Sec.~\ref{sec:nuclear}.

\end{itemize}

Apart from the global analyses, the lattice QCD calculation of non-perturbative TMD information has also seen a lot of progress in recent years. The pioneering lattice efforts were made with the Lorentz-invariant method which mainly focused on the ratios of the $x$-moments of TMDs for various spin and flavor structures~\cite{Hagler:2009mb,Musch:2010ka,Musch:2011er,Engelhardt:2015xja,Yoon:2016dyh,Yoon:2017qzo}. Later on, a breakthrough was made allowing to go beyond TMD ratios and compute individual TMDs, including their $x$-dependence, motivated by the large-momentum effective theory (LaMET)~\cite{Ji:2013dva,Ji:2014gla,Ji:2020ect} that has enabled tremendous progress in the lattice calculation of collinear PDFs~\cite{Lin:2017snn,Constantinou:2020hdm}. In this approach, the TMDs in Drell-Yan and SIDIS factorization theorems can be perturbatively matched from a quasi-TMD calculable on the Euclidean lattice~\cite{Ji:2014hxa,Ji:2018hvs,Ebert:2018gzl,Ebert:2019okf,Ji:2019sxk,Ji:2019ewn,Ebert:2022fmh} with the subtraction of a soft factor that can be extracted from a light-meson form factor~\cite{Ji:2019sxk,Ji:2019ewn,LatticeParton:2020uhz,Li:2021wvl}, up to power corrections suppressed by the large parton momentum. The perturbative matching kernel that relates the quasi- and physical TMDs is diagonal in the $x$-space, independent of the spin structure~\cite{Ebert:2020gxr,Vladimirov:2020ofp}, and free from mixing between quarks and gluons or quarks of different flavors~\cite{Ebert:2022fmh}.
Based on this matching relation, important non-perturbative TMD information can be computed:
\begin{itemize}
    \item The Collins-Soper kernel for TMD evolution. The Collins-Soper kernel plays an important role in the global fitting of TMDs.
    It depends on $k_T=|{\bf k}_T|$ and becomes non-perturbative when $k_T\sim\Lambda_{\rm QCD}$, where it has not been well constrained by the global analyses~\cite{BermudezMartinez:2022ctj}. In this regime, the Collins-Soper kernel can be extracted from the momentum evolution of the quasi-TMD~\cite{Ebert:2018gzl}, which has been applied in several lattice QCD calculations~\cite{Shanahan:2019zcq,Shanahan:2020zxr,Schlemmer:2021aij,LatticeParton:2020uhz,Li:2021wvl,Shanahan:2021tst,LPC:2022ibr,Shu:2023cot}. While the systematic uncertainties need to be improved on both sides, recent global analyses~\cite{Scimemi:2019cmh,Bacchetta:2022awv,Boglione:2022nzq} have shown interesting agreement with the lattice results~\cite{BermudezMartinez:2022ctj}.
    
    \item Ratios of TMDs of different spin and flavor structures. 
    Early efforts with the Lorentz-invariant method calculated ratios of TMD $x$-moments for different spin and flavor structures~\cite{Hagler:2009mb,Musch:2010ka,Musch:2011er,Engelhardt:2015xja,Yoon:2016dyh,Yoon:2017qzo}. These efforts must be extended to include the $x$-dependence, to which purpose also the LaMET approach can be tailored~\cite{Ebert:2019okf,Ebert:2020gxr}. Obtaining the ratios in the full $(x,{\bf k}_T)$ space will provide rich information on the non-perturbative behavior of spin-dependent TMDs such as the helicity, transversity, Sivers and Boer-Mulders TMDs.     Thanks to the non-mixing of the matching for quasi-TMDs, it is straightforward to separate the flavors of quasi-TMDs on the lattice and calculate their ratios.
    
    \item Full kinematic dependence of TMDs in the $(x,{\bf k}_T)$ space. With the soft factor being calculable~\cite{LatticeParton:2020uhz,Li:2021wvl}, one can perform a complete lattice QCD determination of the TMDs, which will provide direct comparison with global analyses for all the spin and (quark and gluon~\cite{Schindler:2022eva,Zhu:2022bja}) flavor structures.
    
    \item Accessing twist-3 PDFs from quasi-TMDs at large $k_T$. For example, at large $k_T$ the quasi Sivers TMD can be related to the twist-3 Qiu-Sterman function through an operator product expansion~\cite{Ji:2020jeb}, which is complementary to the twist-3 quasi-PDF approach~\cite{Bhattacharya:2020cen,Braun:2021gvv}.
    
\item 
Accessing subleading power TMDs~\cite{Chen:2016hgw,Bacchetta:2019qkv,Vladimirov:2021hdn,Rodini:2022wki,Ebert:2022cku,Gao:2022bzi,Gamberg:2022lju},
     which are of great interest in achieving a complete 3-D momentum  tomography of hadrons.

    \item Generalization to GTMD observables that quantify, e.g., parton orbital angular momentum and spin-orbit correlations in the nucleon \cite{Lorce:2011kd,Hatta:2011ku,Lorce:2011ni,Lorce:2015sqe}. Initial studies \cite{Engelhardt:2017miy,Engelhardt:2020qtg,Engelhardt:2021kdo} that include both the Jaffe-Manohar as well as the Ji definitions of these quantities must be extended to understand their scaling properties, power corrections, and other systematics, with the long-term perspective of complementing eventual phenomenological extractions of GTMD observables.
    
\end{itemize}

The EIC Theory Alliance will be an invaluable platform for bringing together talents in both analytical theory and lattice QCD to investigate the above TMD physics in the next two decades. The primary efforts by the EIC Theory Alliance include:

\begin{itemize}

    \item Deeper understanding the QCD factorization relation between lattice and physical TMDs. Factorization theorems have been derived for both quark and gluon TMDs at moderate to large $x$, with the matching coefficients calculated at one-loop order~\cite{Ji:2018hvs,Ebert:2019okf,Schindler:2022eva,Zhu:2022bja} and resummations performed at next-to-leading logarithmic accuracy~\cite{Ebert:2022fmh}. To improve the current lattice calculations at moderate momenta, it is necessary to study the subleading power corrections and higher-order perturbative corrections. It is also worthwhile to explore novel lattice TMDs and their corresponding factorization, which may convergence faster to the physical TMDs. Moreover, the small-$x$ factorization of the lattice TMDs will require a new formalism that needs to be discovered.

    \item Disentangle the systematic effects during the procedure of lattice QCD calculation, which include renormalization, operator mixing, Fourier transform, finite volume effects, and other lattice artifacts~\cite{Shanahan:2021tst}. The solution relies on a profound understanding of the short- and long-distance behaviors of the lattice TMD matrix elements, based on which one can construct proper observables to minimize lattice artifacts, develop an optimal renormalization scheme, and derive the effective theory formulas for physical extrapolations.
    
    \item Development of lattice software to meet the requirement of calculating parton physics, especially in the era of exa-scale computing. The key for all lattice calculation of parton physics is the large hadron momentum, which can only be achieved with smaller lattice spacing, and which is difficult for most lattice ensembles nowadays. This poses challenges for generating the ideal QCD gauge configurations, as well as the corresponding algorithms for simulation on the most advanced exa-scale GPU machines such as Frontier and Aurora.

    \item Precision controlled lattice QCD calculation of TMD physics. By capitalizing other initiatives and awards such as the Computational Nuclear Physics Initiative, SciDAC, INCITE and ALCC, lattice QCD will carry out systematic calculations of TMDs to complement the EIC program. The goals include reliable predictions of the quark Collins-Soper kernel and ratios of quark TMDs of different spin and flavor structures with a 10-20\% level precision, and a 20-40\% level calculation of the full $(x,{\bf k}_T)$ dependence of TMDs, within this decade. The calculation of the gluon Collins-Soper kernel and TMDs are expected to achieve meaningful precisions with further advancement in computing and algorithms in the future. 
    
    \item Synergy between lattice QCD, theory and phenomenology to provide a complete 3D tomography of the nucleon. To realize the full potential of the EIC Theory Alliance, it is also expected to develop a comprehensive program for comparing the lattice QCD predictions and experimental results, and incorporating the non-peturbative lattice inputs for the global analyses to reduce the model uncertainties.
\end{itemize}

Finally, to improve the knowledge and skills required in theoretical analyses, lattice QCD calculations, and phenomenological extractions of TMD physics, the theory alliance will provide valuable training for students and postdocs, who will become the major work force in this field and will also contribute to the other initiatives in Nuclear Physics.

\section{Gluon saturation, small $x$}
\label{sec:gluon_sat}

With its access to high energies, electron and light ion polarization, a wide range of nuclear species, and unprecedented DIS luminosities \cite{Accardi:2012qut,AbdulKhalek:2021gbh}, EIC offers the exciting possibility of uncovering and establishing the properties of gluon saturation, including its effects on proton spin and the physics of strong color fields in QCD with unprecedented precision. Indeed, this discovery and characterization was assessed to be a principal objective of EIC science in a National Academy of Sciences report \cite{NAP25171}. 

At short distances, the proton can be viewed as a collection of weakly interacting quarks and gluons, commonly referred to as partons, which carry each a fraction $x$ of the proton momentum. The parton picture is expected to break down, however, when the probe resolution becomes of order the proton size where confinement forces are at play. On the other hand, at small enough $x$ the proton wave function is characterized by a rapid rise of the number of ``wee'' gluons up to a point where many-body recombination and screening effects become important, which  leading to saturation of the gluon distributions. This novel many-body regime of QCD characterized by strong non-linear gauge fields can be explored in the framework of the Color Glass Condensate effective field theory (CGC)~\cite{Gelis:2010nm}. Such effects are strongly enhanced in nuclei as reflected in the $Q_S^2\propto A^{1/3}\gg \Lambda_{\rm QCD}^2$ dependence of an emergent saturation scale controlling the nonlinear dynamics of saturation. Likewise, there there is a powerful interplay between contributions to the proton's spin at small $x$ and the physics of gluon saturation. 

Ensuring that the EIC can realize its enormous promise and deliver on the discovery of gluon saturation 
requires 
both broad and focused collaborative theoretical research. This effort includes the identification and computation of observables that are sensitive to gluon saturation in DIS off 
polarized protons ($e+p$) and heavy nuclei ($e+A$), along with observables in hadron collider experiments. This effort also requires robust end-to-end calculations that minimize known uncertainties on each of these observables with an ambitious goal of $<10\%$ accuracy, sufficient for unambiguous characterization of the gluon dominated small $x$ regime in protons and nuclei. Not least,  new and potentially transformative ideas (and their empirical consequences) connecting the physics of gluon saturation to the 
intrinsically non-perturbative physics of color confinement and chiral symmetry breaking in QCD  need to be explored.

The ambitious goal of precision in data-theory comparisons at small $x$ is driven by the significant progress achieved in recent years in 
a) next-to-leading order (NLO) computations of process-dependent so-called ``impact factors" for key final states, and 
b) advances in computing the  small-$x$ RG LL BK~\cite{Balitsky:1995ub,Kovchegov:1999yj} and JIMWLK equations~\cite{Jalilian-Marian:1997ubg,Iancu:2000hn} to next-to-leading logarithmic (NLL) accuracy. Combining the developments in impact factors and small $x$ evolution for a number of final states puts us in a position to reach the $O(\alpha_S^2)$ accuracy for cross sections, corresponding to the desired  $10\%$ figure of merit.

In DIS, formal expressions for NLO impact factors have been derived for key observables such as those for inclusive structure functions \cite{Balitsky:2010ze,Beuf:2017bpd}, heavy quark structure functions \cite{Beuf:2021srj},  diffractive dihadron, dijet and exclusive vector meson production \cite{Boussarie:2016ogo,Boussarie:2016bkq,Mantysaari:2022kdm,Fucilla:2022wcg}, single inclusive hadron \cite{Bergabo:2022zhe}, inclusive dijet, dihadron \cite{Caucal:2021ent,Caucal:2022ulg,Bergabo:2022tcu,Bergabo:2023wed,Caucal:2023nci} and hard photon \cite{Roy:2019hwr} final states.  Recently proposed observables such as coherent inelastic dijet production, photon-jet and lepton-jet correlations, as well as nucleon energy correlators, remain to be studied at NLO accuracy~\cite{Iancu:2021rup,Kolbe:2020tlq,Tong:2022zwp,Liu:2023aqb}. However in many cases, even where formal expressions exist, the impact factors have not been explicitly evaluated, primarily due to their numerical complexity. Combining analytical progress with numerical implementation will be an important collaborative task. 

A key aspect of this it to combine advances in computing the various process dependent impact factors with the universal small $x$ evolution in a consistent manner. This matching still has a number of conceptual questions which have spurred a lot of theoretical work. This issues pertain to the poor treatment of the collinear corner of phase space and resummation of higher order terms turned out to be necessary in order to cure the problem. 
Similar to the linear BFKL case~\cite{Fadin:1975cb,Kuraev:1976ge,Kuraev:1977fs,Balitsky:1978ic}, the renormalization group (RG) improved treatment of collinear logs, discussed in the context of the CGC EFT in \cite{Lappi:2016fmu,Ducloue:2019ezk}, is essential for robust predictions. 
Unless these large logarithmic contributions are properly accounted for, NLO level results can lead to unphysical (negative) cross sections \cite{Stasto:2014sea}. It is only recently that first computations that take all these elements into account in the context of back-to-back inclusive dijets have become available~\cite{Caucal:2022ulg}.

Such systematic computations indicate the importance of parton lifetime constraints on RG evolution at small $x$. This is well-known to regulate the behavior of the NLL BFKL~\cite{Fadin:1998py,Ciafaloni:1998gs} equation. A full understanding of its impact on the NLL JIMWLK RG equations~\cite{Balitsky:2013fea,Kovner:2014lca} remains to be understood although there is promising work in this direction~\cite{Ducloue:2019ezk}. At the heart of this is a powerful spacelike-timelike correspondence~\cite{Mueller:2018llt} first discussed in the context of the relation of non-global logarithms in $e^+e^-$ collisions 
to small $x$ evolution~\cite{Hatta:2013iba}, and later exploited to compute key pieces of the next-to-next-to-leading order RG evolution equations~\cite{Caron-Huot:2016tzz}. This connection has led to a powerful synergy between the small $x$ and ``amplitudes" communities in developing precision tools which has the potential to significantly enhance the EIC theory alliance (EIC-TA)~\cite{Travaglini:2022uwo}.

Sub-eikonal corrections to the  eikonal approximation employed in CGC calculations will extend its applicability to higher $p_t$ and improve its accuracy at small $x$~\cite{Jalilian-Marian:2017ttv,Jalilian-Marian:2018iui,Jalilian-Marian:2019kaf,Boussarie:2021wkn,Boussarie:2020fpb}. 
Such sub-eikonal corrections can be calculated systematically either at the level of propagators~\cite{Altinoluk:2020oyd,Altinoluk:2021lvu} or of observables like the DIS dijet cross section~\cite{Altinoluk:2022jkk}. In that context, the study of the back-to-back limit will clarify the matching between the TMD formalism supplemented by higher twist power corrections and the CGC formalism supplemented by sub-eikonal power corrections.   

DGLAP-based fits of helicity PDFs are plagued by extrapolation issues into the small-$x$ regime~\cite{DeFlorian:2019xxt,Borsa:2020lsz}.   
Small-$x$ helicity evolution equations, the Kovchegov-Pitonyak-Sievert--Cougoulic-Tarasov-Tawabutr (KPS-CTT) equations, involving the {\em polarized dipole amplitude}, 
were derived in~\cite{Kovchegov:2015pbl,Kovchegov:2016zex,Kovchegov:2017lsr,Kovchegov:2018znm,Cougoulic:2022gbk}.  The first implementation of the older KPS equations achieved an extraction of the $g_1$ structure function from world's polarized DIS data~\cite{Adamiak:2021ppq}. 
The growth of helicity PDF studies will be aided tremendously by the EIC-TA. For example, compared to previous work 
 on fits of DIS double-spin asymmetries~\cite{Adamiak:2021ppq} with the (unpolarized) denominator determined from  DGLAP  fits, improved self-consistent fits should employ the small $x$ unpolarized RG equations discussed above. 
 Processes such as polarized SIDIS and polarized proton-proton collisions, need to be incorporated into the framework. To distinguish individual flavor contributions,  double-log approximation large-$N_c$ solutions~\cite{Adamiak:2021ppq} employed should be replaced by results in the large-$N_c$\&$N_f$ limit~\cite{Kovchegov:2020hgb}.  
Similarly, improved determination of helicity-dependent initial conditions should replace the {\it ad hoc} fitting procedure in \cite{Adamiak:2021ppq}.  Saturation corrections to the KPS-CTT equations~\cite{Kovchegov:2021lvz} from the unpolarized dipole correlator are required to study their effect on parton spin. Predictions for the proton orbital angular momentum  carried by small-$x$ quarks and gluons are feasible~\cite{Kovchegov:2019rrz}.  The polarized dipole amplitude approach lends itself to systematic improvements within a global analysis framework. 
An important issue which is not addressed by the above analyses is the role of the chiral anomaly. Work in this direction suggests that the quark helicity is proportional to the QCD topological susceptibility~\cite{Narison:1998aq,Tarasov:2020cwl,Tarasov:2021yll}. A similar effect due to the trace anomaly has been uncovered in deeply virtual Compton scattering (DVCS)~\cite{Bhattacharya:2022xxw}. These anomaly studies connect the small $x$ community to researchers working on various aspects of non-perturbative QCD that include lattice QCD and chiral perturbation theory.

Single forward emission in inclusive as well
as exclusive channels gives us  direct access to the unintegrated gluon distribution (UGD) in the proton, and BFKL evolution small-$x$. This BFKL baseed approach to UGDs was previously applied to forward emissions of light vector mesons at HERA ~\cite{Anikin:2009bf,Anikin:2011sa,Besse:2013muy,Bolognino:2018rhb,Bolognino:2019pba,Bolognino:2021niq,Cisek:2022yjj,Luszczak:2022fkf,Celiberto:2019slj,Bolognino:2018mlw,Bolognino:2019bko,Bolognino:2021gjm,Bolognino:2022uty,Bolognino:2022ndh}, to the study of heavy quarkonia~\cite{Bautista:2016xnp,ArroyoGarcia:2019cfl,Hentschinski:2020yfm}; and forward Drell--Yan detections at LHCb~\cite{Motyka:2014lya,Motyka:2016lta,Brzeminski:2016lwh,Celiberto:2018muu}. The EIC can potentially unveil connections between UGDs and the unpolarized and Boer--Mulders gluon TMDs~\cite{Hentschinski:2021lsh,Nefedov:2021vvy,Celiberto:2021zww,Bolognino:2021niq}. Likewise, excellent agreement is obtained between the CGC EFT with HERA data on proton $F_2$ and $F_L$ \cite{Beuf:2020dxl} and heavy quark structure functions~\cite{Hanninen:2022gje}, and for forward single inclusive hadron production data at RHIC and the LHC \cite{Shi:2021hwx}. 

The  onset of gluon saturation should show distinctly different systematics from those seen in the absence of saturation. 
Such differences could be striking in exclusive vector meson production off large nuclei \cite{Mantysaari:2017slo}.
In particular, one expects scaling behavior of vector meson production cross sections in both nuclear mass number, $A$, and photon virtuality, $Q^{2}$, to be strongly modified due to 
saturation effects. Recent simulations with the Sar{\em t}re \cite{Toll:2013gda,Toll:2012mb} Monte Carlo event generator confirm this scaling in a realistic EIC kinematic setup, where the results are obtained via cross-section pseudo-data collected by Sar{\em t}re \cite{Matousek:2022enl}, and then parsed through 
smearing functions, that emulate proposed EIC detector resolutions; theoretical and experimental challenges exist and are detailed in \cite{Matousek:2022enl,AbdulKhalek:2021gbh}. 
 
An interesting signature of  saturation is geometrical scaling (GS) in  exclusive vector meson production  inspired by previous work on inclusive DIS \cite{Stasto:2000er}, which was 
observed  at HERA \cite{Golec-Biernat:1998zce,Golec-Biernat:1999qor,Marquet:2006jb}, both in the proton and pion structure at small $x$ at HERA \cite{Kumar:2022kww}, in hadronic collisions at the LHC \cite{McLerran:2010ex,McLerran:2010wm}, and possibly also in heavy ion collisions at RHIC and LHC \cite{Khachatryan:2019uqn,Khachatryan:2022qrc}. 
Based on these examples, it is anticipated that GS should exist in exclusive vector meson production in $e+A$ (and possibly in $e+p$) at the EIC \cite{Ben:2017xny,Kowalski:2006hc}. In this regard, 
one can use Sar{\em t}re to generate events and perform cross-section calculations as a function of the transverse momentum of the produced vector mesons.

Recent phenomenological studies dedicated to photo-production of vector mesons $J/\Psi$ and $\Psi(2s)$ in ultraperipheral collisions at the LHC indicate that the ratio of their cross-sections can distinguish linear from non-linear low $x$ evolution \cite{Hentschinski:2020yfm}. At the EIC, a similar observation is expected by varying $A$. To explore the full consequences of non-linear low $x$ evolution, it will be  beneficial to complement EIC measurements by LHC forward physics results, which -- albeit limited in precision relative to the EIC -- has a wider range in $x$. See \cite{Hentschinski:2022xnd} for a recent summary of possible measurements.

The Good-Walker paradigm relates coherent photo/electroproduction to the average nuclear configuration, with $d\sigma/dt$ being sensitive to the transverse distribution of gluons in the nuclear target \cite{Klein:2019qfb}.  Incoherent photoproduction is, in turn, sensitive to event-by-event fluctuations in the target, including gluonic hotspots \cite{Miettinen:1978jb}.  Unfortunately, the Good-Walker paradigm is, by nature, a lowest order formalism which fails in some cases \cite {Klein:2023zlf}.  
For large (negative) values of the Mandelstam $t$ variable, the incoherent cross section is nevertheless proportional to the fluctuations of the initial state target wave function. In the dense saturated kinematic regime, these fluctuations are expected to be suppressed \cite{Cepila:2016uku, Mantysaari:2018zdd, Kumar:2022aly}. The EIC 
can therefore trace the onset of saturation in the suppression of the incoherent cross section in exclusive diffraction at intermediate and large values of $|t|$.

Exclusive production of vector mesons, photons (including DVCS) or other final states allows for unique measurements but introduces additional theoretical problems in understanding the Pomeron.  To lowest order, the Pomeron is a color singlet combination of two gluons.  However recent theoretical calculations of exclusive $J/\psi$ photoproduction  found several surprises: a large contribution from quarks (partly due to a significant cancellation in the gluon components) and a strong scale dependence \cite{Eskola:2022vpi}. Additional calculations are needed to understand the implications of this result and to determine how to best use this data in determining nuclear PDFs. 

Quantum coherence \cite{Kopeliovich:2001xj} and color transparency (CT) \cite{Kopeliovich:1993gk} is conveniently studied in diffractive photo- and electro-production of heavy quarkonia \cite{Kopeliovich:2022jwe} 
 within a Green function formalism \cite{Kopeliovich:2001xj,Kopeliovich:2022jwe}; this formalism has also been applied to exclusive production of light vector mesons \cite{Kopeliovich:2001xj}, diffractive DIS and DVCS \cite{Kopeliovich:2010sa}. Since both quantum coherence and color transparency are features of the CGC formalism, it will be interesting to explore further the connections  between apparently different formalisms and and arrive a common understanding. A recent review of the phenomenological status of gluon saturation measurements at colliders can be found in \cite{Morreale:2021pnn}.

The two most widely used initial conditions for small $x$ evolution are those of the Golec-Biernat--W\"usthoff (GBW) \cite{Golec-Biernat:1998zce,Golec-Biernat:1999qor} and the McLerran--Venugopalan (MV) models \cite{McLerran:1993ka,McLerran:1993ni}. The former captures key features of saturation but doesn't match smoothly to LO perturbative QCD. 
The latter is more robust in this sense but its regime of validity is 
strictly for large nuclei.

How to go beyond the GBW and MV models to extract reliable initial conditions for multipole correlators at moderate $x\sim 10^{-2}$ is an outstanding question.  Fluctuation driven non-Gaussian effects derived from analogy of ``BK dipole evolution" to wavefront propagation in statistical mechanics provide useful guidance in going beyond the MV model \cite{Mueller:2014fba,Caucal:2021lgf}, which can be tested in diffractive final states at the EIC \cite{Le:2020zpy}.

A promising first principles approach for light nuclei is to compute color charge correlators on the light front \cite{Dumitru:2018vpr,Dumitru:2020gla,Dumitru:2021tqp}, which can be constrained by DIS data at large $x$ \cite{Dumitru:2019qec,Dumitru:2021tvw}.  Another such promising approach is to employ the techniques of Large Momentum EFT (LaMET) \cite{Ji:2020ect,Zhang:2018diq} to compute the energy dependence of transverse momentum dependent parton distributions from Lattice QCD \cite{Shanahan:2021tst} or extract color charge correlators.  Reaching down to $x\sim 10^{-2}$ is however challenging, since it will require high resolution on the lattice or new ideas to circumvent the need for very small lattice spacing.

Small $x$ physics and gluon saturation have interesting conceptual and practical connections to several subfields of physics. In the high energy Regge limit at large $N_c$, the LL BFKL equation has a rich mathematical structure and can be understood as dual to the XXX spin chain with negative spin which is an integrable model that 
is solvable by the Bethe Ansatz. This mapping of small $x$ dynamics is valuable for providing insight into the role of entanglement in DIS~\cite{Zhang:2021hra}. The role of quantum entanglement and quantum information science (QIS) in DIS at small $x$ has been discussed recently in several works~\cite{Kharzeev:2017qzs,Duan:2021clk,Dumitru:2022tud,Hentschinski:2022rsa}. It has been suggested recently that gluon saturation can be understood  as a maximally occupied state whose microstates saturate the universal Bekenstein bound~\cite{Dvali:2021ooc}. QIS studies are a new direction of EIC physics appropriate for the EIC-TA and shows great promise of productive synergy with other 
areas of physics.

The small $x$ studies undertaken by the EIC-TA are also valuable for QCD studies at the LHC particularly for understanding the role of multi-parton interaction (MPIs) in hard processes. Small $x$ has played a key role in understanding the initial state and the thermalization process in heavy-ion collisions~\cite{Berges:2020fwq}. Prior to the EIC, there will be a significant amount of data  sensitive to small $x$ physics and gluon saturation from RHIC and LHC. The EIC-TA will interact with the heavy-ion community in analyzing this data and incorporating it in global analysis. This is a primary goal of the recently approved SURGE (Saturated Glue) DOE Topical Theory collaboration, which will have considerable overlap with the EIC-TA. 

Small $x$ QCD also places important constraints on the astrophysics of cosmic neutrinos~\cite{Anchordoqui:2006ta,Cooper-Sarkar:2011jtt,Garcia:2020jwr,Candido:2023utz}. Conversely, measurements at cosmic neutrino observatories such as ICECUBE can help distinguish between differing frameworks for QCD evolution at small $x$~\cite{Bhattacharya:2016jce}.

\section{Precision $ep$ physics}

Scattering reactions with polarized or unpolarized electrons and protons are a core component of the physics program at the EIC.
Data for $e+p$ cross sections and spin asymmetries at unprecedented precision and kinematic reach are expected from the EIC measurements. The anticipated experimental precision also sets the bar for theoretical calculations of the corresponding observables. 
The need for an adequate theoretical framework to match the quality of the EIC data has long been recognized. 

A central task of QCD theory for the EIC is to provide precision computations of relevant partonic hard-scattering cross sections and splitting functions to the highest possible orders in perturbation theory. Such computations are vital for the success of the EIC because higher-order corrections are often sizable and strongly reduce the dependence of the theoretical results on the factorization and renormalization scales. The past few years 
have seen tremendous progress in this area. The DGLAP evolution kernels are now fully known through NNLO (or, to three loops), both for the spin-averaged~\cite{Moch:2004pa,Vogt:2004mw} and helicity~\cite{Moch:2014sna} dependent evolution. Parts of the four-loop splitting functions~\cite{Moch:2017uml,Moch:2021qrk} and the lower moments of the five-loop
functions~\cite{Herzog:2018kwj} have become available for unpolarized evolution, elevating evolution of PDFs to an unprecedented level of precision. 
Pertinent partonic cross sections of $e+p$ scattering at NNLO and beyond include inclusive DIS~\cite{Zijlstra:1992qd,Zijlstra:1993sh,Borsa:2022irn,Borsa:2022cap,Blumlein:2022gpp} and
jet production in DIS~\cite{Abelof:2016pby,Currie:2017tpe,Currie:2018fgr,Boughezal:2018azh,Borsa:2020ulb}. The jet calculations use
modern subtraction methods to handle collinear and infrared divergences such as the ``projection to Born''
method or the ``N-jettiness subtraction scheme'', which are also applied to LHC calculations. 
Also at the $e+p$ scattering precision theory frontier are the recent three-loop results  on off-forward evolution equations \cite{Braun:2017cih,Braun:2022byg} and NNLO computations of the DVCS coefficient
functions~\cite{Braun:2022bpn}, pertaining to the study of generalized parton distributions discussed in Sec.~II.
Likewise, as described in Sec.~III, tremendous progress has been made on developing a framework for phenomenological investigations of transverse-momentum dependent parton distributions (TMDs) in $e+p$ scattering at the EIC 
(see, for example, Refs.~\cite{Scimemi:2019cmh,Bury:2020vhj,Bury:2021sue,Ebert:2020dfc,Bacchetta:2022awv}).

These achievements are important first steps toward a new precision era for EIC theory of $e+p$ scattering. 
While much work is ongoing, many tasks and challenges remain. Ideally, by the time the EIC turns on,
it is hoped that the precision of theoretical calculations should be on par with 
that achieved for the LHC, with next-to-next-to-leading order (NNLO) QCD corrections available for 
the relevant observables, along with NNLO extractions standard for PDFs and FFs, using numerically efficient tools. 
The EIC Theory Alliance would be ideal for setting up this framework and for addressing
the associated challenges. 

In order to advance precision $e+p$ theory for the EIC and to set the stage for future consolidated efforts,
a dedicated workshop series ``Precision QCD predictions for $ep$ Physics at the EIC'' was started at the Center for Frontiers in Nuclear Science (CFNS) in Stony Brook. The first workshop was held in 2022 
({\tt https://indico.bnl.gov/event/14374/}); the next edition is planned for September 2023. 
The inaugural workshop brought together some of the theorists studying $e+p$ scattering and addressed higher-order perturbative calculations of EIC observables, resummation,
power corrections, and methods for extracting PDFs, TMDs and fragmentation functions.  At the workshop, it was noted that around 20 years ago, LHC theory was at a similar stage as that of the EIC now: the need for precision theory calculations was recognized and the required higher-order theory calculations were 
identified. This effort culminated in the {\it LHC experimenter's wish list}
that shaped LHC theory for the subsequent decades (see, for example, Ref.~\cite{huston-list}). 
An outcome of the first CFNS workshop was the start of a similar {\it EIC wish list} compiling
calculations and studies that need to be carried out in preparation for the EIC.
We will not present the entire initial EIC wish list here but will only present a few specific
topics, grouped into classes of similar scientific scope that would likely become core activities of the EIC Theory Alliance focused on $e+p$ physics:

\begin{itemize}

\item Computation of higher-order corrections for EIC observables where not yet available (in part at NNLO).
Examples include QCD corrections to semi-inclusive deep inelastic scattering (SIDIS), electro- or photoproduction 
of hadrons and jets, hadron-pair production at the EIC, $\Lambda$-baryon production cross sections and spin 
asymmetries, azimuthal transverse single-spin asymmetries. Regarding the computation of fully-differential fixed-order corrections to (polarized) SIDIS in particular,
we stress that while all ingredients are in place to construct the integrated subtraction terms
for $0$-jettiness subtractions~\cite{Gaunt:2014xga, Ritzmann:2014mka, Boughezal:2017tdd} at NNLO
and $q_T$ subtractions in unpolarized calculations~\cite{Catani:2007vq, Billis:2021ecs} even at N$^3$LO~\cite{Luo:2019szz, Ebert:2020yqt, Luo:2020epw, Ebert:2020qef},
a dedicated and coordinated effort will be required to interface these in publicly available tools
with suitably efficient and precise calculations of the contributions of resolved emission with an identified hadron.

\item QCD resummation studies for the EIC. As is well known, perturbative QCD corrections typically exhibit
single- or double-logarithmic terms that may become important and even dominant in certain regions of
phase space. Examples are threshold logarithms that generically arise when the energy of the incoming
particle is just sufficient to produce the observed final state, or transverse-momentum logarithms that
develop in two-scale situations for a (small) measured transverse momentum of a produced particle
in the presence of an overall hard scale. In many instances it is necessary to resum these large logarithmic
terms to all orders in strong coupling.  In the case of the EIC, such resummations are not yet fully developed,
and many cases still need to be addressed. Examples are threshold resummation studies for final states
produced with large transverse momentum (such as hadrons or jets), studies of resummation for low-$q_T$ jet production
and its matching to NNLO, and resummation for spin asymmetries. High-precision resummation of hadronic event shapes in DIS also promises to lead to new determinations of the strong coupling $\alpha_s$, universal nonperturbative hadronization effects, and nuclear dynamics \cite{Kang:2013wca,Kang:2013nda,Kang:2014qba,Kang:2015swk,Zhu:2021xjn}. 
Resummation may also be used
to derive approximate fixed-order (e.g., NNLO) corrections to observables, as recently shown in~\cite{Abele:2021nyo,Abele:2022wuy}.
These approximate results may be used both as a cross check for full fixed-order calculations and for obtaining
phenomenological results. In addition, the domain of validity of the threshold approximation
may be extended to all collinear initial-state radiation,
as demonstrated for $p+p$ collisions in Ref.~\cite{Lustermans:2019cau},
expected to be relevant to capture the full $x$ and spin dependence in SIDIS.
Power corrections to EIC  observables are closely related to resummation, see, for example, Ref.~\cite{Sterman:2022lki}.

\item Phenomenology of the impact of QED corrections on extractions of (polarized) PDFs. 
Clearly, if NNLO accuracy is to become the standard for QCD, QED and even electroweak corrections should also be considered. The importance
of radiative QED effects in $e+p$ DIS has long been recognized. Recent work~\cite{Liu:2021jfp} presented a new formulation for QED corrections that promises to be applicable to a much broader range of $e+p$ observables such as SIDIS and high-$p_T$ production. Given the importance of understanding the systematics stemming from QED radiative corrections, and given that such corrections 
are among the major sources of the total systematic uncertainty for extracting TMDs from experimental observables, 
a recent analysis \cite{Bishnu:2023} has already compared the QED radiative effects within the factorized \cite{Liu:2021jfp} 
and conventional \cite{Akushevich:2019mbz} approaches for unpolarized beams on unpolarized targets. 
The preliminary results, obtained for three JLab kinematic bins 
as well as three EIC bins, show about 10\% difference between the factorized and conventional 
approaches for the unpolarized case. A similar comparison for an unpolarized beam on a transversely-polarized target is under way. Much more work will be needed to fully explore and understand the role of radiative corrections to EIC observables.
(We note that it will be important for future experimental data to not be released only after correction for QED effects, but also at the uncorrected level.) Another key contribution of precision theory will be the application of insights from QED radiative corrections to the construction of robust experimental lepton/photon recombination schemes,
see e.g.\ Ref.~\cite{Ebert:2020dfc} for a similar effort in precision Drell-Yan studies.

\item 
Monte-Carlo event generators are a cornerstone of analyses at  colliders. 
A generator called \texttt{SIDIS-RC EvGen} has been developed in Ref.~\cite{Byer:2022bqf} for generating 
SIDIS events and calculating cross sections for unpolarized or longitudinally polarized beams and unpolarized, longitudinally 
or transversely polarized targets. The structure and underlying physics of the generator incorporates TMDs and FFs in the 
Gaussian and Wandzura\texttt{-}Wilczek-type approximations \cite{Bastami:2018xqd}, as well as QED calculations of the 
lowest-order radiative effects (using a conventional radiative correction \cite{Akushevich:2019mbz}) 
applied to the leading order Born cross section in the SIDIS process. Thereby, one can obtain multi-dimensional binned 
simulation results, which will help extract essential information about 3D nucleon structure 
from SIDIS measurements. \texttt{SIDIS-RC EvGen} is in the second stage of its development, whereby one can carry out 
high-precision studies of SIDIS cross sections, multiplicities and single-spin asymmetries at the generator level, from 
medium to high lepton beam energies, including studies in EIC kinematics.
In the future, modern parametrizations of TMDs used in most recent phenomenological studies, 
as well as new calculations of exclusive structure functions and hadronic vacuum polarization corrections should be employed in other 
updates of \texttt{SIDIS-RC EvGen} to make more precise comparisons with data. The continued development and improvement of event generators for the EIC will remain an important topic for precision theory in the upcoming decade.

\item New global analyses of hadronic structure (note that Sec.~VIII is also 
dedicated to global analyses). These include global NNLO analyses of helicity PDFs
including lepton scattering and RHIC ($p+p$) data; global analysis of DVCS in terms of GPDs; 
determination of new sets of photon PDFs using HERA data -- which will be vitally important
for studies of photoproduction at the EIC -- and inclusion of threshold resummation in analyses
of PDFs and FFs.
In conjunction with these developments, inclusive measurements and those involving tagged final states can place significant constraints on $\alpha_s$~\cite{dEnterria:2022hzv} itself.  Theoretical improvements to more accurately determine QCD parameters such as $\alpha_s$ and heavy-quark masses in the context of global analyses would also support $e+p$ precision physics~\cite{Alekhin:2017kpj}.
Experience from HERA and the LHC has demonstrated the
importance of combined community efforts to analyze PDFs. These efforts have given 
rise to the widely used LHAPDF ({\tt https://lhapdf.hepforge.org/}, Ref.~\cite{Buckley:2014ana}), PDF4LHC~\cite{DeRoeck:2009zz},
and xFitter ({\tt https://www.xfitter.org/xFitter/}, Ref.~\cite{Alekhin:2014irh}) platforms. A similar setup for the EIC in terms of a new {\it PDF4EIC} collaboration
would be highly beneficial and could best be formed under the auspices of
the EIC Theory Alliance.

\end{itemize}

Clearly, at such an initial stage, this list is still incomplete and will likely be
extended, sharpened and further developed over the next few years. However, the list already lays 
out a set of goals that are vital milestones that 
could only be realized in full via the EIC Theory Alliance.



\section{Global analysis of hadron structure}
\label{sec:glob_ana}

The QCD global analysis program sits at the intersection between theory, experiment and data science and has as a primary goal the extraction of quantum correlation functions (QCFs) from experimental data. These objects synthesize the internal structure of hadrons in terms of their quark and gluon degrees of freedom and provide unique opportunities to understand fundamental questions in nuclear particle physics, such as the origin of spin and mass, nuclear tomography, and the origin of anti-matter asymmetry in hadronic matter, just to mention a few.

At the EIC, an unprecedented amount of data is going to be produced that has a transformational impact in our understanding of nuclear and particle physics, and QCD global analysis will play a key role in delivering the extraction of QCFs by matching sophisticated  theoretical frameworks to observational data. To achieve this, it is necessary to establish coordinated efforts within the EIC theory community that can integrate organically  the available expertise to build the next generation of global analysis tools and to meet the challenges of the EIC science. Specifically, we envision that a theory alliance  will involve:
    i) Training the next generation of scientists from diverse backgrounds (theory, experiment, data science) in QCD theory  that can contribute to the QCD global analysis efforts at the EIC.
    ii) R\&D of reliable and extendable open-source theory libraries using modern programming languages that can be integrated within state-of-the art data science analysis tool kits and be tested and validated  by the community itself with standardized practices.
    iii) Dedicated impact studies to provide reliable projections of the physics outputs  that feedback the EIC experimental  community with continuous updates from the advances in QCD theory.  
    iv) Coordination with the LQCD community to systematically integrate LQCD calculations into the QCD global analysis program~\cite{Constantinou:2020hdm,Lin:2017snn}. 

Coordination of these efforts  will consolidate effectively the link between theory and experiment by preventing  the unnecessary duplication of efforts around the community and  it will deliver the most reliable community driven analysis framework to execute the QCD global analysis program at the EIC. In what follows, we will discuss the opportunities, challenges and needs of the QCD global analysis program within the EIC theory Alliance.

\subsection{Opportunities}

\textbf{\underline{Simultaneous extraction paradigm}}:
The high-quality measurements that will be carried out at the EIC require matching developments from the theory side in order to extract all the relevant information from them. In particular, the EIC will be the first ever experiment for which measurements of unpolarised and polarised lepton-nucleon scattering measurement will be provided alongside with lepton-nucleus scattering and fragmentation function measurements. In particular it will: 
\begin{itemize}
    \item Allow the first joint determination of the quark and gluon structure of protons, deuterons, and heavy nuclei, and in particular illuminate the poorly known behaviour of nuclear corrections for intermediate A values~\cite{AbdulKhalek:2022fyi,Eskola:2021nhw}. This determination will fully account for the correlations between experimental uncertainties provided by the EIC measurements, a unique feature of the EIC that opens the door to many new groundbreaking analyses, see~\cite{Khalek:2021ulf} for an initial study.
    \item Enable to perform such joint analysis for different underlying theory frameworks for the underlying QCD processes, from collinear factorisation to BFKL resummation~\cite{Ball:2017otu,xFitterDevelopersTeam:2018hym} and non-linear QCD dynamics such as those accounted for in the Color Glass Condensate formalism. 
    
    \item Incorporate the information provided by lattice QCD calculations into this universal QCD analysis of hadron structure in ways that minimise the need of intermediate assumptions, e.g. including lattice data at the level of matrix-element calculations alongside the actual EIC measurements.
    
    \item Carry out a joint determination of polarised and unpolarised free-nucleon structure together with that of the hadronic fragmentation functions, which offers an unprecedented window to the mechanisms of mass and spin generation in the proton as well as to the QCD dynamics responsible for the parton-to-hadron transition. Only with the advent of the EIC such analysis would be possible without having to rely on the combination of very different (possibily inconsistent) experiments. 
    
    \item The extended range in $x$ and $Q^2$ covered by the EIC will allow us to study the scale dependence of unpolarized as well as polarized transverse momentum dependent distribution functions (TMDs) to an unprecedented level of precision. In turn, this will enable us to perform simultaneous extractions of collinear and transverse momentum parton degrees of freedom, through unpolarized and transverse-spin sensitive observables, in order to provide the most comprehensive understanding of hadron tomography.
    Moreover, the EIC measurements will provide the theory community with fundamental insight on  the mechanism of QCD factorization. 

\end{itemize}

\textbf{\underline{Going beyond the boundaries of our current knowledge}}:
Measurements at the EIC will traverse a critical region of the quark-to-hadron transition, broadly sampling $W^2$ in $ep$ and $eA$ collisions with significant access to both relatively low and very high $x$. The EIC will also span a range of $Q^2$ and $W^2$ from relatively low values, where power-suppressed corrections ($\sim\! 1 / Q^2$) are prominent, to much higher values, where a purely perturbative, twist-2 description is applicable. The wide purview of the EIC program demands a serious Theory effort to understand and control dynamics in the transition region and to maximize the benefit to analyses of PDFs~\cite{Gao:2017yyd, Hou:2019efy,Bailey:2020ooq,Alekhin:2017kpj,NNPDF:2021njg,Moffat:2021dji,ATLAS:2021vod} and related quantities, including at the TeV scale. This Theory effort encompasses the development of numerical methods to better quantify QCFs and their uncertainties in light of the expected large data sets. In addition, high luminosities at the EIC will provide access to more exclusive processes~\cite{Arratia:2020azl} with small(er) cross sections while, at the same time, the precision of inclusive measurements, such as reduced DIS cross sections, may be systematics-limited. Balancing among studies of inclusive and exclusive processes is non-trivial and requires coordination between theorists with diverse expertise. The Theory Alliance will be positioned to coordinate the development of robust and flexible factorization frameworks for both kinds of analyses, as well as hybrids of these.

\textbf{\underline{Feedback to the experimental community}}:
An EIC Theory Alliance might stimulate the necessary studies of precision $ep$ processes necessary to inform possible extensions of the EIC science program, for instance, related to luminosity upgrades. The Alliance would be a natural setting to weigh the potential of such luminosity improvements, as well as other potential systematic extensions of the EIC program ({\it e.g.}, alternative detector concepts; positron beams). For example, in the context of studies of flavor and $x$ dependence of the projected PDFs, the estimated impacts based on fitting EIC pseudodata can at times be counterintuitive, with projections occasionally depending on the chosen fitting framework, baseline data sets, or varying theory assumptions. As a Theory Alliance initiative, it would be valuable to carry out dedicated benchmarking studies involving multiple groups with agreed-upon methods to identify consistent impact(s) of various EIC run scenarios. Such benchmarking studies~\cite{PDF4LHCWorkingGroup:2022cjn,Ball:2012wy,Rojo:2015acz,Andersen:2016qtm,Andersen:2014efa,Butterworth:2015oua} have proven to be valuable for understanding differences among the conclusions obtained by PDF fitters while analyzing the same data sets.

\textbf{\underline{Next generation of QCFs uncertainty quantification and modeling}}: 
Intepretability, uncertainty quantification, and replicability are generic challenges in multivariate, complex analyses like the global analyses of QCFs. While precise theoretical predictions and experimental measurements will be mandatory, inference from EIC data will also depend on representative explorations~\cite{Courtoy:2022ocu} of various uncertainties and control of approximations in theoretical formalisms and numerical realizations (see also~\cite{Ball:2022uon} for a contrasting perspective). Reliable extraction of QCFs from EIC data will require sustained investments in these issues both in traditional statistical and AI/ML approaches. This involves gaining a systematic understanding of the PDF parameter space, including with respect to the complicated patterns of parametric correlations in fits; prevention of sampling biases in the resulting PDF uncertainties; and quantification of the pulls of specific experiments on the underlying QCFs. 

There is also an associated need to develop numerical tools, such as the $L_2$~\cite{Hobbs:2019gob,Courtoy:2020fex,Accardi:2021ysh} or other sensitivity methods~\cite{Wang:2018heo} or optimized Monte-Carlo techniques~\cite{Courtoy:2022ocu} for PDF sampling, to understand internal aspects of PDF fitting frameworks for EIC analyses and to stress-test ML/AI algorithms which might augment QCF fitting for the EIC. Understanding implications of the bias-variance dilemma for global fits, which affects the flexibility of the functional forms of QCFs in all approaches, will be central. The Theory Alliance could serve as a clearinghouse for this activity.

The EIC program will involve a complicated mix of perturbative and nonperturbative QCD. There is a strong need for insights into nonperturbative dynamics in carrying out next-generation PDF fits of EIC data. Examples include high-$x$ physics, like nucleon structure beyond leading twist, or the structure and dynamics of light and heavy nuclei. The Theory Alliance could serve as a nexus connecting theoretical developments in related areas \cite{Hou:2019efy}.

\textbf{\underline{Synergies with HEP}}: From the HEP vantage, the EIC may enable various PDF improvements connected to phenomenological goals at the LHC~\cite{AbdulKhalek:2022hcn} such as those related to large-$x$ PDFs entering searches for new massive particles~\cite{Ball:2022qtp,Fiaschi:2022wgl} or interpretations of high-$p_T$ tails in the framework of the SM Effective Field Theory~\cite{Greljo:2021kvv,Gao:2022srd}".. The EIC therefore has an important reach toward the TeV scale, but there are no formal community structures or venues to develop this connection beyond {\it ad hoc} collaborations of individual researchers. As a special initiative focused on the QCD program at the EIC, a Theory Alliance would be suited to bridge implications of the EIC program at the GeV and TeV scales while exploring possible overlaps. These overlaps would also extent to commonalities between PDFs for collider physics and the EIC with respect to Monte Carlo event generation~\cite{Campbell:2022qmc} and related computation.

A formal organization would benefit theorists working on EIC science, especially given that the EIC transcends the traditional divide between HEP and NP. A Theory Alliance could therefore ensure that developments on either side of this line are communicated to the other, possibly following the successful mold of previous initiatives like the recent formation of a PDFLattice community. Practically, the Alliance might also strengthen opportunities for young researchers entering the field, providing them with a larger network beyond that more narrowly focused on specific areas in HEP or NP.

\subsection{Challenges}

In order to realise such ambitious program of global analysis of hadron structure at the EIC, some of the main bottlenecks that the Theory community needs to tackle beforehand are:
\begin{itemize}
    \item To carry out the QCD and electroweak higher-order perturbative calculations required to fully exploit the scientific potential of the EIC for global QCD analyses, and implementing these calculations in fast interfaces so that they can be incorporated in existing fitting frameworks.
    \item To develop novel analysis frameworks and techniques, based for instance in ML/AI, to reduce the need for theory and model assumptions in the interpretation of the EIC measurements.
    \item In close connection with the experimental community, to design and formulate non-trivial, out of the box measurements with improved or even optimal sensitivity on the underlying theory parameters, such as for example unbinned multivariate measurements.
    \item All software development should take place under the open source paradigm and fully documented, to ensure the reproducibility and  facilitate the dissemination
    of the results.
    \item To integrate in a cohesive manner all the  different software tools required for global QCD analyses at the LHC, from theory calculations to the fit machinery, into a single "EIC Theory" framework that becomes the battle horse for global QCD analyses in the EIC era and that is supported and maintained by a broad theory community. 
\end{itemize}

\section{Jets at EIC}
\label{sec:jet}

The advent of the ElC with its high luminosity ($\sim 1000$ times higher than HERA) and polarized hadron beams will produce the first-ever jets in polarized electron-hadron scattering and will unlock the full potential of jets as novel tools to probe the structure of nucleons and nuclei.  High energy jets are energetic sprays of particles that are routinely observed in high energy particle colliders. Jet studies have played a key role in the exploration of QCD since its inception~\cite{Ali:2010tw}. Early jet measurements in $e^+e^-$ collisions have confirmed the existence of the gluon and established its spin. With advances in experimental techniques and theory development over time, jets have become powerful tools to explore the fundamental properties of QCD, such as in searches for unexpected phenomena in high-energy collisions~\cite{Larkoski:2017jix,Asquith:2018igt,Marzani:2019hun} and in studying the transport properties of the quark-gluon plasma (QGP) in heavy ion collisions~\cite{Connors:2017ptx,Busza:2018rrf,Cunqueiro:2021wls}. Such work has pushed jet physics to the forefront of phenomenology at the LHC and RHIC.

While jets are familiar in high-energy physics analyses, and appear in many different guises, jets at the EIC can add important pieces of the puzzle on top of insights gained at hadron-hadron machines: Jets at the EIC are naively expected to be very ``clean'', i.e. little energy not associated with the jets~\cite{Aschenauer:2017jsk}. However, the jets themselves contain relatively few particles and the particles have moderate energies~\cite{AbdulKhalek:2021gbh}. This scenario offers unique challenges and opportunities: every particle is precious and differences between jet algorithms or substructure methods can become very apparent, while at the same time, underlying event contamination (that continues to be a major challenge at the LHC) will be much smaller. Thus an assessment of jet properties at the EIC is an exciting theoretical and experimental prospect. On top of that, non-perturbative fragmentation contributions are more pronounced at lower jet masses, which make jets at the EIC a stress test for the universality of jet-based methods in high-energy physics. In addition, the ability to polarize beams and its unmatched versatility, the EIC will catalyze development of new jet-based spin observables. Thus, dedicated studies of jet substructure at the EIC are critical to realize the full potential of the EIC in jet physics. A variety of key measurements include (but are not limited to):
\begin{itemize}

\item Jets for studies of flavor and spin structure of the nucleon, in particular 3D imaging of the nucleon and even 5D Wigner distributions.  Employing jets instead of final-state hadrons reduces the sensitivity to fragmentation. For example, Refs.~\cite{Hatta:2016dxp,Boussarie:2018zwg} demonstrated that diffractive dijet production would access the quantum phase space Wigner distribution of gluons. Refs.~\cite{Liu:2018trl,Liu:2020dct,Arratia:2020nxw} proposed the lepton-jet correlation in deep inelastic scattering as a unique tool for nucleon tomography at the EIC. In particular, the transverse momentum imbalance between the lepton and the jet would probe 3D unpolarized and polarized TMDs. If one further measures the transverse momentum distribution of hadrons inside the jet with respect to the jet axis, one would be able to probe TMD fragmentation functions~\cite{Kang:2020xyq,Kang:2021ffh}. In addition, Ref.~\cite{Kang:2020fka,Lee:2022kdn} proposed utilizing the jet charge for flavor separation, especially critical for spin-dependent PDFs. In the same spirit, Ref.~\cite{Arratia:2022oxd} proposed to use neutrino-tagged jets for flavor separation at the EIC.

\item Transverse momentum measurements using jets to image the 3D structure of the nucleon can benefit from using a recoil-free jet definition. In this case there is an all-order factorization theorem for the cross section~\cite{Gutierrez-Reyes:2018qez} and all ingredients for NNNLL resummation are known~\cite{Gutierrez-Reyes:2019vbx}. (This choice of jet definition avoids the issue of non-global logarithms from correlated soft radiation in- and outside jets, which limits the theoretical accuracy.) 
With this jet definition, the nonvanishing T-odd part of a jet could help probe the chiral-odd nucleon TMDs~\cite{Liu:2021ewb,Lai:2022aly,Lai:2022xox}.
Furthermore, it is possible to consider jets defined by charged-particles only, exploiting the superior angular resolution of the tracking system. This involves only a minimal modification~\cite{Chien:2020hzh} of the function describing the jet, involving moments of the non-perturbative track functions~\cite{Chang:2013rca}.

\item Jets for longitudinal nucleon structure. Observables such as the double longitudinal spin asymmetry in inclusive jet production are sensitive to the collinear partonic structure of both the proton and the polarized photon~\cite{Boughezal:2018azh, Page:2019gbf}. Jet probes of polarized gluon distributions feature different systematic errors than measurements with inclusive DIS and therefore allow for cross checks between the two measurements. The EIC will provide the first window into polarized photon structure through jet measurements. The computation of the hard scattering cross sections for inclusive jet production are amenable to techniques developed for high precision LHC studies~\cite{Abelof:2016pby,Borsa:2020ulb} and therefore provide a natural bridge between the theory efforts of the EIC and LHC programs.

\item Jets for study of 3D evolution equations, which go beyond the well-established DGLAP equations.

\item Jets formed by heavy quarks. Charm quark jets can be produced at the EIC. They can be used to disclose properties of heavy quark systems and light hadron structures~\cite{Zhu:2013yxa,Arratia:2020azl,delCastillo:2020omr,delCastillo:2021znl}. For example, Ref.~\cite{Boer:2016fqd,Kang:2020xgk,Dong:2022xbd} proposed to use heavy flavor dijet production in polarized lepton-nucleon scattering to probe the gluon Sivers function.

\item Jet substructure (such as jet shape, jet mass, jet angularity, etc) in $e+p$ collisions as powerful probes of QCD dynamics.  See, for example, the studies of jet angularity~\cite{Aschenauer:2019uex,Zhu:2021xjn} and jet charge~\cite{Li:2021uww}.

\item Jet algorithms tailored to DIS that can separate target from current fragmentation processes~\cite{Arratia:2020ssx}.

\item Jet-based observables and event shapes (such as 1-jettiness) as  precision probes of fundamental QCD parameters~\cite{Kang:2013nha,Kang:2014qba} such as the running of the strong coupling constant and of nuclear dynamics~\cite{Kang:2012zr,Kang:2013wca,Kang:2013lga}.

Beyond jets, the energy-energy correlator (EEC) event shape observables in $e^+e^-$ annihilation, hadronic collisions, and deep inelastic scattering are precision probes of perturbative and non-perturbative QCD dynamics~\cite{Neill:2022lqx}. These correlators can be calculated to very high precision~\cite{Ebert:2020qef,Ebert:2020sfi} and generalized to DIS by considering the transverse-energy correlation between the lepton and final state hadrons~\cite{Li:2020bub}. EECs provide a complementary way to study TMDs that is, as yet, underexplored and presents an opportunity for the EIC Theory Alliance.    Measurements of QCD observables in DIS are often done in the Breit frame. 
Recently, a new definition of EEC in the Breit frame, a natural frame for the study of TMD physics~\cite{Collins:2011zzd}, was presented~\cite{Li:2021txc}.
In this frame, the target hadron moves along $\hat{z}$ and the  virtual photon  moves in the opposite direction. The Born-level process is described by lepton-parton scattering $e+q_i\to e+q_f$, 
where the outgoing quark $q_f$ backscatters in the direction opposite the proton.  Hadronization of the struck quark will form a collimated spray of radiation close to the $- \hat{z}$ direction. On the other hand, initial state radiation and beam remnants are moving in the opposite direction close to the proton direction of motion.  This feature of the Breit frame, leads to the clean separation of target and current fragmentation utilized to construct and  study novel EEC observables in DIS.

In this spirit, Ref.~\cite{Liu:2022wop}  introduces the concept of nucleon energy correlators, a set of novel objects that encode the
microscopic details of a nucleon, such as the parton angular distribution in a nucleon, collinear
splitting to all orders, as well as the internal transverse dynamics of the nucleon. It was demonstrated that nucleon energy
correlators can be measured in lepton-nucleon DIS and complement the conventional nucleon/nucleus tomography without introducing non-perturbative fragmentation functions or jet clustering algorithms.

\item Modification of jets and jet substructure going from $e+p$ to $e+A$ collisions provides an opportunity to study parton transport through nuclear matter.  The forward proton/nucleus going direction is the optimal region to observe large final-state modifications due to in-medium shower evolution~\cite{Li:2020rqj,Li:2021gjw}. The final-state effects on $R_{eA}$ are large in the relatively low $p_T$ region, whereas initial-state effects, if sizeable, are observed at high $p_T$.  A major advantage of jet measurements relative to those of semi-inclusive hadron production is that, by considering the ratio of cross section modifications for different jet radii, the effects of nuclear PDFs can be strongly suppressed to cleanly  probe the strong interaction between jets and cold nuclear matter.  With a judicious choice of the center-of-mass energy, rapidity interval, and jet radius $R$, the inclusive cross section suppression can be nearly a factor of two -- similar to what is measured to high precision in $A+A$ relative to $p+p$ collisions. The modification of jet substructure is related to jet attenuation in cold nuclear matter. The jet charge modification of individual flavor jets can shed light on the medium-induced scaling violations in QCD,  whereas precision studies of the inclusive jet charge can be used to extract flavor information and constrain the nuclear PDFs~\cite{Li:2020rqj}. Last but not least, first calculations of nuclear-enhanced QED corrections, meriting further investigation, have appeared~\cite{Tomalak:2022kjd}.A broad collaboration among experts on jet physics, global analysis, and nuclear matter, available through the EIC Theory Alliance,  is needed to realize this ambitious program.

\item  Probing gluon saturation with forward dijet production. 
Saturation provides an additional handle on the transverse motion of soft gluons in the target, particularly for inclusive dijet production.  Diffractive dijet production has been shown be sensitive to the Wigner function at small $x$. These observables will probe the singularity of the small $x$ gluon distribution. 

\item
Jets are fundamental for understanding QCD factorization. Current formulas deal mostly with leading-power terms in the perturbative QCD expansion, while investigations into the effects of next-to-leading power terms have become a recent area of interest. Beyond perturbative QCD, higher-type functions enter scattering amplitudes of various jet characteristics. More research is required to apply this theory to EIC data analysis and interpretation, in particular, to extend factorization to higher dimensional symmetric spaces (by adapting modern theory of hypergeometric functions ~\cite{gelfand1994discriminants},  higher L-functions ~\cite{bump2012multiple,gil2017multiple} and modular forms ~\cite{manin2007modular}); develop factorization for resurgent functions (See Refs.~\cite{mitschi2016divergent,balser1979general,olive2003resurgence}); and extend factorization to irregular singularities (\cite{xu2019closure,bridgeland2012stability}).
\end{itemize}

\section{Heavy flavor production and hadronization}
Heavy flavor production in DIS complements the science thrusts outlined in other sections of this white paper and opens a window on new physics inaccessible with light hadrons and inclusive jets.    \\

\subsection{Open heavy flavor} 

Open heavy-flavor  production at the EIC  is an important probe of the partonic content of nucleons and nuclei. In addition to constraining the gluon and sea quark PDFs, feasibility studies suggest that the prospects for constraining unpolarized nucleon strangeness via charge current reactions that produce charm jets in the final state are rather promising~\cite{Arratia:2020azl}. 
Furthermore, the existence of a nonperturbative heavy-quark content in the proton, called intrinsic charm (IC), has long been postulated~\cite{Brodsky:1980pb,Brodsky:1981se}.  
A number of experimental measurements~provided inconclusive evidence of IC, however the recent LHCb $Z + {\rm charm \, jets}$ measurement  
relative to all $Z + {\rm jets}$, is consistent with a 1\% IC
component \cite{LHCb:2021stx}. New work by the NNPDF collaboration has established the existence of IC in their analysis~\cite{Ball:2022qks}, consistent with both the LHCb $Z+$charm results and the EMC $F_2^c$ measurements. 
Data pertinent to IC will become available at the EIC and will shed new light on this exciting topic~\cite{Hobbs:2013bia}. In addition to open heavy flavor, quarkonia can also be used to probe intrinsic charm~\cite{Vogt:2021rcz,Vogt:2021vsc,Vogt:2022glr}. 

In-jet hadron data at the EIC will prove very valuable in the future in analyses of fragmentation functions (FFs). In particular, it can further constrain the detailed momentum dependence of gluon hadronization~\cite{Kneesch:2007ey}. 
The framework developed and applied in~\cite{Anderle:2017cgl} can be straightforwardly generalized to incorporate in-jet data in any future global fit of FFs once such data become available.
The detailed impact of the resummation of logarithms of the jet size parameter $R$ can be further investigated.   
By making use of the results for the in-jet fragmentation of hadrons derived within the SCET formalism~\cite{Kang:2016ehg}, 
it is possible to extract FFs at a combined accuracy of NLO+NLL$_{\rm R}$.

The EIC will provide opportunities to study semi-inclusive $c$-jet and $b$-jet cross sections and substructure in $e+A$ relative to $e+p$ collisions~\cite{Li:2021gjw}.  Heavy flavor-tagged jet production is more sensitive to the gluon and sea quark distributions in nucleons and nuclei compared to light jets. Thus, in kinematic regions where $R_{eA}$ is dominated by initial-state nPDF effects, the modification is expected to be even stronger when compared to inclusive jets. Similar to the case of light jets,  by applying the strategy of studying ratios of the nuclear modification with two different jet radii $R$ we can eliminate nPDF effects, primarily the anti-shadowing and the EMC effect in the forward rapidity region.  The remaining quenching of the jet spectra can be as large as a factor of two for small jet radii, for example $R=0.3$, and can clearly be attributed to final-state interactions and in-medium modification of parton showers containing heavy quarks~\cite{Kang:2016ofv,Sievert:2019cwq,Li:2018xuv}.  These measurements will yield valuable independent constraints on the transport properties of cold nuclear matter.

The EIC Theory Alliance will provide the broad expertise needed to complement the calculation of semi-inclusive jet cross sections with jet substructure. Heavy flavor tagged jets in DIS play a special role since the modifications are expected to be large based on the ``dead cone effect''~\cite{Dokshitzer:2001zm}. First calculations of the groomed, soft-drop momentum-sharing distribution~\cite{Larkoski:2015lea} at the EIC have recently appeared~\cite{Li:2021gjw}. These results show that the substructure modification in $e+A$ relative to $e+p$ reactions is on the order of 10\% or smaller. Still, as in the case of heavy-ion collisions at relatively small $p_T$ the differences in the subjet distribution are most pronounced for $b$-jets, followed by $c$-jets. Heavy-ion collisions have also explored the interplay between the ``dead cone effect'' and nuclear medium dynamics, which should be revisited in the context of the EIC using a model-agnostic framework such as that of the JETSCAPE Collaboration \cite{Putschke:2019yrg}. In contrast to the heavy-ion case, however, there is significant difference between the energy of the parton in the rest frame of the nucleus and the jet scale which determines the available phase space for substructure, even for large radii, $R\sim 1$. Thus the jet  momentum sharing distribution at the EIC probes a different interplay between the heavy quark mass and suppression of small-angle medium-induced radiation  -- a regime that can only be accessed at the EIC and merits detailed investigation in the future. Last but not least, theoretical tools that are becoming available can be used to study how sub-eikonal corrections to in-medium branching, such as the effects of varying matter density~\cite{Sadofyev:2021ohn}, propagate into experimental observables. \\

\subsection{Quarkonia} 

Recent theoretical studies of quarkonia exploit new effective
field theory (EFT) capabilities  that significantly boost the theoretical precision of $J/\psi$ and $\Upsilon$  analyses and propose modern observables~\cite{Lansberg:2019adr,Brambilla:2010cs,Brambilla:2014jmp} that can probe the quarkonium production mechanism. Recent research has enabled reduction of the number of long-distance matrix elements (LDMEs)~\cite{Brambilla:2022rjd,Brambilla:2022ayc}. Based on pNRQCD, the spin-1 S-wave quarkonia (bottomonium and charmonium) LDMEs can be factorized in terms of wave-functions at the origin and $3$ flavor-independent gluon correlators, greatly reducing the nonperturbative unknowns, instrumental for work at the EIC. 
References~\cite{Brambilla:2022rjd,Brambilla:2022ayc} constrain the LDME $\langle {\cal O}^{J/\psi}(^3P_J^{[8]}) \rangle$ to be positive and give a relatively small value of $\langle {\cal O}^{J/\psi}(^1S_0^{[8]}) \rangle$, describing spin-1 S-wave quarkonia production and polarization at large $p_T$ but still overshooting the $J/\psi$ inclusive production rates at HERA and Belle at low $p_T$. The LHC cross section ratios predicted in Refs.~\cite{Brambilla:2022rjd,Brambilla:2022ayc}  (independent of perturbative calculations because the short distance coefficients (SDCs) cancel in the ratios) are in good agreement with the high $p_T$ data but are in conflict with low $p_T$ data, indicating that NRQCD factorization may fail at relatively low $p_T$. 

The EIC will provide new insights in quarkonium factorization and reduce the uncertainties on the LDMEs because, while the predicted cross sections depend dramatically on the LDMEs, the rates are sufficiently high to be measurable even at relatively large $p_T$ \cite{Qiu:2020xum,Brambilla:2022ayc}. At low $p_T$ recent developments have been focused on the TMD description of quarkonium production. This includes a novel factorization formalism: a shape function description of the non-perturbative physics.  This is particularly relevant for the EIC  \cite{Bacchetta:2018ivt,Echevarria:2020qjk,delCastillo:2021znl,Boer:2021ehu}, which will further clarify shape function LDME extraction. Conversely, in the high energy region ($E \gg m_{Q{\overline Q}}$), theoretical advances in understanding quarkonia are also possible based on the picture of parton fragmentation~\cite{Baumgart:2014upa,Bain:2016rrv}. 
Studies of $J/\psi$ and $\Upsilon$ production in jets can better constrain the LDMEs appearing in NRQCD factorization where significant uncertainties still remain.  Another important open question that the EIC Theory Alliance can answer is whether medium-induced radiative processes can contribute significantly to the modification of quarkonium cross sections in $e+A$ reactions. Finally, EIC studies of vector and pseudoscalar quarkonium emission at moderate $p_T$ will be relevant for the transition region from short-distance $(Q \bar Q)$-pair production to the fragmentation mechanism~\cite{Braaten:1993rw,Braaten:1993mp,Zheng:2019dfk,Zheng:2021ylc,Zheng:2021mqr,Bodwin:2014bia,Zhang:2017xoj,Celiberto:2022dyf,Celiberto:2022keu,Silvetti:2022hyc}.

As in the vacuum case, quarkonium dynamics in nuclear matter remain a multi-scale problem accessible to the EFT approach~\cite{Brambilla:2004jw}. The corresponding non-equilibrium evolution in the quark-gluon plasma has been  described recently using pNRQCD (potential NRQCD) \cite{Brambilla:1999xf} at finite temperature \cite{Brambilla:2008cx,Brambilla:2017zei,Brambilla:2021wkt}. Many of these findings are independent of the medium.
Recently, it was explicitly demonstrated how NRQCD  \cite{Bodwin:1994jh} can be generalized to include interactions of non-relativistic heavy quarks with different type of nuclear media~\cite{Sharma:2012dy,Aronson:2017ymv} without loss of generality~\cite{Makris:2019ttx,Makris:2019kap}. This generalization was achieved by incorporating the Glauber and Coulomb gluon exchanges of charm and bottom quarks with different types of scattering centers in nuclear matter. The NRQCD and NRQCD$_{\rm G}$ approachess can facilitate a more robust and accurate theoretical analysis of quarkonium measurements in $e+p$ and $e+A$ reactions at the EIC, presenting the opportunity to investigate modifications of the $Q\bar{Q}$ potential from medium interactions which are Coulomb-like in the vacuum. In addition, interactions with the medium can induce radial excitations that can induce transitions from one quarkonium state to another. Medium-induced transitions from and to exited states can modify the observed relative quarkonia production rates and can be incorporated into a network of rate or master equations.\\

{\em Threshold photo- and electro-production of heavy quarkonia and the mass radius of the proton} 
The mass radius is a fundamental property of the proton that can be rigorously defined through the form factor of the energy-momentum tensor. In the weak gravitational field approximation, it can also be defined through the form factor of the trace of the energy-momentum tensor (EMT). The scale anomaly in QCD enables the extraction of this form factor through measurements of differential heavy quarkonium photoproduction cross sections near threshold~\cite{Kharzeev:1995ij,Kharzeev:1998bz,Hatta:2018ina,Hatta:2019lxo,Mamo:2019mka,Kharzeev:2021qkd,Guo:2021ibg,Lee:2022ymp}. Recent data from the GlueX \cite{GlueX:2019mkq} and $J/\psi$-007 \cite{Duran:2022xag} Collaborations suggested that the mass radius of the proton is significantly smaller than the rms charge radius of the proton.  While this difference has been attributed to the interplay of asymptotic freedom and spontaneous breaking of chiral symmetry in QCD \cite{Kharzeev:2021qkd}, a quantitative QCD-based  theory of the mass distribution inside the proton has yet to be developed. Because high statistics studies of photo- and electro-production of both charmonium and bottomonium are planned at the EIC \cite{Joosten:2018gyo}, a quantitative theory of these processes needs to be developed.

First principles lattice QCD calculations provide valuable information on the gravitational form factors of nucleons \cite{Shanahan:2018pib,Shanahan:2018nnv}.  Collaborations between lattice QCD, phenomenology and experiment to uncover the gravitational and mechanical properties of the proton will be one of research thrusts of the EIC Theory Alliance. 
In addition, lattice calculations of the proton mass radius will be complemented by first principles computations in the continuum. Such calculations begin with the study of the relevant functional differential equations and the methods of their solution. The latter involve deep questions in complex multi-variable functional theory, singularity theory, dynamical systems, and functional analysis. Interpreting the mass distribution inside the proton in terms of the theory of higher transcendental functions is an immediate goal. \\

{\em Hadronization} To interpret the results of  current and future experiments in high energy and nuclear physics, we need a precise understanding of hadron production, especially those composed of both light and heavy quark flavors, and often collected in jets of subatomic particles~\cite{Webber:1999ui,Andersson:1983ia,Andersson:1983jt,Ethier:2017zbq,Bertone:2017tyb}. On general grounds, we expect that hadronization and other non-perturbative effects result in important corrections to quantities that are calculable in perturbation theory. In semi-inclusive deep inelastic scattering (SIDIS)  $e+p/A \rightarrow {\rm jet}/h + X$, the particle production cross sections can be expressed as $ d \sigma = \phi(x) \otimes H \; \otimes D(z)$,  where $\phi(x)$ is the quark or gluon distribution in nucleons and nuclei and $H$ denotes the hard interaction, calculable at high accuracy. The fragmentation function, $D(z)$, describes how partons assemble into an observable bound states carrying a fraction $z$ of the energy of the hard interaction. It was first realized in SIDIS measurements in electron-nucleus scattering by the HERMES experiment~\cite{Airapetian:2000ks,Airapetian:2007vu} that not only the magnitude and shape of $D(z)$, but also the space-time picture of hadronization, plays a critical role in the interpretation of the data. In addition to the picture of parton propagation and energy loss in large nuclei~\cite{Chang:2014fba,Arleo:2003jz}, it is possible that elementary particles themselves can be formed and absorbed inside nuclear matter~\cite{Kopeliovich:2003py,Kopeliovich:2004kq,Kopeliovich:2006xy,Guiot:2020vsf,Adil:2006ra}. 
Relative to the HERMES experiment, the EIC kinematics are subject to a larger medium-induced energy loss, affecting the multiplicity ratios for pions and kaons~\cite{Guiot:2020vsf}. This difference further motivates development of rigorous theoretical approaches based on renormalization group analysis~\cite{Ke:2023ixa} that provide new insights into the resummation of medium-induced radiation and modify fragmentation in reactions with nuclei.    
At present, however, light particle measurements have not provided sufficient discriminating power between those models.

Heavy quark measurements at the EIC will provide the definitive tiebreaker between competing theories of energy loss and in-medium hadronization~\cite{Li:2020sru,Li:2020zbk,Das:2021nqw}. The clean environment and constrained SIDIS kinematics (in contrast to RHIC and LHC) can lead to the first observation of the predicted significant difference in $D(z)$.  First principles evaluations of hadronization times will be an invaluable guide to the interpretation of current and future experimental data.  The distinctly different heavy quark fragmentation functions into $D$ and $B$ mesons provide clear signatures of hadronization dynamics measurable in $e+A$ collisions at the EIC~\cite{Li:2020zbk}.  The EIC Theory Alliance will advance these ideas and perform the first calculations of HF quenching as a function of centrality in DIS~\cite{Chang:2022hkt}. 

Models of hadronization are very important not only for EIC physics, but also for LHC physics. Currently, the interpretation of LHC data relies on string models (often used in event generators), and on elaborate legacy codes with dozens of parameters that have been evolved to fit low energy data. In-depth understanding of hadronization requires development of non perturbative Qcd methods. Factorization formulas take into account only the regular parts of hadronic scattering amplitudes, ignoring higher functional forms (e.g. higher resurgent terms \cite{ecalle1981fonctions,schlomiuk2013bifurcations} are never part of the analysis). Inclusion of these higher functional terms involves development of the corresponding mathematical methods. These methods include studies of functional differential equations (and factorization of their solutions for particular inclusive observables); functional symmetries and embedding collision geometry in function spaces; and extension of resurgence theory to high dimensional symmetric spaces \cite{connes2019noncommutative,braaksma2001differential}.

\section{Hadron Spectrosopy at EIC}


Since the last LRP, hadron spectroscopy has emerged as an area of nuclear and particle physics where new QCD phenomena are regularly being discovered. For over 50 years the quark model of hadrons, much like the nuclear shell model, was capable of explaining the observed symmetry patterns and mass hierarchies of hadrons. The newly observed candidates for multiquark resonances have significantly broadened the hadron landscape beyond the simple quark-model states and into hadronic ``terra incognita''. This second revolution in hadron spectroscopy\footnote{The first revolution of 1974 was marked by the discovery of charmonia, which confirmed QCD beyond doubt as the true theory of strong interactions.} began in 2003 with the discovery of the $X(3872)$ particle~\cite{Belle:2003nnu}, which can be interpreted as containing a large component of a bound state of open-charm mesons, with a radius at least five times larger than that of the deuteron. Charged tetraquark ($Z_c$) and pentaquark ($P_c$) candidates had already been discovered by the time of the 2015 LPR~\cite{BESIII:2013ris,Belle:2013yex,LHCb:2015yax}. A few dozen additional exotic ``$XYZP$'' states have since been observed~\cite{Brambilla:2019esw}. Some recent examples include new structures in the $J/\psi\, J/\psi$ spectrum~\cite{LHCb:2020bwg,CMS:2022yhl,ATLAS:2022hhx} and the doubly charmed tetraquark candidate $T_{cc}^+$~\cite{LHCb:2021auc}. 

This proliferation of hadrons, especially those with heavy quarks and/or gluonic excitations, has fundamental importance for our understanding of QCD in the nonperturbative regime. The implications range from new insights into confinement and other deep features of strongly coupled theories (such as the existence of gravity duals) to the interpretation of new effective degrees of freedom (such as diquarks or constituent gluons)~\cite{Barabanov:2020jvn,Farina:2020slb}, and to new applications of low-energy nonrelativistic effective field theories (EFTs)~\cite{Brambilla:2004jw,Bodwin:1994jh,Brambilla:1999xf}.
 Furthermore, unprecedented advances in algorithms and in theoretical finite-volume formalisms will allow lattice QCD to confirm some of the experimental sightings in the next decade. Such progress will provide a ``theory laboratory'' in which to study the underlying structure of the $XYZP$'s by examining their quark-mass dependence or their response to different external probes. 

This fundamental role of hadron spectroscopy has been highlighted in multiple recent white papers and reviews~\cite{Brambilla:2019esw,Barabanov:2020jvn,Adhikari:2020cvz,JPAC:2021rxu,Arrington:2021alx}, including the recent summary of the Snowmass Planning Exercise white paper~\cite{Lebed:2022vfu}.  The production rates for the $XYZP$ states at the EIC are expected to be quite high~\cite{Klein:2019avl,Albaladejo:2020tzt,Winney:2022tky}, allowing many different states and branching ratios to be studied.  However, because production likely involves Reggeon-like exchange, it is concentrated near the energy threshold and thus at large rapidity, creating an experimental challenge.  For many final states, it may be desirable to take data at lower collision energies. The rates and rapidity distributions would benefit from additional theoretical input.

It is worth recalling that the discovery of QCD exotics had been claimed in the past. For example, the strange pentaquark $\Theta^+$ attracted a lot of attention in the early 2000's after being reported by a dozen experiments, only to disappear after dedicated experiments did not confirm the early signals a couple of years later~\cite{Dzierba:2004db}. Unlike this case, the statistical significance of many $XYZP$ are high, 
reaffirming that structures seen in data are unlikely to be artifacts. On the other hand, it is not always clear that the observations, often made in complicated final states, must be interpreted as QCD resonances. Although the Argand diagrams of some of these structures exhibit a rapid phase motion compatible with resonant behavior, this is insufficient to establish whether or not they correspond to actual excitations of the QCD spectrum. This is why one has to extract resonance information from experimental and lattice data using a variety of reaction amplitudes that fulfill model-independent $S$-matrix principles.
For example, in the case of the light hybrid meson candidate, the $\pi_1(1400)$ and the $\pi_1(1600)$ were previously considered to be different states,  decaying independently into different final states.  However, a simultaneous analysis of $\eta\pi$, $\eta^{\prime}\pi$ final states based on general $S$-matrix principles deduced the existence of a single state able to describe both signals~\cite{JPAC:2018zyd}. Similar cases of apparent duplication of levels might affect the $XYZP$ sector as well, which has immediate consequences for the identification of their multiplets and eventually for understanding their underlying dynamics.
 The Joint Physics Analysis Center (JPAC), formed when JLab12 operations were about to commence, has been developing a reaction theory effort to improve existing methods for analysis and interpretation of spectroscopy data~\cite{jpacwebsite}. The Quarkonium Working Group~\cite{qwgsite} is another example of collaborative efforts between theory and experiment.

Real and virtual photons are some of the most efficient probes for studying the internal structure of hadrons, potentially including the $XYZP$ states.
Resonance properties are directly imprinted in the dependence of their photoproduction observables on momentum exchange and photon virtuality. Measuring these observables to high precision over a large kinematic range will  provide valuable insights into the nature of the exotics. Remarkably, none of these states have yet been unambiguously seen in electro- or photoproduction; such observations would provide complementary information that can further shed light on their composition. 
The theoretical framework to calculate electromagnetic transitions of conventional quarkonia is rather robust, which makes the predictions for photoproduction observables particularly reliable, and thus serve as benchmarks for the $XYZP$ states~\cite{Albaladejo:2020tzt,Winney:2022tky}. 
These studies motivate a spectroscopy program at the EIC, as well as at other future lepton-hadron facilities.
In order for the EIC spectroscopy program to succeed some of the topics that need to be undertaken  include:

\begin{itemize}

\item  A systematic approach to reaction theory for production and decay of heavy-quark resonances that implements $S$-matrix principles, together with the relevant aspects of QCD interactions~\cite{JPAC:2021rxu}.  

\item 
Event generators are required to synergize theoretical and experimental hadron spectroscopy studies at the EIC. Currently, most studies use  \texttt{elSpectro}~\cite{GitHub:elspectro} which is based on the $e+p$ exclusive amplitudes from~\cite{Albaladejo:2020tzt}. The extension to other beam species and to semi-inclusive reactions is needed. 
Moreover, since measurements of polarization observables are unique to the EIC, theoretical studies of how such observables can discriminate among the states are needed.

\item Making use of deep neural networks for interpretation of hadron data, for example 
 using line-shape studies to infer the microscopic nature of the states and learning trainable models of reaction dynamics~\cite{Sombillo:2020ccg,Sombillo:2021rxv,Ng:2021ibr,Gupta:2022vhe,Chen:2023xkz,Zhang:2023czx}. 
\item  Further development of EFTs that exploit
 physical scale separation and factorization will provide systematically improvable descriptions of 
 exotics~\cite{Braaten:2014qka,Guo:2017jvc,Brambilla:2017uyf,Brambilla:2019jfi}.

\item Many of the $XYZP$ states appear close to two-hadron thresholds. Their dynamics can thus be strongly affected by such nearby channels. There are studies~\cite{Eichten:2005ga, Kalashnikova:2005ui, Santopinto:2011zza,  Ferretti:2013faa, Segovia:2016xqb, Ortega:2020tng} that explore how to ``unquench'' the naive quark model to take these dynamics into account. Furthermore, other nonperturbative functional approaches, such as those based on Dyson-Schwinger equations, can bridge the gap between quark models and QCD, providing a more rigorous basis to such frameworks~\cite{Eichmann:2015cra,Eichmann:2016yit,Wallbott:2019dng}. It would be desirable to delve into this topic with the EIC.

\item The recent advances in finite-volume formalism will allow lattice QCD to calculate three-body amplitudes in the channels where exotic candidates are observed. Resonance form factors can be calculated as well and  offer new insights into the nature of exotics~\cite{Briceno:2017max}.

\item Measurements of conventional quarkonium states have shown that nuclear breakup effects are dependent on the radius of the observed state. It is anticipated that similar suppression effects for exotic hadrons can be used to determine their structure.  This suppression has been studied in high multiplicity collisions~\cite{ExHIC:2010gcb,ExHIC:2017smd,Zhang:2020dwn, Wu:2020zbx, Esposito:2020ywk,Braaten:2020iqw} and could continue in $e+A$ collisions with different $A$ at the EIC. Moreover, the impact of the gluon-rich environment at small-$x$ on the production of heavy hybrids must be explored.
\end{itemize}

\section{Opportunities with Nuclei beyond Gluon Saturation}
\label{sec:nuclear}
The investigation of nuclei is important for all aspects of the EIC physics program.  The neutrons within nuclei, in combination with free proton data, permit flavor separation of the partonic substructure of the nucleon.  With free neutron targets unavailable, (polarized) light nuclei function as effective neutron targets.  By measuring coherent exclusive reactions on light nuclei, we can study the tomography of bound nuclear states~\cite{Cano:2003ju,Liuti:2005gi,Liuti:2005qj,Dupre:2015jha,Fucini:2018gso} and connect to their quark and gluon degrees of freedom through the extraction of generalized parton distributions \cite{Berger:2001zb,Cosyn:2018rdm}.  Coherence and saturation effects can be explored by studying interactions of high-energy probes with coherent quark-gluon fields in light and heavy nuclei, see Sec.~\ref{sec:gluon_sat}.  A wealth of nuclear PDF data (see Sec.~\ref{sec:glob_ana}) and diffractive nuclear pdfs~\cite{Armesto:2021fws,Guzey:2020gkk} will become available from inclusive scattering on nuclei.

At the EIC, polarized (spin-1) deuteron beams will be available.  Thus, new aspects of high-energy spin physics can be explored in the 2030s. Spin structure of the spin-1 deuteron is interesting because there are additional structure functions \cite{Frankfurt:1983qs,Hoodbhoy:1988am} associated with its tensor structure that do not exist for spin-1/2 nucleons. These tensor structure functions are appropriate observables for identifying physics beyond a simple bound system of nucleons (non-nucleonic degrees of freedom).  There is parton model sum rule for the leading-twist tensor-polarized structure function $b_1$ \cite{Close:1990zw}, a parametrization for tensor-polarized PDFs \cite{Kumano:2010vz}, and standard theoretical calculations based on convolution models \cite{Cosyn:2017fbo}. Hidden color components could also contribute to $b_1$, together with pion contributions \cite{Miller:2013hla}. Furthermore, transverse-momentum-dependent parton distribution functions (TMDs) and deuteron PDFs were recently obtained up to twist 4 \cite{Bacchetta:2000jk,Kumano:2020ijt,Kumano:2021fem,Kumano:2021xau}, making it possible to investigate spin-1 structure including higher-twist effects. Another interesting deuteron observable is gluon transversity, which corresponds to two units of gluon spin flip~\cite{Jaffe:1989xy,Detmold:2016gpy,Arbuzov:2020cqg}. Consequently it is absent for the nucleon and is sensitive to new non-nucleonic components in nuclei.

Nuclei are used to study the interplay of nuclear interactions and high-energy QCD phenomena.  The EMC effect denotes the medium modification of partonic distributions through residual nucleon-nucleon interactions~\cite{Wang:2021elw,Pace:2022qoj,Kim:2022lng, Guzey:2008fe}.  The EIC will shed light on gluon and polarized EMC effects~\cite{Cloet:2006bq,Wang:2021elw}, the $Q^2$-dependence, and will quantify the EMC effect in processes beyond inclusive scattering~\cite{Liuti:2005qj,Guzey:2008fe,CiofidegliAtti:2010uwl,Fucini:2019xlc,Barry:2023qqh}. At small Bjorken-$x$, medium modifications occur through nuclear shadowing effects~\cite{Brodsky:1989qz,Frankfurt:2011cs,Krelina:2020ipn,Guzey:2022jtv} and coherent power corrections~\cite{Qiu:2003vd,Qiu:2004qk}. The QCD origin of the short-range part of the nuclear force~\cite{Miller:2015tjf,Sargsian:2022rmq}, non-nucleonic degrees of freedom in nuclei~\cite{Bertulani:2022vad} and the role and nature of nuclear short-range correlations~\cite{Hen:2016kwk,West:2020tyo,Hauenstein:2021zql,Lynn:2019vwp,Arrington:2022sov,Cosyn:2021ber} are all topics of great interest.  The phenomenon of color transparency~\cite{Jain:2022xzo} can be studied for various reactions at the EIC. A better theoretical understanding of the transition from nuclear opacity to nuclear transparency is currently needed. 


The role of nuclear structure in EIC physics is twofold.  First, precision nuclear structure input is needed to quantify and understand the role of the nuclear medium.  Second, the EIC enables novel studies of nuclear structure and dynamics~\cite{Miller:2015tjf,Tu:2020ymk,Hauenstein:2021zql,Guzey:2022jtv}.  One possible framework to describe the interaction of a high-energy probe with a nuclear system is light-front quantization.  Nuclear off-shell effects remain finite in the high-energy limit in light-front quantization.  The high-energy scattering event can be separated from the low-energy nuclear structure input which is encoded in objects such as nuclear light-front spectral functions and momentum distributions~\cite{Frankfurt:1981mk, Miller:2000kv, Alessandro:2021cbg}.  The calculation of these nuclear light-front nuclear distributions provide opportunities and demonstrate the need for involvement of the low energy nuclear structure community.  State of the art many-body physics and EFT techniques~\cite{Hammer:2019poc,Maris:2020qne,Tews:2022yfb} can be applied to dedicated calculations of light-front nuclear structure.  The requirement of Poincar\'{e} covariance also places non-trivial constraints on the wave functions and interactions of light nuclei~\cite{Lev:1998qz}.  Lattice QCD calculations of nuclear structure~\cite{Chang:2017eiq,Winter:2017bfs,Detmold:2019ghl,Detmold:2020snb,Sun:2022frr} will provide valuable input to and constraints on non-nucleonic components in nuclei.

Deeper understanding of nuclear properties requires computations to be directly based on first-principles QCD.  First principles QFT shows the necessity of functional differential equations such as the functional Schrodinger equation~\cite{ivanov2020functional,kiefer1994functional,drews2017functional} for the description of bound states. These equations are more general and distinct from the equations used in density functional method \cite{finelli2006relativistic,meng2016relativistic}. Some of these principles on which the study of functional PDEs are built concern the symmetry of underlying functional manifolds, see  Ref.~\cite{hotta2007d}, while others involve insensitivity to the structure of the function space in the form of functional homotopy principles, see Ref.~\cite{ferry2013quantitative,michor2013zoo,hjorth2000classification}. The solution methods for functional equations necessarily involve the development of the theory of higher transcendental functions, extending far beyond current theories of integrable functions (see e.g.  \cite{braaksma2001differential,ecalle1981fonctions,bridgeland2012stability,olive2003resurgence,xu2019closure}). The equations typically encountered in QCD point to a function class that exhibits a dense set of singularities in the analytic continuation of correlation functions, see Refs.~\cite{delabaere1999resurgent,filipuk2009movable} and  \cite{gentile2005degenerate} for the KAM case).  In simple situations \cite{ashok2020aspects}, these functions have a number theory origin~\cite{chakravarty2010parameterizations}. These partial results point to deep connections to transcendental number theory. Truncations of functional equations lead to coupled systems of integral equations with singular kernels, see Ref.~\cite{pachucki1994complete,eides2001theory} for QED case. Methods of solution for such systems must focus on holomorphic aspects distinct from the current focus on geometric methods\cite{constanda2020direct}. The topics of immediate theoretical interest include first principles understanding of the emergence of effective pionic degrees of freedom as carriers of the nuclear force, ab initio spectral calculations beyond effective Hamiltonians, and modeling EIC data for light nuclei. It is essential for both light front quantization and for functional PDE methods to have good control of many body wave functions at spatial infinity (coordinate space methods). This makes it necessary to develop multidimensional versions of the theory of irregular singularities, in particular, extending the 1-dimensional case \cite{harnad2002integrable,toledano2022stokes,bridgeland2013stokes} to multiple dimensions  \cite{barkatou2017formal,katzarkov2015harmonic}. Part of mathematical nuclear physics can be considered as a functional analogy of spectral problems such as spectral theory of variational differential operators. This point of view can be fruitful as it establishes contact with several research directions in spectral theory, operator algebras, functional analysis and dynamical systems. The EIC Theory Alliance can foster interdisciplinary research and mutual enrichment between these areas of science.

The extensive set of forward detectors at the EIC, both in the ePIC detector/interaction region (IR) and a possible second detector/IR), enable precision measurements of specific nuclear breakup channels in the target fragmentation region, also referred to as  ``spectator tagging''.  In the simplest case,  the spectators are treated as nucleons or nuclei.  Such measurements are much more difficult for fixed target experiments than in a collider, where the spectators still have momenta of at least $1/A$ times the ion beam momentum and can be detected in forward detectors.  Compared to measurements where no nuclear fragments are detected and the measurement averages over all initial nuclear configurations, the spectator-tagged measurements provide additional control over the initial nuclear state.  It is possible to extract free neutron structure by using the on-shell extrapolation technique~\cite{Sargsian:2005rm, Cosyn:2020kwu, Jentsch:2021qdp} or to perform a differential study of medium modification effects by varying the kinematics of the detected spectators.  These reactions, however, require much more theoretical input than their inclusive counterparts in the modeling of the reaction mechanisms.  The initial state input requires the calculation of specific overlaps of numerous nuclear states.  One needs to include dynamical descriptions of final-state interactions between the spectators and produced hadrons \cite{Palli:2009it,Strikman:2017koc}.  Current studies have focused on the tagged spectator DIS process (for the deuteron), but extensions are possible for other light nuclei and for more complicated processes including tagged SIDIS, tagged coherent and incoherent exclusive processes such as DVCS and meson production \cite{Goncalves:2015mbf}).

Pre-existing non-nucleonic components in the ground state wave function of nuclei can be explored in tagged measurements.  The observables of interest are correlations between the rapidity of non-nucleonic components in the target fragmentation region and rapidities and transverse momenta of high $p_T$  jets from current fragmentation~\cite{Freese:2014zda}.  Previous studies already demonstrated that it is theoretically possible to handle the final-state interactions of current quark fragments with the residual nucleus~\cite{Cosyn:2011jnm}. In such processes, detected non-nucleon fragments such as $\Delta$ resonances or measuring the relative abundances of kaons and pions will probe unique phenomena such as signatures of hidden-color components in nuclei or the onset of chiral symmetry in nuclear short-range correlations. 

The study of all these topics requires a dedicated and concerted theoretical effort across many areas of expertise (nuclear structure, the hadron/nuclear boundary, high-energy scattering, dynamics of final-state interactions).  This concerted effort is essential to increase our understanding of these phenomena, develop precision theoretical frameworks and provide a meaningful physical interpretation of the EIC data.  This of course cannot happen without the creation of a sizable expert workforce that can guide and support the experimental efforts in nuclear processes at the EIC before and during its operation.
The EIC Theory Alliance will play a critical role by combining expertise from low energy nuclear physicists, QCD practicioners and mathematical physicists and training the next generation of students and postdocs.

\section{Fundamental symmetries at the EIC }

Thanks to its high luminosity and versatile capabilities, the EIC also offers ample opportunities to study physics beyond the Standard Model (BSM). Various scenarios of BSM physics and their experimental signatures at the EIC have already been discussed in the literature, see for example,  Ref.~\cite{AbdulKhalek:2022hcn}. In many cases, the BSM potential of the EIC complements the reach of high-energy physics experiments such as at the LHC, leading to a fruitful synergy between the two programs. In particular, the unique ability of the EIC to polarize both electron and proton beams can provide strong discriminating power for certain observables.
\vspace*{0.3cm} 

{\em Charged Lepton Flavor Violation (CLFV)}
BSM models that explain neutrino masses typically predict new CLFV interactions that can be tested in various experiments including  the EIC~\cite{Barbieri:1995tw,Abada:2007ux,Alonso:2012ji,Cirigliano:2005ck,Raidal:2008jk,deGouvea:2013zba,Calibbi:2017uvl}. 
The $\mu \rightarrow e$ transitions are very strongly constrained by $\mu \rightarrow e \gamma$ and $\mu \rightarrow e$ conversions in nuclei. The constraints on $\tau \rightarrow e$ transitions, however, are much weaker, e.g. ${\rm BR}(\tau \rightarrow e \gamma) < 3.3 \times 10^{-8}$, compared to ${\rm BR}(\mu \rightarrow e \gamma) < 4.2 \times 10^{-13}$. There is, therefore, a competitive
opportunity for the EIC to search for CLFV DIS events with the production of $\tau$ leptons.  
The original study, Ref.~\cite{
Gonderinger:2010yn}, considered specific leptoquark models. More recently, the analysis was repeated in the framework of the Standard Model Effective Field Theory (SMEFT)~\cite{Cirigliano:2021img}, which allows for a straightforward comparison between the EIC, searches of CLFV at the LHC, and low-energy searches in $\tau$ and $B$ meson decays.
To estimate the EIC efficiency,
Ref.~\cite{Cirigliano:2021img} performed preliminary simulations of $e\to \tau$ conversions, with subsequent $\tau$ decays.  More realistic simulations
can be found in Ref.~\cite{Zhang:2022zuz}.
For many SMEFT CLFV parameters, the EIC is as constraining as the LHC. Together, they are complementary to low-energy experiments such as  Belle II. In particular, the EIC will competitively probe CLFV heavy quark couplings that are poorly constrained at low energy. For more details, see
Ref.~\cite{Cirigliano:2021img} and the Snowmass White Papers~\cite{AbdulKhalek:2022hcn,Banerjee:2022xuw}. 
\vspace*{0.3cm}

{\em Complementarity of the EIC with the LHC in exploring the SMEFT}.
%
SMEFT is a powerful theoretical framework for investigating indirect signatures of new interactions and heavy particles in low-energy SM physics. The typical structure of the extension of the SM Lagrangian is $L_{SMEFT} = \sum c_\alpha O_\alpha$, where operators of $O_\alpha$ contain only SM fields. These are typically of dimension $D$ greater than four and thus the couplings $c_\alpha$ (also called Wilson coefficients), which encode the strength of the original BSM interactions, are suppressed by appropriate, $D-4$, powers of the energy scale $\Lambda$ of the new physics. 
 Above $\Lambda$, ultraviolet completion of the theory becomes important and new particles beyond the SM are active degrees of freedom. Below $\Lambda$, phenomena are described by $L_{SMEFT} +L_{SM}$. Considerable effort has been devoted to performing global
analyses of available experimental data within the SMEFT framework (for recent examples of this effort see Refs.~\cite{Ellis:2020unq,Ethier:2021bye}). An issue that arises in these fits is the appearance of
flat directions that occur when the available experimental
measurements cannot disentangle the contributions from different Wilson
coefficients. An example occurs in the semi-leptonic four-fermion sector of the SMEFT (See 
Table 1 in Ref.~\cite{AbdulKhalek:2022hcn}). Low-energy data impose only weak constraints on these operators~\cite{Falkowski:2017pss} while high-energy Drell-Yan data at the LHC probe only a few combinations of the possible Wilson coefficients~\cite{Alte:2018xgc}. 
The EIC, with its ability to polarize both the electron and ion beams, can remove the degeneracies in the semi-leptonic four-fermion Wilson coefficient parameter space that are indistinguishable at the LHC~\cite{Boughezal:2020uwq}. A detailed analysis of the various longitudinal polarization asymmetries that can be measured at the EIC, together with realistic estimates of experimental effects, indicates that ultraviolet scales for semi-leptonic four-fermion Wilson coefficients reaching 4 TeV can be probed at the EIC~\cite{Boughezal:2022pmb}. Furthermore, the flat directions, present after LHC results, can be removed by polarization asymmetry measurements at the EIC. 
This has been demonstrated for the two Wilson coefficients $C_{eu}, C_{lu}$  of certain four-fermion SMEFT couplings~\cite{Boughezal:2022pmb}. While the LHC data are most sensitive  to one linear combination of these parameters, 
simulated EIC data from proton and deuteron runs are sensitive to both linear combinations.

\vspace*{0.3cm}

{\em Nucleon electric dipole moment (EDM)}  A nonzero EDM of a hadron or a nucleus is an unambiguous signal of CP violation necessary to explain the matter-antimatter imbalance of the universe. One of the promising sources of CP violation in certain scenarios of BSM physics is the so-called Weinberg operator~\cite{PhysRevLett.63.2333,Cirigliano:2020msr}
that can be induced in the QCD Lagrangian and contributes to the neutron EDM. A recent study uncovered a novel connection between the hadronic matrix element of the Weinberg operator
and a part of the twist-four corrections in the $g_1$ structure function in polarized DIS~\cite{Hatta:2020riw}. The EIC is capable of constraining such higher twist effects through global analyses due to its large lever arm in $Q^2$.  Another contribution to the nucleon EDM comes from the quark EDM operators whose matrix elements are proportional to the tensor charges of the nucleon. They can be constrained by extracting the transversity distributions in various processes involving a transversely polarized proton 
\cite{Liu:2017olr,Radici:2018iag,Cammarota:2020qcw}. The unprecedented kinematic coverage of the EIC will significantly boost the accuracy of such extractions.  
\vspace*{0.3cm}

{\em Probes of anomalous dipole moments at the EIC}
%
%
Transversely polarized electron or ion beams at the EIC will enable measurements of the beam and target transverse single-spin asymmetries (SSAs). Inclusive DIS SSAs are predicted to be extremely small in the Standard Model. The beam asymmetry is suppressed by a factor of $\alpha_{QED} \times m_e/Q$ where $Q$ the momentum transfer and $m_e$ the electron mass, leading to numerical asymmetries of $10^{-7}$. The target asymmetries are expected to be $\alpha_{QED} \times m_p/Q$, with $m_p$ the proton mass, leading to asymmetries of the order $10^{-4}$. These small SM values make the SSAs a potentially powerful probe of physics beyond the SM. An analysis of both beam and target asymmetries within the SMEFT reveals that they are sensitive probes of the same Wilson coefficients that contribute to electron and quark dipole moments, respectively. The EIC will probe different linear combinations of these Wilson coefficients than measured by experimental determinations of magnetic and electric dipole moments, making EIC measurements complementary to low-energy probes. In particular, the EIC can provide competitive bounds on the magnetic dipole couplings of fermions to the $Z$-boson~\cite{Boughezal:2023ooo}.
\vspace*{0.3cm}

{\em PDF extractions and BSM implications}
As discussed in the section on global analyses of hadron structure, the EIC will yield large
data sets with the potential to stringently constrain PDFs. The resulting PDF improvements,
including EIC-driven refinements in (non)perturbative QCD theory, would increase theoretical
accuracy on SM predictions of TeV-scale processes, for instance, at hadron
colliders like the (HL-)LHC, unlocking a range of BSM investigations in terms of EFT frameworks like SMEFT~\cite{Gao:2022srd}
discussed above or specific UV-complete models. PDF improvements
may underwrite enhanced discovery potential in multiple energy frontier sectors~\cite{AbdulKhalek:2022hcn}, including efforts to measure the
Yukawa couplings of the Higgs, probe possible BSM signatures in high-mass Drell-Yan distributions,
 search for heavy BSM $W'$ and $Z'$ bosons~\cite{Brady_2012}, tails of $p_T$ spectra, or observables such as the forward-backward asymmetry, $A_\mathrm{FB}$~\cite{Ball:2022qtp,Xie:2022tzo}, and to extract
fundamental SM parameters like $M_W$~\cite{Gao:2022wxk}, which can be sensitive to oblique corrections resulting from
BSM contributions.
This sampling of BSM-sensitive observables is variously connected to the $x$ dependence of the PDFs and their
uncertainties~\cite{PDF4LHCWorkingGroup:2022cjn,Hou:2019efy}, including those of the gluon at
small-to-intermediate $x\!\sim\! 10^{-2}$ or the high-$x$ behavior of the nucleon quark sea.
The EIC may also augment extractions of other fundamental SM quantities through broadened global QCD analyses,
including the strong coupling, $\alpha_s$, or heavy quark masses~\cite{Gao:2013wwa}, $m_c$ and $m_b$; as these quantities are generally
fitted alongside PDFs, EIC-based PDF improvements may reciprocally benefit the precision of such QCD parameter
extractions and SM tests.
In addition, the EIC can also facilitate SM tests through measurements of observables like the parity-violating
helicity beam asymmetry on the proton or deuteron, $A^\mathrm{PV}_{p,d}$, which is sensitive to fundamental
electron- and quark-level electroweak couplings, and would extend analogous measurements at JLab
to higher energies while probing the underlying scale dependence.
In addition to electroweak couplings, $A^\mathrm{PV}_{p,d}$ depends on nonperturbative information~\cite{Hobbs:2008mm},
such that control over PDFs (especially for the proton) or partonic charge-symmetry violation and light-nuclear
corrections (for the deuteron) would extend the constraining power of such measurements. 
Finally, precision tests will require a careful and systematic treatment of QED and QCD radiative effects, within a common, factorized framework ~\cite{Liu_2021}. 

\vspace*{0.3cm}

{\it Lattice QCD} 

The search for BSM physics 
at the EIC will greatly benefit from lattice QCD, the best theoretical method available for obtaining the nucleon matrix elements 
of low-energy effective operators in SMEFT with controlled errors 
and steady improvement in precision.    
Conversely, the EIC will pose significant challenges that motivate collaborative efforts within and beyond the lattice QCD community.  
Rare processes at the EIC that are sensitive to BSM physics often probe high-dimension and/or higher-twist operators that are still difficult to calculate using lattice QCD, especially when the operators involve gluon fields and/or mix with lower-dimensional operators under renormalization. For example, it is extremely challenging to evaluate the contribution of  the dimension-six, purely-gluonic Weinberg operator $f_{abc}\tilde{F}^{\mu\nu}_a F^b_{\mu \alpha}F^c_{\nu\beta}$ to the nucleon EDM 
 \cite{Cirigliano:2020msr,Rizik:2020naq}. Equally challenging are dimension-four scalar,  $F^{\mu\nu}_a F_{\mu\nu}^a$, and pseudoscalar, $F^{\mu\nu}_a \tilde{F}_{\mu\nu}^a$~\cite{Syritsyn:2019vvt,Bhattacharya:2021lol}, operators that are of interest in the context of both QCD (origin of hadron mass and spin) and BSM physics (dark matter coupling, CP violation). Novel connections between these operators and  polarized DIS \cite{Tarasov:2021yll} and Deeply Virtual Compton Scattering~\cite{Bhattacharya:2022xxw}  have been recently pointed out and can be explored at the EIC. The quark bilinear operators $\bar{q}_{u,d,s}\Gamma q_{u,d,s}$ with $\Gamma=1,\gamma^\mu,\gamma^\mu\gamma_5,\sigma^{\mu\nu}$ are easier to simulate and still relevant to BSM physics. For example, the strangeness sigma term $\langle N |\bar{s}s|N\rangle$ arises in lepton-number violation processes like $\mu\to e$,  and the tensor charges $\langle N|\bar{q}\sigma^{\mu\nu}q|N\rangle$ are relevant to the quark EDM as mentioned above. High-precision lattice QCD results of these matrix elements can be compared to their phenomenological extraction from the existing and future experimental data~\cite{Courtoy:2015haa,Bhattacharya:2015wna,Gupta:2018qil,Radici:2018iag,Cammarota:2020qcw,FlavourLatticeAveragingGroupFLAG:2021npn,Courtoy:2022kca}. Synergies between lattice QCD, phenomenology, and experiments are  highly valuable, in fact essential, for the development of the EIC community.  The EIC Theory Alliance will encourage and facilitate  all aspects of the involvement of lattice QCD in BSM physics searches at the EIC.  
\vspace*{0.3cm}

In conclusion, BSM physics  at the EIC is a growing subfield that can attract  theorists and experimentalists from many areas of  nuclear and particle physics at both high and low energies.  Lattice QCD and SMEFT are two examples of such cross-cutting disciplines. The EIC theory alliance  can serve as a valuable platform to recruit and promote young talents to tackle these significant challenges.   

\section{Opportunities with AI/ML}
The rapid advances in artificial intelligence (AI) and machine learning (ML) over the last decade have had widespread impact in physics. Various applications of these tools now target almost all facets of QCD theory~\cite{Boehnlein:2021eym}. Naturally, these approaches, and computation more generally, will play an important role in the theory mission of the EIC, supporting goals ranging from efficient data analysis through first-principles theory calculations.

More specifically, ML is a class of tools for optimising the parameters of complex models; ML frameworks can thus be used to describe, model or approximate data or known or unknown functions, and to identify correlations or features in data sets that may be either experimental or simulated data but also ``data" that is the output of theory calculations. In the context of EIC theory, ML applied to data analysis tasks has found applications to global fits of, for example, PDFs or TMDs~\cite{AbdulKhalek:2019mzd}, and to the classification and interpretation of jets and events with the goal of extracting information about underlying physics from simulated or experimental data~\cite{Lai:2020byl,Lee:2022kdn,Pang:2016vdc}. In the context of first-principles theory calculations including applications to lattice QCD, perturbative QCD, EFTs, and nuclear many-body theory, ML can be incorporated as part of more traditional algorithms in such a way that if the ML components are poorly optimised the results are nevertheless correct, but come at a potentially significant computational cost, whereas if the algorithm is well trained or optimised, one might achieve an acceleration which enables otherwise intractable calculations. Examples include the acceleration of sampling processes within lattice QCD~\cite{Kanwar:2020xzo} and the acceleration of multi-loop Feynman integral calculations in perturbative QCD~\cite{Winterhalder:2021ngy}, among many others.

Several applications of AI and ML specific to EIC theory applications have already demonstrated important advantages over more traditional analysis approaches; it will be important to continue to advance and develop this technology as preparation for the EIC continues. For example, since deeply virtual exclusive experiments are characterized by a complex final state with a larger number of kinematic variables and observables, the facility of ML to analyze and interpret these data is particularly important. In this context, it has been found that using a custom deep neural network~\cite{Grigsby:2020auv} uncovers emergent features in the data and learns an accurate approximation of the cross section, outperforming standard baselines. Work has begun to establish frameworks for the benchmarking of both ML and phenomenological analyses of exclusive scattering cross sections~\cite{Almaeen:2022imx}. Critical to this effort is a study of the effects of physics constraints built into ML algorithms; another important aspect is in the treatment of uncertainties~\cite{Almaeen:2022imx}.

As a second definite example, AI/ML algorithms show promise as a means of augmenting extractions of PDFs or related quantum correlation functions, for example, through highly flexible parametrizations of PDFs or TMDs themselves or via enhanced parameter exploration~\cite{AbdulKhalek:2019mzd}. This approach can be applied to experimental data and also to theory calculations; neural network reconstruction of parton distribution functions from lattice QCD correlation functions has been implemented in Ref.~\cite{DelDebbio:2020rgv,Cichy:2019ebf,Gao:2022uhg}. In a recent study, machine learning techniques~\cite{Khan:2022vot} showed a significant advantage for the first determination of gluon helicity distribution from lattice data without relying on any specific model of phenomenological parton distribution functions. 

As such applications continue to develop, it will be critical that  the AI/ML algorithms developed and implemented for EIC physics are benchmarked and tested against classical statistical and computational methods to ensure their proper and effective optimization. This will require an enhanced level of community coordination and support which is well suited to the EIC Theory Alliance. In particular, the Alliance will support AI/ML for QCD by:

\begin{enumerate}
\item
Setting community standards for the benchmarking and stress-testing of novel AI/ML methods while motivating the parallel development of numerical tools to assess the performance of AI/ML algorithms in dedicated theory tasks relevant to the EIC. This work necessarily encompasses the challenge of improving uncertainty quantification obtained through AI/ML methods, with an emphasis on replicability and interpretability.
\item
Cultivating specialized forums devoted to the cross-disciplinary application of AI/ML in distinct areas of QCD theory; these forums might allow expertise developed for hadron collider phenomenology (e.g.,in jet substructure studies) to be extrapolated to EIC use cases and {\it vice versa}.
\item
Developing the workforce of EIC theorists who are expert in AI/ML tools.
\end{enumerate}

Ultimately, the rapid evolution and strengthening of the role of computation in theory, including the development and exploitation of novel algorithmic tools, demands a corresponding evolution of the manner in which the theory community collaborates. At the present time, advances at the intersection of EIC theory and ML remain to be made at every level of complexity and scale, spanning from the application of existing tools through to the development of custom approaches, with computational scales running from a handful of GPU hours through to those requiring exascale hardware. Given the rapidly changing and diverse landscape of this intersection, the strong community coordination, collaboration, and workforce development that the EIC Theory Alliance can provide is essential to fully develop and exploit this opportunity.

\section{Intersections of Quantum Information Science and EIC}


The interface between nuclear theory and quantum information will be an important research thrust of the EIC Theory Alliance. Rapid progress in quantum computing makes it possible to perform quantum simulations of real-time processes in strongly coupled quantum field theories \cite{Bauer:2022hpo}, with a promise to address the real-time dynamics of QCD \cite{Klco:2018kyo,Klco:2019evd,Kharzeev:2020kgc,Ikeda:2020agk,DeJong:2020riy,deJong:2021wsd,Farrell:2022wyt,Farrell:2022vyh}.

Examples of the tight connection between EIC physics and quantum information include the proposed 
relation between structure functions and the entanglement entropy inside the nucleon \cite{Kharzeev:2017qzs,Kharzeev:2021nzh}, and the link between the evolution of parton distributions and momentum space entanglement \cite{Kovner:2018rbf,Armesto:2019mna}. The real-time production of entanglement entropy in deep-inelastic scattering has been evaluated \cite{Zhang:2021hra} by using duality between the effective action of high energy QCD and the ${\rm XXX}$ spin chain with negative spin ~\cite{Lipatov:1995pn}.
Recent experimental analysis of the data from the H1 Collaboration at HERA supports the existence of a link between parton distributions and entanglement \cite{H1:2020zpd,Kharzeev:2021yyf,Hentschinski:2021aux}. 

Quantum computers can potentially provide access to the study of real-time phenomena that cannot be addressed by classical computation. This is because the dimension of Hilbert space spanned by gauge theories is very large and evolving the states in this space in real time requires a huge amount of memory unavailable even in modern supercomputers. 
Since the EIC will study real-time scattering processes, quantum computation can have a transformative effect on the underlying theory. 
First attempts to study DIS on quantum computers can be found in 
\cite{Mueller:2019qqj,Lamm:2019uyc} and a single particle strategy optimized to explore parton dynamics on quantum computers was fleshed out (for a $\phi^4$ scalar field theory) in \cite{Barata:2020jtq}.
Further, quantum computing can advance our understanding of jet evolution inside cold nuclear matter in cases where traditional methods are too difficult to use~\cite{Li:2020uhl,Yao:2022eqm,Barata:2022wim,Barata:2021yri} and address entanglement between  produced jets and the real-time dynamics of vacuum response to their propagation ~\cite{Florio:2023dke}.

Realizing the potential of quantum computing in nuclear physics will, however, require a great deal of work. First, one needs to identify an optimal way to truncate
the infinite-dimensional Hilbert space of quantum field theories to a finite-dimensional one. This
requires both a lattice formulation and replacement of the continuous gauge group by a discrete one. Second, there is a need to develop optimal algorithms for both digital and analog quantum simulations. These algorithms will have to be rooted in the physics needs and theoretical understanding of the process under consideration. 

In particular, one needs an efficient encoding of the relevant QCD degrees of freedom to qubits. Furthermore, protocols for preparing the initial states and for measuring the outcomes of quantum simulations have to be optimized for the processes relevant for the EIC, such as deep-inelastic scattering and jet production. Rigorous control of errors is necessary at each step, especially in the current era of NISC computing \cite{Alexeev:2020xrq}. All of these problems will be studied by the EIC Theory Alliance.

\section{Summary}
\label{NewSummary}

In this document, we presented the objectives and potential impact of an EIC Theory Alliance on both EIC science and the scientific workforce. We reviewed the wide range of scientific opportunities that can be explored at the EIC, and the opportunity that the EIC presents to grow and diversify the field of nuclear theory. We summarized the present state of EIC-related theory and identified the theory progress and workforce development needed for maximizing the impact of EIC science. There are many theoretical challenges that have to be addressed in the coming decades which require close collaboration between various subfields and across many theory groups. Examples include the need for higher-order pQCD calculations and interfacing them with phenomenology as well as lattice QCD for exploring the parton structure of hadrons. Progress is needed to unravel the deeper connections between the TMD and GPD distributions of protons and nuclei with the physics of saturated gluons employing novel jet and heavy flavor observables in deep inelastic scattering. Lattice QCD calculations will play an important role in many aspects of EIC science, including hadron structure, hadron spectroscopy and tests of fundamental symmetries. These calculations in turm will require improvements in algorithms and the development of optimized software as well as access to leadership class computing resources. There are also interesting opportunities at the intersection of EIC science and QIS that should be explored. In addition, AI/ML can aid progress in several areas such as global analyses of hadron structure and improved algorithms for lattice QCD. This many-faceted theory effort within the U.S.\ and internationally will be greatly enhanced by the community-wide effort resulting from the creation of an EIC Theory Alliance.

\section*{Acknowledgements}

A. Baha	Balantekin:	NSF Grant  PHY-2108339.
Joao	Barata: U.S. Department of Energy, Office of Science, National Quantum Information Science Research Centers, Co-design Center for Quantum Advantage (C2QA) under Contract No. DE-SC0012704. U.S. Department of Energy under Contract No. DE-SC0012704.
Carlos	Bertulani:	U.S. DOE grant DE- FG02-08ER41533.
Chiara Bissolotti and Radja	Boughezal: DOE contract DE-AC02-06CH11357.
Nora	Brambilla:	DFG Project-ID 196253076 TRR 110 and the NSFC.
Sino-German: CRC 110 “Symmetries and the Emergence of Structure in QCD”,  by the DFG (Deutsche Forschungs- gemeinschaft, German Research Foundation) Grant No. BR 4058/2-2,  by the DFG cluster of excellence “ORIGINS” under Germany’s Excellence Strategy - EXC-2094 - 390783311.
Vladimir	Braun and Werner Vogelsang:	Deutsche Forschungsgemeinschaft (DFG, German Research Foundation),  grant 409651613.
Duane	Byer:	U.S. Department of Energy under Contract No. DE-FG02-03ER41231.
Jose A. R.	Cembranos:	MICINN (Ministerio de Ciencia e Innovación, Spain) project PID2019-107394GB-I00 (AEI/FEDER, UE).
Martha	Constantinou:	U.S. Department of Energy, Office of Nuclear Physics, Early Career Award, Contract No. DE-SC0020405.
Wim	Cosyn:	National Science Foundation Award No. 2111442.
Aurore	Courtoy:	CONACyT-- Ciencia de Frontera 2019 No.51244 (FORDECYT-PRONACES), UNAM Grant No. DGAPA-PAPIIT IN111222.
J. P. B. C.:  de Melo	INCT-FNA Proc. No. 464898/2014-5, de Melo	CNPq 307131/2020-3, de Melo	FAPESP grants No. 2013/26258-4 and 2019/02923-5.
Debasish Das: The facilities of Saha Institute of Nuclear Physics, Kolkata, India
Adrian	Dumitru: U.S. Department of Energy under Contract No. DE-SC0002307.
Miguel	Echevarria:	Spanish State Agency for Research through grant No. PID2019-106080GB-C21.
Bruno	El-Bennich:	FAPESP grant no. 2018/20218-4.
Michael	Engelhardt:	U.S. Department of Energy, Office of Science, Office of Nuclear Physics, grant No. DE-FG02-96ER40965 and through the TMD Topical Collaboration.
Sean	Fleming: DE-FG02-04ER41338.
Leonard Gamberg:    U.S. Department of Energy, Office of Science under Contract No.~DE-FG02-07ER41460.	
Maria Vittoria	Garzelli:	Bundesministerium f{\"u}r Bildung und Forschung undercontract 05H21GUCCA.
Victor Paulo	Goncalves:	VPG was partially supported by the brazilian support agencies CAPES, CNPq and FAPERGS.
Timothy	Hobbs:	HEP Division of Argonne National Laboratory was supported by the U.S. Department of Energy under contract DE-AC02-06CH11357.
Jamal	Jalilian-Marian:	US DOE Office of Nuclear Physics Grant No. DE-SC0002307 and PSC-CUNY through grant No. 63158-0051.
Chueng-Ryong	Ji:	U.S. Department of Energy Contract No. DE-FG02- 03ER41260.
Hai Tao Li: National Science Foundation of China under grant  No. 12275156.
Zhongbo	Kang:	NSF grant No. PHY-1945471.
Weiyao	Ke:  U.S. Department of Energy, Office of Science, Office of Nuclear Physics through  Contract No. 89233218CNA000001 and by the Laboratory Directed Research and Development Program at LANL.
Spencer	Klein:	DE-AC02-05CH11231 from DOE NP.  This number covers all of LBNL.
Brandon Kriesten: Southeastern Universities Research Association (SURA) Center for Nuclear Femtography

Shunzo 	Kumano: Japan Society for the Promotion of Science (JSPS) Grants-in-Aid  for Scientific Research (KAKENHI) Grant Number 19K03830.
Wai Kin	Lai:	Guangdong Major Project of Basic and Applied Basic Research No. 2020B0301030008, and Science and Technology Program of Guangzhou No. 2019050001.
Heikki	Mäntysaari:	Academy of Finland, the Centre of Excellence in Quark Matter and projects 338263 and 346567.
Nilmani	Mathur:	Department of Atomic Energy, Government of India, under Project Identification No. RTI4002.
Wally	Melnitchouk:	DOE contract No. DE-AC05-06OR23177, under which Jefferson Science Associates, LLC operates Jefferson Lab.
Andreas	Metz:	NSF grant No. PHY-2110472.
Johannes K. L.	Michel:	U.S. Department of Energy under Contract No. DE-SC0011090.
Hamlet Mkrtchyan: The Science Committee of Republic of Armenia, in the frame
of  the research project 21AG-1C028
Asmita 	Mukherjee: 	SERB-MATRICS (MTR/2021/000103).
Pavel	Nadolsky: U.S. Department of Energy under Grant No.~DE-SC0010129 and by the Fermilab URA award, using the resources of the Fermi National Accelerator Laboratory (Fermilab), a U.S. Department of Energy, Office of Science, HEP User Facility. Fermilab is managed by Fermi Research Alliance, LLC (FRA), acting under Contract No. DE-AC02-07CH11359.
Jan	Nemchik: European Regional Development Fund  No. CZ.02.1.01/0.0/0.0/16\_019/0000778 and by the Slovak Funding Agency, Grant No. 2/0020/22.
Emanuele Roberto	Nocera:	Italian Ministry of University and Research (MUR) through the “Rita Levi-Montalcini” Program.
Peter	Petreczky:	 U.S. Department of Energy under Contract No. DE-SC0012704.
Frank	Petriello:	U.S. DOE under the grants DE-FG02-91ER40684 and DE-AC02-06CH11357.
Daniel	Pitonyak:	National Science Foundation under Grant No. PHY-2011763.
Jian-Wei 	Qiu: US Department of Energy (DOE) Contract No. DE-AC05-06OR23177, under which Jefferson Science Associates, LLC operates Jefferson Lab.
Juan	Rojo:	Dutch Research Council (NWO) and Netherlands eScience Center.
Farid Salazar: National Science Foundation under grant No. PHY-1945471,  the UC Southern California Hub, with funding from the UC National Laboratories division of the University of California Office of the President.

Peter	Schweitzer:	National Science Foundation  under Contract No. 1812423 and 2111490.
Ignazio	Scimemi:	Spanish Ministry grant PID2019-106080GB-C21.  European Union Horizon 2020 research and innovation program under grant agreement Num. 824093 (STRONG-2020).
Jorge	Segovia:	Ministerio Espa\~nol de Ciencia e Innovaci\'on under grant No. PID2019-107844GB-C22, Junta de Andaluc\'ia under contract Nos. UHU-1264517 and P18-FR-5057.
Kirill	Semenov-Tian-Shansky:	National Research Foundation of Korea (NRF) under Grants No. NRF-2020R1A2C1007597 and No. NRF-2018R1A6A1A06024970 (Basic Science Research Program); and by the Foundation for the Advancement of Theoretical Physics and Mathematics ``BASIS''.
Rajeev	Singh:	Polish NAWA Bekker program no.: BPN/BEK/2021/1/00342.
Vladimir	Skokov:	U.S. Department of Energy under Contract No. DE-SC0020081.
Iain	Stewart:	U.S.~Department of Energy, Office of Nuclear Physics under DE-SC0011090 and by the Simons Foundation through the Investigator grant 327942.
Sergey Syritsyn: National Science Foundation under CAREER Award PHY-1847893.
Paweł	Sznajder:	Polish National Science Centre with the grant no. 2019/35/D/ST2/00272
Yossathorn	Tawabutr:	The Academy of Finland, the Centre of Excellence in Quark Matter and project 346567; the European Union’s Horizon 2020 research and innovation programme by the European Research Council (ERC, grant agreement No. ERC-2018-ADG-835105 YoctoLHC); the STRONG-2020 project (grant agreement No. 824093). 
Ivan	Vitev:	U.S. Department of Energy under Contract No 89233218CNA000001 and by the Laboratory Directed Research and Development Program at LANL
Ramona Vogt: Office of Nuclear Physics in the U.S. Department of Energy under Contract DE-AC52-07NA27344 at LLNL and the LLNL-LDRD Program. 
Gojko Vujanovic: Natural Sciences and Engineering Research Council of Canada and by the University of Regina President's Tri-Agency Grant Support Program.

Wouter	Waalewijn:	D-ITP consortium, a program of NWO that is funded by the Dutch Ministry of Education, Culture and Science (OCW).
Xiang-Peng	Wang:	 DFG (Deutsche Forschungs- gemeinschaft, German Research Foundation) Grant No. BR 4058/2-2 and DFG cluster of excellence “ORIGINS” under Germany’s Excellence Strategy - EXC-2094 - 390783311
Xiaojun	Yao:	U.S. Department of Energy, Office of Science, Office of Nuclear Physics, InQubator for Quantum Simulation (IQuS) under Award Number DOE (NP) Award DE-SC0020970.
Yong	Zhao:	U.S. Department of Energy, Office of Science, Office of Nuclear Physics through Contract No. DE-AC02-06CH11357, and within the frameworks of Scientific Discovery through Advanced Computing (SciDAC) award \textit{Fundamental Nuclear Physics at the Exascale and Beyond} and the Topical Collaboration in Nuclear Theory \textit{3D quark-gluon structure of hadrons: mass, spin, and tomography}.


\bibliographystyle{apsrev4-1.bst}
\bibliography{ref_tidied.bib}

\begin{thebibliography}{779}%
\makeatletter
\providecommand \@ifxundefined [1]{%
 \@ifx{#1\undefined}
}%
\providecommand \@ifnum [1]{%
 \ifnum #1\expandafter \@firstoftwo
 \else \expandafter \@secondoftwo
 \fi
}%
\providecommand \@ifx [1]{%
 \ifx #1\expandafter \@firstoftwo
 \else \expandafter \@secondoftwo
 \fi
}%
\providecommand \natexlab [1]{#1}%
\providecommand \enquote  [1]{``#1''}%
\providecommand \bibnamefont  [1]{#1}%
\providecommand \bibfnamefont [1]{#1}%
\providecommand \citenamefont [1]{#1}%
\providecommand \href@noop [0]{\@secondoftwo}%
\providecommand \href [0]{\begingroup \@sanitize@url \@href}%
\providecommand \@href[1]{\@@startlink{#1}\@@href}%
\providecommand \@@href[1]{\endgroup#1\@@endlink}%
\providecommand \@sanitize@url [0]{\catcode `\\12\catcode `\$12\catcode
  `\&12\catcode `\#12\catcode `\^12\catcode `\_12\catcode `\%12\relax}%
\providecommand \@@startlink[1]{}%
\providecommand \@@endlink[0]{}%
\providecommand \url  [0]{\begingroup\@sanitize@url \@url }%
\providecommand \@url [1]{\endgroup\@href {#1}{\urlprefix }}%
\providecommand \urlprefix  [0]{URL }%
\providecommand \Eprint [0]{\href }%
\providecommand \doibase [0]{http://dx.doi.org/}%
\providecommand \selectlanguage [0]{\@gobble}%
\providecommand \bibinfo  [0]{\@secondoftwo}%
\providecommand \bibfield  [0]{\@secondoftwo}%
\providecommand \translation [1]{[#1]}%
\providecommand \BibitemOpen [0]{}%
\providecommand \bibitemStop [0]{}%
\providecommand \bibitemNoStop [0]{.\EOS\space}%
\providecommand \EOS [0]{\spacefactor3000\relax}%
\providecommand \BibitemShut  [1]{\csname bibitem#1\endcsname}%
\let\auto@bib@innerbib\@empty
\bibitem [{\citenamefont {{ Ani Aprahamian {\it et al.}, Reaching for the
  horizon: The 2015 long range plan for nuclear science 2015
  (http://science.energy.gov/~/media/np/nsac/pdf/2015LRP/2015\_LRPNS\_091815.pdf)}}(2015)}]{NSAC}%
  \BibitemOpen
  \bibfield  {author} {\bibinfo {author} {\bibnamefont {{ Ani Aprahamian {\it
  et al.}, Reaching for the horizon: The 2015 long range plan for nuclear
  science 2015
  (http://science.energy.gov/~/media/np/nsac/pdf/2015LRP/2015\_LRPNS\_091815.pdf)}}},\
  }\href@noop {} {\  (\bibinfo {year} {2015})}\BibitemShut {NoStop}%
\bibitem [{\citenamefont {of~Sciences Engineering~Medicine}(2018)}]{NAP25171}%
  \BibitemOpen
  \bibfield  {author} {\bibinfo {author} {\bibfnamefont {N.~A.}\ \bibnamefont
  {of~Sciences Engineering~Medicine}},\ }\href
  {https://www.nap.edu/catalog/25171/an-assessment-of-us-based-electron-ion-collider-science}
  {\emph {\bibinfo {title} {{An Assessment of U.S.-Based Electron-Ion Collider
  Science}}}}\ (\bibinfo {year} {2018})\BibitemShut {NoStop}%
\bibitem [{\citenamefont {Society}()}]{APS-stats}%
  \BibitemOpen
  \bibfield  {author} {\bibinfo {author} {\bibfnamefont {A.~P.}\ \bibnamefont
  {Society}},\ }\href {https://www.aps.org/programs/education/statistics/}
  {\enquote {\bibinfo {title} {Physics graphs and statistics},}\ }\BibitemShut
  {NoStop}%
\bibitem [{\citenamefont {Morgan}\ \emph {et~al.}(2022)\citenamefont {Morgan},
  \citenamefont {LaBerge}, \citenamefont {Larremore}, \citenamefont {Galesic},
  \citenamefont {Brand},\ and\ \citenamefont {Clauset}}]{Morgan2022}%
  \BibitemOpen
  \bibfield  {author} {\bibinfo {author} {\bibfnamefont {A.~C.}\ \bibnamefont
  {Morgan}}, \bibinfo {author} {\bibfnamefont {N.}~\bibnamefont {LaBerge}},
  \bibinfo {author} {\bibfnamefont {D.~B.}\ \bibnamefont {Larremore}}, \bibinfo
  {author} {\bibfnamefont {M.}~\bibnamefont {Galesic}}, \bibinfo {author}
  {\bibfnamefont {J.~E.}\ \bibnamefont {Brand}}, \ and\ \bibinfo {author}
  {\bibfnamefont {A.}~\bibnamefont {Clauset}},\ }\href
  {https://www.nature.com/articles/s41562-022-01425-4} {\bibfield  {journal}
  {\bibinfo  {journal} {Nature Human Behaviour}\ }\textbf {\bibinfo {volume}
  {6}} (\bibinfo {year} {2022})}\BibitemShut {NoStop}%
\bibitem [{\citenamefont {Ji}(1997{\natexlab{a}})}]{Ji:1996ek}%
  \BibitemOpen
  \bibfield  {author} {\bibinfo {author} {\bibfnamefont {X.-D.}\ \bibnamefont
  {Ji}},\ }\href {\doibase 10.1103/PhysRevLett.78.610} {\bibfield  {journal}
  {\bibinfo  {journal} {Phys. Rev. Lett.}\ }\textbf {\bibinfo {volume} {78}},\
  \bibinfo {pages} {610} (\bibinfo {year} {1997}{\natexlab{a}})},\ \Eprint
  {http://arxiv.org/abs/hep-ph/9603249} {arXiv:hep-ph/9603249} \BibitemShut
  {NoStop}%
\bibitem [{\citenamefont {Radyushkin}(1997)}]{Radyushkin:1997ki}%
  \BibitemOpen
  \bibfield  {author} {\bibinfo {author} {\bibfnamefont {A.~V.}\ \bibnamefont
  {Radyushkin}},\ }\href {\doibase 10.1103/PhysRevD.56.5524} {\bibfield
  {journal} {\bibinfo  {journal} {Phys. Rev. D}\ }\textbf {\bibinfo {volume}
  {56}},\ \bibinfo {pages} {5524} (\bibinfo {year} {1997})},\ \Eprint
  {http://arxiv.org/abs/hep-ph/9704207} {arXiv:hep-ph/9704207} \BibitemShut
  {NoStop}%
\bibitem [{\citenamefont {M\"uller}\ \emph {et~al.}(1994)\citenamefont
  {M\"uller}, \citenamefont {Robaschik}, \citenamefont {Geyer}, \citenamefont
  {Dittes},\ and\ \citenamefont {Ho\v{r}ej\v{s}i}}]{Muller:1994ses}%
  \BibitemOpen
  \bibfield  {author} {\bibinfo {author} {\bibfnamefont {D.}~\bibnamefont
  {M\"uller}}, \bibinfo {author} {\bibfnamefont {D.}~\bibnamefont {Robaschik}},
  \bibinfo {author} {\bibfnamefont {B.}~\bibnamefont {Geyer}}, \bibinfo
  {author} {\bibfnamefont {F.~M.}\ \bibnamefont {Dittes}}, \ and\ \bibinfo
  {author} {\bibfnamefont {J.}~\bibnamefont {Ho\v{r}ej\v{s}i}},\ }\href
  {\doibase 10.1002/prop.2190420202} {\bibfield  {journal} {\bibinfo  {journal}
  {Fortsch. Phys.}\ }\textbf {\bibinfo {volume} {42}},\ \bibinfo {pages} {101}
  (\bibinfo {year} {1994})},\ \Eprint {http://arxiv.org/abs/hep-ph/9812448}
  {arXiv:hep-ph/9812448} \BibitemShut {NoStop}%
\bibitem [{\citenamefont {Burkardt}(2000)}]{Burkardt:2000za}%
  \BibitemOpen
  \bibfield  {author} {\bibinfo {author} {\bibfnamefont {M.}~\bibnamefont
  {Burkardt}},\ }\href {\doibase 10.1103/PhysRevD.62.071503} {\bibfield
  {journal} {\bibinfo  {journal} {Phys. Rev. D}\ }\textbf {\bibinfo {volume}
  {62}},\ \bibinfo {pages} {071503} (\bibinfo {year} {2000})},\ \bibinfo {note}
  {[Erratum: Phys.Rev.D 66, 119903 (2002)]},\ \Eprint
  {http://arxiv.org/abs/hep-ph/0005108} {arXiv:hep-ph/0005108} \BibitemShut
  {NoStop}%
\bibitem [{\citenamefont {Ralston}\ and\ \citenamefont
  {Pire}(2002)}]{Ralston:2001xs}%
  \BibitemOpen
  \bibfield  {author} {\bibinfo {author} {\bibfnamefont {J.~P.}\ \bibnamefont
  {Ralston}}\ and\ \bibinfo {author} {\bibfnamefont {B.}~\bibnamefont {Pire}},\
  }\href {\doibase 10.1103/PhysRevD.66.111501} {\bibfield  {journal} {\bibinfo
  {journal} {Phys. Rev. D}\ }\textbf {\bibinfo {volume} {66}},\ \bibinfo
  {pages} {111501} (\bibinfo {year} {2002})},\ \Eprint
  {http://arxiv.org/abs/hep-ph/0110075} {arXiv:hep-ph/0110075} \BibitemShut
  {NoStop}%
\bibitem [{\citenamefont {Diehl}(2002)}]{Diehl:2002he}%
  \BibitemOpen
  \bibfield  {author} {\bibinfo {author} {\bibfnamefont {M.}~\bibnamefont
  {Diehl}},\ }\href {\doibase 10.1007/s10052-002-1016-9} {\bibfield  {journal}
  {\bibinfo  {journal} {Eur. Phys. J. C}\ }\textbf {\bibinfo {volume} {25}},\
  \bibinfo {pages} {223} (\bibinfo {year} {2002})},\ \bibinfo {note} {[Erratum:
  Eur.Phys.J.C 31, 277--278 (2003)]},\ \Eprint
  {http://arxiv.org/abs/hep-ph/0205208} {arXiv:hep-ph/0205208} \BibitemShut
  {NoStop}%
\bibitem [{\citenamefont {Burkardt}(2003)}]{Burkardt:2002hr}%
  \BibitemOpen
  \bibfield  {author} {\bibinfo {author} {\bibfnamefont {M.}~\bibnamefont
  {Burkardt}},\ }\href {\doibase 10.1142/S0217751X03012370} {\bibfield
  {journal} {\bibinfo  {journal} {Int. J. Mod. Phys. A}\ }\textbf {\bibinfo
  {volume} {18}},\ \bibinfo {pages} {173} (\bibinfo {year} {2003})},\ \Eprint
  {http://arxiv.org/abs/hep-ph/0207047} {arXiv:hep-ph/0207047} \BibitemShut
  {NoStop}%
\bibitem [{\citenamefont {Goeke}\ \emph {et~al.}(2001)\citenamefont {Goeke},
  \citenamefont {Polyakov},\ and\ \citenamefont
  {Vanderhaeghen}}]{Goeke:2001tz}%
  \BibitemOpen
  \bibfield  {author} {\bibinfo {author} {\bibfnamefont {K.}~\bibnamefont
  {Goeke}}, \bibinfo {author} {\bibfnamefont {M.~V.}\ \bibnamefont {Polyakov}},
  \ and\ \bibinfo {author} {\bibfnamefont {M.}~\bibnamefont {Vanderhaeghen}},\
  }\href {\doibase 10.1016/S0146-6410(01)00158-2} {\bibfield  {journal}
  {\bibinfo  {journal} {Prog. Part. Nucl. Phys.}\ }\textbf {\bibinfo {volume}
  {47}},\ \bibinfo {pages} {401} (\bibinfo {year} {2001})},\ \Eprint
  {http://arxiv.org/abs/hep-ph/0106012} {arXiv:hep-ph/0106012} \BibitemShut
  {NoStop}%
\bibitem [{\citenamefont {Diehl}(2003)}]{Diehl:2003ny}%
  \BibitemOpen
  \bibfield  {author} {\bibinfo {author} {\bibfnamefont {M.}~\bibnamefont
  {Diehl}},\ }\href {\doibase 10.1016/j.physrep.2003.08.002} {\bibfield
  {journal} {\bibinfo  {journal} {Phys. Rept.}\ }\textbf {\bibinfo {volume}
  {388}},\ \bibinfo {pages} {41} (\bibinfo {year} {2003})},\ \Eprint
  {http://arxiv.org/abs/hep-ph/0307382} {arXiv:hep-ph/0307382} \BibitemShut
  {NoStop}%
\bibitem [{\citenamefont {Belitsky}\ and\ \citenamefont
  {Radyushkin}(2005)}]{Belitsky:2005qn}%
  \BibitemOpen
  \bibfield  {author} {\bibinfo {author} {\bibfnamefont {A.~V.}\ \bibnamefont
  {Belitsky}}\ and\ \bibinfo {author} {\bibfnamefont {A.~V.}\ \bibnamefont
  {Radyushkin}},\ }\href {\doibase 10.1016/j.physrep.2005.06.002} {\bibfield
  {journal} {\bibinfo  {journal} {Phys. Rept.}\ }\textbf {\bibinfo {volume}
  {418}},\ \bibinfo {pages} {1} (\bibinfo {year} {2005})},\ \Eprint
  {http://arxiv.org/abs/hep-ph/0504030} {arXiv:hep-ph/0504030} \BibitemShut
  {NoStop}%
\bibitem [{\citenamefont {Goeke}\ \emph {et~al.}(2007)\citenamefont {Goeke}
  \emph {et~al.}}]{Goeke:2007fp}%
  \BibitemOpen
  \bibfield  {author} {\bibinfo {author} {\bibfnamefont {K.}~\bibnamefont
  {Goeke}} \emph {et~al.},\ }\href {\doibase 10.1103/PhysRevD.75.094021}
  {\bibfield  {journal} {\bibinfo  {journal} {Phys. Rev. D}\ }\textbf {\bibinfo
  {volume} {75}},\ \bibinfo {pages} {094021} (\bibinfo {year} {2007})},\
  \Eprint {http://arxiv.org/abs/hep-ph/0702030} {arXiv:hep-ph/0702030}
  \BibitemShut {NoStop}%
\bibitem [{\citenamefont {Boffi}\ and\ \citenamefont
  {Pasquini}(2007)}]{Boffi:2007yc}%
  \BibitemOpen
  \bibfield  {author} {\bibinfo {author} {\bibfnamefont {S.}~\bibnamefont
  {Boffi}}\ and\ \bibinfo {author} {\bibfnamefont {B.}~\bibnamefont
  {Pasquini}},\ }\href {\doibase 10.1393/ncr/i2007-10025-7} {\bibfield
  {journal} {\bibinfo  {journal} {Riv. Nuovo Cim.}\ }\textbf {\bibinfo {volume}
  {30}},\ \bibinfo {pages} {387} (\bibinfo {year} {2007})},\ \Eprint
  {http://arxiv.org/abs/0711.2625} {arXiv:0711.2625 [hep-ph]} \BibitemShut
  {NoStop}%
\bibitem [{\citenamefont {Guidal}\ \emph {et~al.}(2013)\citenamefont {Guidal},
  \citenamefont {Moutarde},\ and\ \citenamefont
  {Vanderhaeghen}}]{Guidal:2013rya}%
  \BibitemOpen
  \bibfield  {author} {\bibinfo {author} {\bibfnamefont {M.}~\bibnamefont
  {Guidal}}, \bibinfo {author} {\bibfnamefont {H.}~\bibnamefont {Moutarde}}, \
  and\ \bibinfo {author} {\bibfnamefont {M.}~\bibnamefont {Vanderhaeghen}},\
  }\href {\doibase 10.1088/0034-4885/76/6/066202} {\bibfield  {journal}
  {\bibinfo  {journal} {Rept. Prog. Phys.}\ }\textbf {\bibinfo {volume} {76}},\
  \bibinfo {pages} {066202} (\bibinfo {year} {2013})},\ \Eprint
  {http://arxiv.org/abs/1303.6600} {arXiv:1303.6600 [hep-ph]} \BibitemShut
  {NoStop}%
\bibitem [{\citenamefont {Mueller}(2014)}]{Mueller:2014hsa}%
  \BibitemOpen
  \bibfield  {author} {\bibinfo {author} {\bibfnamefont {D.}~\bibnamefont
  {Mueller}},\ }\href {\doibase 10.1007/s00601-014-0894-3} {\bibfield
  {journal} {\bibinfo  {journal} {Few Body Syst.}\ }\textbf {\bibinfo {volume}
  {55}},\ \bibinfo {pages} {317} (\bibinfo {year} {2014})},\ \Eprint
  {http://arxiv.org/abs/1405.2817} {arXiv:1405.2817 [hep-ph]} \BibitemShut
  {NoStop}%
\bibitem [{\citenamefont {Kumericki}\ \emph {et~al.}(2016)\citenamefont
  {Kumericki}, \citenamefont {Liuti},\ and\ \citenamefont
  {Moutarde}}]{Kumericki:2016ehc}%
  \BibitemOpen
  \bibfield  {author} {\bibinfo {author} {\bibfnamefont {K.}~\bibnamefont
  {Kumericki}}, \bibinfo {author} {\bibfnamefont {S.}~\bibnamefont {Liuti}}, \
  and\ \bibinfo {author} {\bibfnamefont {H.}~\bibnamefont {Moutarde}},\ }\href
  {\doibase 10.1140/epja/i2016-16157-3} {\bibfield  {journal} {\bibinfo
  {journal} {Eur. Phys. J. A}\ }\textbf {\bibinfo {volume} {52}},\ \bibinfo
  {pages} {157} (\bibinfo {year} {2016})},\ \Eprint
  {http://arxiv.org/abs/1602.02763} {arXiv:1602.02763 [hep-ph]} \BibitemShut
  {NoStop}%
\bibitem [{\citenamefont {Lorc\'e}(2021)}]{Lorce:2021gxs}%
  \BibitemOpen
  \bibfield  {author} {\bibinfo {author} {\bibfnamefont {C.}~\bibnamefont
  {Lorc\'e}},\ }\href {\doibase 10.1140/epjc/s10052-021-09207-4} {\bibfield
  {journal} {\bibinfo  {journal} {Eur. Phys. J. C}\ }\textbf {\bibinfo {volume}
  {81}},\ \bibinfo {pages} {413} (\bibinfo {year} {2021})},\ \Eprint
  {http://arxiv.org/abs/2103.10100} {arXiv:2103.10100 [hep-ph]} \BibitemShut
  {NoStop}%
\bibitem [{\citenamefont {Polyakov}\ and\ \citenamefont
  {Shuvaev}(2002)}]{Polyakov:2002wz}%
  \BibitemOpen
  \bibfield  {author} {\bibinfo {author} {\bibfnamefont {M.~V.}\ \bibnamefont
  {Polyakov}}\ and\ \bibinfo {author} {\bibfnamefont {A.~G.}\ \bibnamefont
  {Shuvaev}},\ }\href@noop {} {\  (\bibinfo {year} {2002})},\ \Eprint
  {http://arxiv.org/abs/hep-ph/0207153} {arXiv:hep-ph/0207153} \BibitemShut
  {NoStop}%
\bibitem [{\citenamefont {Polyakov}(2003)}]{Polyakov:2002yz}%
  \BibitemOpen
  \bibfield  {author} {\bibinfo {author} {\bibfnamefont {M.~V.}\ \bibnamefont
  {Polyakov}},\ }\href {\doibase 10.1016/S0370-2693(03)00036-4} {\bibfield
  {journal} {\bibinfo  {journal} {Phys. Lett. B}\ }\textbf {\bibinfo {volume}
  {555}},\ \bibinfo {pages} {57} (\bibinfo {year} {2003})},\ \Eprint
  {http://arxiv.org/abs/hep-ph/0210165} {arXiv:hep-ph/0210165} \BibitemShut
  {NoStop}%
\bibitem [{\citenamefont {Polyakov}\ and\ \citenamefont
  {Schweitzer}(2018)}]{Polyakov:2018zvc}%
  \BibitemOpen
  \bibfield  {author} {\bibinfo {author} {\bibfnamefont {M.~V.}\ \bibnamefont
  {Polyakov}}\ and\ \bibinfo {author} {\bibfnamefont {P.}~\bibnamefont
  {Schweitzer}},\ }\href {\doibase 10.1142/S0217751X18300259} {\bibfield
  {journal} {\bibinfo  {journal} {Int. J. Mod. Phys. A}\ }\textbf {\bibinfo
  {volume} {33}},\ \bibinfo {pages} {1830025} (\bibinfo {year} {2018})},\
  \Eprint {http://arxiv.org/abs/1805.06596} {arXiv:1805.06596 [hep-ph]}
  \BibitemShut {NoStop}%
\bibitem [{\citenamefont {Lorc\'e}\ \emph {et~al.}(2019)\citenamefont
  {Lorc\'e}, \citenamefont {Moutarde},\ and\ \citenamefont
  {Trawi\'nski}}]{Lorce:2018egm}%
  \BibitemOpen
  \bibfield  {author} {\bibinfo {author} {\bibfnamefont {C.}~\bibnamefont
  {Lorc\'e}}, \bibinfo {author} {\bibfnamefont {H.}~\bibnamefont {Moutarde}}, \
  and\ \bibinfo {author} {\bibfnamefont {A.~P.}\ \bibnamefont {Trawi\'nski}},\
  }\href {\doibase 10.1140/epjc/s10052-019-6572-3} {\bibfield  {journal}
  {\bibinfo  {journal} {Eur. Phys. J. C}\ }\textbf {\bibinfo {volume} {79}},\
  \bibinfo {pages} {89} (\bibinfo {year} {2019})},\ \Eprint
  {http://arxiv.org/abs/1810.09837} {arXiv:1810.09837 [hep-ph]} \BibitemShut
  {NoStop}%
\bibitem [{\citenamefont {Ji}\ and\ \citenamefont {Liu}(2022)}]{Ji:2021mfb}%
  \BibitemOpen
  \bibfield  {author} {\bibinfo {author} {\bibfnamefont {X.}~\bibnamefont
  {Ji}}\ and\ \bibinfo {author} {\bibfnamefont {Y.}~\bibnamefont {Liu}},\
  }\href {\doibase 10.1103/PhysRevD.106.034028} {\bibfield  {journal} {\bibinfo
   {journal} {Phys. Rev. D}\ }\textbf {\bibinfo {volume} {106}},\ \bibinfo
  {pages} {034028} (\bibinfo {year} {2022})},\ \Eprint
  {http://arxiv.org/abs/2110.14781} {arXiv:2110.14781 [hep-ph]} \BibitemShut
  {NoStop}%
\bibitem [{\citenamefont {Burkert}\ \emph {et~al.}(2023)\citenamefont
  {Burkert}, \citenamefont {Elouadrhiri}, \citenamefont {Girod}, \citenamefont
  {Lorc\'e}, \citenamefont {Schweitzer},\ and\ \citenamefont
  {Shanahan}}]{Burkert:2023wzr}%
  \BibitemOpen
  \bibfield  {author} {\bibinfo {author} {\bibfnamefont {V.~D.}\ \bibnamefont
  {Burkert}}, \bibinfo {author} {\bibfnamefont {L.}~\bibnamefont
  {Elouadrhiri}}, \bibinfo {author} {\bibfnamefont {F.~X.}\ \bibnamefont
  {Girod}}, \bibinfo {author} {\bibfnamefont {C.}~\bibnamefont {Lorc\'e}},
  \bibinfo {author} {\bibfnamefont {P.}~\bibnamefont {Schweitzer}}, \ and\
  \bibinfo {author} {\bibfnamefont {P.~E.}\ \bibnamefont {Shanahan}},\
  }\href@noop {} {\  (\bibinfo {year} {2023})},\ \Eprint
  {http://arxiv.org/abs/2303.08347} {arXiv:2303.08347 [hep-ph]} \BibitemShut
  {NoStop}%
\bibitem [{\citenamefont {Radyushkin}(1996{\natexlab{a}})}]{Radyushkin:1996nd}%
  \BibitemOpen
  \bibfield  {author} {\bibinfo {author} {\bibfnamefont {A.~V.}\ \bibnamefont
  {Radyushkin}},\ }\href {\doibase 10.1016/0370-2693(96)00528-X} {\bibfield
  {journal} {\bibinfo  {journal} {Phys. Lett. B}\ }\textbf {\bibinfo {volume}
  {380}},\ \bibinfo {pages} {417} (\bibinfo {year} {1996}{\natexlab{a}})},\
  \Eprint {http://arxiv.org/abs/hep-ph/9604317} {arXiv:hep-ph/9604317}
  \BibitemShut {NoStop}%
\bibitem [{\citenamefont {Ji}(1997{\natexlab{b}})}]{Ji:1996nm}%
  \BibitemOpen
  \bibfield  {author} {\bibinfo {author} {\bibfnamefont {X.-D.}\ \bibnamefont
  {Ji}},\ }\href {\doibase 10.1103/PhysRevD.55.7114} {\bibfield  {journal}
  {\bibinfo  {journal} {Phys. Rev. D}\ }\textbf {\bibinfo {volume} {55}},\
  \bibinfo {pages} {7114} (\bibinfo {year} {1997}{\natexlab{b}})},\ \Eprint
  {http://arxiv.org/abs/hep-ph/9609381} {arXiv:hep-ph/9609381} \BibitemShut
  {NoStop}%
\bibitem [{\citenamefont {Collins}\ and\ \citenamefont
  {Freund}(1999)}]{Collins:1998be}%
  \BibitemOpen
  \bibfield  {author} {\bibinfo {author} {\bibfnamefont {J.~C.}\ \bibnamefont
  {Collins}}\ and\ \bibinfo {author} {\bibfnamefont {A.}~\bibnamefont
  {Freund}},\ }\href {\doibase 10.1103/PhysRevD.59.074009} {\bibfield
  {journal} {\bibinfo  {journal} {Phys. Rev. D}\ }\textbf {\bibinfo {volume}
  {59}},\ \bibinfo {pages} {074009} (\bibinfo {year} {1999})},\ \Eprint
  {http://arxiv.org/abs/hep-ph/9801262} {arXiv:hep-ph/9801262} \BibitemShut
  {NoStop}%
\bibitem [{\citenamefont {Radyushkin}(1996{\natexlab{b}})}]{Radyushkin:1996ru}%
  \BibitemOpen
  \bibfield  {author} {\bibinfo {author} {\bibfnamefont {A.~V.}\ \bibnamefont
  {Radyushkin}},\ }\href {\doibase 10.1016/0370-2693(96)00844-1} {\bibfield
  {journal} {\bibinfo  {journal} {Phys. Lett. B}\ }\textbf {\bibinfo {volume}
  {385}},\ \bibinfo {pages} {333} (\bibinfo {year} {1996}{\natexlab{b}})},\
  \Eprint {http://arxiv.org/abs/hep-ph/9605431} {arXiv:hep-ph/9605431}
  \BibitemShut {NoStop}%
\bibitem [{\citenamefont {Collins}\ \emph {et~al.}(1997)\citenamefont
  {Collins}, \citenamefont {Frankfurt},\ and\ \citenamefont
  {Strikman}}]{Collins:1996fb}%
  \BibitemOpen
  \bibfield  {author} {\bibinfo {author} {\bibfnamefont {J.~C.}\ \bibnamefont
  {Collins}}, \bibinfo {author} {\bibfnamefont {L.}~\bibnamefont {Frankfurt}},
  \ and\ \bibinfo {author} {\bibfnamefont {M.}~\bibnamefont {Strikman}},\
  }\href {\doibase 10.1103/PhysRevD.56.2982} {\bibfield  {journal} {\bibinfo
  {journal} {Phys. Rev. D}\ }\textbf {\bibinfo {volume} {56}},\ \bibinfo
  {pages} {2982} (\bibinfo {year} {1997})},\ \Eprint
  {http://arxiv.org/abs/hep-ph/9611433} {arXiv:hep-ph/9611433} \BibitemShut
  {NoStop}%
\bibitem [{\citenamefont {Mankiewicz}\ \emph {et~al.}(1998)\citenamefont
  {Mankiewicz}, \citenamefont {Piller},\ and\ \citenamefont
  {Weigl}}]{Mankiewicz:1997uy}%
  \BibitemOpen
  \bibfield  {author} {\bibinfo {author} {\bibfnamefont {L.}~\bibnamefont
  {Mankiewicz}}, \bibinfo {author} {\bibfnamefont {G.}~\bibnamefont {Piller}},
  \ and\ \bibinfo {author} {\bibfnamefont {T.}~\bibnamefont {Weigl}},\ }\href
  {\doibase 10.1007/s100520050253} {\bibfield  {journal} {\bibinfo  {journal}
  {Eur. Phys. J. C}\ }\textbf {\bibinfo {volume} {5}},\ \bibinfo {pages} {119}
  (\bibinfo {year} {1998})},\ \Eprint {http://arxiv.org/abs/hep-ph/9711227}
  {arXiv:hep-ph/9711227} \BibitemShut {NoStop}%
\bibitem [{\citenamefont {Abdul~Khalek}\ \emph
  {et~al.}(2022{\natexlab{a}})\citenamefont {Abdul~Khalek} \emph
  {et~al.}}]{AbdulKhalek:2021gbh}%
  \BibitemOpen
  \bibfield  {author} {\bibinfo {author} {\bibfnamefont {R.}~\bibnamefont
  {Abdul~Khalek}} \emph {et~al.},\ }\href {\doibase
  10.1016/j.nuclphysa.2022.122447} {\bibfield  {journal} {\bibinfo  {journal}
  {Nucl. Phys. A}\ }\textbf {\bibinfo {volume} {1026}},\ \bibinfo {pages}
  {122447} (\bibinfo {year} {2022}{\natexlab{a}})},\ \Eprint
  {http://arxiv.org/abs/2103.05419} {arXiv:2103.05419 [physics.ins-det]}
  \BibitemShut {NoStop}%
\bibitem [{\citenamefont {Constantinou}\ \emph {et~al.}(2021)\citenamefont
  {Constantinou} \emph {et~al.}}]{Constantinou:2020hdm}%
  \BibitemOpen
  \bibfield  {author} {\bibinfo {author} {\bibfnamefont {M.}~\bibnamefont
  {Constantinou}} \emph {et~al.},\ }\href {\doibase 10.1016/j.ppnp.2021.103908}
  {\bibfield  {journal} {\bibinfo  {journal} {Prog. Part. Nucl. Phys.}\
  }\textbf {\bibinfo {volume} {121}},\ \bibinfo {pages} {103908} (\bibinfo
  {year} {2021})},\ \Eprint {http://arxiv.org/abs/2006.08636} {arXiv:2006.08636
  [hep-ph]} \BibitemShut {NoStop}%
\bibitem [{\citenamefont {Liu}\ and\ \citenamefont {Dong}(1994)}]{Liu:1993cv}%
  \BibitemOpen
  \bibfield  {author} {\bibinfo {author} {\bibfnamefont {K.-F.}\ \bibnamefont
  {Liu}}\ and\ \bibinfo {author} {\bibfnamefont {S.-J.}\ \bibnamefont {Dong}},\
  }\href {\doibase 10.1103/PhysRevLett.72.1790} {\bibfield  {journal} {\bibinfo
   {journal} {Phys. Rev. Lett.}\ }\textbf {\bibinfo {volume} {72}},\ \bibinfo
  {pages} {1790} (\bibinfo {year} {1994})},\ \Eprint
  {http://arxiv.org/abs/hep-ph/9306299} {arXiv:hep-ph/9306299} \BibitemShut
  {NoStop}%
\bibitem [{\citenamefont {Ji}(2013)}]{Ji:2013dva}%
  \BibitemOpen
  \bibfield  {author} {\bibinfo {author} {\bibfnamefont {X.}~\bibnamefont
  {Ji}},\ }\href {\doibase 10.1103/PhysRevLett.110.262002} {\bibfield
  {journal} {\bibinfo  {journal} {Phys. Rev. Lett.}\ }\textbf {\bibinfo
  {volume} {110}},\ \bibinfo {pages} {262002} (\bibinfo {year} {2013})},\
  \Eprint {http://arxiv.org/abs/1305.1539} {arXiv:1305.1539 [hep-ph]}
  \BibitemShut {NoStop}%
\bibitem [{\citenamefont {Ji}(2014)}]{Ji:2014gla}%
  \BibitemOpen
  \bibfield  {author} {\bibinfo {author} {\bibfnamefont {X.}~\bibnamefont
  {Ji}},\ }\href {\doibase 10.1007/s11433-014-5492-3} {\bibfield  {journal}
  {\bibinfo  {journal} {Sci. China Phys. Mech. Astron.}\ }\textbf {\bibinfo
  {volume} {57}},\ \bibinfo {pages} {1407} (\bibinfo {year} {2014})},\ \Eprint
  {http://arxiv.org/abs/1404.6680} {arXiv:1404.6680 [hep-ph]} \BibitemShut
  {NoStop}%
\bibitem [{\citenamefont {Detmold}\ and\ \citenamefont
  {Lin}(2006)}]{Detmold:2005gg}%
  \BibitemOpen
  \bibfield  {author} {\bibinfo {author} {\bibfnamefont {W.}~\bibnamefont
  {Detmold}}\ and\ \bibinfo {author} {\bibfnamefont {C.~J.~D.}\ \bibnamefont
  {Lin}},\ }\href {\doibase 10.1103/PhysRevD.73.014501} {\bibfield  {journal}
  {\bibinfo  {journal} {Phys. Rev. D}\ }\textbf {\bibinfo {volume} {73}},\
  \bibinfo {pages} {014501} (\bibinfo {year} {2006})},\ \Eprint
  {http://arxiv.org/abs/hep-lat/0507007} {arXiv:hep-lat/0507007} \BibitemShut
  {NoStop}%
\bibitem [{\citenamefont {Braun}\ and\ \citenamefont
  {M\"uller}(2008)}]{Braun:2007wv}%
  \BibitemOpen
  \bibfield  {author} {\bibinfo {author} {\bibfnamefont {V.}~\bibnamefont
  {Braun}}\ and\ \bibinfo {author} {\bibfnamefont {D.}~\bibnamefont
  {M\"uller}},\ }\href {\doibase 10.1140/epjc/s10052-008-0608-4} {\bibfield
  {journal} {\bibinfo  {journal} {Eur. Phys. J. C}\ }\textbf {\bibinfo {volume}
  {55}},\ \bibinfo {pages} {349} (\bibinfo {year} {2008})},\ \Eprint
  {http://arxiv.org/abs/0709.1348} {arXiv:0709.1348 [hep-ph]} \BibitemShut
  {NoStop}%
\bibitem [{\citenamefont {Radyushkin}(2017)}]{Radyushkin:2017cyf}%
  \BibitemOpen
  \bibfield  {author} {\bibinfo {author} {\bibfnamefont {A.~V.}\ \bibnamefont
  {Radyushkin}},\ }\href {\doibase 10.1103/PhysRevD.96.034025} {\bibfield
  {journal} {\bibinfo  {journal} {Phys. Rev. D}\ }\textbf {\bibinfo {volume}
  {96}},\ \bibinfo {pages} {034025} (\bibinfo {year} {2017})},\ \Eprint
  {http://arxiv.org/abs/1705.01488} {arXiv:1705.01488 [hep-ph]} \BibitemShut
  {NoStop}%
\bibitem [{\citenamefont {Orginos}\ \emph {et~al.}(2017)\citenamefont
  {Orginos}, \citenamefont {Radyushkin}, \citenamefont {Karpie},\ and\
  \citenamefont {Zafeiropoulos}}]{Orginos:2017kos}%
  \BibitemOpen
  \bibfield  {author} {\bibinfo {author} {\bibfnamefont {K.}~\bibnamefont
  {Orginos}}, \bibinfo {author} {\bibfnamefont {A.}~\bibnamefont {Radyushkin}},
  \bibinfo {author} {\bibfnamefont {J.}~\bibnamefont {Karpie}}, \ and\ \bibinfo
  {author} {\bibfnamefont {S.}~\bibnamefont {Zafeiropoulos}},\ }\href {\doibase
  10.1103/PhysRevD.96.094503} {\bibfield  {journal} {\bibinfo  {journal} {Phys.
  Rev. D}\ }\textbf {\bibinfo {volume} {96}},\ \bibinfo {pages} {094503}
  (\bibinfo {year} {2017})},\ \Eprint {http://arxiv.org/abs/1706.05373}
  {arXiv:1706.05373 [hep-ph]} \BibitemShut {NoStop}%
\bibitem [{\citenamefont {Chambers}\ \emph {et~al.}(2017)\citenamefont
  {Chambers}, \citenamefont {Horsley}, \citenamefont {Nakamura}, \citenamefont
  {Perlt}, \citenamefont {Rakow}, \citenamefont {Schierholz}, \citenamefont
  {Schiller}, \citenamefont {Somfleth}, \citenamefont {Young},\ and\
  \citenamefont {Zanotti}}]{Chambers:2017dov}%
  \BibitemOpen
  \bibfield  {author} {\bibinfo {author} {\bibfnamefont {A.~J.}\ \bibnamefont
  {Chambers}}, \bibinfo {author} {\bibfnamefont {R.}~\bibnamefont {Horsley}},
  \bibinfo {author} {\bibfnamefont {Y.}~\bibnamefont {Nakamura}}, \bibinfo
  {author} {\bibfnamefont {H.}~\bibnamefont {Perlt}}, \bibinfo {author}
  {\bibfnamefont {P.~E.~L.}\ \bibnamefont {Rakow}}, \bibinfo {author}
  {\bibfnamefont {G.}~\bibnamefont {Schierholz}}, \bibinfo {author}
  {\bibfnamefont {A.}~\bibnamefont {Schiller}}, \bibinfo {author}
  {\bibfnamefont {K.}~\bibnamefont {Somfleth}}, \bibinfo {author}
  {\bibfnamefont {R.~D.}\ \bibnamefont {Young}}, \ and\ \bibinfo {author}
  {\bibfnamefont {J.~M.}\ \bibnamefont {Zanotti}},\ }\href {\doibase
  10.1103/PhysRevLett.118.242001} {\bibfield  {journal} {\bibinfo  {journal}
  {Phys. Rev. Lett.}\ }\textbf {\bibinfo {volume} {118}},\ \bibinfo {pages}
  {242001} (\bibinfo {year} {2017})},\ \Eprint
  {http://arxiv.org/abs/1703.01153} {arXiv:1703.01153 [hep-lat]} \BibitemShut
  {NoStop}%
\bibitem [{\citenamefont {Ma}\ and\ \citenamefont {Qiu}(2018)}]{Ma:2017pxb}%
  \BibitemOpen
  \bibfield  {author} {\bibinfo {author} {\bibfnamefont {Y.-Q.}\ \bibnamefont
  {Ma}}\ and\ \bibinfo {author} {\bibfnamefont {J.-W.}\ \bibnamefont {Qiu}},\
  }\href {\doibase 10.1103/PhysRevLett.120.022003} {\bibfield  {journal}
  {\bibinfo  {journal} {Phys. Rev. Lett.}\ }\textbf {\bibinfo {volume} {120}},\
  \bibinfo {pages} {022003} (\bibinfo {year} {2018})},\ \Eprint
  {http://arxiv.org/abs/1709.03018} {arXiv:1709.03018 [hep-ph]} \BibitemShut
  {NoStop}%
\bibitem [{\citenamefont {Alexandrou}\ \emph {et~al.}(2020)\citenamefont
  {Alexandrou}, \citenamefont {Cichy}, \citenamefont {Constantinou},
  \citenamefont {Hadjiyiannakou}, \citenamefont {Jansen}, \citenamefont
  {Scapellato},\ and\ \citenamefont {Steffens}}]{Alexandrou:2020zbe}%
  \BibitemOpen
  \bibfield  {author} {\bibinfo {author} {\bibfnamefont {C.}~\bibnamefont
  {Alexandrou}}, \bibinfo {author} {\bibfnamefont {K.}~\bibnamefont {Cichy}},
  \bibinfo {author} {\bibfnamefont {M.}~\bibnamefont {Constantinou}}, \bibinfo
  {author} {\bibfnamefont {K.}~\bibnamefont {Hadjiyiannakou}}, \bibinfo
  {author} {\bibfnamefont {K.}~\bibnamefont {Jansen}}, \bibinfo {author}
  {\bibfnamefont {A.}~\bibnamefont {Scapellato}}, \ and\ \bibinfo {author}
  {\bibfnamefont {F.}~\bibnamefont {Steffens}},\ }\href {\doibase
  10.1103/PhysRevLett.125.262001} {\bibfield  {journal} {\bibinfo  {journal}
  {Phys. Rev. Lett.}\ }\textbf {\bibinfo {volume} {125}},\ \bibinfo {pages}
  {262001} (\bibinfo {year} {2020})},\ \Eprint
  {http://arxiv.org/abs/2008.10573} {arXiv:2008.10573 [hep-lat]} \BibitemShut
  {NoStop}%
\bibitem [{\citenamefont {Lin}(2021)}]{Lin:2020rxa}%
  \BibitemOpen
  \bibfield  {author} {\bibinfo {author} {\bibfnamefont {H.-W.}\ \bibnamefont
  {Lin}},\ }\href {\doibase 10.1103/PhysRevLett.127.182001} {\bibfield
  {journal} {\bibinfo  {journal} {Phys. Rev. Lett.}\ }\textbf {\bibinfo
  {volume} {127}},\ \bibinfo {pages} {182001} (\bibinfo {year} {2021})},\
  \Eprint {http://arxiv.org/abs/2008.12474} {arXiv:2008.12474 [hep-ph]}
  \BibitemShut {NoStop}%
\bibitem [{\citenamefont {Alexandrou}\ \emph {et~al.}(2022)\citenamefont
  {Alexandrou}, \citenamefont {Cichy}, \citenamefont {Constantinou},
  \citenamefont {Hadjiyiannakou}, \citenamefont {Jansen}, \citenamefont
  {Scapellato},\ and\ \citenamefont {Steffens}}]{Alexandrou:2021bbo}%
  \BibitemOpen
  \bibfield  {author} {\bibinfo {author} {\bibfnamefont {C.}~\bibnamefont
  {Alexandrou}}, \bibinfo {author} {\bibfnamefont {K.}~\bibnamefont {Cichy}},
  \bibinfo {author} {\bibfnamefont {M.}~\bibnamefont {Constantinou}}, \bibinfo
  {author} {\bibfnamefont {K.}~\bibnamefont {Hadjiyiannakou}}, \bibinfo
  {author} {\bibfnamefont {K.}~\bibnamefont {Jansen}}, \bibinfo {author}
  {\bibfnamefont {A.}~\bibnamefont {Scapellato}}, \ and\ \bibinfo {author}
  {\bibfnamefont {F.}~\bibnamefont {Steffens}},\ }\href {\doibase
  10.1103/PhysRevD.105.034501} {\bibfield  {journal} {\bibinfo  {journal}
  {Phys. Rev. D}\ }\textbf {\bibinfo {volume} {105}},\ \bibinfo {pages}
  {034501} (\bibinfo {year} {2022})},\ \Eprint
  {http://arxiv.org/abs/2108.10789} {arXiv:2108.10789 [hep-lat]} \BibitemShut
  {NoStop}%
\bibitem [{\citenamefont {Hannaford-Gunn}\ \emph {et~al.}(2022)\citenamefont
  {Hannaford-Gunn}, \citenamefont {Can}, \citenamefont {Horsley}, \citenamefont
  {Nakamura}, \citenamefont {Perlt}, \citenamefont {Rakow}, \citenamefont
  {St\"uben}, \citenamefont {Schierholz}, \citenamefont {Young},\ and\
  \citenamefont {Zanotti}}]{CSSMQCDSFUKQCD:2021lkf}%
  \BibitemOpen
  \bibfield  {author} {\bibinfo {author} {\bibfnamefont {A.}~\bibnamefont
  {Hannaford-Gunn}}, \bibinfo {author} {\bibfnamefont {K.~U.}\ \bibnamefont
  {Can}}, \bibinfo {author} {\bibfnamefont {R.}~\bibnamefont {Horsley}},
  \bibinfo {author} {\bibfnamefont {Y.}~\bibnamefont {Nakamura}}, \bibinfo
  {author} {\bibfnamefont {H.}~\bibnamefont {Perlt}}, \bibinfo {author}
  {\bibfnamefont {P.~E.~L.}\ \bibnamefont {Rakow}}, \bibinfo {author}
  {\bibfnamefont {H.}~\bibnamefont {St\"uben}}, \bibinfo {author}
  {\bibfnamefont {G.}~\bibnamefont {Schierholz}}, \bibinfo {author}
  {\bibfnamefont {R.~D.}\ \bibnamefont {Young}}, \ and\ \bibinfo {author}
  {\bibfnamefont {J.~M.}\ \bibnamefont {Zanotti}} (\bibinfo {collaboration}
  {CSSM/QCDSF/UKQCD}),\ }\href {\doibase 10.1103/PhysRevD.105.014502}
  {\bibfield  {journal} {\bibinfo  {journal} {Phys. Rev. D}\ }\textbf {\bibinfo
  {volume} {105}},\ \bibinfo {pages} {014502} (\bibinfo {year} {2022})},\
  \Eprint {http://arxiv.org/abs/2110.11532} {arXiv:2110.11532 [hep-lat]}
  \BibitemShut {NoStop}%
\bibitem [{\citenamefont {Dodson}\ \emph {et~al.}(2022)\citenamefont {Dodson},
  \citenamefont {Bhattacharya}, \citenamefont {Cichy}, \citenamefont
  {Constantinou}, \citenamefont {Metz}, \citenamefont {Scapellato},\ and\
  \citenamefont {Steffens}}]{Bhattacharya:2021oyr}%
  \BibitemOpen
  \bibfield  {author} {\bibinfo {author} {\bibfnamefont {J.}~\bibnamefont
  {Dodson}}, \bibinfo {author} {\bibfnamefont {S.}~\bibnamefont
  {Bhattacharya}}, \bibinfo {author} {\bibfnamefont {K.}~\bibnamefont {Cichy}},
  \bibinfo {author} {\bibfnamefont {M.}~\bibnamefont {Constantinou}}, \bibinfo
  {author} {\bibfnamefont {A.}~\bibnamefont {Metz}}, \bibinfo {author}
  {\bibfnamefont {A.}~\bibnamefont {Scapellato}}, \ and\ \bibinfo {author}
  {\bibfnamefont {F.}~\bibnamefont {Steffens}},\ }\href {\doibase
  10.22323/1.396.0054} {\bibfield  {journal} {\bibinfo  {journal} {PoS}\
  }\textbf {\bibinfo {volume} {LATTICE2021}},\ \bibinfo {pages} {054} (\bibinfo
  {year} {2022})},\ \Eprint {http://arxiv.org/abs/2112.05538} {arXiv:2112.05538
  [hep-lat]} \BibitemShut {NoStop}%
\bibitem [{\citenamefont {Lin}(2022)}]{Lin:2021brq}%
  \BibitemOpen
  \bibfield  {author} {\bibinfo {author} {\bibfnamefont {H.-W.}\ \bibnamefont
  {Lin}},\ }\href {\doibase 10.1016/j.physletb.2021.136821} {\bibfield
  {journal} {\bibinfo  {journal} {Phys. Lett. B}\ }\textbf {\bibinfo {volume}
  {824}},\ \bibinfo {pages} {136821} (\bibinfo {year} {2022})},\ \Eprint
  {http://arxiv.org/abs/2112.07519} {arXiv:2112.07519 [hep-lat]} \BibitemShut
  {NoStop}%
\bibitem [{\citenamefont {Bhattacharya}\ \emph {et~al.}(2022)\citenamefont
  {Bhattacharya}, \citenamefont {Cichy}, \citenamefont {Constantinou},
  \citenamefont {Dodson}, \citenamefont {Gao}, \citenamefont {Metz},
  \citenamefont {Mukherjee}, \citenamefont {Scapellato}, \citenamefont
  {Steffens},\ and\ \citenamefont {Zhao}}]{Bhattacharya:2022aob}%
  \BibitemOpen
  \bibfield  {author} {\bibinfo {author} {\bibfnamefont {S.}~\bibnamefont
  {Bhattacharya}}, \bibinfo {author} {\bibfnamefont {K.}~\bibnamefont {Cichy}},
  \bibinfo {author} {\bibfnamefont {M.}~\bibnamefont {Constantinou}}, \bibinfo
  {author} {\bibfnamefont {J.}~\bibnamefont {Dodson}}, \bibinfo {author}
  {\bibfnamefont {X.}~\bibnamefont {Gao}}, \bibinfo {author} {\bibfnamefont
  {A.}~\bibnamefont {Metz}}, \bibinfo {author} {\bibfnamefont {S.}~\bibnamefont
  {Mukherjee}}, \bibinfo {author} {\bibfnamefont {A.}~\bibnamefont
  {Scapellato}}, \bibinfo {author} {\bibfnamefont {F.}~\bibnamefont
  {Steffens}}, \ and\ \bibinfo {author} {\bibfnamefont {Y.}~\bibnamefont
  {Zhao}},\ }\href {\doibase 10.1103/PhysRevD.106.114512} {\bibfield  {journal}
  {\bibinfo  {journal} {Phys. Rev. D}\ }\textbf {\bibinfo {volume} {106}},\
  \bibinfo {pages} {114512} (\bibinfo {year} {2022})},\ \Eprint
  {http://arxiv.org/abs/2209.05373} {arXiv:2209.05373 [hep-lat]} \BibitemShut
  {NoStop}%
\bibitem [{\citenamefont {Cichy}\ and\ \citenamefont
  {Constantinou}(2019)}]{Cichy:2018mum}%
  \BibitemOpen
  \bibfield  {author} {\bibinfo {author} {\bibfnamefont {K.}~\bibnamefont
  {Cichy}}\ and\ \bibinfo {author} {\bibfnamefont {M.}~\bibnamefont
  {Constantinou}},\ }\href {\doibase 10.1155/2019/3036904} {\bibfield
  {journal} {\bibinfo  {journal} {Adv. High Energy Phys.}\ }\textbf {\bibinfo
  {volume} {2019}},\ \bibinfo {pages} {3036904} (\bibinfo {year} {2019})},\
  \Eprint {http://arxiv.org/abs/1811.07248} {arXiv:1811.07248 [hep-lat]}
  \BibitemShut {NoStop}%
\bibitem [{\citenamefont {Ji}\ \emph {et~al.}(2021{\natexlab{a}})\citenamefont
  {Ji}, \citenamefont {Liu}, \citenamefont {Liu}, \citenamefont {Zhang},\ and\
  \citenamefont {Zhao}}]{Ji:2020ect}%
  \BibitemOpen
  \bibfield  {author} {\bibinfo {author} {\bibfnamefont {X.}~\bibnamefont
  {Ji}}, \bibinfo {author} {\bibfnamefont {Y.-S.}\ \bibnamefont {Liu}},
  \bibinfo {author} {\bibfnamefont {Y.}~\bibnamefont {Liu}}, \bibinfo {author}
  {\bibfnamefont {J.-H.}\ \bibnamefont {Zhang}}, \ and\ \bibinfo {author}
  {\bibfnamefont {Y.}~\bibnamefont {Zhao}},\ }\href {\doibase
  10.1103/RevModPhys.93.035005} {\bibfield  {journal} {\bibinfo  {journal}
  {Rev. Mod. Phys.}\ }\textbf {\bibinfo {volume} {93}},\ \bibinfo {pages}
  {035005} (\bibinfo {year} {2021}{\natexlab{a}})},\ \Eprint
  {http://arxiv.org/abs/2004.03543} {arXiv:2004.03543 [hep-ph]} \BibitemShut
  {NoStop}%
\bibitem [{\citenamefont {Constantinou}(2021)}]{Constantinou:2020pek}%
  \BibitemOpen
  \bibfield  {author} {\bibinfo {author} {\bibfnamefont {M.}~\bibnamefont
  {Constantinou}},\ }\href {\doibase 10.1140/epja/s10050-021-00353-7}
  {\bibfield  {journal} {\bibinfo  {journal} {Eur. Phys. J. A}\ }\textbf
  {\bibinfo {volume} {57}},\ \bibinfo {pages} {77} (\bibinfo {year} {2021})},\
  \Eprint {http://arxiv.org/abs/2010.02445} {arXiv:2010.02445 [hep-lat]}
  \BibitemShut {NoStop}%
\bibitem [{\citenamefont {Cichy}(2022)}]{Cichy:2021lih}%
  \BibitemOpen
  \bibfield  {author} {\bibinfo {author} {\bibfnamefont {K.}~\bibnamefont
  {Cichy}},\ }\href {\doibase 10.22323/1.396.0017} {\bibfield  {journal}
  {\bibinfo  {journal} {PoS}\ }\textbf {\bibinfo {volume} {LATTICE2021}},\
  \bibinfo {pages} {017} (\bibinfo {year} {2022})},\ \Eprint
  {http://arxiv.org/abs/2110.07440} {arXiv:2110.07440 [hep-lat]} \BibitemShut
  {NoStop}%
\bibitem [{\citenamefont {Rodr\'\i{}guez-Quintero}\ \emph
  {et~al.}(2018)\citenamefont {Rodr\'\i{}guez-Quintero}, \citenamefont
  {Binosi}, \citenamefont {Mezrag}, \citenamefont {Papavassiliou},\ and\
  \citenamefont {Roberts}}]{Rodriguez-Quintero:2018wma}%
  \BibitemOpen
  \bibfield  {author} {\bibinfo {author} {\bibfnamefont {J.}~\bibnamefont
  {Rodr\'\i{}guez-Quintero}}, \bibinfo {author} {\bibfnamefont
  {D.}~\bibnamefont {Binosi}}, \bibinfo {author} {\bibfnamefont
  {C.}~\bibnamefont {Mezrag}}, \bibinfo {author} {\bibfnamefont
  {J.}~\bibnamefont {Papavassiliou}}, \ and\ \bibinfo {author} {\bibfnamefont
  {C.~D.}\ \bibnamefont {Roberts}},\ }\href {\doibase
  10.1007/s00601-018-1437-0} {\bibfield  {journal} {\bibinfo  {journal} {Few
  Body Syst.}\ }\textbf {\bibinfo {volume} {59}},\ \bibinfo {pages} {121}
  (\bibinfo {year} {2018})},\ \Eprint {http://arxiv.org/abs/1801.10164}
  {arXiv:1801.10164 [nucl-th]} \BibitemShut {NoStop}%
\bibitem [{\citenamefont {Cui}\ \emph {et~al.}(2020)\citenamefont {Cui},
  \citenamefont {Chen}, \citenamefont {Binosi}, \citenamefont {de~Soto},
  \citenamefont {Roberts}, \citenamefont {Rodr\'\i{}guez-Quintero},
  \citenamefont {Schmidt},\ and\ \citenamefont {Segovia}}]{Cui:2020rmu}%
  \BibitemOpen
  \bibfield  {author} {\bibinfo {author} {\bibfnamefont {Z.-F.}\ \bibnamefont
  {Cui}}, \bibinfo {author} {\bibfnamefont {C.}~\bibnamefont {Chen}}, \bibinfo
  {author} {\bibfnamefont {D.}~\bibnamefont {Binosi}}, \bibinfo {author}
  {\bibfnamefont {F.}~\bibnamefont {de~Soto}}, \bibinfo {author} {\bibfnamefont
  {C.~D.}\ \bibnamefont {Roberts}}, \bibinfo {author} {\bibfnamefont
  {J.}~\bibnamefont {Rodr\'\i{}guez-Quintero}}, \bibinfo {author}
  {\bibfnamefont {S.~M.}\ \bibnamefont {Schmidt}}, \ and\ \bibinfo {author}
  {\bibfnamefont {J.}~\bibnamefont {Segovia}},\ }\href {\doibase
  10.1103/PhysRevD.102.014043} {\bibfield  {journal} {\bibinfo  {journal}
  {Phys. Rev. D}\ }\textbf {\bibinfo {volume} {102}},\ \bibinfo {pages}
  {014043} (\bibinfo {year} {2020})},\ \Eprint
  {http://arxiv.org/abs/2003.11655} {arXiv:2003.11655 [hep-ph]} \BibitemShut
  {NoStop}%
\bibitem [{\citenamefont {Jang}\ \emph {et~al.}(2020)\citenamefont {Jang},
  \citenamefont {Gupta}, \citenamefont {Lin}, \citenamefont {Yoon},\ and\
  \citenamefont {Bhattacharya}}]{Jang:2019jkn}%
  \BibitemOpen
  \bibfield  {author} {\bibinfo {author} {\bibfnamefont {Y.-C.}\ \bibnamefont
  {Jang}}, \bibinfo {author} {\bibfnamefont {R.}~\bibnamefont {Gupta}},
  \bibinfo {author} {\bibfnamefont {H.-W.}\ \bibnamefont {Lin}}, \bibinfo
  {author} {\bibfnamefont {B.}~\bibnamefont {Yoon}}, \ and\ \bibinfo {author}
  {\bibfnamefont {T.}~\bibnamefont {Bhattacharya}},\ }\href {\doibase
  10.1103/PhysRevD.101.014507} {\bibfield  {journal} {\bibinfo  {journal}
  {Phys. Rev. D}\ }\textbf {\bibinfo {volume} {101}},\ \bibinfo {pages}
  {014507} (\bibinfo {year} {2020})},\ \Eprint
  {http://arxiv.org/abs/1906.07217} {arXiv:1906.07217 [hep-lat]} \BibitemShut
  {NoStop}%
\bibitem [{\citenamefont {Park}\ \emph {et~al.}(2022)\citenamefont {Park},
  \citenamefont {Gupta}, \citenamefont {Yoon}, \citenamefont {Mondal},
  \citenamefont {Bhattacharya}, \citenamefont {Jang}, \citenamefont {Jo\'o},\
  and\ \citenamefont {Winter}}]{Park:2021ypf}%
  \BibitemOpen
  \bibfield  {author} {\bibinfo {author} {\bibfnamefont {S.}~\bibnamefont
  {Park}}, \bibinfo {author} {\bibfnamefont {R.}~\bibnamefont {Gupta}},
  \bibinfo {author} {\bibfnamefont {B.}~\bibnamefont {Yoon}}, \bibinfo {author}
  {\bibfnamefont {S.}~\bibnamefont {Mondal}}, \bibinfo {author} {\bibfnamefont
  {T.}~\bibnamefont {Bhattacharya}}, \bibinfo {author} {\bibfnamefont {Y.-C.}\
  \bibnamefont {Jang}}, \bibinfo {author} {\bibfnamefont {B.}~\bibnamefont
  {Jo\'o}}, \ and\ \bibinfo {author} {\bibfnamefont {F.}~\bibnamefont {Winter}}
  (\bibinfo {collaboration} {Nucleon Matrix Elements (NME)}),\ }\href {\doibase
  10.1103/PhysRevD.105.054505} {\bibfield  {journal} {\bibinfo  {journal}
  {Phys. Rev. D}\ }\textbf {\bibinfo {volume} {105}},\ \bibinfo {pages}
  {054505} (\bibinfo {year} {2022})},\ \Eprint
  {http://arxiv.org/abs/2103.05599} {arXiv:2103.05599 [hep-lat]} \BibitemShut
  {NoStop}%
\bibitem [{\citenamefont {Salg}\ \emph {et~al.}(2022)\citenamefont {Salg},
  \citenamefont {Djukanovic}, \citenamefont {von Hippel}, \citenamefont
  {Meyer}, \citenamefont {Ottnad},\ and\ \citenamefont
  {Wittig}}]{Salg:2022poa}%
  \BibitemOpen
  \bibfield  {author} {\bibinfo {author} {\bibfnamefont {M.}~\bibnamefont
  {Salg}}, \bibinfo {author} {\bibfnamefont {D.}~\bibnamefont {Djukanovic}},
  \bibinfo {author} {\bibfnamefont {G.}~\bibnamefont {von Hippel}}, \bibinfo
  {author} {\bibfnamefont {H.~B.}\ \bibnamefont {Meyer}}, \bibinfo {author}
  {\bibfnamefont {K.}~\bibnamefont {Ottnad}}, \ and\ \bibinfo {author}
  {\bibfnamefont {H.}~\bibnamefont {Wittig}},\ }in\ \href {\doibase
  10.22323/1.430.0121} {\emph {\bibinfo {booktitle} {{39th International
  Symposium on Lattice Field Theory}}}},\ Vol.\ \bibinfo {volume}
  {LATTICE2022}\ (\bibinfo {year} {2022})\ p.\ \bibinfo {pages} {121},\ \Eprint
  {http://arxiv.org/abs/2211.17049} {arXiv:2211.17049 [hep-lat]} \BibitemShut
  {NoStop}%
\bibitem [{\citenamefont {Penttinen}\ \emph
  {et~al.}(2000{\natexlab{a}})\citenamefont {Penttinen}, \citenamefont
  {Polyakov}, \citenamefont {Shuvaev},\ and\ \citenamefont
  {Strikman}}]{Penttinen:2000dg}%
  \BibitemOpen
  \bibfield  {author} {\bibinfo {author} {\bibfnamefont {M.}~\bibnamefont
  {Penttinen}}, \bibinfo {author} {\bibfnamefont {M.~V.}\ \bibnamefont
  {Polyakov}}, \bibinfo {author} {\bibfnamefont {A.~G.}\ \bibnamefont
  {Shuvaev}}, \ and\ \bibinfo {author} {\bibfnamefont {M.}~\bibnamefont
  {Strikman}},\ }\href {\doibase 10.1016/S0370-2693(00)01035-2} {\bibfield
  {journal} {\bibinfo  {journal} {Phys. Lett. B}\ }\textbf {\bibinfo {volume}
  {491}},\ \bibinfo {pages} {96} (\bibinfo {year} {2000}{\natexlab{a}})},\
  \Eprint {http://arxiv.org/abs/hep-ph/0006321} {arXiv:hep-ph/0006321}
  \BibitemShut {NoStop}%
\bibitem [{\citenamefont {Kiptily}\ and\ \citenamefont
  {Polyakov}(2004)}]{Kiptily:2002nx}%
  \BibitemOpen
  \bibfield  {author} {\bibinfo {author} {\bibfnamefont {D.~V.}\ \bibnamefont
  {Kiptily}}\ and\ \bibinfo {author} {\bibfnamefont {M.~V.}\ \bibnamefont
  {Polyakov}},\ }\href {\doibase 10.1140/epjc/s2004-01957-3} {\bibfield
  {journal} {\bibinfo  {journal} {Eur. Phys. J. C}\ }\textbf {\bibinfo {volume}
  {37}},\ \bibinfo {pages} {105} (\bibinfo {year} {2004})},\ \Eprint
  {http://arxiv.org/abs/hep-ph/0212372} {arXiv:hep-ph/0212372} \BibitemShut
  {NoStop}%
\bibitem [{\citenamefont {Hatta}\ and\ \citenamefont
  {Yoshida}(2012)}]{Hatta:2012cs}%
  \BibitemOpen
  \bibfield  {author} {\bibinfo {author} {\bibfnamefont {Y.}~\bibnamefont
  {Hatta}}\ and\ \bibinfo {author} {\bibfnamefont {S.}~\bibnamefont
  {Yoshida}},\ }\href {\doibase 10.1007/JHEP10(2012)080} {\bibfield  {journal}
  {\bibinfo  {journal} {JHEP}\ }\textbf {\bibinfo {volume} {10}},\ \bibinfo
  {pages} {080} (\bibinfo {year} {2012})},\ \Eprint
  {http://arxiv.org/abs/1207.5332} {arXiv:1207.5332 [hep-ph]} \BibitemShut
  {NoStop}%
\bibitem [{\citenamefont {Ji}\ \emph {et~al.}(2013)\citenamefont {Ji},
  \citenamefont {Xiong},\ and\ \citenamefont {Yuan}}]{Ji:2012ba}%
  \BibitemOpen
  \bibfield  {author} {\bibinfo {author} {\bibfnamefont {X.}~\bibnamefont
  {Ji}}, \bibinfo {author} {\bibfnamefont {X.}~\bibnamefont {Xiong}}, \ and\
  \bibinfo {author} {\bibfnamefont {F.}~\bibnamefont {Yuan}},\ }\href {\doibase
  10.1103/PhysRevD.88.014041} {\bibfield  {journal} {\bibinfo  {journal} {Phys.
  Rev. D}\ }\textbf {\bibinfo {volume} {88}},\ \bibinfo {pages} {014041}
  (\bibinfo {year} {2013})},\ \Eprint {http://arxiv.org/abs/1207.5221}
  {arXiv:1207.5221 [hep-ph]} \BibitemShut {NoStop}%
\bibitem [{\citenamefont {Leader}\ and\ \citenamefont
  {Lorc\'e}(2014)}]{Leader:2013jra}%
  \BibitemOpen
  \bibfield  {author} {\bibinfo {author} {\bibfnamefont {E.}~\bibnamefont
  {Leader}}\ and\ \bibinfo {author} {\bibfnamefont {C.}~\bibnamefont
  {Lorc\'e}},\ }\href {\doibase 10.1016/j.physrep.2014.02.010} {\bibfield
  {journal} {\bibinfo  {journal} {Phys. Rept.}\ }\textbf {\bibinfo {volume}
  {541}},\ \bibinfo {pages} {163} (\bibinfo {year} {2014})},\ \Eprint
  {http://arxiv.org/abs/1309.4235} {arXiv:1309.4235 [hep-ph]} \BibitemShut
  {NoStop}%
\bibitem [{\citenamefont {Liu}\ and\ \citenamefont
  {Lorc\'e}(2016)}]{Liu:2015xha}%
  \BibitemOpen
  \bibfield  {author} {\bibinfo {author} {\bibfnamefont {K.-F.}\ \bibnamefont
  {Liu}}\ and\ \bibinfo {author} {\bibfnamefont {C.}~\bibnamefont {Lorc\'e}},\
  }\href {\doibase 10.1140/epja/i2016-16160-8} {\bibfield  {journal} {\bibinfo
  {journal} {Eur. Phys. J. A}\ }\textbf {\bibinfo {volume} {52}},\ \bibinfo
  {pages} {160} (\bibinfo {year} {2016})},\ \Eprint
  {http://arxiv.org/abs/1508.00911} {arXiv:1508.00911 [hep-ph]} \BibitemShut
  {NoStop}%
\bibitem [{\citenamefont {Rajan}\ \emph {et~al.}(2016)\citenamefont {Rajan},
  \citenamefont {Courtoy}, \citenamefont {Engelhardt},\ and\ \citenamefont
  {Liuti}}]{Rajan:2016tlg}%
  \BibitemOpen
  \bibfield  {author} {\bibinfo {author} {\bibfnamefont {A.}~\bibnamefont
  {Rajan}}, \bibinfo {author} {\bibfnamefont {A.}~\bibnamefont {Courtoy}},
  \bibinfo {author} {\bibfnamefont {M.}~\bibnamefont {Engelhardt}}, \ and\
  \bibinfo {author} {\bibfnamefont {S.}~\bibnamefont {Liuti}},\ }\href
  {\doibase 10.1103/PhysRevD.94.034041} {\bibfield  {journal} {\bibinfo
  {journal} {Phys. Rev. D}\ }\textbf {\bibinfo {volume} {94}},\ \bibinfo
  {pages} {034041} (\bibinfo {year} {2016})},\ \Eprint
  {http://arxiv.org/abs/1601.06117} {arXiv:1601.06117 [hep-ph]} \BibitemShut
  {NoStop}%
\bibitem [{\citenamefont {Rajan}\ \emph {et~al.}(2018)\citenamefont {Rajan},
  \citenamefont {Engelhardt},\ and\ \citenamefont {Liuti}}]{Rajan:2017cpx}%
  \BibitemOpen
  \bibfield  {author} {\bibinfo {author} {\bibfnamefont {A.}~\bibnamefont
  {Rajan}}, \bibinfo {author} {\bibfnamefont {M.}~\bibnamefont {Engelhardt}}, \
  and\ \bibinfo {author} {\bibfnamefont {S.}~\bibnamefont {Liuti}},\ }\href
  {\doibase 10.1103/PhysRevD.98.074022} {\bibfield  {journal} {\bibinfo
  {journal} {Phys. Rev. D}\ }\textbf {\bibinfo {volume} {98}},\ \bibinfo
  {pages} {074022} (\bibinfo {year} {2018})},\ \Eprint
  {http://arxiv.org/abs/1709.05770} {arXiv:1709.05770 [hep-ph]} \BibitemShut
  {NoStop}%
\bibitem [{\citenamefont {Pire}\ \emph {et~al.}(2021)\citenamefont {Pire},
  \citenamefont {Semenov-Tian-Shansky},\ and\ \citenamefont
  {Szymanowski}}]{Pire:2021hbl}%
  \BibitemOpen
  \bibfield  {author} {\bibinfo {author} {\bibfnamefont {B.}~\bibnamefont
  {Pire}}, \bibinfo {author} {\bibfnamefont {K.}~\bibnamefont
  {Semenov-Tian-Shansky}}, \ and\ \bibinfo {author} {\bibfnamefont
  {L.}~\bibnamefont {Szymanowski}},\ }\href {\doibase
  10.1016/j.physrep.2021.09.002} {\bibfield  {journal} {\bibinfo  {journal}
  {Phys. Rept.}\ }\textbf {\bibinfo {volume} {940}},\ \bibinfo {pages} {1}
  (\bibinfo {year} {2021})},\ \Eprint {http://arxiv.org/abs/2103.01079}
  {arXiv:2103.01079 [hep-ph]} \BibitemShut {NoStop}%
\bibitem [{\citenamefont {Gayoso}\ \emph {et~al.}(2021)\citenamefont {Gayoso}
  \emph {et~al.}}]{Gayoso:2021rzj}%
  \BibitemOpen
  \bibfield  {author} {\bibinfo {author} {\bibfnamefont {C.~A.}\ \bibnamefont
  {Gayoso}} \emph {et~al.},\ }\href {\doibase 10.1140/epja/s10050-021-00625-2}
  {\bibfield  {journal} {\bibinfo  {journal} {Eur. Phys. J. A}\ }\textbf
  {\bibinfo {volume} {57}},\ \bibinfo {pages} {342} (\bibinfo {year} {2021})},\
  \Eprint {http://arxiv.org/abs/2107.06748} {arXiv:2107.06748 [hep-ph]}
  \BibitemShut {NoStop}%
\bibitem [{\citenamefont {Qiu}\ and\ \citenamefont {Yu}(2022)}]{Qiu:2022bpq}%
  \BibitemOpen
  \bibfield  {author} {\bibinfo {author} {\bibfnamefont {J.-W.}\ \bibnamefont
  {Qiu}}\ and\ \bibinfo {author} {\bibfnamefont {Z.}~\bibnamefont {Yu}},\
  }\href {\doibase 10.1007/JHEP08(2022)103} {\bibfield  {journal} {\bibinfo
  {journal} {JHEP}\ }\textbf {\bibinfo {volume} {08}},\ \bibinfo {pages} {103}
  (\bibinfo {year} {2022})},\ \Eprint {http://arxiv.org/abs/2205.07846}
  {arXiv:2205.07846 [hep-ph]} \BibitemShut {NoStop}%
\bibitem [{\citenamefont {Ahmad}\ \emph {et~al.}(2007)\citenamefont {Ahmad},
  \citenamefont {Honkanen}, \citenamefont {Liuti},\ and\ \citenamefont
  {Taneja}}]{Ahmad:2006gn}%
  \BibitemOpen
  \bibfield  {author} {\bibinfo {author} {\bibfnamefont {S.}~\bibnamefont
  {Ahmad}}, \bibinfo {author} {\bibfnamefont {H.}~\bibnamefont {Honkanen}},
  \bibinfo {author} {\bibfnamefont {S.}~\bibnamefont {Liuti}}, \ and\ \bibinfo
  {author} {\bibfnamefont {S.~K.}\ \bibnamefont {Taneja}},\ }\href {\doibase
  10.1103/PhysRevD.75.094003} {\bibfield  {journal} {\bibinfo  {journal} {Phys.
  Rev. D}\ }\textbf {\bibinfo {volume} {75}},\ \bibinfo {pages} {094003}
  (\bibinfo {year} {2007})},\ \Eprint {http://arxiv.org/abs/hep-ph/0611046}
  {arXiv:hep-ph/0611046} \BibitemShut {NoStop}%
\bibitem [{\citenamefont {Ahmad}\ \emph {et~al.}(2009)\citenamefont {Ahmad},
  \citenamefont {Honkanen}, \citenamefont {Liuti},\ and\ \citenamefont
  {Taneja}}]{Ahmad:2009fvg}%
  \BibitemOpen
  \bibfield  {author} {\bibinfo {author} {\bibfnamefont {S.}~\bibnamefont
  {Ahmad}}, \bibinfo {author} {\bibfnamefont {H.}~\bibnamefont {Honkanen}},
  \bibinfo {author} {\bibfnamefont {S.}~\bibnamefont {Liuti}}, \ and\ \bibinfo
  {author} {\bibfnamefont {S.~K.}\ \bibnamefont {Taneja}},\ }\href {\doibase
  10.1140/epjc/s10052-009-1073-4} {\bibfield  {journal} {\bibinfo  {journal}
  {Eur. Phys. J. C}\ }\textbf {\bibinfo {volume} {63}},\ \bibinfo {pages} {407}
  (\bibinfo {year} {2009})},\ \Eprint {http://arxiv.org/abs/0708.0268}
  {arXiv:0708.0268 [hep-ph]} \BibitemShut {NoStop}%
\bibitem [{\citenamefont {Goldstein}\ \emph {et~al.}(2011)\citenamefont
  {Goldstein}, \citenamefont {Hernandez},\ and\ \citenamefont
  {Liuti}}]{Goldstein:2010gu}%
  \BibitemOpen
  \bibfield  {author} {\bibinfo {author} {\bibfnamefont {G.~R.}\ \bibnamefont
  {Goldstein}}, \bibinfo {author} {\bibfnamefont {J.~O.}\ \bibnamefont
  {Hernandez}}, \ and\ \bibinfo {author} {\bibfnamefont {S.}~\bibnamefont
  {Liuti}},\ }\href {\doibase 10.1103/PhysRevD.84.034007} {\bibfield  {journal}
  {\bibinfo  {journal} {Phys. Rev. D}\ }\textbf {\bibinfo {volume} {84}},\
  \bibinfo {pages} {034007} (\bibinfo {year} {2011})},\ \Eprint
  {http://arxiv.org/abs/1012.3776} {arXiv:1012.3776 [hep-ph]} \BibitemShut
  {NoStop}%
\bibitem [{\citenamefont {Gonzalez-Hernandez}\ \emph
  {et~al.}(2013)\citenamefont {Gonzalez-Hernandez}, \citenamefont {Liuti},
  \citenamefont {Goldstein},\ and\ \citenamefont
  {Kathuria}}]{Gonzalez-Hernandez:2012xap}%
  \BibitemOpen
  \bibfield  {author} {\bibinfo {author} {\bibfnamefont {J.~O.}\ \bibnamefont
  {Gonzalez-Hernandez}}, \bibinfo {author} {\bibfnamefont {S.}~\bibnamefont
  {Liuti}}, \bibinfo {author} {\bibfnamefont {G.~R.}\ \bibnamefont
  {Goldstein}}, \ and\ \bibinfo {author} {\bibfnamefont {K.}~\bibnamefont
  {Kathuria}},\ }\href {\doibase 10.1103/PhysRevC.88.065206} {\bibfield
  {journal} {\bibinfo  {journal} {Phys. Rev. C}\ }\textbf {\bibinfo {volume}
  {88}},\ \bibinfo {pages} {065206} (\bibinfo {year} {2013})},\ \Eprint
  {http://arxiv.org/abs/1206.1876} {arXiv:1206.1876 [hep-ph]} \BibitemShut
  {NoStop}%
\bibitem [{\citenamefont {Kriesten}\ \emph {et~al.}(2020)\citenamefont
  {Kriesten}, \citenamefont {Liuti}, \citenamefont {Calero-Diaz}, \citenamefont
  {Keller}, \citenamefont {Meyer}, \citenamefont {Goldstein},\ and\
  \citenamefont {Osvaldo Gonzalez-Hernandez}}]{Kriesten:2019jep}%
  \BibitemOpen
  \bibfield  {author} {\bibinfo {author} {\bibfnamefont {B.}~\bibnamefont
  {Kriesten}}, \bibinfo {author} {\bibfnamefont {S.}~\bibnamefont {Liuti}},
  \bibinfo {author} {\bibfnamefont {L.}~\bibnamefont {Calero-Diaz}}, \bibinfo
  {author} {\bibfnamefont {D.}~\bibnamefont {Keller}}, \bibinfo {author}
  {\bibfnamefont {A.}~\bibnamefont {Meyer}}, \bibinfo {author} {\bibfnamefont
  {G.~R.}\ \bibnamefont {Goldstein}}, \ and\ \bibinfo {author} {\bibfnamefont
  {J.}~\bibnamefont {Osvaldo Gonzalez-Hernandez}},\ }\href {\doibase
  10.1103/PhysRevD.101.054021} {\bibfield  {journal} {\bibinfo  {journal}
  {Phys. Rev. D}\ }\textbf {\bibinfo {volume} {101}},\ \bibinfo {pages}
  {054021} (\bibinfo {year} {2020})},\ \Eprint
  {http://arxiv.org/abs/1903.05742} {arXiv:1903.05742 [hep-ph]} \BibitemShut
  {NoStop}%
\bibitem [{\citenamefont {Kriesten}\ and\ \citenamefont
  {Liuti}(2022)}]{Kriesten:2020wcx}%
  \BibitemOpen
  \bibfield  {author} {\bibinfo {author} {\bibfnamefont {B.}~\bibnamefont
  {Kriesten}}\ and\ \bibinfo {author} {\bibfnamefont {S.}~\bibnamefont
  {Liuti}},\ }\href {\doibase 10.1103/PhysRevD.105.016015} {\bibfield
  {journal} {\bibinfo  {journal} {Phys. Rev. D}\ }\textbf {\bibinfo {volume}
  {105}},\ \bibinfo {pages} {016015} (\bibinfo {year} {2022})},\ \Eprint
  {http://arxiv.org/abs/2004.08890} {arXiv:2004.08890 [hep-ph]} \BibitemShut
  {NoStop}%
\bibitem [{\citenamefont {Kriesten}\ \emph {et~al.}(2022)\citenamefont
  {Kriesten}, \citenamefont {Velie}, \citenamefont {Yeats}, \citenamefont
  {Lopez},\ and\ \citenamefont {Liuti}}]{Kriesten:2021sqc}%
  \BibitemOpen
  \bibfield  {author} {\bibinfo {author} {\bibfnamefont {B.}~\bibnamefont
  {Kriesten}}, \bibinfo {author} {\bibfnamefont {P.}~\bibnamefont {Velie}},
  \bibinfo {author} {\bibfnamefont {E.}~\bibnamefont {Yeats}}, \bibinfo
  {author} {\bibfnamefont {F.~Y.}\ \bibnamefont {Lopez}}, \ and\ \bibinfo
  {author} {\bibfnamefont {S.}~\bibnamefont {Liuti}},\ }\href {\doibase
  10.1103/PhysRevD.105.056022} {\bibfield  {journal} {\bibinfo  {journal}
  {Phys. Rev. D}\ }\textbf {\bibinfo {volume} {105}},\ \bibinfo {pages}
  {056022} (\bibinfo {year} {2022})},\ \Eprint
  {http://arxiv.org/abs/2101.01826} {arXiv:2101.01826 [hep-ph]} \BibitemShut
  {NoStop}%
\bibitem [{\citenamefont {Bertone}\ \emph {et~al.}(2021)\citenamefont
  {Bertone}, \citenamefont {Dutrieux}, \citenamefont {Mezrag}, \citenamefont
  {Moutarde},\ and\ \citenamefont {Sznajder}}]{Bertone:2021yyz}%
  \BibitemOpen
  \bibfield  {author} {\bibinfo {author} {\bibfnamefont {V.}~\bibnamefont
  {Bertone}}, \bibinfo {author} {\bibfnamefont {H.}~\bibnamefont {Dutrieux}},
  \bibinfo {author} {\bibfnamefont {C.}~\bibnamefont {Mezrag}}, \bibinfo
  {author} {\bibfnamefont {H.}~\bibnamefont {Moutarde}}, \ and\ \bibinfo
  {author} {\bibfnamefont {P.}~\bibnamefont {Sznajder}},\ }\href {\doibase
  10.1103/PhysRevD.103.114019} {\bibfield  {journal} {\bibinfo  {journal}
  {Phys. Rev. D}\ }\textbf {\bibinfo {volume} {103}},\ \bibinfo {pages}
  {114019} (\bibinfo {year} {2021})},\ \Eprint
  {http://arxiv.org/abs/2104.03836} {arXiv:2104.03836 [hep-ph]} \BibitemShut
  {NoStop}%
\bibitem [{\citenamefont {El~Beiyad}\ \emph {et~al.}(2010)\citenamefont
  {El~Beiyad}, \citenamefont {Pire}, \citenamefont {Segond}, \citenamefont
  {Szymanowski},\ and\ \citenamefont {Wallon}}]{Beiyad:2010cxa}%
  \BibitemOpen
  \bibfield  {author} {\bibinfo {author} {\bibfnamefont {M.}~\bibnamefont
  {El~Beiyad}}, \bibinfo {author} {\bibfnamefont {B.}~\bibnamefont {Pire}},
  \bibinfo {author} {\bibfnamefont {M.}~\bibnamefont {Segond}}, \bibinfo
  {author} {\bibfnamefont {L.}~\bibnamefont {Szymanowski}}, \ and\ \bibinfo
  {author} {\bibfnamefont {S.}~\bibnamefont {Wallon}},\ }\href {\doibase
  10.1016/j.physletb.2010.02.086} {\bibfield  {journal} {\bibinfo  {journal}
  {Phys. Lett. B}\ }\textbf {\bibinfo {volume} {688}},\ \bibinfo {pages} {154}
  (\bibinfo {year} {2010})},\ \Eprint {http://arxiv.org/abs/hep-ph/1001.4491}
  {arXiv:hep-ph/1001.4491 [hep-ph]} \BibitemShut {NoStop}%
\bibitem [{\citenamefont {Pedrak}\ \emph {et~al.}(2017)\citenamefont {Pedrak},
  \citenamefont {Pire}, \citenamefont {Szymanowski},\ and\ \citenamefont
  {Wagner}}]{Pedrak:2017cpp}%
  \BibitemOpen
  \bibfield  {author} {\bibinfo {author} {\bibfnamefont {A.}~\bibnamefont
  {Pedrak}}, \bibinfo {author} {\bibfnamefont {B.}~\bibnamefont {Pire}},
  \bibinfo {author} {\bibfnamefont {L.}~\bibnamefont {Szymanowski}}, \ and\
  \bibinfo {author} {\bibfnamefont {J.}~\bibnamefont {Wagner}},\ }\href
  {\doibase 10.1103/PhysRevD.96.074008} {\bibfield  {journal} {\bibinfo
  {journal} {Phys. Rev. D}\ }\textbf {\bibinfo {volume} {96}},\ \bibinfo
  {pages} {074008} (\bibinfo {year} {2017})},\ \bibinfo {note} {[Erratum:
  Phys.Rev.D 100, 039901 (2019)]},\ \Eprint
  {http://arxiv.org/abs/hep-ph/1708.01043} {arXiv:hep-ph/1708.01043 [hep-ph]}
  \BibitemShut {NoStop}%
\bibitem [{\citenamefont {Boussarie}\ \emph
  {et~al.}(2017{\natexlab{a}})\citenamefont {Boussarie}, \citenamefont {Pire},
  \citenamefont {Szymanowski},\ and\ \citenamefont
  {Wallon}}]{Boussarie:2016qop}%
  \BibitemOpen
  \bibfield  {author} {\bibinfo {author} {\bibfnamefont {R.}~\bibnamefont
  {Boussarie}}, \bibinfo {author} {\bibfnamefont {B.}~\bibnamefont {Pire}},
  \bibinfo {author} {\bibfnamefont {L.}~\bibnamefont {Szymanowski}}, \ and\
  \bibinfo {author} {\bibfnamefont {S.}~\bibnamefont {Wallon}},\ }\href
  {\doibase 10.1007/JHEP02(2017)054} {\bibfield  {journal} {\bibinfo  {journal}
  {JHEP}\ }\textbf {\bibinfo {volume} {02}},\ \bibinfo {pages} {054} (\bibinfo
  {year} {2017}{\natexlab{a}})},\ \bibinfo {note} {[Erratum: JHEP 10, 029
  (2018)]},\ \Eprint {http://arxiv.org/abs/hep-ph/1609.03830}
  {arXiv:hep-ph/1609.03830 [hep-ph]} \BibitemShut {NoStop}%
\bibitem [{\citenamefont {Pedrak}\ \emph {et~al.}(2020)\citenamefont {Pedrak},
  \citenamefont {Pire}, \citenamefont {Szymanowski},\ and\ \citenamefont
  {Wagner}}]{Pedrak:2020mfm}%
  \BibitemOpen
  \bibfield  {author} {\bibinfo {author} {\bibfnamefont {A.}~\bibnamefont
  {Pedrak}}, \bibinfo {author} {\bibfnamefont {B.}~\bibnamefont {Pire}},
  \bibinfo {author} {\bibfnamefont {L.}~\bibnamefont {Szymanowski}}, \ and\
  \bibinfo {author} {\bibfnamefont {J.}~\bibnamefont {Wagner}},\ }\href
  {\doibase 10.1103/PhysRevD.101.114027} {\bibfield  {journal} {\bibinfo
  {journal} {Phys. Rev. D}\ }\textbf {\bibinfo {volume} {101}},\ \bibinfo
  {pages} {114027} (\bibinfo {year} {2020})},\ \Eprint
  {http://arxiv.org/abs/hep-ph/2003.03263} {arXiv:hep-ph/2003.03263 [hep-ph]}
  \BibitemShut {NoStop}%
\bibitem [{\citenamefont {Grocholski}\ \emph {et~al.}(2021)\citenamefont
  {Grocholski}, \citenamefont {Pire}, \citenamefont {Sznajder}, \citenamefont
  {Szymanowski},\ and\ \citenamefont {Wagner}}]{Grocholski:2021man}%
  \BibitemOpen
  \bibfield  {author} {\bibinfo {author} {\bibfnamefont {O.}~\bibnamefont
  {Grocholski}}, \bibinfo {author} {\bibfnamefont {B.}~\bibnamefont {Pire}},
  \bibinfo {author} {\bibfnamefont {P.}~\bibnamefont {Sznajder}}, \bibinfo
  {author} {\bibfnamefont {L.}~\bibnamefont {Szymanowski}}, \ and\ \bibinfo
  {author} {\bibfnamefont {J.}~\bibnamefont {Wagner}},\ }\href {\doibase
  10.1103/PhysRevD.104.114006} {\bibfield  {journal} {\bibinfo  {journal}
  {Phys. Rev. D}\ }\textbf {\bibinfo {volume} {104}},\ \bibinfo {pages}
  {114006} (\bibinfo {year} {2021})},\ \Eprint
  {http://arxiv.org/abs/2110.00048} {arXiv:2110.00048 [hep-ph]} \BibitemShut
  {NoStop}%
\bibitem [{\citenamefont {Grocholski}\ \emph {et~al.}(2022)\citenamefont
  {Grocholski}, \citenamefont {Pire}, \citenamefont {Sznajder}, \citenamefont
  {Szymanowski},\ and\ \citenamefont {Wagner}}]{Grocholski:2022rqj}%
  \BibitemOpen
  \bibfield  {author} {\bibinfo {author} {\bibfnamefont {O.}~\bibnamefont
  {Grocholski}}, \bibinfo {author} {\bibfnamefont {B.}~\bibnamefont {Pire}},
  \bibinfo {author} {\bibfnamefont {P.}~\bibnamefont {Sznajder}}, \bibinfo
  {author} {\bibfnamefont {L.}~\bibnamefont {Szymanowski}}, \ and\ \bibinfo
  {author} {\bibfnamefont {J.}~\bibnamefont {Wagner}},\ }\href {\doibase
  10.1103/PhysRevD.105.094025} {\bibfield  {journal} {\bibinfo  {journal}
  {Phys. Rev. D}\ }\textbf {\bibinfo {volume} {105}},\ \bibinfo {pages}
  {094025} (\bibinfo {year} {2022})},\ \Eprint
  {http://arxiv.org/abs/2204.00396} {arXiv:2204.00396 [hep-ph]} \BibitemShut
  {NoStop}%
\bibitem [{\citenamefont {Qiu}\ and\ \citenamefont {Yu}(2023)}]{Qiu:2022pla}%
  \BibitemOpen
  \bibfield  {author} {\bibinfo {author} {\bibfnamefont {J.-W.}\ \bibnamefont
  {Qiu}}\ and\ \bibinfo {author} {\bibfnamefont {Z.}~\bibnamefont {Yu}},\
  }\href {\doibase 10.1103/PhysRevD.107.014007} {\bibfield  {journal} {\bibinfo
   {journal} {Phys. Rev. D}\ }\textbf {\bibinfo {volume} {107}},\ \bibinfo
  {pages} {014007} (\bibinfo {year} {2023})},\ \Eprint
  {http://arxiv.org/abs/2210.07995} {arXiv:2210.07995 [hep-ph]} \BibitemShut
  {NoStop}%
\bibitem [{\citenamefont {Polyakov}\ and\ \citenamefont
  {Weiss}(1999)}]{Polyakov:1999gs}%
  \BibitemOpen
  \bibfield  {author} {\bibinfo {author} {\bibfnamefont {M.~V.}\ \bibnamefont
  {Polyakov}}\ and\ \bibinfo {author} {\bibfnamefont {C.}~\bibnamefont
  {Weiss}},\ }\href {\doibase 10.1103/PhysRevD.60.114017} {\bibfield  {journal}
  {\bibinfo  {journal} {Phys. Rev. D}\ }\textbf {\bibinfo {volume} {60}},\
  \bibinfo {pages} {114017} (\bibinfo {year} {1999})},\ \Eprint
  {http://arxiv.org/abs/hep-ph/9902451} {arXiv:hep-ph/9902451} \BibitemShut
  {NoStop}%
\bibitem [{\citenamefont {Theussl}\ \emph {et~al.}(2004)\citenamefont
  {Theussl}, \citenamefont {Noguera},\ and\ \citenamefont
  {Vento}}]{Theussl:2002xp}%
  \BibitemOpen
  \bibfield  {author} {\bibinfo {author} {\bibfnamefont {L.}~\bibnamefont
  {Theussl}}, \bibinfo {author} {\bibfnamefont {S.}~\bibnamefont {Noguera}}, \
  and\ \bibinfo {author} {\bibfnamefont {V.}~\bibnamefont {Vento}},\ }\href
  {\doibase 10.1140/epja/i2003-10174-3} {\bibfield  {journal} {\bibinfo
  {journal} {Eur. Phys. J. A}\ }\textbf {\bibinfo {volume} {20}},\ \bibinfo
  {pages} {483} (\bibinfo {year} {2004})},\ \Eprint
  {http://arxiv.org/abs/nucl-th/0211036} {arXiv:nucl-th/0211036} \BibitemShut
  {NoStop}%
\bibitem [{\citenamefont {Broniowski}\ and\ \citenamefont
  {Ruiz~Arriola}(2003)}]{Broniowski:2003rp}%
  \BibitemOpen
  \bibfield  {author} {\bibinfo {author} {\bibfnamefont {W.}~\bibnamefont
  {Broniowski}}\ and\ \bibinfo {author} {\bibfnamefont {E.}~\bibnamefont
  {Ruiz~Arriola}},\ }\href {\doibase 10.1016/j.physletb.2003.09.009} {\bibfield
   {journal} {\bibinfo  {journal} {Phys. Lett. B}\ }\textbf {\bibinfo {volume}
  {574}},\ \bibinfo {pages} {57} (\bibinfo {year} {2003})},\ \Eprint
  {http://arxiv.org/abs/hep-ph/0307198} {arXiv:hep-ph/0307198} \BibitemShut
  {NoStop}%
\bibitem [{\citenamefont {Aguilar}\ \emph {et~al.}(2019)\citenamefont {Aguilar}
  \emph {et~al.}}]{Aguilar:2019teb}%
  \BibitemOpen
  \bibfield  {author} {\bibinfo {author} {\bibfnamefont {A.~C.}\ \bibnamefont
  {Aguilar}} \emph {et~al.},\ }\href {\doibase 10.1140/epja/i2019-12885-0}
  {\bibfield  {journal} {\bibinfo  {journal} {Eur. Phys. J. A}\ }\textbf
  {\bibinfo {volume} {55}},\ \bibinfo {pages} {190} (\bibinfo {year} {2019})},\
  \Eprint {http://arxiv.org/abs/1907.08218} {arXiv:1907.08218 [nucl-ex]}
  \BibitemShut {NoStop}%
\bibitem [{\citenamefont {Arrington}\ \emph {et~al.}(2021)\citenamefont
  {Arrington} \emph {et~al.}}]{Arrington:2021biu}%
  \BibitemOpen
  \bibfield  {author} {\bibinfo {author} {\bibfnamefont {J.}~\bibnamefont
  {Arrington}} \emph {et~al.},\ }\href {\doibase 10.1088/1361-6471/abf5c3}
  {\bibfield  {journal} {\bibinfo  {journal} {J. Phys. G}\ }\textbf {\bibinfo
  {volume} {48}},\ \bibinfo {pages} {075106} (\bibinfo {year} {2021})},\
  \Eprint {http://arxiv.org/abs/2102.11788} {arXiv:2102.11788 [nucl-ex]}
  \BibitemShut {NoStop}%
\bibitem [{\citenamefont {Burkert}\ \emph {et~al.}(2022)\citenamefont {Burkert}
  \emph {et~al.}}]{Burkert:2022hjz}%
  \BibitemOpen
  \bibfield  {author} {\bibinfo {author} {\bibfnamefont {V.~D.}\ \bibnamefont
  {Burkert}} \emph {et~al.},\ }\href@noop {} {\  (\bibinfo {year} {2022})},\
  \Eprint {http://arxiv.org/abs/2211.15746} {arXiv:2211.15746 [nucl-ex]}
  \BibitemShut {NoStop}%
\bibitem [{\citenamefont {Chavez}\ \emph {et~al.}(2022)\citenamefont {Chavez},
  \citenamefont {Bertone}, \citenamefont {De~Soto~Borrero}, \citenamefont
  {Defurne}, \citenamefont {Mezrag}, \citenamefont {Moutarde}, \citenamefont
  {Rodr\'\i{}guez-Quintero},\ and\ \citenamefont {Segovia}}]{Chavez:2021llq}%
  \BibitemOpen
  \bibfield  {author} {\bibinfo {author} {\bibfnamefont {J.~M.~M.}\
  \bibnamefont {Chavez}}, \bibinfo {author} {\bibfnamefont {V.}~\bibnamefont
  {Bertone}}, \bibinfo {author} {\bibfnamefont {F.}~\bibnamefont
  {De~Soto~Borrero}}, \bibinfo {author} {\bibfnamefont {M.}~\bibnamefont
  {Defurne}}, \bibinfo {author} {\bibfnamefont {C.}~\bibnamefont {Mezrag}},
  \bibinfo {author} {\bibfnamefont {H.}~\bibnamefont {Moutarde}}, \bibinfo
  {author} {\bibfnamefont {J.}~\bibnamefont {Rodr\'\i{}guez-Quintero}}, \ and\
  \bibinfo {author} {\bibfnamefont {J.}~\bibnamefont {Segovia}},\ }\href
  {\doibase 10.1103/PhysRevD.105.094012} {\bibfield  {journal} {\bibinfo
  {journal} {Phys. Rev. D}\ }\textbf {\bibinfo {volume} {105}},\ \bibinfo
  {pages} {094012} (\bibinfo {year} {2022})},\ \Eprint
  {http://arxiv.org/abs/2110.06052} {arXiv:2110.06052 [hep-ph]} \BibitemShut
  {NoStop}%
\bibitem [{\citenamefont {Ch\'avez}\ \emph {et~al.}(2022)\citenamefont
  {Ch\'avez}, \citenamefont {Bertone}, \citenamefont {De~Soto~Borrero},
  \citenamefont {Defurne}, \citenamefont {Mezrag}, \citenamefont {Moutarde},
  \citenamefont {Rodr\'\i{}guez-Quintero},\ and\ \citenamefont
  {Segovia}}]{Chavez:2021koz}%
  \BibitemOpen
  \bibfield  {author} {\bibinfo {author} {\bibfnamefont {J.~M.~M.}\
  \bibnamefont {Ch\'avez}}, \bibinfo {author} {\bibfnamefont {V.}~\bibnamefont
  {Bertone}}, \bibinfo {author} {\bibfnamefont {F.}~\bibnamefont
  {De~Soto~Borrero}}, \bibinfo {author} {\bibfnamefont {M.}~\bibnamefont
  {Defurne}}, \bibinfo {author} {\bibfnamefont {C.}~\bibnamefont {Mezrag}},
  \bibinfo {author} {\bibfnamefont {H.}~\bibnamefont {Moutarde}}, \bibinfo
  {author} {\bibfnamefont {J.}~\bibnamefont {Rodr\'\i{}guez-Quintero}}, \ and\
  \bibinfo {author} {\bibfnamefont {J.}~\bibnamefont {Segovia}},\ }\href
  {\doibase 10.1103/PhysRevLett.128.202501} {\bibfield  {journal} {\bibinfo
  {journal} {Phys. Rev. Lett.}\ }\textbf {\bibinfo {volume} {128}},\ \bibinfo
  {pages} {202501} (\bibinfo {year} {2022})},\ \Eprint
  {http://arxiv.org/abs/2110.09462} {arXiv:2110.09462 [hep-ph]} \BibitemShut
  {NoStop}%
\bibitem [{\citenamefont {Alharazin}\ \emph {et~al.}(2020)\citenamefont
  {Alharazin}, \citenamefont {Djukanovic}, \citenamefont {Gegelia},\ and\
  \citenamefont {Polyakov}}]{Alharazin:2020yjv}%
  \BibitemOpen
  \bibfield  {author} {\bibinfo {author} {\bibfnamefont {H.}~\bibnamefont
  {Alharazin}}, \bibinfo {author} {\bibfnamefont {D.}~\bibnamefont
  {Djukanovic}}, \bibinfo {author} {\bibfnamefont {J.}~\bibnamefont {Gegelia}},
  \ and\ \bibinfo {author} {\bibfnamefont {M.~V.}\ \bibnamefont {Polyakov}},\
  }\href {\doibase 10.1103/PhysRevD.102.076023} {\bibfield  {journal} {\bibinfo
   {journal} {Phys. Rev. D}\ }\textbf {\bibinfo {volume} {102}},\ \bibinfo
  {pages} {076023} (\bibinfo {year} {2020})},\ \Eprint
  {http://arxiv.org/abs/2006.05890} {arXiv:2006.05890 [hep-ph]} \BibitemShut
  {NoStop}%
\bibitem [{\citenamefont {Pasquini}\ \emph {et~al.}(2014)\citenamefont
  {Pasquini}, \citenamefont {Polyakov},\ and\ \citenamefont
  {Vanderhaeghen}}]{Pasquini:2014vua}%
  \BibitemOpen
  \bibfield  {author} {\bibinfo {author} {\bibfnamefont {B.}~\bibnamefont
  {Pasquini}}, \bibinfo {author} {\bibfnamefont {M.~V.}\ \bibnamefont
  {Polyakov}}, \ and\ \bibinfo {author} {\bibfnamefont {M.}~\bibnamefont
  {Vanderhaeghen}},\ }\href {\doibase 10.1016/j.physletb.2014.10.047}
  {\bibfield  {journal} {\bibinfo  {journal} {Phys. Lett. B}\ }\textbf
  {\bibinfo {volume} {739}},\ \bibinfo {pages} {133} (\bibinfo {year}
  {2014})},\ \Eprint {http://arxiv.org/abs/1407.5960} {arXiv:1407.5960
  [hep-ph]} \BibitemShut {NoStop}%
\bibitem [{\citenamefont {Petrov}\ \emph {et~al.}(1998)\citenamefont {Petrov},
  \citenamefont {Pobylitsa}, \citenamefont {Polyakov}, \citenamefont {Bornig},
  \citenamefont {Goeke},\ and\ \citenamefont {Weiss}}]{Petrov:1998kf}%
  \BibitemOpen
  \bibfield  {author} {\bibinfo {author} {\bibfnamefont {V.~Y.}\ \bibnamefont
  {Petrov}}, \bibinfo {author} {\bibfnamefont {P.~V.}\ \bibnamefont
  {Pobylitsa}}, \bibinfo {author} {\bibfnamefont {M.~V.}\ \bibnamefont
  {Polyakov}}, \bibinfo {author} {\bibfnamefont {I.}~\bibnamefont {Bornig}},
  \bibinfo {author} {\bibfnamefont {K.}~\bibnamefont {Goeke}}, \ and\ \bibinfo
  {author} {\bibfnamefont {C.}~\bibnamefont {Weiss}},\ }\href {\doibase
  10.1103/PhysRevD.57.4325} {\bibfield  {journal} {\bibinfo  {journal} {Phys.
  Rev. D}\ }\textbf {\bibinfo {volume} {57}},\ \bibinfo {pages} {4325}
  (\bibinfo {year} {1998})},\ \Eprint {http://arxiv.org/abs/hep-ph/9710270}
  {arXiv:hep-ph/9710270} \BibitemShut {NoStop}%
\bibitem [{\citenamefont {Penttinen}\ \emph
  {et~al.}(2000{\natexlab{b}})\citenamefont {Penttinen}, \citenamefont
  {Polyakov},\ and\ \citenamefont {Goeke}}]{Penttinen:1999th}%
  \BibitemOpen
  \bibfield  {author} {\bibinfo {author} {\bibfnamefont {M.}~\bibnamefont
  {Penttinen}}, \bibinfo {author} {\bibfnamefont {M.~V.}\ \bibnamefont
  {Polyakov}}, \ and\ \bibinfo {author} {\bibfnamefont {K.}~\bibnamefont
  {Goeke}},\ }\href {\doibase 10.1103/PhysRevD.62.014024} {\bibfield  {journal}
  {\bibinfo  {journal} {Phys. Rev. D}\ }\textbf {\bibinfo {volume} {62}},\
  \bibinfo {pages} {014024} (\bibinfo {year} {2000}{\natexlab{b}})},\ \Eprint
  {http://arxiv.org/abs/hep-ph/9909489} {arXiv:hep-ph/9909489} \BibitemShut
  {NoStop}%
\bibitem [{\citenamefont {Ossmann}\ \emph {et~al.}(2005)\citenamefont
  {Ossmann}, \citenamefont {Polyakov}, \citenamefont {Schweitzer},
  \citenamefont {Urbano},\ and\ \citenamefont {Goeke}}]{Ossmann:2004bp}%
  \BibitemOpen
  \bibfield  {author} {\bibinfo {author} {\bibfnamefont {J.}~\bibnamefont
  {Ossmann}}, \bibinfo {author} {\bibfnamefont {M.~V.}\ \bibnamefont
  {Polyakov}}, \bibinfo {author} {\bibfnamefont {P.}~\bibnamefont
  {Schweitzer}}, \bibinfo {author} {\bibfnamefont {D.}~\bibnamefont {Urbano}},
  \ and\ \bibinfo {author} {\bibfnamefont {K.}~\bibnamefont {Goeke}},\ }\href
  {\doibase 10.1103/PhysRevD.71.034011} {\bibfield  {journal} {\bibinfo
  {journal} {Phys. Rev. D}\ }\textbf {\bibinfo {volume} {71}},\ \bibinfo
  {pages} {034011} (\bibinfo {year} {2005})},\ \Eprint
  {http://arxiv.org/abs/hep-ph/0411172} {arXiv:hep-ph/0411172} \BibitemShut
  {NoStop}%
\bibitem [{\citenamefont {Kanatchikov}(2018)}]{Kanatchikov:2018uoy}%
  \BibitemOpen
  \bibfield  {author} {\bibinfo {author} {\bibfnamefont {I.~V.}\ \bibnamefont
  {Kanatchikov}},\ }\href {\doibase 10.1016/S0034-4877(19)30008-4} {\bibfield
  {journal} {\bibinfo  {journal} {Rept. Math. Phys.}\ }\textbf {\bibinfo
  {volume} {82}},\ \bibinfo {pages} {373} (\bibinfo {year} {2018})},\ \Eprint
  {http://arxiv.org/abs/1805.05279} {arXiv:1805.05279 [hep-th]} \BibitemShut
  {NoStop}%
\bibitem [{\citenamefont {Ivanov}\ \emph
  {et~al.}(2020{\natexlab{a}})\citenamefont {Ivanov}, \citenamefont {Kalugin},
  \citenamefont {Ogarkova},\ and\ \citenamefont {Ogarkov}}]{Ivanov:2020zkm}%
  \BibitemOpen
  \bibfield  {author} {\bibinfo {author} {\bibfnamefont {M.~G.}\ \bibnamefont
  {Ivanov}}, \bibinfo {author} {\bibfnamefont {A.~E.}\ \bibnamefont {Kalugin}},
  \bibinfo {author} {\bibfnamefont {A.~A.}\ \bibnamefont {Ogarkova}}, \ and\
  \bibinfo {author} {\bibfnamefont {S.~L.}\ \bibnamefont {Ogarkov}},\ }\href
  {\doibase 10.3390/sym12101657} {\bibfield  {journal} {\bibinfo  {journal}
  {Symmetry}\ }\textbf {\bibinfo {volume} {12}},\ \bibinfo {pages} {1657}
  (\bibinfo {year} {2020}{\natexlab{a}})},\ \Eprint
  {http://arxiv.org/abs/2008.05862} {arXiv:2008.05862 [hep-th]} \BibitemShut
  {NoStop}%
\bibitem [{\citenamefont {Kiefer}(1992)}]{Kiefer:1991xy}%
  \BibitemOpen
  \bibfield  {author} {\bibinfo {author} {\bibfnamefont {C.}~\bibnamefont
  {Kiefer}},\ }\href {\doibase 10.1103/PhysRevD.45.2044} {\bibfield  {journal}
  {\bibinfo  {journal} {Phys. Rev. D}\ }\textbf {\bibinfo {volume} {45}},\
  \bibinfo {pages} {2044} (\bibinfo {year} {1992})}\BibitemShut {NoStop}%
\bibitem [{\citenamefont {de~T{\'e}ramond}\ and\ \citenamefont
  {Brodsky}(2009)}]{de2009light}%
  \BibitemOpen
  \bibfield  {author} {\bibinfo {author} {\bibfnamefont {G.~F.}\ \bibnamefont
  {de~T{\'e}ramond}}\ and\ \bibinfo {author} {\bibfnamefont {S.~J.}\
  \bibnamefont {Brodsky}},\ }\href@noop {} {\bibfield  {journal} {\bibinfo
  {journal} {Physical review letters}\ }\textbf {\bibinfo {volume} {102}},\
  \bibinfo {pages} {081601} (\bibinfo {year} {2009})}\BibitemShut {NoStop}%
\bibitem [{\citenamefont {Braun}\ \emph
  {et~al.}(2022{\natexlab{a}})\citenamefont {Braun}, \citenamefont {Ji},\ and\
  \citenamefont {Schoenleber}}]{Braun:2022bpn}%
  \BibitemOpen
  \bibfield  {author} {\bibinfo {author} {\bibfnamefont {V.~M.}\ \bibnamefont
  {Braun}}, \bibinfo {author} {\bibfnamefont {Y.}~\bibnamefont {Ji}}, \ and\
  \bibinfo {author} {\bibfnamefont {J.}~\bibnamefont {Schoenleber}},\ }\href
  {\doibase 10.1103/PhysRevLett.129.172001} {\bibfield  {journal} {\bibinfo
  {journal} {Phys. Rev. Lett.}\ }\textbf {\bibinfo {volume} {129}},\ \bibinfo
  {pages} {172001} (\bibinfo {year} {2022}{\natexlab{a}})},\ \Eprint
  {http://arxiv.org/abs/2207.06818} {arXiv:2207.06818 [hep-ph]} \BibitemShut
  {NoStop}%
\bibitem [{\citenamefont {Braun}\ \emph {et~al.}(2017)\citenamefont {Braun},
  \citenamefont {Manashov}, \citenamefont {Moch},\ and\ \citenamefont
  {Strohmaier}}]{Braun:2017cih}%
  \BibitemOpen
  \bibfield  {author} {\bibinfo {author} {\bibfnamefont {V.~M.}\ \bibnamefont
  {Braun}}, \bibinfo {author} {\bibfnamefont {A.~N.}\ \bibnamefont {Manashov}},
  \bibinfo {author} {\bibfnamefont {S.}~\bibnamefont {Moch}}, \ and\ \bibinfo
  {author} {\bibfnamefont {M.}~\bibnamefont {Strohmaier}},\ }\href {\doibase
  10.1007/JHEP06(2017)037} {\bibfield  {journal} {\bibinfo  {journal} {JHEP}\
  }\textbf {\bibinfo {volume} {06}},\ \bibinfo {pages} {037} (\bibinfo {year}
  {2017})},\ \Eprint {http://arxiv.org/abs/1703.09532} {arXiv:1703.09532
  [hep-ph]} \BibitemShut {NoStop}%
\bibitem [{\citenamefont {Braun}\ \emph
  {et~al.}(2022{\natexlab{b}})\citenamefont {Braun}, \citenamefont
  {Chetyrkin},\ and\ \citenamefont {Manashov}}]{Braun:2022byg}%
  \BibitemOpen
  \bibfield  {author} {\bibinfo {author} {\bibfnamefont {V.~M.}\ \bibnamefont
  {Braun}}, \bibinfo {author} {\bibfnamefont {K.~G.}\ \bibnamefont
  {Chetyrkin}}, \ and\ \bibinfo {author} {\bibfnamefont {A.~N.}\ \bibnamefont
  {Manashov}},\ }\href {\doibase 10.1016/j.physletb.2022.137409} {\bibfield
  {journal} {\bibinfo  {journal} {Phys. Lett. B}\ }\textbf {\bibinfo {volume}
  {834}},\ \bibinfo {pages} {137409} (\bibinfo {year} {2022}{\natexlab{b}})},\
  \Eprint {http://arxiv.org/abs/2205.08228} {arXiv:2205.08228 [hep-ph]}
  \BibitemShut {NoStop}%
\bibitem [{\citenamefont {Schoenleber}(2023)}]{Schoenleber:2022myb}%
  \BibitemOpen
  \bibfield  {author} {\bibinfo {author} {\bibfnamefont {J.}~\bibnamefont
  {Schoenleber}},\ }\href {\doibase 10.1007/JHEP02(2023)207} {\bibfield
  {journal} {\bibinfo  {journal} {JHEP}\ }\textbf {\bibinfo {volume} {02}},\
  \bibinfo {pages} {207} (\bibinfo {year} {2023})},\ \Eprint
  {http://arxiv.org/abs/2209.09015} {arXiv:2209.09015 [hep-ph]} \BibitemShut
  {NoStop}%
\bibitem [{\citenamefont {Braun}\ \emph {et~al.}(2012)\citenamefont {Braun},
  \citenamefont {Manashov},\ and\ \citenamefont {Pirnay}}]{Braun:2012hq}%
  \BibitemOpen
  \bibfield  {author} {\bibinfo {author} {\bibfnamefont {V.~M.}\ \bibnamefont
  {Braun}}, \bibinfo {author} {\bibfnamefont {A.~N.}\ \bibnamefont {Manashov}},
  \ and\ \bibinfo {author} {\bibfnamefont {B.}~\bibnamefont {Pirnay}},\ }\href
  {\doibase 10.1103/PhysRevLett.109.242001} {\bibfield  {journal} {\bibinfo
  {journal} {Phys. Rev. Lett.}\ }\textbf {\bibinfo {volume} {109}},\ \bibinfo
  {pages} {242001} (\bibinfo {year} {2012})},\ \Eprint
  {http://arxiv.org/abs/1209.2559} {arXiv:1209.2559 [hep-ph]} \BibitemShut
  {NoStop}%
\bibitem [{\citenamefont {Braun}\ \emph {et~al.}(2014)\citenamefont {Braun},
  \citenamefont {Manashov}, \citenamefont {M\"uller},\ and\ \citenamefont
  {Pirnay}}]{Braun:2014sta}%
  \BibitemOpen
  \bibfield  {author} {\bibinfo {author} {\bibfnamefont {V.~M.}\ \bibnamefont
  {Braun}}, \bibinfo {author} {\bibfnamefont {A.~N.}\ \bibnamefont {Manashov}},
  \bibinfo {author} {\bibfnamefont {D.}~\bibnamefont {M\"uller}}, \ and\
  \bibinfo {author} {\bibfnamefont {B.~M.}\ \bibnamefont {Pirnay}},\ }\href
  {\doibase 10.1103/PhysRevD.89.074022} {\bibfield  {journal} {\bibinfo
  {journal} {Phys. Rev. D}\ }\textbf {\bibinfo {volume} {89}},\ \bibinfo
  {pages} {074022} (\bibinfo {year} {2014})},\ \Eprint
  {http://arxiv.org/abs/1401.7621} {arXiv:1401.7621 [hep-ph]} \BibitemShut
  {NoStop}%
\bibitem [{\citenamefont {Braun}\ \emph {et~al.}(2023)\citenamefont {Braun},
  \citenamefont {Ji},\ and\ \citenamefont {Manashov}}]{Braun:2022qly}%
  \BibitemOpen
  \bibfield  {author} {\bibinfo {author} {\bibfnamefont {V.~M.}\ \bibnamefont
  {Braun}}, \bibinfo {author} {\bibfnamefont {Y.}~\bibnamefont {Ji}}, \ and\
  \bibinfo {author} {\bibfnamefont {A.~N.}\ \bibnamefont {Manashov}},\ }\href
  {\doibase 10.1007/JHEP01(2023)078} {\bibfield  {journal} {\bibinfo  {journal}
  {JHEP}\ }\textbf {\bibinfo {volume} {01}},\ \bibinfo {pages} {078} (\bibinfo
  {year} {2023})},\ \Eprint {http://arxiv.org/abs/2211.04902} {arXiv:2211.04902
  [hep-ph]} \BibitemShut {NoStop}%
\bibitem [{\citenamefont {Dutrieux}\ \emph {et~al.}(2022)\citenamefont
  {Dutrieux}, \citenamefont {Dutrieux}, \citenamefont {Grocholski},
  \citenamefont {Grocholski}, \citenamefont {Moutarde}, \citenamefont
  {Moutarde}, \citenamefont {Sznajder},\ and\ \citenamefont
  {Sznajder}}]{Dutrieux:2021wll}%
  \BibitemOpen
  \bibfield  {author} {\bibinfo {author} {\bibfnamefont {H.}~\bibnamefont
  {Dutrieux}}, \bibinfo {author} {\bibfnamefont {H.}~\bibnamefont {Dutrieux}},
  \bibinfo {author} {\bibfnamefont {O.}~\bibnamefont {Grocholski}}, \bibinfo
  {author} {\bibfnamefont {O.}~\bibnamefont {Grocholski}}, \bibinfo {author}
  {\bibfnamefont {H.}~\bibnamefont {Moutarde}}, \bibinfo {author}
  {\bibfnamefont {H.}~\bibnamefont {Moutarde}}, \bibinfo {author}
  {\bibfnamefont {P.}~\bibnamefont {Sznajder}}, \ and\ \bibinfo {author}
  {\bibfnamefont {P.}~\bibnamefont {Sznajder}},\ }\href {\doibase
  10.1140/epjc/s10052-022-10211-5} {\bibfield  {journal} {\bibinfo  {journal}
  {Eur. Phys. J. C}\ }\textbf {\bibinfo {volume} {82}},\ \bibinfo {pages} {252}
  (\bibinfo {year} {2022})},\ \bibinfo {note} {[Erratum: Eur.Phys.J.C 82, 389
  (2022)]},\ \Eprint {http://arxiv.org/abs/2112.10528} {arXiv:2112.10528
  [hep-ph]} \BibitemShut {NoStop}%
\bibitem [{\citenamefont {Berthou}\ \emph {et~al.}(2018)\citenamefont {Berthou}
  \emph {et~al.}}]{Berthou:2015oaw}%
  \BibitemOpen
  \bibfield  {author} {\bibinfo {author} {\bibfnamefont {B.}~\bibnamefont
  {Berthou}} \emph {et~al.},\ }\href {\doibase 10.1140/epjc/s10052-018-5948-0}
  {\bibfield  {journal} {\bibinfo  {journal} {Eur. Phys. J. C}\ }\textbf
  {\bibinfo {volume} {78}},\ \bibinfo {pages} {478} (\bibinfo {year} {2018})},\
  \Eprint {http://arxiv.org/abs/1512.06174} {arXiv:1512.06174 [hep-ph]}
  \BibitemShut {NoStop}%
\bibitem [{\citenamefont {Kumericki}\ \emph {et~al.}(2008)\citenamefont
  {Kumericki}, \citenamefont {Mueller},\ and\ \citenamefont
  {Passek-Kumericki}}]{Kumericki:2007sa}%
  \BibitemOpen
  \bibfield  {author} {\bibinfo {author} {\bibfnamefont {K.}~\bibnamefont
  {Kumericki}}, \bibinfo {author} {\bibfnamefont {D.}~\bibnamefont {Mueller}},
  \ and\ \bibinfo {author} {\bibfnamefont {K.}~\bibnamefont
  {Passek-Kumericki}},\ }\href {\doibase 10.1016/j.nuclphysb.2007.10.029}
  {\bibfield  {journal} {\bibinfo  {journal} {Nucl. Phys. B}\ }\textbf
  {\bibinfo {volume} {794}},\ \bibinfo {pages} {244} (\bibinfo {year}
  {2008})},\ \Eprint {http://arxiv.org/abs/hep-ph/0703179}
  {arXiv:hep-ph/0703179} \BibitemShut {NoStop}%
\bibitem [{\citenamefont {Kotzinian}(1995)}]{Kotzinian:1994dv}%
  \BibitemOpen
  \bibfield  {author} {\bibinfo {author} {\bibfnamefont {A.}~\bibnamefont
  {Kotzinian}},\ }\href {\doibase 10.1016/0550-3213(95)00098-D} {\bibfield
  {journal} {\bibinfo  {journal} {Nucl. Phys.}\ }\textbf {\bibinfo {volume}
  {B441}},\ \bibinfo {pages} {234} (\bibinfo {year} {1995})},\ \Eprint
  {http://arxiv.org/abs/hep-ph/9412283} {arXiv:hep-ph/9412283} \BibitemShut
  {NoStop}%
\bibitem [{\citenamefont {Mulders}\ and\ \citenamefont
  {Tangerman}(1996)}]{Mulders:1995dh}%
  \BibitemOpen
  \bibfield  {author} {\bibinfo {author} {\bibfnamefont {P.~J.}\ \bibnamefont
  {Mulders}}\ and\ \bibinfo {author} {\bibfnamefont {R.~D.}\ \bibnamefont
  {Tangerman}},\ }\href {\doibase 10.1016/0550-3213(95)00632-X} {\bibfield
  {journal} {\bibinfo  {journal} {Nucl. Phys.}\ }\textbf {\bibinfo {volume}
  {B461}},\ \bibinfo {pages} {197} (\bibinfo {year} {1996})},\ \Eprint
  {http://arxiv.org/abs/hep-ph/9510301} {arXiv:hep-ph/9510301} \BibitemShut
  {NoStop}%
\bibitem [{\citenamefont {Bacchetta}\ \emph {et~al.}(2007)\citenamefont
  {Bacchetta}, \citenamefont {Diehl}, \citenamefont {Goeke}, \citenamefont
  {Metz}, \citenamefont {Mulders},\ and\ \citenamefont
  {Schlegel}}]{Bacchetta:2006tn}%
  \BibitemOpen
  \bibfield  {author} {\bibinfo {author} {\bibfnamefont {A.}~\bibnamefont
  {Bacchetta}}, \bibinfo {author} {\bibfnamefont {M.}~\bibnamefont {Diehl}},
  \bibinfo {author} {\bibfnamefont {K.}~\bibnamefont {Goeke}}, \bibinfo
  {author} {\bibfnamefont {A.}~\bibnamefont {Metz}}, \bibinfo {author}
  {\bibfnamefont {P.~J.}\ \bibnamefont {Mulders}}, \ and\ \bibinfo {author}
  {\bibfnamefont {M.}~\bibnamefont {Schlegel}},\ }\href {\doibase
  10.1088/1126-6708/2007/02/093} {\bibfield  {journal} {\bibinfo  {journal}
  {JHEP}\ }\textbf {\bibinfo {volume} {02}},\ \bibinfo {pages} {093} (\bibinfo
  {year} {2007})},\ \Eprint {http://arxiv.org/abs/hep-ph/0611265}
  {arXiv:hep-ph/0611265} \BibitemShut {NoStop}%
\bibitem [{\citenamefont {Bacchetta}\ \emph
  {et~al.}(2019{\natexlab{a}})\citenamefont {Bacchetta}, \citenamefont {Bozzi},
  \citenamefont {Echevarria}, \citenamefont {Pisano}, \citenamefont
  {Prokudin},\ and\ \citenamefont {Radici}}]{Bacchetta:2019qkv}%
  \BibitemOpen
  \bibfield  {author} {\bibinfo {author} {\bibfnamefont {A.}~\bibnamefont
  {Bacchetta}}, \bibinfo {author} {\bibfnamefont {G.}~\bibnamefont {Bozzi}},
  \bibinfo {author} {\bibfnamefont {M.~G.}\ \bibnamefont {Echevarria}},
  \bibinfo {author} {\bibfnamefont {C.}~\bibnamefont {Pisano}}, \bibinfo
  {author} {\bibfnamefont {A.}~\bibnamefont {Prokudin}}, \ and\ \bibinfo
  {author} {\bibfnamefont {M.}~\bibnamefont {Radici}},\ }\href {\doibase
  10.1016/j.physletb.2019.134850} {\bibfield  {journal} {\bibinfo  {journal}
  {Phys. Lett. B}\ }\textbf {\bibinfo {volume} {797}},\ \bibinfo {pages}
  {134850} (\bibinfo {year} {2019}{\natexlab{a}})},\ \Eprint
  {http://arxiv.org/abs/1906.07037} {arXiv:1906.07037 [hep-ph]} \BibitemShut
  {NoStop}%
\bibitem [{\citenamefont {Ebert}\ \emph
  {et~al.}(2021{\natexlab{a}})\citenamefont {Ebert}, \citenamefont {Michel},
  \citenamefont {Stewart},\ and\ \citenamefont {Tackmann}}]{Ebert:2020dfc}%
  \BibitemOpen
  \bibfield  {author} {\bibinfo {author} {\bibfnamefont {M.~A.}\ \bibnamefont
  {Ebert}}, \bibinfo {author} {\bibfnamefont {J.~K.~L.}\ \bibnamefont
  {Michel}}, \bibinfo {author} {\bibfnamefont {I.~W.}\ \bibnamefont {Stewart}},
  \ and\ \bibinfo {author} {\bibfnamefont {F.~J.}\ \bibnamefont {Tackmann}},\
  }\href {\doibase 10.1007/JHEP04(2021)102} {\bibfield  {journal} {\bibinfo
  {journal} {JHEP}\ }\textbf {\bibinfo {volume} {04}},\ \bibinfo {pages} {102}
  (\bibinfo {year} {2021}{\natexlab{a}})},\ \Eprint
  {http://arxiv.org/abs/2006.11382} {arXiv:2006.11382 [hep-ph]} \BibitemShut
  {NoStop}%
\bibitem [{\citenamefont {Ebert}\ \emph
  {et~al.}(2022{\natexlab{a}})\citenamefont {Ebert}, \citenamefont {Gao},\ and\
  \citenamefont {Stewart}}]{Ebert:2021jhy}%
  \BibitemOpen
  \bibfield  {author} {\bibinfo {author} {\bibfnamefont {M.~A.}\ \bibnamefont
  {Ebert}}, \bibinfo {author} {\bibfnamefont {A.}~\bibnamefont {Gao}}, \ and\
  \bibinfo {author} {\bibfnamefont {I.~W.}\ \bibnamefont {Stewart}},\ }\href
  {\doibase 10.1007/JHEP06(2022)007} {\bibfield  {journal} {\bibinfo  {journal}
  {JHEP}\ }\textbf {\bibinfo {volume} {06}},\ \bibinfo {pages} {007} (\bibinfo
  {year} {2022}{\natexlab{a}})},\ \Eprint {http://arxiv.org/abs/2112.07680}
  {arXiv:2112.07680 [hep-ph]} \BibitemShut {NoStop}%
\bibitem [{\citenamefont {Vladimirov}\ \emph {et~al.}(2022)\citenamefont
  {Vladimirov}, \citenamefont {Moos},\ and\ \citenamefont
  {Scimemi}}]{Vladimirov:2021hdn}%
  \BibitemOpen
  \bibfield  {author} {\bibinfo {author} {\bibfnamefont {A.}~\bibnamefont
  {Vladimirov}}, \bibinfo {author} {\bibfnamefont {V.}~\bibnamefont {Moos}}, \
  and\ \bibinfo {author} {\bibfnamefont {I.}~\bibnamefont {Scimemi}},\ }\href
  {\doibase 10.1007/JHEP01(2022)110} {\bibfield  {journal} {\bibinfo  {journal}
  {JHEP}\ }\textbf {\bibinfo {volume} {01}},\ \bibinfo {pages} {110} (\bibinfo
  {year} {2022})},\ \Eprint {http://arxiv.org/abs/2109.09771} {arXiv:2109.09771
  [hep-ph]} \BibitemShut {NoStop}%
\bibitem [{\citenamefont {Rodini}\ and\ \citenamefont
  {Vladimirov}(2022)}]{Rodini:2022wki}%
  \BibitemOpen
  \bibfield  {author} {\bibinfo {author} {\bibfnamefont {S.}~\bibnamefont
  {Rodini}}\ and\ \bibinfo {author} {\bibfnamefont {A.}~\bibnamefont
  {Vladimirov}},\ }\href {\doibase 10.1007/JHEP08(2022)031} {\bibfield
  {journal} {\bibinfo  {journal} {JHEP}\ }\textbf {\bibinfo {volume} {08}},\
  \bibinfo {pages} {031} (\bibinfo {year} {2022})},\ \bibinfo {note} {[Erratum:
  JHEP 12, 048 (2022)]},\ \Eprint {http://arxiv.org/abs/2204.03856}
  {arXiv:2204.03856 [hep-ph]} \BibitemShut {NoStop}%
\bibitem [{\citenamefont {Ebert}\ \emph
  {et~al.}(2022{\natexlab{b}})\citenamefont {Ebert}, \citenamefont {Michel},
  \citenamefont {Stewart},\ and\ \citenamefont {Sun}}]{Ebert:2022cku}%
  \BibitemOpen
  \bibfield  {author} {\bibinfo {author} {\bibfnamefont {M.~A.}\ \bibnamefont
  {Ebert}}, \bibinfo {author} {\bibfnamefont {J.~K.~L.}\ \bibnamefont
  {Michel}}, \bibinfo {author} {\bibfnamefont {I.~W.}\ \bibnamefont {Stewart}},
  \ and\ \bibinfo {author} {\bibfnamefont {Z.}~\bibnamefont {Sun}},\ }\href
  {\doibase 10.1007/JHEP07(2022)129} {\bibfield  {journal} {\bibinfo  {journal}
  {JHEP}\ }\textbf {\bibinfo {volume} {07}},\ \bibinfo {pages} {129} (\bibinfo
  {year} {2022}{\natexlab{b}})},\ \Eprint {http://arxiv.org/abs/2201.07237}
  {arXiv:2201.07237 [hep-ph]} \BibitemShut {NoStop}%
\bibitem [{\citenamefont {Gao}\ \emph {et~al.}(2022{\natexlab{a}})\citenamefont
  {Gao}, \citenamefont {Michel}, \citenamefont {Stewart},\ and\ \citenamefont
  {Sun}}]{Gao:2022bzi}%
  \BibitemOpen
  \bibfield  {author} {\bibinfo {author} {\bibfnamefont {A.}~\bibnamefont
  {Gao}}, \bibinfo {author} {\bibfnamefont {J.~K.~L.}\ \bibnamefont {Michel}},
  \bibinfo {author} {\bibfnamefont {I.~W.}\ \bibnamefont {Stewart}}, \ and\
  \bibinfo {author} {\bibfnamefont {Z.}~\bibnamefont {Sun}},\ }\href@noop {} {\
   (\bibinfo {year} {2022}{\natexlab{a}})},\ \Eprint
  {http://arxiv.org/abs/2209.11211} {arXiv:2209.11211 [hep-ph]} \BibitemShut
  {NoStop}%
\bibitem [{\citenamefont {Gamberg}\ \emph {et~al.}(2022)\citenamefont
  {Gamberg}, \citenamefont {Kang}, \citenamefont {Shao}, \citenamefont
  {Terry},\ and\ \citenamefont {Zhao}}]{Gamberg:2022lju}%
  \BibitemOpen
  \bibfield  {author} {\bibinfo {author} {\bibfnamefont {L.}~\bibnamefont
  {Gamberg}}, \bibinfo {author} {\bibfnamefont {Z.-B.}\ \bibnamefont {Kang}},
  \bibinfo {author} {\bibfnamefont {D.~Y.}\ \bibnamefont {Shao}}, \bibinfo
  {author} {\bibfnamefont {J.}~\bibnamefont {Terry}}, \ and\ \bibinfo {author}
  {\bibfnamefont {F.}~\bibnamefont {Zhao}},\ }\href@noop {} {\  (\bibinfo
  {year} {2022})},\ \Eprint {http://arxiv.org/abs/2211.13209} {arXiv:2211.13209
  [hep-ph]} \BibitemShut {NoStop}%
\bibitem [{\citenamefont {Bacchetta}\ \emph {et~al.}(2017)\citenamefont
  {Bacchetta}, \citenamefont {Delcarro}, \citenamefont {Pisano}, \citenamefont
  {Radici},\ and\ \citenamefont {Signori}}]{Bacchetta:2017gcc}%
  \BibitemOpen
  \bibfield  {author} {\bibinfo {author} {\bibfnamefont {A.}~\bibnamefont
  {Bacchetta}}, \bibinfo {author} {\bibfnamefont {F.}~\bibnamefont {Delcarro}},
  \bibinfo {author} {\bibfnamefont {C.}~\bibnamefont {Pisano}}, \bibinfo
  {author} {\bibfnamefont {M.}~\bibnamefont {Radici}}, \ and\ \bibinfo {author}
  {\bibfnamefont {A.}~\bibnamefont {Signori}},\ }\href {\doibase
  10.1007/JHEP06(2017)081} {\bibfield  {journal} {\bibinfo  {journal} {JHEP}\
  }\textbf {\bibinfo {volume} {06}},\ \bibinfo {pages} {081} (\bibinfo {year}
  {2017})},\ \bibinfo {note} {[Erratum: JHEP 06, 051 (2019)]},\ \Eprint
  {http://arxiv.org/abs/1703.10157} {arXiv:1703.10157 [hep-ph]} \BibitemShut
  {NoStop}%
\bibitem [{\citenamefont {Scimemi}\ and\ \citenamefont
  {Vladimirov}(2020)}]{Scimemi:2019cmh}%
  \BibitemOpen
  \bibfield  {author} {\bibinfo {author} {\bibfnamefont {I.}~\bibnamefont
  {Scimemi}}\ and\ \bibinfo {author} {\bibfnamefont {A.}~\bibnamefont
  {Vladimirov}},\ }\href {\doibase 10.1007/JHEP06(2020)137} {\bibfield
  {journal} {\bibinfo  {journal} {JHEP}\ }\textbf {\bibinfo {volume} {06}},\
  \bibinfo {pages} {137} (\bibinfo {year} {2020})},\ \Eprint
  {http://arxiv.org/abs/1912.06532} {arXiv:1912.06532 [hep-ph]} \BibitemShut
  {NoStop}%
\bibitem [{\citenamefont {Echevarria}\ \emph {et~al.}(2021)\citenamefont
  {Echevarria}, \citenamefont {Kang},\ and\ \citenamefont
  {Terry}}]{Echevarria:2020hpy}%
  \BibitemOpen
  \bibfield  {author} {\bibinfo {author} {\bibfnamefont {M.~G.}\ \bibnamefont
  {Echevarria}}, \bibinfo {author} {\bibfnamefont {Z.-B.}\ \bibnamefont
  {Kang}}, \ and\ \bibinfo {author} {\bibfnamefont {J.}~\bibnamefont {Terry}},\
  }\href {\doibase 10.1007/JHEP01(2021)126} {\bibfield  {journal} {\bibinfo
  {journal} {JHEP}\ }\textbf {\bibinfo {volume} {01}},\ \bibinfo {pages} {126}
  (\bibinfo {year} {2021})},\ \Eprint {http://arxiv.org/abs/2009.10710}
  {arXiv:2009.10710 [hep-ph]} \BibitemShut {NoStop}%
\bibitem [{\citenamefont {Bury}\ \emph
  {et~al.}(2021{\natexlab{a}})\citenamefont {Bury}, \citenamefont {Prokudin},\
  and\ \citenamefont {Vladimirov}}]{Bury:2020vhj}%
  \BibitemOpen
  \bibfield  {author} {\bibinfo {author} {\bibfnamefont {M.}~\bibnamefont
  {Bury}}, \bibinfo {author} {\bibfnamefont {A.}~\bibnamefont {Prokudin}}, \
  and\ \bibinfo {author} {\bibfnamefont {A.}~\bibnamefont {Vladimirov}},\
  }\href {\doibase 10.1103/PhysRevLett.126.112002} {\bibfield  {journal}
  {\bibinfo  {journal} {Phys. Rev. Lett.}\ }\textbf {\bibinfo {volume} {126}},\
  \bibinfo {pages} {112002} (\bibinfo {year} {2021}{\natexlab{a}})},\ \Eprint
  {http://arxiv.org/abs/2012.05135} {arXiv:2012.05135 [hep-ph]} \BibitemShut
  {NoStop}%
\bibitem [{\citenamefont {Bury}\ \emph
  {et~al.}(2021{\natexlab{b}})\citenamefont {Bury}, \citenamefont {Prokudin},\
  and\ \citenamefont {Vladimirov}}]{Bury:2021sue}%
  \BibitemOpen
  \bibfield  {author} {\bibinfo {author} {\bibfnamefont {M.}~\bibnamefont
  {Bury}}, \bibinfo {author} {\bibfnamefont {A.}~\bibnamefont {Prokudin}}, \
  and\ \bibinfo {author} {\bibfnamefont {A.}~\bibnamefont {Vladimirov}},\
  }\href {\doibase 10.1007/JHEP05(2021)151} {\bibfield  {journal} {\bibinfo
  {journal} {JHEP}\ }\textbf {\bibinfo {volume} {05}},\ \bibinfo {pages} {151}
  (\bibinfo {year} {2021}{\natexlab{b}})},\ \Eprint
  {http://arxiv.org/abs/2103.03270} {arXiv:2103.03270 [hep-ph]} \BibitemShut
  {NoStop}%
\bibitem [{\citenamefont {Alrashed}\ \emph {et~al.}(2022)\citenamefont
  {Alrashed}, \citenamefont {Anderle}, \citenamefont {Kang}, \citenamefont
  {Terry},\ and\ \citenamefont {Xing}}]{Alrashed:2021csd}%
  \BibitemOpen
  \bibfield  {author} {\bibinfo {author} {\bibfnamefont {M.}~\bibnamefont
  {Alrashed}}, \bibinfo {author} {\bibfnamefont {D.}~\bibnamefont {Anderle}},
  \bibinfo {author} {\bibfnamefont {Z.-B.}\ \bibnamefont {Kang}}, \bibinfo
  {author} {\bibfnamefont {J.}~\bibnamefont {Terry}}, \ and\ \bibinfo {author}
  {\bibfnamefont {H.}~\bibnamefont {Xing}},\ }\href {\doibase
  10.1103/PhysRevLett.129.242001} {\bibfield  {journal} {\bibinfo  {journal}
  {Phys. Rev. Lett.}\ }\textbf {\bibinfo {volume} {129}},\ \bibinfo {pages}
  {242001} (\bibinfo {year} {2022})},\ \Eprint
  {http://arxiv.org/abs/2107.12401} {arXiv:2107.12401 [hep-ph]} \BibitemShut
  {NoStop}%
\bibitem [{\citenamefont {Bacchetta}\ \emph
  {et~al.}(2022{\natexlab{a}})\citenamefont {Bacchetta}, \citenamefont
  {Delcarro}, \citenamefont {Pisano},\ and\ \citenamefont
  {Radici}}]{Bacchetta:2020gko}%
  \BibitemOpen
  \bibfield  {author} {\bibinfo {author} {\bibfnamefont {A.}~\bibnamefont
  {Bacchetta}}, \bibinfo {author} {\bibfnamefont {F.}~\bibnamefont {Delcarro}},
  \bibinfo {author} {\bibfnamefont {C.}~\bibnamefont {Pisano}}, \ and\ \bibinfo
  {author} {\bibfnamefont {M.}~\bibnamefont {Radici}},\ }\href {\doibase
  10.1016/j.physletb.2022.136961} {\bibfield  {journal} {\bibinfo  {journal}
  {Phys. Lett. B}\ }\textbf {\bibinfo {volume} {827}},\ \bibinfo {pages}
  {136961} (\bibinfo {year} {2022}{\natexlab{a}})},\ \Eprint
  {http://arxiv.org/abs/2004.14278} {arXiv:2004.14278 [hep-ph]} \BibitemShut
  {NoStop}%
\bibitem [{\citenamefont {Bacchetta}\ \emph
  {et~al.}(2022{\natexlab{b}})\citenamefont {Bacchetta}, \citenamefont
  {Bertone}, \citenamefont {Bissolotti}, \citenamefont {Bozzi}, \citenamefont
  {Cerutti}, \citenamefont {Piacenza}, \citenamefont {Radici},\ and\
  \citenamefont {Signori}}]{Bacchetta:2022awv}%
  \BibitemOpen
  \bibfield  {author} {\bibinfo {author} {\bibfnamefont {A.}~\bibnamefont
  {Bacchetta}}, \bibinfo {author} {\bibfnamefont {V.}~\bibnamefont {Bertone}},
  \bibinfo {author} {\bibfnamefont {C.}~\bibnamefont {Bissolotti}}, \bibinfo
  {author} {\bibfnamefont {G.}~\bibnamefont {Bozzi}}, \bibinfo {author}
  {\bibfnamefont {M.}~\bibnamefont {Cerutti}}, \bibinfo {author} {\bibfnamefont
  {F.}~\bibnamefont {Piacenza}}, \bibinfo {author} {\bibfnamefont
  {M.}~\bibnamefont {Radici}}, \ and\ \bibinfo {author} {\bibfnamefont
  {A.}~\bibnamefont {Signori}} (\bibinfo {collaboration} {WMAP}),\ }\href
  {\doibase 10.1007/JHEP10(2022)127} {\bibfield  {journal} {\bibinfo  {journal}
  {JHEP}\ }\textbf {\bibinfo {volume} {10}},\ \bibinfo {pages} {127} (\bibinfo
  {year} {2022}{\natexlab{b}})},\ \Eprint {http://arxiv.org/abs/2206.07598}
  {arXiv:2206.07598 [hep-ph]} \BibitemShut {NoStop}%
\bibitem [{\citenamefont {Barry}\ \emph {et~al.}(2023)\citenamefont {Barry},
  \citenamefont {Gamberg}, \citenamefont {Melnitchouk}, \citenamefont {Moffat},
  \citenamefont {Pitonyak}, \citenamefont {Prokudin},\ and\ \citenamefont
  {Sato}}]{Barry:2023qqh}%
  \BibitemOpen
  \bibfield  {author} {\bibinfo {author} {\bibfnamefont {P.~C.}\ \bibnamefont
  {Barry}}, \bibinfo {author} {\bibfnamefont {L.}~\bibnamefont {Gamberg}},
  \bibinfo {author} {\bibfnamefont {W.}~\bibnamefont {Melnitchouk}}, \bibinfo
  {author} {\bibfnamefont {E.}~\bibnamefont {Moffat}}, \bibinfo {author}
  {\bibfnamefont {D.}~\bibnamefont {Pitonyak}}, \bibinfo {author}
  {\bibfnamefont {A.}~\bibnamefont {Prokudin}}, \ and\ \bibinfo {author}
  {\bibfnamefont {N.}~\bibnamefont {Sato}},\ }\href@noop {} {\  (\bibinfo
  {year} {2023})},\ \Eprint {http://arxiv.org/abs/2302.01192} {arXiv:2302.01192
  [hep-ph]} \BibitemShut {NoStop}%
\bibitem [{\citenamefont {Amoroso}\ \emph {et~al.}(2022)\citenamefont {Amoroso}
  \emph {et~al.}}]{Amoroso:2022eow}%
  \BibitemOpen
  \bibfield  {author} {\bibinfo {author} {\bibfnamefont {S.}~\bibnamefont
  {Amoroso}} \emph {et~al.},\ }\href@noop {} {\  (\bibinfo {year} {2022})},\
  \Eprint {http://arxiv.org/abs/2203.13923} {arXiv:2203.13923 [hep-ph]}
  \BibitemShut {NoStop}%
\bibitem [{\citenamefont {Abe}\ \emph {et~al.}(1988)\citenamefont {Abe} \emph
  {et~al.}}]{CDF:1988lbl}%
  \BibitemOpen
  \bibfield  {author} {\bibinfo {author} {\bibfnamefont {F.}~\bibnamefont
  {Abe}} \emph {et~al.} (\bibinfo {collaboration} {CDF}),\ }\href {\doibase
  10.1016/0168-9002(88)90298-7} {\bibfield  {journal} {\bibinfo  {journal}
  {Nucl. Instrum. Meth. A}\ }\textbf {\bibinfo {volume} {271}},\ \bibinfo
  {pages} {387} (\bibinfo {year} {1988})}\BibitemShut {NoStop}%
\bibitem [{\citenamefont {Airapetian}\ \emph {et~al.}(2005)\citenamefont
  {Airapetian} \emph {et~al.}}]{HERMES:2004vsf}%
  \BibitemOpen
  \bibfield  {author} {\bibinfo {author} {\bibfnamefont {A.}~\bibnamefont
  {Airapetian}} \emph {et~al.} (\bibinfo {collaboration} {HERMES}),\ }\href
  {\doibase 10.1016/j.nima.2004.11.020} {\bibfield  {journal} {\bibinfo
  {journal} {Nucl. Instrum. Meth. A}\ }\textbf {\bibinfo {volume} {540}},\
  \bibinfo {pages} {68} (\bibinfo {year} {2005})},\ \Eprint
  {http://arxiv.org/abs/physics/0408137} {arXiv:physics/0408137} \BibitemShut
  {NoStop}%
\bibitem [{\citenamefont {Jaeckel}\ \emph {et~al.}(2020)\citenamefont
  {Jaeckel}, \citenamefont {Lamont},\ and\ \citenamefont
  {Vall\'ee}}]{Jaeckel:2020dxj}%
  \BibitemOpen
  \bibfield  {author} {\bibinfo {author} {\bibfnamefont {J.}~\bibnamefont
  {Jaeckel}}, \bibinfo {author} {\bibfnamefont {M.}~\bibnamefont {Lamont}}, \
  and\ \bibinfo {author} {\bibfnamefont {C.}~\bibnamefont {Vall\'ee}},\ }\href
  {\doibase 10.1038/s41567-020-0838-4} {\bibfield  {journal} {\bibinfo
  {journal} {Nature Phys.}\ }\textbf {\bibinfo {volume} {16}},\ \bibinfo
  {pages} {393} (\bibinfo {year} {2020})}\BibitemShut {NoStop}%
\bibitem [{\citenamefont {Hadjidakis}\ \emph {et~al.}(2021)\citenamefont
  {Hadjidakis} \emph {et~al.}}]{Hadjidakis:2018ifr}%
  \BibitemOpen
  \bibfield  {author} {\bibinfo {author} {\bibfnamefont {C.}~\bibnamefont
  {Hadjidakis}} \emph {et~al.},\ }\href {\doibase
  10.1016/j.physrep.2021.01.002} {\bibfield  {journal} {\bibinfo  {journal}
  {Phys. Rept.}\ }\textbf {\bibinfo {volume} {911}},\ \bibinfo {pages} {1}
  (\bibinfo {year} {2021})},\ \Eprint {http://arxiv.org/abs/1807.00603}
  {arXiv:1807.00603 [hep-ex]} \BibitemShut {NoStop}%
\bibitem [{\citenamefont {Gautheron}\ \emph {et~al.}(2010)\citenamefont
  {Gautheron} \emph {et~al.}}]{COMPASS:2010shj}%
  \BibitemOpen
  \bibfield  {author} {\bibinfo {author} {\bibfnamefont {F.}~\bibnamefont
  {Gautheron}} \emph {et~al.} (\bibinfo {collaboration} {COMPASS}),\
  }\href@noop {} {\  (\bibinfo {year} {2010})}\BibitemShut {NoStop}%
\bibitem [{\citenamefont {Aschenauer}\ \emph {et~al.}(2015)\citenamefont
  {Aschenauer} \emph {et~al.}}]{Aschenauer:2015eha}%
  \BibitemOpen
  \bibfield  {author} {\bibinfo {author} {\bibfnamefont {E.-C.}\ \bibnamefont
  {Aschenauer}} \emph {et~al.},\ }\href@noop {} {\  (\bibinfo {year} {2015})},\
  \Eprint {http://arxiv.org/abs/1501.01220} {arXiv:1501.01220 [nucl-ex]}
  \BibitemShut {NoStop}%
\bibitem [{\citenamefont {Aschenauer}\ \emph {et~al.}(2016)\citenamefont
  {Aschenauer} \emph {et~al.}}]{Aschenauer:2016our}%
  \BibitemOpen
  \bibfield  {author} {\bibinfo {author} {\bibfnamefont {E.-C.}\ \bibnamefont
  {Aschenauer}} \emph {et~al.},\ }\href@noop {} {\  (\bibinfo {year} {2016})},\
  \Eprint {http://arxiv.org/abs/1602.03922} {arXiv:1602.03922 [nucl-ex]}
  \BibitemShut {NoStop}%
\bibitem [{\citenamefont {Dudek}\ \emph {et~al.}(2012)\citenamefont {Dudek}
  \emph {et~al.}}]{Dudek:2012vr}%
  \BibitemOpen
  \bibfield  {author} {\bibinfo {author} {\bibfnamefont {J.}~\bibnamefont
  {Dudek}} \emph {et~al.},\ }\href {\doibase 10.1140/epja/i2012-12187-1}
  {\bibfield  {journal} {\bibinfo  {journal} {Eur. Phys. J. A}\ }\textbf
  {\bibinfo {volume} {48}},\ \bibinfo {pages} {187} (\bibinfo {year} {2012})},\
  \Eprint {http://arxiv.org/abs/1208.1244} {arXiv:1208.1244 [hep-ex]}
  \BibitemShut {NoStop}%
\bibitem [{\citenamefont {Abe}\ \emph {et~al.}(2010)\citenamefont {Abe} \emph
  {et~al.}}]{Belle-II:2010dht}%
  \BibitemOpen
  \bibfield  {author} {\bibinfo {author} {\bibfnamefont {T.}~\bibnamefont
  {Abe}} \emph {et~al.} (\bibinfo {collaboration} {Belle-II}),\ }\href@noop {}
  {\  (\bibinfo {year} {2010})},\ \Eprint {http://arxiv.org/abs/1011.0352}
  {arXiv:1011.0352 [physics.ins-det]} \BibitemShut {NoStop}%
\bibitem [{\citenamefont {Anderle}\ \emph {et~al.}(2021)\citenamefont {Anderle}
  \emph {et~al.}}]{Anderle:2021wcy}%
  \BibitemOpen
  \bibfield  {author} {\bibinfo {author} {\bibfnamefont {D.~P.}\ \bibnamefont
  {Anderle}} \emph {et~al.},\ }\href {\doibase 10.1007/s11467-021-1062-0}
  {\bibfield  {journal} {\bibinfo  {journal} {Front. Phys. (Beijing)}\ }\textbf
  {\bibinfo {volume} {16}},\ \bibinfo {pages} {64701} (\bibinfo {year}
  {2021})},\ \Eprint {http://arxiv.org/abs/2102.09222} {arXiv:2102.09222
  [nucl-ex]} \BibitemShut {NoStop}%
\bibitem [{\citenamefont {Boer}\ \emph {et~al.}(2011)\citenamefont {Boer} \emph
  {et~al.}}]{Boer:2011fh}%
  \BibitemOpen
  \bibfield  {author} {\bibinfo {author} {\bibfnamefont {D.}~\bibnamefont
  {Boer}} \emph {et~al.},\ }\href@noop {} {\  (\bibinfo {year} {2011})},\
  \Eprint {http://arxiv.org/abs/1108.1713} {arXiv:1108.1713 [nucl-th]}
  \BibitemShut {NoStop}%
\bibitem [{\citenamefont {Accardi}\ \emph {et~al.}(2016)\citenamefont {Accardi}
  \emph {et~al.}}]{Accardi:2012qut}%
  \BibitemOpen
  \bibfield  {author} {\bibinfo {author} {\bibfnamefont {A.}~\bibnamefont
  {Accardi}} \emph {et~al.},\ }\href {\doibase 10.1140/epja/i2016-16268-9}
  {\bibfield  {journal} {\bibinfo  {journal} {Eur. Phys. J. A}\ }\textbf
  {\bibinfo {volume} {52}},\ \bibinfo {pages} {268} (\bibinfo {year} {2016})},\
  \Eprint {http://arxiv.org/abs/1212.1701} {arXiv:1212.1701 [nucl-ex]}
  \BibitemShut {NoStop}%
\bibitem [{\citenamefont {Borsa}\ \emph
  {et~al.}(2020{\natexlab{a}})\citenamefont {Borsa}, \citenamefont {Lucero},
  \citenamefont {Sassot}, \citenamefont {Aschenauer},\ and\ \citenamefont
  {Nunes}}]{Borsa:2020lsz}%
  \BibitemOpen
  \bibfield  {author} {\bibinfo {author} {\bibfnamefont {I.}~\bibnamefont
  {Borsa}}, \bibinfo {author} {\bibfnamefont {G.}~\bibnamefont {Lucero}},
  \bibinfo {author} {\bibfnamefont {R.}~\bibnamefont {Sassot}}, \bibinfo
  {author} {\bibfnamefont {E.~C.}\ \bibnamefont {Aschenauer}}, \ and\ \bibinfo
  {author} {\bibfnamefont {A.~S.}\ \bibnamefont {Nunes}},\ }\href {\doibase
  10.1103/PhysRevD.102.094018} {\bibfield  {journal} {\bibinfo  {journal}
  {Phys. Rev. D}\ }\textbf {\bibinfo {volume} {102}},\ \bibinfo {pages}
  {094018} (\bibinfo {year} {2020}{\natexlab{a}})},\ \Eprint
  {http://arxiv.org/abs/2007.08300} {arXiv:2007.08300 [hep-ph]} \BibitemShut
  {NoStop}%
\bibitem [{\citenamefont {Kanazawa}\ \emph {et~al.}(2014)\citenamefont
  {Kanazawa}, \citenamefont {Lorc\'e}, \citenamefont {Metz}, \citenamefont
  {Pasquini},\ and\ \citenamefont {Schlegel}}]{Kanazawa:2014nha}%
  \BibitemOpen
  \bibfield  {author} {\bibinfo {author} {\bibfnamefont {K.}~\bibnamefont
  {Kanazawa}}, \bibinfo {author} {\bibfnamefont {C.}~\bibnamefont {Lorc\'e}},
  \bibinfo {author} {\bibfnamefont {A.}~\bibnamefont {Metz}}, \bibinfo {author}
  {\bibfnamefont {B.}~\bibnamefont {Pasquini}}, \ and\ \bibinfo {author}
  {\bibfnamefont {M.}~\bibnamefont {Schlegel}},\ }\href {\doibase
  10.1103/PhysRevD.90.014028} {\bibfield  {journal} {\bibinfo  {journal} {Phys.
  Rev. D}\ }\textbf {\bibinfo {volume} {90}},\ \bibinfo {pages} {014028}
  (\bibinfo {year} {2014})},\ \Eprint {http://arxiv.org/abs/1403.5226}
  {arXiv:1403.5226 [hep-ph]} \BibitemShut {NoStop}%
\bibitem [{\citenamefont {Echevarria}\ \emph {et~al.}(2016)\citenamefont
  {Echevarria}, \citenamefont {Idilbi}, \citenamefont {Kanazawa}, \citenamefont
  {Lorc\'e}, \citenamefont {Metz}, \citenamefont {Pasquini},\ and\
  \citenamefont {Schlegel}}]{Echevarria:2016mrc}%
  \BibitemOpen
  \bibfield  {author} {\bibinfo {author} {\bibfnamefont {M.~G.}\ \bibnamefont
  {Echevarria}}, \bibinfo {author} {\bibfnamefont {A.}~\bibnamefont {Idilbi}},
  \bibinfo {author} {\bibfnamefont {K.}~\bibnamefont {Kanazawa}}, \bibinfo
  {author} {\bibfnamefont {C.}~\bibnamefont {Lorc\'e}}, \bibinfo {author}
  {\bibfnamefont {A.}~\bibnamefont {Metz}}, \bibinfo {author} {\bibfnamefont
  {B.}~\bibnamefont {Pasquini}}, \ and\ \bibinfo {author} {\bibfnamefont
  {M.}~\bibnamefont {Schlegel}},\ }\href {\doibase
  10.1016/j.physletb.2016.05.086} {\bibfield  {journal} {\bibinfo  {journal}
  {Phys. Lett. B}\ }\textbf {\bibinfo {volume} {759}},\ \bibinfo {pages} {336}
  (\bibinfo {year} {2016})},\ \Eprint {http://arxiv.org/abs/1602.06953}
  {arXiv:1602.06953 [hep-ph]} \BibitemShut {NoStop}%
\bibitem [{\citenamefont {Bertone}(2022)}]{Bertone:2022awq}%
  \BibitemOpen
  \bibfield  {author} {\bibinfo {author} {\bibfnamefont {V.}~\bibnamefont
  {Bertone}},\ }\href {\doibase 10.1140/epjc/s10052-022-10863-3} {\bibfield
  {journal} {\bibinfo  {journal} {Eur. Phys. J. C}\ }\textbf {\bibinfo {volume}
  {82}},\ \bibinfo {pages} {941} (\bibinfo {year} {2022})},\ \Eprint
  {http://arxiv.org/abs/2207.09526} {arXiv:2207.09526 [hep-ph]} \BibitemShut
  {NoStop}%
\bibitem [{\citenamefont {Echevarria}\ \emph {et~al.}(2022)\citenamefont
  {Echevarria}, \citenamefont {Gutierrez~Garcia},\ and\ \citenamefont
  {Scimemi}}]{Echevarria:2022ztg}%
  \BibitemOpen
  \bibfield  {author} {\bibinfo {author} {\bibfnamefont {M.~G.}\ \bibnamefont
  {Echevarria}}, \bibinfo {author} {\bibfnamefont {P.~A.}\ \bibnamefont
  {Gutierrez~Garcia}}, \ and\ \bibinfo {author} {\bibfnamefont
  {I.}~\bibnamefont {Scimemi}},\ }\href@noop {} {\  (\bibinfo {year} {2022})},\
  \Eprint {http://arxiv.org/abs/2208.00021} {arXiv:2208.00021 [hep-ph]}
  \BibitemShut {NoStop}%
\bibitem [{\citenamefont {Hatta}\ \emph {et~al.}(2016)\citenamefont {Hatta},
  \citenamefont {Xiao},\ and\ \citenamefont {Yuan}}]{Hatta:2016dxp}%
  \BibitemOpen
  \bibfield  {author} {\bibinfo {author} {\bibfnamefont {Y.}~\bibnamefont
  {Hatta}}, \bibinfo {author} {\bibfnamefont {B.-W.}\ \bibnamefont {Xiao}}, \
  and\ \bibinfo {author} {\bibfnamefont {F.}~\bibnamefont {Yuan}},\ }\href
  {\doibase 10.1103/PhysRevLett.116.202301} {\bibfield  {journal} {\bibinfo
  {journal} {Phys. Rev. Lett.}\ }\textbf {\bibinfo {volume} {116}},\ \bibinfo
  {pages} {202301} (\bibinfo {year} {2016})},\ \Eprint
  {http://arxiv.org/abs/1601.01585} {arXiv:1601.01585 [hep-ph]} \BibitemShut
  {NoStop}%
\bibitem [{\citenamefont {M\"antysaari}\ \emph {et~al.}(2019)\citenamefont
  {M\"antysaari}, \citenamefont {Mueller},\ and\ \citenamefont
  {Schenke}}]{Mantysaari:2019csc}%
  \BibitemOpen
  \bibfield  {author} {\bibinfo {author} {\bibfnamefont {H.}~\bibnamefont
  {M\"antysaari}}, \bibinfo {author} {\bibfnamefont {N.}~\bibnamefont
  {Mueller}}, \ and\ \bibinfo {author} {\bibfnamefont {B.}~\bibnamefont
  {Schenke}},\ }\href {\doibase 10.1103/PhysRevD.99.074004} {\bibfield
  {journal} {\bibinfo  {journal} {Phys. Rev. D}\ }\textbf {\bibinfo {volume}
  {99}},\ \bibinfo {pages} {074004} (\bibinfo {year} {2019})},\ \Eprint
  {http://arxiv.org/abs/1902.05087} {arXiv:1902.05087 [hep-ph]} \BibitemShut
  {NoStop}%
\bibitem [{\citenamefont {Boer}\ and\ \citenamefont
  {Setyadi}(2021)}]{Boer:2021upt}%
  \BibitemOpen
  \bibfield  {author} {\bibinfo {author} {\bibfnamefont {D.}~\bibnamefont
  {Boer}}\ and\ \bibinfo {author} {\bibfnamefont {C.}~\bibnamefont {Setyadi}},\
  }\href {\doibase 10.1103/PhysRevD.104.074006} {\bibfield  {journal} {\bibinfo
   {journal} {Phys. Rev. D}\ }\textbf {\bibinfo {volume} {104}},\ \bibinfo
  {pages} {074006} (\bibinfo {year} {2021})},\ \Eprint
  {http://arxiv.org/abs/2106.15148} {arXiv:2106.15148 [hep-ph]} \BibitemShut
  {NoStop}%
\bibitem [{\citenamefont {Byer}\ \emph {et~al.}(2023)\citenamefont {Byer},
  \citenamefont {Khachatryan}, \citenamefont {Gao}, \citenamefont {Akushevich},
  \citenamefont {Ilyichev}, \citenamefont {Peng}, \citenamefont {Prokudin},
  \citenamefont {Srednyak},\ and\ \citenamefont {Zhao}}]{Byer:2022bqf}%
  \BibitemOpen
  \bibfield  {author} {\bibinfo {author} {\bibfnamefont {D.}~\bibnamefont
  {Byer}}, \bibinfo {author} {\bibfnamefont {V.}~\bibnamefont {Khachatryan}},
  \bibinfo {author} {\bibfnamefont {H.}~\bibnamefont {Gao}}, \bibinfo {author}
  {\bibfnamefont {I.}~\bibnamefont {Akushevich}}, \bibinfo {author}
  {\bibfnamefont {A.}~\bibnamefont {Ilyichev}}, \bibinfo {author}
  {\bibfnamefont {C.}~\bibnamefont {Peng}}, \bibinfo {author} {\bibfnamefont
  {A.}~\bibnamefont {Prokudin}}, \bibinfo {author} {\bibfnamefont
  {S.}~\bibnamefont {Srednyak}}, \ and\ \bibinfo {author} {\bibfnamefont
  {Z.}~\bibnamefont {Zhao}},\ }\href {\doibase 10.1016/j.cpc.2023.108702}
  {\bibfield  {journal} {\bibinfo  {journal} {Comput. Phys. Commun.}\ }\textbf
  {\bibinfo {volume} {287}},\ \bibinfo {pages} {108702} (\bibinfo {year}
  {2023})},\ \Eprint {http://arxiv.org/abs/2210.03785} {arXiv:2210.03785
  [hep-ph]} \BibitemShut {NoStop}%
\bibitem [{\citenamefont {Liu}\ \emph {et~al.}(2021{\natexlab{a}})\citenamefont
  {Liu}, \citenamefont {Melnitchouk}, \citenamefont {Qiu},\ and\ \citenamefont
  {Sato}}]{Liu:2021jfp}%
  \BibitemOpen
  \bibfield  {author} {\bibinfo {author} {\bibfnamefont {T.}~\bibnamefont
  {Liu}}, \bibinfo {author} {\bibfnamefont {W.}~\bibnamefont {Melnitchouk}},
  \bibinfo {author} {\bibfnamefont {J.-W.}\ \bibnamefont {Qiu}}, \ and\
  \bibinfo {author} {\bibfnamefont {N.}~\bibnamefont {Sato}},\ }\href {\doibase
  10.1007/JHEP11(2021)157} {\bibfield  {journal} {\bibinfo  {journal} {JHEP}\
  }\textbf {\bibinfo {volume} {11}},\ \bibinfo {pages} {157} (\bibinfo {year}
  {2021}{\natexlab{a}})},\ \Eprint {http://arxiv.org/abs/2108.13371}
  {arXiv:2108.13371 [hep-ph]} \BibitemShut {NoStop}%
\bibitem [{\citenamefont {Vladimirov}(2019)}]{Vladimirov:2019bfa}%
  \BibitemOpen
  \bibfield  {author} {\bibinfo {author} {\bibfnamefont {A.}~\bibnamefont
  {Vladimirov}},\ }\href {\doibase 10.1007/JHEP10(2019)090} {\bibfield
  {journal} {\bibinfo  {journal} {JHEP}\ }\textbf {\bibinfo {volume} {10}},\
  \bibinfo {pages} {090} (\bibinfo {year} {2019})},\ \Eprint
  {http://arxiv.org/abs/1907.10356} {arXiv:1907.10356 [hep-ph]} \BibitemShut
  {NoStop}%
\bibitem [{\citenamefont {Cerutti}\ \emph {et~al.}(2023)\citenamefont
  {Cerutti}, \citenamefont {Rossi}, \citenamefont {Venturini}, \citenamefont
  {Bacchetta}, \citenamefont {Bertone}, \citenamefont {Bissolotti},\ and\
  \citenamefont {Radici}}]{Cerutti:2022lmb}%
  \BibitemOpen
  \bibfield  {author} {\bibinfo {author} {\bibfnamefont {M.}~\bibnamefont
  {Cerutti}}, \bibinfo {author} {\bibfnamefont {L.}~\bibnamefont {Rossi}},
  \bibinfo {author} {\bibfnamefont {S.}~\bibnamefont {Venturini}}, \bibinfo
  {author} {\bibfnamefont {A.}~\bibnamefont {Bacchetta}}, \bibinfo {author}
  {\bibfnamefont {V.}~\bibnamefont {Bertone}}, \bibinfo {author} {\bibfnamefont
  {C.}~\bibnamefont {Bissolotti}}, \ and\ \bibinfo {author} {\bibfnamefont
  {M.}~\bibnamefont {Radici}} (\bibinfo {collaboration} {(MAP
  Collaboration)\textdagger{}\textdagger{}, WMAP}),\ }\href {\doibase
  10.1103/PhysRevD.107.014014} {\bibfield  {journal} {\bibinfo  {journal}
  {Phys. Rev. D}\ }\textbf {\bibinfo {volume} {107}},\ \bibinfo {pages}
  {014014} (\bibinfo {year} {2023})},\ \Eprint
  {http://arxiv.org/abs/2210.01733} {arXiv:2210.01733 [hep-ph]} \BibitemShut
  {NoStop}%
\bibitem [{\citenamefont {Bacchetta}\ \emph
  {et~al.}(2020{\natexlab{a}})\citenamefont {Bacchetta}, \citenamefont
  {Celiberto}, \citenamefont {Radici},\ and\ \citenamefont
  {Taels}}]{Bacchetta:2020vty}%
  \BibitemOpen
  \bibfield  {author} {\bibinfo {author} {\bibfnamefont {A.}~\bibnamefont
  {Bacchetta}}, \bibinfo {author} {\bibfnamefont {F.~G.}\ \bibnamefont
  {Celiberto}}, \bibinfo {author} {\bibfnamefont {M.}~\bibnamefont {Radici}}, \
  and\ \bibinfo {author} {\bibfnamefont {P.}~\bibnamefont {Taels}},\ }\href
  {\doibase 10.1140/epjc/s10052-020-8327-6} {\bibfield  {journal} {\bibinfo
  {journal} {Eur. Phys. J. C}\ }\textbf {\bibinfo {volume} {80}},\ \bibinfo
  {pages} {733} (\bibinfo {year} {2020}{\natexlab{a}})},\ \Eprint
  {http://arxiv.org/abs/2005.02288} {arXiv:2005.02288 [hep-ph]} \BibitemShut
  {NoStop}%
\bibitem [{\citenamefont {Bacchetta}\ \emph
  {et~al.}(2022{\natexlab{c}})\citenamefont {Bacchetta}, \citenamefont
  {Celiberto},\ and\ \citenamefont {Radici}}]{Bacchetta:2022esb}%
  \BibitemOpen
  \bibfield  {author} {\bibinfo {author} {\bibfnamefont {A.}~\bibnamefont
  {Bacchetta}}, \bibinfo {author} {\bibfnamefont {F.~G.}\ \bibnamefont
  {Celiberto}}, \ and\ \bibinfo {author} {\bibfnamefont {M.}~\bibnamefont
  {Radici}},\ }\href {\doibase 10.7566/JPSCP.37.020124} {\bibfield  {journal}
  {\bibinfo  {journal} {JPS Conf. Proc.}\ }\textbf {\bibinfo {volume} {37}},\
  \bibinfo {pages} {020124} (\bibinfo {year} {2022}{\natexlab{c}})},\ \Eprint
  {http://arxiv.org/abs/2201.10508} {arXiv:2201.10508 [hep-ph]} \BibitemShut
  {NoStop}%
\bibitem [{\citenamefont {Bacchetta}\ \emph
  {et~al.}(2019{\natexlab{b}})\citenamefont {Bacchetta}, \citenamefont {Bozzi},
  \citenamefont {Radici}, \citenamefont {Ritzmann},\ and\ \citenamefont
  {Signori}}]{Bacchetta:2018lna}%
  \BibitemOpen
  \bibfield  {author} {\bibinfo {author} {\bibfnamefont {A.}~\bibnamefont
  {Bacchetta}}, \bibinfo {author} {\bibfnamefont {G.}~\bibnamefont {Bozzi}},
  \bibinfo {author} {\bibfnamefont {M.}~\bibnamefont {Radici}}, \bibinfo
  {author} {\bibfnamefont {M.}~\bibnamefont {Ritzmann}}, \ and\ \bibinfo
  {author} {\bibfnamefont {A.}~\bibnamefont {Signori}},\ }\href {\doibase
  10.1016/j.physletb.2018.11.002} {\bibfield  {journal} {\bibinfo  {journal}
  {Phys. Lett. B}\ }\textbf {\bibinfo {volume} {788}},\ \bibinfo {pages} {542}
  (\bibinfo {year} {2019}{\natexlab{b}})},\ \Eprint
  {http://arxiv.org/abs/1807.02101} {arXiv:1807.02101 [hep-ph]} \BibitemShut
  {NoStop}%
\bibitem [{\citenamefont {Collins}\ and\ \citenamefont
  {Soper}(1981)}]{Collins:1981uk}%
  \BibitemOpen
  \bibfield  {author} {\bibinfo {author} {\bibfnamefont {J.~C.}\ \bibnamefont
  {Collins}}\ and\ \bibinfo {author} {\bibfnamefont {D.~E.}\ \bibnamefont
  {Soper}},\ }\href {\doibase 10.1016/0550-3213(81)90339-4} {\bibfield
  {journal} {\bibinfo  {journal} {Nucl. Phys. B}\ }\textbf {\bibinfo {volume}
  {193}},\ \bibinfo {pages} {381} (\bibinfo {year} {1981})},\ \bibinfo {note}
  {[Erratum: Nucl.Phys.B 213, 545 (1983)]}\BibitemShut {NoStop}%
\bibitem [{\citenamefont {Collins}\ and\ \citenamefont
  {Soper}(1982)}]{Collins:1981va}%
  \BibitemOpen
  \bibfield  {author} {\bibinfo {author} {\bibfnamefont {J.~C.}\ \bibnamefont
  {Collins}}\ and\ \bibinfo {author} {\bibfnamefont {D.~E.}\ \bibnamefont
  {Soper}},\ }\href {\doibase 10.1016/0550-3213(82)90453-9} {\bibfield
  {journal} {\bibinfo  {journal} {Nucl. Phys. B}\ }\textbf {\bibinfo {volume}
  {197}},\ \bibinfo {pages} {446} (\bibinfo {year} {1982})}\BibitemShut
  {NoStop}%
\bibitem [{\citenamefont {Collins}\ \emph {et~al.}(1985)\citenamefont
  {Collins}, \citenamefont {Soper},\ and\ \citenamefont
  {Sterman}}]{Collins:1984kg}%
  \BibitemOpen
  \bibfield  {author} {\bibinfo {author} {\bibfnamefont {J.~C.}\ \bibnamefont
  {Collins}}, \bibinfo {author} {\bibfnamefont {D.~E.}\ \bibnamefont {Soper}},
  \ and\ \bibinfo {author} {\bibfnamefont {G.~F.}\ \bibnamefont {Sterman}},\
  }\href {\doibase 10.1016/0550-3213(85)90479-1} {\bibfield  {journal}
  {\bibinfo  {journal} {Nucl. Phys. B}\ }\textbf {\bibinfo {volume} {250}},\
  \bibinfo {pages} {199} (\bibinfo {year} {1985})}\BibitemShut {NoStop}%
\bibitem [{\citenamefont {Collins}\ and\ \citenamefont
  {Rogers}(2015)}]{Collins:2014jpa}%
  \BibitemOpen
  \bibfield  {author} {\bibinfo {author} {\bibfnamefont {J.}~\bibnamefont
  {Collins}}\ and\ \bibinfo {author} {\bibfnamefont {T.}~\bibnamefont
  {Rogers}},\ }\href {\doibase 10.1103/PhysRevD.91.074020} {\bibfield
  {journal} {\bibinfo  {journal} {Phys. Rev. D}\ }\textbf {\bibinfo {volume}
  {91}},\ \bibinfo {pages} {074020} (\bibinfo {year} {2015})},\ \Eprint
  {http://arxiv.org/abs/1412.3820} {arXiv:1412.3820 [hep-ph]} \BibitemShut
  {NoStop}%
\bibitem [{\citenamefont {Liang}\ \emph {et~al.}(2008)\citenamefont {Liang},
  \citenamefont {Wang},\ and\ \citenamefont {Zhou}}]{Liang:2008vz}%
  \BibitemOpen
  \bibfield  {author} {\bibinfo {author} {\bibfnamefont {Z.-t.}\ \bibnamefont
  {Liang}}, \bibinfo {author} {\bibfnamefont {X.-N.}\ \bibnamefont {Wang}}, \
  and\ \bibinfo {author} {\bibfnamefont {J.}~\bibnamefont {Zhou}},\ }\href
  {\doibase 10.1103/PhysRevD.77.125010} {\bibfield  {journal} {\bibinfo
  {journal} {Phys. Rev. D}\ }\textbf {\bibinfo {volume} {77}},\ \bibinfo
  {pages} {125010} (\bibinfo {year} {2008})},\ \Eprint
  {http://arxiv.org/abs/0801.0434} {arXiv:0801.0434 [hep-ph]} \BibitemShut
  {NoStop}%
\bibitem [{\citenamefont {Sch\"afer}\ and\ \citenamefont
  {Zhou}(2013)}]{Schafer:2013mza}%
  \BibitemOpen
  \bibfield  {author} {\bibinfo {author} {\bibfnamefont {A.}~\bibnamefont
  {Sch\"afer}}\ and\ \bibinfo {author} {\bibfnamefont {J.}~\bibnamefont
  {Zhou}},\ }\href {\doibase 10.1103/PhysRevD.88.074012} {\bibfield  {journal}
  {\bibinfo  {journal} {Phys. Rev. D}\ }\textbf {\bibinfo {volume} {88}},\
  \bibinfo {pages} {074012} (\bibinfo {year} {2013})},\ \Eprint
  {http://arxiv.org/abs/1305.5042} {arXiv:1305.5042 [hep-ph]} \BibitemShut
  {NoStop}%
\bibitem [{\citenamefont {Boer}\ \emph {et~al.}(2015)\citenamefont {Boer},
  \citenamefont {Buffing},\ and\ \citenamefont {Mulders}}]{Boer:2015kxa}%
  \BibitemOpen
  \bibfield  {author} {\bibinfo {author} {\bibfnamefont {D.}~\bibnamefont
  {Boer}}, \bibinfo {author} {\bibfnamefont {M.~G.~A.}\ \bibnamefont
  {Buffing}}, \ and\ \bibinfo {author} {\bibfnamefont {P.~J.}\ \bibnamefont
  {Mulders}},\ }\href {\doibase 10.1007/JHEP08(2015)053} {\bibfield  {journal}
  {\bibinfo  {journal} {JHEP}\ }\textbf {\bibinfo {volume} {08}},\ \bibinfo
  {pages} {053} (\bibinfo {year} {2015})},\ \Eprint
  {http://arxiv.org/abs/1503.03760} {arXiv:1503.03760 [hep-ph]} \BibitemShut
  {NoStop}%
\bibitem [{\citenamefont {Guo}(1998)}]{Guo:1998rd}%
  \BibitemOpen
  \bibfield  {author} {\bibinfo {author} {\bibfnamefont {X.-f.}\ \bibnamefont
  {Guo}},\ }\href {\doibase 10.1103/PhysRevD.58.114033} {\bibfield  {journal}
  {\bibinfo  {journal} {Phys. Rev. D}\ }\textbf {\bibinfo {volume} {58}},\
  \bibinfo {pages} {114033} (\bibinfo {year} {1998})},\ \Eprint
  {http://arxiv.org/abs/hep-ph/9804234} {arXiv:hep-ph/9804234} \BibitemShut
  {NoStop}%
\bibitem [{\citenamefont {Kang}\ and\ \citenamefont {Qiu}(2008)}]{Kang:2008us}%
  \BibitemOpen
  \bibfield  {author} {\bibinfo {author} {\bibfnamefont {Z.-B.}\ \bibnamefont
  {Kang}}\ and\ \bibinfo {author} {\bibfnamefont {J.-W.}\ \bibnamefont {Qiu}},\
  }\href {\doibase 10.1103/PhysRevD.77.114027} {\bibfield  {journal} {\bibinfo
  {journal} {Phys. Rev. D}\ }\textbf {\bibinfo {volume} {77}},\ \bibinfo
  {pages} {114027} (\bibinfo {year} {2008})},\ \Eprint
  {http://arxiv.org/abs/0802.2904} {arXiv:0802.2904 [hep-ph]} \BibitemShut
  {NoStop}%
\bibitem [{\citenamefont {Kang}\ \emph
  {et~al.}(2012{\natexlab{a}})\citenamefont {Kang}, \citenamefont {Vitev},\
  and\ \citenamefont {Xing}}]{Kang:2011bp}%
  \BibitemOpen
  \bibfield  {author} {\bibinfo {author} {\bibfnamefont {Z.-B.}\ \bibnamefont
  {Kang}}, \bibinfo {author} {\bibfnamefont {I.}~\bibnamefont {Vitev}}, \ and\
  \bibinfo {author} {\bibfnamefont {H.}~\bibnamefont {Xing}},\ }\href {\doibase
  10.1103/PhysRevD.85.054024} {\bibfield  {journal} {\bibinfo  {journal} {Phys.
  Rev. D}\ }\textbf {\bibinfo {volume} {85}},\ \bibinfo {pages} {054024}
  (\bibinfo {year} {2012}{\natexlab{a}})},\ \Eprint
  {http://arxiv.org/abs/1112.6021} {arXiv:1112.6021 [hep-ph]} \BibitemShut
  {NoStop}%
\bibitem [{\citenamefont {Xing}\ \emph {et~al.}(2012)\citenamefont {Xing},
  \citenamefont {Kang}, \citenamefont {Vitev},\ and\ \citenamefont
  {Wang}}]{Xing:2012ii}%
  \BibitemOpen
  \bibfield  {author} {\bibinfo {author} {\bibfnamefont {H.}~\bibnamefont
  {Xing}}, \bibinfo {author} {\bibfnamefont {Z.-B.}\ \bibnamefont {Kang}},
  \bibinfo {author} {\bibfnamefont {I.}~\bibnamefont {Vitev}}, \ and\ \bibinfo
  {author} {\bibfnamefont {E.}~\bibnamefont {Wang}},\ }\href {\doibase
  10.1103/PhysRevD.86.094010} {\bibfield  {journal} {\bibinfo  {journal} {Phys.
  Rev. D}\ }\textbf {\bibinfo {volume} {86}},\ \bibinfo {pages} {094010}
  (\bibinfo {year} {2012})},\ \Eprint {http://arxiv.org/abs/1206.1826}
  {arXiv:1206.1826 [hep-ph]} \BibitemShut {NoStop}%
\bibitem [{\citenamefont {Kang}\ \emph
  {et~al.}(2014{\natexlab{a}})\citenamefont {Kang}, \citenamefont {Wang},
  \citenamefont {Wang},\ and\ \citenamefont {Xing}}]{Kang:2013raa}%
  \BibitemOpen
  \bibfield  {author} {\bibinfo {author} {\bibfnamefont {Z.-B.}\ \bibnamefont
  {Kang}}, \bibinfo {author} {\bibfnamefont {E.}~\bibnamefont {Wang}}, \bibinfo
  {author} {\bibfnamefont {X.-N.}\ \bibnamefont {Wang}}, \ and\ \bibinfo
  {author} {\bibfnamefont {H.}~\bibnamefont {Xing}},\ }\href {\doibase
  10.1103/PhysRevLett.112.102001} {\bibfield  {journal} {\bibinfo  {journal}
  {Phys. Rev. Lett.}\ }\textbf {\bibinfo {volume} {112}},\ \bibinfo {pages}
  {102001} (\bibinfo {year} {2014}{\natexlab{a}})},\ \Eprint
  {http://arxiv.org/abs/1310.6759} {arXiv:1310.6759 [hep-ph]} \BibitemShut
  {NoStop}%
\bibitem [{\citenamefont {Ru}\ \emph {et~al.}(2021)\citenamefont {Ru},
  \citenamefont {Kang}, \citenamefont {Wang}, \citenamefont {Xing},\ and\
  \citenamefont {Zhang}}]{Ru:2019qvz}%
  \BibitemOpen
  \bibfield  {author} {\bibinfo {author} {\bibfnamefont {P.}~\bibnamefont
  {Ru}}, \bibinfo {author} {\bibfnamefont {Z.-B.}\ \bibnamefont {Kang}},
  \bibinfo {author} {\bibfnamefont {E.}~\bibnamefont {Wang}}, \bibinfo {author}
  {\bibfnamefont {H.}~\bibnamefont {Xing}}, \ and\ \bibinfo {author}
  {\bibfnamefont {B.-W.}\ \bibnamefont {Zhang}},\ }\href {\doibase
  10.1103/PhysRevD.103.L031901} {\bibfield  {journal} {\bibinfo  {journal}
  {Phys. Rev. D}\ }\textbf {\bibinfo {volume} {103}},\ \bibinfo {pages}
  {L031901} (\bibinfo {year} {2021})},\ \Eprint
  {http://arxiv.org/abs/1907.11808} {arXiv:1907.11808 [hep-ph]} \BibitemShut
  {NoStop}%
\bibitem [{\citenamefont {Ru}\ \emph {et~al.}(2023)\citenamefont {Ru},
  \citenamefont {Kang}, \citenamefont {Wang}, \citenamefont {Xing},\ and\
  \citenamefont {Zhang}}]{Ru:2023ars}%
  \BibitemOpen
  \bibfield  {author} {\bibinfo {author} {\bibfnamefont {P.}~\bibnamefont
  {Ru}}, \bibinfo {author} {\bibfnamefont {Z.-B.}\ \bibnamefont {Kang}},
  \bibinfo {author} {\bibfnamefont {E.}~\bibnamefont {Wang}}, \bibinfo {author}
  {\bibfnamefont {H.}~\bibnamefont {Xing}}, \ and\ \bibinfo {author}
  {\bibfnamefont {B.-W.}\ \bibnamefont {Zhang}},\ }\href@noop {} {\  (\bibinfo
  {year} {2023})},\ \Eprint {http://arxiv.org/abs/2302.02329} {arXiv:2302.02329
  [nucl-th]} \BibitemShut {NoStop}%
\bibitem [{\citenamefont {Bomhof}\ and\ \citenamefont
  {Mulders}(2008)}]{Bomhof:2007xt}%
  \BibitemOpen
  \bibfield  {author} {\bibinfo {author} {\bibfnamefont {C.~J.}\ \bibnamefont
  {Bomhof}}\ and\ \bibinfo {author} {\bibfnamefont {P.~J.}\ \bibnamefont
  {Mulders}},\ }\href {\doibase 10.1016/j.nuclphysb.2007.11.024} {\bibfield
  {journal} {\bibinfo  {journal} {Nucl. Phys. B}\ }\textbf {\bibinfo {volume}
  {795}},\ \bibinfo {pages} {409} (\bibinfo {year} {2008})},\ \Eprint
  {http://arxiv.org/abs/0709.1390} {arXiv:0709.1390 [hep-ph]} \BibitemShut
  {NoStop}%
\bibitem [{\citenamefont {Dominguez}\ \emph {et~al.}(2011)\citenamefont
  {Dominguez}, \citenamefont {Marquet}, \citenamefont {Xiao},\ and\
  \citenamefont {Yuan}}]{Dominguez:2011wm}%
  \BibitemOpen
  \bibfield  {author} {\bibinfo {author} {\bibfnamefont {F.}~\bibnamefont
  {Dominguez}}, \bibinfo {author} {\bibfnamefont {C.}~\bibnamefont {Marquet}},
  \bibinfo {author} {\bibfnamefont {B.-W.}\ \bibnamefont {Xiao}}, \ and\
  \bibinfo {author} {\bibfnamefont {F.}~\bibnamefont {Yuan}},\ }\href {\doibase
  10.1103/PhysRevD.83.105005} {\bibfield  {journal} {\bibinfo  {journal} {Phys.
  Rev. D}\ }\textbf {\bibinfo {volume} {83}},\ \bibinfo {pages} {105005}
  (\bibinfo {year} {2011})},\ \Eprint {http://arxiv.org/abs/1101.0715}
  {arXiv:1101.0715 [hep-ph]} \BibitemShut {NoStop}%
\bibitem [{\citenamefont {Scheihing-Hitschfeld}\ and\ \citenamefont
  {Yao}(2023)}]{Scheihing-Hitschfeld:2022xqx}%
  \BibitemOpen
  \bibfield  {author} {\bibinfo {author} {\bibfnamefont {B.}~\bibnamefont
  {Scheihing-Hitschfeld}}\ and\ \bibinfo {author} {\bibfnamefont
  {X.}~\bibnamefont {Yao}},\ }\href {\doibase 10.1103/PhysRevLett.130.052302}
  {\bibfield  {journal} {\bibinfo  {journal} {Phys. Rev. Lett.}\ }\textbf
  {\bibinfo {volume} {130}},\ \bibinfo {pages} {052302} (\bibinfo {year}
  {2023})},\ \Eprint {http://arxiv.org/abs/2205.04477} {arXiv:2205.04477
  [hep-ph]} \BibitemShut {NoStop}%
\bibitem [{\citenamefont {Hagler}\ \emph {et~al.}(2009)\citenamefont {Hagler},
  \citenamefont {Musch}, \citenamefont {Negele},\ and\ \citenamefont
  {Schafer}}]{Hagler:2009mb}%
  \BibitemOpen
  \bibfield  {author} {\bibinfo {author} {\bibfnamefont {P.}~\bibnamefont
  {Hagler}}, \bibinfo {author} {\bibfnamefont {B.~U.}\ \bibnamefont {Musch}},
  \bibinfo {author} {\bibfnamefont {J.~W.}\ \bibnamefont {Negele}}, \ and\
  \bibinfo {author} {\bibfnamefont {A.}~\bibnamefont {Schafer}},\ }\href
  {\doibase 10.1209/0295-5075/88/61001} {\bibfield  {journal} {\bibinfo
  {journal} {EPL}\ }\textbf {\bibinfo {volume} {88}},\ \bibinfo {pages} {61001}
  (\bibinfo {year} {2009})},\ \Eprint {http://arxiv.org/abs/0908.1283}
  {arXiv:0908.1283 [hep-lat]} \BibitemShut {NoStop}%
\bibitem [{\citenamefont {Musch}\ \emph {et~al.}(2011)\citenamefont {Musch},
  \citenamefont {Hagler}, \citenamefont {Negele},\ and\ \citenamefont
  {Schafer}}]{Musch:2010ka}%
  \BibitemOpen
  \bibfield  {author} {\bibinfo {author} {\bibfnamefont {B.~U.}\ \bibnamefont
  {Musch}}, \bibinfo {author} {\bibfnamefont {P.}~\bibnamefont {Hagler}},
  \bibinfo {author} {\bibfnamefont {J.~W.}\ \bibnamefont {Negele}}, \ and\
  \bibinfo {author} {\bibfnamefont {A.}~\bibnamefont {Schafer}},\ }\href
  {\doibase 10.1103/PhysRevD.83.094507} {\bibfield  {journal} {\bibinfo
  {journal} {Phys. Rev. D}\ }\textbf {\bibinfo {volume} {83}},\ \bibinfo
  {pages} {094507} (\bibinfo {year} {2011})},\ \Eprint
  {http://arxiv.org/abs/1011.1213} {arXiv:1011.1213 [hep-lat]} \BibitemShut
  {NoStop}%
\bibitem [{\citenamefont {Musch}\ \emph {et~al.}(2012)\citenamefont {Musch},
  \citenamefont {Hagler}, \citenamefont {Engelhardt}, \citenamefont {Negele},\
  and\ \citenamefont {Schafer}}]{Musch:2011er}%
  \BibitemOpen
  \bibfield  {author} {\bibinfo {author} {\bibfnamefont {B.~U.}\ \bibnamefont
  {Musch}}, \bibinfo {author} {\bibfnamefont {P.}~\bibnamefont {Hagler}},
  \bibinfo {author} {\bibfnamefont {M.}~\bibnamefont {Engelhardt}}, \bibinfo
  {author} {\bibfnamefont {J.~W.}\ \bibnamefont {Negele}}, \ and\ \bibinfo
  {author} {\bibfnamefont {A.}~\bibnamefont {Schafer}},\ }\href {\doibase
  10.1103/PhysRevD.85.094510} {\bibfield  {journal} {\bibinfo  {journal} {Phys.
  Rev. D}\ }\textbf {\bibinfo {volume} {85}},\ \bibinfo {pages} {094510}
  (\bibinfo {year} {2012})},\ \Eprint {http://arxiv.org/abs/1111.4249}
  {arXiv:1111.4249 [hep-lat]} \BibitemShut {NoStop}%
\bibitem [{\citenamefont {Engelhardt}\ \emph {et~al.}(2016)\citenamefont
  {Engelhardt}, \citenamefont {H\"agler}, \citenamefont {Musch}, \citenamefont
  {Negele},\ and\ \citenamefont {Sch\"afer}}]{Engelhardt:2015xja}%
  \BibitemOpen
  \bibfield  {author} {\bibinfo {author} {\bibfnamefont {M.}~\bibnamefont
  {Engelhardt}}, \bibinfo {author} {\bibfnamefont {P.}~\bibnamefont
  {H\"agler}}, \bibinfo {author} {\bibfnamefont {B.}~\bibnamefont {Musch}},
  \bibinfo {author} {\bibfnamefont {J.}~\bibnamefont {Negele}}, \ and\ \bibinfo
  {author} {\bibfnamefont {A.}~\bibnamefont {Sch\"afer}},\ }\href {\doibase
  10.1103/PhysRevD.93.054501} {\bibfield  {journal} {\bibinfo  {journal} {Phys.
  Rev. D}\ }\textbf {\bibinfo {volume} {93}},\ \bibinfo {pages} {054501}
  (\bibinfo {year} {2016})},\ \Eprint {http://arxiv.org/abs/1506.07826}
  {arXiv:1506.07826 [hep-lat]} \BibitemShut {NoStop}%
\bibitem [{\citenamefont {Yoon}\ \emph {et~al.}(2015)\citenamefont {Yoon},
  \citenamefont {Bhattacharya}, \citenamefont {Engelhardt}, \citenamefont
  {Green}, \citenamefont {Gupta}, \citenamefont {H\"agler}, \citenamefont
  {Musch}, \citenamefont {Negele}, \citenamefont {Pochinsky},\ and\
  \citenamefont {Syritsyn}}]{Yoon:2016dyh}%
  \BibitemOpen
  \bibfield  {author} {\bibinfo {author} {\bibfnamefont {B.}~\bibnamefont
  {Yoon}}, \bibinfo {author} {\bibfnamefont {T.}~\bibnamefont {Bhattacharya}},
  \bibinfo {author} {\bibfnamefont {M.}~\bibnamefont {Engelhardt}}, \bibinfo
  {author} {\bibfnamefont {J.}~\bibnamefont {Green}}, \bibinfo {author}
  {\bibfnamefont {R.}~\bibnamefont {Gupta}}, \bibinfo {author} {\bibfnamefont
  {P.}~\bibnamefont {H\"agler}}, \bibinfo {author} {\bibfnamefont
  {B.}~\bibnamefont {Musch}}, \bibinfo {author} {\bibfnamefont
  {J.}~\bibnamefont {Negele}}, \bibinfo {author} {\bibfnamefont
  {A.}~\bibnamefont {Pochinsky}}, \ and\ \bibinfo {author} {\bibfnamefont
  {S.}~\bibnamefont {Syritsyn}},\ }in\ \href@noop {} {\emph {\bibinfo
  {booktitle} {{33rd International Symposium on Lattice Field Theory}}}}\
  (\bibinfo  {publisher} {SISSA},\ \bibinfo {year} {2015})\ \Eprint
  {http://arxiv.org/abs/1601.05717} {arXiv:1601.05717 [hep-lat]} \BibitemShut
  {NoStop}%
\bibitem [{\citenamefont {Yoon}\ \emph {et~al.}(2017)\citenamefont {Yoon},
  \citenamefont {Engelhardt}, \citenamefont {Gupta}, \citenamefont
  {Bhattacharya}, \citenamefont {Green}, \citenamefont {Musch}, \citenamefont
  {Negele}, \citenamefont {Pochinsky}, \citenamefont {Sch\"afer},\ and\
  \citenamefont {Syritsyn}}]{Yoon:2017qzo}%
  \BibitemOpen
  \bibfield  {author} {\bibinfo {author} {\bibfnamefont {B.}~\bibnamefont
  {Yoon}}, \bibinfo {author} {\bibfnamefont {M.}~\bibnamefont {Engelhardt}},
  \bibinfo {author} {\bibfnamefont {R.}~\bibnamefont {Gupta}}, \bibinfo
  {author} {\bibfnamefont {T.}~\bibnamefont {Bhattacharya}}, \bibinfo {author}
  {\bibfnamefont {J.~R.}\ \bibnamefont {Green}}, \bibinfo {author}
  {\bibfnamefont {B.~U.}\ \bibnamefont {Musch}}, \bibinfo {author}
  {\bibfnamefont {J.~W.}\ \bibnamefont {Negele}}, \bibinfo {author}
  {\bibfnamefont {A.~V.}\ \bibnamefont {Pochinsky}}, \bibinfo {author}
  {\bibfnamefont {A.}~\bibnamefont {Sch\"afer}}, \ and\ \bibinfo {author}
  {\bibfnamefont {S.~N.}\ \bibnamefont {Syritsyn}},\ }\href {\doibase
  10.1103/PhysRevD.96.094508} {\bibfield  {journal} {\bibinfo  {journal} {Phys.
  Rev. D}\ }\textbf {\bibinfo {volume} {96}},\ \bibinfo {pages} {094508}
  (\bibinfo {year} {2017})},\ \Eprint {http://arxiv.org/abs/1706.03406}
  {arXiv:1706.03406 [hep-lat]} \BibitemShut {NoStop}%
\bibitem [{\citenamefont {Lin}\ \emph {et~al.}(2018)\citenamefont {Lin} \emph
  {et~al.}}]{Lin:2017snn}%
  \BibitemOpen
  \bibfield  {author} {\bibinfo {author} {\bibfnamefont {H.-W.}\ \bibnamefont
  {Lin}} \emph {et~al.},\ }\href {\doibase 10.1016/j.ppnp.2018.01.007}
  {\bibfield  {journal} {\bibinfo  {journal} {Prog. Part. Nucl. Phys.}\
  }\textbf {\bibinfo {volume} {100}},\ \bibinfo {pages} {107} (\bibinfo {year}
  {2018})},\ \Eprint {http://arxiv.org/abs/1711.07916} {arXiv:1711.07916
  [hep-ph]} \BibitemShut {NoStop}%
\bibitem [{\citenamefont {Ji}\ \emph {et~al.}(2015)\citenamefont {Ji},
  \citenamefont {Sun}, \citenamefont {Xiong},\ and\ \citenamefont
  {Yuan}}]{Ji:2014hxa}%
  \BibitemOpen
  \bibfield  {author} {\bibinfo {author} {\bibfnamefont {X.}~\bibnamefont
  {Ji}}, \bibinfo {author} {\bibfnamefont {P.}~\bibnamefont {Sun}}, \bibinfo
  {author} {\bibfnamefont {X.}~\bibnamefont {Xiong}}, \ and\ \bibinfo {author}
  {\bibfnamefont {F.}~\bibnamefont {Yuan}},\ }\href {\doibase
  10.1103/PhysRevD.91.074009} {\bibfield  {journal} {\bibinfo  {journal} {Phys.
  Rev. D}\ }\textbf {\bibinfo {volume} {91}},\ \bibinfo {pages} {074009}
  (\bibinfo {year} {2015})},\ \Eprint {http://arxiv.org/abs/1405.7640}
  {arXiv:1405.7640 [hep-ph]} \BibitemShut {NoStop}%
\bibitem [{\citenamefont {Ji}\ \emph {et~al.}(2019)\citenamefont {Ji},
  \citenamefont {Jin}, \citenamefont {Yuan}, \citenamefont {Zhang},\ and\
  \citenamefont {Zhao}}]{Ji:2018hvs}%
  \BibitemOpen
  \bibfield  {author} {\bibinfo {author} {\bibfnamefont {X.}~\bibnamefont
  {Ji}}, \bibinfo {author} {\bibfnamefont {L.-C.}\ \bibnamefont {Jin}},
  \bibinfo {author} {\bibfnamefont {F.}~\bibnamefont {Yuan}}, \bibinfo {author}
  {\bibfnamefont {J.-H.}\ \bibnamefont {Zhang}}, \ and\ \bibinfo {author}
  {\bibfnamefont {Y.}~\bibnamefont {Zhao}},\ }\href {\doibase
  10.1103/PhysRevD.99.114006} {\bibfield  {journal} {\bibinfo  {journal} {Phys.
  Rev. D}\ }\textbf {\bibinfo {volume} {99}},\ \bibinfo {pages} {114006}
  (\bibinfo {year} {2019})},\ \Eprint {http://arxiv.org/abs/1801.05930}
  {arXiv:1801.05930 [hep-ph]} \BibitemShut {NoStop}%
\bibitem [{\citenamefont {Ebert}\ \emph
  {et~al.}(2019{\natexlab{a}})\citenamefont {Ebert}, \citenamefont {Stewart},\
  and\ \citenamefont {Zhao}}]{Ebert:2018gzl}%
  \BibitemOpen
  \bibfield  {author} {\bibinfo {author} {\bibfnamefont {M.~A.}\ \bibnamefont
  {Ebert}}, \bibinfo {author} {\bibfnamefont {I.~W.}\ \bibnamefont {Stewart}},
  \ and\ \bibinfo {author} {\bibfnamefont {Y.}~\bibnamefont {Zhao}},\ }\href
  {\doibase 10.1103/PhysRevD.99.034505} {\bibfield  {journal} {\bibinfo
  {journal} {Phys. Rev. D}\ }\textbf {\bibinfo {volume} {99}},\ \bibinfo
  {pages} {034505} (\bibinfo {year} {2019}{\natexlab{a}})},\ \Eprint
  {http://arxiv.org/abs/1811.00026} {arXiv:1811.00026 [hep-ph]} \BibitemShut
  {NoStop}%
\bibitem [{\citenamefont {Ebert}\ \emph
  {et~al.}(2019{\natexlab{b}})\citenamefont {Ebert}, \citenamefont {Stewart},\
  and\ \citenamefont {Zhao}}]{Ebert:2019okf}%
  \BibitemOpen
  \bibfield  {author} {\bibinfo {author} {\bibfnamefont {M.~A.}\ \bibnamefont
  {Ebert}}, \bibinfo {author} {\bibfnamefont {I.~W.}\ \bibnamefont {Stewart}},
  \ and\ \bibinfo {author} {\bibfnamefont {Y.}~\bibnamefont {Zhao}},\ }\href
  {\doibase 10.1007/JHEP09(2019)037} {\bibfield  {journal} {\bibinfo  {journal}
  {JHEP}\ }\textbf {\bibinfo {volume} {09}},\ \bibinfo {pages} {037} (\bibinfo
  {year} {2019}{\natexlab{b}})},\ \Eprint {http://arxiv.org/abs/1901.03685}
  {arXiv:1901.03685 [hep-ph]} \BibitemShut {NoStop}%
\bibitem [{\citenamefont {Ji}\ \emph {et~al.}(2020{\natexlab{a}})\citenamefont
  {Ji}, \citenamefont {Liu},\ and\ \citenamefont {Liu}}]{Ji:2019sxk}%
  \BibitemOpen
  \bibfield  {author} {\bibinfo {author} {\bibfnamefont {X.}~\bibnamefont
  {Ji}}, \bibinfo {author} {\bibfnamefont {Y.}~\bibnamefont {Liu}}, \ and\
  \bibinfo {author} {\bibfnamefont {Y.-S.}\ \bibnamefont {Liu}},\ }\href
  {\doibase 10.1016/j.nuclphysb.2020.115054} {\bibfield  {journal} {\bibinfo
  {journal} {Nucl. Phys. B}\ }\textbf {\bibinfo {volume} {955}},\ \bibinfo
  {pages} {115054} (\bibinfo {year} {2020}{\natexlab{a}})},\ \Eprint
  {http://arxiv.org/abs/1910.11415} {arXiv:1910.11415 [hep-ph]} \BibitemShut
  {NoStop}%
\bibitem [{\citenamefont {Ji}\ \emph {et~al.}(2020{\natexlab{b}})\citenamefont
  {Ji}, \citenamefont {Liu},\ and\ \citenamefont {Liu}}]{Ji:2019ewn}%
  \BibitemOpen
  \bibfield  {author} {\bibinfo {author} {\bibfnamefont {X.}~\bibnamefont
  {Ji}}, \bibinfo {author} {\bibfnamefont {Y.}~\bibnamefont {Liu}}, \ and\
  \bibinfo {author} {\bibfnamefont {Y.-S.}\ \bibnamefont {Liu}},\ }\href
  {\doibase 10.1016/j.physletb.2020.135946} {\bibfield  {journal} {\bibinfo
  {journal} {Phys. Lett. B}\ }\textbf {\bibinfo {volume} {811}},\ \bibinfo
  {pages} {135946} (\bibinfo {year} {2020}{\natexlab{b}})},\ \Eprint
  {http://arxiv.org/abs/1911.03840} {arXiv:1911.03840 [hep-ph]} \BibitemShut
  {NoStop}%
\bibitem [{\citenamefont {Ebert}\ \emph
  {et~al.}(2022{\natexlab{c}})\citenamefont {Ebert}, \citenamefont {Schindler},
  \citenamefont {Stewart},\ and\ \citenamefont {Zhao}}]{Ebert:2022fmh}%
  \BibitemOpen
  \bibfield  {author} {\bibinfo {author} {\bibfnamefont {M.~A.}\ \bibnamefont
  {Ebert}}, \bibinfo {author} {\bibfnamefont {S.~T.}\ \bibnamefont
  {Schindler}}, \bibinfo {author} {\bibfnamefont {I.~W.}\ \bibnamefont
  {Stewart}}, \ and\ \bibinfo {author} {\bibfnamefont {Y.}~\bibnamefont
  {Zhao}},\ }\href {\doibase 10.1007/JHEP04(2022)178} {\bibfield  {journal}
  {\bibinfo  {journal} {JHEP}\ }\textbf {\bibinfo {volume} {04}},\ \bibinfo
  {pages} {178} (\bibinfo {year} {2022}{\natexlab{c}})},\ \Eprint
  {http://arxiv.org/abs/2201.08401} {arXiv:2201.08401 [hep-ph]} \BibitemShut
  {NoStop}%
\bibitem [{\citenamefont {Zhang}\ \emph {et~al.}(2020)\citenamefont {Zhang}
  \emph {et~al.}}]{LatticeParton:2020uhz}%
  \BibitemOpen
  \bibfield  {author} {\bibinfo {author} {\bibfnamefont {Q.-A.}\ \bibnamefont
  {Zhang}} \emph {et~al.} (\bibinfo {collaboration} {Lattice Parton}),\ }\href
  {\doibase 10.22323/1.396.0477} {\bibfield  {journal} {\bibinfo  {journal}
  {Phys. Rev. Lett.}\ }\textbf {\bibinfo {volume} {125}},\ \bibinfo {pages}
  {192001} (\bibinfo {year} {2020})},\ \Eprint
  {http://arxiv.org/abs/2005.14572} {arXiv:2005.14572 [hep-lat]} \BibitemShut
  {NoStop}%
\bibitem [{\citenamefont {Li}\ \emph {et~al.}(2022{\natexlab{a}})\citenamefont
  {Li} \emph {et~al.}}]{Li:2021wvl}%
  \BibitemOpen
  \bibfield  {author} {\bibinfo {author} {\bibfnamefont {Y.}~\bibnamefont {Li}}
  \emph {et~al.},\ }\href {\doibase 10.1103/PhysRevLett.128.062002} {\bibfield
  {journal} {\bibinfo  {journal} {Phys. Rev. Lett.}\ }\textbf {\bibinfo
  {volume} {128}},\ \bibinfo {pages} {062002} (\bibinfo {year}
  {2022}{\natexlab{a}})},\ \Eprint {http://arxiv.org/abs/2106.13027}
  {arXiv:2106.13027 [hep-lat]} \BibitemShut {NoStop}%
\bibitem [{\citenamefont {Ebert}\ \emph
  {et~al.}(2020{\natexlab{a}})\citenamefont {Ebert}, \citenamefont {Schindler},
  \citenamefont {Stewart},\ and\ \citenamefont {Zhao}}]{Ebert:2020gxr}%
  \BibitemOpen
  \bibfield  {author} {\bibinfo {author} {\bibfnamefont {M.~A.}\ \bibnamefont
  {Ebert}}, \bibinfo {author} {\bibfnamefont {S.~T.}\ \bibnamefont
  {Schindler}}, \bibinfo {author} {\bibfnamefont {I.~W.}\ \bibnamefont
  {Stewart}}, \ and\ \bibinfo {author} {\bibfnamefont {Y.}~\bibnamefont
  {Zhao}},\ }\href {\doibase 10.1007/JHEP09(2020)099} {\bibfield  {journal}
  {\bibinfo  {journal} {JHEP}\ }\textbf {\bibinfo {volume} {09}},\ \bibinfo
  {pages} {099} (\bibinfo {year} {2020}{\natexlab{a}})},\ \Eprint
  {http://arxiv.org/abs/2004.14831} {arXiv:2004.14831 [hep-ph]} \BibitemShut
  {NoStop}%
\bibitem [{\citenamefont {Vladimirov}\ and\ \citenamefont
  {Sch\"afer}(2020)}]{Vladimirov:2020ofp}%
  \BibitemOpen
  \bibfield  {author} {\bibinfo {author} {\bibfnamefont {A.~A.}\ \bibnamefont
  {Vladimirov}}\ and\ \bibinfo {author} {\bibfnamefont {A.}~\bibnamefont
  {Sch\"afer}},\ }\href {\doibase 10.1103/PhysRevD.101.074517} {\bibfield
  {journal} {\bibinfo  {journal} {Phys. Rev. D}\ }\textbf {\bibinfo {volume}
  {101}},\ \bibinfo {pages} {074517} (\bibinfo {year} {2020})},\ \Eprint
  {http://arxiv.org/abs/2002.07527} {arXiv:2002.07527 [hep-ph]} \BibitemShut
  {NoStop}%
\bibitem [{\citenamefont {Bermudez~Martinez}\ and\ \citenamefont
  {Vladimirov}(2022)}]{BermudezMartinez:2022ctj}%
  \BibitemOpen
  \bibfield  {author} {\bibinfo {author} {\bibfnamefont {A.}~\bibnamefont
  {Bermudez~Martinez}}\ and\ \bibinfo {author} {\bibfnamefont {A.}~\bibnamefont
  {Vladimirov}},\ }\href {\doibase 10.1103/PhysRevD.106.L091501} {\bibfield
  {journal} {\bibinfo  {journal} {Phys. Rev. D}\ }\textbf {\bibinfo {volume}
  {106}},\ \bibinfo {pages} {L091501} (\bibinfo {year} {2022})},\ \Eprint
  {http://arxiv.org/abs/2206.01105} {arXiv:2206.01105 [hep-ph]} \BibitemShut
  {NoStop}%
\bibitem [{\citenamefont {Shanahan}\ \emph
  {et~al.}(2020{\natexlab{a}})\citenamefont {Shanahan}, \citenamefont
  {Wagman},\ and\ \citenamefont {Zhao}}]{Shanahan:2019zcq}%
  \BibitemOpen
  \bibfield  {author} {\bibinfo {author} {\bibfnamefont {P.}~\bibnamefont
  {Shanahan}}, \bibinfo {author} {\bibfnamefont {M.~L.}\ \bibnamefont
  {Wagman}}, \ and\ \bibinfo {author} {\bibfnamefont {Y.}~\bibnamefont
  {Zhao}},\ }\href {\doibase 10.1103/PhysRevD.101.074505} {\bibfield  {journal}
  {\bibinfo  {journal} {Phys. Rev. D}\ }\textbf {\bibinfo {volume} {101}},\
  \bibinfo {pages} {074505} (\bibinfo {year} {2020}{\natexlab{a}})},\ \Eprint
  {http://arxiv.org/abs/1911.00800} {arXiv:1911.00800 [hep-lat]} \BibitemShut
  {NoStop}%
\bibitem [{\citenamefont {Shanahan}\ \emph
  {et~al.}(2020{\natexlab{b}})\citenamefont {Shanahan}, \citenamefont
  {Wagman},\ and\ \citenamefont {Zhao}}]{Shanahan:2020zxr}%
  \BibitemOpen
  \bibfield  {author} {\bibinfo {author} {\bibfnamefont {P.}~\bibnamefont
  {Shanahan}}, \bibinfo {author} {\bibfnamefont {M.}~\bibnamefont {Wagman}}, \
  and\ \bibinfo {author} {\bibfnamefont {Y.}~\bibnamefont {Zhao}},\ }\href
  {\doibase 10.1103/PhysRevD.102.014511} {\bibfield  {journal} {\bibinfo
  {journal} {Phys. Rev. D}\ }\textbf {\bibinfo {volume} {102}},\ \bibinfo
  {pages} {014511} (\bibinfo {year} {2020}{\natexlab{b}})},\ \Eprint
  {http://arxiv.org/abs/2003.06063} {arXiv:2003.06063 [hep-lat]} \BibitemShut
  {NoStop}%
\bibitem [{\citenamefont {Schlemmer}\ \emph {et~al.}(2021)\citenamefont
  {Schlemmer}, \citenamefont {Vladimirov}, \citenamefont {Zimmermann},
  \citenamefont {Engelhardt},\ and\ \citenamefont
  {Sch\"afer}}]{Schlemmer:2021aij}%
  \BibitemOpen
  \bibfield  {author} {\bibinfo {author} {\bibfnamefont {M.}~\bibnamefont
  {Schlemmer}}, \bibinfo {author} {\bibfnamefont {A.}~\bibnamefont
  {Vladimirov}}, \bibinfo {author} {\bibfnamefont {C.}~\bibnamefont
  {Zimmermann}}, \bibinfo {author} {\bibfnamefont {M.}~\bibnamefont
  {Engelhardt}}, \ and\ \bibinfo {author} {\bibfnamefont {A.}~\bibnamefont
  {Sch\"afer}},\ }\href {\doibase 10.1007/JHEP08(2021)004} {\bibfield
  {journal} {\bibinfo  {journal} {JHEP}\ }\textbf {\bibinfo {volume} {08}},\
  \bibinfo {pages} {004} (\bibinfo {year} {2021})},\ \Eprint
  {http://arxiv.org/abs/2103.16991} {arXiv:2103.16991 [hep-lat]} \BibitemShut
  {NoStop}%
\bibitem [{\citenamefont {Shanahan}\ \emph {et~al.}(2021)\citenamefont
  {Shanahan}, \citenamefont {Wagman},\ and\ \citenamefont
  {Zhao}}]{Shanahan:2021tst}%
  \BibitemOpen
  \bibfield  {author} {\bibinfo {author} {\bibfnamefont {P.}~\bibnamefont
  {Shanahan}}, \bibinfo {author} {\bibfnamefont {M.}~\bibnamefont {Wagman}}, \
  and\ \bibinfo {author} {\bibfnamefont {Y.}~\bibnamefont {Zhao}},\ }\href
  {\doibase 10.1103/PhysRevD.104.114502} {\bibfield  {journal} {\bibinfo
  {journal} {Phys. Rev. D}\ }\textbf {\bibinfo {volume} {104}},\ \bibinfo
  {pages} {114502} (\bibinfo {year} {2021})},\ \Eprint
  {http://arxiv.org/abs/2107.11930} {arXiv:2107.11930 [hep-lat]} \BibitemShut
  {NoStop}%
\bibitem [{\citenamefont {Chu}\ \emph {et~al.}(2022)\citenamefont {Chu} \emph
  {et~al.}}]{LPC:2022ibr}%
  \BibitemOpen
  \bibfield  {author} {\bibinfo {author} {\bibfnamefont {M.-H.}\ \bibnamefont
  {Chu}} \emph {et~al.} (\bibinfo {collaboration} {LPC}),\ }\href {\doibase
  10.1103/PhysRevD.106.034509} {\bibfield  {journal} {\bibinfo  {journal}
  {Phys. Rev. D}\ }\textbf {\bibinfo {volume} {106}},\ \bibinfo {pages}
  {034509} (\bibinfo {year} {2022})},\ \Eprint
  {http://arxiv.org/abs/2204.00200} {arXiv:2204.00200 [hep-lat]} \BibitemShut
  {NoStop}%
\bibitem [{\citenamefont {Shu}\ \emph {et~al.}(2023)\citenamefont {Shu},
  \citenamefont {Schlemmer}, \citenamefont {Sizmann}, \citenamefont
  {Vladimirov}, \citenamefont {Walter}, \citenamefont {Engelhardt},
  \citenamefont {Sch\"afer},\ and\ \citenamefont {Yang}}]{Shu:2023cot}%
  \BibitemOpen
  \bibfield  {author} {\bibinfo {author} {\bibfnamefont {H.-T.}\ \bibnamefont
  {Shu}}, \bibinfo {author} {\bibfnamefont {M.}~\bibnamefont {Schlemmer}},
  \bibinfo {author} {\bibfnamefont {T.}~\bibnamefont {Sizmann}}, \bibinfo
  {author} {\bibfnamefont {A.}~\bibnamefont {Vladimirov}}, \bibinfo {author}
  {\bibfnamefont {L.}~\bibnamefont {Walter}}, \bibinfo {author} {\bibfnamefont
  {M.}~\bibnamefont {Engelhardt}}, \bibinfo {author} {\bibfnamefont
  {A.}~\bibnamefont {Sch\"afer}}, \ and\ \bibinfo {author} {\bibfnamefont
  {Y.-B.}\ \bibnamefont {Yang}},\ }\href@noop {} {\  (\bibinfo {year}
  {2023})},\ \Eprint {http://arxiv.org/abs/2302.06502} {arXiv:2302.06502
  [hep-lat]} \BibitemShut {NoStop}%
\bibitem [{\citenamefont {Boglione}\ \emph {et~al.}(2022)\citenamefont
  {Boglione}, \citenamefont {Gonzalez-Hernandez},\ and\ \citenamefont
  {Simonelli}}]{Boglione:2022nzq}%
  \BibitemOpen
  \bibfield  {author} {\bibinfo {author} {\bibfnamefont {M.}~\bibnamefont
  {Boglione}}, \bibinfo {author} {\bibfnamefont {J.~O.}\ \bibnamefont
  {Gonzalez-Hernandez}}, \ and\ \bibinfo {author} {\bibfnamefont
  {A.}~\bibnamefont {Simonelli}},\ }\href {\doibase
  10.1103/PhysRevD.106.074024} {\bibfield  {journal} {\bibinfo  {journal}
  {Phys. Rev. D}\ }\textbf {\bibinfo {volume} {106}},\ \bibinfo {pages}
  {074024} (\bibinfo {year} {2022})},\ \Eprint
  {http://arxiv.org/abs/2206.08876} {arXiv:2206.08876 [hep-ph]} \BibitemShut
  {NoStop}%
\bibitem [{\citenamefont {Schindler}\ \emph {et~al.}(2022)\citenamefont
  {Schindler}, \citenamefont {Stewart},\ and\ \citenamefont
  {Zhao}}]{Schindler:2022eva}%
  \BibitemOpen
  \bibfield  {author} {\bibinfo {author} {\bibfnamefont {S.~T.}\ \bibnamefont
  {Schindler}}, \bibinfo {author} {\bibfnamefont {I.~W.}\ \bibnamefont
  {Stewart}}, \ and\ \bibinfo {author} {\bibfnamefont {Y.}~\bibnamefont
  {Zhao}},\ }\href {\doibase 10.1007/JHEP08(2022)084} {\bibfield  {journal}
  {\bibinfo  {journal} {JHEP}\ }\textbf {\bibinfo {volume} {08}},\ \bibinfo
  {pages} {084} (\bibinfo {year} {2022})},\ \Eprint
  {http://arxiv.org/abs/2205.12369} {arXiv:2205.12369 [hep-ph]} \BibitemShut
  {NoStop}%
\bibitem [{\citenamefont {Zhu}\ \emph {et~al.}(2023)\citenamefont {Zhu},
  \citenamefont {Ji}, \citenamefont {Zhang},\ and\ \citenamefont
  {Zhao}}]{Zhu:2022bja}%
  \BibitemOpen
  \bibfield  {author} {\bibinfo {author} {\bibfnamefont {R.}~\bibnamefont
  {Zhu}}, \bibinfo {author} {\bibfnamefont {Y.}~\bibnamefont {Ji}}, \bibinfo
  {author} {\bibfnamefont {J.-H.}\ \bibnamefont {Zhang}}, \ and\ \bibinfo
  {author} {\bibfnamefont {S.}~\bibnamefont {Zhao}},\ }\href {\doibase
  10.1007/JHEP02(2023)114} {\bibfield  {journal} {\bibinfo  {journal} {JHEP}\
  }\textbf {\bibinfo {volume} {02}},\ \bibinfo {pages} {114} (\bibinfo {year}
  {2023})},\ \Eprint {http://arxiv.org/abs/2209.05443} {arXiv:2209.05443
  [hep-ph]} \BibitemShut {NoStop}%
\bibitem [{\citenamefont {Ji}\ \emph {et~al.}(2021{\natexlab{b}})\citenamefont
  {Ji}, \citenamefont {Liu}, \citenamefont {Sch\"afer},\ and\ \citenamefont
  {Yuan}}]{Ji:2020jeb}%
  \BibitemOpen
  \bibfield  {author} {\bibinfo {author} {\bibfnamefont {X.}~\bibnamefont
  {Ji}}, \bibinfo {author} {\bibfnamefont {Y.}~\bibnamefont {Liu}}, \bibinfo
  {author} {\bibfnamefont {A.}~\bibnamefont {Sch\"afer}}, \ and\ \bibinfo
  {author} {\bibfnamefont {F.}~\bibnamefont {Yuan}},\ }\href {\doibase
  10.1103/PhysRevD.103.074005} {\bibfield  {journal} {\bibinfo  {journal}
  {Phys. Rev. D}\ }\textbf {\bibinfo {volume} {103}},\ \bibinfo {pages}
  {074005} (\bibinfo {year} {2021}{\natexlab{b}})},\ \Eprint
  {http://arxiv.org/abs/2011.13397} {arXiv:2011.13397 [hep-ph]} \BibitemShut
  {NoStop}%
\bibitem [{\citenamefont {Bhattacharya}\ \emph {et~al.}(2020)\citenamefont
  {Bhattacharya}, \citenamefont {Cichy}, \citenamefont {Constantinou},
  \citenamefont {Metz}, \citenamefont {Scapellato},\ and\ \citenamefont
  {Steffens}}]{Bhattacharya:2020cen}%
  \BibitemOpen
  \bibfield  {author} {\bibinfo {author} {\bibfnamefont {S.}~\bibnamefont
  {Bhattacharya}}, \bibinfo {author} {\bibfnamefont {K.}~\bibnamefont {Cichy}},
  \bibinfo {author} {\bibfnamefont {M.}~\bibnamefont {Constantinou}}, \bibinfo
  {author} {\bibfnamefont {A.}~\bibnamefont {Metz}}, \bibinfo {author}
  {\bibfnamefont {A.}~\bibnamefont {Scapellato}}, \ and\ \bibinfo {author}
  {\bibfnamefont {F.}~\bibnamefont {Steffens}},\ }\href {\doibase
  10.1103/PhysRevD.102.111501} {\bibfield  {journal} {\bibinfo  {journal}
  {Phys. Rev. D}\ }\textbf {\bibinfo {volume} {102}},\ \bibinfo {pages}
  {111501} (\bibinfo {year} {2020})},\ \Eprint
  {http://arxiv.org/abs/2004.04130} {arXiv:2004.04130 [hep-lat]} \BibitemShut
  {NoStop}%
\bibitem [{\citenamefont {Braun}\ \emph {et~al.}(2021)\citenamefont {Braun},
  \citenamefont {Ji},\ and\ \citenamefont {Vladimirov}}]{Braun:2021gvv}%
  \BibitemOpen
  \bibfield  {author} {\bibinfo {author} {\bibfnamefont {V.~M.}\ \bibnamefont
  {Braun}}, \bibinfo {author} {\bibfnamefont {Y.}~\bibnamefont {Ji}}, \ and\
  \bibinfo {author} {\bibfnamefont {A.}~\bibnamefont {Vladimirov}},\ }\href
  {\doibase 10.1007/JHEP10(2021)087} {\bibfield  {journal} {\bibinfo  {journal}
  {JHEP}\ }\textbf {\bibinfo {volume} {10}},\ \bibinfo {pages} {087} (\bibinfo
  {year} {2021})},\ \Eprint {http://arxiv.org/abs/2108.03065} {arXiv:2108.03065
  [hep-ph]} \BibitemShut {NoStop}%
\bibitem [{\citenamefont {Chen}\ and\ \citenamefont {Ma}(2017)}]{Chen:2016hgw}%
  \BibitemOpen
  \bibfield  {author} {\bibinfo {author} {\bibfnamefont {A.~P.}\ \bibnamefont
  {Chen}}\ and\ \bibinfo {author} {\bibfnamefont {J.~P.}\ \bibnamefont {Ma}},\
  }\href {\doibase 10.1016/j.physletb.2017.03.015} {\bibfield  {journal}
  {\bibinfo  {journal} {Phys. Lett. B}\ }\textbf {\bibinfo {volume} {768}},\
  \bibinfo {pages} {380} (\bibinfo {year} {2017})},\ \Eprint
  {http://arxiv.org/abs/1610.08634} {arXiv:1610.08634 [hep-ph]} \BibitemShut
  {NoStop}%
\bibitem [{\citenamefont {Lorc\'e}\ and\ \citenamefont
  {Pasquini}(2011)}]{Lorce:2011kd}%
  \BibitemOpen
  \bibfield  {author} {\bibinfo {author} {\bibfnamefont {C.}~\bibnamefont
  {Lorc\'e}}\ and\ \bibinfo {author} {\bibfnamefont {B.}~\bibnamefont
  {Pasquini}},\ }\href {\doibase 10.1103/PhysRevD.84.014015} {\bibfield
  {journal} {\bibinfo  {journal} {Phys. Rev. D}\ }\textbf {\bibinfo {volume}
  {84}},\ \bibinfo {pages} {014015} (\bibinfo {year} {2011})},\ \Eprint
  {http://arxiv.org/abs/1106.0139} {arXiv:1106.0139 [hep-ph]} \BibitemShut
  {NoStop}%
\bibitem [{\citenamefont {Hatta}(2012)}]{Hatta:2011ku}%
  \BibitemOpen
  \bibfield  {author} {\bibinfo {author} {\bibfnamefont {Y.}~\bibnamefont
  {Hatta}},\ }\href {\doibase 10.1016/j.physletb.2012.01.024} {\bibfield
  {journal} {\bibinfo  {journal} {Phys. Lett. B}\ }\textbf {\bibinfo {volume}
  {708}},\ \bibinfo {pages} {186} (\bibinfo {year} {2012})},\ \Eprint
  {http://arxiv.org/abs/1111.3547} {arXiv:1111.3547 [hep-ph]} \BibitemShut
  {NoStop}%
\bibitem [{\citenamefont {Lorc\'e}\ \emph {et~al.}(2012)\citenamefont
  {Lorc\'e}, \citenamefont {Pasquini}, \citenamefont {Xiong},\ and\
  \citenamefont {Yuan}}]{Lorce:2011ni}%
  \BibitemOpen
  \bibfield  {author} {\bibinfo {author} {\bibfnamefont {C.}~\bibnamefont
  {Lorc\'e}}, \bibinfo {author} {\bibfnamefont {B.}~\bibnamefont {Pasquini}},
  \bibinfo {author} {\bibfnamefont {X.}~\bibnamefont {Xiong}}, \ and\ \bibinfo
  {author} {\bibfnamefont {F.}~\bibnamefont {Yuan}},\ }\href {\doibase
  10.1103/PhysRevD.85.114006} {\bibfield  {journal} {\bibinfo  {journal} {Phys.
  Rev. D}\ }\textbf {\bibinfo {volume} {85}},\ \bibinfo {pages} {114006}
  (\bibinfo {year} {2012})},\ \Eprint {http://arxiv.org/abs/1111.4827}
  {arXiv:1111.4827 [hep-ph]} \BibitemShut {NoStop}%
\bibitem [{\citenamefont {Lorc\'e}\ and\ \citenamefont
  {Pasquini}(2016)}]{Lorce:2015sqe}%
  \BibitemOpen
  \bibfield  {author} {\bibinfo {author} {\bibfnamefont {C.}~\bibnamefont
  {Lorc\'e}}\ and\ \bibinfo {author} {\bibfnamefont {B.}~\bibnamefont
  {Pasquini}},\ }\href {\doibase 10.1103/PhysRevD.93.034040} {\bibfield
  {journal} {\bibinfo  {journal} {Phys. Rev. D}\ }\textbf {\bibinfo {volume}
  {93}},\ \bibinfo {pages} {034040} (\bibinfo {year} {2016})},\ \Eprint
  {http://arxiv.org/abs/1512.06744} {arXiv:1512.06744 [hep-ph]} \BibitemShut
  {NoStop}%
\bibitem [{\citenamefont {Engelhardt}(2017)}]{Engelhardt:2017miy}%
  \BibitemOpen
  \bibfield  {author} {\bibinfo {author} {\bibfnamefont {M.}~\bibnamefont
  {Engelhardt}},\ }\href {\doibase 10.1103/PhysRevD.95.094505} {\bibfield
  {journal} {\bibinfo  {journal} {Phys. Rev. D}\ }\textbf {\bibinfo {volume}
  {95}},\ \bibinfo {pages} {094505} (\bibinfo {year} {2017})},\ \Eprint
  {http://arxiv.org/abs/1701.01536} {arXiv:1701.01536 [hep-lat]} \BibitemShut
  {NoStop}%
\bibitem [{\citenamefont {Engelhardt}\ \emph {et~al.}(2020)\citenamefont
  {Engelhardt}, \citenamefont {Green}, \citenamefont {Hasan}, \citenamefont
  {Krieg}, \citenamefont {Meinel}, \citenamefont {Negele}, \citenamefont
  {Pochinsky},\ and\ \citenamefont {Syritsyn}}]{Engelhardt:2020qtg}%
  \BibitemOpen
  \bibfield  {author} {\bibinfo {author} {\bibfnamefont {M.}~\bibnamefont
  {Engelhardt}}, \bibinfo {author} {\bibfnamefont {J.~R.}\ \bibnamefont
  {Green}}, \bibinfo {author} {\bibfnamefont {N.}~\bibnamefont {Hasan}},
  \bibinfo {author} {\bibfnamefont {S.}~\bibnamefont {Krieg}}, \bibinfo
  {author} {\bibfnamefont {S.}~\bibnamefont {Meinel}}, \bibinfo {author}
  {\bibfnamefont {J.}~\bibnamefont {Negele}}, \bibinfo {author} {\bibfnamefont
  {A.}~\bibnamefont {Pochinsky}}, \ and\ \bibinfo {author} {\bibfnamefont
  {S.}~\bibnamefont {Syritsyn}},\ }\href {\doibase 10.1103/PhysRevD.102.074505}
  {\bibfield  {journal} {\bibinfo  {journal} {Phys. Rev. D}\ }\textbf {\bibinfo
  {volume} {102}},\ \bibinfo {pages} {074505} (\bibinfo {year} {2020})},\
  \Eprint {http://arxiv.org/abs/2008.03660} {arXiv:2008.03660 [hep-lat]}
  \BibitemShut {NoStop}%
\bibitem [{\citenamefont {Engelhardt}\ \emph {et~al.}(2022)\citenamefont
  {Engelhardt} \emph {et~al.}}]{Engelhardt:2021kdo}%
  \BibitemOpen
  \bibfield  {author} {\bibinfo {author} {\bibfnamefont {M.}~\bibnamefont
  {Engelhardt}} \emph {et~al.},\ }\href {\doibase 10.22323/1.396.0413}
  {\bibfield  {journal} {\bibinfo  {journal} {PoS}\ }\textbf {\bibinfo {volume}
  {LATTICE2021}},\ \bibinfo {pages} {413} (\bibinfo {year} {2022})},\ \Eprint
  {http://arxiv.org/abs/2112.13464} {arXiv:2112.13464 [hep-lat]} \BibitemShut
  {NoStop}%
\bibitem [{\citenamefont {Gelis}\ \emph {et~al.}(2010)\citenamefont {Gelis},
  \citenamefont {Iancu}, \citenamefont {Jalilian-Marian},\ and\ \citenamefont
  {Venugopalan}}]{Gelis:2010nm}%
  \BibitemOpen
  \bibfield  {author} {\bibinfo {author} {\bibfnamefont {F.}~\bibnamefont
  {Gelis}}, \bibinfo {author} {\bibfnamefont {E.}~\bibnamefont {Iancu}},
  \bibinfo {author} {\bibfnamefont {J.}~\bibnamefont {Jalilian-Marian}}, \ and\
  \bibinfo {author} {\bibfnamefont {R.}~\bibnamefont {Venugopalan}},\ }\href
  {\doibase 10.1146/annurev.nucl.010909.083629} {\bibfield  {journal} {\bibinfo
   {journal} {Ann. Rev. Nucl. Part. Sci.}\ }\textbf {\bibinfo {volume} {60}},\
  \bibinfo {pages} {463} (\bibinfo {year} {2010})},\ \Eprint
  {http://arxiv.org/abs/1002.0333} {arXiv:1002.0333 [hep-ph]} \BibitemShut
  {NoStop}%
\bibitem [{\citenamefont {Balitsky}(1996)}]{Balitsky:1995ub}%
  \BibitemOpen
  \bibfield  {author} {\bibinfo {author} {\bibfnamefont {I.}~\bibnamefont
  {Balitsky}},\ }\href {\doibase 10.1016/0550-3213(95)00638-9} {\bibfield
  {journal} {\bibinfo  {journal} {Nucl. Phys. B}\ }\textbf {\bibinfo {volume}
  {463}},\ \bibinfo {pages} {99} (\bibinfo {year} {1996})},\ \Eprint
  {http://arxiv.org/abs/hep-ph/9509348} {arXiv:hep-ph/9509348} \BibitemShut
  {NoStop}%
\bibitem [{\citenamefont {Kovchegov}(1999)}]{Kovchegov:1999yj}%
  \BibitemOpen
  \bibfield  {author} {\bibinfo {author} {\bibfnamefont {Y.~V.}\ \bibnamefont
  {Kovchegov}},\ }\href {\doibase 10.1103/PhysRevD.60.034008} {\bibfield
  {journal} {\bibinfo  {journal} {Phys. Rev. D}\ }\textbf {\bibinfo {volume}
  {60}},\ \bibinfo {pages} {034008} (\bibinfo {year} {1999})},\ \Eprint
  {http://arxiv.org/abs/hep-ph/9901281} {arXiv:hep-ph/9901281} \BibitemShut
  {NoStop}%
\bibitem [{\citenamefont {Jalilian-Marian}\ \emph {et~al.}(1998)\citenamefont
  {Jalilian-Marian}, \citenamefont {Kovner},\ and\ \citenamefont
  {Weigert}}]{Jalilian-Marian:1997ubg}%
  \BibitemOpen
  \bibfield  {author} {\bibinfo {author} {\bibfnamefont {J.}~\bibnamefont
  {Jalilian-Marian}}, \bibinfo {author} {\bibfnamefont {A.}~\bibnamefont
  {Kovner}}, \ and\ \bibinfo {author} {\bibfnamefont {H.}~\bibnamefont
  {Weigert}},\ }\href {\doibase 10.1103/PhysRevD.59.014015} {\bibfield
  {journal} {\bibinfo  {journal} {Phys. Rev. D}\ }\textbf {\bibinfo {volume}
  {59}},\ \bibinfo {pages} {014015} (\bibinfo {year} {1998})},\ \Eprint
  {http://arxiv.org/abs/hep-ph/9709432} {arXiv:hep-ph/9709432} \BibitemShut
  {NoStop}%
\bibitem [{\citenamefont {Iancu}\ \emph {et~al.}(2001)\citenamefont {Iancu},
  \citenamefont {Leonidov},\ and\ \citenamefont {McLerran}}]{Iancu:2000hn}%
  \BibitemOpen
  \bibfield  {author} {\bibinfo {author} {\bibfnamefont {E.}~\bibnamefont
  {Iancu}}, \bibinfo {author} {\bibfnamefont {A.}~\bibnamefont {Leonidov}}, \
  and\ \bibinfo {author} {\bibfnamefont {L.~D.}\ \bibnamefont {McLerran}},\
  }\href {\doibase 10.1016/S0375-9474(01)00642-X} {\bibfield  {journal}
  {\bibinfo  {journal} {Nucl. Phys. A}\ }\textbf {\bibinfo {volume} {692}},\
  \bibinfo {pages} {583} (\bibinfo {year} {2001})},\ \Eprint
  {http://arxiv.org/abs/hep-ph/0011241} {arXiv:hep-ph/0011241} \BibitemShut
  {NoStop}%
\bibitem [{\citenamefont {Balitsky}\ and\ \citenamefont
  {Chirilli}(2011)}]{Balitsky:2010ze}%
  \BibitemOpen
  \bibfield  {author} {\bibinfo {author} {\bibfnamefont {I.}~\bibnamefont
  {Balitsky}}\ and\ \bibinfo {author} {\bibfnamefont {G.~A.}\ \bibnamefont
  {Chirilli}},\ }\href {\doibase 10.1103/PhysRevD.83.031502} {\bibfield
  {journal} {\bibinfo  {journal} {Phys. Rev. D}\ }\textbf {\bibinfo {volume}
  {83}},\ \bibinfo {pages} {031502} (\bibinfo {year} {2011})},\ \Eprint
  {http://arxiv.org/abs/1009.4729} {arXiv:1009.4729 [hep-ph]} \BibitemShut
  {NoStop}%
\bibitem [{\citenamefont {Beuf}(2017)}]{Beuf:2017bpd}%
  \BibitemOpen
  \bibfield  {author} {\bibinfo {author} {\bibfnamefont {G.}~\bibnamefont
  {Beuf}},\ }\href {\doibase 10.1103/PhysRevD.96.074033} {\bibfield  {journal}
  {\bibinfo  {journal} {Phys. Rev. D}\ }\textbf {\bibinfo {volume} {96}},\
  \bibinfo {pages} {074033} (\bibinfo {year} {2017})},\ \Eprint
  {http://arxiv.org/abs/1708.06557} {arXiv:1708.06557 [hep-ph]} \BibitemShut
  {NoStop}%
\bibitem [{\citenamefont {Beuf}\ \emph {et~al.}(2022)\citenamefont {Beuf},
  \citenamefont {Lappi},\ and\ \citenamefont {Paatelainen}}]{Beuf:2021srj}%
  \BibitemOpen
  \bibfield  {author} {\bibinfo {author} {\bibfnamefont {G.}~\bibnamefont
  {Beuf}}, \bibinfo {author} {\bibfnamefont {T.}~\bibnamefont {Lappi}}, \ and\
  \bibinfo {author} {\bibfnamefont {R.}~\bibnamefont {Paatelainen}},\ }\href
  {\doibase 10.1103/PhysRevLett.129.072001} {\bibfield  {journal} {\bibinfo
  {journal} {Phys. Rev. Lett.}\ }\textbf {\bibinfo {volume} {129}},\ \bibinfo
  {pages} {072001} (\bibinfo {year} {2022})},\ \Eprint
  {http://arxiv.org/abs/2112.03158} {arXiv:2112.03158 [hep-ph]} \BibitemShut
  {NoStop}%
\bibitem [{\citenamefont {Boussarie}\ \emph {et~al.}(2016)\citenamefont
  {Boussarie}, \citenamefont {Grabovsky}, \citenamefont {Szymanowski},\ and\
  \citenamefont {Wallon}}]{Boussarie:2016ogo}%
  \BibitemOpen
  \bibfield  {author} {\bibinfo {author} {\bibfnamefont {R.}~\bibnamefont
  {Boussarie}}, \bibinfo {author} {\bibfnamefont {A.~V.}\ \bibnamefont
  {Grabovsky}}, \bibinfo {author} {\bibfnamefont {L.}~\bibnamefont
  {Szymanowski}}, \ and\ \bibinfo {author} {\bibfnamefont {S.}~\bibnamefont
  {Wallon}},\ }\href {\doibase 10.1007/JHEP11(2016)149} {\bibfield  {journal}
  {\bibinfo  {journal} {JHEP}\ }\textbf {\bibinfo {volume} {11}},\ \bibinfo
  {pages} {149} (\bibinfo {year} {2016})},\ \Eprint
  {http://arxiv.org/abs/1606.00419} {arXiv:1606.00419 [hep-ph]} \BibitemShut
  {NoStop}%
\bibitem [{\citenamefont {Boussarie}\ \emph
  {et~al.}(2017{\natexlab{b}})\citenamefont {Boussarie}, \citenamefont
  {Grabovsky}, \citenamefont {Ivanov}, \citenamefont {Szymanowski},\ and\
  \citenamefont {Wallon}}]{Boussarie:2016bkq}%
  \BibitemOpen
  \bibfield  {author} {\bibinfo {author} {\bibfnamefont {R.}~\bibnamefont
  {Boussarie}}, \bibinfo {author} {\bibfnamefont {A.~V.}\ \bibnamefont
  {Grabovsky}}, \bibinfo {author} {\bibfnamefont {D.~Y.}\ \bibnamefont
  {Ivanov}}, \bibinfo {author} {\bibfnamefont {L.}~\bibnamefont {Szymanowski}},
  \ and\ \bibinfo {author} {\bibfnamefont {S.}~\bibnamefont {Wallon}},\ }\href
  {\doibase 10.1103/PhysRevLett.119.072002} {\bibfield  {journal} {\bibinfo
  {journal} {Phys. Rev. Lett.}\ }\textbf {\bibinfo {volume} {119}},\ \bibinfo
  {pages} {072002} (\bibinfo {year} {2017}{\natexlab{b}})},\ \Eprint
  {http://arxiv.org/abs/1612.08026} {arXiv:1612.08026 [hep-ph]} \BibitemShut
  {NoStop}%
\bibitem [{\citenamefont {M\"antysaari}\ and\ \citenamefont
  {Penttala}(2022)}]{Mantysaari:2022kdm}%
  \BibitemOpen
  \bibfield  {author} {\bibinfo {author} {\bibfnamefont {H.}~\bibnamefont
  {M\"antysaari}}\ and\ \bibinfo {author} {\bibfnamefont {J.}~\bibnamefont
  {Penttala}},\ }\href {\doibase 10.1007/JHEP08(2022)247} {\bibfield  {journal}
  {\bibinfo  {journal} {JHEP}\ }\textbf {\bibinfo {volume} {08}},\ \bibinfo
  {pages} {247} (\bibinfo {year} {2022})},\ \Eprint
  {http://arxiv.org/abs/2204.14031} {arXiv:2204.14031 [hep-ph]} \BibitemShut
  {NoStop}%
\bibitem [{\citenamefont {Fucilla}\ \emph {et~al.}(2022)\citenamefont
  {Fucilla}, \citenamefont {Grabovsky}, \citenamefont {Li}, \citenamefont
  {Szymanowski},\ and\ \citenamefont {Wallon}}]{Fucilla:2022wcg}%
  \BibitemOpen
  \bibfield  {author} {\bibinfo {author} {\bibfnamefont {M.}~\bibnamefont
  {Fucilla}}, \bibinfo {author} {\bibfnamefont {A.~V.}\ \bibnamefont
  {Grabovsky}}, \bibinfo {author} {\bibfnamefont {E.}~\bibnamefont {Li}},
  \bibinfo {author} {\bibfnamefont {L.}~\bibnamefont {Szymanowski}}, \ and\
  \bibinfo {author} {\bibfnamefont {S.}~\bibnamefont {Wallon}},\ }\href@noop {}
  {\  (\bibinfo {year} {2022})},\ \Eprint {http://arxiv.org/abs/2211.05774}
  {arXiv:2211.05774 [hep-ph]} \BibitemShut {NoStop}%
\bibitem [{\citenamefont {Bergabo}\ and\ \citenamefont
  {Jalilian-Marian}(2023{\natexlab{a}})}]{Bergabo:2022zhe}%
  \BibitemOpen
  \bibfield  {author} {\bibinfo {author} {\bibfnamefont {F.}~\bibnamefont
  {Bergabo}}\ and\ \bibinfo {author} {\bibfnamefont {J.}~\bibnamefont
  {Jalilian-Marian}},\ }\href {\doibase 10.1007/JHEP01(2023)095} {\bibfield
  {journal} {\bibinfo  {journal} {JHEP}\ }\textbf {\bibinfo {volume} {01}},\
  \bibinfo {pages} {095} (\bibinfo {year} {2023}{\natexlab{a}})},\ \Eprint
  {http://arxiv.org/abs/2210.03208} {arXiv:2210.03208 [hep-ph]} \BibitemShut
  {NoStop}%
\bibitem [{\citenamefont {Caucal}\ \emph {et~al.}(2021)\citenamefont {Caucal},
  \citenamefont {Salazar},\ and\ \citenamefont {Venugopalan}}]{Caucal:2021ent}%
  \BibitemOpen
  \bibfield  {author} {\bibinfo {author} {\bibfnamefont {P.}~\bibnamefont
  {Caucal}}, \bibinfo {author} {\bibfnamefont {F.}~\bibnamefont {Salazar}}, \
  and\ \bibinfo {author} {\bibfnamefont {R.}~\bibnamefont {Venugopalan}},\
  }\href {\doibase 10.1007/JHEP11(2021)222} {\bibfield  {journal} {\bibinfo
  {journal} {JHEP}\ }\textbf {\bibinfo {volume} {11}},\ \bibinfo {pages} {222}
  (\bibinfo {year} {2021})},\ \Eprint {http://arxiv.org/abs/2108.06347}
  {arXiv:2108.06347 [hep-ph]} \BibitemShut {NoStop}%
\bibitem [{\citenamefont {Caucal}\ \emph {et~al.}(2022)\citenamefont {Caucal},
  \citenamefont {Salazar}, \citenamefont {Schenke},\ and\ \citenamefont
  {Venugopalan}}]{Caucal:2022ulg}%
  \BibitemOpen
  \bibfield  {author} {\bibinfo {author} {\bibfnamefont {P.}~\bibnamefont
  {Caucal}}, \bibinfo {author} {\bibfnamefont {F.}~\bibnamefont {Salazar}},
  \bibinfo {author} {\bibfnamefont {B.}~\bibnamefont {Schenke}}, \ and\
  \bibinfo {author} {\bibfnamefont {R.}~\bibnamefont {Venugopalan}},\ }\href
  {\doibase 10.1007/JHEP11(2022)169} {\bibfield  {journal} {\bibinfo  {journal}
  {JHEP}\ }\textbf {\bibinfo {volume} {11}},\ \bibinfo {pages} {169} (\bibinfo
  {year} {2022})},\ \Eprint {http://arxiv.org/abs/2208.13872} {arXiv:2208.13872
  [hep-ph]} \BibitemShut {NoStop}%
\bibitem [{\citenamefont {Bergabo}\ and\ \citenamefont
  {Jalilian-Marian}(2022)}]{Bergabo:2022tcu}%
  \BibitemOpen
  \bibfield  {author} {\bibinfo {author} {\bibfnamefont {F.}~\bibnamefont
  {Bergabo}}\ and\ \bibinfo {author} {\bibfnamefont {J.}~\bibnamefont
  {Jalilian-Marian}},\ }\href {\doibase 10.1103/PhysRevD.106.054035} {\bibfield
   {journal} {\bibinfo  {journal} {Phys. Rev. D}\ }\textbf {\bibinfo {volume}
  {106}},\ \bibinfo {pages} {054035} (\bibinfo {year} {2022})},\ \Eprint
  {http://arxiv.org/abs/2207.03606} {arXiv:2207.03606 [hep-ph]} \BibitemShut
  {NoStop}%
\bibitem [{\citenamefont {Bergabo}\ and\ \citenamefont
  {Jalilian-Marian}(2023{\natexlab{b}})}]{Bergabo:2023wed}%
  \BibitemOpen
  \bibfield  {author} {\bibinfo {author} {\bibfnamefont {F.}~\bibnamefont
  {Bergabo}}\ and\ \bibinfo {author} {\bibfnamefont {J.}~\bibnamefont
  {Jalilian-Marian}},\ }\href@noop {} {\  (\bibinfo {year}
  {2023}{\natexlab{b}})},\ \Eprint {http://arxiv.org/abs/2301.03117}
  {arXiv:2301.03117 [hep-ph]} \BibitemShut {NoStop}%
\bibitem [{\citenamefont {Caucal}\ \emph {et~al.}(2023)\citenamefont {Caucal},
  \citenamefont {Salazar}, \citenamefont {Schenke}, \citenamefont {Stebel},\
  and\ \citenamefont {Venugopalan}}]{Caucal:2023nci}%
  \BibitemOpen
  \bibfield  {author} {\bibinfo {author} {\bibfnamefont {P.}~\bibnamefont
  {Caucal}}, \bibinfo {author} {\bibfnamefont {F.}~\bibnamefont {Salazar}},
  \bibinfo {author} {\bibfnamefont {B.}~\bibnamefont {Schenke}}, \bibinfo
  {author} {\bibfnamefont {T.}~\bibnamefont {Stebel}}, \ and\ \bibinfo {author}
  {\bibfnamefont {R.}~\bibnamefont {Venugopalan}},\ }\href@noop {} {\
  (\bibinfo {year} {2023})},\ \Eprint {http://arxiv.org/abs/2304.03304}
  {arXiv:2304.03304 [hep-ph]} \BibitemShut {NoStop}%
\bibitem [{\citenamefont {Roy}\ and\ \citenamefont
  {Venugopalan}(2020)}]{Roy:2019hwr}%
  \BibitemOpen
  \bibfield  {author} {\bibinfo {author} {\bibfnamefont {K.}~\bibnamefont
  {Roy}}\ and\ \bibinfo {author} {\bibfnamefont {R.}~\bibnamefont
  {Venugopalan}},\ }\href {\doibase 10.1103/PhysRevD.101.034028} {\bibfield
  {journal} {\bibinfo  {journal} {Phys. Rev. D}\ }\textbf {\bibinfo {volume}
  {101}},\ \bibinfo {pages} {034028} (\bibinfo {year} {2020})},\ \Eprint
  {http://arxiv.org/abs/1911.04530} {arXiv:1911.04530 [hep-ph]} \BibitemShut
  {NoStop}%
\bibitem [{\citenamefont {Iancu}\ \emph {et~al.}(2022)\citenamefont {Iancu},
  \citenamefont {Mueller},\ and\ \citenamefont
  {Triantafyllopoulos}}]{Iancu:2021rup}%
  \BibitemOpen
  \bibfield  {author} {\bibinfo {author} {\bibfnamefont {E.}~\bibnamefont
  {Iancu}}, \bibinfo {author} {\bibfnamefont {A.~H.}\ \bibnamefont {Mueller}},
  \ and\ \bibinfo {author} {\bibfnamefont {D.~N.}\ \bibnamefont
  {Triantafyllopoulos}},\ }\href {\doibase 10.1103/PhysRevLett.128.202001}
  {\bibfield  {journal} {\bibinfo  {journal} {Phys. Rev. Lett.}\ }\textbf
  {\bibinfo {volume} {128}},\ \bibinfo {pages} {202001} (\bibinfo {year}
  {2022})},\ \Eprint {http://arxiv.org/abs/2112.06353} {arXiv:2112.06353
  [hep-ph]} \BibitemShut {NoStop}%
\bibitem [{\citenamefont {Kolb\'e}\ \emph {et~al.}(2021)\citenamefont
  {Kolb\'e}, \citenamefont {Roy}, \citenamefont {Salazar}, \citenamefont
  {Schenke},\ and\ \citenamefont {Venugopalan}}]{Kolbe:2020tlq}%
  \BibitemOpen
  \bibfield  {author} {\bibinfo {author} {\bibfnamefont {I.}~\bibnamefont
  {Kolb\'e}}, \bibinfo {author} {\bibfnamefont {K.}~\bibnamefont {Roy}},
  \bibinfo {author} {\bibfnamefont {F.}~\bibnamefont {Salazar}}, \bibinfo
  {author} {\bibfnamefont {B.}~\bibnamefont {Schenke}}, \ and\ \bibinfo
  {author} {\bibfnamefont {R.}~\bibnamefont {Venugopalan}},\ }\href {\doibase
  10.1007/JHEP01(2021)052} {\bibfield  {journal} {\bibinfo  {journal} {JHEP}\
  }\textbf {\bibinfo {volume} {01}},\ \bibinfo {pages} {052} (\bibinfo {year}
  {2021})},\ \Eprint {http://arxiv.org/abs/2008.04372} {arXiv:2008.04372
  [hep-ph]} \BibitemShut {NoStop}%
\bibitem [{\citenamefont {Tong}\ \emph {et~al.}(2022)\citenamefont {Tong},
  \citenamefont {Xiao},\ and\ \citenamefont {Zhang}}]{Tong:2022zwp}%
  \BibitemOpen
  \bibfield  {author} {\bibinfo {author} {\bibfnamefont {X.-B.}\ \bibnamefont
  {Tong}}, \bibinfo {author} {\bibfnamefont {B.-W.}\ \bibnamefont {Xiao}}, \
  and\ \bibinfo {author} {\bibfnamefont {Y.-Y.}\ \bibnamefont {Zhang}},\
  }\href@noop {} {\  (\bibinfo {year} {2022})},\ \Eprint
  {http://arxiv.org/abs/2211.01647} {arXiv:2211.01647 [hep-ph]} \BibitemShut
  {NoStop}%
\bibitem [{\citenamefont {Liu}\ \emph {et~al.}(2023)\citenamefont {Liu},
  \citenamefont {Liu}, \citenamefont {Pan}, \citenamefont {Yuan},\ and\
  \citenamefont {Zhu}}]{Liu:2023aqb}%
  \BibitemOpen
  \bibfield  {author} {\bibinfo {author} {\bibfnamefont {H.-Y.}\ \bibnamefont
  {Liu}}, \bibinfo {author} {\bibfnamefont {X.}~\bibnamefont {Liu}}, \bibinfo
  {author} {\bibfnamefont {J.-C.}\ \bibnamefont {Pan}}, \bibinfo {author}
  {\bibfnamefont {F.}~\bibnamefont {Yuan}}, \ and\ \bibinfo {author}
  {\bibfnamefont {H.~X.}\ \bibnamefont {Zhu}},\ }\href@noop {} {\  (\bibinfo
  {year} {2023})},\ \Eprint {http://arxiv.org/abs/2301.01788} {arXiv:2301.01788
  [hep-ph]} \BibitemShut {NoStop}%
\bibitem [{\citenamefont {Fadin}\ \emph {et~al.}(1975)\citenamefont {Fadin},
  \citenamefont {Kuraev},\ and\ \citenamefont {Lipatov}}]{Fadin:1975cb}%
  \BibitemOpen
  \bibfield  {author} {\bibinfo {author} {\bibfnamefont {V.~S.}\ \bibnamefont
  {Fadin}}, \bibinfo {author} {\bibfnamefont {E.~A.}\ \bibnamefont {Kuraev}}, \
  and\ \bibinfo {author} {\bibfnamefont {L.~N.}\ \bibnamefont {Lipatov}},\
  }\href {\doibase 10.1016/0370-2693(75)90524-9} {\bibfield  {journal}
  {\bibinfo  {journal} {Phys. Lett. B}\ }\textbf {\bibinfo {volume} {60}},\
  \bibinfo {pages} {50} (\bibinfo {year} {1975})}\BibitemShut {NoStop}%
\bibitem [{\citenamefont {Kuraev}\ \emph {et~al.}(1976)\citenamefont {Kuraev},
  \citenamefont {Lipatov},\ and\ \citenamefont {Fadin}}]{Kuraev:1976ge}%
  \BibitemOpen
  \bibfield  {author} {\bibinfo {author} {\bibfnamefont {E.~A.}\ \bibnamefont
  {Kuraev}}, \bibinfo {author} {\bibfnamefont {L.~N.}\ \bibnamefont {Lipatov}},
  \ and\ \bibinfo {author} {\bibfnamefont {V.~S.}\ \bibnamefont {Fadin}},\
  }\href@noop {} {\bibfield  {journal} {\bibinfo  {journal} {Sov. Phys. JETP}\
  }\textbf {\bibinfo {volume} {44}},\ \bibinfo {pages} {443} (\bibinfo {year}
  {1976})}\BibitemShut {NoStop}%
\bibitem [{\citenamefont {Kuraev}\ \emph {et~al.}(1977)\citenamefont {Kuraev},
  \citenamefont {Lipatov},\ and\ \citenamefont {Fadin}}]{Kuraev:1977fs}%
  \BibitemOpen
  \bibfield  {author} {\bibinfo {author} {\bibfnamefont {E.~A.}\ \bibnamefont
  {Kuraev}}, \bibinfo {author} {\bibfnamefont {L.~N.}\ \bibnamefont {Lipatov}},
  \ and\ \bibinfo {author} {\bibfnamefont {V.~S.}\ \bibnamefont {Fadin}},\
  }\href@noop {} {\bibfield  {journal} {\bibinfo  {journal} {Sov. Phys. JETP}\
  }\textbf {\bibinfo {volume} {45}},\ \bibinfo {pages} {199} (\bibinfo {year}
  {1977})}\BibitemShut {NoStop}%
\bibitem [{\citenamefont {Balitsky}\ and\ \citenamefont
  {Lipatov}(1978)}]{Balitsky:1978ic}%
  \BibitemOpen
  \bibfield  {author} {\bibinfo {author} {\bibfnamefont {I.~I.}\ \bibnamefont
  {Balitsky}}\ and\ \bibinfo {author} {\bibfnamefont {L.~N.}\ \bibnamefont
  {Lipatov}},\ }\href@noop {} {\bibfield  {journal} {\bibinfo  {journal} {Sov.
  J. Nucl. Phys.}\ }\textbf {\bibinfo {volume} {28}},\ \bibinfo {pages} {822}
  (\bibinfo {year} {1978})}\BibitemShut {NoStop}%
\bibitem [{\citenamefont {Lappi}\ and\ \citenamefont
  {M\"antysaari}(2016)}]{Lappi:2016fmu}%
  \BibitemOpen
  \bibfield  {author} {\bibinfo {author} {\bibfnamefont {T.}~\bibnamefont
  {Lappi}}\ and\ \bibinfo {author} {\bibfnamefont {H.}~\bibnamefont
  {M\"antysaari}},\ }\href {\doibase 10.1103/PhysRevD.93.094004} {\bibfield
  {journal} {\bibinfo  {journal} {Phys. Rev. D}\ }\textbf {\bibinfo {volume}
  {93}},\ \bibinfo {pages} {094004} (\bibinfo {year} {2016})},\ \Eprint
  {http://arxiv.org/abs/1601.06598} {arXiv:1601.06598 [hep-ph]} \BibitemShut
  {NoStop}%
\bibitem [{\citenamefont {Duclou\'e}\ \emph {et~al.}(2019)\citenamefont
  {Duclou\'e}, \citenamefont {Iancu}, \citenamefont {Mueller}, \citenamefont
  {Soyez},\ and\ \citenamefont {Triantafyllopoulos}}]{Ducloue:2019ezk}%
  \BibitemOpen
  \bibfield  {author} {\bibinfo {author} {\bibfnamefont {B.}~\bibnamefont
  {Duclou\'e}}, \bibinfo {author} {\bibfnamefont {E.}~\bibnamefont {Iancu}},
  \bibinfo {author} {\bibfnamefont {A.~H.}\ \bibnamefont {Mueller}}, \bibinfo
  {author} {\bibfnamefont {G.}~\bibnamefont {Soyez}}, \ and\ \bibinfo {author}
  {\bibfnamefont {D.~N.}\ \bibnamefont {Triantafyllopoulos}},\ }\href {\doibase
  10.1007/JHEP04(2019)081} {\bibfield  {journal} {\bibinfo  {journal} {JHEP}\
  }\textbf {\bibinfo {volume} {04}},\ \bibinfo {pages} {081} (\bibinfo {year}
  {2019})},\ \Eprint {http://arxiv.org/abs/1902.06637} {arXiv:1902.06637
  [hep-ph]} \BibitemShut {NoStop}%
\bibitem [{\citenamefont {Sta\'sto}\ \emph {et~al.}(2014)\citenamefont
  {Sta\'sto}, \citenamefont {Xiao}, \citenamefont {Yuan},\ and\ \citenamefont
  {Zaslavsky}}]{Stasto:2014sea}%
  \BibitemOpen
  \bibfield  {author} {\bibinfo {author} {\bibfnamefont {A.~M.}\ \bibnamefont
  {Sta\'sto}}, \bibinfo {author} {\bibfnamefont {B.-W.}\ \bibnamefont {Xiao}},
  \bibinfo {author} {\bibfnamefont {F.}~\bibnamefont {Yuan}}, \ and\ \bibinfo
  {author} {\bibfnamefont {D.}~\bibnamefont {Zaslavsky}},\ }\href {\doibase
  10.1103/PhysRevD.90.014047} {\bibfield  {journal} {\bibinfo  {journal} {Phys.
  Rev. D}\ }\textbf {\bibinfo {volume} {90}},\ \bibinfo {pages} {014047}
  (\bibinfo {year} {2014})},\ \Eprint {http://arxiv.org/abs/1405.6311}
  {arXiv:1405.6311 [hep-ph]} \BibitemShut {NoStop}%
\bibitem [{\citenamefont {Fadin}\ and\ \citenamefont
  {Lipatov}(1998)}]{Fadin:1998py}%
  \BibitemOpen
  \bibfield  {author} {\bibinfo {author} {\bibfnamefont {V.~S.}\ \bibnamefont
  {Fadin}}\ and\ \bibinfo {author} {\bibfnamefont {L.~N.}\ \bibnamefont
  {Lipatov}},\ }\href {\doibase 10.1016/S0370-2693(98)00473-0} {\bibfield
  {journal} {\bibinfo  {journal} {Phys. Lett. B}\ }\textbf {\bibinfo {volume}
  {429}},\ \bibinfo {pages} {127} (\bibinfo {year} {1998})},\ \Eprint
  {http://arxiv.org/abs/hep-ph/9802290} {arXiv:hep-ph/9802290} \BibitemShut
  {NoStop}%
\bibitem [{\citenamefont {Ciafaloni}\ and\ \citenamefont
  {Camici}(1998)}]{Ciafaloni:1998gs}%
  \BibitemOpen
  \bibfield  {author} {\bibinfo {author} {\bibfnamefont {M.}~\bibnamefont
  {Ciafaloni}}\ and\ \bibinfo {author} {\bibfnamefont {G.}~\bibnamefont
  {Camici}},\ }\href {\doibase 10.1016/S0370-2693(98)00551-6} {\bibfield
  {journal} {\bibinfo  {journal} {Phys. Lett. B}\ }\textbf {\bibinfo {volume}
  {430}},\ \bibinfo {pages} {349} (\bibinfo {year} {1998})},\ \Eprint
  {http://arxiv.org/abs/hep-ph/9803389} {arXiv:hep-ph/9803389} \BibitemShut
  {NoStop}%
\bibitem [{\citenamefont {Balitsky}\ and\ \citenamefont
  {Chirilli}(2013)}]{Balitsky:2013fea}%
  \BibitemOpen
  \bibfield  {author} {\bibinfo {author} {\bibfnamefont {I.}~\bibnamefont
  {Balitsky}}\ and\ \bibinfo {author} {\bibfnamefont {G.~A.}\ \bibnamefont
  {Chirilli}},\ }\href {\doibase 10.1103/PhysRevD.88.111501} {\bibfield
  {journal} {\bibinfo  {journal} {Phys. Rev. D}\ }\textbf {\bibinfo {volume}
  {88}},\ \bibinfo {pages} {111501} (\bibinfo {year} {2013})},\ \Eprint
  {http://arxiv.org/abs/1309.7644} {arXiv:1309.7644 [hep-ph]} \BibitemShut
  {NoStop}%
\bibitem [{\citenamefont {Kovner}\ \emph {et~al.}(2014)\citenamefont {Kovner},
  \citenamefont {Lublinsky},\ and\ \citenamefont {Mulian}}]{Kovner:2014lca}%
  \BibitemOpen
  \bibfield  {author} {\bibinfo {author} {\bibfnamefont {A.}~\bibnamefont
  {Kovner}}, \bibinfo {author} {\bibfnamefont {M.}~\bibnamefont {Lublinsky}}, \
  and\ \bibinfo {author} {\bibfnamefont {Y.}~\bibnamefont {Mulian}},\ }\href
  {\doibase 10.1007/JHEP08(2014)114} {\bibfield  {journal} {\bibinfo  {journal}
  {JHEP}\ }\textbf {\bibinfo {volume} {08}},\ \bibinfo {pages} {114} (\bibinfo
  {year} {2014})},\ \Eprint {http://arxiv.org/abs/1405.0418} {arXiv:1405.0418
  [hep-ph]} \BibitemShut {NoStop}%
\bibitem [{\citenamefont {Mueller}(2018)}]{Mueller:2018llt}%
  \BibitemOpen
  \bibfield  {author} {\bibinfo {author} {\bibfnamefont {A.~H.}\ \bibnamefont
  {Mueller}},\ }\href {\doibase 10.1007/JHEP08(2018)139} {\bibfield  {journal}
  {\bibinfo  {journal} {JHEP}\ }\textbf {\bibinfo {volume} {08}},\ \bibinfo
  {pages} {139} (\bibinfo {year} {2018})},\ \Eprint
  {http://arxiv.org/abs/1804.07249} {arXiv:1804.07249 [hep-th]} \BibitemShut
  {NoStop}%
\bibitem [{\citenamefont {Hatta}\ and\ \citenamefont
  {Ueda}(2013)}]{Hatta:2013iba}%
  \BibitemOpen
  \bibfield  {author} {\bibinfo {author} {\bibfnamefont {Y.}~\bibnamefont
  {Hatta}}\ and\ \bibinfo {author} {\bibfnamefont {T.}~\bibnamefont {Ueda}},\
  }\href {\doibase 10.1016/j.nuclphysb.2013.06.021} {\bibfield  {journal}
  {\bibinfo  {journal} {Nucl. Phys. B}\ }\textbf {\bibinfo {volume} {874}},\
  \bibinfo {pages} {808} (\bibinfo {year} {2013})},\ \Eprint
  {http://arxiv.org/abs/1304.6930} {arXiv:1304.6930 [hep-ph]} \BibitemShut
  {NoStop}%
\bibitem [{\citenamefont {Caron-Huot}\ and\ \citenamefont
  {Herranen}(2018)}]{Caron-Huot:2016tzz}%
  \BibitemOpen
  \bibfield  {author} {\bibinfo {author} {\bibfnamefont {S.}~\bibnamefont
  {Caron-Huot}}\ and\ \bibinfo {author} {\bibfnamefont {M.}~\bibnamefont
  {Herranen}},\ }\href {\doibase 10.1007/JHEP02(2018)058} {\bibfield  {journal}
  {\bibinfo  {journal} {JHEP}\ }\textbf {\bibinfo {volume} {02}},\ \bibinfo
  {pages} {058} (\bibinfo {year} {2018})},\ \Eprint
  {http://arxiv.org/abs/1604.07417} {arXiv:1604.07417 [hep-ph]} \BibitemShut
  {NoStop}%
\bibitem [{\citenamefont {Travaglini}\ \emph {et~al.}(2022)\citenamefont
  {Travaglini} \emph {et~al.}}]{Travaglini:2022uwo}%
  \BibitemOpen
  \bibfield  {author} {\bibinfo {author} {\bibfnamefont {G.}~\bibnamefont
  {Travaglini}} \emph {et~al.},\ }\href {\doibase 10.1088/1751-8121/ac8380}
  {\bibfield  {journal} {\bibinfo  {journal} {J. Phys. A}\ }\textbf {\bibinfo
  {volume} {55}},\ \bibinfo {pages} {443001} (\bibinfo {year} {2022})},\
  \Eprint {http://arxiv.org/abs/2203.13011} {arXiv:2203.13011 [hep-th]}
  \BibitemShut {NoStop}%
\bibitem [{\citenamefont {Jalilian-Marian}(2017)}]{Jalilian-Marian:2017ttv}%
  \BibitemOpen
  \bibfield  {author} {\bibinfo {author} {\bibfnamefont {J.}~\bibnamefont
  {Jalilian-Marian}},\ }\href {\doibase 10.1103/PhysRevD.96.074020} {\bibfield
  {journal} {\bibinfo  {journal} {Phys. Rev. D}\ }\textbf {\bibinfo {volume}
  {96}},\ \bibinfo {pages} {074020} (\bibinfo {year} {2017})},\ \Eprint
  {http://arxiv.org/abs/1708.07533} {arXiv:1708.07533 [hep-ph]} \BibitemShut
  {NoStop}%
\bibitem [{\citenamefont {Jalilian-Marian}(2019)}]{Jalilian-Marian:2018iui}%
  \BibitemOpen
  \bibfield  {author} {\bibinfo {author} {\bibfnamefont {J.}~\bibnamefont
  {Jalilian-Marian}},\ }\href {\doibase 10.1103/PhysRevD.99.014043} {\bibfield
  {journal} {\bibinfo  {journal} {Phys. Rev. D}\ }\textbf {\bibinfo {volume}
  {99}},\ \bibinfo {pages} {014043} (\bibinfo {year} {2019})},\ \Eprint
  {http://arxiv.org/abs/1809.04625} {arXiv:1809.04625 [hep-ph]} \BibitemShut
  {NoStop}%
\bibitem [{\citenamefont {Jalilian-Marian}(2020)}]{Jalilian-Marian:2019kaf}%
  \BibitemOpen
  \bibfield  {author} {\bibinfo {author} {\bibfnamefont {J.}~\bibnamefont
  {Jalilian-Marian}},\ }\href {\doibase 10.1103/PhysRevD.102.014008} {\bibfield
   {journal} {\bibinfo  {journal} {Phys. Rev. D}\ }\textbf {\bibinfo {volume}
  {102}},\ \bibinfo {pages} {014008} (\bibinfo {year} {2020})},\ \Eprint
  {http://arxiv.org/abs/1912.08878} {arXiv:1912.08878 [hep-ph]} \BibitemShut
  {NoStop}%
\bibitem [{\citenamefont {Boussarie}\ and\ \citenamefont
  {Mehtar-Tani}(2022{\natexlab{a}})}]{Boussarie:2021wkn}%
  \BibitemOpen
  \bibfield  {author} {\bibinfo {author} {\bibfnamefont {R.}~\bibnamefont
  {Boussarie}}\ and\ \bibinfo {author} {\bibfnamefont {Y.}~\bibnamefont
  {Mehtar-Tani}},\ }\href {\doibase 10.1007/JHEP07(2022)080} {\bibfield
  {journal} {\bibinfo  {journal} {JHEP}\ }\textbf {\bibinfo {volume} {07}},\
  \bibinfo {pages} {080} (\bibinfo {year} {2022}{\natexlab{a}})},\ \Eprint
  {http://arxiv.org/abs/2112.01412} {arXiv:2112.01412 [hep-ph]} \BibitemShut
  {NoStop}%
\bibitem [{\citenamefont {Boussarie}\ and\ \citenamefont
  {Mehtar-Tani}(2022{\natexlab{b}})}]{Boussarie:2020fpb}%
  \BibitemOpen
  \bibfield  {author} {\bibinfo {author} {\bibfnamefont {R.}~\bibnamefont
  {Boussarie}}\ and\ \bibinfo {author} {\bibfnamefont {Y.}~\bibnamefont
  {Mehtar-Tani}},\ }\href {\doibase 10.1016/j.physletb.2022.137125} {\bibfield
  {journal} {\bibinfo  {journal} {Phys. Lett. B}\ }\textbf {\bibinfo {volume}
  {831}},\ \bibinfo {pages} {137125} (\bibinfo {year} {2022}{\natexlab{b}})},\
  \Eprint {http://arxiv.org/abs/2006.14569} {arXiv:2006.14569 [hep-ph]}
  \BibitemShut {NoStop}%
\bibitem [{\citenamefont {Altinoluk}\ \emph {et~al.}(2021)\citenamefont
  {Altinoluk}, \citenamefont {Beuf}, \citenamefont {Czajka},\ and\
  \citenamefont {Tymowska}}]{Altinoluk:2020oyd}%
  \BibitemOpen
  \bibfield  {author} {\bibinfo {author} {\bibfnamefont {T.}~\bibnamefont
  {Altinoluk}}, \bibinfo {author} {\bibfnamefont {G.}~\bibnamefont {Beuf}},
  \bibinfo {author} {\bibfnamefont {A.}~\bibnamefont {Czajka}}, \ and\ \bibinfo
  {author} {\bibfnamefont {A.}~\bibnamefont {Tymowska}},\ }\href {\doibase
  10.1103/PhysRevD.104.014019} {\bibfield  {journal} {\bibinfo  {journal}
  {Phys. Rev. D}\ }\textbf {\bibinfo {volume} {104}},\ \bibinfo {pages}
  {014019} (\bibinfo {year} {2021})},\ \Eprint
  {http://arxiv.org/abs/2012.03886} {arXiv:2012.03886 [hep-ph]} \BibitemShut
  {NoStop}%
\bibitem [{\citenamefont {Altinoluk}\ and\ \citenamefont
  {Beuf}(2022)}]{Altinoluk:2021lvu}%
  \BibitemOpen
  \bibfield  {author} {\bibinfo {author} {\bibfnamefont {T.}~\bibnamefont
  {Altinoluk}}\ and\ \bibinfo {author} {\bibfnamefont {G.}~\bibnamefont
  {Beuf}},\ }\href {\doibase 10.1103/PhysRevD.105.074026} {\bibfield  {journal}
  {\bibinfo  {journal} {Phys. Rev. D}\ }\textbf {\bibinfo {volume} {105}},\
  \bibinfo {pages} {074026} (\bibinfo {year} {2022})},\ \Eprint
  {http://arxiv.org/abs/2109.01620} {arXiv:2109.01620 [hep-ph]} \BibitemShut
  {NoStop}%
\bibitem [{\citenamefont {Altinoluk}\ \emph {et~al.}(2022)\citenamefont
  {Altinoluk}, \citenamefont {Beuf}, \citenamefont {Czajka},\ and\
  \citenamefont {Tymowska}}]{Altinoluk:2022jkk}%
  \BibitemOpen
  \bibfield  {author} {\bibinfo {author} {\bibfnamefont {T.}~\bibnamefont
  {Altinoluk}}, \bibinfo {author} {\bibfnamefont {G.}~\bibnamefont {Beuf}},
  \bibinfo {author} {\bibfnamefont {A.}~\bibnamefont {Czajka}}, \ and\ \bibinfo
  {author} {\bibfnamefont {A.}~\bibnamefont {Tymowska}},\ }\href@noop {} {\
  (\bibinfo {year} {2022})},\ \Eprint {http://arxiv.org/abs/2212.10484}
  {arXiv:2212.10484 [hep-ph]} \BibitemShut {NoStop}%
\bibitem [{\citenamefont {De~Florian}\ \emph {et~al.}(2019)\citenamefont
  {De~Florian}, \citenamefont {Lucero}, \citenamefont {Sassot}, \citenamefont
  {Stratmann},\ and\ \citenamefont {Vogelsang}}]{DeFlorian:2019xxt}%
  \BibitemOpen
  \bibfield  {author} {\bibinfo {author} {\bibfnamefont {D.}~\bibnamefont
  {De~Florian}}, \bibinfo {author} {\bibfnamefont {G.~A.}\ \bibnamefont
  {Lucero}}, \bibinfo {author} {\bibfnamefont {R.}~\bibnamefont {Sassot}},
  \bibinfo {author} {\bibfnamefont {M.}~\bibnamefont {Stratmann}}, \ and\
  \bibinfo {author} {\bibfnamefont {W.}~\bibnamefont {Vogelsang}},\ }\href
  {\doibase 10.1103/PhysRevD.100.114027} {\bibfield  {journal} {\bibinfo
  {journal} {Phys. Rev. D}\ }\textbf {\bibinfo {volume} {100}},\ \bibinfo
  {pages} {114027} (\bibinfo {year} {2019})},\ \Eprint
  {http://arxiv.org/abs/1902.10548} {arXiv:1902.10548 [hep-ph]} \BibitemShut
  {NoStop}%
\bibitem [{\citenamefont {Kovchegov}\ \emph {et~al.}(2016)\citenamefont
  {Kovchegov}, \citenamefont {Pitonyak},\ and\ \citenamefont
  {Sievert}}]{Kovchegov:2015pbl}%
  \BibitemOpen
  \bibfield  {author} {\bibinfo {author} {\bibfnamefont {Y.~V.}\ \bibnamefont
  {Kovchegov}}, \bibinfo {author} {\bibfnamefont {D.}~\bibnamefont {Pitonyak}},
  \ and\ \bibinfo {author} {\bibfnamefont {M.~D.}\ \bibnamefont {Sievert}},\
  }\href {\doibase 10.1007/JHEP01(2016)072} {\bibfield  {journal} {\bibinfo
  {journal} {JHEP}\ }\textbf {\bibinfo {volume} {01}},\ \bibinfo {pages} {072}
  (\bibinfo {year} {2016})},\ \bibinfo {note} {[Erratum: JHEP 10, 148
  (2016)]},\ \Eprint {http://arxiv.org/abs/1511.06737} {arXiv:1511.06737
  [hep-ph]} \BibitemShut {NoStop}%
\bibitem [{\citenamefont {Kovchegov}\ \emph
  {et~al.}(2017{\natexlab{a}})\citenamefont {Kovchegov}, \citenamefont
  {Pitonyak},\ and\ \citenamefont {Sievert}}]{Kovchegov:2016zex}%
  \BibitemOpen
  \bibfield  {author} {\bibinfo {author} {\bibfnamefont {Y.~V.}\ \bibnamefont
  {Kovchegov}}, \bibinfo {author} {\bibfnamefont {D.}~\bibnamefont {Pitonyak}},
  \ and\ \bibinfo {author} {\bibfnamefont {M.~D.}\ \bibnamefont {Sievert}},\
  }\href {\doibase 10.1103/PhysRevD.95.014033} {\bibfield  {journal} {\bibinfo
  {journal} {Phys. Rev. D}\ }\textbf {\bibinfo {volume} {95}},\ \bibinfo
  {pages} {014033} (\bibinfo {year} {2017}{\natexlab{a}})},\ \Eprint
  {http://arxiv.org/abs/1610.06197} {arXiv:1610.06197 [hep-ph]} \BibitemShut
  {NoStop}%
\bibitem [{\citenamefont {Kovchegov}\ \emph
  {et~al.}(2017{\natexlab{b}})\citenamefont {Kovchegov}, \citenamefont
  {Pitonyak},\ and\ \citenamefont {Sievert}}]{Kovchegov:2017lsr}%
  \BibitemOpen
  \bibfield  {author} {\bibinfo {author} {\bibfnamefont {Y.~V.}\ \bibnamefont
  {Kovchegov}}, \bibinfo {author} {\bibfnamefont {D.}~\bibnamefont {Pitonyak}},
  \ and\ \bibinfo {author} {\bibfnamefont {M.~D.}\ \bibnamefont {Sievert}},\
  }\href {\doibase 10.1007/JHEP10(2017)198} {\bibfield  {journal} {\bibinfo
  {journal} {JHEP}\ }\textbf {\bibinfo {volume} {10}},\ \bibinfo {pages} {198}
  (\bibinfo {year} {2017}{\natexlab{b}})},\ \Eprint
  {http://arxiv.org/abs/1706.04236} {arXiv:1706.04236 [nucl-th]} \BibitemShut
  {NoStop}%
\bibitem [{\citenamefont {Kovchegov}\ and\ \citenamefont
  {Sievert}(2019)}]{Kovchegov:2018znm}%
  \BibitemOpen
  \bibfield  {author} {\bibinfo {author} {\bibfnamefont {Y.~V.}\ \bibnamefont
  {Kovchegov}}\ and\ \bibinfo {author} {\bibfnamefont {M.~D.}\ \bibnamefont
  {Sievert}},\ }\href {\doibase 10.1103/PhysRevD.99.054032} {\bibfield
  {journal} {\bibinfo  {journal} {Phys. Rev. D}\ }\textbf {\bibinfo {volume}
  {99}},\ \bibinfo {pages} {054032} (\bibinfo {year} {2019})},\ \Eprint
  {http://arxiv.org/abs/1808.09010} {arXiv:1808.09010 [hep-ph]} \BibitemShut
  {NoStop}%
\bibitem [{\citenamefont {Cougoulic}\ \emph {et~al.}(2022)\citenamefont
  {Cougoulic}, \citenamefont {Kovchegov}, \citenamefont {Tarasov},\ and\
  \citenamefont {Tawabutr}}]{Cougoulic:2022gbk}%
  \BibitemOpen
  \bibfield  {author} {\bibinfo {author} {\bibfnamefont {F.}~\bibnamefont
  {Cougoulic}}, \bibinfo {author} {\bibfnamefont {Y.~V.}\ \bibnamefont
  {Kovchegov}}, \bibinfo {author} {\bibfnamefont {A.}~\bibnamefont {Tarasov}},
  \ and\ \bibinfo {author} {\bibfnamefont {Y.}~\bibnamefont {Tawabutr}},\
  }\href {\doibase 10.1007/JHEP07(2022)095} {\bibfield  {journal} {\bibinfo
  {journal} {JHEP}\ }\textbf {\bibinfo {volume} {07}},\ \bibinfo {pages} {095}
  (\bibinfo {year} {2022})},\ \Eprint {http://arxiv.org/abs/2204.11898}
  {arXiv:2204.11898 [hep-ph]} \BibitemShut {NoStop}%
\bibitem [{\citenamefont {Adamiak}\ \emph {et~al.}(2021)\citenamefont
  {Adamiak}, \citenamefont {Kovchegov}, \citenamefont {Melnitchouk},
  \citenamefont {Pitonyak}, \citenamefont {Sato},\ and\ \citenamefont
  {Sievert}}]{Adamiak:2021ppq}%
  \BibitemOpen
  \bibfield  {author} {\bibinfo {author} {\bibfnamefont {D.}~\bibnamefont
  {Adamiak}}, \bibinfo {author} {\bibfnamefont {Y.~V.}\ \bibnamefont
  {Kovchegov}}, \bibinfo {author} {\bibfnamefont {W.}~\bibnamefont
  {Melnitchouk}}, \bibinfo {author} {\bibfnamefont {D.}~\bibnamefont
  {Pitonyak}}, \bibinfo {author} {\bibfnamefont {N.}~\bibnamefont {Sato}}, \
  and\ \bibinfo {author} {\bibfnamefont {M.~D.}\ \bibnamefont {Sievert}}
  (\bibinfo {collaboration} {Jefferson Lab Angular Momentum}),\ }\href
  {\doibase 10.1103/PhysRevD.104.L031501} {\bibfield  {journal} {\bibinfo
  {journal} {Phys. Rev. D}\ }\textbf {\bibinfo {volume} {104}},\ \bibinfo
  {pages} {L031501} (\bibinfo {year} {2021})},\ \Eprint
  {http://arxiv.org/abs/2102.06159} {arXiv:2102.06159 [hep-ph]} \BibitemShut
  {NoStop}%
\bibitem [{\citenamefont {Kovchegov}\ and\ \citenamefont
  {Tawabutr}(2020)}]{Kovchegov:2020hgb}%
  \BibitemOpen
  \bibfield  {author} {\bibinfo {author} {\bibfnamefont {Y.~V.}\ \bibnamefont
  {Kovchegov}}\ and\ \bibinfo {author} {\bibfnamefont {Y.}~\bibnamefont
  {Tawabutr}},\ }\href {\doibase 10.1007/JHEP08(2020)014} {\bibfield  {journal}
  {\bibinfo  {journal} {JHEP}\ }\textbf {\bibinfo {volume} {08}},\ \bibinfo
  {pages} {014} (\bibinfo {year} {2020})},\ \Eprint
  {http://arxiv.org/abs/2005.07285} {arXiv:2005.07285 [hep-ph]} \BibitemShut
  {NoStop}%
\bibitem [{\citenamefont {Kovchegov}\ \emph {et~al.}(2022)\citenamefont
  {Kovchegov}, \citenamefont {Tarasov},\ and\ \citenamefont
  {Tawabutr}}]{Kovchegov:2021lvz}%
  \BibitemOpen
  \bibfield  {author} {\bibinfo {author} {\bibfnamefont {Y.~V.}\ \bibnamefont
  {Kovchegov}}, \bibinfo {author} {\bibfnamefont {A.}~\bibnamefont {Tarasov}},
  \ and\ \bibinfo {author} {\bibfnamefont {Y.}~\bibnamefont {Tawabutr}},\
  }\href {\doibase 10.1007/JHEP03(2022)184} {\bibfield  {journal} {\bibinfo
  {journal} {JHEP}\ }\textbf {\bibinfo {volume} {03}},\ \bibinfo {pages} {184}
  (\bibinfo {year} {2022})},\ \Eprint {http://arxiv.org/abs/2104.11765}
  {arXiv:2104.11765 [hep-ph]} \BibitemShut {NoStop}%
\bibitem [{\citenamefont {Kovchegov}(2019)}]{Kovchegov:2019rrz}%
  \BibitemOpen
  \bibfield  {author} {\bibinfo {author} {\bibfnamefont {Y.~V.}\ \bibnamefont
  {Kovchegov}},\ }\href {\doibase 10.1007/JHEP03(2019)174} {\bibfield
  {journal} {\bibinfo  {journal} {JHEP}\ }\textbf {\bibinfo {volume} {03}},\
  \bibinfo {pages} {174} (\bibinfo {year} {2019})},\ \Eprint
  {http://arxiv.org/abs/1901.07453} {arXiv:1901.07453 [hep-ph]} \BibitemShut
  {NoStop}%
\bibitem [{\citenamefont {Narison}\ \emph {et~al.}(1999)\citenamefont
  {Narison}, \citenamefont {Shore},\ and\ \citenamefont
  {Veneziano}}]{Narison:1998aq}%
  \BibitemOpen
  \bibfield  {author} {\bibinfo {author} {\bibfnamefont {S.}~\bibnamefont
  {Narison}}, \bibinfo {author} {\bibfnamefont {G.~M.}\ \bibnamefont {Shore}},
  \ and\ \bibinfo {author} {\bibfnamefont {G.}~\bibnamefont {Veneziano}},\
  }\href {\doibase 10.1016/S0550-3213(99)00061-9} {\bibfield  {journal}
  {\bibinfo  {journal} {Nucl. Phys. B}\ }\textbf {\bibinfo {volume} {546}},\
  \bibinfo {pages} {235} (\bibinfo {year} {1999})},\ \Eprint
  {http://arxiv.org/abs/hep-ph/9812333} {arXiv:hep-ph/9812333} \BibitemShut
  {NoStop}%
\bibitem [{\citenamefont {Tarasov}\ and\ \citenamefont
  {Venugopalan}(2020)}]{Tarasov:2020cwl}%
  \BibitemOpen
  \bibfield  {author} {\bibinfo {author} {\bibfnamefont {A.}~\bibnamefont
  {Tarasov}}\ and\ \bibinfo {author} {\bibfnamefont {R.}~\bibnamefont
  {Venugopalan}},\ }\href {\doibase 10.1103/PhysRevD.102.114022} {\bibfield
  {journal} {\bibinfo  {journal} {Phys. Rev. D}\ }\textbf {\bibinfo {volume}
  {102}},\ \bibinfo {pages} {114022} (\bibinfo {year} {2020})},\ \Eprint
  {http://arxiv.org/abs/2008.08104} {arXiv:2008.08104 [hep-ph]} \BibitemShut
  {NoStop}%
\bibitem [{\citenamefont {Tarasov}\ and\ \citenamefont
  {Venugopalan}(2022)}]{Tarasov:2021yll}%
  \BibitemOpen
  \bibfield  {author} {\bibinfo {author} {\bibfnamefont {A.}~\bibnamefont
  {Tarasov}}\ and\ \bibinfo {author} {\bibfnamefont {R.}~\bibnamefont
  {Venugopalan}},\ }\href {\doibase 10.1103/PhysRevD.105.014020} {\bibfield
  {journal} {\bibinfo  {journal} {Phys. Rev. D}\ }\textbf {\bibinfo {volume}
  {105}},\ \bibinfo {pages} {014020} (\bibinfo {year} {2022})},\ \Eprint
  {http://arxiv.org/abs/2109.10370} {arXiv:2109.10370 [hep-ph]} \BibitemShut
  {NoStop}%
\bibitem [{\citenamefont {Bhattacharya}\ \emph {et~al.}(2023)\citenamefont
  {Bhattacharya}, \citenamefont {Hatta},\ and\ \citenamefont
  {Vogelsang}}]{Bhattacharya:2022xxw}%
  \BibitemOpen
  \bibfield  {author} {\bibinfo {author} {\bibfnamefont {S.}~\bibnamefont
  {Bhattacharya}}, \bibinfo {author} {\bibfnamefont {Y.}~\bibnamefont {Hatta}},
  \ and\ \bibinfo {author} {\bibfnamefont {W.}~\bibnamefont {Vogelsang}},\
  }\href {\doibase 10.1103/PhysRevD.107.014026} {\bibfield  {journal} {\bibinfo
   {journal} {Phys. Rev. D}\ }\textbf {\bibinfo {volume} {107}},\ \bibinfo
  {pages} {014026} (\bibinfo {year} {2023})},\ \Eprint
  {http://arxiv.org/abs/2210.13419} {arXiv:2210.13419 [hep-ph]} \BibitemShut
  {NoStop}%
\bibitem [{\citenamefont {Anikin}\ \emph {et~al.}(2010)\citenamefont {Anikin},
  \citenamefont {Ivanov}, \citenamefont {Pire}, \citenamefont {Szymanowski},\
  and\ \citenamefont {Wallon}}]{Anikin:2009bf}%
  \BibitemOpen
  \bibfield  {author} {\bibinfo {author} {\bibfnamefont {I.~V.}\ \bibnamefont
  {Anikin}}, \bibinfo {author} {\bibfnamefont {D.~Y.}\ \bibnamefont {Ivanov}},
  \bibinfo {author} {\bibfnamefont {B.}~\bibnamefont {Pire}}, \bibinfo {author}
  {\bibfnamefont {L.}~\bibnamefont {Szymanowski}}, \ and\ \bibinfo {author}
  {\bibfnamefont {S.}~\bibnamefont {Wallon}},\ }\href {\doibase
  10.1016/j.nuclphysb.2009.10.022} {\bibfield  {journal} {\bibinfo  {journal}
  {Nucl. Phys. B}\ }\textbf {\bibinfo {volume} {828}},\ \bibinfo {pages} {1}
  (\bibinfo {year} {2010})},\ \Eprint {http://arxiv.org/abs/0909.4090}
  {arXiv:0909.4090 [hep-ph]} \BibitemShut {NoStop}%
\bibitem [{\citenamefont {Anikin}\ \emph {et~al.}(2011)\citenamefont {Anikin},
  \citenamefont {Besse}, \citenamefont {Ivanov}, \citenamefont {Pire},
  \citenamefont {Szymanowski},\ and\ \citenamefont {Wallon}}]{Anikin:2011sa}%
  \BibitemOpen
  \bibfield  {author} {\bibinfo {author} {\bibfnamefont {I.~V.}\ \bibnamefont
  {Anikin}}, \bibinfo {author} {\bibfnamefont {A.}~\bibnamefont {Besse}},
  \bibinfo {author} {\bibfnamefont {D.~Y.}\ \bibnamefont {Ivanov}}, \bibinfo
  {author} {\bibfnamefont {B.}~\bibnamefont {Pire}}, \bibinfo {author}
  {\bibfnamefont {L.}~\bibnamefont {Szymanowski}}, \ and\ \bibinfo {author}
  {\bibfnamefont {S.}~\bibnamefont {Wallon}},\ }\href {\doibase
  10.1103/PhysRevD.84.054004} {\bibfield  {journal} {\bibinfo  {journal} {Phys.
  Rev. D}\ }\textbf {\bibinfo {volume} {84}},\ \bibinfo {pages} {054004}
  (\bibinfo {year} {2011})},\ \Eprint {http://arxiv.org/abs/1105.1761}
  {arXiv:1105.1761 [hep-ph]} \BibitemShut {NoStop}%
\bibitem [{\citenamefont {Besse}\ \emph {et~al.}(2013)\citenamefont {Besse},
  \citenamefont {Szymanowski},\ and\ \citenamefont {Wallon}}]{Besse:2013muy}%
  \BibitemOpen
  \bibfield  {author} {\bibinfo {author} {\bibfnamefont {A.}~\bibnamefont
  {Besse}}, \bibinfo {author} {\bibfnamefont {L.}~\bibnamefont {Szymanowski}},
  \ and\ \bibinfo {author} {\bibfnamefont {S.}~\bibnamefont {Wallon}},\ }\href
  {\doibase 10.1007/JHEP11(2013)062} {\bibfield  {journal} {\bibinfo  {journal}
  {JHEP}\ }\textbf {\bibinfo {volume} {11}},\ \bibinfo {pages} {062} (\bibinfo
  {year} {2013})},\ \Eprint {http://arxiv.org/abs/1302.1766} {arXiv:1302.1766
  [hep-ph]} \BibitemShut {NoStop}%
\bibitem [{\citenamefont {Bolognino}\ \emph
  {et~al.}(2018{\natexlab{a}})\citenamefont {Bolognino}, \citenamefont
  {Celiberto}, \citenamefont {Ivanov},\ and\ \citenamefont
  {Papa}}]{Bolognino:2018rhb}%
  \BibitemOpen
  \bibfield  {author} {\bibinfo {author} {\bibfnamefont {A.~D.}\ \bibnamefont
  {Bolognino}}, \bibinfo {author} {\bibfnamefont {F.~G.}\ \bibnamefont
  {Celiberto}}, \bibinfo {author} {\bibfnamefont {D.~Y.}\ \bibnamefont
  {Ivanov}}, \ and\ \bibinfo {author} {\bibfnamefont {A.}~\bibnamefont
  {Papa}},\ }\href {\doibase 10.1140/epjc/s10052-018-6493-6} {\bibfield
  {journal} {\bibinfo  {journal} {Eur. Phys. J. C}\ }\textbf {\bibinfo {volume}
  {78}},\ \bibinfo {pages} {1023} (\bibinfo {year} {2018}{\natexlab{a}})},\
  \Eprint {http://arxiv.org/abs/1808.02395} {arXiv:1808.02395 [hep-ph]}
  \BibitemShut {NoStop}%
\bibitem [{\citenamefont {Bolognino}\ \emph {et~al.}(2020)\citenamefont
  {Bolognino}, \citenamefont {Szczurek},\ and\ \citenamefont
  {Sch\"afer}}]{Bolognino:2019pba}%
  \BibitemOpen
  \bibfield  {author} {\bibinfo {author} {\bibfnamefont {A.~D.}\ \bibnamefont
  {Bolognino}}, \bibinfo {author} {\bibfnamefont {A.}~\bibnamefont {Szczurek}},
  \ and\ \bibinfo {author} {\bibfnamefont {W.}~\bibnamefont {Sch\"afer}},\
  }\href {\doibase 10.1103/PhysRevD.101.054041} {\bibfield  {journal} {\bibinfo
   {journal} {Phys. Rev. D}\ }\textbf {\bibinfo {volume} {101}},\ \bibinfo
  {pages} {054041} (\bibinfo {year} {2020})},\ \Eprint
  {http://arxiv.org/abs/1912.06507} {arXiv:1912.06507 [hep-ph]} \BibitemShut
  {NoStop}%
\bibitem [{\citenamefont {Bolognino}\ \emph {et~al.}(2021)\citenamefont
  {Bolognino}, \citenamefont {Celiberto}, \citenamefont {Ivanov}, \citenamefont
  {Papa}, \citenamefont {Sch\"afer},\ and\ \citenamefont
  {Szczurek}}]{Bolognino:2021niq}%
  \BibitemOpen
  \bibfield  {author} {\bibinfo {author} {\bibfnamefont {A.~D.}\ \bibnamefont
  {Bolognino}}, \bibinfo {author} {\bibfnamefont {F.~G.}\ \bibnamefont
  {Celiberto}}, \bibinfo {author} {\bibfnamefont {D.~Y.}\ \bibnamefont
  {Ivanov}}, \bibinfo {author} {\bibfnamefont {A.}~\bibnamefont {Papa}},
  \bibinfo {author} {\bibfnamefont {W.}~\bibnamefont {Sch\"afer}}, \ and\
  \bibinfo {author} {\bibfnamefont {A.}~\bibnamefont {Szczurek}},\ }\href
  {\doibase 10.1140/epjc/s10052-021-09593-9} {\bibfield  {journal} {\bibinfo
  {journal} {Eur. Phys. J. C}\ }\textbf {\bibinfo {volume} {81}},\ \bibinfo
  {pages} {846} (\bibinfo {year} {2021})},\ \Eprint
  {http://arxiv.org/abs/2107.13415} {arXiv:2107.13415 [hep-ph]} \BibitemShut
  {NoStop}%
\bibitem [{\citenamefont {Cisek}\ \emph {et~al.}(2023)\citenamefont {Cisek},
  \citenamefont {Sch\"afer},\ and\ \citenamefont {Szczurek}}]{Cisek:2022yjj}%
  \BibitemOpen
  \bibfield  {author} {\bibinfo {author} {\bibfnamefont {A.}~\bibnamefont
  {Cisek}}, \bibinfo {author} {\bibfnamefont {W.}~\bibnamefont {Sch\"afer}}, \
  and\ \bibinfo {author} {\bibfnamefont {A.}~\bibnamefont {Szczurek}},\ }\href
  {\doibase 10.1016/j.physletb.2022.137595} {\bibfield  {journal} {\bibinfo
  {journal} {Phys. Lett. B}\ }\textbf {\bibinfo {volume} {836}},\ \bibinfo
  {pages} {137595} (\bibinfo {year} {2023})},\ \Eprint
  {http://arxiv.org/abs/2209.06578} {arXiv:2209.06578 [hep-ph]} \BibitemShut
  {NoStop}%
\bibitem [{\citenamefont {\L{}uszczak}\ \emph {et~al.}(2022)\citenamefont
  {\L{}uszczak}, \citenamefont {\L{}uszczak},\ and\ \citenamefont
  {Sch\"afer}}]{Luszczak:2022fkf}%
  \BibitemOpen
  \bibfield  {author} {\bibinfo {author} {\bibfnamefont {A.}~\bibnamefont
  {\L{}uszczak}}, \bibinfo {author} {\bibfnamefont {M.}~\bibnamefont
  {\L{}uszczak}}, \ and\ \bibinfo {author} {\bibfnamefont {W.}~\bibnamefont
  {Sch\"afer}},\ }\href {\doibase 10.1016/j.physletb.2022.137582} {\bibfield
  {journal} {\bibinfo  {journal} {Phys. Lett. B}\ }\textbf {\bibinfo {volume}
  {835}},\ \bibinfo {pages} {137582} (\bibinfo {year} {2022})},\ \Eprint
  {http://arxiv.org/abs/2210.02877} {arXiv:2210.02877 [hep-ph]} \BibitemShut
  {NoStop}%
\bibitem [{\citenamefont {Celiberto}(2019)}]{Celiberto:2019slj}%
  \BibitemOpen
  \bibfield  {author} {\bibinfo {author} {\bibfnamefont {F.~G.}\ \bibnamefont
  {Celiberto}},\ }\href {\doibase 10.1393/ncc/i2019-19220-9} {\bibfield
  {journal} {\bibinfo  {journal} {Nuovo Cim. C}\ }\textbf {\bibinfo {volume}
  {42}},\ \bibinfo {pages} {220} (\bibinfo {year} {2019})},\ \Eprint
  {http://arxiv.org/abs/1912.11313} {arXiv:1912.11313 [hep-ph]} \BibitemShut
  {NoStop}%
\bibitem [{\citenamefont {Bolognino}\ \emph
  {et~al.}(2018{\natexlab{b}})\citenamefont {Bolognino}, \citenamefont
  {Celiberto}, \citenamefont {Ivanov},\ and\ \citenamefont
  {Papa}}]{Bolognino:2018mlw}%
  \BibitemOpen
  \bibfield  {author} {\bibinfo {author} {\bibfnamefont {A.~D.}\ \bibnamefont
  {Bolognino}}, \bibinfo {author} {\bibfnamefont {F.~G.}\ \bibnamefont
  {Celiberto}}, \bibinfo {author} {\bibfnamefont {D.~Y.}\ \bibnamefont
  {Ivanov}}, \ and\ \bibinfo {author} {\bibfnamefont {A.}~\bibnamefont
  {Papa}},\ }\href@noop {} {\bibfield  {journal} {\bibinfo  {journal} {Frascati
  Phys. Ser.}\ }\textbf {\bibinfo {volume} {67}},\ \bibinfo {pages} {76}
  (\bibinfo {year} {2018}{\natexlab{b}})},\ \Eprint
  {http://arxiv.org/abs/1808.02958} {arXiv:1808.02958 [hep-ph]} \BibitemShut
  {NoStop}%
\bibitem [{\citenamefont {Bolognino}\ \emph {et~al.}(2019)\citenamefont
  {Bolognino}, \citenamefont {Celiberto}, \citenamefont {Ivanov},\ and\
  \citenamefont {Papa}}]{Bolognino:2019bko}%
  \BibitemOpen
  \bibfield  {author} {\bibinfo {author} {\bibfnamefont {A.~D.}\ \bibnamefont
  {Bolognino}}, \bibinfo {author} {\bibfnamefont {F.~G.}\ \bibnamefont
  {Celiberto}}, \bibinfo {author} {\bibfnamefont {D.~Y.}\ \bibnamefont
  {Ivanov}}, \ and\ \bibinfo {author} {\bibfnamefont {A.}~\bibnamefont
  {Papa}},\ }\href {\doibase 10.5506/APhysPolBSupp.12.891} {\bibfield
  {journal} {\bibinfo  {journal} {Acta Phys. Polon. Supp.}\ }\textbf {\bibinfo
  {volume} {12}},\ \bibinfo {pages} {891} (\bibinfo {year} {2019})},\ \Eprint
  {http://arxiv.org/abs/1902.04520} {arXiv:1902.04520 [hep-ph]} \BibitemShut
  {NoStop}%
\bibitem [{\citenamefont {Bolognino}\ \emph
  {et~al.}(2022{\natexlab{a}})\citenamefont {Bolognino}, \citenamefont
  {Celiberto}, \citenamefont {Ivanov},\ and\ \citenamefont
  {Papa}}]{Bolognino:2021gjm}%
  \BibitemOpen
  \bibfield  {author} {\bibinfo {author} {\bibfnamefont {A.~D.}\ \bibnamefont
  {Bolognino}}, \bibinfo {author} {\bibfnamefont {F.~G.}\ \bibnamefont
  {Celiberto}}, \bibinfo {author} {\bibfnamefont {D.~Y.}\ \bibnamefont
  {Ivanov}}, \ and\ \bibinfo {author} {\bibfnamefont {A.}~\bibnamefont
  {Papa}},\ }\href {\doibase 10.21468/SciPostPhysProc.8.089} {\bibfield
  {journal} {\bibinfo  {journal} {SciPost Phys. Proc.}\ }\textbf {\bibinfo
  {volume} {8}},\ \bibinfo {pages} {089} (\bibinfo {year}
  {2022}{\natexlab{a}})},\ \Eprint {http://arxiv.org/abs/2107.12725}
  {arXiv:2107.12725 [hep-ph]} \BibitemShut {NoStop}%
\bibitem [{\citenamefont {Bolognino}\ \emph
  {et~al.}(2022{\natexlab{b}})\citenamefont {Bolognino}, \citenamefont
  {Celiberto}, \citenamefont {Fucilla}, \citenamefont {Ivanov}, \citenamefont
  {Papa}, \citenamefont {Sch\"afer},\ and\ \citenamefont
  {Szczurek}}]{Bolognino:2022uty}%
  \BibitemOpen
  \bibfield  {author} {\bibinfo {author} {\bibfnamefont {A.~D.}\ \bibnamefont
  {Bolognino}}, \bibinfo {author} {\bibfnamefont {F.~G.}\ \bibnamefont
  {Celiberto}}, \bibinfo {author} {\bibfnamefont {M.}~\bibnamefont {Fucilla}},
  \bibinfo {author} {\bibfnamefont {D.~Y.}\ \bibnamefont {Ivanov}}, \bibinfo
  {author} {\bibfnamefont {A.}~\bibnamefont {Papa}}, \bibinfo {author}
  {\bibfnamefont {W.}~\bibnamefont {Sch\"afer}}, \ and\ \bibinfo {author}
  {\bibfnamefont {A.}~\bibnamefont {Szczurek}},\ }\href {\doibase
  10.31349/SuplRevMexFis.3.0308109} {\bibfield  {journal} {\bibinfo  {journal}
  {Rev. Mex. Fis. Suppl.}\ }\textbf {\bibinfo {volume} {3}},\ \bibinfo {pages}
  {0308109} (\bibinfo {year} {2022}{\natexlab{b}})},\ \Eprint
  {http://arxiv.org/abs/2202.02513} {arXiv:2202.02513 [hep-ph]} \BibitemShut
  {NoStop}%
\bibitem [{\citenamefont {Bolognino}\ \emph
  {et~al.}(2022{\natexlab{c}})\citenamefont {Bolognino}, \citenamefont
  {Celiberto}, \citenamefont {Ivanov}, \citenamefont {Papa}, \citenamefont
  {Sch\"afer},\ and\ \citenamefont {Szczurek}}]{Bolognino:2022ndh}%
  \BibitemOpen
  \bibfield  {author} {\bibinfo {author} {\bibfnamefont {A.~D.}\ \bibnamefont
  {Bolognino}}, \bibinfo {author} {\bibfnamefont {F.~G.}\ \bibnamefont
  {Celiberto}}, \bibinfo {author} {\bibfnamefont {D.~Y.}\ \bibnamefont
  {Ivanov}}, \bibinfo {author} {\bibfnamefont {A.}~\bibnamefont {Papa}},
  \bibinfo {author} {\bibfnamefont {W.}~\bibnamefont {Sch\"afer}}, \ and\
  \bibinfo {author} {\bibfnamefont {A.}~\bibnamefont {Szczurek}},\ }in\
  \href@noop {} {\emph {\bibinfo {booktitle} {{29th International Workshop on
  Deep-Inelastic Scattering and Related Subjects}}}}\ (\bibinfo {year} {2022})\
  \Eprint {http://arxiv.org/abs/2207.05726} {arXiv:2207.05726 [hep-ph]}
  \BibitemShut {NoStop}%
\bibitem [{\citenamefont {Bautista}\ \emph {et~al.}(2016)\citenamefont
  {Bautista}, \citenamefont {Fernandez~Tellez},\ and\ \citenamefont
  {Hentschinski}}]{Bautista:2016xnp}%
  \BibitemOpen
  \bibfield  {author} {\bibinfo {author} {\bibfnamefont {I.}~\bibnamefont
  {Bautista}}, \bibinfo {author} {\bibfnamefont {A.}~\bibnamefont
  {Fernandez~Tellez}}, \ and\ \bibinfo {author} {\bibfnamefont
  {M.}~\bibnamefont {Hentschinski}},\ }\href {\doibase
  10.1103/PhysRevD.94.054002} {\bibfield  {journal} {\bibinfo  {journal} {Phys.
  Rev. D}\ }\textbf {\bibinfo {volume} {94}},\ \bibinfo {pages} {054002}
  (\bibinfo {year} {2016})},\ \Eprint {http://arxiv.org/abs/1607.05203}
  {arXiv:1607.05203 [hep-ph]} \BibitemShut {NoStop}%
\bibitem [{\citenamefont {Arroyo~Garcia}\ \emph {et~al.}(2019)\citenamefont
  {Arroyo~Garcia}, \citenamefont {Hentschinski},\ and\ \citenamefont
  {Kutak}}]{ArroyoGarcia:2019cfl}%
  \BibitemOpen
  \bibfield  {author} {\bibinfo {author} {\bibfnamefont {A.}~\bibnamefont
  {Arroyo~Garcia}}, \bibinfo {author} {\bibfnamefont {M.}~\bibnamefont
  {Hentschinski}}, \ and\ \bibinfo {author} {\bibfnamefont {K.}~\bibnamefont
  {Kutak}},\ }\href {\doibase 10.1016/j.physletb.2019.06.061} {\bibfield
  {journal} {\bibinfo  {journal} {Phys. Lett. B}\ }\textbf {\bibinfo {volume}
  {795}},\ \bibinfo {pages} {569} (\bibinfo {year} {2019})},\ \Eprint
  {http://arxiv.org/abs/1904.04394} {arXiv:1904.04394 [hep-ph]} \BibitemShut
  {NoStop}%
\bibitem [{\citenamefont {Hentschinski}\ and\ \citenamefont
  {Padr\'on~Molina}(2021)}]{Hentschinski:2020yfm}%
  \BibitemOpen
  \bibfield  {author} {\bibinfo {author} {\bibfnamefont {M.}~\bibnamefont
  {Hentschinski}}\ and\ \bibinfo {author} {\bibfnamefont {E.}~\bibnamefont
  {Padr\'on~Molina}},\ }\href {\doibase 10.1103/PhysRevD.103.074008} {\bibfield
   {journal} {\bibinfo  {journal} {Phys. Rev. D}\ }\textbf {\bibinfo {volume}
  {103}},\ \bibinfo {pages} {074008} (\bibinfo {year} {2021})},\ \Eprint
  {http://arxiv.org/abs/2011.02640} {arXiv:2011.02640 [hep-ph]} \BibitemShut
  {NoStop}%
\bibitem [{\citenamefont {Motyka}\ \emph {et~al.}(2015)\citenamefont {Motyka},
  \citenamefont {Sadzikowski},\ and\ \citenamefont {Stebel}}]{Motyka:2014lya}%
  \BibitemOpen
  \bibfield  {author} {\bibinfo {author} {\bibfnamefont {L.}~\bibnamefont
  {Motyka}}, \bibinfo {author} {\bibfnamefont {M.}~\bibnamefont {Sadzikowski}},
  \ and\ \bibinfo {author} {\bibfnamefont {T.}~\bibnamefont {Stebel}},\ }\href
  {\doibase 10.1007/JHEP05(2015)087} {\bibfield  {journal} {\bibinfo  {journal}
  {JHEP}\ }\textbf {\bibinfo {volume} {05}},\ \bibinfo {pages} {087} (\bibinfo
  {year} {2015})},\ \Eprint {http://arxiv.org/abs/1412.4675} {arXiv:1412.4675
  [hep-ph]} \BibitemShut {NoStop}%
\bibitem [{\citenamefont {Motyka}\ \emph {et~al.}(2017)\citenamefont {Motyka},
  \citenamefont {Sadzikowski},\ and\ \citenamefont {Stebel}}]{Motyka:2016lta}%
  \BibitemOpen
  \bibfield  {author} {\bibinfo {author} {\bibfnamefont {L.}~\bibnamefont
  {Motyka}}, \bibinfo {author} {\bibfnamefont {M.}~\bibnamefont {Sadzikowski}},
  \ and\ \bibinfo {author} {\bibfnamefont {T.}~\bibnamefont {Stebel}},\ }\href
  {\doibase 10.1103/PhysRevD.95.114025} {\bibfield  {journal} {\bibinfo
  {journal} {Phys. Rev. D}\ }\textbf {\bibinfo {volume} {95}},\ \bibinfo
  {pages} {114025} (\bibinfo {year} {2017})},\ \Eprint
  {http://arxiv.org/abs/1609.04300} {arXiv:1609.04300 [hep-ph]} \BibitemShut
  {NoStop}%
\bibitem [{\citenamefont {Brzeminski}\ \emph {et~al.}(2017)\citenamefont
  {Brzeminski}, \citenamefont {Motyka}, \citenamefont {Sadzikowski},\ and\
  \citenamefont {Stebel}}]{Brzeminski:2016lwh}%
  \BibitemOpen
  \bibfield  {author} {\bibinfo {author} {\bibfnamefont {D.}~\bibnamefont
  {Brzeminski}}, \bibinfo {author} {\bibfnamefont {L.}~\bibnamefont {Motyka}},
  \bibinfo {author} {\bibfnamefont {M.}~\bibnamefont {Sadzikowski}}, \ and\
  \bibinfo {author} {\bibfnamefont {T.}~\bibnamefont {Stebel}},\ }\href
  {\doibase 10.1007/JHEP01(2017)005} {\bibfield  {journal} {\bibinfo  {journal}
  {JHEP}\ }\textbf {\bibinfo {volume} {01}},\ \bibinfo {pages} {005} (\bibinfo
  {year} {2017})},\ \Eprint {http://arxiv.org/abs/1611.04449} {arXiv:1611.04449
  [hep-ph]} \BibitemShut {NoStop}%
\bibitem [{\citenamefont {Celiberto}\ \emph {et~al.}(2018)\citenamefont
  {Celiberto}, \citenamefont {Gordo~G\'omez},\ and\ \citenamefont
  {Sabio~Vera}}]{Celiberto:2018muu}%
  \BibitemOpen
  \bibfield  {author} {\bibinfo {author} {\bibfnamefont {F.~G.}\ \bibnamefont
  {Celiberto}}, \bibinfo {author} {\bibfnamefont {D.}~\bibnamefont
  {Gordo~G\'omez}}, \ and\ \bibinfo {author} {\bibfnamefont {A.}~\bibnamefont
  {Sabio~Vera}},\ }\href {\doibase 10.1016/j.physletb.2018.09.045} {\bibfield
  {journal} {\bibinfo  {journal} {Phys. Lett. B}\ }\textbf {\bibinfo {volume}
  {786}},\ \bibinfo {pages} {201} (\bibinfo {year} {2018})},\ \Eprint
  {http://arxiv.org/abs/1808.09511} {arXiv:1808.09511 [hep-ph]} \BibitemShut
  {NoStop}%
\bibitem [{\citenamefont {Hentschinski}(2021)}]{Hentschinski:2021lsh}%
  \BibitemOpen
  \bibfield  {author} {\bibinfo {author} {\bibfnamefont {M.}~\bibnamefont
  {Hentschinski}},\ }\href {\doibase 10.1103/PhysRevD.104.054014} {\bibfield
  {journal} {\bibinfo  {journal} {Phys. Rev. D}\ }\textbf {\bibinfo {volume}
  {104}},\ \bibinfo {pages} {054014} (\bibinfo {year} {2021})},\ \Eprint
  {http://arxiv.org/abs/2107.06203} {arXiv:2107.06203 [hep-ph]} \BibitemShut
  {NoStop}%
\bibitem [{\citenamefont {Nefedov}(2021)}]{Nefedov:2021vvy}%
  \BibitemOpen
  \bibfield  {author} {\bibinfo {author} {\bibfnamefont {M.}~\bibnamefont
  {Nefedov}},\ }\href {\doibase 10.1103/PhysRevD.104.054039} {\bibfield
  {journal} {\bibinfo  {journal} {Phys. Rev. D}\ }\textbf {\bibinfo {volume}
  {104}},\ \bibinfo {pages} {054039} (\bibinfo {year} {2021})},\ \Eprint
  {http://arxiv.org/abs/2105.13915} {arXiv:2105.13915 [hep-ph]} \BibitemShut
  {NoStop}%
\bibitem [{\citenamefont {Celiberto}(2021)}]{Celiberto:2021zww}%
  \BibitemOpen
  \bibfield  {author} {\bibinfo {author} {\bibfnamefont {F.~G.}\ \bibnamefont
  {Celiberto}},\ }\href {\doibase 10.1393/ncc/i2021-21036-3} {\bibfield
  {journal} {\bibinfo  {journal} {Nuovo Cim. C}\ }\textbf {\bibinfo {volume}
  {44}},\ \bibinfo {pages} {36} (\bibinfo {year} {2021})},\ \Eprint
  {http://arxiv.org/abs/2101.04630} {arXiv:2101.04630 [hep-ph]} \BibitemShut
  {NoStop}%
\bibitem [{\citenamefont {Beuf}\ \emph {et~al.}(2020)\citenamefont {Beuf},
  \citenamefont {H\"anninen}, \citenamefont {Lappi},\ and\ \citenamefont
  {M\"antysaari}}]{Beuf:2020dxl}%
  \BibitemOpen
  \bibfield  {author} {\bibinfo {author} {\bibfnamefont {G.}~\bibnamefont
  {Beuf}}, \bibinfo {author} {\bibfnamefont {H.}~\bibnamefont {H\"anninen}},
  \bibinfo {author} {\bibfnamefont {T.}~\bibnamefont {Lappi}}, \ and\ \bibinfo
  {author} {\bibfnamefont {H.}~\bibnamefont {M\"antysaari}},\ }\href {\doibase
  10.1103/PhysRevD.102.074028} {\bibfield  {journal} {\bibinfo  {journal}
  {Phys. Rev. D}\ }\textbf {\bibinfo {volume} {102}},\ \bibinfo {pages}
  {074028} (\bibinfo {year} {2020})},\ \Eprint
  {http://arxiv.org/abs/2007.01645} {arXiv:2007.01645 [hep-ph]} \BibitemShut
  {NoStop}%
\bibitem [{\citenamefont {H\"anninen}\ \emph {et~al.}(2022)\citenamefont
  {H\"anninen}, \citenamefont {M\"antysaari}, \citenamefont {Paatelainen},\
  and\ \citenamefont {Penttala}}]{Hanninen:2022gje}%
  \BibitemOpen
  \bibfield  {author} {\bibinfo {author} {\bibfnamefont {H.}~\bibnamefont
  {H\"anninen}}, \bibinfo {author} {\bibfnamefont {H.}~\bibnamefont
  {M\"antysaari}}, \bibinfo {author} {\bibfnamefont {R.}~\bibnamefont
  {Paatelainen}}, \ and\ \bibinfo {author} {\bibfnamefont {J.}~\bibnamefont
  {Penttala}},\ }\href@noop {} {\  (\bibinfo {year} {2022})},\ \Eprint
  {http://arxiv.org/abs/2211.03504} {arXiv:2211.03504 [hep-ph]} \BibitemShut
  {NoStop}%
\bibitem [{\citenamefont {Shi}\ \emph {et~al.}(2022)\citenamefont {Shi},
  \citenamefont {Wang}, \citenamefont {Wei},\ and\ \citenamefont
  {Xiao}}]{Shi:2021hwx}%
  \BibitemOpen
  \bibfield  {author} {\bibinfo {author} {\bibfnamefont {Y.}~\bibnamefont
  {Shi}}, \bibinfo {author} {\bibfnamefont {L.}~\bibnamefont {Wang}}, \bibinfo
  {author} {\bibfnamefont {S.-Y.}\ \bibnamefont {Wei}}, \ and\ \bibinfo
  {author} {\bibfnamefont {B.-W.}\ \bibnamefont {Xiao}},\ }\href {\doibase
  10.1103/PhysRevLett.128.202302} {\bibfield  {journal} {\bibinfo  {journal}
  {Phys. Rev. Lett.}\ }\textbf {\bibinfo {volume} {128}},\ \bibinfo {pages}
  {202302} (\bibinfo {year} {2022})},\ \Eprint
  {http://arxiv.org/abs/2112.06975} {arXiv:2112.06975 [hep-ph]} \BibitemShut
  {NoStop}%
\bibitem [{\citenamefont {M\"antysaari}\ and\ \citenamefont
  {Venugopalan}(2018)}]{Mantysaari:2017slo}%
  \BibitemOpen
  \bibfield  {author} {\bibinfo {author} {\bibfnamefont {H.}~\bibnamefont
  {M\"antysaari}}\ and\ \bibinfo {author} {\bibfnamefont {R.}~\bibnamefont
  {Venugopalan}},\ }\href {\doibase 10.1016/j.physletb.2018.04.044} {\bibfield
  {journal} {\bibinfo  {journal} {Phys. Lett. B}\ }\textbf {\bibinfo {volume}
  {781}},\ \bibinfo {pages} {664} (\bibinfo {year} {2018})},\ \Eprint
  {http://arxiv.org/abs/1712.02508} {arXiv:1712.02508 [nucl-th]} \BibitemShut
  {NoStop}%
\bibitem [{\citenamefont {Toll}\ and\ \citenamefont
  {Ullrich}(2014)}]{Toll:2013gda}%
  \BibitemOpen
  \bibfield  {author} {\bibinfo {author} {\bibfnamefont {T.}~\bibnamefont
  {Toll}}\ and\ \bibinfo {author} {\bibfnamefont {T.}~\bibnamefont {Ullrich}},\
  }\href {\doibase 10.1016/j.cpc.2014.03.010} {\bibfield  {journal} {\bibinfo
  {journal} {Comput. Phys. Commun.}\ }\textbf {\bibinfo {volume} {185}},\
  \bibinfo {pages} {1835} (\bibinfo {year} {2014})},\ \Eprint
  {http://arxiv.org/abs/1307.8059} {arXiv:1307.8059 [hep-ph]} \BibitemShut
  {NoStop}%
\bibitem [{\citenamefont {Toll}\ and\ \citenamefont
  {Ullrich}(2013)}]{Toll:2012mb}%
  \BibitemOpen
  \bibfield  {author} {\bibinfo {author} {\bibfnamefont {T.}~\bibnamefont
  {Toll}}\ and\ \bibinfo {author} {\bibfnamefont {T.}~\bibnamefont {Ullrich}},\
  }\href {\doibase 10.1103/PhysRevC.87.024913} {\bibfield  {journal} {\bibinfo
  {journal} {Phys. Rev. C}\ }\textbf {\bibinfo {volume} {87}},\ \bibinfo
  {pages} {024913} (\bibinfo {year} {2013})},\ \Eprint
  {http://arxiv.org/abs/1211.3048} {arXiv:1211.3048 [hep-ph]} \BibitemShut
  {NoStop}%
\bibitem [{\citenamefont {Matousek}\ \emph {et~al.}(2023)\citenamefont
  {Matousek}, \citenamefont {Khachatryan},\ and\ \citenamefont
  {Zhang}}]{Matousek:2022enl}%
  \BibitemOpen
  \bibfield  {author} {\bibinfo {author} {\bibfnamefont {G.}~\bibnamefont
  {Matousek}}, \bibinfo {author} {\bibfnamefont {V.}~\bibnamefont
  {Khachatryan}}, \ and\ \bibinfo {author} {\bibfnamefont {J.}~\bibnamefont
  {Zhang}},\ }\href {\doibase 10.1140/epjp/s13360-023-03729-4} {\bibfield
  {journal} {\bibinfo  {journal} {Eur. Phys. J. Plus}\ }\textbf {\bibinfo
  {volume} {138}},\ \bibinfo {pages} {113} (\bibinfo {year} {2023})},\ \Eprint
  {http://arxiv.org/abs/2202.05981} {arXiv:2202.05981 [nucl-th]} \BibitemShut
  {NoStop}%
\bibitem [{\citenamefont {Stasto}\ \emph {et~al.}(2001)\citenamefont {Stasto},
  \citenamefont {Golec-Biernat},\ and\ \citenamefont
  {Kwiecinski}}]{Stasto:2000er}%
  \BibitemOpen
  \bibfield  {author} {\bibinfo {author} {\bibfnamefont {A.~M.}\ \bibnamefont
  {Stasto}}, \bibinfo {author} {\bibfnamefont {K.~J.}\ \bibnamefont
  {Golec-Biernat}}, \ and\ \bibinfo {author} {\bibfnamefont {J.}~\bibnamefont
  {Kwiecinski}},\ }\href {\doibase 10.1103/PhysRevLett.86.596} {\bibfield
  {journal} {\bibinfo  {journal} {Phys. Rev. Lett.}\ }\textbf {\bibinfo
  {volume} {86}},\ \bibinfo {pages} {596} (\bibinfo {year} {2001})},\ \Eprint
  {http://arxiv.org/abs/hep-ph/0007192} {arXiv:hep-ph/0007192} \BibitemShut
  {NoStop}%
\bibitem [{\citenamefont {Golec-Biernat}\ and\ \citenamefont
  {Wusthoff}(1998)}]{Golec-Biernat:1998zce}%
  \BibitemOpen
  \bibfield  {author} {\bibinfo {author} {\bibfnamefont {K.~J.}\ \bibnamefont
  {Golec-Biernat}}\ and\ \bibinfo {author} {\bibfnamefont {M.}~\bibnamefont
  {Wusthoff}},\ }\href {\doibase 10.1103/PhysRevD.59.014017} {\bibfield
  {journal} {\bibinfo  {journal} {Phys. Rev. D}\ }\textbf {\bibinfo {volume}
  {59}},\ \bibinfo {pages} {014017} (\bibinfo {year} {1998})},\ \Eprint
  {http://arxiv.org/abs/hep-ph/9807513} {arXiv:hep-ph/9807513} \BibitemShut
  {NoStop}%
\bibitem [{\citenamefont {Golec-Biernat}\ and\ \citenamefont
  {Wusthoff}(1999)}]{Golec-Biernat:1999qor}%
  \BibitemOpen
  \bibfield  {author} {\bibinfo {author} {\bibfnamefont {K.~J.}\ \bibnamefont
  {Golec-Biernat}}\ and\ \bibinfo {author} {\bibfnamefont {M.}~\bibnamefont
  {Wusthoff}},\ }\href {\doibase 10.1103/PhysRevD.60.114023} {\bibfield
  {journal} {\bibinfo  {journal} {Phys. Rev. D}\ }\textbf {\bibinfo {volume}
  {60}},\ \bibinfo {pages} {114023} (\bibinfo {year} {1999})},\ \Eprint
  {http://arxiv.org/abs/hep-ph/9903358} {arXiv:hep-ph/9903358} \BibitemShut
  {NoStop}%
\bibitem [{\citenamefont {Marquet}\ and\ \citenamefont
  {Schoeffel}(2006)}]{Marquet:2006jb}%
  \BibitemOpen
  \bibfield  {author} {\bibinfo {author} {\bibfnamefont {C.}~\bibnamefont
  {Marquet}}\ and\ \bibinfo {author} {\bibfnamefont {L.}~\bibnamefont
  {Schoeffel}},\ }\href {\doibase 10.1016/j.physletb.2006.07.004} {\bibfield
  {journal} {\bibinfo  {journal} {Phys. Lett. B}\ }\textbf {\bibinfo {volume}
  {639}},\ \bibinfo {pages} {471} (\bibinfo {year} {2006})},\ \Eprint
  {http://arxiv.org/abs/hep-ph/0606079} {arXiv:hep-ph/0606079} \BibitemShut
  {NoStop}%
\bibitem [{\citenamefont {Kumar}(2023)}]{Kumar:2022kww}%
  \BibitemOpen
  \bibfield  {author} {\bibinfo {author} {\bibfnamefont {A.}~\bibnamefont
  {Kumar}},\ }\href {\doibase 10.1103/PhysRevD.107.034005} {\bibfield
  {journal} {\bibinfo  {journal} {Phys. Rev. D}\ }\textbf {\bibinfo {volume}
  {107}},\ \bibinfo {pages} {034005} (\bibinfo {year} {2023})},\ \Eprint
  {http://arxiv.org/abs/2208.14200} {arXiv:2208.14200 [hep-ph]} \BibitemShut
  {NoStop}%
\bibitem [{\citenamefont {McLerran}\ and\ \citenamefont
  {Praszalowicz}(2010)}]{McLerran:2010ex}%
  \BibitemOpen
  \bibfield  {author} {\bibinfo {author} {\bibfnamefont {L.}~\bibnamefont
  {McLerran}}\ and\ \bibinfo {author} {\bibfnamefont {M.}~\bibnamefont
  {Praszalowicz}},\ }\href@noop {} {\bibfield  {journal} {\bibinfo  {journal}
  {Acta Phys. Polon. B}\ }\textbf {\bibinfo {volume} {41}},\ \bibinfo {pages}
  {1917} (\bibinfo {year} {2010})},\ \Eprint {http://arxiv.org/abs/1006.4293}
  {arXiv:1006.4293 [hep-ph]} \BibitemShut {NoStop}%
\bibitem [{\citenamefont {McLerran}\ and\ \citenamefont
  {Praszalowicz}(2011)}]{McLerran:2010wm}%
  \BibitemOpen
  \bibfield  {author} {\bibinfo {author} {\bibfnamefont {L.}~\bibnamefont
  {McLerran}}\ and\ \bibinfo {author} {\bibfnamefont {M.}~\bibnamefont
  {Praszalowicz}},\ }\href {\doibase 10.5506/APhysPolB.42.99} {\bibfield
  {journal} {\bibinfo  {journal} {Acta Phys. Polon. B}\ }\textbf {\bibinfo
  {volume} {42}},\ \bibinfo {pages} {99} (\bibinfo {year} {2011})},\ \Eprint
  {http://arxiv.org/abs/1011.3403} {arXiv:1011.3403 [hep-ph]} \BibitemShut
  {NoStop}%
\bibitem [{\citenamefont {Khachatryan}\ and\ \citenamefont
  {Prasza\l{}owicz}(2020)}]{Khachatryan:2019uqn}%
  \BibitemOpen
  \bibfield  {author} {\bibinfo {author} {\bibfnamefont {V.}~\bibnamefont
  {Khachatryan}}\ and\ \bibinfo {author} {\bibfnamefont {M.}~\bibnamefont
  {Prasza\l{}owicz}},\ }\href {\doibase 10.1140/epjc/s10052-020-8219-9}
  {\bibfield  {journal} {\bibinfo  {journal} {Eur. Phys. J. C}\ }\textbf
  {\bibinfo {volume} {80}},\ \bibinfo {pages} {670} (\bibinfo {year} {2020})},\
  \Eprint {http://arxiv.org/abs/1907.03815} {arXiv:1907.03815 [nucl-th]}
  \BibitemShut {NoStop}%
\bibitem [{\citenamefont {Khachatryan}\ and\ \citenamefont
  {Praszalowicz}(2022)}]{Khachatryan:2022qrc}%
  \BibitemOpen
  \bibfield  {author} {\bibinfo {author} {\bibfnamefont {V.}~\bibnamefont
  {Khachatryan}}\ and\ \bibinfo {author} {\bibfnamefont {M.}~\bibnamefont
  {Praszalowicz}},\ }\href@noop {} {\  (\bibinfo {year} {2022})},\ \Eprint
  {http://arxiv.org/abs/2203.16204} {arXiv:2203.16204 [nucl-th]} \BibitemShut
  {NoStop}%
\bibitem [{\citenamefont {Ben}\ \emph {et~al.}(2017)\citenamefont {Ben},
  \citenamefont {Machado},\ and\ \citenamefont {Sauter}}]{Ben:2017xny}%
  \BibitemOpen
  \bibfield  {author} {\bibinfo {author} {\bibfnamefont {F.~G.}\ \bibnamefont
  {Ben}}, \bibinfo {author} {\bibfnamefont {M.~V.~T.}\ \bibnamefont {Machado}},
  \ and\ \bibinfo {author} {\bibfnamefont {W.~K.}\ \bibnamefont {Sauter}},\
  }\href {\doibase 10.1103/PhysRevD.96.054015} {\bibfield  {journal} {\bibinfo
  {journal} {Phys. Rev. D}\ }\textbf {\bibinfo {volume} {96}},\ \bibinfo
  {pages} {054015} (\bibinfo {year} {2017})},\ \Eprint
  {http://arxiv.org/abs/1701.01141} {arXiv:1701.01141 [hep-ph]} \BibitemShut
  {NoStop}%
\bibitem [{\citenamefont {Kowalski}\ \emph {et~al.}(2006)\citenamefont
  {Kowalski}, \citenamefont {Motyka},\ and\ \citenamefont
  {Watt}}]{Kowalski:2006hc}%
  \BibitemOpen
  \bibfield  {author} {\bibinfo {author} {\bibfnamefont {H.}~\bibnamefont
  {Kowalski}}, \bibinfo {author} {\bibfnamefont {L.}~\bibnamefont {Motyka}}, \
  and\ \bibinfo {author} {\bibfnamefont {G.}~\bibnamefont {Watt}},\ }\href
  {\doibase 10.1103/PhysRevD.74.074016} {\bibfield  {journal} {\bibinfo
  {journal} {Phys. Rev. D}\ }\textbf {\bibinfo {volume} {74}},\ \bibinfo
  {pages} {074016} (\bibinfo {year} {2006})},\ \Eprint
  {http://arxiv.org/abs/hep-ph/0606272} {arXiv:hep-ph/0606272} \BibitemShut
  {NoStop}%
\bibitem [{\citenamefont {Hentschinski}\ \emph
  {et~al.}(2022{\natexlab{a}})\citenamefont {Hentschinski} \emph
  {et~al.}}]{Hentschinski:2022xnd}%
  \BibitemOpen
  \bibfield  {author} {\bibinfo {author} {\bibfnamefont {M.}~\bibnamefont
  {Hentschinski}} \emph {et~al.},\ }\href@noop {} {\  (\bibinfo {year}
  {2022}{\natexlab{a}})},\ \Eprint {http://arxiv.org/abs/2203.08129}
  {arXiv:2203.08129 [hep-ph]} \BibitemShut {NoStop}%
\bibitem [{\citenamefont {Klein}\ and\ \citenamefont
  {M\"antysaari}(2019)}]{Klein:2019qfb}%
  \BibitemOpen
  \bibfield  {author} {\bibinfo {author} {\bibfnamefont {S.~R.}\ \bibnamefont
  {Klein}}\ and\ \bibinfo {author} {\bibfnamefont {H.}~\bibnamefont
  {M\"antysaari}},\ }\href {\doibase 10.1038/s42254-019-0107-6} {\bibfield
  {journal} {\bibinfo  {journal} {Nature Rev. Phys.}\ }\textbf {\bibinfo
  {volume} {1}},\ \bibinfo {pages} {662} (\bibinfo {year} {2019})},\ \Eprint
  {http://arxiv.org/abs/1910.10858} {arXiv:1910.10858 [hep-ex]} \BibitemShut
  {NoStop}%
\bibitem [{\citenamefont {Miettinen}\ and\ \citenamefont
  {Pumplin}(1978)}]{Miettinen:1978jb}%
  \BibitemOpen
  \bibfield  {author} {\bibinfo {author} {\bibfnamefont {H.~I.}\ \bibnamefont
  {Miettinen}}\ and\ \bibinfo {author} {\bibfnamefont {J.}~\bibnamefont
  {Pumplin}},\ }\href {\doibase 10.1103/PhysRevD.18.1696} {\bibfield  {journal}
  {\bibinfo  {journal} {Phys. Rev. D}\ }\textbf {\bibinfo {volume} {18}},\
  \bibinfo {pages} {1696} (\bibinfo {year} {1978})}\BibitemShut {NoStop}%
\bibitem [{\citenamefont {Klein}(2023)}]{Klein:2023zlf}%
  \BibitemOpen
  \bibfield  {author} {\bibinfo {author} {\bibfnamefont {S.~R.}\ \bibnamefont
  {Klein}},\ }\href@noop {} {\  (\bibinfo {year} {2023})},\ \Eprint
  {http://arxiv.org/abs/2301.01408} {arXiv:2301.01408 [hep-ph]} \BibitemShut
  {NoStop}%
\bibitem [{\citenamefont {Cepila}\ \emph {et~al.}(2017)\citenamefont {Cepila},
  \citenamefont {Contreras},\ and\ \citenamefont
  {Tapia~Takaki}}]{Cepila:2016uku}%
  \BibitemOpen
  \bibfield  {author} {\bibinfo {author} {\bibfnamefont {J.}~\bibnamefont
  {Cepila}}, \bibinfo {author} {\bibfnamefont {J.~G.}\ \bibnamefont
  {Contreras}}, \ and\ \bibinfo {author} {\bibfnamefont {J.~D.}\ \bibnamefont
  {Tapia~Takaki}},\ }\href {\doibase 10.1016/j.physletb.2016.12.063} {\bibfield
   {journal} {\bibinfo  {journal} {Phys. Lett. B}\ }\textbf {\bibinfo {volume}
  {766}},\ \bibinfo {pages} {186} (\bibinfo {year} {2017})},\ \Eprint
  {http://arxiv.org/abs/1608.07559} {arXiv:1608.07559 [hep-ph]} \BibitemShut
  {NoStop}%
\bibitem [{\citenamefont {M\"antysaari}\ and\ \citenamefont
  {Schenke}(2018)}]{Mantysaari:2018zdd}%
  \BibitemOpen
  \bibfield  {author} {\bibinfo {author} {\bibfnamefont {H.}~\bibnamefont
  {M\"antysaari}}\ and\ \bibinfo {author} {\bibfnamefont {B.}~\bibnamefont
  {Schenke}},\ }\href {\doibase 10.1103/PhysRevD.98.034013} {\bibfield
  {journal} {\bibinfo  {journal} {Phys. Rev. D}\ }\textbf {\bibinfo {volume}
  {98}},\ \bibinfo {pages} {034013} (\bibinfo {year} {2018})},\ \Eprint
  {http://arxiv.org/abs/1806.06783} {arXiv:1806.06783 [hep-ph]} \BibitemShut
  {NoStop}%
\bibitem [{\citenamefont {Kumar}\ and\ \citenamefont
  {Toll}(2022)}]{Kumar:2022aly}%
  \BibitemOpen
  \bibfield  {author} {\bibinfo {author} {\bibfnamefont {A.}~\bibnamefont
  {Kumar}}\ and\ \bibinfo {author} {\bibfnamefont {T.}~\bibnamefont {Toll}},\
  }\href {\doibase 10.1103/PhysRevD.105.114011} {\bibfield  {journal} {\bibinfo
   {journal} {Phys. Rev. D}\ }\textbf {\bibinfo {volume} {105}},\ \bibinfo
  {pages} {114011} (\bibinfo {year} {2022})},\ \Eprint
  {http://arxiv.org/abs/2202.06631} {arXiv:2202.06631 [hep-ph]} \BibitemShut
  {NoStop}%
\bibitem [{\citenamefont {Eskola}\ \emph
  {et~al.}(2022{\natexlab{a}})\citenamefont {Eskola}, \citenamefont {Flett},
  \citenamefont {Guzey}, \citenamefont {L\"oyt\"ainen},\ and\ \citenamefont
  {Paukkunen}}]{Eskola:2022vpi}%
  \BibitemOpen
  \bibfield  {author} {\bibinfo {author} {\bibfnamefont {K.~J.}\ \bibnamefont
  {Eskola}}, \bibinfo {author} {\bibfnamefont {C.~A.}\ \bibnamefont {Flett}},
  \bibinfo {author} {\bibfnamefont {V.}~\bibnamefont {Guzey}}, \bibinfo
  {author} {\bibfnamefont {T.}~\bibnamefont {L\"oyt\"ainen}}, \ and\ \bibinfo
  {author} {\bibfnamefont {H.}~\bibnamefont {Paukkunen}},\ }\href {\doibase
  10.1103/PhysRevC.106.035202} {\bibfield  {journal} {\bibinfo  {journal}
  {Phys. Rev. C}\ }\textbf {\bibinfo {volume} {106}},\ \bibinfo {pages}
  {035202} (\bibinfo {year} {2022}{\natexlab{a}})},\ \Eprint
  {http://arxiv.org/abs/2203.11613} {arXiv:2203.11613 [hep-ph]} \BibitemShut
  {NoStop}%
\bibitem [{\citenamefont {Kopeliovich}\ \emph {et~al.}(2002)\citenamefont
  {Kopeliovich}, \citenamefont {Nemchik}, \citenamefont {Schafer},\ and\
  \citenamefont {Tarasov}}]{Kopeliovich:2001xj}%
  \BibitemOpen
  \bibfield  {author} {\bibinfo {author} {\bibfnamefont {B.~Z.}\ \bibnamefont
  {Kopeliovich}}, \bibinfo {author} {\bibfnamefont {J.}~\bibnamefont
  {Nemchik}}, \bibinfo {author} {\bibfnamefont {A.}~\bibnamefont {Schafer}}, \
  and\ \bibinfo {author} {\bibfnamefont {A.~V.}\ \bibnamefont {Tarasov}},\
  }\href {\doibase 10.1103/PhysRevC.65.035201} {\bibfield  {journal} {\bibinfo
  {journal} {Phys. Rev. C}\ }\textbf {\bibinfo {volume} {65}},\ \bibinfo
  {pages} {035201} (\bibinfo {year} {2002})},\ \Eprint
  {http://arxiv.org/abs/hep-ph/0107227} {arXiv:hep-ph/0107227} \BibitemShut
  {NoStop}%
\bibitem [{\citenamefont {Kopeliovich}\ \emph {et~al.}(1993)\citenamefont
  {Kopeliovich}, \citenamefont {Nemchick}, \citenamefont {Nikolaev},\ and\
  \citenamefont {Zakharov}}]{Kopeliovich:1993gk}%
  \BibitemOpen
  \bibfield  {author} {\bibinfo {author} {\bibfnamefont {B.~Z.}\ \bibnamefont
  {Kopeliovich}}, \bibinfo {author} {\bibfnamefont {J.}~\bibnamefont
  {Nemchick}}, \bibinfo {author} {\bibfnamefont {N.~N.}\ \bibnamefont
  {Nikolaev}}, \ and\ \bibinfo {author} {\bibfnamefont {B.~G.}\ \bibnamefont
  {Zakharov}},\ }\href {\doibase 10.1016/0370-2693(93)91523-P} {\bibfield
  {journal} {\bibinfo  {journal} {Phys. Lett. B}\ }\textbf {\bibinfo {volume}
  {309}},\ \bibinfo {pages} {179} (\bibinfo {year} {1993})},\ \Eprint
  {http://arxiv.org/abs/hep-ph/9305225} {arXiv:hep-ph/9305225} \BibitemShut
  {NoStop}%
\bibitem [{\citenamefont {Kopeliovich}\ \emph {et~al.}(2022)\citenamefont
  {Kopeliovich}, \citenamefont {Krelina}, \citenamefont {Nemchik},\ and\
  \citenamefont {Potashnikova}}]{Kopeliovich:2022jwe}%
  \BibitemOpen
  \bibfield  {author} {\bibinfo {author} {\bibfnamefont {B.~Z.}\ \bibnamefont
  {Kopeliovich}}, \bibinfo {author} {\bibfnamefont {M.}~\bibnamefont
  {Krelina}}, \bibinfo {author} {\bibfnamefont {J.}~\bibnamefont {Nemchik}}, \
  and\ \bibinfo {author} {\bibfnamefont {I.~K.}\ \bibnamefont {Potashnikova}},\
  }\href {\doibase 10.1103/PhysRevD.105.054023} {\bibfield  {journal} {\bibinfo
   {journal} {Phys. Rev. D}\ }\textbf {\bibinfo {volume} {105}},\ \bibinfo
  {pages} {054023} (\bibinfo {year} {2022})},\ \Eprint
  {http://arxiv.org/abs/2201.13021} {arXiv:2201.13021 [hep-ph]} \BibitemShut
  {NoStop}%
\bibitem [{\citenamefont {Kopeliovich}\ \emph {et~al.}(2010)\citenamefont
  {Kopeliovich}, \citenamefont {Schmidt},\ and\ \citenamefont
  {Siddikov}}]{Kopeliovich:2010sa}%
  \BibitemOpen
  \bibfield  {author} {\bibinfo {author} {\bibfnamefont {B.~Z.}\ \bibnamefont
  {Kopeliovich}}, \bibinfo {author} {\bibfnamefont {I.}~\bibnamefont
  {Schmidt}}, \ and\ \bibinfo {author} {\bibfnamefont {M.}~\bibnamefont
  {Siddikov}},\ }\href {\doibase 10.1103/PhysRevD.81.094013} {\bibfield
  {journal} {\bibinfo  {journal} {Phys. Rev. D}\ }\textbf {\bibinfo {volume}
  {81}},\ \bibinfo {pages} {094013} (\bibinfo {year} {2010})},\ \Eprint
  {http://arxiv.org/abs/1003.4188} {arXiv:1003.4188 [hep-ph]} \BibitemShut
  {NoStop}%
\bibitem [{\citenamefont {Morreale}\ and\ \citenamefont
  {Salazar}(2021)}]{Morreale:2021pnn}%
  \BibitemOpen
  \bibfield  {author} {\bibinfo {author} {\bibfnamefont {A.}~\bibnamefont
  {Morreale}}\ and\ \bibinfo {author} {\bibfnamefont {F.}~\bibnamefont
  {Salazar}},\ }\href {\doibase 10.3390/universe7080312} {\bibfield  {journal}
  {\bibinfo  {journal} {Universe}\ }\textbf {\bibinfo {volume} {7}},\ \bibinfo
  {pages} {312} (\bibinfo {year} {2021})},\ \Eprint
  {http://arxiv.org/abs/2108.08254} {arXiv:2108.08254 [hep-ph]} \BibitemShut
  {NoStop}%
\bibitem [{\citenamefont {McLerran}\ and\ \citenamefont
  {Venugopalan}(1994{\natexlab{a}})}]{McLerran:1993ka}%
  \BibitemOpen
  \bibfield  {author} {\bibinfo {author} {\bibfnamefont {L.~D.}\ \bibnamefont
  {McLerran}}\ and\ \bibinfo {author} {\bibfnamefont {R.}~\bibnamefont
  {Venugopalan}},\ }\href {\doibase 10.1103/PhysRevD.49.3352} {\bibfield
  {journal} {\bibinfo  {journal} {Phys. Rev. D}\ }\textbf {\bibinfo {volume}
  {49}},\ \bibinfo {pages} {3352} (\bibinfo {year} {1994}{\natexlab{a}})},\
  \Eprint {http://arxiv.org/abs/hep-ph/9311205} {arXiv:hep-ph/9311205}
  \BibitemShut {NoStop}%
\bibitem [{\citenamefont {McLerran}\ and\ \citenamefont
  {Venugopalan}(1994{\natexlab{b}})}]{McLerran:1993ni}%
  \BibitemOpen
  \bibfield  {author} {\bibinfo {author} {\bibfnamefont {L.~D.}\ \bibnamefont
  {McLerran}}\ and\ \bibinfo {author} {\bibfnamefont {R.}~\bibnamefont
  {Venugopalan}},\ }\href {\doibase 10.1103/PhysRevD.49.2233} {\bibfield
  {journal} {\bibinfo  {journal} {Phys. Rev. D}\ }\textbf {\bibinfo {volume}
  {49}},\ \bibinfo {pages} {2233} (\bibinfo {year} {1994}{\natexlab{b}})},\
  \Eprint {http://arxiv.org/abs/hep-ph/9309289} {arXiv:hep-ph/9309289}
  \BibitemShut {NoStop}%
\bibitem [{\citenamefont {Mueller}\ and\ \citenamefont
  {Munier}(2014)}]{Mueller:2014fba}%
  \BibitemOpen
  \bibfield  {author} {\bibinfo {author} {\bibfnamefont {A.~H.}\ \bibnamefont
  {Mueller}}\ and\ \bibinfo {author} {\bibfnamefont {S.}~\bibnamefont
  {Munier}},\ }\href {\doibase 10.1016/j.physletb.2014.08.058} {\bibfield
  {journal} {\bibinfo  {journal} {Phys. Lett. B}\ }\textbf {\bibinfo {volume}
  {737}},\ \bibinfo {pages} {303} (\bibinfo {year} {2014})},\ \Eprint
  {http://arxiv.org/abs/1405.3131} {arXiv:1405.3131 [hep-ph]} \BibitemShut
  {NoStop}%
\bibitem [{\citenamefont {Caucal}\ and\ \citenamefont
  {Mehtar-Tani}(2022)}]{Caucal:2021lgf}%
  \BibitemOpen
  \bibfield  {author} {\bibinfo {author} {\bibfnamefont {P.}~\bibnamefont
  {Caucal}}\ and\ \bibinfo {author} {\bibfnamefont {Y.}~\bibnamefont
  {Mehtar-Tani}},\ }\href {\doibase 10.1103/PhysRevD.106.L051501} {\bibfield
  {journal} {\bibinfo  {journal} {Phys. Rev. D}\ }\textbf {\bibinfo {volume}
  {106}},\ \bibinfo {pages} {L051501} (\bibinfo {year} {2022})},\ \Eprint
  {http://arxiv.org/abs/2109.12041} {arXiv:2109.12041 [hep-ph]} \BibitemShut
  {NoStop}%
\bibitem [{\citenamefont {Le}\ \emph {et~al.}(2021)\citenamefont {Le},
  \citenamefont {Mueller},\ and\ \citenamefont {Munier}}]{Le:2020zpy}%
  \BibitemOpen
  \bibfield  {author} {\bibinfo {author} {\bibfnamefont {A.~D.}\ \bibnamefont
  {Le}}, \bibinfo {author} {\bibfnamefont {A.~H.}\ \bibnamefont {Mueller}}, \
  and\ \bibinfo {author} {\bibfnamefont {S.}~\bibnamefont {Munier}},\ }\href
  {\doibase 10.1103/PhysRevD.103.054031} {\bibfield  {journal} {\bibinfo
  {journal} {Phys. Rev. D}\ }\textbf {\bibinfo {volume} {103}},\ \bibinfo
  {pages} {054031} (\bibinfo {year} {2021})},\ \Eprint
  {http://arxiv.org/abs/2010.15546} {arXiv:2010.15546 [hep-ph]} \BibitemShut
  {NoStop}%
\bibitem [{\citenamefont {Dumitru}\ \emph {et~al.}(2018)\citenamefont
  {Dumitru}, \citenamefont {Miller},\ and\ \citenamefont
  {Venugopalan}}]{Dumitru:2018vpr}%
  \BibitemOpen
  \bibfield  {author} {\bibinfo {author} {\bibfnamefont {A.}~\bibnamefont
  {Dumitru}}, \bibinfo {author} {\bibfnamefont {G.~A.}\ \bibnamefont {Miller}},
  \ and\ \bibinfo {author} {\bibfnamefont {R.}~\bibnamefont {Venugopalan}},\
  }\href {\doibase 10.1103/PhysRevD.98.094004} {\bibfield  {journal} {\bibinfo
  {journal} {Phys. Rev. D}\ }\textbf {\bibinfo {volume} {98}},\ \bibinfo
  {pages} {094004} (\bibinfo {year} {2018})},\ \Eprint
  {http://arxiv.org/abs/1808.02501} {arXiv:1808.02501 [hep-ph]} \BibitemShut
  {NoStop}%
\bibitem [{\citenamefont {Dumitru}\ and\ \citenamefont
  {Paatelainen}(2021)}]{Dumitru:2020gla}%
  \BibitemOpen
  \bibfield  {author} {\bibinfo {author} {\bibfnamefont {A.}~\bibnamefont
  {Dumitru}}\ and\ \bibinfo {author} {\bibfnamefont {R.}~\bibnamefont
  {Paatelainen}},\ }\href {\doibase 10.1103/PhysRevD.103.034026} {\bibfield
  {journal} {\bibinfo  {journal} {Phys. Rev. D}\ }\textbf {\bibinfo {volume}
  {103}},\ \bibinfo {pages} {034026} (\bibinfo {year} {2021})},\ \Eprint
  {http://arxiv.org/abs/2010.11245} {arXiv:2010.11245 [hep-ph]} \BibitemShut
  {NoStop}%
\bibitem [{\citenamefont {Dumitru}\ \emph {et~al.}(2022)\citenamefont
  {Dumitru}, \citenamefont {M\"antysaari},\ and\ \citenamefont
  {Paatelainen}}]{Dumitru:2021tqp}%
  \BibitemOpen
  \bibfield  {author} {\bibinfo {author} {\bibfnamefont {A.}~\bibnamefont
  {Dumitru}}, \bibinfo {author} {\bibfnamefont {H.}~\bibnamefont
  {M\"antysaari}}, \ and\ \bibinfo {author} {\bibfnamefont {R.}~\bibnamefont
  {Paatelainen}},\ }\href {\doibase 10.1103/PhysRevD.105.036007} {\bibfield
  {journal} {\bibinfo  {journal} {Phys. Rev. D}\ }\textbf {\bibinfo {volume}
  {105}},\ \bibinfo {pages} {036007} (\bibinfo {year} {2022})},\ \Eprint
  {http://arxiv.org/abs/2106.12623} {arXiv:2106.12623 [hep-ph]} \BibitemShut
  {NoStop}%
\bibitem [{\citenamefont {Dumitru}\ and\ \citenamefont
  {Stebel}(2019)}]{Dumitru:2019qec}%
  \BibitemOpen
  \bibfield  {author} {\bibinfo {author} {\bibfnamefont {A.}~\bibnamefont
  {Dumitru}}\ and\ \bibinfo {author} {\bibfnamefont {T.}~\bibnamefont
  {Stebel}},\ }\href {\doibase 10.1103/PhysRevD.99.094038} {\bibfield
  {journal} {\bibinfo  {journal} {Phys. Rev. D}\ }\textbf {\bibinfo {volume}
  {99}},\ \bibinfo {pages} {094038} (\bibinfo {year} {2019})},\ \Eprint
  {http://arxiv.org/abs/1903.07660} {arXiv:1903.07660 [hep-ph]} \BibitemShut
  {NoStop}%
\bibitem [{\citenamefont {Dumitru}\ \emph {et~al.}(2021)\citenamefont
  {Dumitru}, \citenamefont {M\"antysaari},\ and\ \citenamefont
  {Paatelainen}}]{Dumitru:2021tvw}%
  \BibitemOpen
  \bibfield  {author} {\bibinfo {author} {\bibfnamefont {A.}~\bibnamefont
  {Dumitru}}, \bibinfo {author} {\bibfnamefont {H.}~\bibnamefont
  {M\"antysaari}}, \ and\ \bibinfo {author} {\bibfnamefont {R.}~\bibnamefont
  {Paatelainen}},\ }\href {\doibase 10.1016/j.physletb.2021.136560} {\bibfield
  {journal} {\bibinfo  {journal} {Phys. Lett. B}\ }\textbf {\bibinfo {volume}
  {820}},\ \bibinfo {pages} {136560} (\bibinfo {year} {2021})},\ \Eprint
  {http://arxiv.org/abs/2103.11682} {arXiv:2103.11682 [hep-ph]} \BibitemShut
  {NoStop}%
\bibitem [{\citenamefont {Zhang}\ \emph {et~al.}(2019)\citenamefont {Zhang},
  \citenamefont {Ji}, \citenamefont {Sch\"afer}, \citenamefont {Wang},\ and\
  \citenamefont {Zhao}}]{Zhang:2018diq}%
  \BibitemOpen
  \bibfield  {author} {\bibinfo {author} {\bibfnamefont {J.-H.}\ \bibnamefont
  {Zhang}}, \bibinfo {author} {\bibfnamefont {X.}~\bibnamefont {Ji}}, \bibinfo
  {author} {\bibfnamefont {A.}~\bibnamefont {Sch\"afer}}, \bibinfo {author}
  {\bibfnamefont {W.}~\bibnamefont {Wang}}, \ and\ \bibinfo {author}
  {\bibfnamefont {S.}~\bibnamefont {Zhao}},\ }\href {\doibase
  10.1103/PhysRevLett.122.142001} {\bibfield  {journal} {\bibinfo  {journal}
  {Phys. Rev. Lett.}\ }\textbf {\bibinfo {volume} {122}},\ \bibinfo {pages}
  {142001} (\bibinfo {year} {2019})},\ \Eprint
  {http://arxiv.org/abs/1808.10824} {arXiv:1808.10824 [hep-ph]} \BibitemShut
  {NoStop}%
\bibitem [{\citenamefont {Zhang}\ \emph
  {et~al.}(2022{\natexlab{a}})\citenamefont {Zhang}, \citenamefont {Hao},
  \citenamefont {Kharzeev},\ and\ \citenamefont {Korepin}}]{Zhang:2021hra}%
  \BibitemOpen
  \bibfield  {author} {\bibinfo {author} {\bibfnamefont {K.}~\bibnamefont
  {Zhang}}, \bibinfo {author} {\bibfnamefont {K.}~\bibnamefont {Hao}}, \bibinfo
  {author} {\bibfnamefont {D.}~\bibnamefont {Kharzeev}}, \ and\ \bibinfo
  {author} {\bibfnamefont {V.}~\bibnamefont {Korepin}},\ }\href {\doibase
  10.1103/PhysRevD.105.014002} {\bibfield  {journal} {\bibinfo  {journal}
  {Phys. Rev. D}\ }\textbf {\bibinfo {volume} {105}},\ \bibinfo {pages}
  {014002} (\bibinfo {year} {2022}{\natexlab{a}})},\ \Eprint
  {http://arxiv.org/abs/2110.04881} {arXiv:2110.04881 [quant-ph]} \BibitemShut
  {NoStop}%
\bibitem [{\citenamefont {Kharzeev}\ and\ \citenamefont
  {Levin}(2017)}]{Kharzeev:2017qzs}%
  \BibitemOpen
  \bibfield  {author} {\bibinfo {author} {\bibfnamefont {D.~E.}\ \bibnamefont
  {Kharzeev}}\ and\ \bibinfo {author} {\bibfnamefont {E.~M.}\ \bibnamefont
  {Levin}},\ }\href {\doibase 10.1103/PhysRevD.95.114008} {\bibfield  {journal}
  {\bibinfo  {journal} {Phys. Rev. D}\ }\textbf {\bibinfo {volume} {95}},\
  \bibinfo {pages} {114008} (\bibinfo {year} {2017})},\ \Eprint
  {http://arxiv.org/abs/1702.03489} {arXiv:1702.03489 [hep-ph]} \BibitemShut
  {NoStop}%
\bibitem [{\citenamefont {Duan}\ \emph {et~al.}(2022)\citenamefont {Duan},
  \citenamefont {Kovner},\ and\ \citenamefont {Skokov}}]{Duan:2021clk}%
  \BibitemOpen
  \bibfield  {author} {\bibinfo {author} {\bibfnamefont {H.}~\bibnamefont
  {Duan}}, \bibinfo {author} {\bibfnamefont {A.}~\bibnamefont {Kovner}}, \ and\
  \bibinfo {author} {\bibfnamefont {V.~V.}\ \bibnamefont {Skokov}},\ }\href
  {\doibase 10.1103/PhysRevD.105.056009} {\bibfield  {journal} {\bibinfo
  {journal} {Phys. Rev. D}\ }\textbf {\bibinfo {volume} {105}},\ \bibinfo
  {pages} {056009} (\bibinfo {year} {2022})},\ \Eprint
  {http://arxiv.org/abs/2111.06475} {arXiv:2111.06475 [hep-ph]} \BibitemShut
  {NoStop}%
\bibitem [{\citenamefont {Dumitru}\ and\ \citenamefont
  {Kolbusz}(2022)}]{Dumitru:2022tud}%
  \BibitemOpen
  \bibfield  {author} {\bibinfo {author} {\bibfnamefont {A.}~\bibnamefont
  {Dumitru}}\ and\ \bibinfo {author} {\bibfnamefont {E.}~\bibnamefont
  {Kolbusz}},\ }\href {\doibase 10.1103/PhysRevD.105.074030} {\bibfield
  {journal} {\bibinfo  {journal} {Phys. Rev. D}\ }\textbf {\bibinfo {volume}
  {105}},\ \bibinfo {pages} {074030} (\bibinfo {year} {2022})},\ \Eprint
  {http://arxiv.org/abs/2202.01803} {arXiv:2202.01803 [hep-ph]} \BibitemShut
  {NoStop}%
\bibitem [{\citenamefont {Hentschinski}\ \emph
  {et~al.}(2022{\natexlab{b}})\citenamefont {Hentschinski}, \citenamefont
  {Kutak},\ and\ \citenamefont {Straka}}]{Hentschinski:2022rsa}%
  \BibitemOpen
  \bibfield  {author} {\bibinfo {author} {\bibfnamefont {M.}~\bibnamefont
  {Hentschinski}}, \bibinfo {author} {\bibfnamefont {K.}~\bibnamefont {Kutak}},
  \ and\ \bibinfo {author} {\bibfnamefont {R.}~\bibnamefont {Straka}},\ }\href
  {\doibase 10.1140/epjc/s10052-022-11122-1} {\bibfield  {journal} {\bibinfo
  {journal} {Eur. Phys. J. C}\ }\textbf {\bibinfo {volume} {82}},\ \bibinfo
  {pages} {1147} (\bibinfo {year} {2022}{\natexlab{b}})},\ \Eprint
  {http://arxiv.org/abs/2207.09430} {arXiv:2207.09430 [hep-ph]} \BibitemShut
  {NoStop}%
\bibitem [{\citenamefont {Dvali}\ and\ \citenamefont
  {Venugopalan}(2022)}]{Dvali:2021ooc}%
  \BibitemOpen
  \bibfield  {author} {\bibinfo {author} {\bibfnamefont {G.}~\bibnamefont
  {Dvali}}\ and\ \bibinfo {author} {\bibfnamefont {R.}~\bibnamefont
  {Venugopalan}},\ }\href {\doibase 10.1103/PhysRevD.105.056026} {\bibfield
  {journal} {\bibinfo  {journal} {Phys. Rev. D}\ }\textbf {\bibinfo {volume}
  {105}},\ \bibinfo {pages} {056026} (\bibinfo {year} {2022})},\ \Eprint
  {http://arxiv.org/abs/2106.11989} {arXiv:2106.11989 [hep-th]} \BibitemShut
  {NoStop}%
\bibitem [{\citenamefont {Berges}\ \emph {et~al.}(2021)\citenamefont {Berges},
  \citenamefont {Heller}, \citenamefont {Mazeliauskas},\ and\ \citenamefont
  {Venugopalan}}]{Berges:2020fwq}%
  \BibitemOpen
  \bibfield  {author} {\bibinfo {author} {\bibfnamefont {J.}~\bibnamefont
  {Berges}}, \bibinfo {author} {\bibfnamefont {M.~P.}\ \bibnamefont {Heller}},
  \bibinfo {author} {\bibfnamefont {A.}~\bibnamefont {Mazeliauskas}}, \ and\
  \bibinfo {author} {\bibfnamefont {R.}~\bibnamefont {Venugopalan}},\ }\href
  {\doibase 10.1103/RevModPhys.93.035003} {\bibfield  {journal} {\bibinfo
  {journal} {Rev. Mod. Phys.}\ }\textbf {\bibinfo {volume} {93}},\ \bibinfo
  {pages} {035003} (\bibinfo {year} {2021})},\ \Eprint
  {http://arxiv.org/abs/2005.12299} {arXiv:2005.12299 [hep-th]} \BibitemShut
  {NoStop}%
\bibitem [{\citenamefont {Anchordoqui}\ \emph {et~al.}(2006)\citenamefont
  {Anchordoqui}, \citenamefont {Cooper-Sarkar}, \citenamefont {Hooper},\ and\
  \citenamefont {Sarkar}}]{Anchordoqui:2006ta}%
  \BibitemOpen
  \bibfield  {author} {\bibinfo {author} {\bibfnamefont {L.~A.}\ \bibnamefont
  {Anchordoqui}}, \bibinfo {author} {\bibfnamefont {A.~M.}\ \bibnamefont
  {Cooper-Sarkar}}, \bibinfo {author} {\bibfnamefont {D.}~\bibnamefont
  {Hooper}}, \ and\ \bibinfo {author} {\bibfnamefont {S.}~\bibnamefont
  {Sarkar}},\ }\href {\doibase 10.1103/PhysRevD.74.043008} {\bibfield
  {journal} {\bibinfo  {journal} {Phys. Rev. D}\ }\textbf {\bibinfo {volume}
  {74}},\ \bibinfo {pages} {043008} (\bibinfo {year} {2006})},\ \Eprint
  {http://arxiv.org/abs/hep-ph/0605086} {arXiv:hep-ph/0605086} \BibitemShut
  {NoStop}%
\bibitem [{\citenamefont {Cooper-Sarkar}\ \emph {et~al.}(2011)\citenamefont
  {Cooper-Sarkar}, \citenamefont {Mertsch},\ and\ \citenamefont
  {Sarkar}}]{Cooper-Sarkar:2011jtt}%
  \BibitemOpen
  \bibfield  {author} {\bibinfo {author} {\bibfnamefont {A.}~\bibnamefont
  {Cooper-Sarkar}}, \bibinfo {author} {\bibfnamefont {P.}~\bibnamefont
  {Mertsch}}, \ and\ \bibinfo {author} {\bibfnamefont {S.}~\bibnamefont
  {Sarkar}},\ }\href {\doibase 10.1007/JHEP08(2011)042} {\bibfield  {journal}
  {\bibinfo  {journal} {JHEP}\ }\textbf {\bibinfo {volume} {08}},\ \bibinfo
  {pages} {042} (\bibinfo {year} {2011})},\ \Eprint
  {http://arxiv.org/abs/1106.3723} {arXiv:1106.3723 [hep-ph]} \BibitemShut
  {NoStop}%
\bibitem [{\citenamefont {Garcia}\ \emph {et~al.}(2020)\citenamefont {Garcia},
  \citenamefont {Gauld}, \citenamefont {Heijboer},\ and\ \citenamefont
  {Rojo}}]{Garcia:2020jwr}%
  \BibitemOpen
  \bibfield  {author} {\bibinfo {author} {\bibfnamefont {A.}~\bibnamefont
  {Garcia}}, \bibinfo {author} {\bibfnamefont {R.}~\bibnamefont {Gauld}},
  \bibinfo {author} {\bibfnamefont {A.}~\bibnamefont {Heijboer}}, \ and\
  \bibinfo {author} {\bibfnamefont {J.}~\bibnamefont {Rojo}},\ }\href {\doibase
  10.1088/1475-7516/2020/09/025} {\bibfield  {journal} {\bibinfo  {journal}
  {JCAP}\ }\textbf {\bibinfo {volume} {09}},\ \bibinfo {pages} {025} (\bibinfo
  {year} {2020})},\ \Eprint {http://arxiv.org/abs/2004.04756} {arXiv:2004.04756
  [hep-ph]} \BibitemShut {NoStop}%
\bibitem [{\citenamefont {Candido}\ \emph {et~al.}(2023)\citenamefont
  {Candido}, \citenamefont {Garcia}, \citenamefont {Magni}, \citenamefont
  {Rabemananjara}, \citenamefont {Rojo},\ and\ \citenamefont
  {Stegeman}}]{Candido:2023utz}%
  \BibitemOpen
  \bibfield  {author} {\bibinfo {author} {\bibfnamefont {A.}~\bibnamefont
  {Candido}}, \bibinfo {author} {\bibfnamefont {A.}~\bibnamefont {Garcia}},
  \bibinfo {author} {\bibfnamefont {G.}~\bibnamefont {Magni}}, \bibinfo
  {author} {\bibfnamefont {T.}~\bibnamefont {Rabemananjara}}, \bibinfo {author}
  {\bibfnamefont {J.}~\bibnamefont {Rojo}}, \ and\ \bibinfo {author}
  {\bibfnamefont {R.}~\bibnamefont {Stegeman}},\ }\href@noop {} {\  (\bibinfo
  {year} {2023})},\ \Eprint {http://arxiv.org/abs/2302.08527} {arXiv:2302.08527
  [hep-ph]} \BibitemShut {NoStop}%
\bibitem [{\citenamefont {Bhattacharya}\ \emph {et~al.}(2016)\citenamefont
  {Bhattacharya}, \citenamefont {Enberg}, \citenamefont {Jeong}, \citenamefont
  {Kim}, \citenamefont {Reno}, \citenamefont {Sarcevic},\ and\ \citenamefont
  {Stasto}}]{Bhattacharya:2016jce}%
  \BibitemOpen
  \bibfield  {author} {\bibinfo {author} {\bibfnamefont {A.}~\bibnamefont
  {Bhattacharya}}, \bibinfo {author} {\bibfnamefont {R.}~\bibnamefont
  {Enberg}}, \bibinfo {author} {\bibfnamefont {Y.~S.}\ \bibnamefont {Jeong}},
  \bibinfo {author} {\bibfnamefont {C.~S.}\ \bibnamefont {Kim}}, \bibinfo
  {author} {\bibfnamefont {M.~H.}\ \bibnamefont {Reno}}, \bibinfo {author}
  {\bibfnamefont {I.}~\bibnamefont {Sarcevic}}, \ and\ \bibinfo {author}
  {\bibfnamefont {A.}~\bibnamefont {Stasto}},\ }\href {\doibase
  10.1007/JHEP11(2016)167} {\bibfield  {journal} {\bibinfo  {journal} {JHEP}\
  }\textbf {\bibinfo {volume} {11}},\ \bibinfo {pages} {167} (\bibinfo {year}
  {2016})},\ \Eprint {http://arxiv.org/abs/1607.00193} {arXiv:1607.00193
  [hep-ph]} \BibitemShut {NoStop}%
\bibitem [{\citenamefont {Moch}\ \emph {et~al.}(2004)\citenamefont {Moch},
  \citenamefont {Vermaseren},\ and\ \citenamefont {Vogt}}]{Moch:2004pa}%
  \BibitemOpen
  \bibfield  {author} {\bibinfo {author} {\bibfnamefont {S.}~\bibnamefont
  {Moch}}, \bibinfo {author} {\bibfnamefont {J.~A.~M.}\ \bibnamefont
  {Vermaseren}}, \ and\ \bibinfo {author} {\bibfnamefont {A.}~\bibnamefont
  {Vogt}},\ }\href {\doibase 10.1016/j.nuclphysb.2004.03.030} {\bibfield
  {journal} {\bibinfo  {journal} {Nucl. Phys. B}\ }\textbf {\bibinfo {volume}
  {688}},\ \bibinfo {pages} {101} (\bibinfo {year} {2004})},\ \Eprint
  {http://arxiv.org/abs/hep-ph/0403192} {arXiv:hep-ph/0403192} \BibitemShut
  {NoStop}%
\bibitem [{\citenamefont {Vogt}\ \emph {et~al.}(2004)\citenamefont {Vogt},
  \citenamefont {Moch},\ and\ \citenamefont {Vermaseren}}]{Vogt:2004mw}%
  \BibitemOpen
  \bibfield  {author} {\bibinfo {author} {\bibfnamefont {A.}~\bibnamefont
  {Vogt}}, \bibinfo {author} {\bibfnamefont {S.}~\bibnamefont {Moch}}, \ and\
  \bibinfo {author} {\bibfnamefont {J.~A.~M.}\ \bibnamefont {Vermaseren}},\
  }\href {\doibase 10.1016/j.nuclphysb.2004.04.024} {\bibfield  {journal}
  {\bibinfo  {journal} {Nucl. Phys. B}\ }\textbf {\bibinfo {volume} {691}},\
  \bibinfo {pages} {129} (\bibinfo {year} {2004})},\ \Eprint
  {http://arxiv.org/abs/hep-ph/0404111} {arXiv:hep-ph/0404111} \BibitemShut
  {NoStop}%
\bibitem [{\citenamefont {Moch}\ \emph {et~al.}(2014)\citenamefont {Moch},
  \citenamefont {Vermaseren},\ and\ \citenamefont {Vogt}}]{Moch:2014sna}%
  \BibitemOpen
  \bibfield  {author} {\bibinfo {author} {\bibfnamefont {S.}~\bibnamefont
  {Moch}}, \bibinfo {author} {\bibfnamefont {J.~A.~M.}\ \bibnamefont
  {Vermaseren}}, \ and\ \bibinfo {author} {\bibfnamefont {A.}~\bibnamefont
  {Vogt}},\ }\href {\doibase 10.1016/j.nuclphysb.2014.10.016} {\bibfield
  {journal} {\bibinfo  {journal} {Nucl. Phys. B}\ }\textbf {\bibinfo {volume}
  {889}},\ \bibinfo {pages} {351} (\bibinfo {year} {2014})},\ \Eprint
  {http://arxiv.org/abs/1409.5131} {arXiv:1409.5131 [hep-ph]} \BibitemShut
  {NoStop}%
\bibitem [{\citenamefont {Moch}\ \emph {et~al.}(2017)\citenamefont {Moch},
  \citenamefont {Ruijl}, \citenamefont {Ueda}, \citenamefont {Vermaseren},\
  and\ \citenamefont {Vogt}}]{Moch:2017uml}%
  \BibitemOpen
  \bibfield  {author} {\bibinfo {author} {\bibfnamefont {S.}~\bibnamefont
  {Moch}}, \bibinfo {author} {\bibfnamefont {B.}~\bibnamefont {Ruijl}},
  \bibinfo {author} {\bibfnamefont {T.}~\bibnamefont {Ueda}}, \bibinfo {author}
  {\bibfnamefont {J.~A.~M.}\ \bibnamefont {Vermaseren}}, \ and\ \bibinfo
  {author} {\bibfnamefont {A.}~\bibnamefont {Vogt}},\ }\href {\doibase
  10.1007/JHEP10(2017)041} {\bibfield  {journal} {\bibinfo  {journal} {JHEP}\
  }\textbf {\bibinfo {volume} {10}},\ \bibinfo {pages} {041} (\bibinfo {year}
  {2017})},\ \Eprint {http://arxiv.org/abs/1707.08315} {arXiv:1707.08315
  [hep-ph]} \BibitemShut {NoStop}%
\bibitem [{\citenamefont {Moch}\ \emph {et~al.}(2022)\citenamefont {Moch},
  \citenamefont {Ruijl}, \citenamefont {Ueda}, \citenamefont {Vermaseren},\
  and\ \citenamefont {Vogt}}]{Moch:2021qrk}%
  \BibitemOpen
  \bibfield  {author} {\bibinfo {author} {\bibfnamefont {S.}~\bibnamefont
  {Moch}}, \bibinfo {author} {\bibfnamefont {B.}~\bibnamefont {Ruijl}},
  \bibinfo {author} {\bibfnamefont {T.}~\bibnamefont {Ueda}}, \bibinfo {author}
  {\bibfnamefont {J.~A.~M.}\ \bibnamefont {Vermaseren}}, \ and\ \bibinfo
  {author} {\bibfnamefont {A.}~\bibnamefont {Vogt}},\ }\href {\doibase
  10.1016/j.physletb.2021.136853} {\bibfield  {journal} {\bibinfo  {journal}
  {Phys. Lett. B}\ }\textbf {\bibinfo {volume} {825}},\ \bibinfo {pages}
  {136853} (\bibinfo {year} {2022})},\ \Eprint
  {http://arxiv.org/abs/2111.15561} {arXiv:2111.15561 [hep-ph]} \BibitemShut
  {NoStop}%
\bibitem [{\citenamefont {Herzog}\ \emph {et~al.}(2019)\citenamefont {Herzog},
  \citenamefont {Moch}, \citenamefont {Ruijl}, \citenamefont {Ueda},
  \citenamefont {Vermaseren},\ and\ \citenamefont {Vogt}}]{Herzog:2018kwj}%
  \BibitemOpen
  \bibfield  {author} {\bibinfo {author} {\bibfnamefont {F.}~\bibnamefont
  {Herzog}}, \bibinfo {author} {\bibfnamefont {S.}~\bibnamefont {Moch}},
  \bibinfo {author} {\bibfnamefont {B.}~\bibnamefont {Ruijl}}, \bibinfo
  {author} {\bibfnamefont {T.}~\bibnamefont {Ueda}}, \bibinfo {author}
  {\bibfnamefont {J.~A.~M.}\ \bibnamefont {Vermaseren}}, \ and\ \bibinfo
  {author} {\bibfnamefont {A.}~\bibnamefont {Vogt}},\ }\href {\doibase
  10.1016/j.physletb.2019.01.060} {\bibfield  {journal} {\bibinfo  {journal}
  {Phys. Lett. B}\ }\textbf {\bibinfo {volume} {790}},\ \bibinfo {pages} {436}
  (\bibinfo {year} {2019})},\ \Eprint {http://arxiv.org/abs/1812.11818}
  {arXiv:1812.11818 [hep-ph]} \BibitemShut {NoStop}%
\bibitem [{\citenamefont {Zijlstra}\ and\ \citenamefont {van
  Neerven}(1992)}]{Zijlstra:1992qd}%
  \BibitemOpen
  \bibfield  {author} {\bibinfo {author} {\bibfnamefont {E.~B.}\ \bibnamefont
  {Zijlstra}}\ and\ \bibinfo {author} {\bibfnamefont {W.~L.}\ \bibnamefont {van
  Neerven}},\ }\href {\doibase 10.1016/0550-3213(92)90087-R} {\bibfield
  {journal} {\bibinfo  {journal} {Nucl. Phys. B}\ }\textbf {\bibinfo {volume}
  {383}},\ \bibinfo {pages} {525} (\bibinfo {year} {1992})}\BibitemShut
  {NoStop}%
\bibitem [{\citenamefont {Zijlstra}\ and\ \citenamefont {van
  Neerven}(1994)}]{Zijlstra:1993sh}%
  \BibitemOpen
  \bibfield  {author} {\bibinfo {author} {\bibfnamefont {E.~B.}\ \bibnamefont
  {Zijlstra}}\ and\ \bibinfo {author} {\bibfnamefont {W.~L.}\ \bibnamefont {van
  Neerven}},\ }\href {\doibase 10.1016/0550-3213(94)90538-X} {\bibfield
  {journal} {\bibinfo  {journal} {Nucl. Phys. B}\ }\textbf {\bibinfo {volume}
  {417}},\ \bibinfo {pages} {61} (\bibinfo {year} {1994})},\ \bibinfo {note}
  {[Erratum: Nucl.Phys.B 426, 245 (1994), Erratum: Nucl.Phys.B 773, 105--106
  (2007), Erratum: Nucl.Phys.B 501, 599--599 (1997)]}\BibitemShut {NoStop}%
\bibitem [{\citenamefont {Borsa}\ \emph
  {et~al.}(2022{\natexlab{a}})\citenamefont {Borsa}, \citenamefont
  {de~Florian},\ and\ \citenamefont {Pedron}}]{Borsa:2022irn}%
  \BibitemOpen
  \bibfield  {author} {\bibinfo {author} {\bibfnamefont {I.}~\bibnamefont
  {Borsa}}, \bibinfo {author} {\bibfnamefont {D.}~\bibnamefont {de~Florian}}, \
  and\ \bibinfo {author} {\bibfnamefont {I.}~\bibnamefont {Pedron}},\ }\href
  {\doibase 10.1140/epjc/s10052-022-11140-z} {\bibfield  {journal} {\bibinfo
  {journal} {Eur. Phys. J. C}\ }\textbf {\bibinfo {volume} {82}},\ \bibinfo
  {pages} {1167} (\bibinfo {year} {2022}{\natexlab{a}})},\ \Eprint
  {http://arxiv.org/abs/2210.12014} {arXiv:2210.12014 [hep-ph]} \BibitemShut
  {NoStop}%
\bibitem [{\citenamefont {Borsa}\ \emph
  {et~al.}(2022{\natexlab{b}})\citenamefont {Borsa}, \citenamefont
  {de~Florian},\ and\ \citenamefont {Pedron}}]{Borsa:2022cap}%
  \BibitemOpen
  \bibfield  {author} {\bibinfo {author} {\bibfnamefont {I.}~\bibnamefont
  {Borsa}}, \bibinfo {author} {\bibfnamefont {D.}~\bibnamefont {de~Florian}}, \
  and\ \bibinfo {author} {\bibfnamefont {I.}~\bibnamefont {Pedron}},\
  }\href@noop {} {\  (\bibinfo {year} {2022}{\natexlab{b}})},\ \Eprint
  {http://arxiv.org/abs/2212.06625} {arXiv:2212.06625 [hep-ph]} \BibitemShut
  {NoStop}%
\bibitem [{\citenamefont {Bl\"umlein}\ \emph {et~al.}(2022)\citenamefont
  {Bl\"umlein}, \citenamefont {Marquard}, \citenamefont {Schneider},\ and\
  \citenamefont {Sch\"onwald}}]{Blumlein:2022gpp}%
  \BibitemOpen
  \bibfield  {author} {\bibinfo {author} {\bibfnamefont {J.}~\bibnamefont
  {Bl\"umlein}}, \bibinfo {author} {\bibfnamefont {P.}~\bibnamefont
  {Marquard}}, \bibinfo {author} {\bibfnamefont {C.}~\bibnamefont {Schneider}},
  \ and\ \bibinfo {author} {\bibfnamefont {K.}~\bibnamefont {Sch\"onwald}},\
  }\href {\doibase 10.1007/JHEP11(2022)156} {\bibfield  {journal} {\bibinfo
  {journal} {JHEP}\ }\textbf {\bibinfo {volume} {11}},\ \bibinfo {pages} {156}
  (\bibinfo {year} {2022})},\ \Eprint {http://arxiv.org/abs/2208.14325}
  {arXiv:2208.14325 [hep-ph]} \BibitemShut {NoStop}%
\bibitem [{\citenamefont {Abelof}\ \emph {et~al.}(2016)\citenamefont {Abelof},
  \citenamefont {Boughezal}, \citenamefont {Liu},\ and\ \citenamefont
  {Petriello}}]{Abelof:2016pby}%
  \BibitemOpen
  \bibfield  {author} {\bibinfo {author} {\bibfnamefont {G.}~\bibnamefont
  {Abelof}}, \bibinfo {author} {\bibfnamefont {R.}~\bibnamefont {Boughezal}},
  \bibinfo {author} {\bibfnamefont {X.}~\bibnamefont {Liu}}, \ and\ \bibinfo
  {author} {\bibfnamefont {F.}~\bibnamefont {Petriello}},\ }\href {\doibase
  10.1016/j.physletb.2016.10.022} {\bibfield  {journal} {\bibinfo  {journal}
  {Phys. Lett. B}\ }\textbf {\bibinfo {volume} {763}},\ \bibinfo {pages} {52}
  (\bibinfo {year} {2016})},\ \Eprint {http://arxiv.org/abs/1607.04921}
  {arXiv:1607.04921 [hep-ph]} \BibitemShut {NoStop}%
\bibitem [{\citenamefont {Currie}\ \emph {et~al.}(2017)\citenamefont {Currie},
  \citenamefont {Gehrmann}, \citenamefont {Huss},\ and\ \citenamefont
  {Niehues}}]{Currie:2017tpe}%
  \BibitemOpen
  \bibfield  {author} {\bibinfo {author} {\bibfnamefont {J.}~\bibnamefont
  {Currie}}, \bibinfo {author} {\bibfnamefont {T.}~\bibnamefont {Gehrmann}},
  \bibinfo {author} {\bibfnamefont {A.}~\bibnamefont {Huss}}, \ and\ \bibinfo
  {author} {\bibfnamefont {J.}~\bibnamefont {Niehues}},\ }\href {\doibase
  10.1007/JHEP07(2017)018} {\bibfield  {journal} {\bibinfo  {journal} {JHEP}\
  }\textbf {\bibinfo {volume} {07}},\ \bibinfo {pages} {018} (\bibinfo {year}
  {2017})},\ \bibinfo {note} {[Erratum: JHEP 12, 042 (2020)]},\ \Eprint
  {http://arxiv.org/abs/1703.05977} {arXiv:1703.05977 [hep-ph]} \BibitemShut
  {NoStop}%
\bibitem [{\citenamefont {Currie}\ \emph {et~al.}(2018)\citenamefont {Currie},
  \citenamefont {Gehrmann}, \citenamefont {Glover}, \citenamefont {Huss},
  \citenamefont {Niehues},\ and\ \citenamefont {Vogt}}]{Currie:2018fgr}%
  \BibitemOpen
  \bibfield  {author} {\bibinfo {author} {\bibfnamefont {J.}~\bibnamefont
  {Currie}}, \bibinfo {author} {\bibfnamefont {T.}~\bibnamefont {Gehrmann}},
  \bibinfo {author} {\bibfnamefont {E.~W.~N.}\ \bibnamefont {Glover}}, \bibinfo
  {author} {\bibfnamefont {A.}~\bibnamefont {Huss}}, \bibinfo {author}
  {\bibfnamefont {J.}~\bibnamefont {Niehues}}, \ and\ \bibinfo {author}
  {\bibfnamefont {A.}~\bibnamefont {Vogt}},\ }\href {\doibase
  10.1007/JHEP05(2018)209} {\bibfield  {journal} {\bibinfo  {journal} {JHEP}\
  }\textbf {\bibinfo {volume} {05}},\ \bibinfo {pages} {209} (\bibinfo {year}
  {2018})},\ \Eprint {http://arxiv.org/abs/1803.09973} {arXiv:1803.09973
  [hep-ph]} \BibitemShut {NoStop}%
\bibitem [{\citenamefont {Boughezal}\ \emph {et~al.}(2018)\citenamefont
  {Boughezal}, \citenamefont {Petriello},\ and\ \citenamefont
  {Xing}}]{Boughezal:2018azh}%
  \BibitemOpen
  \bibfield  {author} {\bibinfo {author} {\bibfnamefont {R.}~\bibnamefont
  {Boughezal}}, \bibinfo {author} {\bibfnamefont {F.}~\bibnamefont
  {Petriello}}, \ and\ \bibinfo {author} {\bibfnamefont {H.}~\bibnamefont
  {Xing}},\ }\href {\doibase 10.1103/PhysRevD.98.054031} {\bibfield  {journal}
  {\bibinfo  {journal} {Phys. Rev. D}\ }\textbf {\bibinfo {volume} {98}},\
  \bibinfo {pages} {054031} (\bibinfo {year} {2018})},\ \Eprint
  {http://arxiv.org/abs/1806.07311} {arXiv:1806.07311 [hep-ph]} \BibitemShut
  {NoStop}%
\bibitem [{\citenamefont {Borsa}\ \emph
  {et~al.}(2020{\natexlab{b}})\citenamefont {Borsa}, \citenamefont
  {de~Florian},\ and\ \citenamefont {Pedron}}]{Borsa:2020ulb}%
  \BibitemOpen
  \bibfield  {author} {\bibinfo {author} {\bibfnamefont {I.}~\bibnamefont
  {Borsa}}, \bibinfo {author} {\bibfnamefont {D.}~\bibnamefont {de~Florian}}, \
  and\ \bibinfo {author} {\bibfnamefont {I.}~\bibnamefont {Pedron}},\ }\href
  {\doibase 10.1103/PhysRevLett.125.082001} {\bibfield  {journal} {\bibinfo
  {journal} {Phys. Rev. Lett.}\ }\textbf {\bibinfo {volume} {125}},\ \bibinfo
  {pages} {082001} (\bibinfo {year} {2020}{\natexlab{b}})},\ \Eprint
  {http://arxiv.org/abs/2005.10705} {arXiv:2005.10705 [hep-ph]} \BibitemShut
  {NoStop}%
\bibitem [{\citenamefont {Huston}(2005)}]{huston-list}%
  \BibitemOpen
  \bibfield  {author} {\bibinfo {author} {\bibfnamefont {J.}~\bibnamefont
  {Huston}},\ }\href
  {{https://indico.cern.ch/event/334268/attachments/652642/897410/huston\_lpc\_wishlist.pdf}}
  {\emph {\bibinfo {title} {{The new Les Houches high precision wishlist}}}}\
  (\bibinfo {year} {2005})\BibitemShut {NoStop}%
\bibitem [{\citenamefont {Gaunt}\ \emph {et~al.}(2014)\citenamefont {Gaunt},
  \citenamefont {Stahlhofen},\ and\ \citenamefont {Tackmann}}]{Gaunt:2014xga}%
  \BibitemOpen
  \bibfield  {author} {\bibinfo {author} {\bibfnamefont {J.~R.}\ \bibnamefont
  {Gaunt}}, \bibinfo {author} {\bibfnamefont {M.}~\bibnamefont {Stahlhofen}}, \
  and\ \bibinfo {author} {\bibfnamefont {F.~J.}\ \bibnamefont {Tackmann}},\
  }\href {\doibase 10.1007/JHEP04(2014)113} {\bibfield  {journal} {\bibinfo
  {journal} {JHEP}\ }\textbf {\bibinfo {volume} {04}},\ \bibinfo {pages} {113}
  (\bibinfo {year} {2014})},\ \Eprint {http://arxiv.org/abs/1401.5478}
  {arXiv:1401.5478 [hep-ph]} \BibitemShut {NoStop}%
\bibitem [{\citenamefont {Ritzmann}\ and\ \citenamefont
  {Waalewijn}(2014)}]{Ritzmann:2014mka}%
  \BibitemOpen
  \bibfield  {author} {\bibinfo {author} {\bibfnamefont {M.}~\bibnamefont
  {Ritzmann}}\ and\ \bibinfo {author} {\bibfnamefont {W.~J.}\ \bibnamefont
  {Waalewijn}},\ }\href {\doibase 10.1103/PhysRevD.90.054029} {\bibfield
  {journal} {\bibinfo  {journal} {Phys. Rev. D}\ }\textbf {\bibinfo {volume}
  {90}},\ \bibinfo {pages} {054029} (\bibinfo {year} {2014})},\ \Eprint
  {http://arxiv.org/abs/1407.3272} {arXiv:1407.3272 [hep-ph]} \BibitemShut
  {NoStop}%
\bibitem [{\citenamefont {Boughezal}\ \emph {et~al.}(2017)\citenamefont
  {Boughezal}, \citenamefont {Petriello}, \citenamefont {Schubert},\ and\
  \citenamefont {Xing}}]{Boughezal:2017tdd}%
  \BibitemOpen
  \bibfield  {author} {\bibinfo {author} {\bibfnamefont {R.}~\bibnamefont
  {Boughezal}}, \bibinfo {author} {\bibfnamefont {F.}~\bibnamefont
  {Petriello}}, \bibinfo {author} {\bibfnamefont {U.}~\bibnamefont {Schubert}},
  \ and\ \bibinfo {author} {\bibfnamefont {H.}~\bibnamefont {Xing}},\ }\href
  {\doibase 10.1103/PhysRevD.96.034001} {\bibfield  {journal} {\bibinfo
  {journal} {Phys. Rev. D}\ }\textbf {\bibinfo {volume} {96}},\ \bibinfo
  {pages} {034001} (\bibinfo {year} {2017})},\ \Eprint
  {http://arxiv.org/abs/1704.05457} {arXiv:1704.05457 [hep-ph]} \BibitemShut
  {NoStop}%
\bibitem [{\citenamefont {Catani}\ and\ \citenamefont
  {Grazzini}(2007)}]{Catani:2007vq}%
  \BibitemOpen
  \bibfield  {author} {\bibinfo {author} {\bibfnamefont {S.}~\bibnamefont
  {Catani}}\ and\ \bibinfo {author} {\bibfnamefont {M.}~\bibnamefont
  {Grazzini}},\ }\href {\doibase 10.1103/PhysRevLett.98.222002} {\bibfield
  {journal} {\bibinfo  {journal} {Phys. Rev. Lett.}\ }\textbf {\bibinfo
  {volume} {98}},\ \bibinfo {pages} {222002} (\bibinfo {year} {2007})},\
  \Eprint {http://arxiv.org/abs/hep-ph/0703012} {arXiv:hep-ph/0703012}
  \BibitemShut {NoStop}%
\bibitem [{\citenamefont {Billis}\ \emph {et~al.}(2021)\citenamefont {Billis},
  \citenamefont {Dehnadi}, \citenamefont {Ebert}, \citenamefont {Michel},\ and\
  \citenamefont {Tackmann}}]{Billis:2021ecs}%
  \BibitemOpen
  \bibfield  {author} {\bibinfo {author} {\bibfnamefont {G.}~\bibnamefont
  {Billis}}, \bibinfo {author} {\bibfnamefont {B.}~\bibnamefont {Dehnadi}},
  \bibinfo {author} {\bibfnamefont {M.~A.}\ \bibnamefont {Ebert}}, \bibinfo
  {author} {\bibfnamefont {J.~K.~L.}\ \bibnamefont {Michel}}, \ and\ \bibinfo
  {author} {\bibfnamefont {F.~J.}\ \bibnamefont {Tackmann}},\ }\href {\doibase
  10.1103/PhysRevLett.127.072001} {\bibfield  {journal} {\bibinfo  {journal}
  {Phys. Rev. Lett.}\ }\textbf {\bibinfo {volume} {127}},\ \bibinfo {pages}
  {072001} (\bibinfo {year} {2021})},\ \Eprint
  {http://arxiv.org/abs/2102.08039} {arXiv:2102.08039 [hep-ph]} \BibitemShut
  {NoStop}%
\bibitem [{\citenamefont {Luo}\ \emph {et~al.}(2020)\citenamefont {Luo},
  \citenamefont {Yang}, \citenamefont {Zhu},\ and\ \citenamefont
  {Zhu}}]{Luo:2019szz}%
  \BibitemOpen
  \bibfield  {author} {\bibinfo {author} {\bibfnamefont {M.-x.}\ \bibnamefont
  {Luo}}, \bibinfo {author} {\bibfnamefont {T.-Z.}\ \bibnamefont {Yang}},
  \bibinfo {author} {\bibfnamefont {H.~X.}\ \bibnamefont {Zhu}}, \ and\
  \bibinfo {author} {\bibfnamefont {Y.~J.}\ \bibnamefont {Zhu}},\ }\href
  {\doibase 10.1103/PhysRevLett.124.092001} {\bibfield  {journal} {\bibinfo
  {journal} {Phys. Rev. Lett.}\ }\textbf {\bibinfo {volume} {124}},\ \bibinfo
  {pages} {092001} (\bibinfo {year} {2020})},\ \Eprint
  {http://arxiv.org/abs/1912.05778} {arXiv:1912.05778 [hep-ph]} \BibitemShut
  {NoStop}%
\bibitem [{\citenamefont {Ebert}\ \emph
  {et~al.}(2020{\natexlab{b}})\citenamefont {Ebert}, \citenamefont
  {Mistlberger},\ and\ \citenamefont {Vita}}]{Ebert:2020yqt}%
  \BibitemOpen
  \bibfield  {author} {\bibinfo {author} {\bibfnamefont {M.~A.}\ \bibnamefont
  {Ebert}}, \bibinfo {author} {\bibfnamefont {B.}~\bibnamefont {Mistlberger}},
  \ and\ \bibinfo {author} {\bibfnamefont {G.}~\bibnamefont {Vita}},\ }\href
  {\doibase 10.1007/JHEP09(2020)146} {\bibfield  {journal} {\bibinfo  {journal}
  {JHEP}\ }\textbf {\bibinfo {volume} {09}},\ \bibinfo {pages} {146} (\bibinfo
  {year} {2020}{\natexlab{b}})},\ \Eprint {http://arxiv.org/abs/2006.05329}
  {arXiv:2006.05329 [hep-ph]} \BibitemShut {NoStop}%
\bibitem [{\citenamefont {Luo}\ \emph {et~al.}(2021)\citenamefont {Luo},
  \citenamefont {Yang}, \citenamefont {Zhu},\ and\ \citenamefont
  {Zhu}}]{Luo:2020epw}%
  \BibitemOpen
  \bibfield  {author} {\bibinfo {author} {\bibfnamefont {M.-x.}\ \bibnamefont
  {Luo}}, \bibinfo {author} {\bibfnamefont {T.-Z.}\ \bibnamefont {Yang}},
  \bibinfo {author} {\bibfnamefont {H.~X.}\ \bibnamefont {Zhu}}, \ and\
  \bibinfo {author} {\bibfnamefont {Y.~J.}\ \bibnamefont {Zhu}},\ }\href
  {\doibase 10.1007/JHEP06(2021)115} {\bibfield  {journal} {\bibinfo  {journal}
  {JHEP}\ }\textbf {\bibinfo {volume} {06}},\ \bibinfo {pages} {115} (\bibinfo
  {year} {2021})},\ \Eprint {http://arxiv.org/abs/2012.03256} {arXiv:2012.03256
  [hep-ph]} \BibitemShut {NoStop}%
\bibitem [{\citenamefont {Ebert}\ \emph
  {et~al.}(2021{\natexlab{b}})\citenamefont {Ebert}, \citenamefont
  {Mistlberger},\ and\ \citenamefont {Vita}}]{Ebert:2020qef}%
  \BibitemOpen
  \bibfield  {author} {\bibinfo {author} {\bibfnamefont {M.~A.}\ \bibnamefont
  {Ebert}}, \bibinfo {author} {\bibfnamefont {B.}~\bibnamefont {Mistlberger}},
  \ and\ \bibinfo {author} {\bibfnamefont {G.}~\bibnamefont {Vita}},\ }\href
  {\doibase 10.1007/JHEP07(2021)121} {\bibfield  {journal} {\bibinfo  {journal}
  {JHEP}\ }\textbf {\bibinfo {volume} {07}},\ \bibinfo {pages} {121} (\bibinfo
  {year} {2021}{\natexlab{b}})},\ \Eprint {http://arxiv.org/abs/2012.07853}
  {arXiv:2012.07853 [hep-ph]} \BibitemShut {NoStop}%
\bibitem [{\citenamefont {Kang}\ \emph
  {et~al.}(2013{\natexlab{a}})\citenamefont {Kang}, \citenamefont {Liu},
  \citenamefont {Mantry},\ and\ \citenamefont {Qiu}}]{Kang:2013wca}%
  \BibitemOpen
  \bibfield  {author} {\bibinfo {author} {\bibfnamefont {Z.-B.}\ \bibnamefont
  {Kang}}, \bibinfo {author} {\bibfnamefont {X.}~\bibnamefont {Liu}}, \bibinfo
  {author} {\bibfnamefont {S.}~\bibnamefont {Mantry}}, \ and\ \bibinfo {author}
  {\bibfnamefont {J.-W.}\ \bibnamefont {Qiu}},\ }\href {\doibase
  10.1103/PhysRevD.88.074020} {\bibfield  {journal} {\bibinfo  {journal} {Phys.
  Rev. D}\ }\textbf {\bibinfo {volume} {88}},\ \bibinfo {pages} {074020}
  (\bibinfo {year} {2013}{\natexlab{a}})},\ \Eprint
  {http://arxiv.org/abs/1303.3063} {arXiv:1303.3063 [hep-ph]} \BibitemShut
  {NoStop}%
\bibitem [{\citenamefont {Kang}\ \emph
  {et~al.}(2013{\natexlab{b}})\citenamefont {Kang}, \citenamefont {Lee},\ and\
  \citenamefont {Stewart}}]{Kang:2013nda}%
  \BibitemOpen
  \bibfield  {author} {\bibinfo {author} {\bibfnamefont {D.}~\bibnamefont
  {Kang}}, \bibinfo {author} {\bibfnamefont {C.}~\bibnamefont {Lee}}, \ and\
  \bibinfo {author} {\bibfnamefont {I.~W.}\ \bibnamefont {Stewart}},\ }\href
  {\doibase 10.22323/1.191.0158} {\bibfield  {journal} {\bibinfo  {journal}
  {PoS}\ }\textbf {\bibinfo {volume} {DIS2013}},\ \bibinfo {pages} {158}
  (\bibinfo {year} {2013}{\natexlab{b}})},\ \Eprint
  {http://arxiv.org/abs/1308.4473} {arXiv:1308.4473 [hep-ph]} \BibitemShut
  {NoStop}%
\bibitem [{\citenamefont {Kang}\ \emph
  {et~al.}(2014{\natexlab{b}})\citenamefont {Kang}, \citenamefont {Lee},\ and\
  \citenamefont {Stewart}}]{Kang:2014qba}%
  \BibitemOpen
  \bibfield  {author} {\bibinfo {author} {\bibfnamefont {D.}~\bibnamefont
  {Kang}}, \bibinfo {author} {\bibfnamefont {C.}~\bibnamefont {Lee}}, \ and\
  \bibinfo {author} {\bibfnamefont {I.~W.}\ \bibnamefont {Stewart}},\ }\href
  {\doibase 10.1007/JHEP11(2014)132} {\bibfield  {journal} {\bibinfo  {journal}
  {JHEP}\ }\textbf {\bibinfo {volume} {11}},\ \bibinfo {pages} {132} (\bibinfo
  {year} {2014}{\natexlab{b}})},\ \Eprint {http://arxiv.org/abs/1407.6706}
  {arXiv:1407.6706 [hep-ph]} \BibitemShut {NoStop}%
\bibitem [{\citenamefont {Kang}\ \emph {et~al.}(2015)\citenamefont {Kang},
  \citenamefont {Lee},\ and\ \citenamefont {Stewart}}]{Kang:2015swk}%
  \BibitemOpen
  \bibfield  {author} {\bibinfo {author} {\bibfnamefont {D.}~\bibnamefont
  {Kang}}, \bibinfo {author} {\bibfnamefont {C.}~\bibnamefont {Lee}}, \ and\
  \bibinfo {author} {\bibfnamefont {I.~W.}\ \bibnamefont {Stewart}},\ }\href
  {\doibase 10.22323/1.247.0142} {\bibfield  {journal} {\bibinfo  {journal}
  {PoS}\ }\textbf {\bibinfo {volume} {DIS2015}},\ \bibinfo {pages} {142}
  (\bibinfo {year} {2015})}\BibitemShut {NoStop}%
\bibitem [{\citenamefont {Zhu}\ \emph {et~al.}(2021)\citenamefont {Zhu},
  \citenamefont {Kang},\ and\ \citenamefont {Maji}}]{Zhu:2021xjn}%
  \BibitemOpen
  \bibfield  {author} {\bibinfo {author} {\bibfnamefont {J.}~\bibnamefont
  {Zhu}}, \bibinfo {author} {\bibfnamefont {D.}~\bibnamefont {Kang}}, \ and\
  \bibinfo {author} {\bibfnamefont {T.}~\bibnamefont {Maji}},\ }\href {\doibase
  10.1007/JHEP11(2021)026} {\bibfield  {journal} {\bibinfo  {journal} {JHEP}\
  }\textbf {\bibinfo {volume} {11}},\ \bibinfo {pages} {026} (\bibinfo {year}
  {2021})},\ \Eprint {http://arxiv.org/abs/2106.14429} {arXiv:2106.14429
  [hep-ph]} \BibitemShut {NoStop}%
\bibitem [{\citenamefont {Abele}\ \emph {et~al.}(2021)\citenamefont {Abele},
  \citenamefont {de~Florian},\ and\ \citenamefont {Vogelsang}}]{Abele:2021nyo}%
  \BibitemOpen
  \bibfield  {author} {\bibinfo {author} {\bibfnamefont {M.}~\bibnamefont
  {Abele}}, \bibinfo {author} {\bibfnamefont {D.}~\bibnamefont {de~Florian}}, \
  and\ \bibinfo {author} {\bibfnamefont {W.}~\bibnamefont {Vogelsang}},\ }\href
  {\doibase 10.1103/PhysRevD.104.094046} {\bibfield  {journal} {\bibinfo
  {journal} {Phys. Rev. D}\ }\textbf {\bibinfo {volume} {104}},\ \bibinfo
  {pages} {094046} (\bibinfo {year} {2021})},\ \Eprint
  {http://arxiv.org/abs/2109.00847} {arXiv:2109.00847 [hep-ph]} \BibitemShut
  {NoStop}%
\bibitem [{\citenamefont {Abele}\ \emph {et~al.}(2022)\citenamefont {Abele},
  \citenamefont {de~Florian},\ and\ \citenamefont {Vogelsang}}]{Abele:2022wuy}%
  \BibitemOpen
  \bibfield  {author} {\bibinfo {author} {\bibfnamefont {M.}~\bibnamefont
  {Abele}}, \bibinfo {author} {\bibfnamefont {D.}~\bibnamefont {de~Florian}}, \
  and\ \bibinfo {author} {\bibfnamefont {W.}~\bibnamefont {Vogelsang}},\ }\href
  {\doibase 10.1103/PhysRevD.106.014015} {\bibfield  {journal} {\bibinfo
  {journal} {Phys. Rev. D}\ }\textbf {\bibinfo {volume} {106}},\ \bibinfo
  {pages} {014015} (\bibinfo {year} {2022})},\ \Eprint
  {http://arxiv.org/abs/2203.07928} {arXiv:2203.07928 [hep-ph]} \BibitemShut
  {NoStop}%
\bibitem [{\citenamefont {Lustermans}\ \emph {et~al.}(2019)\citenamefont
  {Lustermans}, \citenamefont {Michel},\ and\ \citenamefont
  {Tackmann}}]{Lustermans:2019cau}%
  \BibitemOpen
  \bibfield  {author} {\bibinfo {author} {\bibfnamefont {G.}~\bibnamefont
  {Lustermans}}, \bibinfo {author} {\bibfnamefont {J.~K.~L.}\ \bibnamefont
  {Michel}}, \ and\ \bibinfo {author} {\bibfnamefont {F.~J.}\ \bibnamefont
  {Tackmann}},\ }\href@noop {} {\  (\bibinfo {year} {2019})},\ \Eprint
  {http://arxiv.org/abs/1908.00985} {arXiv:1908.00985 [hep-ph]} \BibitemShut
  {NoStop}%
\bibitem [{\citenamefont {Sterman}\ and\ \citenamefont
  {Vogelsang}(2022)}]{Sterman:2022lki}%
  \BibitemOpen
  \bibfield  {author} {\bibinfo {author} {\bibfnamefont {G.}~\bibnamefont
  {Sterman}}\ and\ \bibinfo {author} {\bibfnamefont {W.}~\bibnamefont
  {Vogelsang}},\ }\href {\doibase 10.1103/PhysRevD.107.014009} {\bibfield
  {journal} {\bibinfo  {journal} {Phys. Rev. D}\ }\textbf {\bibinfo {volume}
  {107}},\ \bibinfo {pages} {014009} (\bibinfo {year} {2022})},\ \Eprint
  {http://arxiv.org/abs/2208.00937} {arXiv:2208.00937 [hep-ph]} \BibitemShut
  {NoStop}%
\bibitem [{\citenamefont {Karki}\ \emph {et~al.}(2023)\citenamefont {Karki},
  \citenamefont {Byer},\ and\ \citenamefont {Gao}}]{Bishnu:2023}%
  \BibitemOpen
  \bibfield  {author} {\bibinfo {author} {\bibfnamefont {B.}~\bibnamefont
  {Karki}}, \bibinfo {author} {\bibfnamefont {D.}~\bibnamefont {Byer}}, \ and\
  \bibinfo {author} {\bibfnamefont {H.}~\bibnamefont {Gao}},\ }\href@noop {}
  {\bibfield  {journal} {\bibinfo  {journal} {Manuscript in preparation}\ }
  (\bibinfo {year} {2023})}\BibitemShut {NoStop}%
\bibitem [{\citenamefont {Akushevich}\ and\ \citenamefont
  {Ilyichev}(2019)}]{Akushevich:2019mbz}%
  \BibitemOpen
  \bibfield  {author} {\bibinfo {author} {\bibfnamefont {I.}~\bibnamefont
  {Akushevich}}\ and\ \bibinfo {author} {\bibfnamefont {A.}~\bibnamefont
  {Ilyichev}},\ }\href {\doibase 10.1103/PhysRevD.100.033005} {\bibfield
  {journal} {\bibinfo  {journal} {Phys. Rev. D}\ }\textbf {\bibinfo {volume}
  {100}},\ \bibinfo {pages} {033005} (\bibinfo {year} {2019})},\ \Eprint
  {http://arxiv.org/abs/1905.09232} {arXiv:1905.09232 [hep-ph]} \BibitemShut
  {NoStop}%
\bibitem [{\citenamefont {Bastami}\ \emph {et~al.}(2019)\citenamefont {Bastami}
  \emph {et~al.}}]{Bastami:2018xqd}%
  \BibitemOpen
  \bibfield  {author} {\bibinfo {author} {\bibfnamefont {S.}~\bibnamefont
  {Bastami}} \emph {et~al.},\ }\href {\doibase 10.1007/JHEP06(2019)007}
  {\bibfield  {journal} {\bibinfo  {journal} {JHEP}\ }\textbf {\bibinfo
  {volume} {06}},\ \bibinfo {pages} {007} (\bibinfo {year} {2019})},\ \Eprint
  {http://arxiv.org/abs/1807.10606} {arXiv:1807.10606 [hep-ph]} \BibitemShut
  {NoStop}%
\bibitem [{\citenamefont {d'Enterria}\ \emph {et~al.}(2022)\citenamefont
  {d'Enterria} \emph {et~al.}}]{dEnterria:2022hzv}%
  \BibitemOpen
  \bibfield  {author} {\bibinfo {author} {\bibfnamefont {D.}~\bibnamefont
  {d'Enterria}} \emph {et~al.},\ }\href@noop {} {\  (\bibinfo {year} {2022})},\
  \Eprint {http://arxiv.org/abs/2203.08271} {arXiv:2203.08271 [hep-ph]}
  \BibitemShut {NoStop}%
\bibitem [{\citenamefont {Alekhin}\ \emph {et~al.}(2017)\citenamefont
  {Alekhin}, \citenamefont {Bl\"umlein}, \citenamefont {Moch},\ and\
  \citenamefont {Placakyte}}]{Alekhin:2017kpj}%
  \BibitemOpen
  \bibfield  {author} {\bibinfo {author} {\bibfnamefont {S.}~\bibnamefont
  {Alekhin}}, \bibinfo {author} {\bibfnamefont {J.}~\bibnamefont {Bl\"umlein}},
  \bibinfo {author} {\bibfnamefont {S.}~\bibnamefont {Moch}}, \ and\ \bibinfo
  {author} {\bibfnamefont {R.}~\bibnamefont {Placakyte}},\ }\href {\doibase
  10.1103/PhysRevD.96.014011} {\bibfield  {journal} {\bibinfo  {journal} {Phys.
  Rev. D}\ }\textbf {\bibinfo {volume} {96}},\ \bibinfo {pages} {014011}
  (\bibinfo {year} {2017})},\ \Eprint {http://arxiv.org/abs/1701.05838}
  {arXiv:1701.05838 [hep-ph]} \BibitemShut {NoStop}%
\bibitem [{\citenamefont {Buckley}\ \emph {et~al.}(2015)\citenamefont
  {Buckley}, \citenamefont {Ferrando}, \citenamefont {Lloyd}, \citenamefont
  {Nordstr\"om}, \citenamefont {Page}, \citenamefont {R\"ufenacht},
  \citenamefont {Sch\"onherr},\ and\ \citenamefont {Watt}}]{Buckley:2014ana}%
  \BibitemOpen
  \bibfield  {author} {\bibinfo {author} {\bibfnamefont {A.}~\bibnamefont
  {Buckley}}, \bibinfo {author} {\bibfnamefont {J.}~\bibnamefont {Ferrando}},
  \bibinfo {author} {\bibfnamefont {S.}~\bibnamefont {Lloyd}}, \bibinfo
  {author} {\bibfnamefont {K.}~\bibnamefont {Nordstr\"om}}, \bibinfo {author}
  {\bibfnamefont {B.}~\bibnamefont {Page}}, \bibinfo {author} {\bibfnamefont
  {M.}~\bibnamefont {R\"ufenacht}}, \bibinfo {author} {\bibfnamefont
  {M.}~\bibnamefont {Sch\"onherr}}, \ and\ \bibinfo {author} {\bibfnamefont
  {G.}~\bibnamefont {Watt}},\ }\href {\doibase 10.1140/epjc/s10052-015-3318-8}
  {\bibfield  {journal} {\bibinfo  {journal} {Eur. Phys. J. C}\ }\textbf
  {\bibinfo {volume} {75}},\ \bibinfo {pages} {132} (\bibinfo {year} {2015})},\
  \Eprint {http://arxiv.org/abs/1412.7420} {arXiv:1412.7420 [hep-ph]}
  \BibitemShut {NoStop}%
\bibitem [{\citenamefont {De~Roeck}(2009)}]{DeRoeck:2009zz}%
  \BibitemOpen
  \bibfield  {author} {\bibinfo {author} {\bibfnamefont {A.}~\bibnamefont
  {De~Roeck}},\ }in\ \href {\doibase 10.3204/DESY-PROC-2009-02/63} {\emph
  {\bibinfo {booktitle} {{HERA and the LHC: 4th Workshop on the Implications of
  HERA for LHC Physics}}}}\ (\bibinfo {year} {2009})\ pp.\ \bibinfo {pages}
  {125--126}\BibitemShut {NoStop}%
\bibitem [{\citenamefont {Alekhin}\ \emph {et~al.}(2015)\citenamefont {Alekhin}
  \emph {et~al.}}]{Alekhin:2014irh}%
  \BibitemOpen
  \bibfield  {author} {\bibinfo {author} {\bibfnamefont {S.}~\bibnamefont
  {Alekhin}} \emph {et~al.},\ }\href {\doibase 10.1140/epjc/s10052-015-3480-z}
  {\bibfield  {journal} {\bibinfo  {journal} {Eur. Phys. J. C}\ }\textbf
  {\bibinfo {volume} {75}},\ \bibinfo {pages} {304} (\bibinfo {year} {2015})},\
  \Eprint {http://arxiv.org/abs/1410.4412} {arXiv:1410.4412 [hep-ph]}
  \BibitemShut {NoStop}%
\bibitem [{\citenamefont {Abdul~Khalek}\ \emph
  {et~al.}(2022{\natexlab{b}})\citenamefont {Abdul~Khalek}, \citenamefont
  {Gauld}, \citenamefont {Giani}, \citenamefont {Nocera}, \citenamefont
  {Rabemananjara},\ and\ \citenamefont {Rojo}}]{AbdulKhalek:2022fyi}%
  \BibitemOpen
  \bibfield  {author} {\bibinfo {author} {\bibfnamefont {R.}~\bibnamefont
  {Abdul~Khalek}}, \bibinfo {author} {\bibfnamefont {R.}~\bibnamefont {Gauld}},
  \bibinfo {author} {\bibfnamefont {T.}~\bibnamefont {Giani}}, \bibinfo
  {author} {\bibfnamefont {E.~R.}\ \bibnamefont {Nocera}}, \bibinfo {author}
  {\bibfnamefont {T.~R.}\ \bibnamefont {Rabemananjara}}, \ and\ \bibinfo
  {author} {\bibfnamefont {J.}~\bibnamefont {Rojo}},\ }\href {\doibase
  10.1140/epjc/s10052-022-10417-7} {\bibfield  {journal} {\bibinfo  {journal}
  {Eur. Phys. J. C}\ }\textbf {\bibinfo {volume} {82}},\ \bibinfo {pages} {507}
  (\bibinfo {year} {2022}{\natexlab{b}})},\ \Eprint
  {http://arxiv.org/abs/2201.12363} {arXiv:2201.12363 [hep-ph]} \BibitemShut
  {NoStop}%
\bibitem [{\citenamefont {Eskola}\ \emph
  {et~al.}(2022{\natexlab{b}})\citenamefont {Eskola}, \citenamefont
  {Paakkinen}, \citenamefont {Paukkunen},\ and\ \citenamefont
  {Salgado}}]{Eskola:2021nhw}%
  \BibitemOpen
  \bibfield  {author} {\bibinfo {author} {\bibfnamefont {K.~J.}\ \bibnamefont
  {Eskola}}, \bibinfo {author} {\bibfnamefont {P.}~\bibnamefont {Paakkinen}},
  \bibinfo {author} {\bibfnamefont {H.}~\bibnamefont {Paukkunen}}, \ and\
  \bibinfo {author} {\bibfnamefont {C.~A.}\ \bibnamefont {Salgado}},\ }\href
  {\doibase 10.1140/epjc/s10052-022-10359-0} {\bibfield  {journal} {\bibinfo
  {journal} {Eur. Phys. J. C}\ }\textbf {\bibinfo {volume} {82}},\ \bibinfo
  {pages} {413} (\bibinfo {year} {2022}{\natexlab{b}})},\ \Eprint
  {http://arxiv.org/abs/2112.12462} {arXiv:2112.12462 [hep-ph]} \BibitemShut
  {NoStop}%
\bibitem [{\citenamefont {Khalek}\ \emph {et~al.}(2021)\citenamefont {Khalek},
  \citenamefont {Ethier}, \citenamefont {Nocera},\ and\ \citenamefont
  {Rojo}}]{Khalek:2021ulf}%
  \BibitemOpen
  \bibfield  {author} {\bibinfo {author} {\bibfnamefont {R.~A.}\ \bibnamefont
  {Khalek}}, \bibinfo {author} {\bibfnamefont {J.~J.}\ \bibnamefont {Ethier}},
  \bibinfo {author} {\bibfnamefont {E.~R.}\ \bibnamefont {Nocera}}, \ and\
  \bibinfo {author} {\bibfnamefont {J.}~\bibnamefont {Rojo}},\ }\href {\doibase
  10.1103/PhysRevD.103.096005} {\bibfield  {journal} {\bibinfo  {journal}
  {Phys. Rev. D}\ }\textbf {\bibinfo {volume} {103}},\ \bibinfo {pages}
  {096005} (\bibinfo {year} {2021})},\ \Eprint
  {http://arxiv.org/abs/2102.00018} {arXiv:2102.00018 [hep-ph]} \BibitemShut
  {NoStop}%
\bibitem [{\citenamefont {Ball}\ \emph {et~al.}(2018)\citenamefont {Ball},
  \citenamefont {Bertone}, \citenamefont {Bonvini}, \citenamefont {Marzani},
  \citenamefont {Rojo},\ and\ \citenamefont {Rottoli}}]{Ball:2017otu}%
  \BibitemOpen
  \bibfield  {author} {\bibinfo {author} {\bibfnamefont {R.~D.}\ \bibnamefont
  {Ball}}, \bibinfo {author} {\bibfnamefont {V.}~\bibnamefont {Bertone}},
  \bibinfo {author} {\bibfnamefont {M.}~\bibnamefont {Bonvini}}, \bibinfo
  {author} {\bibfnamefont {S.}~\bibnamefont {Marzani}}, \bibinfo {author}
  {\bibfnamefont {J.}~\bibnamefont {Rojo}}, \ and\ \bibinfo {author}
  {\bibfnamefont {L.}~\bibnamefont {Rottoli}},\ }\href {\doibase
  10.1140/epjc/s10052-018-5774-4} {\bibfield  {journal} {\bibinfo  {journal}
  {Eur. Phys. J. C}\ }\textbf {\bibinfo {volume} {78}},\ \bibinfo {pages} {321}
  (\bibinfo {year} {2018})},\ \Eprint {http://arxiv.org/abs/1710.05935}
  {arXiv:1710.05935 [hep-ph]} \BibitemShut {NoStop}%
\bibitem [{\citenamefont {Abdolmaleki}\ \emph {et~al.}(2018)\citenamefont
  {Abdolmaleki} \emph {et~al.}}]{xFitterDevelopersTeam:2018hym}%
  \BibitemOpen
  \bibfield  {author} {\bibinfo {author} {\bibfnamefont {H.}~\bibnamefont
  {Abdolmaleki}} \emph {et~al.} (\bibinfo {collaboration} {xFitter Developers'
  Team}),\ }\href {\doibase 10.1140/epjc/s10052-018-6090-8} {\bibfield
  {journal} {\bibinfo  {journal} {Eur. Phys. J. C}\ }\textbf {\bibinfo {volume}
  {78}},\ \bibinfo {pages} {621} (\bibinfo {year} {2018})},\ \Eprint
  {http://arxiv.org/abs/1802.00064} {arXiv:1802.00064 [hep-ph]} \BibitemShut
  {NoStop}%
\bibitem [{\citenamefont {Gao}\ \emph {et~al.}(2018)\citenamefont {Gao},
  \citenamefont {Harland-Lang},\ and\ \citenamefont {Rojo}}]{Gao:2017yyd}%
  \BibitemOpen
  \bibfield  {author} {\bibinfo {author} {\bibfnamefont {J.}~\bibnamefont
  {Gao}}, \bibinfo {author} {\bibfnamefont {L.}~\bibnamefont {Harland-Lang}}, \
  and\ \bibinfo {author} {\bibfnamefont {J.}~\bibnamefont {Rojo}},\ }\href
  {\doibase 10.1016/j.physrep.2018.03.002} {\bibfield  {journal} {\bibinfo
  {journal} {Phys. Rept.}\ }\textbf {\bibinfo {volume} {742}},\ \bibinfo
  {pages} {1} (\bibinfo {year} {2018})},\ \Eprint
  {http://arxiv.org/abs/1709.04922} {arXiv:1709.04922 [hep-ph]} \BibitemShut
  {NoStop}%
\bibitem [{\citenamefont {Hou}\ \emph {et~al.}(2021)\citenamefont {Hou} \emph
  {et~al.}}]{Hou:2019efy}%
  \BibitemOpen
  \bibfield  {author} {\bibinfo {author} {\bibfnamefont {T.-J.}\ \bibnamefont
  {Hou}} \emph {et~al.},\ }\href {\doibase 10.1103/PhysRevD.103.014013}
  {\bibfield  {journal} {\bibinfo  {journal} {Phys. Rev. D}\ }\textbf {\bibinfo
  {volume} {103}},\ \bibinfo {pages} {014013} (\bibinfo {year} {2021})},\
  \Eprint {http://arxiv.org/abs/1912.10053} {arXiv:1912.10053 [hep-ph]}
  \BibitemShut {NoStop}%
\bibitem [{\citenamefont {Bailey}\ \emph {et~al.}(2021)\citenamefont {Bailey},
  \citenamefont {Cridge}, \citenamefont {Harland-Lang}, \citenamefont
  {Martin},\ and\ \citenamefont {Thorne}}]{Bailey:2020ooq}%
  \BibitemOpen
  \bibfield  {author} {\bibinfo {author} {\bibfnamefont {S.}~\bibnamefont
  {Bailey}}, \bibinfo {author} {\bibfnamefont {T.}~\bibnamefont {Cridge}},
  \bibinfo {author} {\bibfnamefont {L.~A.}\ \bibnamefont {Harland-Lang}},
  \bibinfo {author} {\bibfnamefont {A.~D.}\ \bibnamefont {Martin}}, \ and\
  \bibinfo {author} {\bibfnamefont {R.~S.}\ \bibnamefont {Thorne}},\ }\href
  {\doibase 10.1140/epjc/s10052-021-09057-0} {\bibfield  {journal} {\bibinfo
  {journal} {Eur. Phys. J. C}\ }\textbf {\bibinfo {volume} {81}},\ \bibinfo
  {pages} {341} (\bibinfo {year} {2021})},\ \Eprint
  {http://arxiv.org/abs/2012.04684} {arXiv:2012.04684 [hep-ph]} \BibitemShut
  {NoStop}%
\bibitem [{\citenamefont {Ball}\ \emph
  {et~al.}(2022{\natexlab{a}})\citenamefont {Ball} \emph
  {et~al.}}]{NNPDF:2021njg}%
  \BibitemOpen
  \bibfield  {author} {\bibinfo {author} {\bibfnamefont {R.~D.}\ \bibnamefont
  {Ball}} \emph {et~al.} (\bibinfo {collaboration} {NNPDF}),\ }\href {\doibase
  10.1140/epjc/s10052-022-10328-7} {\bibfield  {journal} {\bibinfo  {journal}
  {Eur. Phys. J. C}\ }\textbf {\bibinfo {volume} {82}},\ \bibinfo {pages} {428}
  (\bibinfo {year} {2022}{\natexlab{a}})},\ \Eprint
  {http://arxiv.org/abs/2109.02653} {arXiv:2109.02653 [hep-ph]} \BibitemShut
  {NoStop}%
\bibitem [{\citenamefont {Moffat}\ \emph {et~al.}(2021)\citenamefont {Moffat},
  \citenamefont {Melnitchouk}, \citenamefont {Rogers},\ and\ \citenamefont
  {Sato}}]{Moffat:2021dji}%
  \BibitemOpen
  \bibfield  {author} {\bibinfo {author} {\bibfnamefont {E.}~\bibnamefont
  {Moffat}}, \bibinfo {author} {\bibfnamefont {W.}~\bibnamefont {Melnitchouk}},
  \bibinfo {author} {\bibfnamefont {T.~C.}\ \bibnamefont {Rogers}}, \ and\
  \bibinfo {author} {\bibfnamefont {N.}~\bibnamefont {Sato}} (\bibinfo
  {collaboration} {Jefferson Lab Angular Momentum (JAM)}),\ }\href {\doibase
  10.1103/PhysRevD.104.016015} {\bibfield  {journal} {\bibinfo  {journal}
  {Phys. Rev. D}\ }\textbf {\bibinfo {volume} {104}},\ \bibinfo {pages}
  {016015} (\bibinfo {year} {2021})},\ \Eprint
  {http://arxiv.org/abs/2101.04664} {arXiv:2101.04664 [hep-ph]} \BibitemShut
  {NoStop}%
\bibitem [{\citenamefont {Aad}\ \emph {et~al.}(2022)\citenamefont {Aad} \emph
  {et~al.}}]{ATLAS:2021vod}%
  \BibitemOpen
  \bibfield  {author} {\bibinfo {author} {\bibfnamefont {G.}~\bibnamefont
  {Aad}} \emph {et~al.} (\bibinfo {collaboration} {ATLAS}),\ }\href {\doibase
  10.1140/epjc/s10052-022-10217-z} {\bibfield  {journal} {\bibinfo  {journal}
  {Eur. Phys. J. C}\ }\textbf {\bibinfo {volume} {82}},\ \bibinfo {pages} {438}
  (\bibinfo {year} {2022})},\ \Eprint {http://arxiv.org/abs/2112.11266}
  {arXiv:2112.11266 [hep-ex]} \BibitemShut {NoStop}%
\bibitem [{\citenamefont {Arratia}\ \emph
  {et~al.}(2021{\natexlab{a}})\citenamefont {Arratia}, \citenamefont
  {Furletova}, \citenamefont {Hobbs}, \citenamefont {Olness},\ and\
  \citenamefont {Sekula}}]{Arratia:2020azl}%
  \BibitemOpen
  \bibfield  {author} {\bibinfo {author} {\bibfnamefont {M.}~\bibnamefont
  {Arratia}}, \bibinfo {author} {\bibfnamefont {Y.}~\bibnamefont {Furletova}},
  \bibinfo {author} {\bibfnamefont {T.~J.}\ \bibnamefont {Hobbs}}, \bibinfo
  {author} {\bibfnamefont {F.}~\bibnamefont {Olness}}, \ and\ \bibinfo {author}
  {\bibfnamefont {S.~J.}\ \bibnamefont {Sekula}},\ }\href {\doibase
  10.1103/PhysRevD.103.074023} {\bibfield  {journal} {\bibinfo  {journal}
  {Phys. Rev. D}\ }\textbf {\bibinfo {volume} {103}},\ \bibinfo {pages}
  {074023} (\bibinfo {year} {2021}{\natexlab{a}})},\ \Eprint
  {http://arxiv.org/abs/2006.12520} {arXiv:2006.12520 [hep-ph]} \BibitemShut
  {NoStop}%
\bibitem [{\citenamefont {Ball}\ \emph
  {et~al.}(2022{\natexlab{b}})\citenamefont {Ball} \emph
  {et~al.}}]{PDF4LHCWorkingGroup:2022cjn}%
  \BibitemOpen
  \bibfield  {author} {\bibinfo {author} {\bibfnamefont {R.~D.}\ \bibnamefont
  {Ball}} \emph {et~al.} (\bibinfo {collaboration} {PDF4LHC Working Group}),\
  }\href {\doibase 10.1088/1361-6471/ac7216} {\bibfield  {journal} {\bibinfo
  {journal} {J. Phys. G}\ }\textbf {\bibinfo {volume} {49}},\ \bibinfo {pages}
  {080501} (\bibinfo {year} {2022}{\natexlab{b}})},\ \Eprint
  {http://arxiv.org/abs/2203.05506} {arXiv:2203.05506 [hep-ph]} \BibitemShut
  {NoStop}%
\bibitem [{\citenamefont {Ball}\ \emph {et~al.}(2013)\citenamefont {Ball} \emph
  {et~al.}}]{Ball:2012wy}%
  \BibitemOpen
  \bibfield  {author} {\bibinfo {author} {\bibfnamefont {R.~D.}\ \bibnamefont
  {Ball}} \emph {et~al.},\ }\href {\doibase 10.1007/JHEP04(2013)125} {\bibfield
   {journal} {\bibinfo  {journal} {JHEP}\ }\textbf {\bibinfo {volume} {04}},\
  \bibinfo {pages} {125} (\bibinfo {year} {2013})},\ \Eprint
  {http://arxiv.org/abs/1211.5142} {arXiv:1211.5142 [hep-ph]} \BibitemShut
  {NoStop}%
\bibitem [{\citenamefont {Rojo}\ \emph {et~al.}(2015)\citenamefont {Rojo} \emph
  {et~al.}}]{Rojo:2015acz}%
  \BibitemOpen
  \bibfield  {author} {\bibinfo {author} {\bibfnamefont {J.}~\bibnamefont
  {Rojo}} \emph {et~al.},\ }\href {\doibase 10.1088/0954-3899/42/10/103103}
  {\bibfield  {journal} {\bibinfo  {journal} {J. Phys. G}\ }\textbf {\bibinfo
  {volume} {42}},\ \bibinfo {pages} {103103} (\bibinfo {year} {2015})},\
  \Eprint {http://arxiv.org/abs/1507.00556} {arXiv:1507.00556 [hep-ph]}
  \BibitemShut {NoStop}%
\bibitem [{\citenamefont {Andersen}\ \emph {et~al.}(2016)\citenamefont
  {Andersen} \emph {et~al.}}]{Andersen:2016qtm}%
  \BibitemOpen
  \bibfield  {author} {\bibinfo {author} {\bibfnamefont {J.~R.}\ \bibnamefont
  {Andersen}} \emph {et~al.},\ }in\ \href@noop {} {\emph {\bibinfo {booktitle}
  {{9th Les Houches Workshop on Physics at TeV Colliders}}}}\ (\bibinfo {year}
  {2016})\ \Eprint {http://arxiv.org/abs/1605.04692} {arXiv:1605.04692
  [hep-ph]} \BibitemShut {NoStop}%
\bibitem [{\citenamefont {Andersen}\ \emph {et~al.}(2014)\citenamefont
  {Andersen} \emph {et~al.}}]{Andersen:2014efa}%
  \BibitemOpen
  \bibfield  {author} {\bibinfo {author} {\bibfnamefont {J.~R.}\ \bibnamefont
  {Andersen}} \emph {et~al.},\ }\href@noop {} {\  (\bibinfo {year} {2014})},\
  \Eprint {http://arxiv.org/abs/1405.1067} {arXiv:1405.1067 [hep-ph]}
  \BibitemShut {NoStop}%
\bibitem [{\citenamefont {Butterworth}\ \emph {et~al.}(2016)\citenamefont
  {Butterworth} \emph {et~al.}}]{Butterworth:2015oua}%
  \BibitemOpen
  \bibfield  {author} {\bibinfo {author} {\bibfnamefont {J.}~\bibnamefont
  {Butterworth}} \emph {et~al.},\ }\href {\doibase
  10.1088/0954-3899/43/2/023001} {\bibfield  {journal} {\bibinfo  {journal} {J.
  Phys. G}\ }\textbf {\bibinfo {volume} {43}},\ \bibinfo {pages} {023001}
  (\bibinfo {year} {2016})},\ \Eprint {http://arxiv.org/abs/1510.03865}
  {arXiv:1510.03865 [hep-ph]} \BibitemShut {NoStop}%
\bibitem [{\citenamefont {Courtoy}\ \emph {et~al.}(2023)\citenamefont
  {Courtoy}, \citenamefont {Huston}, \citenamefont {Nadolsky}, \citenamefont
  {Xie}, \citenamefont {Yan},\ and\ \citenamefont {Yuan}}]{Courtoy:2022ocu}%
  \BibitemOpen
  \bibfield  {author} {\bibinfo {author} {\bibfnamefont {A.}~\bibnamefont
  {Courtoy}}, \bibinfo {author} {\bibfnamefont {J.}~\bibnamefont {Huston}},
  \bibinfo {author} {\bibfnamefont {P.}~\bibnamefont {Nadolsky}}, \bibinfo
  {author} {\bibfnamefont {K.}~\bibnamefont {Xie}}, \bibinfo {author}
  {\bibfnamefont {M.}~\bibnamefont {Yan}}, \ and\ \bibinfo {author}
  {\bibfnamefont {C.~P.}\ \bibnamefont {Yuan}},\ }\href {\doibase
  10.1103/PhysRevD.107.034008} {\bibfield  {journal} {\bibinfo  {journal}
  {Phys. Rev. D}\ }\textbf {\bibinfo {volume} {107}},\ \bibinfo {pages}
  {034008} (\bibinfo {year} {2023})},\ \Eprint
  {http://arxiv.org/abs/2205.10444} {arXiv:2205.10444 [hep-ph]} \BibitemShut
  {NoStop}%
\bibitem [{\citenamefont {Ball}\ \emph
  {et~al.}(2022{\natexlab{c}})\citenamefont {Ball}, \citenamefont
  {Cruz-Martinez}, \citenamefont {Del~Debbio}, \citenamefont {Forte},
  \citenamefont {Kassabov}, \citenamefont {Nocera}, \citenamefont {Rojo},
  \citenamefont {Stegeman},\ and\ \citenamefont {Ubiali}}]{Ball:2022uon}%
  \BibitemOpen
  \bibfield  {author} {\bibinfo {author} {\bibfnamefont {R.~D.}\ \bibnamefont
  {Ball}}, \bibinfo {author} {\bibfnamefont {J.}~\bibnamefont {Cruz-Martinez}},
  \bibinfo {author} {\bibfnamefont {L.}~\bibnamefont {Del~Debbio}}, \bibinfo
  {author} {\bibfnamefont {S.}~\bibnamefont {Forte}}, \bibinfo {author}
  {\bibfnamefont {Z.}~\bibnamefont {Kassabov}}, \bibinfo {author}
  {\bibfnamefont {E.~R.}\ \bibnamefont {Nocera}}, \bibinfo {author}
  {\bibfnamefont {J.}~\bibnamefont {Rojo}}, \bibinfo {author} {\bibfnamefont
  {R.}~\bibnamefont {Stegeman}}, \ and\ \bibinfo {author} {\bibfnamefont
  {M.}~\bibnamefont {Ubiali}} (\bibinfo {collaboration} {NNPDF}),\ }\href@noop
  {} {\  (\bibinfo {year} {2022}{\natexlab{c}})},\ \Eprint
  {http://arxiv.org/abs/2211.12961} {arXiv:2211.12961 [hep-ph]} \BibitemShut
  {NoStop}%
\bibitem [{\citenamefont {Hobbs}\ \emph {et~al.}(2019)\citenamefont {Hobbs},
  \citenamefont {Wang}, \citenamefont {Nadolsky},\ and\ \citenamefont
  {Olness}}]{Hobbs:2019gob}%
  \BibitemOpen
  \bibfield  {author} {\bibinfo {author} {\bibfnamefont {T.~J.}\ \bibnamefont
  {Hobbs}}, \bibinfo {author} {\bibfnamefont {B.-T.}\ \bibnamefont {Wang}},
  \bibinfo {author} {\bibfnamefont {P.~M.}\ \bibnamefont {Nadolsky}}, \ and\
  \bibinfo {author} {\bibfnamefont {F.~I.}\ \bibnamefont {Olness}},\ }\href
  {\doibase 10.1103/PhysRevD.100.094040} {\bibfield  {journal} {\bibinfo
  {journal} {Phys. Rev. D}\ }\textbf {\bibinfo {volume} {100}},\ \bibinfo
  {pages} {094040} (\bibinfo {year} {2019})},\ \Eprint
  {http://arxiv.org/abs/1904.00022} {arXiv:1904.00022 [hep-ph]} \BibitemShut
  {NoStop}%
\bibitem [{\citenamefont {Courtoy}\ and\ \citenamefont
  {Nadolsky}(2021)}]{Courtoy:2020fex}%
  \BibitemOpen
  \bibfield  {author} {\bibinfo {author} {\bibfnamefont {A.}~\bibnamefont
  {Courtoy}}\ and\ \bibinfo {author} {\bibfnamefont {P.~M.}\ \bibnamefont
  {Nadolsky}},\ }\href {\doibase 10.1103/PhysRevD.103.054029} {\bibfield
  {journal} {\bibinfo  {journal} {Phys. Rev. D}\ }\textbf {\bibinfo {volume}
  {103}},\ \bibinfo {pages} {054029} (\bibinfo {year} {2021})},\ \Eprint
  {http://arxiv.org/abs/2011.10078} {arXiv:2011.10078 [hep-ph]} \BibitemShut
  {NoStop}%
\bibitem [{\citenamefont {Accardi}\ \emph {et~al.}(2021)\citenamefont
  {Accardi}, \citenamefont {Hobbs}, \citenamefont {Jing},\ and\ \citenamefont
  {Nadolsky}}]{Accardi:2021ysh}%
  \BibitemOpen
  \bibfield  {author} {\bibinfo {author} {\bibfnamefont {A.}~\bibnamefont
  {Accardi}}, \bibinfo {author} {\bibfnamefont {T.~J.}\ \bibnamefont {Hobbs}},
  \bibinfo {author} {\bibfnamefont {X.}~\bibnamefont {Jing}}, \ and\ \bibinfo
  {author} {\bibfnamefont {P.~M.}\ \bibnamefont {Nadolsky}},\ }\href {\doibase
  10.1140/epjc/s10052-021-09318-y} {\bibfield  {journal} {\bibinfo  {journal}
  {Eur. Phys. J. C}\ }\textbf {\bibinfo {volume} {81}},\ \bibinfo {pages} {603}
  (\bibinfo {year} {2021})},\ \Eprint {http://arxiv.org/abs/2102.01107}
  {arXiv:2102.01107 [hep-ph]} \BibitemShut {NoStop}%
\bibitem [{\citenamefont {Wang}\ \emph {et~al.}(2018)\citenamefont {Wang},
  \citenamefont {Hobbs}, \citenamefont {Doyle}, \citenamefont {Gao},
  \citenamefont {Hou}, \citenamefont {Nadolsky},\ and\ \citenamefont
  {Olness}}]{Wang:2018heo}%
  \BibitemOpen
  \bibfield  {author} {\bibinfo {author} {\bibfnamefont {B.-T.}\ \bibnamefont
  {Wang}}, \bibinfo {author} {\bibfnamefont {T.~J.}\ \bibnamefont {Hobbs}},
  \bibinfo {author} {\bibfnamefont {S.}~\bibnamefont {Doyle}}, \bibinfo
  {author} {\bibfnamefont {J.}~\bibnamefont {Gao}}, \bibinfo {author}
  {\bibfnamefont {T.-J.}\ \bibnamefont {Hou}}, \bibinfo {author} {\bibfnamefont
  {P.~M.}\ \bibnamefont {Nadolsky}}, \ and\ \bibinfo {author} {\bibfnamefont
  {F.~I.}\ \bibnamefont {Olness}},\ }\href {\doibase
  10.1103/PhysRevD.98.094030} {\bibfield  {journal} {\bibinfo  {journal} {Phys.
  Rev. D}\ }\textbf {\bibinfo {volume} {98}},\ \bibinfo {pages} {094030}
  (\bibinfo {year} {2018})},\ \Eprint {http://arxiv.org/abs/1803.02777}
  {arXiv:1803.02777 [hep-ph]} \BibitemShut {NoStop}%
\bibitem [{\citenamefont {Abdul~Khalek}\ \emph
  {et~al.}(2022{\natexlab{c}})\citenamefont {Abdul~Khalek} \emph
  {et~al.}}]{AbdulKhalek:2022hcn}%
  \BibitemOpen
  \bibfield  {author} {\bibinfo {author} {\bibfnamefont {R.}~\bibnamefont
  {Abdul~Khalek}} \emph {et~al.},\ }\href@noop {} {\  (\bibinfo {year}
  {2022}{\natexlab{c}})},\ \Eprint {http://arxiv.org/abs/2203.13199}
  {arXiv:2203.13199 [hep-ph]} \BibitemShut {NoStop}%
\bibitem [{\citenamefont {Ball}\ \emph
  {et~al.}(2022{\natexlab{d}})\citenamefont {Ball}, \citenamefont {Candido},
  \citenamefont {Forte}, \citenamefont {Hekhorn}, \citenamefont {Nocera},
  \citenamefont {Rojo},\ and\ \citenamefont {Schwan}}]{Ball:2022qtp}%
  \BibitemOpen
  \bibfield  {author} {\bibinfo {author} {\bibfnamefont {R.~D.}\ \bibnamefont
  {Ball}}, \bibinfo {author} {\bibfnamefont {A.}~\bibnamefont {Candido}},
  \bibinfo {author} {\bibfnamefont {S.}~\bibnamefont {Forte}}, \bibinfo
  {author} {\bibfnamefont {F.}~\bibnamefont {Hekhorn}}, \bibinfo {author}
  {\bibfnamefont {E.~R.}\ \bibnamefont {Nocera}}, \bibinfo {author}
  {\bibfnamefont {J.}~\bibnamefont {Rojo}}, \ and\ \bibinfo {author}
  {\bibfnamefont {C.}~\bibnamefont {Schwan}},\ }\href {\doibase
  10.1140/epjc/s10052-022-11133-y} {\bibfield  {journal} {\bibinfo  {journal}
  {Eur. Phys. J. C}\ }\textbf {\bibinfo {volume} {82}},\ \bibinfo {pages}
  {1160} (\bibinfo {year} {2022}{\natexlab{d}})},\ \Eprint
  {http://arxiv.org/abs/2209.08115} {arXiv:2209.08115 [hep-ph]} \BibitemShut
  {NoStop}%
\bibitem [{\citenamefont {Fiaschi}\ \emph {et~al.}(2022)\citenamefont
  {Fiaschi}, \citenamefont {Giuli}, \citenamefont {Hautmann}, \citenamefont
  {Moch},\ and\ \citenamefont {Moretti}}]{Fiaschi:2022wgl}%
  \BibitemOpen
  \bibfield  {author} {\bibinfo {author} {\bibfnamefont {J.}~\bibnamefont
  {Fiaschi}}, \bibinfo {author} {\bibfnamefont {F.}~\bibnamefont {Giuli}},
  \bibinfo {author} {\bibfnamefont {F.}~\bibnamefont {Hautmann}}, \bibinfo
  {author} {\bibfnamefont {S.}~\bibnamefont {Moch}}, \ and\ \bibinfo {author}
  {\bibfnamefont {S.}~\bibnamefont {Moretti}},\ }\href@noop {} {\  (\bibinfo
  {year} {2022})},\ \Eprint {http://arxiv.org/abs/2211.06188} {arXiv:2211.06188
  [hep-ph]} \BibitemShut {NoStop}%
\bibitem [{\citenamefont {Greljo}\ \emph {et~al.}(2021)\citenamefont {Greljo},
  \citenamefont {Iranipour}, \citenamefont {Kassabov}, \citenamefont {Madigan},
  \citenamefont {Moore}, \citenamefont {Rojo}, \citenamefont {Ubiali},\ and\
  \citenamefont {Voisey}}]{Greljo:2021kvv}%
  \BibitemOpen
  \bibfield  {author} {\bibinfo {author} {\bibfnamefont {A.}~\bibnamefont
  {Greljo}}, \bibinfo {author} {\bibfnamefont {S.}~\bibnamefont {Iranipour}},
  \bibinfo {author} {\bibfnamefont {Z.}~\bibnamefont {Kassabov}}, \bibinfo
  {author} {\bibfnamefont {M.}~\bibnamefont {Madigan}}, \bibinfo {author}
  {\bibfnamefont {J.}~\bibnamefont {Moore}}, \bibinfo {author} {\bibfnamefont
  {J.}~\bibnamefont {Rojo}}, \bibinfo {author} {\bibfnamefont {M.}~\bibnamefont
  {Ubiali}}, \ and\ \bibinfo {author} {\bibfnamefont {C.}~\bibnamefont
  {Voisey}},\ }\href {\doibase 10.1007/JHEP07(2021)122} {\bibfield  {journal}
  {\bibinfo  {journal} {JHEP}\ }\textbf {\bibinfo {volume} {07}},\ \bibinfo
  {pages} {122} (\bibinfo {year} {2021})},\ \Eprint
  {http://arxiv.org/abs/2104.02723} {arXiv:2104.02723 [hep-ph]} \BibitemShut
  {NoStop}%
\bibitem [{\citenamefont {Gao}\ \emph {et~al.}(2022{\natexlab{b}})\citenamefont
  {Gao}, \citenamefont {Gao}, \citenamefont {Hobbs}, \citenamefont {Liu},\ and\
  \citenamefont {Shen}}]{Gao:2022srd}%
  \BibitemOpen
  \bibfield  {author} {\bibinfo {author} {\bibfnamefont {J.}~\bibnamefont
  {Gao}}, \bibinfo {author} {\bibfnamefont {M.}~\bibnamefont {Gao}}, \bibinfo
  {author} {\bibfnamefont {T.~J.}\ \bibnamefont {Hobbs}}, \bibinfo {author}
  {\bibfnamefont {D.}~\bibnamefont {Liu}}, \ and\ \bibinfo {author}
  {\bibfnamefont {X.}~\bibnamefont {Shen}},\ }\href@noop {} {\  (\bibinfo
  {year} {2022}{\natexlab{b}})},\ \Eprint {http://arxiv.org/abs/2211.01094}
  {arXiv:2211.01094 [hep-ph]} \BibitemShut {NoStop}%
\bibitem [{\citenamefont {Campbell}\ \emph {et~al.}(2022)\citenamefont
  {Campbell} \emph {et~al.}}]{Campbell:2022qmc}%
  \BibitemOpen
  \bibfield  {author} {\bibinfo {author} {\bibfnamefont {J.~M.}\ \bibnamefont
  {Campbell}} \emph {et~al.},\ }in\ \href@noop {} {\emph {\bibinfo {booktitle}
  {{2022 Snowmass Summer Study}}}}\ (\bibinfo {year} {2022})\ \Eprint
  {http://arxiv.org/abs/2203.11110} {arXiv:2203.11110 [hep-ph]} \BibitemShut
  {NoStop}%
\bibitem [{\citenamefont {Ali}\ and\ \citenamefont
  {Kramer}(2011)}]{Ali:2010tw}%
  \BibitemOpen
  \bibfield  {author} {\bibinfo {author} {\bibfnamefont {A.}~\bibnamefont
  {Ali}}\ and\ \bibinfo {author} {\bibfnamefont {G.}~\bibnamefont {Kramer}},\
  }\href {\doibase 10.1140/epjh/e2011-10047-1} {\bibfield  {journal} {\bibinfo
  {journal} {Eur. Phys. J. H}\ }\textbf {\bibinfo {volume} {36}},\ \bibinfo
  {pages} {245} (\bibinfo {year} {2011})},\ \Eprint
  {http://arxiv.org/abs/1012.2288} {arXiv:1012.2288 [hep-ph]} \BibitemShut
  {NoStop}%
\bibitem [{\citenamefont {Larkoski}\ \emph {et~al.}(2020)\citenamefont
  {Larkoski}, \citenamefont {Moult},\ and\ \citenamefont
  {Nachman}}]{Larkoski:2017jix}%
  \BibitemOpen
  \bibfield  {author} {\bibinfo {author} {\bibfnamefont {A.~J.}\ \bibnamefont
  {Larkoski}}, \bibinfo {author} {\bibfnamefont {I.}~\bibnamefont {Moult}}, \
  and\ \bibinfo {author} {\bibfnamefont {B.}~\bibnamefont {Nachman}},\ }\href
  {\doibase 10.1016/j.physrep.2019.11.001} {\bibfield  {journal} {\bibinfo
  {journal} {Phys. Rept.}\ }\textbf {\bibinfo {volume} {841}},\ \bibinfo
  {pages} {1} (\bibinfo {year} {2020})},\ \Eprint
  {http://arxiv.org/abs/1709.04464} {arXiv:1709.04464 [hep-ph]} \BibitemShut
  {NoStop}%
\bibitem [{\citenamefont {Kogler}\ \emph {et~al.}(2019)\citenamefont {Kogler}
  \emph {et~al.}}]{Asquith:2018igt}%
  \BibitemOpen
  \bibfield  {author} {\bibinfo {author} {\bibfnamefont {R.}~\bibnamefont
  {Kogler}} \emph {et~al.},\ }\href {\doibase 10.1103/RevModPhys.91.045003}
  {\bibfield  {journal} {\bibinfo  {journal} {Rev. Mod. Phys.}\ }\textbf
  {\bibinfo {volume} {91}},\ \bibinfo {pages} {045003} (\bibinfo {year}
  {2019})},\ \Eprint {http://arxiv.org/abs/1803.06991} {arXiv:1803.06991
  [hep-ex]} \BibitemShut {NoStop}%
\bibitem [{\citenamefont {Marzani}\ \emph {et~al.}(2019)\citenamefont
  {Marzani}, \citenamefont {Soyez},\ and\ \citenamefont
  {Spannowsky}}]{Marzani:2019hun}%
  \BibitemOpen
  \bibfield  {author} {\bibinfo {author} {\bibfnamefont {S.}~\bibnamefont
  {Marzani}}, \bibinfo {author} {\bibfnamefont {G.}~\bibnamefont {Soyez}}, \
  and\ \bibinfo {author} {\bibfnamefont {M.}~\bibnamefont {Spannowsky}},\
  }\href {\doibase 10.1007/978-3-030-15709-8} {\emph {\bibinfo {title}
  {{Looking inside jets: an introduction to jet substructure and boosted-object
  phenomenology}}}},\ Vol.\ \bibinfo {volume} {958}\ (\bibinfo  {publisher}
  {Springer},\ \bibinfo {year} {2019})\ \Eprint
  {http://arxiv.org/abs/1901.10342} {arXiv:1901.10342 [hep-ph]} \BibitemShut
  {NoStop}%
\bibitem [{\citenamefont {Connors}\ \emph {et~al.}(2018)\citenamefont
  {Connors}, \citenamefont {Nattrass}, \citenamefont {Reed},\ and\
  \citenamefont {Salur}}]{Connors:2017ptx}%
  \BibitemOpen
  \bibfield  {author} {\bibinfo {author} {\bibfnamefont {M.}~\bibnamefont
  {Connors}}, \bibinfo {author} {\bibfnamefont {C.}~\bibnamefont {Nattrass}},
  \bibinfo {author} {\bibfnamefont {R.}~\bibnamefont {Reed}}, \ and\ \bibinfo
  {author} {\bibfnamefont {S.}~\bibnamefont {Salur}},\ }\href {\doibase
  10.1103/RevModPhys.90.025005} {\bibfield  {journal} {\bibinfo  {journal}
  {Rev. Mod. Phys.}\ }\textbf {\bibinfo {volume} {90}},\ \bibinfo {pages}
  {025005} (\bibinfo {year} {2018})},\ \Eprint
  {http://arxiv.org/abs/1705.01974} {arXiv:1705.01974 [nucl-ex]} \BibitemShut
  {NoStop}%
\bibitem [{\citenamefont {Busza}\ \emph {et~al.}(2018)\citenamefont {Busza},
  \citenamefont {Rajagopal},\ and\ \citenamefont {van~der
  Schee}}]{Busza:2018rrf}%
  \BibitemOpen
  \bibfield  {author} {\bibinfo {author} {\bibfnamefont {W.}~\bibnamefont
  {Busza}}, \bibinfo {author} {\bibfnamefont {K.}~\bibnamefont {Rajagopal}}, \
  and\ \bibinfo {author} {\bibfnamefont {W.}~\bibnamefont {van~der Schee}},\
  }\href {\doibase 10.1146/annurev-nucl-101917-020852} {\bibfield  {journal}
  {\bibinfo  {journal} {Ann. Rev. Nucl. Part. Sci.}\ }\textbf {\bibinfo
  {volume} {68}},\ \bibinfo {pages} {339} (\bibinfo {year} {2018})},\ \Eprint
  {http://arxiv.org/abs/1802.04801} {arXiv:1802.04801 [hep-ph]} \BibitemShut
  {NoStop}%
\bibitem [{\citenamefont {Cunqueiro}\ and\ \citenamefont
  {Sickles}(2022)}]{Cunqueiro:2021wls}%
  \BibitemOpen
  \bibfield  {author} {\bibinfo {author} {\bibfnamefont {L.}~\bibnamefont
  {Cunqueiro}}\ and\ \bibinfo {author} {\bibfnamefont {A.~M.}\ \bibnamefont
  {Sickles}},\ }\href {\doibase 10.1016/j.ppnp.2022.103940} {\bibfield
  {journal} {\bibinfo  {journal} {Prog. Part. Nucl. Phys.}\ }\textbf {\bibinfo
  {volume} {124}},\ \bibinfo {pages} {103940} (\bibinfo {year} {2022})},\
  \Eprint {http://arxiv.org/abs/2110.14490} {arXiv:2110.14490 [nucl-ex]}
  \BibitemShut {NoStop}%
\bibitem [{\citenamefont {Aschenauer}\ \emph {et~al.}(2019)\citenamefont
  {Aschenauer}, \citenamefont {Fazio}, \citenamefont {Lee}, \citenamefont
  {Mantysaari}, \citenamefont {Page}, \citenamefont {Schenke}, \citenamefont
  {Ullrich}, \citenamefont {Venugopalan},\ and\ \citenamefont
  {Zurita}}]{Aschenauer:2017jsk}%
  \BibitemOpen
  \bibfield  {author} {\bibinfo {author} {\bibfnamefont {E.~C.}\ \bibnamefont
  {Aschenauer}}, \bibinfo {author} {\bibfnamefont {S.}~\bibnamefont {Fazio}},
  \bibinfo {author} {\bibfnamefont {J.~H.}\ \bibnamefont {Lee}}, \bibinfo
  {author} {\bibfnamefont {H.}~\bibnamefont {Mantysaari}}, \bibinfo {author}
  {\bibfnamefont {B.~S.}\ \bibnamefont {Page}}, \bibinfo {author}
  {\bibfnamefont {B.}~\bibnamefont {Schenke}}, \bibinfo {author} {\bibfnamefont
  {T.}~\bibnamefont {Ullrich}}, \bibinfo {author} {\bibfnamefont
  {R.}~\bibnamefont {Venugopalan}}, \ and\ \bibinfo {author} {\bibfnamefont
  {P.}~\bibnamefont {Zurita}},\ }\href {\doibase 10.1088/1361-6633/aaf216}
  {\bibfield  {journal} {\bibinfo  {journal} {Rept. Prog. Phys.}\ }\textbf
  {\bibinfo {volume} {82}},\ \bibinfo {pages} {024301} (\bibinfo {year}
  {2019})},\ \Eprint {http://arxiv.org/abs/1708.01527} {arXiv:1708.01527
  [nucl-ex]} \BibitemShut {NoStop}%
\bibitem [{\citenamefont {Boussarie}\ \emph {et~al.}(2018)\citenamefont
  {Boussarie}, \citenamefont {Hatta}, \citenamefont {Xiao},\ and\ \citenamefont
  {Yuan}}]{Boussarie:2018zwg}%
  \BibitemOpen
  \bibfield  {author} {\bibinfo {author} {\bibfnamefont {R.}~\bibnamefont
  {Boussarie}}, \bibinfo {author} {\bibfnamefont {Y.}~\bibnamefont {Hatta}},
  \bibinfo {author} {\bibfnamefont {B.-W.}\ \bibnamefont {Xiao}}, \ and\
  \bibinfo {author} {\bibfnamefont {F.}~\bibnamefont {Yuan}},\ }\href {\doibase
  10.1103/PhysRevD.98.074015} {\bibfield  {journal} {\bibinfo  {journal} {Phys.
  Rev. D}\ }\textbf {\bibinfo {volume} {98}},\ \bibinfo {pages} {074015}
  (\bibinfo {year} {2018})},\ \Eprint {http://arxiv.org/abs/1807.08697}
  {arXiv:1807.08697 [hep-ph]} \BibitemShut {NoStop}%
\bibitem [{\citenamefont {Liu}\ \emph {et~al.}(2019)\citenamefont {Liu},
  \citenamefont {Ringer}, \citenamefont {Vogelsang},\ and\ \citenamefont
  {Yuan}}]{Liu:2018trl}%
  \BibitemOpen
  \bibfield  {author} {\bibinfo {author} {\bibfnamefont {X.}~\bibnamefont
  {Liu}}, \bibinfo {author} {\bibfnamefont {F.}~\bibnamefont {Ringer}},
  \bibinfo {author} {\bibfnamefont {W.}~\bibnamefont {Vogelsang}}, \ and\
  \bibinfo {author} {\bibfnamefont {F.}~\bibnamefont {Yuan}},\ }\href {\doibase
  10.1103/PhysRevLett.122.192003} {\bibfield  {journal} {\bibinfo  {journal}
  {Phys. Rev. Lett.}\ }\textbf {\bibinfo {volume} {122}},\ \bibinfo {pages}
  {192003} (\bibinfo {year} {2019})},\ \Eprint
  {http://arxiv.org/abs/1812.08077} {arXiv:1812.08077 [hep-ph]} \BibitemShut
  {NoStop}%
\bibitem [{\citenamefont {Liu}\ \emph {et~al.}(2020)\citenamefont {Liu},
  \citenamefont {Ringer}, \citenamefont {Vogelsang},\ and\ \citenamefont
  {Yuan}}]{Liu:2020dct}%
  \BibitemOpen
  \bibfield  {author} {\bibinfo {author} {\bibfnamefont {X.}~\bibnamefont
  {Liu}}, \bibinfo {author} {\bibfnamefont {F.}~\bibnamefont {Ringer}},
  \bibinfo {author} {\bibfnamefont {W.}~\bibnamefont {Vogelsang}}, \ and\
  \bibinfo {author} {\bibfnamefont {F.}~\bibnamefont {Yuan}},\ }\href {\doibase
  10.1103/PhysRevD.102.094022} {\bibfield  {journal} {\bibinfo  {journal}
  {Phys. Rev. D}\ }\textbf {\bibinfo {volume} {102}},\ \bibinfo {pages}
  {094022} (\bibinfo {year} {2020})},\ \Eprint
  {http://arxiv.org/abs/2007.12866} {arXiv:2007.12866 [hep-ph]} \BibitemShut
  {NoStop}%
\bibitem [{\citenamefont {Arratia}\ \emph {et~al.}(2020)\citenamefont
  {Arratia}, \citenamefont {Kang}, \citenamefont {Prokudin},\ and\
  \citenamefont {Ringer}}]{Arratia:2020nxw}%
  \BibitemOpen
  \bibfield  {author} {\bibinfo {author} {\bibfnamefont {M.}~\bibnamefont
  {Arratia}}, \bibinfo {author} {\bibfnamefont {Z.-B.}\ \bibnamefont {Kang}},
  \bibinfo {author} {\bibfnamefont {A.}~\bibnamefont {Prokudin}}, \ and\
  \bibinfo {author} {\bibfnamefont {F.}~\bibnamefont {Ringer}},\ }\href
  {\doibase 10.1103/PhysRevD.102.074015} {\bibfield  {journal} {\bibinfo
  {journal} {Phys. Rev. D}\ }\textbf {\bibinfo {volume} {102}},\ \bibinfo
  {pages} {074015} (\bibinfo {year} {2020})},\ \Eprint
  {http://arxiv.org/abs/2007.07281} {arXiv:2007.07281 [hep-ph]} \BibitemShut
  {NoStop}%
\bibitem [{\citenamefont {Kang}\ \emph
  {et~al.}(2020{\natexlab{a}})\citenamefont {Kang}, \citenamefont {Lee},\ and\
  \citenamefont {Zhao}}]{Kang:2020xyq}%
  \BibitemOpen
  \bibfield  {author} {\bibinfo {author} {\bibfnamefont {Z.-B.}\ \bibnamefont
  {Kang}}, \bibinfo {author} {\bibfnamefont {K.}~\bibnamefont {Lee}}, \ and\
  \bibinfo {author} {\bibfnamefont {F.}~\bibnamefont {Zhao}},\ }\href {\doibase
  10.1016/j.physletb.2020.135756} {\bibfield  {journal} {\bibinfo  {journal}
  {Phys. Lett. B}\ }\textbf {\bibinfo {volume} {809}},\ \bibinfo {pages}
  {135756} (\bibinfo {year} {2020}{\natexlab{a}})},\ \Eprint
  {http://arxiv.org/abs/2005.02398} {arXiv:2005.02398 [hep-ph]} \BibitemShut
  {NoStop}%
\bibitem [{\citenamefont {Kang}\ \emph
  {et~al.}(2021{\natexlab{a}})\citenamefont {Kang}, \citenamefont {Lee},
  \citenamefont {Shao},\ and\ \citenamefont {Zhao}}]{Kang:2021ffh}%
  \BibitemOpen
  \bibfield  {author} {\bibinfo {author} {\bibfnamefont {Z.-B.}\ \bibnamefont
  {Kang}}, \bibinfo {author} {\bibfnamefont {K.}~\bibnamefont {Lee}}, \bibinfo
  {author} {\bibfnamefont {D.~Y.}\ \bibnamefont {Shao}}, \ and\ \bibinfo
  {author} {\bibfnamefont {F.}~\bibnamefont {Zhao}},\ }\href {\doibase
  10.1007/JHEP11(2021)005} {\bibfield  {journal} {\bibinfo  {journal} {JHEP}\
  }\textbf {\bibinfo {volume} {11}},\ \bibinfo {pages} {005} (\bibinfo {year}
  {2021}{\natexlab{a}})},\ \Eprint {http://arxiv.org/abs/2106.15624}
  {arXiv:2106.15624 [hep-ph]} \BibitemShut {NoStop}%
\bibitem [{\citenamefont {Kang}\ \emph
  {et~al.}(2020{\natexlab{b}})\citenamefont {Kang}, \citenamefont {Liu},
  \citenamefont {Mantry},\ and\ \citenamefont {Shao}}]{Kang:2020fka}%
  \BibitemOpen
  \bibfield  {author} {\bibinfo {author} {\bibfnamefont {Z.-B.}\ \bibnamefont
  {Kang}}, \bibinfo {author} {\bibfnamefont {X.}~\bibnamefont {Liu}}, \bibinfo
  {author} {\bibfnamefont {S.}~\bibnamefont {Mantry}}, \ and\ \bibinfo {author}
  {\bibfnamefont {D.~Y.}\ \bibnamefont {Shao}},\ }\href {\doibase
  10.1103/PhysRevLett.125.242003} {\bibfield  {journal} {\bibinfo  {journal}
  {Phys. Rev. Lett.}\ }\textbf {\bibinfo {volume} {125}},\ \bibinfo {pages}
  {242003} (\bibinfo {year} {2020}{\natexlab{b}})},\ \Eprint
  {http://arxiv.org/abs/2008.00655} {arXiv:2008.00655 [hep-ph]} \BibitemShut
  {NoStop}%
\bibitem [{\citenamefont {Lee}\ \emph {et~al.}(2022{\natexlab{a}})\citenamefont
  {Lee}, \citenamefont {Mulligan}, \citenamefont {P\l{}osko\'n}, \citenamefont
  {Ringer},\ and\ \citenamefont {Yuan}}]{Lee:2022kdn}%
  \BibitemOpen
  \bibfield  {author} {\bibinfo {author} {\bibfnamefont {K.}~\bibnamefont
  {Lee}}, \bibinfo {author} {\bibfnamefont {J.}~\bibnamefont {Mulligan}},
  \bibinfo {author} {\bibfnamefont {M.}~\bibnamefont {P\l{}osko\'n}}, \bibinfo
  {author} {\bibfnamefont {F.}~\bibnamefont {Ringer}}, \ and\ \bibinfo {author}
  {\bibfnamefont {F.}~\bibnamefont {Yuan}},\ }\href@noop {} {\  (\bibinfo
  {year} {2022}{\natexlab{a}})},\ \Eprint {http://arxiv.org/abs/2210.06450}
  {arXiv:2210.06450 [hep-ph]} \BibitemShut {NoStop}%
\bibitem [{\citenamefont {Arratia}\ \emph {et~al.}(2022)\citenamefont
  {Arratia}, \citenamefont {Kang}, \citenamefont {Paul}, \citenamefont
  {Prokudin}, \citenamefont {Ringer},\ and\ \citenamefont
  {Zhao}}]{Arratia:2022oxd}%
  \BibitemOpen
  \bibfield  {author} {\bibinfo {author} {\bibfnamefont {M.}~\bibnamefont
  {Arratia}}, \bibinfo {author} {\bibfnamefont {Z.-B.}\ \bibnamefont {Kang}},
  \bibinfo {author} {\bibfnamefont {S.~J.}\ \bibnamefont {Paul}}, \bibinfo
  {author} {\bibfnamefont {A.}~\bibnamefont {Prokudin}}, \bibinfo {author}
  {\bibfnamefont {F.}~\bibnamefont {Ringer}}, \ and\ \bibinfo {author}
  {\bibfnamefont {F.}~\bibnamefont {Zhao}},\ }\href@noop {} {\  (\bibinfo
  {year} {2022})},\ \Eprint {http://arxiv.org/abs/2212.02432} {arXiv:2212.02432
  [hep-ph]} \BibitemShut {NoStop}%
\bibitem [{\citenamefont {Gutierrez-Reyes}\ \emph {et~al.}(2018)\citenamefont
  {Gutierrez-Reyes}, \citenamefont {Scimemi}, \citenamefont {Waalewijn},\ and\
  \citenamefont {Zoppi}}]{Gutierrez-Reyes:2018qez}%
  \BibitemOpen
  \bibfield  {author} {\bibinfo {author} {\bibfnamefont {D.}~\bibnamefont
  {Gutierrez-Reyes}}, \bibinfo {author} {\bibfnamefont {I.}~\bibnamefont
  {Scimemi}}, \bibinfo {author} {\bibfnamefont {W.~J.}\ \bibnamefont
  {Waalewijn}}, \ and\ \bibinfo {author} {\bibfnamefont {L.}~\bibnamefont
  {Zoppi}},\ }\href {\doibase 10.1103/PhysRevLett.121.162001} {\bibfield
  {journal} {\bibinfo  {journal} {Phys. Rev. Lett.}\ }\textbf {\bibinfo
  {volume} {121}},\ \bibinfo {pages} {162001} (\bibinfo {year} {2018})},\
  \Eprint {http://arxiv.org/abs/1807.07573} {arXiv:1807.07573 [hep-ph]}
  \BibitemShut {NoStop}%
\bibitem [{\citenamefont {Gutierrez-Reyes}\ \emph {et~al.}(2019)\citenamefont
  {Gutierrez-Reyes}, \citenamefont {Scimemi}, \citenamefont {Waalewijn},\ and\
  \citenamefont {Zoppi}}]{Gutierrez-Reyes:2019vbx}%
  \BibitemOpen
  \bibfield  {author} {\bibinfo {author} {\bibfnamefont {D.}~\bibnamefont
  {Gutierrez-Reyes}}, \bibinfo {author} {\bibfnamefont {I.}~\bibnamefont
  {Scimemi}}, \bibinfo {author} {\bibfnamefont {W.~J.}\ \bibnamefont
  {Waalewijn}}, \ and\ \bibinfo {author} {\bibfnamefont {L.}~\bibnamefont
  {Zoppi}},\ }\href {\doibase 10.1007/JHEP10(2019)031} {\bibfield  {journal}
  {\bibinfo  {journal} {JHEP}\ }\textbf {\bibinfo {volume} {10}},\ \bibinfo
  {pages} {031} (\bibinfo {year} {2019})},\ \Eprint
  {http://arxiv.org/abs/1904.04259} {arXiv:1904.04259 [hep-ph]} \BibitemShut
  {NoStop}%
\bibitem [{\citenamefont {Liu}\ and\ \citenamefont {Xing}(2021)}]{Liu:2021ewb}%
  \BibitemOpen
  \bibfield  {author} {\bibinfo {author} {\bibfnamefont {X.}~\bibnamefont
  {Liu}}\ and\ \bibinfo {author} {\bibfnamefont {H.}~\bibnamefont {Xing}},\
  }\href@noop {} {\  (\bibinfo {year} {2021})},\ \Eprint
  {http://arxiv.org/abs/2104.03328} {arXiv:2104.03328 [hep-ph]} \BibitemShut
  {NoStop}%
\bibitem [{\citenamefont {Lai}\ \emph {et~al.}(2022{\natexlab{a}})\citenamefont
  {Lai}, \citenamefont {Liu}, \citenamefont {Wang},\ and\ \citenamefont
  {Xing}}]{Lai:2022aly}%
  \BibitemOpen
  \bibfield  {author} {\bibinfo {author} {\bibfnamefont {W.~K.}\ \bibnamefont
  {Lai}}, \bibinfo {author} {\bibfnamefont {X.}~\bibnamefont {Liu}}, \bibinfo
  {author} {\bibfnamefont {M.}~\bibnamefont {Wang}}, \ and\ \bibinfo {author}
  {\bibfnamefont {H.}~\bibnamefont {Xing}},\ }\href@noop {} {\  (\bibinfo
  {year} {2022}{\natexlab{a}})},\ \Eprint {http://arxiv.org/abs/2205.04570}
  {arXiv:2205.04570 [hep-ph]} \BibitemShut {NoStop}%
\bibitem [{\citenamefont {Lai}\ \emph {et~al.}(2022{\natexlab{b}})\citenamefont
  {Lai}, \citenamefont {Liu}, \citenamefont {Wang},\ and\ \citenamefont
  {Xing}}]{Lai:2022xox}%
  \BibitemOpen
  \bibfield  {author} {\bibinfo {author} {\bibfnamefont {W.~K.}\ \bibnamefont
  {Lai}}, \bibinfo {author} {\bibfnamefont {X.}~\bibnamefont {Liu}}, \bibinfo
  {author} {\bibfnamefont {M.}~\bibnamefont {Wang}}, \ and\ \bibinfo {author}
  {\bibfnamefont {H.}~\bibnamefont {Xing}},\ }\href {\doibase
  10.7566/JPSCP.37.020204} {\bibfield  {journal} {\bibinfo  {journal} {JPS
  Conf. Proc.}\ }\textbf {\bibinfo {volume} {37}},\ \bibinfo {pages} {020204}
  (\bibinfo {year} {2022}{\natexlab{b}})}\BibitemShut {NoStop}%
\bibitem [{\citenamefont {Chien}\ \emph {et~al.}(2021)\citenamefont {Chien},
  \citenamefont {Rahn}, \citenamefont {Schrijnder~van Velzen}, \citenamefont
  {Shao}, \citenamefont {Waalewijn},\ and\ \citenamefont {Wu}}]{Chien:2020hzh}%
  \BibitemOpen
  \bibfield  {author} {\bibinfo {author} {\bibfnamefont {Y.-T.}\ \bibnamefont
  {Chien}}, \bibinfo {author} {\bibfnamefont {R.}~\bibnamefont {Rahn}},
  \bibinfo {author} {\bibfnamefont {S.}~\bibnamefont {Schrijnder~van Velzen}},
  \bibinfo {author} {\bibfnamefont {D.~Y.}\ \bibnamefont {Shao}}, \bibinfo
  {author} {\bibfnamefont {W.~J.}\ \bibnamefont {Waalewijn}}, \ and\ \bibinfo
  {author} {\bibfnamefont {B.}~\bibnamefont {Wu}},\ }\href {\doibase
  10.1016/j.physletb.2021.136124} {\bibfield  {journal} {\bibinfo  {journal}
  {Phys. Lett. B}\ }\textbf {\bibinfo {volume} {815}},\ \bibinfo {pages}
  {136124} (\bibinfo {year} {2021})},\ \Eprint
  {http://arxiv.org/abs/2005.12279} {arXiv:2005.12279 [hep-ph]} \BibitemShut
  {NoStop}%
\bibitem [{\citenamefont {Chang}\ \emph {et~al.}(2013)\citenamefont {Chang},
  \citenamefont {Procura}, \citenamefont {Thaler},\ and\ \citenamefont
  {Waalewijn}}]{Chang:2013rca}%
  \BibitemOpen
  \bibfield  {author} {\bibinfo {author} {\bibfnamefont {H.-M.}\ \bibnamefont
  {Chang}}, \bibinfo {author} {\bibfnamefont {M.}~\bibnamefont {Procura}},
  \bibinfo {author} {\bibfnamefont {J.}~\bibnamefont {Thaler}}, \ and\ \bibinfo
  {author} {\bibfnamefont {W.~J.}\ \bibnamefont {Waalewijn}},\ }\href {\doibase
  10.1103/PhysRevLett.111.102002} {\bibfield  {journal} {\bibinfo  {journal}
  {Phys. Rev. Lett.}\ }\textbf {\bibinfo {volume} {111}},\ \bibinfo {pages}
  {102002} (\bibinfo {year} {2013})},\ \Eprint {http://arxiv.org/abs/1303.6637}
  {arXiv:1303.6637 [hep-ph]} \BibitemShut {NoStop}%
\bibitem [{\citenamefont {Page}\ \emph {et~al.}(2020)\citenamefont {Page},
  \citenamefont {Chu},\ and\ \citenamefont {Aschenauer}}]{Page:2019gbf}%
  \BibitemOpen
  \bibfield  {author} {\bibinfo {author} {\bibfnamefont {B.~S.}\ \bibnamefont
  {Page}}, \bibinfo {author} {\bibfnamefont {X.}~\bibnamefont {Chu}}, \ and\
  \bibinfo {author} {\bibfnamefont {E.~C.}\ \bibnamefont {Aschenauer}},\ }\href
  {\doibase 10.1103/PhysRevD.101.072003} {\bibfield  {journal} {\bibinfo
  {journal} {Phys. Rev. D}\ }\textbf {\bibinfo {volume} {101}},\ \bibinfo
  {pages} {072003} (\bibinfo {year} {2020})},\ \Eprint
  {http://arxiv.org/abs/1911.00657} {arXiv:1911.00657 [hep-ph]} \BibitemShut
  {NoStop}%
\bibitem [{\citenamefont {Zhu}\ \emph {et~al.}(2013)\citenamefont {Zhu},
  \citenamefont {Sun},\ and\ \citenamefont {Yuan}}]{Zhu:2013yxa}%
  \BibitemOpen
  \bibfield  {author} {\bibinfo {author} {\bibfnamefont {R.}~\bibnamefont
  {Zhu}}, \bibinfo {author} {\bibfnamefont {P.}~\bibnamefont {Sun}}, \ and\
  \bibinfo {author} {\bibfnamefont {F.}~\bibnamefont {Yuan}},\ }\href {\doibase
  10.1016/j.physletb.2013.11.002} {\bibfield  {journal} {\bibinfo  {journal}
  {Phys. Lett. B}\ }\textbf {\bibinfo {volume} {727}},\ \bibinfo {pages} {474}
  (\bibinfo {year} {2013})},\ \Eprint {http://arxiv.org/abs/1309.0780}
  {arXiv:1309.0780 [hep-ph]} \BibitemShut {NoStop}%
\bibitem [{\citenamefont {del Castillo}\ \emph {et~al.}(2021)\citenamefont {del
  Castillo}, \citenamefont {Echevarria}, \citenamefont {Makris},\ and\
  \citenamefont {Scimemi}}]{delCastillo:2020omr}%
  \BibitemOpen
  \bibfield  {author} {\bibinfo {author} {\bibfnamefont {R.~F.}\ \bibnamefont
  {del Castillo}}, \bibinfo {author} {\bibfnamefont {M.~G.}\ \bibnamefont
  {Echevarria}}, \bibinfo {author} {\bibfnamefont {Y.}~\bibnamefont {Makris}},
  \ and\ \bibinfo {author} {\bibfnamefont {I.}~\bibnamefont {Scimemi}},\ }\href
  {\doibase 10.1007/JHEP01(2021)088} {\bibfield  {journal} {\bibinfo  {journal}
  {JHEP}\ }\textbf {\bibinfo {volume} {01}},\ \bibinfo {pages} {088} (\bibinfo
  {year} {2021})},\ \Eprint {http://arxiv.org/abs/2008.07531} {arXiv:2008.07531
  [hep-ph]} \BibitemShut {NoStop}%
\bibitem [{\citenamefont {del Castillo}\ \emph {et~al.}(2022)\citenamefont {del
  Castillo}, \citenamefont {Echevarria}, \citenamefont {Makris},\ and\
  \citenamefont {Scimemi}}]{delCastillo:2021znl}%
  \BibitemOpen
  \bibfield  {author} {\bibinfo {author} {\bibfnamefont {R.~F.}\ \bibnamefont
  {del Castillo}}, \bibinfo {author} {\bibfnamefont {M.~G.}\ \bibnamefont
  {Echevarria}}, \bibinfo {author} {\bibfnamefont {Y.}~\bibnamefont {Makris}},
  \ and\ \bibinfo {author} {\bibfnamefont {I.}~\bibnamefont {Scimemi}},\ }\href
  {\doibase 10.1007/JHEP03(2022)047} {\bibfield  {journal} {\bibinfo  {journal}
  {JHEP}\ }\textbf {\bibinfo {volume} {03}},\ \bibinfo {pages} {047} (\bibinfo
  {year} {2022})},\ \Eprint {http://arxiv.org/abs/2111.03703} {arXiv:2111.03703
  [hep-ph]} \BibitemShut {NoStop}%
\bibitem [{\citenamefont {Boer}\ \emph {et~al.}(2016)\citenamefont {Boer},
  \citenamefont {Mulders}, \citenamefont {Pisano},\ and\ \citenamefont
  {Zhou}}]{Boer:2016fqd}%
  \BibitemOpen
  \bibfield  {author} {\bibinfo {author} {\bibfnamefont {D.}~\bibnamefont
  {Boer}}, \bibinfo {author} {\bibfnamefont {P.~J.}\ \bibnamefont {Mulders}},
  \bibinfo {author} {\bibfnamefont {C.}~\bibnamefont {Pisano}}, \ and\ \bibinfo
  {author} {\bibfnamefont {J.}~\bibnamefont {Zhou}},\ }\href {\doibase
  10.1007/JHEP08(2016)001} {\bibfield  {journal} {\bibinfo  {journal} {JHEP}\
  }\textbf {\bibinfo {volume} {08}},\ \bibinfo {pages} {001} (\bibinfo {year}
  {2016})},\ \Eprint {http://arxiv.org/abs/1605.07934} {arXiv:1605.07934
  [hep-ph]} \BibitemShut {NoStop}%
\bibitem [{\citenamefont {Kang}\ \emph
  {et~al.}(2021{\natexlab{b}})\citenamefont {Kang}, \citenamefont {Reiten},
  \citenamefont {Shao},\ and\ \citenamefont {Terry}}]{Kang:2020xgk}%
  \BibitemOpen
  \bibfield  {author} {\bibinfo {author} {\bibfnamefont {Z.-B.}\ \bibnamefont
  {Kang}}, \bibinfo {author} {\bibfnamefont {J.}~\bibnamefont {Reiten}},
  \bibinfo {author} {\bibfnamefont {D.~Y.}\ \bibnamefont {Shao}}, \ and\
  \bibinfo {author} {\bibfnamefont {J.}~\bibnamefont {Terry}},\ }\href
  {\doibase 10.1007/JHEP05(2021)286} {\bibfield  {journal} {\bibinfo  {journal}
  {JHEP}\ }\textbf {\bibinfo {volume} {05}},\ \bibinfo {pages} {286} (\bibinfo
  {year} {2021}{\natexlab{b}})},\ \Eprint {http://arxiv.org/abs/2012.01756}
  {arXiv:2012.01756 [hep-ph]} \BibitemShut {NoStop}%
\bibitem [{\citenamefont {Dong}\ \emph {et~al.}(2022)\citenamefont {Dong},
  \citenamefont {Ji}, \citenamefont {Kelsey}, \citenamefont {Radhakrishnan},
  \citenamefont {Sichtermann},\ and\ \citenamefont {Zhao}}]{Dong:2022xbd}%
  \BibitemOpen
  \bibfield  {author} {\bibinfo {author} {\bibfnamefont {X.}~\bibnamefont
  {Dong}}, \bibinfo {author} {\bibfnamefont {Y.}~\bibnamefont {Ji}}, \bibinfo
  {author} {\bibfnamefont {M.}~\bibnamefont {Kelsey}}, \bibinfo {author}
  {\bibfnamefont {S.}~\bibnamefont {Radhakrishnan}}, \bibinfo {author}
  {\bibfnamefont {E.}~\bibnamefont {Sichtermann}}, \ and\ \bibinfo {author}
  {\bibfnamefont {Y.}~\bibnamefont {Zhao}},\ }\href@noop {} {\  (\bibinfo
  {year} {2022})},\ \Eprint {http://arxiv.org/abs/2210.08609} {arXiv:2210.08609
  [hep-ph]} \BibitemShut {NoStop}%
\bibitem [{\citenamefont {Aschenauer}\ \emph {et~al.}(2020)\citenamefont
  {Aschenauer}, \citenamefont {Lee}, \citenamefont {Page},\ and\ \citenamefont
  {Ringer}}]{Aschenauer:2019uex}%
  \BibitemOpen
  \bibfield  {author} {\bibinfo {author} {\bibfnamefont {E.-C.}\ \bibnamefont
  {Aschenauer}}, \bibinfo {author} {\bibfnamefont {K.}~\bibnamefont {Lee}},
  \bibinfo {author} {\bibfnamefont {B.~S.}\ \bibnamefont {Page}}, \ and\
  \bibinfo {author} {\bibfnamefont {F.}~\bibnamefont {Ringer}},\ }\href
  {\doibase 10.1103/PhysRevD.101.054028} {\bibfield  {journal} {\bibinfo
  {journal} {Phys. Rev. D}\ }\textbf {\bibinfo {volume} {101}},\ \bibinfo
  {pages} {054028} (\bibinfo {year} {2020})},\ \Eprint
  {http://arxiv.org/abs/1910.11460} {arXiv:1910.11460 [hep-ph]} \BibitemShut
  {NoStop}%
\bibitem [{\citenamefont {Li}\ \emph {et~al.}(2022{\natexlab{b}})\citenamefont
  {Li}, \citenamefont {Yan},\ and\ \citenamefont {Yuan}}]{Li:2021uww}%
  \BibitemOpen
  \bibfield  {author} {\bibinfo {author} {\bibfnamefont {H.~T.}\ \bibnamefont
  {Li}}, \bibinfo {author} {\bibfnamefont {B.}~\bibnamefont {Yan}}, \ and\
  \bibinfo {author} {\bibfnamefont {C.~P.}\ \bibnamefont {Yuan}},\ }\href
  {\doibase 10.1016/j.physletb.2022.137300} {\bibfield  {journal} {\bibinfo
  {journal} {Phys. Lett. B}\ }\textbf {\bibinfo {volume} {833}},\ \bibinfo
  {pages} {137300} (\bibinfo {year} {2022}{\natexlab{b}})},\ \Eprint
  {http://arxiv.org/abs/2112.07747} {arXiv:2112.07747 [hep-ph]} \BibitemShut
  {NoStop}%
\bibitem [{\citenamefont {Arratia}\ \emph
  {et~al.}(2021{\natexlab{b}})\citenamefont {Arratia}, \citenamefont {Makris},
  \citenamefont {Neill}, \citenamefont {Ringer},\ and\ \citenamefont
  {Sato}}]{Arratia:2020ssx}%
  \BibitemOpen
  \bibfield  {author} {\bibinfo {author} {\bibfnamefont {M.}~\bibnamefont
  {Arratia}}, \bibinfo {author} {\bibfnamefont {Y.}~\bibnamefont {Makris}},
  \bibinfo {author} {\bibfnamefont {D.}~\bibnamefont {Neill}}, \bibinfo
  {author} {\bibfnamefont {F.}~\bibnamefont {Ringer}}, \ and\ \bibinfo {author}
  {\bibfnamefont {N.}~\bibnamefont {Sato}},\ }\href {\doibase
  10.1103/PhysRevD.104.034005} {\bibfield  {journal} {\bibinfo  {journal}
  {Phys. Rev. D}\ }\textbf {\bibinfo {volume} {104}},\ \bibinfo {pages}
  {034005} (\bibinfo {year} {2021}{\natexlab{b}})},\ \Eprint
  {http://arxiv.org/abs/2006.10751} {arXiv:2006.10751 [hep-ph]} \BibitemShut
  {NoStop}%
\bibitem [{\citenamefont {Kang}\ \emph
  {et~al.}(2013{\natexlab{c}})\citenamefont {Kang}, \citenamefont {Lee},\ and\
  \citenamefont {Stewart}}]{Kang:2013nha}%
  \BibitemOpen
  \bibfield  {author} {\bibinfo {author} {\bibfnamefont {D.}~\bibnamefont
  {Kang}}, \bibinfo {author} {\bibfnamefont {C.}~\bibnamefont {Lee}}, \ and\
  \bibinfo {author} {\bibfnamefont {I.~W.}\ \bibnamefont {Stewart}},\ }\href
  {\doibase 10.1103/PhysRevD.88.054004} {\bibfield  {journal} {\bibinfo
  {journal} {Phys. Rev. D}\ }\textbf {\bibinfo {volume} {88}},\ \bibinfo
  {pages} {054004} (\bibinfo {year} {2013}{\natexlab{c}})},\ \Eprint
  {http://arxiv.org/abs/1303.6952} {arXiv:1303.6952 [hep-ph]} \BibitemShut
  {NoStop}%
\bibitem [{\citenamefont {Kang}\ \emph
  {et~al.}(2012{\natexlab{b}})\citenamefont {Kang}, \citenamefont {Mantry},\
  and\ \citenamefont {Qiu}}]{Kang:2012zr}%
  \BibitemOpen
  \bibfield  {author} {\bibinfo {author} {\bibfnamefont {Z.-B.}\ \bibnamefont
  {Kang}}, \bibinfo {author} {\bibfnamefont {S.}~\bibnamefont {Mantry}}, \ and\
  \bibinfo {author} {\bibfnamefont {J.-W.}\ \bibnamefont {Qiu}},\ }\href
  {\doibase 10.1103/PhysRevD.86.114011} {\bibfield  {journal} {\bibinfo
  {journal} {Phys. Rev. D}\ }\textbf {\bibinfo {volume} {86}},\ \bibinfo
  {pages} {114011} (\bibinfo {year} {2012}{\natexlab{b}})},\ \Eprint
  {http://arxiv.org/abs/1204.5469} {arXiv:1204.5469 [hep-ph]} \BibitemShut
  {NoStop}%
\bibitem [{\citenamefont {Kang}\ \emph
  {et~al.}(2014{\natexlab{c}})\citenamefont {Kang}, \citenamefont {Liu},\ and\
  \citenamefont {Mantry}}]{Kang:2013lga}%
  \BibitemOpen
  \bibfield  {author} {\bibinfo {author} {\bibfnamefont {Z.-B.}\ \bibnamefont
  {Kang}}, \bibinfo {author} {\bibfnamefont {X.}~\bibnamefont {Liu}}, \ and\
  \bibinfo {author} {\bibfnamefont {S.}~\bibnamefont {Mantry}},\ }\href
  {\doibase 10.1103/PhysRevD.90.014041} {\bibfield  {journal} {\bibinfo
  {journal} {Phys. Rev. D}\ }\textbf {\bibinfo {volume} {90}},\ \bibinfo
  {pages} {014041} (\bibinfo {year} {2014}{\natexlab{c}})},\ \Eprint
  {http://arxiv.org/abs/1312.0301} {arXiv:1312.0301 [hep-ph]} \BibitemShut
  {NoStop}%
\bibitem [{\citenamefont {Neill}\ \emph {et~al.}(2022)\citenamefont {Neill},
  \citenamefont {Vita}, \citenamefont {Vitev},\ and\ \citenamefont
  {Zhu}}]{Neill:2022lqx}%
  \BibitemOpen
  \bibfield  {author} {\bibinfo {author} {\bibfnamefont {D.}~\bibnamefont
  {Neill}}, \bibinfo {author} {\bibfnamefont {G.}~\bibnamefont {Vita}},
  \bibinfo {author} {\bibfnamefont {I.}~\bibnamefont {Vitev}}, \ and\ \bibinfo
  {author} {\bibfnamefont {H.~X.}\ \bibnamefont {Zhu}},\ }in\ \href@noop {}
  {\emph {\bibinfo {booktitle} {{2022 Snowmass Summer Study}}}}\ (\bibinfo
  {year} {2022})\ \Eprint {http://arxiv.org/abs/2203.07113} {arXiv:2203.07113
  [hep-ph]} \BibitemShut {NoStop}%
\bibitem [{\citenamefont {Ebert}\ \emph
  {et~al.}(2021{\natexlab{c}})\citenamefont {Ebert}, \citenamefont
  {Mistlberger},\ and\ \citenamefont {Vita}}]{Ebert:2020sfi}%
  \BibitemOpen
  \bibfield  {author} {\bibinfo {author} {\bibfnamefont {M.~A.}\ \bibnamefont
  {Ebert}}, \bibinfo {author} {\bibfnamefont {B.}~\bibnamefont {Mistlberger}},
  \ and\ \bibinfo {author} {\bibfnamefont {G.}~\bibnamefont {Vita}},\ }\href
  {\doibase 10.1007/JHEP08(2021)022} {\bibfield  {journal} {\bibinfo  {journal}
  {JHEP}\ }\textbf {\bibinfo {volume} {08}},\ \bibinfo {pages} {022} (\bibinfo
  {year} {2021}{\natexlab{c}})},\ \Eprint {http://arxiv.org/abs/2012.07859}
  {arXiv:2012.07859 [hep-ph]} \BibitemShut {NoStop}%
\bibitem [{\citenamefont {Li}\ \emph {et~al.}(2020{\natexlab{a}})\citenamefont
  {Li}, \citenamefont {Vitev},\ and\ \citenamefont {Zhu}}]{Li:2020bub}%
  \BibitemOpen
  \bibfield  {author} {\bibinfo {author} {\bibfnamefont {H.~T.}\ \bibnamefont
  {Li}}, \bibinfo {author} {\bibfnamefont {I.}~\bibnamefont {Vitev}}, \ and\
  \bibinfo {author} {\bibfnamefont {Y.~J.}\ \bibnamefont {Zhu}},\ }\href
  {\doibase 10.1007/JHEP11(2020)051} {\bibfield  {journal} {\bibinfo  {journal}
  {JHEP}\ }\textbf {\bibinfo {volume} {11}},\ \bibinfo {pages} {051} (\bibinfo
  {year} {2020}{\natexlab{a}})},\ \Eprint {http://arxiv.org/abs/2006.02437}
  {arXiv:2006.02437 [hep-ph]} \BibitemShut {NoStop}%
\bibitem [{\citenamefont {Collins}(2013)}]{Collins:2011zzd}%
  \BibitemOpen
  \bibfield  {author} {\bibinfo {author} {\bibfnamefont {J.}~\bibnamefont
  {Collins}},\ }\href@noop {} {\emph {\bibinfo {title} {{Foundations of
  perturbative QCD}}}},\ Vol.~\bibinfo {volume} {32}\ (\bibinfo  {publisher}
  {Cambridge University Press},\ \bibinfo {year} {2013})\BibitemShut {NoStop}%
\bibitem [{\citenamefont {Li}\ \emph {et~al.}(2021{\natexlab{a}})\citenamefont
  {Li}, \citenamefont {Makris},\ and\ \citenamefont {Vitev}}]{Li:2021txc}%
  \BibitemOpen
  \bibfield  {author} {\bibinfo {author} {\bibfnamefont {H.~T.}\ \bibnamefont
  {Li}}, \bibinfo {author} {\bibfnamefont {Y.}~\bibnamefont {Makris}}, \ and\
  \bibinfo {author} {\bibfnamefont {I.}~\bibnamefont {Vitev}},\ }\href
  {\doibase 10.1103/PhysRevD.103.094005} {\bibfield  {journal} {\bibinfo
  {journal} {Phys. Rev. D}\ }\textbf {\bibinfo {volume} {103}},\ \bibinfo
  {pages} {094005} (\bibinfo {year} {2021}{\natexlab{a}})},\ \Eprint
  {http://arxiv.org/abs/2102.05669} {arXiv:2102.05669 [hep-ph]} \BibitemShut
  {NoStop}%
\bibitem [{\citenamefont {Liu}\ and\ \citenamefont {Zhu}(2023)}]{Liu:2022wop}%
  \BibitemOpen
  \bibfield  {author} {\bibinfo {author} {\bibfnamefont {X.}~\bibnamefont
  {Liu}}\ and\ \bibinfo {author} {\bibfnamefont {H.~X.}\ \bibnamefont {Zhu}},\
  }\href {\doibase 10.1103/PhysRevLett.130.091901} {\bibfield  {journal}
  {\bibinfo  {journal} {Phys. Rev. Lett.}\ }\textbf {\bibinfo {volume} {130}},\
  \bibinfo {pages} {091901} (\bibinfo {year} {2023})},\ \Eprint
  {http://arxiv.org/abs/2209.02080} {arXiv:2209.02080 [hep-ph]} \BibitemShut
  {NoStop}%
\bibitem [{\citenamefont {Li}\ and\ \citenamefont {Vitev}(2021)}]{Li:2020rqj}%
  \BibitemOpen
  \bibfield  {author} {\bibinfo {author} {\bibfnamefont {H.~T.}\ \bibnamefont
  {Li}}\ and\ \bibinfo {author} {\bibfnamefont {I.}~\bibnamefont {Vitev}},\
  }\href {\doibase 10.1103/PhysRevLett.126.252001} {\bibfield  {journal}
  {\bibinfo  {journal} {Phys. Rev. Lett.}\ }\textbf {\bibinfo {volume} {126}},\
  \bibinfo {pages} {252001} (\bibinfo {year} {2021})},\ \Eprint
  {http://arxiv.org/abs/2010.05912} {arXiv:2010.05912 [hep-ph]} \BibitemShut
  {NoStop}%
\bibitem [{\citenamefont {Li}\ \emph {et~al.}(2022{\natexlab{c}})\citenamefont
  {Li}, \citenamefont {Liu},\ and\ \citenamefont {Vitev}}]{Li:2021gjw}%
  \BibitemOpen
  \bibfield  {author} {\bibinfo {author} {\bibfnamefont {H.~T.}\ \bibnamefont
  {Li}}, \bibinfo {author} {\bibfnamefont {Z.~L.}\ \bibnamefont {Liu}}, \ and\
  \bibinfo {author} {\bibfnamefont {I.}~\bibnamefont {Vitev}},\ }\href
  {\doibase 10.1016/j.physletb.2022.137007} {\bibfield  {journal} {\bibinfo
  {journal} {Phys. Lett. B}\ }\textbf {\bibinfo {volume} {827}},\ \bibinfo
  {pages} {137007} (\bibinfo {year} {2022}{\natexlab{c}})},\ \Eprint
  {http://arxiv.org/abs/2108.07809} {arXiv:2108.07809 [hep-ph]} \BibitemShut
  {NoStop}%
\bibitem [{\citenamefont {Tomalak}\ and\ \citenamefont
  {Vitev}(2022)}]{Tomalak:2022kjd}%
  \BibitemOpen
  \bibfield  {author} {\bibinfo {author} {\bibfnamefont {O.}~\bibnamefont
  {Tomalak}}\ and\ \bibinfo {author} {\bibfnamefont {I.}~\bibnamefont
  {Vitev}},\ }\href {\doibase 10.1016/j.physletb.2022.137492} {\bibfield
  {journal} {\bibinfo  {journal} {Phys. Lett. B}\ }\textbf {\bibinfo {volume}
  {835}},\ \bibinfo {pages} {137492} (\bibinfo {year} {2022})},\ \Eprint
  {http://arxiv.org/abs/2206.10637} {arXiv:2206.10637 [nucl-th]} \BibitemShut
  {NoStop}%
\bibitem [{\citenamefont {Gelfand}\ \emph {et~al.}(1994)\citenamefont
  {Gelfand}, \citenamefont {Kapranov},\ and\ \citenamefont
  {Zelevinsky}}]{gelfand1994discriminants}%
  \BibitemOpen
  \bibfield  {author} {\bibinfo {author} {\bibfnamefont {I.~M.}\ \bibnamefont
  {Gelfand}}, \bibinfo {author} {\bibfnamefont {M.~M.}\ \bibnamefont
  {Kapranov}}, \ and\ \bibinfo {author} {\bibfnamefont {A.~V.}\ \bibnamefont
  {Zelevinsky}},\ }\href@noop {} {\emph {\bibinfo {title} {Discriminants,
  Resultants, and Multidimensional Determinants}}}\ (\bibinfo  {publisher}
  {Springer},\ \bibinfo {year} {1994})\ pp.\ \bibinfo {pages}
  {271--296}\BibitemShut {NoStop}%
\bibitem [{\citenamefont {Bump}\ \emph {et~al.}(2012)\citenamefont {Bump},
  \citenamefont {Friedberg},\ and\ \citenamefont
  {Goldfeld}}]{bump2012multiple}%
  \BibitemOpen
  \bibfield  {author} {\bibinfo {author} {\bibfnamefont {D.}~\bibnamefont
  {Bump}}, \bibinfo {author} {\bibfnamefont {S.}~\bibnamefont {Friedberg}}, \
  and\ \bibinfo {author} {\bibfnamefont {D.}~\bibnamefont {Goldfeld}},\
  }\href@noop {} {\emph {\bibinfo {title} {Multiple Dirichlet series,
  L-functions and automorphic forms}}},\ Vol.\ \bibinfo {volume} {300}\
  (\bibinfo  {publisher} {Springer},\ \bibinfo {year} {2012})\BibitemShut
  {NoStop}%
\bibitem [{\citenamefont {Gil}\ and\ \citenamefont
  {Fresan}(2017)}]{gil2017multiple}%
  \BibitemOpen
  \bibfield  {author} {\bibinfo {author} {\bibfnamefont {J.~B.}\ \bibnamefont
  {Gil}}\ and\ \bibinfo {author} {\bibfnamefont {J.}~\bibnamefont {Fresan}},\
  }\href@noop {} {\bibfield  {journal} {\bibinfo  {journal} {Clay Mathematics
  Proceedings, to appear}\ } (\bibinfo {year} {2017})}\BibitemShut {NoStop}%
\bibitem [{\citenamefont {Manin}\ and\ \citenamefont
  {Marcolli}(2007)}]{manin2007modular}%
  \BibitemOpen
  \bibfield  {author} {\bibinfo {author} {\bibfnamefont {Y.}~\bibnamefont
  {Manin}}\ and\ \bibinfo {author} {\bibfnamefont {M.}~\bibnamefont
  {Marcolli}},\ }\href@noop {} {\bibfield  {journal} {\bibinfo  {journal}
  {arXiv preprint math/0703718}\ } (\bibinfo {year} {2007})}\BibitemShut
  {NoStop}%
\bibitem [{\citenamefont {Mitschi}\ \emph {et~al.}(2016)\citenamefont
  {Mitschi}, \citenamefont {Sauzin}, \citenamefont {Loday-Richaud},\ and\
  \citenamefont {Delabaere}}]{mitschi2016divergent}%
  \BibitemOpen
  \bibfield  {author} {\bibinfo {author} {\bibfnamefont {C.}~\bibnamefont
  {Mitschi}}, \bibinfo {author} {\bibfnamefont {D.}~\bibnamefont {Sauzin}},
  \bibinfo {author} {\bibfnamefont {M.}~\bibnamefont {Loday-Richaud}}, \ and\
  \bibinfo {author} {\bibfnamefont {{\'E}.}~\bibnamefont {Delabaere}},\
  }\href@noop {} {\  (\bibinfo {year} {2016})}\BibitemShut {NoStop}%
\bibitem [{\citenamefont {Balser}\ \emph {et~al.}(1979)\citenamefont {Balser},
  \citenamefont {Jurkat},\ and\ \citenamefont {Lutz}}]{balser1979general}%
  \BibitemOpen
  \bibfield  {author} {\bibinfo {author} {\bibfnamefont {W.}~\bibnamefont
  {Balser}}, \bibinfo {author} {\bibfnamefont {W.}~\bibnamefont {Jurkat}}, \
  and\ \bibinfo {author} {\bibfnamefont {D.}~\bibnamefont {Lutz}},\ }\href@noop
  {} {\bibfield  {journal} {\bibinfo  {journal} {Funkcial. Ekvac}\ }\textbf
  {\bibinfo {volume} {22}},\ \bibinfo {pages} {257} (\bibinfo {year}
  {1979})}\BibitemShut {NoStop}%
\bibitem [{\citenamefont {Oliv{\'e}}\ \emph {et~al.}(2003)\citenamefont
  {Oliv{\'e}}, \citenamefont {Sauzin},\ and\ \citenamefont
  {Seara}}]{olive2003resurgence}%
  \BibitemOpen
  \bibfield  {author} {\bibinfo {author} {\bibfnamefont {C.}~\bibnamefont
  {Oliv{\'e}}}, \bibinfo {author} {\bibfnamefont {D.}~\bibnamefont {Sauzin}}, \
  and\ \bibinfo {author} {\bibfnamefont {T.~M.}\ \bibnamefont {Seara}},\ }in\
  \href@noop {} {\emph {\bibinfo {booktitle} {Annales de l'institut
  Fourier}}},\ Vol.~\bibinfo {volume} {53}\ (\bibinfo {year} {2003})\ pp.\
  \bibinfo {pages} {1185--1235}\BibitemShut {NoStop}%
\bibitem [{\citenamefont {Xu}(2019)}]{xu2019closure}%
  \BibitemOpen
  \bibfield  {author} {\bibinfo {author} {\bibfnamefont {X.}~\bibnamefont
  {Xu}},\ }\href@noop {} {\bibfield  {journal} {\bibinfo  {journal} {arXiv
  preprint arXiv:1912.07196}\ } (\bibinfo {year} {2019})}\BibitemShut {NoStop}%
\bibitem [{\citenamefont {Bridgeland}\ and\ \citenamefont
  {Toledano~Laredo}(2012)}]{bridgeland2012stability}%
  \BibitemOpen
  \bibfield  {author} {\bibinfo {author} {\bibfnamefont {T.}~\bibnamefont
  {Bridgeland}}\ and\ \bibinfo {author} {\bibfnamefont {V.}~\bibnamefont
  {Toledano~Laredo}},\ }\href@noop {} {\bibfield  {journal} {\bibinfo
  {journal} {Inventiones mathematicae}\ }\textbf {\bibinfo {volume} {187}},\
  \bibinfo {pages} {61} (\bibinfo {year} {2012})}\BibitemShut {NoStop}%
\bibitem [{\citenamefont {Brodsky}\ \emph {et~al.}(1980)\citenamefont
  {Brodsky}, \citenamefont {Hoyer}, \citenamefont {Peterson},\ and\
  \citenamefont {Sakai}}]{Brodsky:1980pb}%
  \BibitemOpen
  \bibfield  {author} {\bibinfo {author} {\bibfnamefont {S.~J.}\ \bibnamefont
  {Brodsky}}, \bibinfo {author} {\bibfnamefont {P.}~\bibnamefont {Hoyer}},
  \bibinfo {author} {\bibfnamefont {C.}~\bibnamefont {Peterson}}, \ and\
  \bibinfo {author} {\bibfnamefont {N.}~\bibnamefont {Sakai}},\ }\href
  {\doibase 10.1016/0370-2693(80)90364-0} {\bibfield  {journal} {\bibinfo
  {journal} {Phys. Lett. B}\ }\textbf {\bibinfo {volume} {93}},\ \bibinfo
  {pages} {451} (\bibinfo {year} {1980})}\BibitemShut {NoStop}%
\bibitem [{\citenamefont {Brodsky}\ \emph {et~al.}(1981)\citenamefont
  {Brodsky}, \citenamefont {Peterson},\ and\ \citenamefont
  {Sakai}}]{Brodsky:1981se}%
  \BibitemOpen
  \bibfield  {author} {\bibinfo {author} {\bibfnamefont {S.~J.}\ \bibnamefont
  {Brodsky}}, \bibinfo {author} {\bibfnamefont {C.}~\bibnamefont {Peterson}}, \
  and\ \bibinfo {author} {\bibfnamefont {N.}~\bibnamefont {Sakai}},\ }\href
  {\doibase 10.1103/PhysRevD.23.2745} {\bibfield  {journal} {\bibinfo
  {journal} {Phys. Rev. D}\ }\textbf {\bibinfo {volume} {23}},\ \bibinfo
  {pages} {2745} (\bibinfo {year} {1981})}\BibitemShut {NoStop}%
\bibitem [{\citenamefont {Aaij}\ \emph
  {et~al.}(2022{\natexlab{a}})\citenamefont {Aaij} \emph
  {et~al.}}]{LHCb:2021stx}%
  \BibitemOpen
  \bibfield  {author} {\bibinfo {author} {\bibfnamefont {R.}~\bibnamefont
  {Aaij}} \emph {et~al.} (\bibinfo {collaboration} {LHCb}),\ }\href {\doibase
  10.1103/PhysRevLett.128.082001} {\bibfield  {journal} {\bibinfo  {journal}
  {Phys. Rev. Lett.}\ }\textbf {\bibinfo {volume} {128}},\ \bibinfo {pages}
  {082001} (\bibinfo {year} {2022}{\natexlab{a}})},\ \Eprint
  {http://arxiv.org/abs/2109.08084} {arXiv:2109.08084 [hep-ex]} \BibitemShut
  {NoStop}%
\bibitem [{\citenamefont {Ball}\ \emph
  {et~al.}(2022{\natexlab{e}})\citenamefont {Ball}, \citenamefont {Candido},
  \citenamefont {Cruz-Martinez}, \citenamefont {Forte}, \citenamefont {Giani},
  \citenamefont {Hekhorn}, \citenamefont {Kudashkin}, \citenamefont {Magni},\
  and\ \citenamefont {Rojo}}]{Ball:2022qks}%
  \BibitemOpen
  \bibfield  {author} {\bibinfo {author} {\bibfnamefont {R.~D.}\ \bibnamefont
  {Ball}}, \bibinfo {author} {\bibfnamefont {A.}~\bibnamefont {Candido}},
  \bibinfo {author} {\bibfnamefont {J.}~\bibnamefont {Cruz-Martinez}}, \bibinfo
  {author} {\bibfnamefont {S.}~\bibnamefont {Forte}}, \bibinfo {author}
  {\bibfnamefont {T.}~\bibnamefont {Giani}}, \bibinfo {author} {\bibfnamefont
  {F.}~\bibnamefont {Hekhorn}}, \bibinfo {author} {\bibfnamefont
  {K.}~\bibnamefont {Kudashkin}}, \bibinfo {author} {\bibfnamefont
  {G.}~\bibnamefont {Magni}}, \ and\ \bibinfo {author} {\bibfnamefont
  {J.}~\bibnamefont {Rojo}} (\bibinfo {collaboration} {NNPDF}),\ }\href
  {\doibase 10.1038/s41586-022-04998-2} {\bibfield  {journal} {\bibinfo
  {journal} {Nature}\ }\textbf {\bibinfo {volume} {608}},\ \bibinfo {pages}
  {483} (\bibinfo {year} {2022}{\natexlab{e}})},\ \Eprint
  {http://arxiv.org/abs/2208.08372} {arXiv:2208.08372 [hep-ph]} \BibitemShut
  {NoStop}%
\bibitem [{\citenamefont {Hobbs}\ \emph {et~al.}(2014)\citenamefont {Hobbs},
  \citenamefont {Londergan},\ and\ \citenamefont
  {Melnitchouk}}]{Hobbs:2013bia}%
  \BibitemOpen
  \bibfield  {author} {\bibinfo {author} {\bibfnamefont {T.~J.}\ \bibnamefont
  {Hobbs}}, \bibinfo {author} {\bibfnamefont {J.~T.}\ \bibnamefont
  {Londergan}}, \ and\ \bibinfo {author} {\bibfnamefont {W.}~\bibnamefont
  {Melnitchouk}},\ }\href {\doibase 10.1103/PhysRevD.89.074008} {\bibfield
  {journal} {\bibinfo  {journal} {Phys. Rev. D}\ }\textbf {\bibinfo {volume}
  {89}},\ \bibinfo {pages} {074008} (\bibinfo {year} {2014})},\ \Eprint
  {http://arxiv.org/abs/1311.1578} {arXiv:1311.1578 [hep-ph]} \BibitemShut
  {NoStop}%
\bibitem [{\citenamefont {Vogt}(2021{\natexlab{a}})}]{Vogt:2021rcz}%
  \BibitemOpen
  \bibfield  {author} {\bibinfo {author} {\bibfnamefont {R.}~\bibnamefont
  {Vogt}},\ }\href {\doibase 10.22323/1.385.0038} {\bibfield  {journal}
  {\bibinfo  {journal} {PoS}\ }\textbf {\bibinfo {volume} {CHARM2020}},\
  \bibinfo {pages} {038} (\bibinfo {year} {2021}{\natexlab{a}})}\BibitemShut
  {NoStop}%
\bibitem [{\citenamefont {Vogt}(2021{\natexlab{b}})}]{Vogt:2021vsc}%
  \BibitemOpen
  \bibfield  {author} {\bibinfo {author} {\bibfnamefont {R.}~\bibnamefont
  {Vogt}},\ }\href {\doibase 10.1103/PhysRevC.103.035204} {\bibfield  {journal}
  {\bibinfo  {journal} {Phys. Rev. C}\ }\textbf {\bibinfo {volume} {103}},\
  \bibinfo {pages} {035204} (\bibinfo {year} {2021}{\natexlab{b}})},\ \Eprint
  {http://arxiv.org/abs/2101.02858} {arXiv:2101.02858 [hep-ph]} \BibitemShut
  {NoStop}%
\bibitem [{\citenamefont {Vogt}(2022)}]{Vogt:2022glr}%
  \BibitemOpen
  \bibfield  {author} {\bibinfo {author} {\bibfnamefont {R.}~\bibnamefont
  {Vogt}},\ }\href {\doibase 10.1103/PhysRevC.106.025201} {\bibfield  {journal}
  {\bibinfo  {journal} {Phys. Rev. C}\ }\textbf {\bibinfo {volume} {106}},\
  \bibinfo {pages} {025201} (\bibinfo {year} {2022})},\ \Eprint
  {http://arxiv.org/abs/2207.04347} {arXiv:2207.04347 [hep-ph]} \BibitemShut
  {NoStop}%
\bibitem [{\citenamefont {Kneesch}\ \emph {et~al.}(2008)\citenamefont
  {Kneesch}, \citenamefont {Kniehl}, \citenamefont {Kramer},\ and\
  \citenamefont {Schienbein}}]{Kneesch:2007ey}%
  \BibitemOpen
  \bibfield  {author} {\bibinfo {author} {\bibfnamefont {T.}~\bibnamefont
  {Kneesch}}, \bibinfo {author} {\bibfnamefont {B.~A.}\ \bibnamefont {Kniehl}},
  \bibinfo {author} {\bibfnamefont {G.}~\bibnamefont {Kramer}}, \ and\ \bibinfo
  {author} {\bibfnamefont {I.}~\bibnamefont {Schienbein}},\ }\href {\doibase
  10.1016/j.nuclphysb.2008.02.015} {\bibfield  {journal} {\bibinfo  {journal}
  {Nucl. Phys. B}\ }\textbf {\bibinfo {volume} {799}},\ \bibinfo {pages} {34}
  (\bibinfo {year} {2008})},\ \Eprint {http://arxiv.org/abs/0712.0481}
  {arXiv:0712.0481 [hep-ph]} \BibitemShut {NoStop}%
\bibitem [{\citenamefont {Anderle}\ \emph {et~al.}(2017)\citenamefont
  {Anderle}, \citenamefont {Kaufmann}, \citenamefont {Stratmann}, \citenamefont
  {Ringer},\ and\ \citenamefont {Vitev}}]{Anderle:2017cgl}%
  \BibitemOpen
  \bibfield  {author} {\bibinfo {author} {\bibfnamefont {D.~P.}\ \bibnamefont
  {Anderle}}, \bibinfo {author} {\bibfnamefont {T.}~\bibnamefont {Kaufmann}},
  \bibinfo {author} {\bibfnamefont {M.}~\bibnamefont {Stratmann}}, \bibinfo
  {author} {\bibfnamefont {F.}~\bibnamefont {Ringer}}, \ and\ \bibinfo {author}
  {\bibfnamefont {I.}~\bibnamefont {Vitev}},\ }\href {\doibase
  10.1103/PhysRevD.96.034028} {\bibfield  {journal} {\bibinfo  {journal} {Phys.
  Rev. D}\ }\textbf {\bibinfo {volume} {96}},\ \bibinfo {pages} {034028}
  (\bibinfo {year} {2017})},\ \Eprint {http://arxiv.org/abs/1706.09857}
  {arXiv:1706.09857 [hep-ph]} \BibitemShut {NoStop}%
\bibitem [{\citenamefont {Kang}\ \emph {et~al.}(2016)\citenamefont {Kang},
  \citenamefont {Ringer},\ and\ \citenamefont {Vitev}}]{Kang:2016ehg}%
  \BibitemOpen
  \bibfield  {author} {\bibinfo {author} {\bibfnamefont {Z.-B.}\ \bibnamefont
  {Kang}}, \bibinfo {author} {\bibfnamefont {F.}~\bibnamefont {Ringer}}, \ and\
  \bibinfo {author} {\bibfnamefont {I.}~\bibnamefont {Vitev}},\ }\href
  {\doibase 10.1007/JHEP11(2016)155} {\bibfield  {journal} {\bibinfo  {journal}
  {JHEP}\ }\textbf {\bibinfo {volume} {11}},\ \bibinfo {pages} {155} (\bibinfo
  {year} {2016})},\ \Eprint {http://arxiv.org/abs/1606.07063} {arXiv:1606.07063
  [hep-ph]} \BibitemShut {NoStop}%
\bibitem [{\citenamefont {Kang}\ \emph {et~al.}(2017)\citenamefont {Kang},
  \citenamefont {Ringer},\ and\ \citenamefont {Vitev}}]{Kang:2016ofv}%
  \BibitemOpen
  \bibfield  {author} {\bibinfo {author} {\bibfnamefont {Z.-B.}\ \bibnamefont
  {Kang}}, \bibinfo {author} {\bibfnamefont {F.}~\bibnamefont {Ringer}}, \ and\
  \bibinfo {author} {\bibfnamefont {I.}~\bibnamefont {Vitev}},\ }\href
  {\doibase 10.1007/JHEP03(2017)146} {\bibfield  {journal} {\bibinfo  {journal}
  {JHEP}\ }\textbf {\bibinfo {volume} {03}},\ \bibinfo {pages} {146} (\bibinfo
  {year} {2017})},\ \Eprint {http://arxiv.org/abs/1610.02043} {arXiv:1610.02043
  [hep-ph]} \BibitemShut {NoStop}%
\bibitem [{\citenamefont {Sievert}\ \emph {et~al.}(2019)\citenamefont
  {Sievert}, \citenamefont {Vitev},\ and\ \citenamefont
  {Yoon}}]{Sievert:2019cwq}%
  \BibitemOpen
  \bibfield  {author} {\bibinfo {author} {\bibfnamefont {M.~D.}\ \bibnamefont
  {Sievert}}, \bibinfo {author} {\bibfnamefont {I.}~\bibnamefont {Vitev}}, \
  and\ \bibinfo {author} {\bibfnamefont {B.}~\bibnamefont {Yoon}},\ }\href
  {\doibase 10.1016/j.physletb.2019.06.019} {\bibfield  {journal} {\bibinfo
  {journal} {Phys. Lett. B}\ }\textbf {\bibinfo {volume} {795}},\ \bibinfo
  {pages} {502} (\bibinfo {year} {2019})},\ \Eprint
  {http://arxiv.org/abs/1903.06170} {arXiv:1903.06170 [hep-ph]} \BibitemShut
  {NoStop}%
\bibitem [{\citenamefont {Li}\ and\ \citenamefont {Vitev}(2019)}]{Li:2018xuv}%
  \BibitemOpen
  \bibfield  {author} {\bibinfo {author} {\bibfnamefont {H.~T.}\ \bibnamefont
  {Li}}\ and\ \bibinfo {author} {\bibfnamefont {I.}~\bibnamefont {Vitev}},\
  }\href {\doibase 10.1007/JHEP07(2019)148} {\bibfield  {journal} {\bibinfo
  {journal} {JHEP}\ }\textbf {\bibinfo {volume} {07}},\ \bibinfo {pages} {148}
  (\bibinfo {year} {2019})},\ \Eprint {http://arxiv.org/abs/1811.07905}
  {arXiv:1811.07905 [hep-ph]} \BibitemShut {NoStop}%
\bibitem [{\citenamefont {Dokshitzer}\ and\ \citenamefont
  {Kharzeev}(2001)}]{Dokshitzer:2001zm}%
  \BibitemOpen
  \bibfield  {author} {\bibinfo {author} {\bibfnamefont {Y.~L.}\ \bibnamefont
  {Dokshitzer}}\ and\ \bibinfo {author} {\bibfnamefont {D.~E.}\ \bibnamefont
  {Kharzeev}},\ }\href {\doibase 10.1016/S0370-2693(01)01130-3} {\bibfield
  {journal} {\bibinfo  {journal} {Phys. Lett. B}\ }\textbf {\bibinfo {volume}
  {519}},\ \bibinfo {pages} {199} (\bibinfo {year} {2001})},\ \Eprint
  {http://arxiv.org/abs/hep-ph/0106202} {arXiv:hep-ph/0106202} \BibitemShut
  {NoStop}%
\bibitem [{\citenamefont {Larkoski}\ \emph {et~al.}(2015)\citenamefont
  {Larkoski}, \citenamefont {Marzani},\ and\ \citenamefont
  {Thaler}}]{Larkoski:2015lea}%
  \BibitemOpen
  \bibfield  {author} {\bibinfo {author} {\bibfnamefont {A.~J.}\ \bibnamefont
  {Larkoski}}, \bibinfo {author} {\bibfnamefont {S.}~\bibnamefont {Marzani}}, \
  and\ \bibinfo {author} {\bibfnamefont {J.}~\bibnamefont {Thaler}},\ }\href
  {\doibase 10.1103/PhysRevD.91.111501} {\bibfield  {journal} {\bibinfo
  {journal} {Phys. Rev. D}\ }\textbf {\bibinfo {volume} {91}},\ \bibinfo
  {pages} {111501} (\bibinfo {year} {2015})},\ \Eprint
  {http://arxiv.org/abs/1502.01719} {arXiv:1502.01719 [hep-ph]} \BibitemShut
  {NoStop}%
\bibitem [{\citenamefont {Putschke}\ \emph {et~al.}(2019)\citenamefont
  {Putschke} \emph {et~al.}}]{Putschke:2019yrg}%
  \BibitemOpen
  \bibfield  {author} {\bibinfo {author} {\bibfnamefont {J.~H.}\ \bibnamefont
  {Putschke}} \emph {et~al.},\ }\href@noop {} {\  (\bibinfo {year} {2019})},\
  \Eprint {http://arxiv.org/abs/1903.07706} {arXiv:1903.07706 [nucl-th]}
  \BibitemShut {NoStop}%
\bibitem [{\citenamefont {Sadofyev}\ \emph {et~al.}(2021)\citenamefont
  {Sadofyev}, \citenamefont {Sievert},\ and\ \citenamefont
  {Vitev}}]{Sadofyev:2021ohn}%
  \BibitemOpen
  \bibfield  {author} {\bibinfo {author} {\bibfnamefont {A.~V.}\ \bibnamefont
  {Sadofyev}}, \bibinfo {author} {\bibfnamefont {M.~D.}\ \bibnamefont
  {Sievert}}, \ and\ \bibinfo {author} {\bibfnamefont {I.}~\bibnamefont
  {Vitev}},\ }\href {\doibase 10.1103/PhysRevD.104.094044} {\bibfield
  {journal} {\bibinfo  {journal} {Phys. Rev. D}\ }\textbf {\bibinfo {volume}
  {104}},\ \bibinfo {pages} {094044} (\bibinfo {year} {2021})},\ \Eprint
  {http://arxiv.org/abs/2104.09513} {arXiv:2104.09513 [hep-ph]} \BibitemShut
  {NoStop}%
\bibitem [{\citenamefont {Lansberg}(2020)}]{Lansberg:2019adr}%
  \BibitemOpen
  \bibfield  {author} {\bibinfo {author} {\bibfnamefont {J.-P.}\ \bibnamefont
  {Lansberg}},\ }\href {\doibase 10.1016/j.physrep.2020.08.007} {\bibfield
  {journal} {\bibinfo  {journal} {Phys. Rept.}\ }\textbf {\bibinfo {volume}
  {889}},\ \bibinfo {pages} {1} (\bibinfo {year} {2020})},\ \Eprint
  {http://arxiv.org/abs/1903.09185} {arXiv:1903.09185 [hep-ph]} \BibitemShut
  {NoStop}%
\bibitem [{\citenamefont {Brambilla}\ \emph {et~al.}(2011)\citenamefont
  {Brambilla} \emph {et~al.}}]{Brambilla:2010cs}%
  \BibitemOpen
  \bibfield  {author} {\bibinfo {author} {\bibfnamefont {N.}~\bibnamefont
  {Brambilla}} \emph {et~al.},\ }\href {\doibase
  10.1140/epjc/s10052-010-1534-9} {\bibfield  {journal} {\bibinfo  {journal}
  {Eur. Phys. J. C}\ }\textbf {\bibinfo {volume} {71}},\ \bibinfo {pages}
  {1534} (\bibinfo {year} {2011})},\ \Eprint {http://arxiv.org/abs/1010.5827}
  {arXiv:1010.5827 [hep-ph]} \BibitemShut {NoStop}%
\bibitem [{\citenamefont {Brambilla}\ \emph {et~al.}(2014)\citenamefont
  {Brambilla} \emph {et~al.}}]{Brambilla:2014jmp}%
  \BibitemOpen
  \bibfield  {author} {\bibinfo {author} {\bibfnamefont {N.}~\bibnamefont
  {Brambilla}} \emph {et~al.},\ }\href {\doibase
  10.1140/epjc/s10052-014-2981-5} {\bibfield  {journal} {\bibinfo  {journal}
  {Eur. Phys. J. C}\ }\textbf {\bibinfo {volume} {74}},\ \bibinfo {pages}
  {2981} (\bibinfo {year} {2014})},\ \Eprint {http://arxiv.org/abs/1404.3723}
  {arXiv:1404.3723 [hep-ph]} \BibitemShut {NoStop}%
\bibitem [{\citenamefont {Brambilla}\ \emph
  {et~al.}(2022{\natexlab{a}})\citenamefont {Brambilla}, \citenamefont {Chung},
  \citenamefont {Vairo},\ and\ \citenamefont {Wang}}]{Brambilla:2022rjd}%
  \BibitemOpen
  \bibfield  {author} {\bibinfo {author} {\bibfnamefont {N.}~\bibnamefont
  {Brambilla}}, \bibinfo {author} {\bibfnamefont {H.~S.}\ \bibnamefont
  {Chung}}, \bibinfo {author} {\bibfnamefont {A.}~\bibnamefont {Vairo}}, \ and\
  \bibinfo {author} {\bibfnamefont {X.-P.}\ \bibnamefont {Wang}},\ }\href
  {\doibase 10.1103/PhysRevD.105.L111503} {\bibfield  {journal} {\bibinfo
  {journal} {Phys. Rev. D}\ }\textbf {\bibinfo {volume} {105}},\ \bibinfo
  {pages} {L111503} (\bibinfo {year} {2022}{\natexlab{a}})},\ \Eprint
  {http://arxiv.org/abs/2203.07778} {arXiv:2203.07778 [hep-ph]} \BibitemShut
  {NoStop}%
\bibitem [{\citenamefont {Brambilla}\ \emph
  {et~al.}(2022{\natexlab{b}})\citenamefont {Brambilla}, \citenamefont {Chung},
  \citenamefont {Vairo},\ and\ \citenamefont {Wang}}]{Brambilla:2022ayc}%
  \BibitemOpen
  \bibfield  {author} {\bibinfo {author} {\bibfnamefont {N.}~\bibnamefont
  {Brambilla}}, \bibinfo {author} {\bibfnamefont {H.~S.}\ \bibnamefont
  {Chung}}, \bibinfo {author} {\bibfnamefont {A.}~\bibnamefont {Vairo}}, \ and\
  \bibinfo {author} {\bibfnamefont {X.-P.}\ \bibnamefont {Wang}},\ }\href@noop
  {} {\  (\bibinfo {year} {2022}{\natexlab{b}})},\ \Eprint
  {http://arxiv.org/abs/2210.17345} {arXiv:2210.17345 [hep-ph]} \BibitemShut
  {NoStop}%
\bibitem [{\citenamefont {Qiu}\ \emph {et~al.}(2021)\citenamefont {Qiu},
  \citenamefont {Wang},\ and\ \citenamefont {Xing}}]{Qiu:2020xum}%
  \BibitemOpen
  \bibfield  {author} {\bibinfo {author} {\bibfnamefont {J.-W.}\ \bibnamefont
  {Qiu}}, \bibinfo {author} {\bibfnamefont {X.-P.}\ \bibnamefont {Wang}}, \
  and\ \bibinfo {author} {\bibfnamefont {H.}~\bibnamefont {Xing}},\ }\href
  {\doibase 10.1088/0256-307X/38/4/041201} {\bibfield  {journal} {\bibinfo
  {journal} {Chin. Phys. Lett.}\ }\textbf {\bibinfo {volume} {38}},\ \bibinfo
  {pages} {041201} (\bibinfo {year} {2021})},\ \Eprint
  {http://arxiv.org/abs/2005.10832} {arXiv:2005.10832 [hep-ph]} \BibitemShut
  {NoStop}%
\bibitem [{\citenamefont {Bacchetta}\ \emph
  {et~al.}(2020{\natexlab{b}})\citenamefont {Bacchetta}, \citenamefont {Boer},
  \citenamefont {Pisano},\ and\ \citenamefont {Taels}}]{Bacchetta:2018ivt}%
  \BibitemOpen
  \bibfield  {author} {\bibinfo {author} {\bibfnamefont {A.}~\bibnamefont
  {Bacchetta}}, \bibinfo {author} {\bibfnamefont {D.}~\bibnamefont {Boer}},
  \bibinfo {author} {\bibfnamefont {C.}~\bibnamefont {Pisano}}, \ and\ \bibinfo
  {author} {\bibfnamefont {P.}~\bibnamefont {Taels}},\ }\href {\doibase
  10.1140/epjc/s10052-020-7620-8} {\bibfield  {journal} {\bibinfo  {journal}
  {Eur. Phys. J. C}\ }\textbf {\bibinfo {volume} {80}},\ \bibinfo {pages} {72}
  (\bibinfo {year} {2020}{\natexlab{b}})},\ \Eprint
  {http://arxiv.org/abs/1809.02056} {arXiv:1809.02056 [hep-ph]} \BibitemShut
  {NoStop}%
\bibitem [{\citenamefont {Echevarria}\ \emph {et~al.}(2020)\citenamefont
  {Echevarria}, \citenamefont {Makris},\ and\ \citenamefont
  {Scimemi}}]{Echevarria:2020qjk}%
  \BibitemOpen
  \bibfield  {author} {\bibinfo {author} {\bibfnamefont {M.~G.}\ \bibnamefont
  {Echevarria}}, \bibinfo {author} {\bibfnamefont {Y.}~\bibnamefont {Makris}},
  \ and\ \bibinfo {author} {\bibfnamefont {I.}~\bibnamefont {Scimemi}},\ }\href
  {\doibase 10.1007/JHEP10(2020)164} {\bibfield  {journal} {\bibinfo  {journal}
  {JHEP}\ }\textbf {\bibinfo {volume} {10}},\ \bibinfo {pages} {164} (\bibinfo
  {year} {2020})},\ \Eprint {http://arxiv.org/abs/2007.05547} {arXiv:2007.05547
  [hep-ph]} \BibitemShut {NoStop}%
\bibitem [{\citenamefont {Boer}\ \emph {et~al.}(2021)\citenamefont {Boer},
  \citenamefont {Pisano},\ and\ \citenamefont {Taels}}]{Boer:2021ehu}%
  \BibitemOpen
  \bibfield  {author} {\bibinfo {author} {\bibfnamefont {D.}~\bibnamefont
  {Boer}}, \bibinfo {author} {\bibfnamefont {C.}~\bibnamefont {Pisano}}, \ and\
  \bibinfo {author} {\bibfnamefont {P.}~\bibnamefont {Taels}},\ }\href
  {\doibase 10.1103/PhysRevD.103.074012} {\bibfield  {journal} {\bibinfo
  {journal} {Phys. Rev. D}\ }\textbf {\bibinfo {volume} {103}},\ \bibinfo
  {pages} {074012} (\bibinfo {year} {2021})},\ \Eprint
  {http://arxiv.org/abs/2102.00003} {arXiv:2102.00003 [hep-ph]} \BibitemShut
  {NoStop}%
\bibitem [{\citenamefont {Baumgart}\ \emph {et~al.}(2014)\citenamefont
  {Baumgart}, \citenamefont {Leibovich}, \citenamefont {Mehen},\ and\
  \citenamefont {Rothstein}}]{Baumgart:2014upa}%
  \BibitemOpen
  \bibfield  {author} {\bibinfo {author} {\bibfnamefont {M.}~\bibnamefont
  {Baumgart}}, \bibinfo {author} {\bibfnamefont {A.~K.}\ \bibnamefont
  {Leibovich}}, \bibinfo {author} {\bibfnamefont {T.}~\bibnamefont {Mehen}}, \
  and\ \bibinfo {author} {\bibfnamefont {I.~Z.}\ \bibnamefont {Rothstein}},\
  }\href {\doibase 10.1007/JHEP11(2014)003} {\bibfield  {journal} {\bibinfo
  {journal} {JHEP}\ }\textbf {\bibinfo {volume} {11}},\ \bibinfo {pages} {003}
  (\bibinfo {year} {2014})},\ \Eprint {http://arxiv.org/abs/1406.2295}
  {arXiv:1406.2295 [hep-ph]} \BibitemShut {NoStop}%
\bibitem [{\citenamefont {Bain}\ \emph {et~al.}(2016)\citenamefont {Bain},
  \citenamefont {Makris},\ and\ \citenamefont {Mehen}}]{Bain:2016rrv}%
  \BibitemOpen
  \bibfield  {author} {\bibinfo {author} {\bibfnamefont {R.}~\bibnamefont
  {Bain}}, \bibinfo {author} {\bibfnamefont {Y.}~\bibnamefont {Makris}}, \ and\
  \bibinfo {author} {\bibfnamefont {T.}~\bibnamefont {Mehen}},\ }\href
  {\doibase 10.1007/JHEP11(2016)144} {\bibfield  {journal} {\bibinfo  {journal}
  {JHEP}\ }\textbf {\bibinfo {volume} {11}},\ \bibinfo {pages} {144} (\bibinfo
  {year} {2016})},\ \Eprint {http://arxiv.org/abs/1610.06508} {arXiv:1610.06508
  [hep-ph]} \BibitemShut {NoStop}%
\bibitem [{\citenamefont {Braaten}\ and\ \citenamefont
  {Yuan}(1993)}]{Braaten:1993rw}%
  \BibitemOpen
  \bibfield  {author} {\bibinfo {author} {\bibfnamefont {E.}~\bibnamefont
  {Braaten}}\ and\ \bibinfo {author} {\bibfnamefont {T.~C.}\ \bibnamefont
  {Yuan}},\ }\href {\doibase 10.1103/PhysRevLett.71.1673} {\bibfield  {journal}
  {\bibinfo  {journal} {Phys. Rev. Lett.}\ }\textbf {\bibinfo {volume} {71}},\
  \bibinfo {pages} {1673} (\bibinfo {year} {1993})},\ \Eprint
  {http://arxiv.org/abs/hep-ph/9303205} {arXiv:hep-ph/9303205} \BibitemShut
  {NoStop}%
\bibitem [{\citenamefont {Braaten}\ \emph {et~al.}(1993)\citenamefont
  {Braaten}, \citenamefont {Cheung},\ and\ \citenamefont
  {Yuan}}]{Braaten:1993mp}%
  \BibitemOpen
  \bibfield  {author} {\bibinfo {author} {\bibfnamefont {E.}~\bibnamefont
  {Braaten}}, \bibinfo {author} {\bibfnamefont {K.-m.}\ \bibnamefont {Cheung}},
  \ and\ \bibinfo {author} {\bibfnamefont {T.~C.}\ \bibnamefont {Yuan}},\
  }\href {\doibase 10.1103/PhysRevD.48.4230} {\bibfield  {journal} {\bibinfo
  {journal} {Phys. Rev. D}\ }\textbf {\bibinfo {volume} {48}},\ \bibinfo
  {pages} {4230} (\bibinfo {year} {1993})},\ \Eprint
  {http://arxiv.org/abs/hep-ph/9302307} {arXiv:hep-ph/9302307} \BibitemShut
  {NoStop}%
\bibitem [{\citenamefont {Zheng}\ \emph {et~al.}(2019)\citenamefont {Zheng},
  \citenamefont {Chang},\ and\ \citenamefont {Wu}}]{Zheng:2019dfk}%
  \BibitemOpen
  \bibfield  {author} {\bibinfo {author} {\bibfnamefont {X.-C.}\ \bibnamefont
  {Zheng}}, \bibinfo {author} {\bibfnamefont {C.-H.}\ \bibnamefont {Chang}}, \
  and\ \bibinfo {author} {\bibfnamefont {X.-G.}\ \bibnamefont {Wu}},\ }\href
  {\doibase 10.1103/PhysRevD.100.014005} {\bibfield  {journal} {\bibinfo
  {journal} {Phys. Rev. D}\ }\textbf {\bibinfo {volume} {100}},\ \bibinfo
  {pages} {014005} (\bibinfo {year} {2019})},\ \Eprint
  {http://arxiv.org/abs/1905.09171} {arXiv:1905.09171 [hep-ph]} \BibitemShut
  {NoStop}%
\bibitem [{\citenamefont {Zheng}\ \emph
  {et~al.}(2021{\natexlab{a}})\citenamefont {Zheng}, \citenamefont {Wu},\ and\
  \citenamefont {Huang}}]{Zheng:2021ylc}%
  \BibitemOpen
  \bibfield  {author} {\bibinfo {author} {\bibfnamefont {X.-C.}\ \bibnamefont
  {Zheng}}, \bibinfo {author} {\bibfnamefont {X.-G.}\ \bibnamefont {Wu}}, \
  and\ \bibinfo {author} {\bibfnamefont {X.-D.}\ \bibnamefont {Huang}},\ }\href
  {\doibase 10.1007/JHEP07(2021)014} {\bibfield  {journal} {\bibinfo  {journal}
  {JHEP}\ }\textbf {\bibinfo {volume} {07}},\ \bibinfo {pages} {014} (\bibinfo
  {year} {2021}{\natexlab{a}})},\ \Eprint {http://arxiv.org/abs/2105.14580}
  {arXiv:2105.14580 [hep-ph]} \BibitemShut {NoStop}%
\bibitem [{\citenamefont {Zheng}\ \emph
  {et~al.}(2021{\natexlab{b}})\citenamefont {Zheng}, \citenamefont {Zhang},\
  and\ \citenamefont {Wu}}]{Zheng:2021mqr}%
  \BibitemOpen
  \bibfield  {author} {\bibinfo {author} {\bibfnamefont {X.-C.}\ \bibnamefont
  {Zheng}}, \bibinfo {author} {\bibfnamefont {Z.-Y.}\ \bibnamefont {Zhang}}, \
  and\ \bibinfo {author} {\bibfnamefont {X.-G.}\ \bibnamefont {Wu}},\ }\href
  {\doibase 10.1103/PhysRevD.103.074004} {\bibfield  {journal} {\bibinfo
  {journal} {Phys. Rev. D}\ }\textbf {\bibinfo {volume} {103}},\ \bibinfo
  {pages} {074004} (\bibinfo {year} {2021}{\natexlab{b}})},\ \Eprint
  {http://arxiv.org/abs/2101.01527} {arXiv:2101.01527 [hep-ph]} \BibitemShut
  {NoStop}%
\bibitem [{\citenamefont {Bodwin}\ \emph {et~al.}(2015)\citenamefont {Bodwin},
  \citenamefont {Chung}, \citenamefont {Kim},\ and\ \citenamefont
  {Lee}}]{Bodwin:2014bia}%
  \BibitemOpen
  \bibfield  {author} {\bibinfo {author} {\bibfnamefont {G.~T.}\ \bibnamefont
  {Bodwin}}, \bibinfo {author} {\bibfnamefont {H.~S.}\ \bibnamefont {Chung}},
  \bibinfo {author} {\bibfnamefont {U.-R.}\ \bibnamefont {Kim}}, \ and\
  \bibinfo {author} {\bibfnamefont {J.}~\bibnamefont {Lee}},\ }\href {\doibase
  10.1103/PhysRevD.91.074013} {\bibfield  {journal} {\bibinfo  {journal} {Phys.
  Rev. D}\ }\textbf {\bibinfo {volume} {91}},\ \bibinfo {pages} {074013}
  (\bibinfo {year} {2015})},\ \Eprint {http://arxiv.org/abs/1412.7106}
  {arXiv:1412.7106 [hep-ph]} \BibitemShut {NoStop}%
\bibitem [{\citenamefont {Zhang}\ \emph {et~al.}(2017)\citenamefont {Zhang},
  \citenamefont {Ma}, \citenamefont {Chen},\ and\ \citenamefont
  {Chao}}]{Zhang:2017xoj}%
  \BibitemOpen
  \bibfield  {author} {\bibinfo {author} {\bibfnamefont {P.}~\bibnamefont
  {Zhang}}, \bibinfo {author} {\bibfnamefont {Y.-Q.}\ \bibnamefont {Ma}},
  \bibinfo {author} {\bibfnamefont {Q.}~\bibnamefont {Chen}}, \ and\ \bibinfo
  {author} {\bibfnamefont {K.-T.}\ \bibnamefont {Chao}},\ }\href {\doibase
  10.1103/PhysRevD.96.094016} {\bibfield  {journal} {\bibinfo  {journal} {Phys.
  Rev. D}\ }\textbf {\bibinfo {volume} {96}},\ \bibinfo {pages} {094016}
  (\bibinfo {year} {2017})},\ \Eprint {http://arxiv.org/abs/1708.01129}
  {arXiv:1708.01129 [hep-ph]} \BibitemShut {NoStop}%
\bibitem [{\citenamefont {Celiberto}\ and\ \citenamefont
  {Fucilla}(2022)}]{Celiberto:2022dyf}%
  \BibitemOpen
  \bibfield  {author} {\bibinfo {author} {\bibfnamefont {F.~G.}\ \bibnamefont
  {Celiberto}}\ and\ \bibinfo {author} {\bibfnamefont {M.}~\bibnamefont
  {Fucilla}},\ }\href {\doibase 10.1140/epjc/s10052-022-10818-8} {\bibfield
  {journal} {\bibinfo  {journal} {Eur. Phys. J. C}\ }\textbf {\bibinfo {volume}
  {82}},\ \bibinfo {pages} {929} (\bibinfo {year} {2022})},\ \Eprint
  {http://arxiv.org/abs/2202.12227} {arXiv:2202.12227 [hep-ph]} \BibitemShut
  {NoStop}%
\bibitem [{\citenamefont {Celiberto}(2022)}]{Celiberto:2022keu}%
  \BibitemOpen
  \bibfield  {author} {\bibinfo {author} {\bibfnamefont {F.~G.}\ \bibnamefont
  {Celiberto}},\ }\href {\doibase 10.1016/j.physletb.2022.137554} {\bibfield
  {journal} {\bibinfo  {journal} {Phys. Lett. B}\ }\textbf {\bibinfo {volume}
  {835}},\ \bibinfo {pages} {137554} (\bibinfo {year} {2022})},\ \Eprint
  {http://arxiv.org/abs/2206.09413} {arXiv:2206.09413 [hep-ph]} \BibitemShut
  {NoStop}%
\bibitem [{\citenamefont {Silvetti}\ and\ \citenamefont
  {Bonvini}(2022)}]{Silvetti:2022hyc}%
  \BibitemOpen
  \bibfield  {author} {\bibinfo {author} {\bibfnamefont {F.}~\bibnamefont
  {Silvetti}}\ and\ \bibinfo {author} {\bibfnamefont {M.}~\bibnamefont
  {Bonvini}},\ }\href@noop {} {\  (\bibinfo {year} {2022})},\ \Eprint
  {http://arxiv.org/abs/2211.10142} {arXiv:2211.10142 [hep-ph]} \BibitemShut
  {NoStop}%
\bibitem [{\citenamefont {Brambilla}\ \emph {et~al.}(2005)\citenamefont
  {Brambilla}, \citenamefont {Pineda}, \citenamefont {Soto},\ and\
  \citenamefont {Vairo}}]{Brambilla:2004jw}%
  \BibitemOpen
  \bibfield  {author} {\bibinfo {author} {\bibfnamefont {N.}~\bibnamefont
  {Brambilla}}, \bibinfo {author} {\bibfnamefont {A.}~\bibnamefont {Pineda}},
  \bibinfo {author} {\bibfnamefont {J.}~\bibnamefont {Soto}}, \ and\ \bibinfo
  {author} {\bibfnamefont {A.}~\bibnamefont {Vairo}},\ }\href {\doibase
  10.1103/RevModPhys.77.1423} {\bibfield  {journal} {\bibinfo  {journal} {Rev.
  Mod. Phys.}\ }\textbf {\bibinfo {volume} {77}},\ \bibinfo {pages} {1423}
  (\bibinfo {year} {2005})},\ \Eprint {http://arxiv.org/abs/hep-ph/0410047}
  {arXiv:hep-ph/0410047} \BibitemShut {NoStop}%
\bibitem [{\citenamefont {Brambilla}\ \emph {et~al.}(2000)\citenamefont
  {Brambilla}, \citenamefont {Pineda}, \citenamefont {Soto},\ and\
  \citenamefont {Vairo}}]{Brambilla:1999xf}%
  \BibitemOpen
  \bibfield  {author} {\bibinfo {author} {\bibfnamefont {N.}~\bibnamefont
  {Brambilla}}, \bibinfo {author} {\bibfnamefont {A.}~\bibnamefont {Pineda}},
  \bibinfo {author} {\bibfnamefont {J.}~\bibnamefont {Soto}}, \ and\ \bibinfo
  {author} {\bibfnamefont {A.}~\bibnamefont {Vairo}},\ }\href {\doibase
  10.1016/S0550-3213(99)00693-8} {\bibfield  {journal} {\bibinfo  {journal}
  {Nucl. Phys. B}\ }\textbf {\bibinfo {volume} {566}},\ \bibinfo {pages} {275}
  (\bibinfo {year} {2000})},\ \Eprint {http://arxiv.org/abs/hep-ph/9907240}
  {arXiv:hep-ph/9907240} \BibitemShut {NoStop}%
\bibitem [{\citenamefont {Brambilla}\ \emph {et~al.}(2008)\citenamefont
  {Brambilla}, \citenamefont {Ghiglieri}, \citenamefont {Vairo},\ and\
  \citenamefont {Petreczky}}]{Brambilla:2008cx}%
  \BibitemOpen
  \bibfield  {author} {\bibinfo {author} {\bibfnamefont {N.}~\bibnamefont
  {Brambilla}}, \bibinfo {author} {\bibfnamefont {J.}~\bibnamefont
  {Ghiglieri}}, \bibinfo {author} {\bibfnamefont {A.}~\bibnamefont {Vairo}}, \
  and\ \bibinfo {author} {\bibfnamefont {P.}~\bibnamefont {Petreczky}},\ }\href
  {\doibase 10.1103/PhysRevD.78.014017} {\bibfield  {journal} {\bibinfo
  {journal} {Phys. Rev. D}\ }\textbf {\bibinfo {volume} {78}},\ \bibinfo
  {pages} {014017} (\bibinfo {year} {2008})},\ \Eprint
  {http://arxiv.org/abs/0804.0993} {arXiv:0804.0993 [hep-ph]} \BibitemShut
  {NoStop}%
\bibitem [{\citenamefont {Brambilla}\ \emph
  {et~al.}(2018{\natexlab{a}})\citenamefont {Brambilla}, \citenamefont
  {Escobedo}, \citenamefont {Soto},\ and\ \citenamefont
  {Vairo}}]{Brambilla:2017zei}%
  \BibitemOpen
  \bibfield  {author} {\bibinfo {author} {\bibfnamefont {N.}~\bibnamefont
  {Brambilla}}, \bibinfo {author} {\bibfnamefont {M.~A.}\ \bibnamefont
  {Escobedo}}, \bibinfo {author} {\bibfnamefont {J.}~\bibnamefont {Soto}}, \
  and\ \bibinfo {author} {\bibfnamefont {A.}~\bibnamefont {Vairo}},\ }\href
  {\doibase 10.1103/PhysRevD.97.074009} {\bibfield  {journal} {\bibinfo
  {journal} {Phys. Rev. D}\ }\textbf {\bibinfo {volume} {97}},\ \bibinfo
  {pages} {074009} (\bibinfo {year} {2018}{\natexlab{a}})},\ \Eprint
  {http://arxiv.org/abs/1711.04515} {arXiv:1711.04515 [hep-ph]} \BibitemShut
  {NoStop}%
\bibitem [{\citenamefont {Brambilla}\ \emph {et~al.}(2021)\citenamefont
  {Brambilla}, \citenamefont {Escobedo}, \citenamefont {Strickland},
  \citenamefont {Vairo}, \citenamefont {Vander~Griend},\ and\ \citenamefont
  {Weber}}]{Brambilla:2021wkt}%
  \BibitemOpen
  \bibfield  {author} {\bibinfo {author} {\bibfnamefont {N.}~\bibnamefont
  {Brambilla}}, \bibinfo {author} {\bibfnamefont {M.~A.}\ \bibnamefont
  {Escobedo}}, \bibinfo {author} {\bibfnamefont {M.}~\bibnamefont
  {Strickland}}, \bibinfo {author} {\bibfnamefont {A.}~\bibnamefont {Vairo}},
  \bibinfo {author} {\bibfnamefont {P.}~\bibnamefont {Vander~Griend}}, \ and\
  \bibinfo {author} {\bibfnamefont {J.~H.}\ \bibnamefont {Weber}},\ }\href
  {\doibase 10.1103/PhysRevD.104.094049} {\bibfield  {journal} {\bibinfo
  {journal} {Phys. Rev. D}\ }\textbf {\bibinfo {volume} {104}},\ \bibinfo
  {pages} {094049} (\bibinfo {year} {2021})},\ \Eprint
  {http://arxiv.org/abs/2107.06222} {arXiv:2107.06222 [hep-ph]} \BibitemShut
  {NoStop}%
\bibitem [{\citenamefont {Bodwin}\ \emph {et~al.}(1995)\citenamefont {Bodwin},
  \citenamefont {Braaten},\ and\ \citenamefont {Lepage}}]{Bodwin:1994jh}%
  \BibitemOpen
  \bibfield  {author} {\bibinfo {author} {\bibfnamefont {G.~T.}\ \bibnamefont
  {Bodwin}}, \bibinfo {author} {\bibfnamefont {E.}~\bibnamefont {Braaten}}, \
  and\ \bibinfo {author} {\bibfnamefont {G.~P.}\ \bibnamefont {Lepage}},\
  }\href {\doibase 10.1103/PhysRevD.55.5853} {\bibfield  {journal} {\bibinfo
  {journal} {Phys. Rev. D}\ }\textbf {\bibinfo {volume} {51}},\ \bibinfo
  {pages} {1125} (\bibinfo {year} {1995})},\ \bibinfo {note} {[Erratum:
  Phys.Rev.D 55, 5853 (1997)]},\ \Eprint {http://arxiv.org/abs/hep-ph/9407339}
  {arXiv:hep-ph/9407339} \BibitemShut {NoStop}%
\bibitem [{\citenamefont {Sharma}\ and\ \citenamefont
  {Vitev}(2013)}]{Sharma:2012dy}%
  \BibitemOpen
  \bibfield  {author} {\bibinfo {author} {\bibfnamefont {R.}~\bibnamefont
  {Sharma}}\ and\ \bibinfo {author} {\bibfnamefont {I.}~\bibnamefont {Vitev}},\
  }\href {\doibase 10.1103/PhysRevC.87.044905} {\bibfield  {journal} {\bibinfo
  {journal} {Phys. Rev. C}\ }\textbf {\bibinfo {volume} {87}},\ \bibinfo
  {pages} {044905} (\bibinfo {year} {2013})},\ \Eprint
  {http://arxiv.org/abs/1203.0329} {arXiv:1203.0329 [hep-ph]} \BibitemShut
  {NoStop}%
\bibitem [{\citenamefont {Aronson}\ \emph {et~al.}(2018)\citenamefont
  {Aronson}, \citenamefont {Borras}, \citenamefont {Odegard}, \citenamefont
  {Sharma},\ and\ \citenamefont {Vitev}}]{Aronson:2017ymv}%
  \BibitemOpen
  \bibfield  {author} {\bibinfo {author} {\bibfnamefont {S.}~\bibnamefont
  {Aronson}}, \bibinfo {author} {\bibfnamefont {E.}~\bibnamefont {Borras}},
  \bibinfo {author} {\bibfnamefont {B.}~\bibnamefont {Odegard}}, \bibinfo
  {author} {\bibfnamefont {R.}~\bibnamefont {Sharma}}, \ and\ \bibinfo {author}
  {\bibfnamefont {I.}~\bibnamefont {Vitev}},\ }\href {\doibase
  10.1016/j.physletb.2018.01.038} {\bibfield  {journal} {\bibinfo  {journal}
  {Phys. Lett. B}\ }\textbf {\bibinfo {volume} {778}},\ \bibinfo {pages} {384}
  (\bibinfo {year} {2018})},\ \Eprint {http://arxiv.org/abs/1709.02372}
  {arXiv:1709.02372 [hep-ph]} \BibitemShut {NoStop}%
\bibitem [{\citenamefont {Makris}\ and\ \citenamefont
  {Vitev}(2019)}]{Makris:2019ttx}%
  \BibitemOpen
  \bibfield  {author} {\bibinfo {author} {\bibfnamefont {Y.}~\bibnamefont
  {Makris}}\ and\ \bibinfo {author} {\bibfnamefont {I.}~\bibnamefont {Vitev}},\
  }\href {\doibase 10.1007/JHEP10(2019)111} {\bibfield  {journal} {\bibinfo
  {journal} {JHEP}\ }\textbf {\bibinfo {volume} {10}},\ \bibinfo {pages} {111}
  (\bibinfo {year} {2019})},\ \Eprint {http://arxiv.org/abs/1906.04186}
  {arXiv:1906.04186 [hep-ph]} \BibitemShut {NoStop}%
\bibitem [{\citenamefont {Makris}\ and\ \citenamefont
  {Vitev}(2021)}]{Makris:2019kap}%
  \BibitemOpen
  \bibfield  {author} {\bibinfo {author} {\bibfnamefont {Y.}~\bibnamefont
  {Makris}}\ and\ \bibinfo {author} {\bibfnamefont {I.}~\bibnamefont {Vitev}},\
  }\href {\doibase 10.1016/j.nuclphysa.2020.121848} {\bibfield  {journal}
  {\bibinfo  {journal} {Nucl. Phys. A}\ }\textbf {\bibinfo {volume} {1005}},\
  \bibinfo {pages} {121848} (\bibinfo {year} {2021})},\ \Eprint
  {http://arxiv.org/abs/1912.08008} {arXiv:1912.08008 [hep-ph]} \BibitemShut
  {NoStop}%
\bibitem [{\citenamefont {Kharzeev}(1996)}]{Kharzeev:1995ij}%
  \BibitemOpen
  \bibfield  {author} {\bibinfo {author} {\bibfnamefont {D.}~\bibnamefont
  {Kharzeev}},\ }\href {\doibase 10.3254/978-1-61499-215-8-105} {\bibfield
  {journal} {\bibinfo  {journal} {Proc. Int. Sch. Phys. Fermi}\ }\textbf
  {\bibinfo {volume} {130}},\ \bibinfo {pages} {105} (\bibinfo {year}
  {1996})},\ \Eprint {http://arxiv.org/abs/nucl-th/9601029}
  {arXiv:nucl-th/9601029} \BibitemShut {NoStop}%
\bibitem [{\citenamefont {Kharzeev}\ \emph {et~al.}(1999)\citenamefont
  {Kharzeev}, \citenamefont {Satz}, \citenamefont {Syamtomov},\ and\
  \citenamefont {Zinovjev}}]{Kharzeev:1998bz}%
  \BibitemOpen
  \bibfield  {author} {\bibinfo {author} {\bibfnamefont {D.}~\bibnamefont
  {Kharzeev}}, \bibinfo {author} {\bibfnamefont {H.}~\bibnamefont {Satz}},
  \bibinfo {author} {\bibfnamefont {A.}~\bibnamefont {Syamtomov}}, \ and\
  \bibinfo {author} {\bibfnamefont {G.}~\bibnamefont {Zinovjev}},\ }\href
  {\doibase 10.1007/s100529900047} {\bibfield  {journal} {\bibinfo  {journal}
  {Eur. Phys. J. C}\ }\textbf {\bibinfo {volume} {9}},\ \bibinfo {pages} {459}
  (\bibinfo {year} {1999})},\ \Eprint {http://arxiv.org/abs/hep-ph/9901375}
  {arXiv:hep-ph/9901375} \BibitemShut {NoStop}%
\bibitem [{\citenamefont {Hatta}\ and\ \citenamefont
  {Yang}(2018)}]{Hatta:2018ina}%
  \BibitemOpen
  \bibfield  {author} {\bibinfo {author} {\bibfnamefont {Y.}~\bibnamefont
  {Hatta}}\ and\ \bibinfo {author} {\bibfnamefont {D.-L.}\ \bibnamefont
  {Yang}},\ }\href {\doibase 10.1103/PhysRevD.98.074003} {\bibfield  {journal}
  {\bibinfo  {journal} {Phys. Rev. D}\ }\textbf {\bibinfo {volume} {98}},\
  \bibinfo {pages} {074003} (\bibinfo {year} {2018})},\ \Eprint
  {http://arxiv.org/abs/1808.02163} {arXiv:1808.02163 [hep-ph]} \BibitemShut
  {NoStop}%
\bibitem [{\citenamefont {Hatta}\ \emph {et~al.}(2019)\citenamefont {Hatta},
  \citenamefont {Rajan},\ and\ \citenamefont {Yang}}]{Hatta:2019lxo}%
  \BibitemOpen
  \bibfield  {author} {\bibinfo {author} {\bibfnamefont {Y.}~\bibnamefont
  {Hatta}}, \bibinfo {author} {\bibfnamefont {A.}~\bibnamefont {Rajan}}, \ and\
  \bibinfo {author} {\bibfnamefont {D.-L.}\ \bibnamefont {Yang}},\ }\href
  {\doibase 10.1103/PhysRevD.100.014032} {\bibfield  {journal} {\bibinfo
  {journal} {Phys. Rev. D}\ }\textbf {\bibinfo {volume} {100}},\ \bibinfo
  {pages} {014032} (\bibinfo {year} {2019})},\ \Eprint
  {http://arxiv.org/abs/1906.00894} {arXiv:1906.00894 [hep-ph]} \BibitemShut
  {NoStop}%
\bibitem [{\citenamefont {Mamo}\ and\ \citenamefont
  {Zahed}(2020)}]{Mamo:2019mka}%
  \BibitemOpen
  \bibfield  {author} {\bibinfo {author} {\bibfnamefont {K.~A.}\ \bibnamefont
  {Mamo}}\ and\ \bibinfo {author} {\bibfnamefont {I.}~\bibnamefont {Zahed}},\
  }\href {\doibase 10.1103/PhysRevD.101.086003} {\bibfield  {journal} {\bibinfo
   {journal} {Phys. Rev. D}\ }\textbf {\bibinfo {volume} {101}},\ \bibinfo
  {pages} {086003} (\bibinfo {year} {2020})},\ \Eprint
  {http://arxiv.org/abs/1910.04707} {arXiv:1910.04707 [hep-ph]} \BibitemShut
  {NoStop}%
\bibitem [{\citenamefont {Kharzeev}(2021{\natexlab{a}})}]{Kharzeev:2021qkd}%
  \BibitemOpen
  \bibfield  {author} {\bibinfo {author} {\bibfnamefont {D.~E.}\ \bibnamefont
  {Kharzeev}},\ }\href {\doibase 10.1103/PhysRevD.104.054015} {\bibfield
  {journal} {\bibinfo  {journal} {Phys. Rev. D}\ }\textbf {\bibinfo {volume}
  {104}},\ \bibinfo {pages} {054015} (\bibinfo {year} {2021}{\natexlab{a}})},\
  \Eprint {http://arxiv.org/abs/2102.00110} {arXiv:2102.00110 [hep-ph]}
  \BibitemShut {NoStop}%
\bibitem [{\citenamefont {Guo}\ \emph {et~al.}(2021)\citenamefont {Guo},
  \citenamefont {Ji},\ and\ \citenamefont {Liu}}]{Guo:2021ibg}%
  \BibitemOpen
  \bibfield  {author} {\bibinfo {author} {\bibfnamefont {Y.}~\bibnamefont
  {Guo}}, \bibinfo {author} {\bibfnamefont {X.}~\bibnamefont {Ji}}, \ and\
  \bibinfo {author} {\bibfnamefont {Y.}~\bibnamefont {Liu}},\ }\href {\doibase
  10.1103/PhysRevD.103.096010} {\bibfield  {journal} {\bibinfo  {journal}
  {Phys. Rev. D}\ }\textbf {\bibinfo {volume} {103}},\ \bibinfo {pages}
  {096010} (\bibinfo {year} {2021})},\ \Eprint
  {http://arxiv.org/abs/2103.11506} {arXiv:2103.11506 [hep-ph]} \BibitemShut
  {NoStop}%
\bibitem [{\citenamefont {Lee}\ \emph {et~al.}(2022{\natexlab{b}})\citenamefont
  {Lee}, \citenamefont {Sakinah},\ and\ \citenamefont {Oh}}]{Lee:2022ymp}%
  \BibitemOpen
  \bibfield  {author} {\bibinfo {author} {\bibfnamefont {T.~S.~H.}\
  \bibnamefont {Lee}}, \bibinfo {author} {\bibfnamefont {S.}~\bibnamefont
  {Sakinah}}, \ and\ \bibinfo {author} {\bibfnamefont {Y.}~\bibnamefont {Oh}},\
  }\href {\doibase 10.1140/epja/s10050-022-00901-9} {\bibfield  {journal}
  {\bibinfo  {journal} {Eur. Phys. J. A}\ }\textbf {\bibinfo {volume} {58}},\
  \bibinfo {pages} {252} (\bibinfo {year} {2022}{\natexlab{b}})},\ \Eprint
  {http://arxiv.org/abs/2210.02154} {arXiv:2210.02154 [hep-ph]} \BibitemShut
  {NoStop}%
\bibitem [{\citenamefont {Ali}\ \emph {et~al.}(2019)\citenamefont {Ali} \emph
  {et~al.}}]{GlueX:2019mkq}%
  \BibitemOpen
  \bibfield  {author} {\bibinfo {author} {\bibfnamefont {A.}~\bibnamefont
  {Ali}} \emph {et~al.} (\bibinfo {collaboration} {GlueX}),\ }\href {\doibase
  10.1103/PhysRevLett.123.072001} {\bibfield  {journal} {\bibinfo  {journal}
  {Phys. Rev. Lett.}\ }\textbf {\bibinfo {volume} {123}},\ \bibinfo {pages}
  {072001} (\bibinfo {year} {2019})},\ \Eprint
  {http://arxiv.org/abs/1905.10811} {arXiv:1905.10811 [nucl-ex]} \BibitemShut
  {NoStop}%
\bibitem [{\citenamefont {Duran}\ \emph {et~al.}(2022)\citenamefont {Duran}
  \emph {et~al.}}]{Duran:2022xag}%
  \BibitemOpen
  \bibfield  {author} {\bibinfo {author} {\bibfnamefont {B.}~\bibnamefont
  {Duran}} \emph {et~al.},\ }\href@noop {} {\  (\bibinfo {year} {2022})},\
  \Eprint {http://arxiv.org/abs/2207.05212} {arXiv:2207.05212 [nucl-ex]}
  \BibitemShut {NoStop}%
\bibitem [{\citenamefont {Joosten}\ and\ \citenamefont
  {Meziani}(2018)}]{Joosten:2018gyo}%
  \BibitemOpen
  \bibfield  {author} {\bibinfo {author} {\bibfnamefont {S.}~\bibnamefont
  {Joosten}}\ and\ \bibinfo {author} {\bibfnamefont {Z.~E.}\ \bibnamefont
  {Meziani}},\ }\href {\doibase 10.22323/1.308.0017} {\bibfield  {journal}
  {\bibinfo  {journal} {PoS}\ }\textbf {\bibinfo {volume} {QCDEV2017}},\
  \bibinfo {pages} {017} (\bibinfo {year} {2018})},\ \Eprint
  {http://arxiv.org/abs/1802.02616} {arXiv:1802.02616 [hep-ex]} \BibitemShut
  {NoStop}%
\bibitem [{\citenamefont {Shanahan}\ and\ \citenamefont
  {Detmold}(2019{\natexlab{a}})}]{Shanahan:2018pib}%
  \BibitemOpen
  \bibfield  {author} {\bibinfo {author} {\bibfnamefont {P.~E.}\ \bibnamefont
  {Shanahan}}\ and\ \bibinfo {author} {\bibfnamefont {W.}~\bibnamefont
  {Detmold}},\ }\href {\doibase 10.1103/PhysRevD.99.014511} {\bibfield
  {journal} {\bibinfo  {journal} {Phys. Rev. D}\ }\textbf {\bibinfo {volume}
  {99}},\ \bibinfo {pages} {014511} (\bibinfo {year} {2019}{\natexlab{a}})},\
  \Eprint {http://arxiv.org/abs/1810.04626} {arXiv:1810.04626 [hep-lat]}
  \BibitemShut {NoStop}%
\bibitem [{\citenamefont {Shanahan}\ and\ \citenamefont
  {Detmold}(2019{\natexlab{b}})}]{Shanahan:2018nnv}%
  \BibitemOpen
  \bibfield  {author} {\bibinfo {author} {\bibfnamefont {P.~E.}\ \bibnamefont
  {Shanahan}}\ and\ \bibinfo {author} {\bibfnamefont {W.}~\bibnamefont
  {Detmold}},\ }\href {\doibase 10.1103/PhysRevLett.122.072003} {\bibfield
  {journal} {\bibinfo  {journal} {Phys. Rev. Lett.}\ }\textbf {\bibinfo
  {volume} {122}},\ \bibinfo {pages} {072003} (\bibinfo {year}
  {2019}{\natexlab{b}})},\ \Eprint {http://arxiv.org/abs/1810.07589}
  {arXiv:1810.07589 [nucl-th]} \BibitemShut {NoStop}%
\bibitem [{\citenamefont {Webber}(2000)}]{Webber:1999ui}%
  \BibitemOpen
  \bibfield  {author} {\bibinfo {author} {\bibfnamefont {B.~R.}\ \bibnamefont
  {Webber}},\ }\bibfield  {booktitle} {\emph {\bibinfo {booktitle} {{Lepton and
  photon interactions at high energies. Proceedings, 19th International
  Symposium, LP'99, Stanford, USA, August 9-14, 1999}}},\ }\href {\doibase
  10.1142/S0217751X00005334} {\bibfield  {journal} {\bibinfo  {journal} {Int.
  J. Mod. Phys.}\ }\textbf {\bibinfo {volume} {A15S1}},\ \bibinfo {pages} {577}
  (\bibinfo {year} {2000})},\ \bibinfo {note} {[eConfC990809,577(2000)]},\
  \Eprint {http://arxiv.org/abs/hep-ph/9912292} {arXiv:hep-ph/9912292 [hep-ph]}
  \BibitemShut {NoStop}%
\bibitem [{\citenamefont {Andersson}\ \emph
  {et~al.}(1983{\natexlab{a}})\citenamefont {Andersson}, \citenamefont
  {Gustafson}, \citenamefont {Ingelman},\ and\ \citenamefont
  {Sjostrand}}]{Andersson:1983ia}%
  \BibitemOpen
  \bibfield  {author} {\bibinfo {author} {\bibfnamefont {B.}~\bibnamefont
  {Andersson}}, \bibinfo {author} {\bibfnamefont {G.}~\bibnamefont
  {Gustafson}}, \bibinfo {author} {\bibfnamefont {G.}~\bibnamefont {Ingelman}},
  \ and\ \bibinfo {author} {\bibfnamefont {T.}~\bibnamefont {Sjostrand}},\
  }\href {\doibase 10.1016/0370-1573(83)90080-7} {\bibfield  {journal}
  {\bibinfo  {journal} {Phys. Rept.}\ }\textbf {\bibinfo {volume} {97}},\
  \bibinfo {pages} {31} (\bibinfo {year} {1983}{\natexlab{a}})}\BibitemShut
  {NoStop}%
\bibitem [{\citenamefont {Andersson}\ \emph
  {et~al.}(1983{\natexlab{b}})\citenamefont {Andersson}, \citenamefont
  {Gustafson},\ and\ \citenamefont {Soderberg}}]{Andersson:1983jt}%
  \BibitemOpen
  \bibfield  {author} {\bibinfo {author} {\bibfnamefont {B.}~\bibnamefont
  {Andersson}}, \bibinfo {author} {\bibfnamefont {G.}~\bibnamefont
  {Gustafson}}, \ and\ \bibinfo {author} {\bibfnamefont {B.}~\bibnamefont
  {Soderberg}},\ }\href {\doibase 10.1007/BF01407824} {\bibfield  {journal}
  {\bibinfo  {journal} {Z. Phys.}\ }\textbf {\bibinfo {volume} {C20}},\
  \bibinfo {pages} {317} (\bibinfo {year} {1983}{\natexlab{b}})}\BibitemShut
  {NoStop}%
\bibitem [{\citenamefont {Ethier}\ \emph {et~al.}(2017)\citenamefont {Ethier},
  \citenamefont {Sato},\ and\ \citenamefont {Melnitchouk}}]{Ethier:2017zbq}%
  \BibitemOpen
  \bibfield  {author} {\bibinfo {author} {\bibfnamefont {J.~J.}\ \bibnamefont
  {Ethier}}, \bibinfo {author} {\bibfnamefont {N.}~\bibnamefont {Sato}}, \ and\
  \bibinfo {author} {\bibfnamefont {W.}~\bibnamefont {Melnitchouk}},\ }\href
  {\doibase 10.1103/PhysRevLett.119.132001} {\bibfield  {journal} {\bibinfo
  {journal} {Phys. Rev. Lett.}\ }\textbf {\bibinfo {volume} {119}},\ \bibinfo
  {pages} {132001} (\bibinfo {year} {2017})},\ \Eprint
  {http://arxiv.org/abs/1705.05889} {arXiv:1705.05889 [hep-ph]} \BibitemShut
  {NoStop}%
\bibitem [{\citenamefont {Bertone}\ \emph {et~al.}(2017)\citenamefont
  {Bertone}, \citenamefont {Carrazza}, \citenamefont {Hartland}, \citenamefont
  {Nocera},\ and\ \citenamefont {Rojo}}]{Bertone:2017tyb}%
  \BibitemOpen
  \bibfield  {author} {\bibinfo {author} {\bibfnamefont {V.}~\bibnamefont
  {Bertone}}, \bibinfo {author} {\bibfnamefont {S.}~\bibnamefont {Carrazza}},
  \bibinfo {author} {\bibfnamefont {N.~P.}\ \bibnamefont {Hartland}}, \bibinfo
  {author} {\bibfnamefont {E.~R.}\ \bibnamefont {Nocera}}, \ and\ \bibinfo
  {author} {\bibfnamefont {J.}~\bibnamefont {Rojo}} (\bibinfo {collaboration}
  {NNPDF}),\ }\href {\doibase 10.1140/epjc/s10052-017-5088-y} {\bibfield
  {journal} {\bibinfo  {journal} {Eur. Phys. J. C}\ }\textbf {\bibinfo {volume}
  {77}},\ \bibinfo {pages} {516} (\bibinfo {year} {2017})},\ \Eprint
  {http://arxiv.org/abs/1706.07049} {arXiv:1706.07049 [hep-ph]} \BibitemShut
  {NoStop}%
\bibitem [{\citenamefont {Airapetian}\ \emph {et~al.}(2001)\citenamefont
  {Airapetian} \emph {et~al.}}]{Airapetian:2000ks}%
  \BibitemOpen
  \bibfield  {author} {\bibinfo {author} {\bibfnamefont {A.}~\bibnamefont
  {Airapetian}} \emph {et~al.} (\bibinfo {collaboration} {HERMES}),\ }\href
  {\doibase 10.1007/s100520100697} {\bibfield  {journal} {\bibinfo  {journal}
  {Eur. Phys. J.}\ }\textbf {\bibinfo {volume} {C20}},\ \bibinfo {pages} {479}
  (\bibinfo {year} {2001})},\ \Eprint {http://arxiv.org/abs/hep-ex/0012049}
  {arXiv:hep-ex/0012049 [hep-ex]} \BibitemShut {NoStop}%
\bibitem [{\citenamefont {Airapetian}\ \emph {et~al.}(2007)\citenamefont
  {Airapetian} \emph {et~al.}}]{Airapetian:2007vu}%
  \BibitemOpen
  \bibfield  {author} {\bibinfo {author} {\bibfnamefont {A.}~\bibnamefont
  {Airapetian}} \emph {et~al.} (\bibinfo {collaboration} {HERMES}),\ }\href
  {\doibase 10.1016/j.nuclphysb.2007.06.004} {\bibfield  {journal} {\bibinfo
  {journal} {Nucl. Phys. B}\ }\textbf {\bibinfo {volume} {780}},\ \bibinfo
  {pages} {1} (\bibinfo {year} {2007})},\ \Eprint
  {http://arxiv.org/abs/0704.3270} {arXiv:0704.3270 [hep-ex]} \BibitemShut
  {NoStop}%
\bibitem [{\citenamefont {Chang}\ \emph {et~al.}(2014)\citenamefont {Chang},
  \citenamefont {Deng},\ and\ \citenamefont {Wang}}]{Chang:2014fba}%
  \BibitemOpen
  \bibfield  {author} {\bibinfo {author} {\bibfnamefont {N.-B.}\ \bibnamefont
  {Chang}}, \bibinfo {author} {\bibfnamefont {W.-T.}\ \bibnamefont {Deng}}, \
  and\ \bibinfo {author} {\bibfnamefont {X.-N.}\ \bibnamefont {Wang}},\ }\href
  {\doibase 10.1103/PhysRevC.89.034911} {\bibfield  {journal} {\bibinfo
  {journal} {Phys. Rev.}\ }\textbf {\bibinfo {volume} {C89}},\ \bibinfo {pages}
  {034911} (\bibinfo {year} {2014})},\ \Eprint {http://arxiv.org/abs/1401.5109}
  {arXiv:1401.5109 [nucl-th]} \BibitemShut {NoStop}%
\bibitem [{\citenamefont {Arleo}(2003)}]{Arleo:2003jz}%
  \BibitemOpen
  \bibfield  {author} {\bibinfo {author} {\bibfnamefont {F.}~\bibnamefont
  {Arleo}},\ }\href {\doibase 10.1140/epjc/s2003-01289-x} {\bibfield  {journal}
  {\bibinfo  {journal} {Eur. Phys. J.}\ }\textbf {\bibinfo {volume} {C30}},\
  \bibinfo {pages} {213} (\bibinfo {year} {2003})},\ \Eprint
  {http://arxiv.org/abs/hep-ph/0306235} {arXiv:hep-ph/0306235 [hep-ph]}
  \BibitemShut {NoStop}%
\bibitem [{\citenamefont {Kopeliovich}\ \emph
  {et~al.}(2004{\natexlab{a}})\citenamefont {Kopeliovich}, \citenamefont
  {Nemchik}, \citenamefont {Predazzi},\ and\ \citenamefont
  {Hayashigaki}}]{Kopeliovich:2003py}%
  \BibitemOpen
  \bibfield  {author} {\bibinfo {author} {\bibfnamefont {B.~Z.}\ \bibnamefont
  {Kopeliovich}}, \bibinfo {author} {\bibfnamefont {J.}~\bibnamefont
  {Nemchik}}, \bibinfo {author} {\bibfnamefont {E.}~\bibnamefont {Predazzi}}, \
  and\ \bibinfo {author} {\bibfnamefont {A.}~\bibnamefont {Hayashigaki}},\
  }\bibfield  {booktitle} {\emph {\bibinfo {booktitle} {{EURESCO Conference on
  Hadron Structure Viewed with Electromagnetic Probes Santorini, Greece,
  October 7-12, 2003}}},\ }\href {\doibase 10.1016/j.nuclphysa.2004.04.110}
  {\bibfield  {journal} {\bibinfo  {journal} {Nucl. Phys.}\ }\textbf {\bibinfo
  {volume} {A740}},\ \bibinfo {pages} {211} (\bibinfo {year}
  {2004}{\natexlab{a}})},\ \Eprint {http://arxiv.org/abs/hep-ph/0311220}
  {arXiv:hep-ph/0311220 [hep-ph]} \BibitemShut {NoStop}%
\bibitem [{\citenamefont {Kopeliovich}\ \emph
  {et~al.}(2004{\natexlab{b}})\citenamefont {Kopeliovich}, \citenamefont
  {Nemchik}, \citenamefont {Predazzi},\ and\ \citenamefont
  {Hayashigaki}}]{Kopeliovich:2004kq}%
  \BibitemOpen
  \bibfield  {author} {\bibinfo {author} {\bibfnamefont {B.~Z.}\ \bibnamefont
  {Kopeliovich}}, \bibinfo {author} {\bibfnamefont {J.}~\bibnamefont
  {Nemchik}}, \bibinfo {author} {\bibfnamefont {E.}~\bibnamefont {Predazzi}}, \
  and\ \bibinfo {author} {\bibfnamefont {A.}~\bibnamefont {Hayashigaki}},\
  }\href {\doibase 10.1140/epjad/s2004-03-019-7} {\bibfield  {journal}
  {\bibinfo  {journal} {Eur. Phys. J. A}\ }\textbf {\bibinfo {volume} {19S1}},\
  \bibinfo {pages} {111} (\bibinfo {year} {2004}{\natexlab{b}})}\BibitemShut
  {NoStop}%
\bibitem [{\citenamefont {Kopeliovich}\ \emph {et~al.}(2007)\citenamefont
  {Kopeliovich}, \citenamefont {Nemchik},\ and\ \citenamefont
  {Schmidt}}]{Kopeliovich:2006xy}%
  \BibitemOpen
  \bibfield  {author} {\bibinfo {author} {\bibfnamefont {B.~Z.}\ \bibnamefont
  {Kopeliovich}}, \bibinfo {author} {\bibfnamefont {J.}~\bibnamefont
  {Nemchik}}, \ and\ \bibinfo {author} {\bibfnamefont {I.}~\bibnamefont
  {Schmidt}},\ }\href {\doibase 10.1016/j.nuclphysa.2006.10.059} {\bibfield
  {journal} {\bibinfo  {journal} {Nucl. Phys. A}\ }\textbf {\bibinfo {volume}
  {782}},\ \bibinfo {pages} {224} (\bibinfo {year} {2007})},\ \Eprint
  {http://arxiv.org/abs/hep-ph/0608044} {arXiv:hep-ph/0608044} \BibitemShut
  {NoStop}%
\bibitem [{\citenamefont {Guiot}\ and\ \citenamefont
  {Kopeliovich}(2020)}]{Guiot:2020vsf}%
  \BibitemOpen
  \bibfield  {author} {\bibinfo {author} {\bibfnamefont {B.}~\bibnamefont
  {Guiot}}\ and\ \bibinfo {author} {\bibfnamefont {B.~Z.}\ \bibnamefont
  {Kopeliovich}},\ }\href {\doibase 10.1103/PhysRevC.102.045201} {\bibfield
  {journal} {\bibinfo  {journal} {Phys. Rev. C}\ }\textbf {\bibinfo {volume}
  {102}},\ \bibinfo {pages} {045201} (\bibinfo {year} {2020})},\ \Eprint
  {http://arxiv.org/abs/2001.00974} {arXiv:2001.00974 [hep-ph]} \BibitemShut
  {NoStop}%
\bibitem [{\citenamefont {Adil}\ and\ \citenamefont
  {Vitev}(2007)}]{Adil:2006ra}%
  \BibitemOpen
  \bibfield  {author} {\bibinfo {author} {\bibfnamefont {A.}~\bibnamefont
  {Adil}}\ and\ \bibinfo {author} {\bibfnamefont {I.}~\bibnamefont {Vitev}},\
  }\href {\doibase 10.1016/j.physletb.2007.03.050} {\bibfield  {journal}
  {\bibinfo  {journal} {Phys. Lett.}\ }\textbf {\bibinfo {volume} {B649}},\
  \bibinfo {pages} {139} (\bibinfo {year} {2007})},\ \Eprint
  {http://arxiv.org/abs/hep-ph/0611109} {arXiv:hep-ph/0611109 [hep-ph]}
  \BibitemShut {NoStop}%
\bibitem [{\citenamefont {Ke}\ and\ \citenamefont {Vitev}(2023)}]{Ke:2023ixa}%
  \BibitemOpen
  \bibfield  {author} {\bibinfo {author} {\bibfnamefont {W.}~\bibnamefont
  {Ke}}\ and\ \bibinfo {author} {\bibfnamefont {I.}~\bibnamefont {Vitev}},\
  }\href@noop {} {\  (\bibinfo {year} {2023})},\ \Eprint
  {http://arxiv.org/abs/2301.11940} {arXiv:2301.11940 [hep-ph]} \BibitemShut
  {NoStop}%
\bibitem [{\citenamefont {Li}\ \emph {et~al.}(2020{\natexlab{b}})\citenamefont
  {Li} \emph {et~al.}}]{Li:2020sru}%
  \BibitemOpen
  \bibfield  {author} {\bibinfo {author} {\bibfnamefont {X.}~\bibnamefont {Li}}
  \emph {et~al.},\ }\href {\doibase 10.1051/epjconf/202023504002} {\bibfield
  {journal} {\bibinfo  {journal} {EPJ Web Conf.}\ }\textbf {\bibinfo {volume}
  {235}},\ \bibinfo {pages} {04002} (\bibinfo {year} {2020}{\natexlab{b}})},\
  \Eprint {http://arxiv.org/abs/2002.05880} {arXiv:2002.05880 [nucl-ex]}
  \BibitemShut {NoStop}%
\bibitem [{\citenamefont {Li}\ \emph {et~al.}(2021{\natexlab{b}})\citenamefont
  {Li}, \citenamefont {Liu},\ and\ \citenamefont {Vitev}}]{Li:2020zbk}%
  \BibitemOpen
  \bibfield  {author} {\bibinfo {author} {\bibfnamefont {H.~T.}\ \bibnamefont
  {Li}}, \bibinfo {author} {\bibfnamefont {Z.~L.}\ \bibnamefont {Liu}}, \ and\
  \bibinfo {author} {\bibfnamefont {I.}~\bibnamefont {Vitev}},\ }\href
  {\doibase 10.1016/j.physletb.2021.136261} {\bibfield  {journal} {\bibinfo
  {journal} {Phys. Lett. B}\ }\textbf {\bibinfo {volume} {816}},\ \bibinfo
  {pages} {136261} (\bibinfo {year} {2021}{\natexlab{b}})},\ \Eprint
  {http://arxiv.org/abs/2007.10994} {arXiv:2007.10994 [hep-ph]} \BibitemShut
  {NoStop}%
\bibitem [{\citenamefont {Das}(2021)}]{Das:2021nqw}%
  \BibitemOpen
  \bibfield  {author} {\bibinfo {author} {\bibfnamefont {D.}~\bibnamefont
  {Das}},\ }\href {\doibase 10.1016/j.nuclphysa.2020.122132} {\bibfield
  {journal} {\bibinfo  {journal} {Nucl. Phys. A}\ }\textbf {\bibinfo {volume}
  {1007}},\ \bibinfo {pages} {122132} (\bibinfo {year} {2021})},\ \Eprint
  {http://arxiv.org/abs/2111.03926} {arXiv:2111.03926 [nucl-ex]} \BibitemShut
  {NoStop}%
\bibitem [{\citenamefont {Chang}\ \emph {et~al.}(2022)\citenamefont {Chang},
  \citenamefont {Aschenauer}, \citenamefont {Baker}, \citenamefont {Jentsch},
  \citenamefont {Lee}, \citenamefont {Tu}, \citenamefont {Yin},\ and\
  \citenamefont {Zheng}}]{Chang:2022hkt}%
  \BibitemOpen
  \bibfield  {author} {\bibinfo {author} {\bibfnamefont {W.}~\bibnamefont
  {Chang}}, \bibinfo {author} {\bibfnamefont {E.-C.}\ \bibnamefont
  {Aschenauer}}, \bibinfo {author} {\bibfnamefont {M.~D.}\ \bibnamefont
  {Baker}}, \bibinfo {author} {\bibfnamefont {A.}~\bibnamefont {Jentsch}},
  \bibinfo {author} {\bibfnamefont {J.-H.}\ \bibnamefont {Lee}}, \bibinfo
  {author} {\bibfnamefont {Z.}~\bibnamefont {Tu}}, \bibinfo {author}
  {\bibfnamefont {Z.}~\bibnamefont {Yin}}, \ and\ \bibinfo {author}
  {\bibfnamefont {L.}~\bibnamefont {Zheng}},\ }\href {\doibase
  10.1103/PhysRevD.106.012007} {\bibfield  {journal} {\bibinfo  {journal}
  {Phys. Rev. D}\ }\textbf {\bibinfo {volume} {106}},\ \bibinfo {pages}
  {012007} (\bibinfo {year} {2022})},\ \Eprint
  {http://arxiv.org/abs/2204.11998} {arXiv:2204.11998 [physics.comp-ph]}
  \BibitemShut {NoStop}%
\bibitem [{\citenamefont {{\'E}calle}(1981)}]{ecalle1981fonctions}%
  \BibitemOpen
  \bibfield  {author} {\bibinfo {author} {\bibfnamefont {J.}~\bibnamefont
  {{\'E}calle}},\ }\href@noop {} {\emph {\bibinfo {title} {Les fonctions
  r{\'e}surgentes:(en trois parties)}}},\ Vol.~\bibinfo {volume} {1}\ (\bibinfo
   {publisher} {Universit{\'e} de Paris-Sud, D{\'e}partement de
  Math{\'e}matique, B{\^a}t. 425},\ \bibinfo {year} {1981})\BibitemShut
  {NoStop}%
\bibitem [{\citenamefont {Schlomiuk}(2013)}]{schlomiuk2013bifurcations}%
  \BibitemOpen
  \bibfield  {author} {\bibinfo {author} {\bibfnamefont {D.}~\bibnamefont
  {Schlomiuk}},\ }\href@noop {} {\emph {\bibinfo {title} {Bifurcations and
  periodic orbits of Vector Fields}}},\ Vol.\ \bibinfo {volume} {408}\
  (\bibinfo  {publisher} {Springer Science \& Business Media},\ \bibinfo {year}
  {2013})\BibitemShut {NoStop}%
\bibitem [{\citenamefont {Connes}\ and\ \citenamefont
  {Marcolli}(2019)}]{connes2019noncommutative}%
  \BibitemOpen
  \bibfield  {author} {\bibinfo {author} {\bibfnamefont {A.}~\bibnamefont
  {Connes}}\ and\ \bibinfo {author} {\bibfnamefont {M.}~\bibnamefont
  {Marcolli}},\ }\href@noop {} {\emph {\bibinfo {title} {Noncommutative
  geometry, quantum fields and motives}}},\ Vol.~\bibinfo {volume} {55}\
  (\bibinfo  {publisher} {American Mathematical Soc.},\ \bibinfo {year}
  {2019})\BibitemShut {NoStop}%
\bibitem [{\citenamefont {Braaksma}\ \emph {et~al.}(2001)\citenamefont
  {Braaksma}, \citenamefont {Immink}, \citenamefont {Van~der Put},\ and\
  \citenamefont {Top}}]{braaksma2001differential}%
  \BibitemOpen
  \bibfield  {author} {\bibinfo {author} {\bibfnamefont {B.}~\bibnamefont
  {Braaksma}}, \bibinfo {author} {\bibfnamefont {G.}~\bibnamefont {Immink}},
  \bibinfo {author} {\bibfnamefont {M.}~\bibnamefont {Van~der Put}}, \ and\
  \bibinfo {author} {\bibfnamefont {J.}~\bibnamefont {Top}},\ }in\ \href@noop
  {} {\emph {\bibinfo {booktitle} {The Conference on Differential Equations and
  the Stokes Phenomenon}}},\ Vol.~\bibinfo {volume} {28}\ (\bibinfo
  {organization} {World Scientific},\ \bibinfo {year} {2001})\ p.~\bibinfo
  {pages} {30}\BibitemShut {NoStop}%
\bibitem [{\citenamefont {Choi}\ \emph {et~al.}(2003)\citenamefont {Choi} \emph
  {et~al.}}]{Belle:2003nnu}%
  \BibitemOpen
  \bibfield  {author} {\bibinfo {author} {\bibfnamefont {S.~K.}\ \bibnamefont
  {Choi}} \emph {et~al.} (\bibinfo {collaboration} {Belle}),\ }\href {\doibase
  10.1103/PhysRevLett.91.262001} {\bibfield  {journal} {\bibinfo  {journal}
  {Phys. Rev. Lett.}\ }\textbf {\bibinfo {volume} {91}},\ \bibinfo {pages}
  {262001} (\bibinfo {year} {2003})},\ \Eprint
  {http://arxiv.org/abs/hep-ex/0309032} {arXiv:hep-ex/0309032} \BibitemShut
  {NoStop}%
\bibitem [{\citenamefont {Ablikim}\ \emph {et~al.}(2013)\citenamefont {Ablikim}
  \emph {et~al.}}]{BESIII:2013ris}%
  \BibitemOpen
  \bibfield  {author} {\bibinfo {author} {\bibfnamefont {M.}~\bibnamefont
  {Ablikim}} \emph {et~al.} (\bibinfo {collaboration} {BESIII}),\ }\href
  {\doibase 10.1103/PhysRevLett.110.252001} {\bibfield  {journal} {\bibinfo
  {journal} {Phys. Rev. Lett.}\ }\textbf {\bibinfo {volume} {110}},\ \bibinfo
  {pages} {252001} (\bibinfo {year} {2013})},\ \Eprint
  {http://arxiv.org/abs/1303.5949} {arXiv:1303.5949 [hep-ex]} \BibitemShut
  {NoStop}%
\bibitem [{\citenamefont {Liu}\ \emph {et~al.}(2013)\citenamefont {Liu} \emph
  {et~al.}}]{Belle:2013yex}%
  \BibitemOpen
  \bibfield  {author} {\bibinfo {author} {\bibfnamefont {Z.~Q.}\ \bibnamefont
  {Liu}} \emph {et~al.} (\bibinfo {collaboration} {Belle}),\ }\href {\doibase
  10.1103/PhysRevLett.110.252002} {\bibfield  {journal} {\bibinfo  {journal}
  {Phys. Rev. Lett.}\ }\textbf {\bibinfo {volume} {110}},\ \bibinfo {pages}
  {252002} (\bibinfo {year} {2013})},\ \bibinfo {note} {[Erratum:
  Phys.Rev.Lett. 111, 019901 (2013)]},\ \Eprint
  {http://arxiv.org/abs/1304.0121} {arXiv:1304.0121 [hep-ex]} \BibitemShut
  {NoStop}%
\bibitem [{\citenamefont {Aaij}\ \emph {et~al.}(2015)\citenamefont {Aaij} \emph
  {et~al.}}]{LHCb:2015yax}%
  \BibitemOpen
  \bibfield  {author} {\bibinfo {author} {\bibfnamefont {R.}~\bibnamefont
  {Aaij}} \emph {et~al.} (\bibinfo {collaboration} {LHCb}),\ }\href {\doibase
  10.1103/PhysRevLett.115.072001} {\bibfield  {journal} {\bibinfo  {journal}
  {Phys. Rev. Lett.}\ }\textbf {\bibinfo {volume} {115}},\ \bibinfo {pages}
  {072001} (\bibinfo {year} {2015})},\ \Eprint
  {http://arxiv.org/abs/1507.03414} {arXiv:1507.03414 [hep-ex]} \BibitemShut
  {NoStop}%
\bibitem [{\citenamefont {Brambilla}\ \emph
  {et~al.}(2020{\natexlab{a}})\citenamefont {Brambilla}, \citenamefont
  {Eidelman}, \citenamefont {Hanhart}, \citenamefont {Nefediev}, \citenamefont
  {Shen}, \citenamefont {Thomas}, \citenamefont {Vairo},\ and\ \citenamefont
  {Yuan}}]{Brambilla:2019esw}%
  \BibitemOpen
  \bibfield  {author} {\bibinfo {author} {\bibfnamefont {N.}~\bibnamefont
  {Brambilla}}, \bibinfo {author} {\bibfnamefont {S.}~\bibnamefont {Eidelman}},
  \bibinfo {author} {\bibfnamefont {C.}~\bibnamefont {Hanhart}}, \bibinfo
  {author} {\bibfnamefont {A.}~\bibnamefont {Nefediev}}, \bibinfo {author}
  {\bibfnamefont {C.-P.}\ \bibnamefont {Shen}}, \bibinfo {author}
  {\bibfnamefont {C.~E.}\ \bibnamefont {Thomas}}, \bibinfo {author}
  {\bibfnamefont {A.}~\bibnamefont {Vairo}}, \ and\ \bibinfo {author}
  {\bibfnamefont {C.-Z.}\ \bibnamefont {Yuan}},\ }\href {\doibase
  10.1016/j.physrep.2020.05.001} {\bibfield  {journal} {\bibinfo  {journal}
  {Phys. Rept.}\ }\textbf {\bibinfo {volume} {873}},\ \bibinfo {pages} {1}
  (\bibinfo {year} {2020}{\natexlab{a}})},\ \Eprint
  {http://arxiv.org/abs/1907.07583} {arXiv:1907.07583 [hep-ex]} \BibitemShut
  {NoStop}%
\bibitem [{\citenamefont {Aaij}\ \emph {et~al.}(2020)\citenamefont {Aaij} \emph
  {et~al.}}]{LHCb:2020bwg}%
  \BibitemOpen
  \bibfield  {author} {\bibinfo {author} {\bibfnamefont {R.}~\bibnamefont
  {Aaij}} \emph {et~al.} (\bibinfo {collaboration} {LHCb}),\ }\href {\doibase
  10.1016/j.scib.2020.08.032} {\bibfield  {journal} {\bibinfo  {journal} {Sci.
  Bull.}\ }\textbf {\bibinfo {volume} {65}},\ \bibinfo {pages} {1983} (\bibinfo
  {year} {2020})},\ \Eprint {http://arxiv.org/abs/2006.16957} {arXiv:2006.16957
  [hep-ex]} \BibitemShut {NoStop}%
\bibitem [{CMS(2022)}]{CMS:2022yhl}%
  \BibitemOpen
  \href@noop {} {\  (\bibinfo {year} {2022})}\BibitemShut {NoStop}%
\bibitem [{ATL(2022)}]{ATLAS:2022hhx}%
  \BibitemOpen
  \href@noop {} {\  (\bibinfo {year} {2022})}\BibitemShut {NoStop}%
\bibitem [{\citenamefont {Aaij}\ \emph
  {et~al.}(2022{\natexlab{b}})\citenamefont {Aaij} \emph
  {et~al.}}]{LHCb:2021auc}%
  \BibitemOpen
  \bibfield  {author} {\bibinfo {author} {\bibfnamefont {R.}~\bibnamefont
  {Aaij}} \emph {et~al.} (\bibinfo {collaboration} {LHCb}),\ }\href {\doibase
  10.1038/s41467-022-30206-w} {\bibfield  {journal} {\bibinfo  {journal}
  {Nature Commun.}\ }\textbf {\bibinfo {volume} {13}},\ \bibinfo {pages} {3351}
  (\bibinfo {year} {2022}{\natexlab{b}})},\ \Eprint
  {http://arxiv.org/abs/2109.01056} {arXiv:2109.01056 [hep-ex]} \BibitemShut
  {NoStop}%
\bibitem [{\citenamefont {Barabanov}\ \emph {et~al.}(2021)\citenamefont
  {Barabanov} \emph {et~al.}}]{Barabanov:2020jvn}%
  \BibitemOpen
  \bibfield  {author} {\bibinfo {author} {\bibfnamefont {M.~Y.}\ \bibnamefont
  {Barabanov}} \emph {et~al.},\ }\href {\doibase 10.1016/j.ppnp.2020.103835}
  {\bibfield  {journal} {\bibinfo  {journal} {Prog. Part. Nucl. Phys.}\
  }\textbf {\bibinfo {volume} {116}},\ \bibinfo {pages} {103835} (\bibinfo
  {year} {2021})},\ \Eprint {http://arxiv.org/abs/2008.07630} {arXiv:2008.07630
  [hep-ph]} \BibitemShut {NoStop}%
\bibitem [{\citenamefont {Farina}\ \emph {et~al.}(2020)\citenamefont {Farina},
  \citenamefont {Garcia~Tecocoatzi}, \citenamefont {Giachino}, \citenamefont
  {Santopinto},\ and\ \citenamefont {Swanson}}]{Farina:2020slb}%
  \BibitemOpen
  \bibfield  {author} {\bibinfo {author} {\bibfnamefont {C.}~\bibnamefont
  {Farina}}, \bibinfo {author} {\bibfnamefont {H.}~\bibnamefont
  {Garcia~Tecocoatzi}}, \bibinfo {author} {\bibfnamefont {A.}~\bibnamefont
  {Giachino}}, \bibinfo {author} {\bibfnamefont {E.}~\bibnamefont
  {Santopinto}}, \ and\ \bibinfo {author} {\bibfnamefont {E.~S.}\ \bibnamefont
  {Swanson}},\ }\href {\doibase 10.1103/PhysRevD.102.014023} {\bibfield
  {journal} {\bibinfo  {journal} {Phys. Rev. D}\ }\textbf {\bibinfo {volume}
  {102}},\ \bibinfo {pages} {014023} (\bibinfo {year} {2020})},\ \Eprint
  {http://arxiv.org/abs/2005.10850} {arXiv:2005.10850 [hep-ph]} \BibitemShut
  {NoStop}%
\bibitem [{\citenamefont {Adhikari}\ \emph {et~al.}(2021)\citenamefont
  {Adhikari} \emph {et~al.}}]{Adhikari:2020cvz}%
  \BibitemOpen
  \bibfield  {author} {\bibinfo {author} {\bibfnamefont {S.}~\bibnamefont
  {Adhikari}} \emph {et~al.} (\bibinfo {collaboration} {GlueX}),\ }\href
  {\doibase 10.1016/j.nima.2020.164807} {\bibfield  {journal} {\bibinfo
  {journal} {Nucl. Instrum. Meth. A}\ }\textbf {\bibinfo {volume} {987}},\
  \bibinfo {pages} {164807} (\bibinfo {year} {2021})},\ \Eprint
  {http://arxiv.org/abs/2005.14272} {arXiv:2005.14272 [physics.ins-det]}
  \BibitemShut {NoStop}%
\bibitem [{\citenamefont {Albaladejo}\ \emph {et~al.}(2022)\citenamefont
  {Albaladejo} \emph {et~al.}}]{JPAC:2021rxu}%
  \BibitemOpen
  \bibfield  {author} {\bibinfo {author} {\bibfnamefont {M.}~\bibnamefont
  {Albaladejo}} \emph {et~al.} (\bibinfo {collaboration} {JPAC}),\ }\href
  {\doibase 10.1016/j.ppnp.2022.103981} {\bibfield  {journal} {\bibinfo
  {journal} {Prog. Part. Nucl. Phys.}\ }\textbf {\bibinfo {volume} {127}},\
  \bibinfo {pages} {103981} (\bibinfo {year} {2022})},\ \Eprint
  {http://arxiv.org/abs/2112.13436} {arXiv:2112.13436 [hep-ph]} \BibitemShut
  {NoStop}%
\bibitem [{\citenamefont {Arrington}\ \emph
  {et~al.}(2022{\natexlab{a}})\citenamefont {Arrington} \emph
  {et~al.}}]{Arrington:2021alx}%
  \BibitemOpen
  \bibfield  {author} {\bibinfo {author} {\bibfnamefont {J.}~\bibnamefont
  {Arrington}} \emph {et~al.},\ }\href {\doibase 10.1016/j.ppnp.2022.103985}
  {\bibfield  {journal} {\bibinfo  {journal} {Prog. Part. Nucl. Phys.}\
  }\textbf {\bibinfo {volume} {127}},\ \bibinfo {pages} {103985} (\bibinfo
  {year} {2022}{\natexlab{a}})},\ \Eprint {http://arxiv.org/abs/2112.00060}
  {arXiv:2112.00060 [nucl-ex]} \BibitemShut {NoStop}%
\bibitem [{\citenamefont {Lebed}\ \emph {et~al.}(2022)\citenamefont {Lebed}
  \emph {et~al.}}]{Lebed:2022vfu}%
  \BibitemOpen
  \bibfield  {author} {\bibinfo {author} {\bibfnamefont {R.~F.}\ \bibnamefont
  {Lebed}} \emph {et~al.},\ }in\ \href@noop {} {\emph {\bibinfo {booktitle}
  {{2022 Snowmass Summer Study}}}},\ \bibinfo {editor} {edited by\ \bibinfo
  {editor} {\bibfnamefont {R.~F.}\ \bibnamefont {Lebed}}\ and\ \bibinfo
  {editor} {\bibfnamefont {T.}~\bibnamefont {Skwarnicki}}}\ (\bibinfo {year}
  {2022})\ \Eprint {http://arxiv.org/abs/2207.14594} {arXiv:2207.14594
  [hep-ph]} \BibitemShut {NoStop}%
\bibitem [{\citenamefont {Klein}\ and\ \citenamefont
  {Xie}(2019)}]{Klein:2019avl}%
  \BibitemOpen
  \bibfield  {author} {\bibinfo {author} {\bibfnamefont {S.~R.}\ \bibnamefont
  {Klein}}\ and\ \bibinfo {author} {\bibfnamefont {Y.-P.}\ \bibnamefont
  {Xie}},\ }\href {\doibase 10.1103/PhysRevC.100.024620} {\bibfield  {journal}
  {\bibinfo  {journal} {Phys. Rev. C}\ }\textbf {\bibinfo {volume} {100}},\
  \bibinfo {pages} {024620} (\bibinfo {year} {2019})},\ \Eprint
  {http://arxiv.org/abs/1903.02680} {arXiv:1903.02680 [nucl-th]} \BibitemShut
  {NoStop}%
\bibitem [{\citenamefont {Albaladejo}\ \emph {et~al.}(2020)\citenamefont
  {Albaladejo}, \citenamefont {Blin}, \citenamefont {Pilloni}, \citenamefont
  {Winney}, \citenamefont {Fern\'andez-Ram\'\i{}rez}, \citenamefont {Mathieu},\
  and\ \citenamefont {Szczepaniak}}]{Albaladejo:2020tzt}%
  \BibitemOpen
  \bibfield  {author} {\bibinfo {author} {\bibfnamefont {M.}~\bibnamefont
  {Albaladejo}}, \bibinfo {author} {\bibfnamefont {A.~N.~H.}\ \bibnamefont
  {Blin}}, \bibinfo {author} {\bibfnamefont {A.}~\bibnamefont {Pilloni}},
  \bibinfo {author} {\bibfnamefont {D.}~\bibnamefont {Winney}}, \bibinfo
  {author} {\bibfnamefont {C.}~\bibnamefont {Fern\'andez-Ram\'\i{}rez}},
  \bibinfo {author} {\bibfnamefont {V.}~\bibnamefont {Mathieu}}, \ and\
  \bibinfo {author} {\bibfnamefont {A.}~\bibnamefont {Szczepaniak}} (\bibinfo
  {collaboration} {JPAC}),\ }\href {\doibase 10.1103/PhysRevD.102.114010}
  {\bibfield  {journal} {\bibinfo  {journal} {Phys. Rev. D}\ }\textbf {\bibinfo
  {volume} {102}},\ \bibinfo {pages} {114010} (\bibinfo {year} {2020})},\
  \Eprint {http://arxiv.org/abs/2008.01001} {arXiv:2008.01001 [hep-ph]}
  \BibitemShut {NoStop}%
\bibitem [{\citenamefont {Winney}\ \emph {et~al.}(2022)\citenamefont {Winney},
  \citenamefont {Pilloni}, \citenamefont {Mathieu}, \citenamefont
  {Hiller~Blin}, \citenamefont {Albaladejo}, \citenamefont {Smith},\ and\
  \citenamefont {Szczepaniak}}]{Winney:2022tky}%
  \BibitemOpen
  \bibfield  {author} {\bibinfo {author} {\bibfnamefont {D.}~\bibnamefont
  {Winney}}, \bibinfo {author} {\bibfnamefont {A.}~\bibnamefont {Pilloni}},
  \bibinfo {author} {\bibfnamefont {V.}~\bibnamefont {Mathieu}}, \bibinfo
  {author} {\bibfnamefont {A.~N.}\ \bibnamefont {Hiller~Blin}}, \bibinfo
  {author} {\bibfnamefont {M.}~\bibnamefont {Albaladejo}}, \bibinfo {author}
  {\bibfnamefont {W.~A.}\ \bibnamefont {Smith}}, \ and\ \bibinfo {author}
  {\bibfnamefont {A.}~\bibnamefont {Szczepaniak}} (\bibinfo {collaboration}
  {JPAC}),\ }\href {\doibase 10.1103/PhysRevD.106.094009} {\bibfield  {journal}
  {\bibinfo  {journal} {Phys. Rev. D}\ }\textbf {\bibinfo {volume} {106}},\
  \bibinfo {pages} {094009} (\bibinfo {year} {2022})},\ \Eprint
  {http://arxiv.org/abs/2209.05882} {arXiv:2209.05882 [hep-ph]} \BibitemShut
  {NoStop}%
\bibitem [{\citenamefont {Dzierba}\ \emph {et~al.}(2005)\citenamefont
  {Dzierba}, \citenamefont {Meyer},\ and\ \citenamefont
  {Szczepaniak}}]{Dzierba:2004db}%
  \BibitemOpen
  \bibfield  {author} {\bibinfo {author} {\bibfnamefont {A.~R.}\ \bibnamefont
  {Dzierba}}, \bibinfo {author} {\bibfnamefont {C.~A.}\ \bibnamefont {Meyer}},
  \ and\ \bibinfo {author} {\bibfnamefont {A.~P.}\ \bibnamefont
  {Szczepaniak}},\ }\href {\doibase 10.1088/1742-6596/9/1/036} {\bibfield
  {journal} {\bibinfo  {journal} {J. Phys. Conf. Ser.}\ }\textbf {\bibinfo
  {volume} {9}},\ \bibinfo {pages} {192} (\bibinfo {year} {2005})},\ \Eprint
  {http://arxiv.org/abs/hep-ex/0412077} {arXiv:hep-ex/0412077} \BibitemShut
  {NoStop}%
\bibitem [{\citenamefont {Rodas}\ \emph {et~al.}(2019)\citenamefont {Rodas}
  \emph {et~al.}}]{JPAC:2018zyd}%
  \BibitemOpen
  \bibfield  {author} {\bibinfo {author} {\bibfnamefont {A.}~\bibnamefont
  {Rodas}} \emph {et~al.} (\bibinfo {collaboration} {JPAC}),\ }\href {\doibase
  10.1103/PhysRevLett.122.042002} {\bibfield  {journal} {\bibinfo  {journal}
  {Phys. Rev. Lett.}\ }\textbf {\bibinfo {volume} {122}},\ \bibinfo {pages}
  {042002} (\bibinfo {year} {2019})},\ \Eprint
  {http://arxiv.org/abs/1810.04171} {arXiv:1810.04171 [hep-ph]} \BibitemShut
  {NoStop}%
\bibitem [{jpa()}]{jpacwebsite}%
  \BibitemOpen
  \href@noop {} {\enquote {\bibinfo {title} {Joint physics analysis center},}\
  }\bibinfo {howpublished} {\url{http://jpac-physics.org}}\BibitemShut
  {NoStop}%
\bibitem [{qwg()}]{qwgsite}%
  \BibitemOpen
  \href@noop {} {\enquote {\bibinfo {title} {Quarkonium working group},}\
  }\bibinfo {howpublished} {\url{https://qwg.ph.nat.tum.de/}}\BibitemShut
  {NoStop}%
\bibitem [{\citenamefont {Glazier}\ \emph {et~al.}(2020)\citenamefont {Glazier}
  \emph {et~al.}}]{GitHub:elspectro}%
  \BibitemOpen
  \bibfield  {author} {\bibinfo {author} {\bibfnamefont {D.}~\bibnamefont
  {Glazier}} \emph {et~al.},\ }\href@noop {} {\enquote {\bibinfo {title}
  {{elSpectro: an event generator framework for incorporating spectroscopy into
  electro/photoproduction reactions}},}\ }\bibinfo {howpublished}
  {\url{https://github.com/dglazier/elSpectro}} (\bibinfo {year}
  {2020})\BibitemShut {NoStop}%
\bibitem [{\citenamefont {Sombillo}\ \emph {et~al.}(2020)\citenamefont
  {Sombillo}, \citenamefont {Ikeda}, \citenamefont {Sato},\ and\ \citenamefont
  {Hosaka}}]{Sombillo:2020ccg}%
  \BibitemOpen
  \bibfield  {author} {\bibinfo {author} {\bibfnamefont {D.~L.~B.}\
  \bibnamefont {Sombillo}}, \bibinfo {author} {\bibfnamefont {Y.}~\bibnamefont
  {Ikeda}}, \bibinfo {author} {\bibfnamefont {T.}~\bibnamefont {Sato}}, \ and\
  \bibinfo {author} {\bibfnamefont {A.}~\bibnamefont {Hosaka}},\ }\href
  {\doibase 10.1103/PhysRevD.102.016024} {\bibfield  {journal} {\bibinfo
  {journal} {Phys. Rev. D}\ }\textbf {\bibinfo {volume} {102}},\ \bibinfo
  {pages} {016024} (\bibinfo {year} {2020})},\ \Eprint
  {http://arxiv.org/abs/2003.10770} {arXiv:2003.10770 [hep-ph]} \BibitemShut
  {NoStop}%
\bibitem [{\citenamefont {Sombillo}\ \emph {et~al.}(2021)\citenamefont
  {Sombillo}, \citenamefont {Ikeda}, \citenamefont {Sato},\ and\ \citenamefont
  {Hosaka}}]{Sombillo:2021rxv}%
  \BibitemOpen
  \bibfield  {author} {\bibinfo {author} {\bibfnamefont {D.~L.~B.}\
  \bibnamefont {Sombillo}}, \bibinfo {author} {\bibfnamefont {Y.}~\bibnamefont
  {Ikeda}}, \bibinfo {author} {\bibfnamefont {T.}~\bibnamefont {Sato}}, \ and\
  \bibinfo {author} {\bibfnamefont {A.}~\bibnamefont {Hosaka}},\ }\href
  {\doibase 10.1103/PhysRevD.104.036001} {\bibfield  {journal} {\bibinfo
  {journal} {Phys. Rev. D}\ }\textbf {\bibinfo {volume} {104}},\ \bibinfo
  {pages} {036001} (\bibinfo {year} {2021})},\ \Eprint
  {http://arxiv.org/abs/2105.04898} {arXiv:2105.04898 [hep-ph]} \BibitemShut
  {NoStop}%
\bibitem [{\citenamefont {Ng}\ \emph {et~al.}(2022)\citenamefont {Ng},
  \citenamefont {Bibrzycki}, \citenamefont {Nys}, \citenamefont
  {Fern\'andez-Ram\'irez}, \citenamefont {Pilloni}, \citenamefont {Mathieu},
  \citenamefont {Rasmusson},\ and\ \citenamefont {Szczepaniak}}]{Ng:2021ibr}%
  \BibitemOpen
  \bibfield  {author} {\bibinfo {author} {\bibfnamefont {L.}~\bibnamefont
  {Ng}}, \bibinfo {author} {\bibfnamefont {L.}~\bibnamefont {Bibrzycki}},
  \bibinfo {author} {\bibfnamefont {J.}~\bibnamefont {Nys}}, \bibinfo {author}
  {\bibfnamefont {C.}~\bibnamefont {Fern\'andez-Ram\'irez}}, \bibinfo {author}
  {\bibfnamefont {A.}~\bibnamefont {Pilloni}}, \bibinfo {author} {\bibfnamefont
  {V.}~\bibnamefont {Mathieu}}, \bibinfo {author} {\bibfnamefont {A.~J.}\
  \bibnamefont {Rasmusson}}, \ and\ \bibinfo {author} {\bibfnamefont {A.~P.}\
  \bibnamefont {Szczepaniak}} (\bibinfo {collaboration} {Joint Physics Analysis
  Center, JPAC}),\ }\href {\doibase 10.1103/PhysRevD.105.L091501} {\bibfield
  {journal} {\bibinfo  {journal} {Phys. Rev. D}\ }\textbf {\bibinfo {volume}
  {105}},\ \bibinfo {pages} {L091501} (\bibinfo {year} {2022})},\ \Eprint
  {http://arxiv.org/abs/2110.13742} {arXiv:2110.13742 [hep-ph]} \BibitemShut
  {NoStop}%
\bibitem [{\citenamefont {Gupta}\ \emph {et~al.}(2022)\citenamefont {Gupta},
  \citenamefont {Bhattacharya},\ and\ \citenamefont {Yoon}}]{Gupta:2022vhe}%
  \BibitemOpen
  \bibfield  {author} {\bibinfo {author} {\bibfnamefont {R.}~\bibnamefont
  {Gupta}}, \bibinfo {author} {\bibfnamefont {T.}~\bibnamefont {Bhattacharya}},
  \ and\ \bibinfo {author} {\bibfnamefont {B.}~\bibnamefont {Yoon}},\
  }\href@noop {} {\  (\bibinfo {year} {2022})},\ \Eprint
  {http://arxiv.org/abs/2205.05803} {arXiv:2205.05803 [hep-lat]} \BibitemShut
  {NoStop}%
\bibitem [{\citenamefont {Chen}\ \emph {et~al.}(2023)\citenamefont {Chen},
  \citenamefont {Chen}, \citenamefont {Niu},\ and\ \citenamefont
  {Zheng}}]{Chen:2023xkz}%
  \BibitemOpen
  \bibfield  {author} {\bibinfo {author} {\bibfnamefont {C.}~\bibnamefont
  {Chen}}, \bibinfo {author} {\bibfnamefont {H.}~\bibnamefont {Chen}}, \bibinfo
  {author} {\bibfnamefont {W.-Q.}\ \bibnamefont {Niu}}, \ and\ \bibinfo
  {author} {\bibfnamefont {H.-Q.}\ \bibnamefont {Zheng}},\ }\href {\doibase
  10.1140/epjc/s10052-023-11170-1} {\bibfield  {journal} {\bibinfo  {journal}
  {Eur. Phys. J. C}\ }\textbf {\bibinfo {volume} {83}},\ \bibinfo {pages} {52}
  (\bibinfo {year} {2023})}\BibitemShut {NoStop}%
\bibitem [{\citenamefont {Zhang}\ \emph {et~al.}(2023)\citenamefont {Zhang},
  \citenamefont {Liu}, \citenamefont {Hu}, \citenamefont {Wang},\ and\
  \citenamefont {Mei\ss{}ner}}]{Zhang:2023czx}%
  \BibitemOpen
  \bibfield  {author} {\bibinfo {author} {\bibfnamefont {Z.}~\bibnamefont
  {Zhang}}, \bibinfo {author} {\bibfnamefont {J.}~\bibnamefont {Liu}}, \bibinfo
  {author} {\bibfnamefont {J.}~\bibnamefont {Hu}}, \bibinfo {author}
  {\bibfnamefont {Q.}~\bibnamefont {Wang}}, \ and\ \bibinfo {author}
  {\bibfnamefont {U.-G.}\ \bibnamefont {Mei\ss{}ner}},\ }\href@noop {} {\
  (\bibinfo {year} {2023})},\ \Eprint {http://arxiv.org/abs/2301.05364}
  {arXiv:2301.05364 [hep-ph]} \BibitemShut {NoStop}%
\bibitem [{\citenamefont {Braaten}\ \emph {et~al.}(2014)\citenamefont
  {Braaten}, \citenamefont {Langmack},\ and\ \citenamefont
  {Smith}}]{Braaten:2014qka}%
  \BibitemOpen
  \bibfield  {author} {\bibinfo {author} {\bibfnamefont {E.}~\bibnamefont
  {Braaten}}, \bibinfo {author} {\bibfnamefont {C.}~\bibnamefont {Langmack}}, \
  and\ \bibinfo {author} {\bibfnamefont {D.~H.}\ \bibnamefont {Smith}},\ }\href
  {\doibase 10.1103/PhysRevD.90.014044} {\bibfield  {journal} {\bibinfo
  {journal} {Phys. Rev. D}\ }\textbf {\bibinfo {volume} {90}},\ \bibinfo
  {pages} {014044} (\bibinfo {year} {2014})},\ \Eprint
  {http://arxiv.org/abs/1402.0438} {arXiv:1402.0438 [hep-ph]} \BibitemShut
  {NoStop}%
\bibitem [{\citenamefont {Guo}\ \emph {et~al.}(2018)\citenamefont {Guo},
  \citenamefont {Hanhart}, \citenamefont {Mei\ss{}ner}, \citenamefont {Wang},
  \citenamefont {Zhao},\ and\ \citenamefont {Zou}}]{Guo:2017jvc}%
  \BibitemOpen
  \bibfield  {author} {\bibinfo {author} {\bibfnamefont {F.-K.}\ \bibnamefont
  {Guo}}, \bibinfo {author} {\bibfnamefont {C.}~\bibnamefont {Hanhart}},
  \bibinfo {author} {\bibfnamefont {U.-G.}\ \bibnamefont {Mei\ss{}ner}},
  \bibinfo {author} {\bibfnamefont {Q.}~\bibnamefont {Wang}}, \bibinfo {author}
  {\bibfnamefont {Q.}~\bibnamefont {Zhao}}, \ and\ \bibinfo {author}
  {\bibfnamefont {B.-S.}\ \bibnamefont {Zou}},\ }\href {\doibase
  10.1103/RevModPhys.90.015004} {\bibfield  {journal} {\bibinfo  {journal}
  {Rev. Mod. Phys.}\ }\textbf {\bibinfo {volume} {90}},\ \bibinfo {pages}
  {015004} (\bibinfo {year} {2018})},\ \bibinfo {note} {[Erratum: Rev.Mod.Phys.
  94, 029901 (2022)]},\ \Eprint {http://arxiv.org/abs/1705.00141}
  {arXiv:1705.00141 [hep-ph]} \BibitemShut {NoStop}%
\bibitem [{\citenamefont {Brambilla}\ \emph
  {et~al.}(2018{\natexlab{b}})\citenamefont {Brambilla}, \citenamefont {Krein},
  \citenamefont {Tarr\'us~Castell\`a},\ and\ \citenamefont
  {Vairo}}]{Brambilla:2017uyf}%
  \BibitemOpen
  \bibfield  {author} {\bibinfo {author} {\bibfnamefont {N.}~\bibnamefont
  {Brambilla}}, \bibinfo {author} {\bibfnamefont {G.~a.}\ \bibnamefont
  {Krein}}, \bibinfo {author} {\bibfnamefont {J.}~\bibnamefont
  {Tarr\'us~Castell\`a}}, \ and\ \bibinfo {author} {\bibfnamefont
  {A.}~\bibnamefont {Vairo}},\ }\href {\doibase 10.1103/PhysRevD.97.016016}
  {\bibfield  {journal} {\bibinfo  {journal} {Phys. Rev. D}\ }\textbf {\bibinfo
  {volume} {97}},\ \bibinfo {pages} {016016} (\bibinfo {year}
  {2018}{\natexlab{b}})},\ \Eprint {http://arxiv.org/abs/1707.09647}
  {arXiv:1707.09647 [hep-ph]} \BibitemShut {NoStop}%
\bibitem [{\citenamefont {Brambilla}\ \emph
  {et~al.}(2020{\natexlab{b}})\citenamefont {Brambilla}, \citenamefont {Lai},
  \citenamefont {Segovia},\ and\ \citenamefont
  {Tarr\'us~Castell\`a}}]{Brambilla:2019jfi}%
  \BibitemOpen
  \bibfield  {author} {\bibinfo {author} {\bibfnamefont {N.}~\bibnamefont
  {Brambilla}}, \bibinfo {author} {\bibfnamefont {W.~K.}\ \bibnamefont {Lai}},
  \bibinfo {author} {\bibfnamefont {J.}~\bibnamefont {Segovia}}, \ and\
  \bibinfo {author} {\bibfnamefont {J.}~\bibnamefont {Tarr\'us~Castell\`a}},\
  }\href {\doibase 10.1103/PhysRevD.101.054040} {\bibfield  {journal} {\bibinfo
   {journal} {Phys. Rev. D}\ }\textbf {\bibinfo {volume} {101}},\ \bibinfo
  {pages} {054040} (\bibinfo {year} {2020}{\natexlab{b}})},\ \Eprint
  {http://arxiv.org/abs/1908.11699} {arXiv:1908.11699 [hep-ph]} \BibitemShut
  {NoStop}%
\bibitem [{\citenamefont {Eichten}\ \emph {et~al.}(2006)\citenamefont
  {Eichten}, \citenamefont {Lane},\ and\ \citenamefont
  {Quigg}}]{Eichten:2005ga}%
  \BibitemOpen
  \bibfield  {author} {\bibinfo {author} {\bibfnamefont {E.~J.}\ \bibnamefont
  {Eichten}}, \bibinfo {author} {\bibfnamefont {K.}~\bibnamefont {Lane}}, \
  and\ \bibinfo {author} {\bibfnamefont {C.}~\bibnamefont {Quigg}},\ }\href
  {\doibase 10.1103/PhysRevD.73.014014} {\bibfield  {journal} {\bibinfo
  {journal} {Phys. Rev. D}\ }\textbf {\bibinfo {volume} {73}},\ \bibinfo
  {pages} {014014} (\bibinfo {year} {2006})},\ \bibinfo {note} {[Erratum:
  Phys.Rev.D 73, 079903 (2006)]},\ \Eprint
  {http://arxiv.org/abs/hep-ph/0511179} {arXiv:hep-ph/0511179} \BibitemShut
  {NoStop}%
\bibitem [{\citenamefont {Kalashnikova}(2005)}]{Kalashnikova:2005ui}%
  \BibitemOpen
  \bibfield  {author} {\bibinfo {author} {\bibfnamefont {Y.~S.}\ \bibnamefont
  {Kalashnikova}},\ }\href {\doibase 10.1103/PhysRevD.72.034010} {\bibfield
  {journal} {\bibinfo  {journal} {Phys. Rev. D}\ }\textbf {\bibinfo {volume}
  {72}},\ \bibinfo {pages} {034010} (\bibinfo {year} {2005})},\ \Eprint
  {http://arxiv.org/abs/hep-ph/0506270} {arXiv:hep-ph/0506270} \BibitemShut
  {NoStop}%
\bibitem [{\citenamefont {Santopinto}\ \emph {et~al.}(2011)\citenamefont
  {Santopinto}, \citenamefont {Bijker},\ and\ \citenamefont
  {Ferretti}}]{Santopinto:2011zza}%
  \BibitemOpen
  \bibfield  {author} {\bibinfo {author} {\bibfnamefont {E.}~\bibnamefont
  {Santopinto}}, \bibinfo {author} {\bibfnamefont {R.}~\bibnamefont {Bijker}},
  \ and\ \bibinfo {author} {\bibfnamefont {J.}~\bibnamefont {Ferretti}},\
  }\href {\doibase 10.1007/s00601-010-0167-8} {\bibfield  {journal} {\bibinfo
  {journal} {Few Body Syst.}\ }\textbf {\bibinfo {volume} {50}},\ \bibinfo
  {pages} {199} (\bibinfo {year} {2011})}\BibitemShut {NoStop}%
\bibitem [{\citenamefont {Ferretti}\ \emph {et~al.}(2013)\citenamefont
  {Ferretti}, \citenamefont {Galat\`a},\ and\ \citenamefont
  {Santopinto}}]{Ferretti:2013faa}%
  \BibitemOpen
  \bibfield  {author} {\bibinfo {author} {\bibfnamefont {J.}~\bibnamefont
  {Ferretti}}, \bibinfo {author} {\bibfnamefont {G.}~\bibnamefont {Galat\`a}},
  \ and\ \bibinfo {author} {\bibfnamefont {E.}~\bibnamefont {Santopinto}},\
  }\href {\doibase 10.1103/PhysRevC.88.015207} {\bibfield  {journal} {\bibinfo
  {journal} {Phys. Rev. C}\ }\textbf {\bibinfo {volume} {88}},\ \bibinfo
  {pages} {015207} (\bibinfo {year} {2013})},\ \Eprint
  {http://arxiv.org/abs/1302.6857} {arXiv:1302.6857 [hep-ph]} \BibitemShut
  {NoStop}%
\bibitem [{\citenamefont {Segovia}\ \emph {et~al.}(2016)\citenamefont
  {Segovia}, \citenamefont {Ortega}, \citenamefont {Entem},\ and\ \citenamefont
  {Fern\'andez}}]{Segovia:2016xqb}%
  \BibitemOpen
  \bibfield  {author} {\bibinfo {author} {\bibfnamefont {J.}~\bibnamefont
  {Segovia}}, \bibinfo {author} {\bibfnamefont {P.~G.}\ \bibnamefont {Ortega}},
  \bibinfo {author} {\bibfnamefont {D.~R.}\ \bibnamefont {Entem}}, \ and\
  \bibinfo {author} {\bibfnamefont {F.}~\bibnamefont {Fern\'andez}},\ }\href
  {\doibase 10.1103/PhysRevD.93.074027} {\bibfield  {journal} {\bibinfo
  {journal} {Phys. Rev. D}\ }\textbf {\bibinfo {volume} {93}},\ \bibinfo
  {pages} {074027} (\bibinfo {year} {2016})},\ \Eprint
  {http://arxiv.org/abs/1601.05093} {arXiv:1601.05093 [hep-ph]} \BibitemShut
  {NoStop}%
\bibitem [{\citenamefont {Ortega}\ and\ \citenamefont
  {Entem}(2021)}]{Ortega:2020tng}%
  \BibitemOpen
  \bibfield  {author} {\bibinfo {author} {\bibfnamefont {P.~G.}\ \bibnamefont
  {Ortega}}\ and\ \bibinfo {author} {\bibfnamefont {D.~R.}\ \bibnamefont
  {Entem}},\ }\href {\doibase 10.3390/sym13020279} {\bibfield  {journal}
  {\bibinfo  {journal} {Symmetry}\ }\textbf {\bibinfo {volume} {13}},\ \bibinfo
  {pages} {279} (\bibinfo {year} {2021})},\ \Eprint
  {http://arxiv.org/abs/2012.10105} {arXiv:2012.10105 [hep-ph]} \BibitemShut
  {NoStop}%
\bibitem [{\citenamefont {Eichmann}\ \emph
  {et~al.}(2016{\natexlab{a}})\citenamefont {Eichmann}, \citenamefont
  {Fischer},\ and\ \citenamefont {Heupel}}]{Eichmann:2015cra}%
  \BibitemOpen
  \bibfield  {author} {\bibinfo {author} {\bibfnamefont {G.}~\bibnamefont
  {Eichmann}}, \bibinfo {author} {\bibfnamefont {C.~S.}\ \bibnamefont
  {Fischer}}, \ and\ \bibinfo {author} {\bibfnamefont {W.}~\bibnamefont
  {Heupel}},\ }\href {\doibase 10.1016/j.physletb.2015.12.036} {\bibfield
  {journal} {\bibinfo  {journal} {Phys. Lett. B}\ }\textbf {\bibinfo {volume}
  {753}},\ \bibinfo {pages} {282} (\bibinfo {year} {2016}{\natexlab{a}})},\
  \Eprint {http://arxiv.org/abs/1508.07178} {arXiv:1508.07178 [hep-ph]}
  \BibitemShut {NoStop}%
\bibitem [{\citenamefont {Eichmann}\ \emph
  {et~al.}(2016{\natexlab{b}})\citenamefont {Eichmann}, \citenamefont
  {Sanchis-Alepuz}, \citenamefont {Williams}, \citenamefont {Alkofer},\ and\
  \citenamefont {Fischer}}]{Eichmann:2016yit}%
  \BibitemOpen
  \bibfield  {author} {\bibinfo {author} {\bibfnamefont {G.}~\bibnamefont
  {Eichmann}}, \bibinfo {author} {\bibfnamefont {H.}~\bibnamefont
  {Sanchis-Alepuz}}, \bibinfo {author} {\bibfnamefont {R.}~\bibnamefont
  {Williams}}, \bibinfo {author} {\bibfnamefont {R.}~\bibnamefont {Alkofer}}, \
  and\ \bibinfo {author} {\bibfnamefont {C.~S.}\ \bibnamefont {Fischer}},\
  }\href {\doibase 10.1016/j.ppnp.2016.07.001} {\bibfield  {journal} {\bibinfo
  {journal} {Prog. Part. Nucl. Phys.}\ }\textbf {\bibinfo {volume} {91}},\
  \bibinfo {pages} {1} (\bibinfo {year} {2016}{\natexlab{b}})},\ \Eprint
  {http://arxiv.org/abs/1606.09602} {arXiv:1606.09602 [hep-ph]} \BibitemShut
  {NoStop}%
\bibitem [{\citenamefont {Wallbott}\ \emph {et~al.}(2019)\citenamefont
  {Wallbott}, \citenamefont {Eichmann},\ and\ \citenamefont
  {Fischer}}]{Wallbott:2019dng}%
  \BibitemOpen
  \bibfield  {author} {\bibinfo {author} {\bibfnamefont {P.~C.}\ \bibnamefont
  {Wallbott}}, \bibinfo {author} {\bibfnamefont {G.}~\bibnamefont {Eichmann}},
  \ and\ \bibinfo {author} {\bibfnamefont {C.~S.}\ \bibnamefont {Fischer}},\
  }\href {\doibase 10.1103/PhysRevD.100.014033} {\bibfield  {journal} {\bibinfo
   {journal} {Phys. Rev. D}\ }\textbf {\bibinfo {volume} {100}},\ \bibinfo
  {pages} {014033} (\bibinfo {year} {2019})},\ \Eprint
  {http://arxiv.org/abs/1905.02615} {arXiv:1905.02615 [hep-ph]} \BibitemShut
  {NoStop}%
\bibitem [{\citenamefont {Briceno}\ \emph {et~al.}(2018)\citenamefont
  {Briceno}, \citenamefont {Dudek},\ and\ \citenamefont
  {Young}}]{Briceno:2017max}%
  \BibitemOpen
  \bibfield  {author} {\bibinfo {author} {\bibfnamefont {R.~A.}\ \bibnamefont
  {Briceno}}, \bibinfo {author} {\bibfnamefont {J.~J.}\ \bibnamefont {Dudek}},
  \ and\ \bibinfo {author} {\bibfnamefont {R.~D.}\ \bibnamefont {Young}},\
  }\href {\doibase 10.1103/RevModPhys.90.025001} {\bibfield  {journal}
  {\bibinfo  {journal} {Rev. Mod. Phys.}\ }\textbf {\bibinfo {volume} {90}},\
  \bibinfo {pages} {025001} (\bibinfo {year} {2018})},\ \Eprint
  {http://arxiv.org/abs/1706.06223} {arXiv:1706.06223 [hep-lat]} \BibitemShut
  {NoStop}%
\bibitem [{\citenamefont {Cho}\ \emph {et~al.}(2011)\citenamefont {Cho} \emph
  {et~al.}}]{ExHIC:2010gcb}%
  \BibitemOpen
  \bibfield  {author} {\bibinfo {author} {\bibfnamefont {S.}~\bibnamefont
  {Cho}} \emph {et~al.} (\bibinfo {collaboration} {ExHIC}),\ }\href {\doibase
  10.1103/PhysRevLett.106.212001} {\bibfield  {journal} {\bibinfo  {journal}
  {Phys. Rev. Lett.}\ }\textbf {\bibinfo {volume} {106}},\ \bibinfo {pages}
  {212001} (\bibinfo {year} {2011})},\ \Eprint {http://arxiv.org/abs/1011.0852}
  {arXiv:1011.0852 [nucl-th]} \BibitemShut {NoStop}%
\bibitem [{\citenamefont {Cho}\ \emph {et~al.}(2017)\citenamefont {Cho} \emph
  {et~al.}}]{ExHIC:2017smd}%
  \BibitemOpen
  \bibfield  {author} {\bibinfo {author} {\bibfnamefont {S.}~\bibnamefont
  {Cho}} \emph {et~al.} (\bibinfo {collaboration} {ExHIC}),\ }\href {\doibase
  10.1016/j.ppnp.2017.02.002} {\bibfield  {journal} {\bibinfo  {journal} {Prog.
  Part. Nucl. Phys.}\ }\textbf {\bibinfo {volume} {95}},\ \bibinfo {pages}
  {279} (\bibinfo {year} {2017})},\ \Eprint {http://arxiv.org/abs/1702.00486}
  {arXiv:1702.00486 [nucl-th]} \BibitemShut {NoStop}%
\bibitem [{\citenamefont {Zhang}\ \emph {et~al.}(2021)\citenamefont {Zhang},
  \citenamefont {Liao}, \citenamefont {Wang}, \citenamefont {Wang},\ and\
  \citenamefont {Xing}}]{Zhang:2020dwn}%
  \BibitemOpen
  \bibfield  {author} {\bibinfo {author} {\bibfnamefont {H.}~\bibnamefont
  {Zhang}}, \bibinfo {author} {\bibfnamefont {J.}~\bibnamefont {Liao}},
  \bibinfo {author} {\bibfnamefont {E.}~\bibnamefont {Wang}}, \bibinfo {author}
  {\bibfnamefont {Q.}~\bibnamefont {Wang}}, \ and\ \bibinfo {author}
  {\bibfnamefont {H.}~\bibnamefont {Xing}},\ }\href {\doibase
  10.1103/PhysRevLett.126.012301} {\bibfield  {journal} {\bibinfo  {journal}
  {Phys. Rev. Lett.}\ }\textbf {\bibinfo {volume} {126}},\ \bibinfo {pages}
  {012301} (\bibinfo {year} {2021})},\ \Eprint
  {http://arxiv.org/abs/2004.00024} {arXiv:2004.00024 [hep-ph]} \BibitemShut
  {NoStop}%
\bibitem [{\citenamefont {Wu}\ \emph {et~al.}(2021)\citenamefont {Wu},
  \citenamefont {Du}, \citenamefont {Sibila},\ and\ \citenamefont
  {Rapp}}]{Wu:2020zbx}%
  \BibitemOpen
  \bibfield  {author} {\bibinfo {author} {\bibfnamefont {B.}~\bibnamefont
  {Wu}}, \bibinfo {author} {\bibfnamefont {X.}~\bibnamefont {Du}}, \bibinfo
  {author} {\bibfnamefont {M.}~\bibnamefont {Sibila}}, \ and\ \bibinfo {author}
  {\bibfnamefont {R.}~\bibnamefont {Rapp}},\ }\href {\doibase
  10.1140/epja/s10050-021-00623-4} {\bibfield  {journal} {\bibinfo  {journal}
  {Eur. Phys. J. A}\ }\textbf {\bibinfo {volume} {57}},\ \bibinfo {pages} {122}
  (\bibinfo {year} {2021})},\ \bibinfo {note} {[Erratum: Eur.Phys.J.A 57, 314
  (2021)]},\ \Eprint {http://arxiv.org/abs/2006.09945} {arXiv:2006.09945
  [nucl-th]} \BibitemShut {NoStop}%
\bibitem [{\citenamefont {Esposito}\ \emph {et~al.}(2021)\citenamefont
  {Esposito}, \citenamefont {Ferreiro}, \citenamefont {Pilloni}, \citenamefont
  {Polosa},\ and\ \citenamefont {Salgado}}]{Esposito:2020ywk}%
  \BibitemOpen
  \bibfield  {author} {\bibinfo {author} {\bibfnamefont {A.}~\bibnamefont
  {Esposito}}, \bibinfo {author} {\bibfnamefont {E.~G.}\ \bibnamefont
  {Ferreiro}}, \bibinfo {author} {\bibfnamefont {A.}~\bibnamefont {Pilloni}},
  \bibinfo {author} {\bibfnamefont {A.~D.}\ \bibnamefont {Polosa}}, \ and\
  \bibinfo {author} {\bibfnamefont {C.~A.}\ \bibnamefont {Salgado}},\ }\href
  {\doibase 10.1140/epjc/s10052-021-09425-w} {\bibfield  {journal} {\bibinfo
  {journal} {Eur. Phys. J. C}\ }\textbf {\bibinfo {volume} {81}},\ \bibinfo
  {pages} {669} (\bibinfo {year} {2021})},\ \Eprint
  {http://arxiv.org/abs/2006.15044} {arXiv:2006.15044 [hep-ph]} \BibitemShut
  {NoStop}%
\bibitem [{\citenamefont {Braaten}\ \emph {et~al.}(2021)\citenamefont
  {Braaten}, \citenamefont {He}, \citenamefont {Ingles},\ and\ \citenamefont
  {Jiang}}]{Braaten:2020iqw}%
  \BibitemOpen
  \bibfield  {author} {\bibinfo {author} {\bibfnamefont {E.}~\bibnamefont
  {Braaten}}, \bibinfo {author} {\bibfnamefont {L.-P.}\ \bibnamefont {He}},
  \bibinfo {author} {\bibfnamefont {K.}~\bibnamefont {Ingles}}, \ and\ \bibinfo
  {author} {\bibfnamefont {J.}~\bibnamefont {Jiang}},\ }\href {\doibase
  10.1103/PhysRevD.103.L071901} {\bibfield  {journal} {\bibinfo  {journal}
  {Phys. Rev. D}\ }\textbf {\bibinfo {volume} {103}},\ \bibinfo {pages}
  {L071901} (\bibinfo {year} {2021})},\ \Eprint
  {http://arxiv.org/abs/2012.13499} {arXiv:2012.13499 [hep-ph]} \BibitemShut
  {NoStop}%
\bibitem [{\citenamefont {Cano}\ and\ \citenamefont
  {Pire}(2004)}]{Cano:2003ju}%
  \BibitemOpen
  \bibfield  {author} {\bibinfo {author} {\bibfnamefont {F.}~\bibnamefont
  {Cano}}\ and\ \bibinfo {author} {\bibfnamefont {B.}~\bibnamefont {Pire}},\
  }\href {\doibase 10.1140/epja/i2003-10127-x} {\bibfield  {journal} {\bibinfo
  {journal} {Eur. Phys. J. A}\ }\textbf {\bibinfo {volume} {19}},\ \bibinfo
  {pages} {423} (\bibinfo {year} {2004})},\ \Eprint
  {http://arxiv.org/abs/hep-ph/0307231} {arXiv:hep-ph/0307231} \BibitemShut
  {NoStop}%
\bibitem [{\citenamefont {Liuti}\ and\ \citenamefont
  {Taneja}(2005{\natexlab{a}})}]{Liuti:2005gi}%
  \BibitemOpen
  \bibfield  {author} {\bibinfo {author} {\bibfnamefont {S.}~\bibnamefont
  {Liuti}}\ and\ \bibinfo {author} {\bibfnamefont {S.~K.}\ \bibnamefont
  {Taneja}},\ }\href {\doibase 10.1103/PhysRevC.72.032201} {\bibfield
  {journal} {\bibinfo  {journal} {Phys. Rev. C}\ }\textbf {\bibinfo {volume}
  {72}},\ \bibinfo {pages} {032201} (\bibinfo {year} {2005}{\natexlab{a}})},\
  \Eprint {http://arxiv.org/abs/hep-ph/0505123} {arXiv:hep-ph/0505123}
  \BibitemShut {NoStop}%
\bibitem [{\citenamefont {Liuti}\ and\ \citenamefont
  {Taneja}(2005{\natexlab{b}})}]{Liuti:2005qj}%
  \BibitemOpen
  \bibfield  {author} {\bibinfo {author} {\bibfnamefont {S.}~\bibnamefont
  {Liuti}}\ and\ \bibinfo {author} {\bibfnamefont {S.~K.}\ \bibnamefont
  {Taneja}},\ }\href {\doibase 10.1103/PhysRevC.72.034902} {\bibfield
  {journal} {\bibinfo  {journal} {Phys. Rev. C}\ }\textbf {\bibinfo {volume}
  {72}},\ \bibinfo {pages} {034902} (\bibinfo {year} {2005}{\natexlab{b}})},\
  \Eprint {http://arxiv.org/abs/hep-ph/0504027} {arXiv:hep-ph/0504027}
  \BibitemShut {NoStop}%
\bibitem [{\citenamefont {Dupr\'e}\ and\ \citenamefont
  {Scopetta}(2016)}]{Dupre:2015jha}%
  \BibitemOpen
  \bibfield  {author} {\bibinfo {author} {\bibfnamefont {R.}~\bibnamefont
  {Dupr\'e}}\ and\ \bibinfo {author} {\bibfnamefont {S.}~\bibnamefont
  {Scopetta}},\ }\href {\doibase 10.1140/epja/i2016-16159-1} {\bibfield
  {journal} {\bibinfo  {journal} {Eur. Phys. J. A}\ }\textbf {\bibinfo {volume}
  {52}},\ \bibinfo {pages} {159} (\bibinfo {year} {2016})},\ \Eprint
  {http://arxiv.org/abs/1510.00794} {arXiv:1510.00794 [nucl-ex]} \BibitemShut
  {NoStop}%
\bibitem [{\citenamefont {Fucini}\ \emph {et~al.}(2018)\citenamefont {Fucini},
  \citenamefont {Scopetta},\ and\ \citenamefont {Viviani}}]{Fucini:2018gso}%
  \BibitemOpen
  \bibfield  {author} {\bibinfo {author} {\bibfnamefont {S.}~\bibnamefont
  {Fucini}}, \bibinfo {author} {\bibfnamefont {S.}~\bibnamefont {Scopetta}}, \
  and\ \bibinfo {author} {\bibfnamefont {M.}~\bibnamefont {Viviani}},\ }\href
  {\doibase 10.1103/PhysRevC.98.015203} {\bibfield  {journal} {\bibinfo
  {journal} {Phys. Rev. C}\ }\textbf {\bibinfo {volume} {98}},\ \bibinfo
  {pages} {015203} (\bibinfo {year} {2018})},\ \Eprint
  {http://arxiv.org/abs/1805.05877} {arXiv:1805.05877 [nucl-th]} \BibitemShut
  {NoStop}%
\bibitem [{\citenamefont {Berger}\ \emph {et~al.}(2001)\citenamefont {Berger},
  \citenamefont {Cano}, \citenamefont {Diehl},\ and\ \citenamefont
  {Pire}}]{Berger:2001zb}%
  \BibitemOpen
  \bibfield  {author} {\bibinfo {author} {\bibfnamefont {E.~R.}\ \bibnamefont
  {Berger}}, \bibinfo {author} {\bibfnamefont {F.}~\bibnamefont {Cano}},
  \bibinfo {author} {\bibfnamefont {M.}~\bibnamefont {Diehl}}, \ and\ \bibinfo
  {author} {\bibfnamefont {B.}~\bibnamefont {Pire}},\ }\href {\doibase
  10.1103/PhysRevLett.87.142302} {\bibfield  {journal} {\bibinfo  {journal}
  {Phys. Rev. Lett.}\ }\textbf {\bibinfo {volume} {87}},\ \bibinfo {pages}
  {142302} (\bibinfo {year} {2001})},\ \Eprint
  {http://arxiv.org/abs/hep-ph/0106192} {arXiv:hep-ph/0106192} \BibitemShut
  {NoStop}%
\bibitem [{\citenamefont {Cosyn}\ and\ \citenamefont
  {Pire}(2018)}]{Cosyn:2018rdm}%
  \BibitemOpen
  \bibfield  {author} {\bibinfo {author} {\bibfnamefont {W.}~\bibnamefont
  {Cosyn}}\ and\ \bibinfo {author} {\bibfnamefont {B.}~\bibnamefont {Pire}},\
  }\href {\doibase 10.1103/PhysRevD.98.074020} {\bibfield  {journal} {\bibinfo
  {journal} {Phys. Rev. D}\ }\textbf {\bibinfo {volume} {98}},\ \bibinfo
  {pages} {074020} (\bibinfo {year} {2018})},\ \Eprint
  {http://arxiv.org/abs/1806.01177} {arXiv:1806.01177 [hep-ph]} \BibitemShut
  {NoStop}%
\bibitem [{\citenamefont {Armesto}\ \emph {et~al.}(2022)\citenamefont
  {Armesto}, \citenamefont {Newman}, \citenamefont {Slominski},\ and\
  \citenamefont {Stasto}}]{Armesto:2021fws}%
  \BibitemOpen
  \bibfield  {author} {\bibinfo {author} {\bibfnamefont {N.}~\bibnamefont
  {Armesto}}, \bibinfo {author} {\bibfnamefont {P.~R.}\ \bibnamefont {Newman}},
  \bibinfo {author} {\bibfnamefont {W.}~\bibnamefont {Slominski}}, \ and\
  \bibinfo {author} {\bibfnamefont {A.~M.}\ \bibnamefont {Stasto}},\ }\href
  {\doibase 10.1103/PhysRevD.105.074006} {\bibfield  {journal} {\bibinfo
  {journal} {Phys. Rev. D}\ }\textbf {\bibinfo {volume} {105}},\ \bibinfo
  {pages} {074006} (\bibinfo {year} {2022})},\ \Eprint
  {http://arxiv.org/abs/2112.06839} {arXiv:2112.06839 [hep-ph]} \BibitemShut
  {NoStop}%
\bibitem [{\citenamefont {Guzey}\ and\ \citenamefont
  {Klasen}(2020)}]{Guzey:2020gkk}%
  \BibitemOpen
  \bibfield  {author} {\bibinfo {author} {\bibfnamefont {V.}~\bibnamefont
  {Guzey}}\ and\ \bibinfo {author} {\bibfnamefont {M.}~\bibnamefont {Klasen}},\
  }\href {\doibase 10.1007/JHEP05(2020)074} {\bibfield  {journal} {\bibinfo
  {journal} {JHEP}\ }\textbf {\bibinfo {volume} {05}},\ \bibinfo {pages} {074}
  (\bibinfo {year} {2020})},\ \Eprint {http://arxiv.org/abs/2004.06972}
  {arXiv:2004.06972 [hep-ph]} \BibitemShut {NoStop}%
\bibitem [{\citenamefont {Frankfurt}\ and\ \citenamefont
  {Strikman}(1983)}]{Frankfurt:1983qs}%
  \BibitemOpen
  \bibfield  {author} {\bibinfo {author} {\bibfnamefont {L.~L.}\ \bibnamefont
  {Frankfurt}}\ and\ \bibinfo {author} {\bibfnamefont {M.~I.}\ \bibnamefont
  {Strikman}},\ }\href {\doibase 10.1016/0375-9474(83)90518-3} {\bibfield
  {journal} {\bibinfo  {journal} {Nucl. Phys. A}\ }\textbf {\bibinfo {volume}
  {405}},\ \bibinfo {pages} {557} (\bibinfo {year} {1983})}\BibitemShut
  {NoStop}%
\bibitem [{\citenamefont {Hoodbhoy}\ \emph {et~al.}(1989)\citenamefont
  {Hoodbhoy}, \citenamefont {Jaffe},\ and\ \citenamefont
  {Manohar}}]{Hoodbhoy:1988am}%
  \BibitemOpen
  \bibfield  {author} {\bibinfo {author} {\bibfnamefont {P.}~\bibnamefont
  {Hoodbhoy}}, \bibinfo {author} {\bibfnamefont {R.~L.}\ \bibnamefont {Jaffe}},
  \ and\ \bibinfo {author} {\bibfnamefont {A.}~\bibnamefont {Manohar}},\ }\href
  {\doibase 10.1016/0550-3213(89)90572-5} {\bibfield  {journal} {\bibinfo
  {journal} {Nucl. Phys. B}\ }\textbf {\bibinfo {volume} {312}},\ \bibinfo
  {pages} {571} (\bibinfo {year} {1989})}\BibitemShut {NoStop}%
\bibitem [{\citenamefont {Close}\ and\ \citenamefont
  {Kumano}(1990)}]{Close:1990zw}%
  \BibitemOpen
  \bibfield  {author} {\bibinfo {author} {\bibfnamefont {F.~E.}\ \bibnamefont
  {Close}}\ and\ \bibinfo {author} {\bibfnamefont {S.}~\bibnamefont {Kumano}},\
  }\href {\doibase 10.1103/PhysRevD.42.2377} {\bibfield  {journal} {\bibinfo
  {journal} {Phys. Rev. D}\ }\textbf {\bibinfo {volume} {42}},\ \bibinfo
  {pages} {2377} (\bibinfo {year} {1990})}\BibitemShut {NoStop}%
\bibitem [{\citenamefont {Kumano}(2010)}]{Kumano:2010vz}%
  \BibitemOpen
  \bibfield  {author} {\bibinfo {author} {\bibfnamefont {S.}~\bibnamefont
  {Kumano}},\ }\href {\doibase 10.1103/PhysRevD.82.017501} {\bibfield
  {journal} {\bibinfo  {journal} {Phys. Rev. D}\ }\textbf {\bibinfo {volume}
  {82}},\ \bibinfo {pages} {017501} (\bibinfo {year} {2010})},\ \Eprint
  {http://arxiv.org/abs/1005.4524} {arXiv:1005.4524 [hep-ph]} \BibitemShut
  {NoStop}%
\bibitem [{\citenamefont {Cosyn}\ \emph {et~al.}(2017)\citenamefont {Cosyn},
  \citenamefont {Dong}, \citenamefont {Kumano},\ and\ \citenamefont
  {Sargsian}}]{Cosyn:2017fbo}%
  \BibitemOpen
  \bibfield  {author} {\bibinfo {author} {\bibfnamefont {W.}~\bibnamefont
  {Cosyn}}, \bibinfo {author} {\bibfnamefont {Y.-B.}\ \bibnamefont {Dong}},
  \bibinfo {author} {\bibfnamefont {S.}~\bibnamefont {Kumano}}, \ and\ \bibinfo
  {author} {\bibfnamefont {M.}~\bibnamefont {Sargsian}},\ }\href {\doibase
  10.1103/PhysRevD.95.074036} {\bibfield  {journal} {\bibinfo  {journal} {Phys.
  Rev. D}\ }\textbf {\bibinfo {volume} {95}},\ \bibinfo {pages} {074036}
  (\bibinfo {year} {2017})},\ \Eprint {http://arxiv.org/abs/1702.05337}
  {arXiv:1702.05337 [hep-ph]} \BibitemShut {NoStop}%
\bibitem [{\citenamefont {Miller}(2014)}]{Miller:2013hla}%
  \BibitemOpen
  \bibfield  {author} {\bibinfo {author} {\bibfnamefont {G.~A.}\ \bibnamefont
  {Miller}},\ }\href {\doibase 10.1103/PhysRevC.89.045203} {\bibfield
  {journal} {\bibinfo  {journal} {Phys. Rev. C}\ }\textbf {\bibinfo {volume}
  {89}},\ \bibinfo {pages} {045203} (\bibinfo {year} {2014})},\ \Eprint
  {http://arxiv.org/abs/1311.4561} {arXiv:1311.4561 [nucl-th]} \BibitemShut
  {NoStop}%
\bibitem [{\citenamefont {Bacchetta}\ and\ \citenamefont
  {Mulders}(2000)}]{Bacchetta:2000jk}%
  \BibitemOpen
  \bibfield  {author} {\bibinfo {author} {\bibfnamefont {A.}~\bibnamefont
  {Bacchetta}}\ and\ \bibinfo {author} {\bibfnamefont {P.~J.}\ \bibnamefont
  {Mulders}},\ }\href {\doibase 10.1103/PhysRevD.62.114004} {\bibfield
  {journal} {\bibinfo  {journal} {Phys. Rev. D}\ }\textbf {\bibinfo {volume}
  {62}},\ \bibinfo {pages} {114004} (\bibinfo {year} {2000})},\ \Eprint
  {http://arxiv.org/abs/hep-ph/0007120} {arXiv:hep-ph/0007120} \BibitemShut
  {NoStop}%
\bibitem [{\citenamefont {Kumano}\ and\ \citenamefont
  {Song}(2021{\natexlab{a}})}]{Kumano:2020ijt}%
  \BibitemOpen
  \bibfield  {author} {\bibinfo {author} {\bibfnamefont {S.}~\bibnamefont
  {Kumano}}\ and\ \bibinfo {author} {\bibfnamefont {Q.-T.}\ \bibnamefont
  {Song}},\ }\href {\doibase 10.1103/PhysRevD.103.014025} {\bibfield  {journal}
  {\bibinfo  {journal} {Phys. Rev. D}\ }\textbf {\bibinfo {volume} {103}},\
  \bibinfo {pages} {014025} (\bibinfo {year} {2021}{\natexlab{a}})},\ \Eprint
  {http://arxiv.org/abs/2011.08583} {arXiv:2011.08583 [hep-ph]} \BibitemShut
  {NoStop}%
\bibitem [{\citenamefont {Kumano}\ and\ \citenamefont
  {Song}(2021{\natexlab{b}})}]{Kumano:2021fem}%
  \BibitemOpen
  \bibfield  {author} {\bibinfo {author} {\bibfnamefont {S.}~\bibnamefont
  {Kumano}}\ and\ \bibinfo {author} {\bibfnamefont {Q.-T.}\ \bibnamefont
  {Song}},\ }\href {\doibase 10.1007/JHEP09(2021)141} {\bibfield  {journal}
  {\bibinfo  {journal} {JHEP}\ }\textbf {\bibinfo {volume} {09}},\ \bibinfo
  {pages} {141} (\bibinfo {year} {2021}{\natexlab{b}})},\ \Eprint
  {http://arxiv.org/abs/2106.15849} {arXiv:2106.15849 [hep-ph]} \BibitemShut
  {NoStop}%
\bibitem [{\citenamefont {Kumano}\ and\ \citenamefont
  {Song}(2022)}]{Kumano:2021xau}%
  \BibitemOpen
  \bibfield  {author} {\bibinfo {author} {\bibfnamefont {S.}~\bibnamefont
  {Kumano}}\ and\ \bibinfo {author} {\bibfnamefont {Q.-T.}\ \bibnamefont
  {Song}},\ }\href {\doibase 10.1016/j.physletb.2022.136908} {\bibfield
  {journal} {\bibinfo  {journal} {Phys. Lett. B}\ }\textbf {\bibinfo {volume}
  {826}},\ \bibinfo {pages} {136908} (\bibinfo {year} {2022})},\ \Eprint
  {http://arxiv.org/abs/2112.13218} {arXiv:2112.13218 [hep-ph]} \BibitemShut
  {NoStop}%
\bibitem [{\citenamefont {Jaffe}\ and\ \citenamefont
  {Manohar}(1989)}]{Jaffe:1989xy}%
  \BibitemOpen
  \bibfield  {author} {\bibinfo {author} {\bibfnamefont {R.~L.}\ \bibnamefont
  {Jaffe}}\ and\ \bibinfo {author} {\bibfnamefont {A.}~\bibnamefont
  {Manohar}},\ }\href {\doibase 10.1016/0370-2693(89)90242-6} {\bibfield
  {journal} {\bibinfo  {journal} {Phys. Lett. B}\ }\textbf {\bibinfo {volume}
  {223}},\ \bibinfo {pages} {218} (\bibinfo {year} {1989})}\BibitemShut
  {NoStop}%
\bibitem [{\citenamefont {Detmold}\ and\ \citenamefont
  {Shanahan}(2016)}]{Detmold:2016gpy}%
  \BibitemOpen
  \bibfield  {author} {\bibinfo {author} {\bibfnamefont {W.}~\bibnamefont
  {Detmold}}\ and\ \bibinfo {author} {\bibfnamefont {P.~E.}\ \bibnamefont
  {Shanahan}},\ }\href {\doibase 10.1103/PhysRevD.94.014507} {\bibfield
  {journal} {\bibinfo  {journal} {Phys. Rev. D}\ }\textbf {\bibinfo {volume}
  {94}},\ \bibinfo {pages} {014507} (\bibinfo {year} {2016})},\ \bibinfo {note}
  {[Erratum: Phys.Rev.D 95, 079902 (2017)]},\ \Eprint
  {http://arxiv.org/abs/1606.04505} {arXiv:1606.04505 [hep-lat]} \BibitemShut
  {NoStop}%
\bibitem [{\citenamefont {Arbuzov}\ \emph {et~al.}(2021)\citenamefont {Arbuzov}
  \emph {et~al.}}]{Arbuzov:2020cqg}%
  \BibitemOpen
  \bibfield  {author} {\bibinfo {author} {\bibfnamefont {A.}~\bibnamefont
  {Arbuzov}} \emph {et~al.},\ }\href {\doibase 10.1016/j.ppnp.2021.103858}
  {\bibfield  {journal} {\bibinfo  {journal} {Prog. Part. Nucl. Phys.}\
  }\textbf {\bibinfo {volume} {119}},\ \bibinfo {pages} {103858} (\bibinfo
  {year} {2021})},\ \Eprint {http://arxiv.org/abs/2011.15005} {arXiv:2011.15005
  [hep-ex]} \BibitemShut {NoStop}%
\bibitem [{\citenamefont {Wang}\ \emph {et~al.}(2022)\citenamefont {Wang},
  \citenamefont {Bentz}, \citenamefont {Clo\"et},\ and\ \citenamefont
  {Thomas}}]{Wang:2021elw}%
  \BibitemOpen
  \bibfield  {author} {\bibinfo {author} {\bibfnamefont {X.-G.}\ \bibnamefont
  {Wang}}, \bibinfo {author} {\bibfnamefont {W.}~\bibnamefont {Bentz}},
  \bibinfo {author} {\bibfnamefont {I.~C.}\ \bibnamefont {Clo\"et}}, \ and\
  \bibinfo {author} {\bibfnamefont {A.~W.}\ \bibnamefont {Thomas}},\ }\href
  {\doibase 10.1088/1361-6471/ac4c90} {\bibfield  {journal} {\bibinfo
  {journal} {J. Phys. G}\ }\textbf {\bibinfo {volume} {49}},\ \bibinfo {pages}
  {03LT01} (\bibinfo {year} {2022})},\ \Eprint
  {http://arxiv.org/abs/2109.03591} {arXiv:2109.03591 [hep-ph]} \BibitemShut
  {NoStop}%
\bibitem [{\citenamefont {Pace}\ \emph {et~al.}(2022)\citenamefont {Pace},
  \citenamefont {Rinaldi}, \citenamefont {Salm\`e},\ and\ \citenamefont
  {Scopetta}}]{Pace:2022qoj}%
  \BibitemOpen
  \bibfield  {author} {\bibinfo {author} {\bibfnamefont {E.}~\bibnamefont
  {Pace}}, \bibinfo {author} {\bibfnamefont {M.}~\bibnamefont {Rinaldi}},
  \bibinfo {author} {\bibfnamefont {G.}~\bibnamefont {Salm\`e}}, \ and\
  \bibinfo {author} {\bibfnamefont {S.}~\bibnamefont {Scopetta}},\ }\href@noop
  {} {\  (\bibinfo {year} {2022})},\ \Eprint {http://arxiv.org/abs/2206.05485}
  {arXiv:2206.05485 [nucl-th]} \BibitemShut {NoStop}%
\bibitem [{\citenamefont {Kim}\ and\ \citenamefont
  {Miller}(2022)}]{Kim:2022lng}%
  \BibitemOpen
  \bibfield  {author} {\bibinfo {author} {\bibfnamefont {D.~N.}\ \bibnamefont
  {Kim}}\ and\ \bibinfo {author} {\bibfnamefont {G.~A.}\ \bibnamefont
  {Miller}},\ }\href {\doibase 10.1103/PhysRevC.106.055202} {\bibfield
  {journal} {\bibinfo  {journal} {Phys. Rev. C}\ }\textbf {\bibinfo {volume}
  {106}},\ \bibinfo {pages} {055202} (\bibinfo {year} {2022})},\ \Eprint
  {http://arxiv.org/abs/2209.13753} {arXiv:2209.13753 [nucl-th]} \BibitemShut
  {NoStop}%
\bibitem [{\citenamefont {Guzey}\ \emph {et~al.}(2009)\citenamefont {Guzey},
  \citenamefont {Thomas},\ and\ \citenamefont {Tsushima}}]{Guzey:2008fe}%
  \BibitemOpen
  \bibfield  {author} {\bibinfo {author} {\bibfnamefont {V.}~\bibnamefont
  {Guzey}}, \bibinfo {author} {\bibfnamefont {A.~W.}\ \bibnamefont {Thomas}}, \
  and\ \bibinfo {author} {\bibfnamefont {K.}~\bibnamefont {Tsushima}},\ }\href
  {\doibase 10.1016/j.physletb.2009.01.064} {\bibfield  {journal} {\bibinfo
  {journal} {Phys. Lett. B}\ }\textbf {\bibinfo {volume} {673}},\ \bibinfo
  {pages} {9} (\bibinfo {year} {2009})},\ \Eprint
  {http://arxiv.org/abs/0806.3288} {arXiv:0806.3288 [hep-ph]} \BibitemShut
  {NoStop}%
\bibitem [{\citenamefont {Cloet}\ \emph {et~al.}(2006)\citenamefont {Cloet},
  \citenamefont {Bentz},\ and\ \citenamefont {Thomas}}]{Cloet:2006bq}%
  \BibitemOpen
  \bibfield  {author} {\bibinfo {author} {\bibfnamefont {I.~C.}\ \bibnamefont
  {Cloet}}, \bibinfo {author} {\bibfnamefont {W.}~\bibnamefont {Bentz}}, \ and\
  \bibinfo {author} {\bibfnamefont {A.~W.}\ \bibnamefont {Thomas}},\ }\href
  {\doibase 10.1016/j.physletb.2006.08.076} {\bibfield  {journal} {\bibinfo
  {journal} {Phys. Lett. B}\ }\textbf {\bibinfo {volume} {642}},\ \bibinfo
  {pages} {210} (\bibinfo {year} {2006})},\ \Eprint
  {http://arxiv.org/abs/nucl-th/0605061} {arXiv:nucl-th/0605061} \BibitemShut
  {NoStop}%
\bibitem [{\citenamefont {Ciofi~degli Atti}\ and\ \citenamefont
  {Kaptari}(2011)}]{CiofidegliAtti:2010uwl}%
  \BibitemOpen
  \bibfield  {author} {\bibinfo {author} {\bibfnamefont {C.}~\bibnamefont
  {Ciofi~degli Atti}}\ and\ \bibinfo {author} {\bibfnamefont {L.~P.}\
  \bibnamefont {Kaptari}},\ }\href {\doibase 10.1103/PhysRevC.83.044602}
  {\bibfield  {journal} {\bibinfo  {journal} {Phys. Rev. C}\ }\textbf {\bibinfo
  {volume} {83}},\ \bibinfo {pages} {044602} (\bibinfo {year} {2011})},\
  \Eprint {http://arxiv.org/abs/1011.5960} {arXiv:1011.5960 [nucl-th]}
  \BibitemShut {NoStop}%
\bibitem [{\citenamefont {Fucini}\ \emph {et~al.}(2020)\citenamefont {Fucini},
  \citenamefont {Scopetta},\ and\ \citenamefont {Viviani}}]{Fucini:2019xlc}%
  \BibitemOpen
  \bibfield  {author} {\bibinfo {author} {\bibfnamefont {S.}~\bibnamefont
  {Fucini}}, \bibinfo {author} {\bibfnamefont {S.}~\bibnamefont {Scopetta}}, \
  and\ \bibinfo {author} {\bibfnamefont {M.}~\bibnamefont {Viviani}},\ }\href
  {\doibase 10.1103/PhysRevD.101.071501} {\bibfield  {journal} {\bibinfo
  {journal} {Phys. Rev. D}\ }\textbf {\bibinfo {volume} {101}},\ \bibinfo
  {pages} {071501} (\bibinfo {year} {2020})},\ \Eprint
  {http://arxiv.org/abs/1909.12261} {arXiv:1909.12261 [nucl-th]} \BibitemShut
  {NoStop}%
\bibitem [{\citenamefont {Brodsky}\ and\ \citenamefont
  {Lu}(1990)}]{Brodsky:1989qz}%
  \BibitemOpen
  \bibfield  {author} {\bibinfo {author} {\bibfnamefont {S.~J.}\ \bibnamefont
  {Brodsky}}\ and\ \bibinfo {author} {\bibfnamefont {H.~J.}\ \bibnamefont
  {Lu}},\ }\href {\doibase 10.1103/PhysRevLett.64.1342} {\bibfield  {journal}
  {\bibinfo  {journal} {Phys. Rev. Lett.}\ }\textbf {\bibinfo {volume} {64}},\
  \bibinfo {pages} {1342} (\bibinfo {year} {1990})}\BibitemShut {NoStop}%
\bibitem [{\citenamefont {Frankfurt}\ \emph {et~al.}(2012)\citenamefont
  {Frankfurt}, \citenamefont {Guzey},\ and\ \citenamefont
  {Strikman}}]{Frankfurt:2011cs}%
  \BibitemOpen
  \bibfield  {author} {\bibinfo {author} {\bibfnamefont {L.}~\bibnamefont
  {Frankfurt}}, \bibinfo {author} {\bibfnamefont {V.}~\bibnamefont {Guzey}}, \
  and\ \bibinfo {author} {\bibfnamefont {M.}~\bibnamefont {Strikman}},\ }\href
  {\doibase 10.1016/j.physrep.2011.12.002} {\bibfield  {journal} {\bibinfo
  {journal} {Phys. Rept.}\ }\textbf {\bibinfo {volume} {512}},\ \bibinfo
  {pages} {255} (\bibinfo {year} {2012})},\ \Eprint
  {http://arxiv.org/abs/1106.2091} {arXiv:1106.2091 [hep-ph]} \BibitemShut
  {NoStop}%
\bibitem [{\citenamefont {Krelina}\ and\ \citenamefont
  {Nemchik}(2020)}]{Krelina:2020ipn}%
  \BibitemOpen
  \bibfield  {author} {\bibinfo {author} {\bibfnamefont {M.}~\bibnamefont
  {Krelina}}\ and\ \bibinfo {author} {\bibfnamefont {J.}~\bibnamefont
  {Nemchik}},\ }\href {\doibase 10.1140/epjp/s13360-020-00498-2} {\bibfield
  {journal} {\bibinfo  {journal} {Eur. Phys. J. Plus}\ }\textbf {\bibinfo
  {volume} {135}},\ \bibinfo {pages} {444} (\bibinfo {year} {2020})},\ \Eprint
  {http://arxiv.org/abs/2003.04156} {arXiv:2003.04156 [hep-ph]} \BibitemShut
  {NoStop}%
\bibitem [{\citenamefont {Guzey}\ \emph {et~al.}(2022)\citenamefont {Guzey},
  \citenamefont {Rinaldi}, \citenamefont {Scopetta}, \citenamefont {Strikman},\
  and\ \citenamefont {Viviani}}]{Guzey:2022jtv}%
  \BibitemOpen
  \bibfield  {author} {\bibinfo {author} {\bibfnamefont {V.}~\bibnamefont
  {Guzey}}, \bibinfo {author} {\bibfnamefont {M.}~\bibnamefont {Rinaldi}},
  \bibinfo {author} {\bibfnamefont {S.}~\bibnamefont {Scopetta}}, \bibinfo
  {author} {\bibfnamefont {M.}~\bibnamefont {Strikman}}, \ and\ \bibinfo
  {author} {\bibfnamefont {M.}~\bibnamefont {Viviani}},\ }\href {\doibase
  10.1103/PhysRevLett.129.242503} {\bibfield  {journal} {\bibinfo  {journal}
  {Phys. Rev. Lett.}\ }\textbf {\bibinfo {volume} {129}},\ \bibinfo {pages}
  {242503} (\bibinfo {year} {2022})},\ \Eprint
  {http://arxiv.org/abs/2202.12200} {arXiv:2202.12200 [hep-ph]} \BibitemShut
  {NoStop}%
\bibitem [{\citenamefont {Qiu}\ and\ \citenamefont
  {Vitev}(2004{\natexlab{a}})}]{Qiu:2003vd}%
  \BibitemOpen
  \bibfield  {author} {\bibinfo {author} {\bibfnamefont {J.-w.}\ \bibnamefont
  {Qiu}}\ and\ \bibinfo {author} {\bibfnamefont {I.}~\bibnamefont {Vitev}},\
  }\href {\doibase 10.1103/PhysRevLett.93.262301} {\bibfield  {journal}
  {\bibinfo  {journal} {Phys. Rev. Lett.}\ }\textbf {\bibinfo {volume} {93}},\
  \bibinfo {pages} {262301} (\bibinfo {year} {2004}{\natexlab{a}})},\ \Eprint
  {http://arxiv.org/abs/hep-ph/0309094} {arXiv:hep-ph/0309094} \BibitemShut
  {NoStop}%
\bibitem [{\citenamefont {Qiu}\ and\ \citenamefont
  {Vitev}(2004{\natexlab{b}})}]{Qiu:2004qk}%
  \BibitemOpen
  \bibfield  {author} {\bibinfo {author} {\bibfnamefont {J.-W.}\ \bibnamefont
  {Qiu}}\ and\ \bibinfo {author} {\bibfnamefont {I.}~\bibnamefont {Vitev}},\
  }\href {\doibase 10.1016/j.physletb.2004.02.065} {\bibfield  {journal}
  {\bibinfo  {journal} {Phys. Lett. B}\ }\textbf {\bibinfo {volume} {587}},\
  \bibinfo {pages} {52} (\bibinfo {year} {2004}{\natexlab{b}})},\ \Eprint
  {http://arxiv.org/abs/hep-ph/0401062} {arXiv:hep-ph/0401062} \BibitemShut
  {NoStop}%
\bibitem [{\citenamefont {Miller}\ \emph {et~al.}(2016)\citenamefont {Miller},
  \citenamefont {Sievert},\ and\ \citenamefont {Venugopalan}}]{Miller:2015tjf}%
  \BibitemOpen
  \bibfield  {author} {\bibinfo {author} {\bibfnamefont {G.~A.}\ \bibnamefont
  {Miller}}, \bibinfo {author} {\bibfnamefont {M.~D.}\ \bibnamefont {Sievert}},
  \ and\ \bibinfo {author} {\bibfnamefont {R.}~\bibnamefont {Venugopalan}},\
  }\href {\doibase 10.1103/PhysRevC.93.045202} {\bibfield  {journal} {\bibinfo
  {journal} {Phys. Rev. C}\ }\textbf {\bibinfo {volume} {93}},\ \bibinfo
  {pages} {045202} (\bibinfo {year} {2016})},\ \Eprint
  {http://arxiv.org/abs/1512.03111} {arXiv:1512.03111 [nucl-th]} \BibitemShut
  {NoStop}%
\bibitem [{\citenamefont {Sargsian}\ and\ \citenamefont
  {Vera}(2022)}]{Sargsian:2022rmq}%
  \BibitemOpen
  \bibfield  {author} {\bibinfo {author} {\bibfnamefont {M.~M.}\ \bibnamefont
  {Sargsian}}\ and\ \bibinfo {author} {\bibfnamefont {F.}~\bibnamefont
  {Vera}},\ }\href@noop {} {\  (\bibinfo {year} {2022})},\ \Eprint
  {http://arxiv.org/abs/2208.00501} {arXiv:2208.00501 [nucl-th]} \BibitemShut
  {NoStop}%
\bibitem [{\citenamefont {Bertulani}(2023)}]{Bertulani:2022vad}%
  \BibitemOpen
  \bibfield  {author} {\bibinfo {author} {\bibfnamefont {C.~A.}\ \bibnamefont
  {Bertulani}},\ }\href {\doibase 10.1016/j.physletb.2022.137639} {\bibfield
  {journal} {\bibinfo  {journal} {Phys. Lett. B}\ }\textbf {\bibinfo {volume}
  {837}},\ \bibinfo {pages} {137639} (\bibinfo {year} {2023})},\ \Eprint
  {http://arxiv.org/abs/2211.12643} {arXiv:2211.12643 [nucl-th]} \BibitemShut
  {NoStop}%
\bibitem [{\citenamefont {Hen}\ \emph {et~al.}(2017)\citenamefont {Hen},
  \citenamefont {Miller}, \citenamefont {Piasetzky},\ and\ \citenamefont
  {Weinstein}}]{Hen:2016kwk}%
  \BibitemOpen
  \bibfield  {author} {\bibinfo {author} {\bibfnamefont {O.}~\bibnamefont
  {Hen}}, \bibinfo {author} {\bibfnamefont {G.~A.}\ \bibnamefont {Miller}},
  \bibinfo {author} {\bibfnamefont {E.}~\bibnamefont {Piasetzky}}, \ and\
  \bibinfo {author} {\bibfnamefont {L.~B.}\ \bibnamefont {Weinstein}},\ }\href
  {\doibase 10.1103/RevModPhys.89.045002} {\bibfield  {journal} {\bibinfo
  {journal} {Rev. Mod. Phys.}\ }\textbf {\bibinfo {volume} {89}},\ \bibinfo
  {pages} {045002} (\bibinfo {year} {2017})},\ \Eprint
  {http://arxiv.org/abs/1611.09748} {arXiv:1611.09748 [nucl-ex]} \BibitemShut
  {NoStop}%
\bibitem [{\citenamefont {West}(2023)}]{West:2020tyo}%
  \BibitemOpen
  \bibfield  {author} {\bibinfo {author} {\bibfnamefont {J.~R.}\ \bibnamefont
  {West}},\ }\href {\doibase 10.1016/j.nuclphysa.2022.122563} {\bibfield
  {journal} {\bibinfo  {journal} {Nucl. Phys. A}\ }\textbf {\bibinfo {volume}
  {1029}},\ \bibinfo {pages} {122563} (\bibinfo {year} {2023})},\ \Eprint
  {http://arxiv.org/abs/2009.06968} {arXiv:2009.06968 [hep-ph]} \BibitemShut
  {NoStop}%
\bibitem [{\citenamefont {Hauenstein}\ \emph {et~al.}(2022)\citenamefont
  {Hauenstein} \emph {et~al.}}]{Hauenstein:2021zql}%
  \BibitemOpen
  \bibfield  {author} {\bibinfo {author} {\bibfnamefont {F.}~\bibnamefont
  {Hauenstein}} \emph {et~al.},\ }\href {\doibase 10.1103/PhysRevC.105.034001}
  {\bibfield  {journal} {\bibinfo  {journal} {Phys. Rev. C}\ }\textbf {\bibinfo
  {volume} {105}},\ \bibinfo {pages} {034001} (\bibinfo {year} {2022})},\
  \Eprint {http://arxiv.org/abs/2109.09509} {arXiv:2109.09509
  [physics.ins-det]} \BibitemShut {NoStop}%
\bibitem [{\citenamefont {Lynn}\ \emph {et~al.}(2020)\citenamefont {Lynn},
  \citenamefont {Lonardoni}, \citenamefont {Carlson}, \citenamefont {Chen},
  \citenamefont {Detmold}, \citenamefont {Gandolfi},\ and\ \citenamefont
  {Schwenk}}]{Lynn:2019vwp}%
  \BibitemOpen
  \bibfield  {author} {\bibinfo {author} {\bibfnamefont {J.~E.}\ \bibnamefont
  {Lynn}}, \bibinfo {author} {\bibfnamefont {D.}~\bibnamefont {Lonardoni}},
  \bibinfo {author} {\bibfnamefont {J.}~\bibnamefont {Carlson}}, \bibinfo
  {author} {\bibfnamefont {J.~W.}\ \bibnamefont {Chen}}, \bibinfo {author}
  {\bibfnamefont {W.}~\bibnamefont {Detmold}}, \bibinfo {author} {\bibfnamefont
  {S.}~\bibnamefont {Gandolfi}}, \ and\ \bibinfo {author} {\bibfnamefont
  {A.}~\bibnamefont {Schwenk}},\ }\href {\doibase 10.1088/1361-6471/ab6af7}
  {\bibfield  {journal} {\bibinfo  {journal} {J. Phys. G}\ }\textbf {\bibinfo
  {volume} {47}},\ \bibinfo {pages} {045109} (\bibinfo {year} {2020})},\
  \Eprint {http://arxiv.org/abs/1903.12587} {arXiv:1903.12587 [nucl-th]}
  \BibitemShut {NoStop}%
\bibitem [{\citenamefont {Arrington}\ \emph
  {et~al.}(2022{\natexlab{b}})\citenamefont {Arrington}, \citenamefont
  {Fomin},\ and\ \citenamefont {Schmidt}}]{Arrington:2022sov}%
  \BibitemOpen
  \bibfield  {author} {\bibinfo {author} {\bibfnamefont {J.}~\bibnamefont
  {Arrington}}, \bibinfo {author} {\bibfnamefont {N.}~\bibnamefont {Fomin}}, \
  and\ \bibinfo {author} {\bibfnamefont {A.}~\bibnamefont {Schmidt}},\ }\href
  {\doibase 10.1146/annurev-nucl-102020-022253} {\  (\bibinfo {year}
  {2022}{\natexlab{b}}),\ 10.1146/annurev-nucl-102020-022253},\ \Eprint
  {http://arxiv.org/abs/2203.02608} {arXiv:2203.02608 [nucl-ex]} \BibitemShut
  {NoStop}%
\bibitem [{\citenamefont {Cosyn}\ and\ \citenamefont
  {Ryckebusch}(2021)}]{Cosyn:2021ber}%
  \BibitemOpen
  \bibfield  {author} {\bibinfo {author} {\bibfnamefont {W.}~\bibnamefont
  {Cosyn}}\ and\ \bibinfo {author} {\bibfnamefont {J.}~\bibnamefont
  {Ryckebusch}},\ }\href {\doibase 10.1016/j.physletb.2021.136526} {\bibfield
  {journal} {\bibinfo  {journal} {Phys. Lett. B}\ }\textbf {\bibinfo {volume}
  {820}},\ \bibinfo {pages} {136526} (\bibinfo {year} {2021})},\ \Eprint
  {http://arxiv.org/abs/2106.01249} {arXiv:2106.01249 [nucl-th]} \BibitemShut
  {NoStop}%
\bibitem [{\citenamefont {Jain}\ \emph {et~al.}(2022)\citenamefont {Jain},
  \citenamefont {Pire},\ and\ \citenamefont {Ralston}}]{Jain:2022xzo}%
  \BibitemOpen
  \bibfield  {author} {\bibinfo {author} {\bibfnamefont {P.}~\bibnamefont
  {Jain}}, \bibinfo {author} {\bibfnamefont {B.}~\bibnamefont {Pire}}, \ and\
  \bibinfo {author} {\bibfnamefont {J.~P.}\ \bibnamefont {Ralston}},\ }\href
  {\doibase 10.3390/physics4020038} {\bibfield  {journal} {\bibinfo  {journal}
  {MDPI Physics}\ }\textbf {\bibinfo {volume} {4}},\ \bibinfo {pages} {578}
  (\bibinfo {year} {2022})},\ \Eprint {http://arxiv.org/abs/2203.02579}
  {arXiv:2203.02579 [hep-ph]} \BibitemShut {NoStop}%
\bibitem [{\citenamefont {Tu}\ \emph {et~al.}(2020)\citenamefont {Tu},
  \citenamefont {Jentsch}, \citenamefont {Baker}, \citenamefont {Zheng},
  \citenamefont {Lee}, \citenamefont {Venugopalan}, \citenamefont {Hen},
  \citenamefont {Higinbotham}, \citenamefont {Aschenauer},\ and\ \citenamefont
  {Ullrich}}]{Tu:2020ymk}%
  \BibitemOpen
  \bibfield  {author} {\bibinfo {author} {\bibfnamefont {Z.}~\bibnamefont
  {Tu}}, \bibinfo {author} {\bibfnamefont {A.}~\bibnamefont {Jentsch}},
  \bibinfo {author} {\bibfnamefont {M.}~\bibnamefont {Baker}}, \bibinfo
  {author} {\bibfnamefont {L.}~\bibnamefont {Zheng}}, \bibinfo {author}
  {\bibfnamefont {J.-H.}\ \bibnamefont {Lee}}, \bibinfo {author} {\bibfnamefont
  {R.}~\bibnamefont {Venugopalan}}, \bibinfo {author} {\bibfnamefont
  {O.}~\bibnamefont {Hen}}, \bibinfo {author} {\bibfnamefont {D.}~\bibnamefont
  {Higinbotham}}, \bibinfo {author} {\bibfnamefont {E.-C.}\ \bibnamefont
  {Aschenauer}}, \ and\ \bibinfo {author} {\bibfnamefont {T.}~\bibnamefont
  {Ullrich}},\ }\href {\doibase 10.1016/j.physletb.2020.135877} {\bibfield
  {journal} {\bibinfo  {journal} {Phys. Lett. B}\ }\textbf {\bibinfo {volume}
  {811}},\ \bibinfo {pages} {135877} (\bibinfo {year} {2020})},\ \Eprint
  {http://arxiv.org/abs/2005.14706} {arXiv:2005.14706 [nucl-ex]} \BibitemShut
  {NoStop}%
\bibitem [{\citenamefont {Frankfurt}\ and\ \citenamefont
  {Strikman}(1981)}]{Frankfurt:1981mk}%
  \BibitemOpen
  \bibfield  {author} {\bibinfo {author} {\bibfnamefont {L.~L.}\ \bibnamefont
  {Frankfurt}}\ and\ \bibinfo {author} {\bibfnamefont {M.~I.}\ \bibnamefont
  {Strikman}},\ }\href {\doibase 10.1016/0370-1573(81)90129-0} {\bibfield
  {journal} {\bibinfo  {journal} {Phys. Rept.}\ }\textbf {\bibinfo {volume}
  {76}},\ \bibinfo {pages} {215} (\bibinfo {year} {1981})}\BibitemShut
  {NoStop}%
\bibitem [{\citenamefont {Miller}(2000)}]{Miller:2000kv}%
  \BibitemOpen
  \bibfield  {author} {\bibinfo {author} {\bibfnamefont {G.~A.}\ \bibnamefont
  {Miller}},\ }\href {\doibase 10.1016/S0146-6410(00)00103-4} {\bibfield
  {journal} {\bibinfo  {journal} {Prog. Part. Nucl. Phys.}\ }\textbf {\bibinfo
  {volume} {45}},\ \bibinfo {pages} {83} (\bibinfo {year} {2000})},\ \Eprint
  {http://arxiv.org/abs/nucl-th/0002059} {arXiv:nucl-th/0002059} \BibitemShut
  {NoStop}%
\bibitem [{\citenamefont {Alessandro}\ \emph {et~al.}(2021)\citenamefont
  {Alessandro}, \citenamefont {Del~Dotto}, \citenamefont {Pace}, \citenamefont
  {Perna}, \citenamefont {Salm\`e},\ and\ \citenamefont
  {Scopetta}}]{Alessandro:2021cbg}%
  \BibitemOpen
  \bibfield  {author} {\bibinfo {author} {\bibfnamefont {R.}~\bibnamefont
  {Alessandro}}, \bibinfo {author} {\bibfnamefont {A.}~\bibnamefont
  {Del~Dotto}}, \bibinfo {author} {\bibfnamefont {E.}~\bibnamefont {Pace}},
  \bibinfo {author} {\bibfnamefont {G.}~\bibnamefont {Perna}}, \bibinfo
  {author} {\bibfnamefont {G.}~\bibnamefont {Salm\`e}}, \ and\ \bibinfo
  {author} {\bibfnamefont {S.}~\bibnamefont {Scopetta}},\ }\href {\doibase
  10.1103/PhysRevC.104.065204} {\bibfield  {journal} {\bibinfo  {journal}
  {Phys. Rev. C}\ }\textbf {\bibinfo {volume} {104}},\ \bibinfo {pages}
  {065204} (\bibinfo {year} {2021})},\ \Eprint
  {http://arxiv.org/abs/2107.10187} {arXiv:2107.10187 [nucl-th]} \BibitemShut
  {NoStop}%
\bibitem [{\citenamefont {Hammer}\ \emph {et~al.}(2020)\citenamefont {Hammer},
  \citenamefont {K\"onig},\ and\ \citenamefont {van Kolck}}]{Hammer:2019poc}%
  \BibitemOpen
  \bibfield  {author} {\bibinfo {author} {\bibfnamefont {H.~W.}\ \bibnamefont
  {Hammer}}, \bibinfo {author} {\bibfnamefont {S.}~\bibnamefont {K\"onig}}, \
  and\ \bibinfo {author} {\bibfnamefont {U.}~\bibnamefont {van Kolck}},\ }\href
  {\doibase 10.1103/RevModPhys.92.025004} {\bibfield  {journal} {\bibinfo
  {journal} {Rev. Mod. Phys.}\ }\textbf {\bibinfo {volume} {92}},\ \bibinfo
  {pages} {025004} (\bibinfo {year} {2020})},\ \Eprint
  {http://arxiv.org/abs/1906.12122} {arXiv:1906.12122 [nucl-th]} \BibitemShut
  {NoStop}%
\bibitem [{\citenamefont {Maris}\ \emph {et~al.}(2021)\citenamefont {Maris}
  \emph {et~al.}}]{Maris:2020qne}%
  \BibitemOpen
  \bibfield  {author} {\bibinfo {author} {\bibfnamefont {P.}~\bibnamefont
  {Maris}} \emph {et~al.},\ }\href {\doibase 10.1103/PhysRevC.103.054001}
  {\bibfield  {journal} {\bibinfo  {journal} {Phys. Rev. C}\ }\textbf {\bibinfo
  {volume} {103}},\ \bibinfo {pages} {054001} (\bibinfo {year} {2021})},\
  \Eprint {http://arxiv.org/abs/2012.12396} {arXiv:2012.12396 [nucl-th]}
  \BibitemShut {NoStop}%
\bibitem [{\citenamefont {Tews}\ \emph {et~al.}(2022)\citenamefont {Tews} \emph
  {et~al.}}]{Tews:2022yfb}%
  \BibitemOpen
  \bibfield  {author} {\bibinfo {author} {\bibfnamefont {I.}~\bibnamefont
  {Tews}} \emph {et~al.},\ }\href {\doibase 10.1007/s00601-022-01749-x}
  {\bibfield  {journal} {\bibinfo  {journal} {Few Body Syst.}\ }\textbf
  {\bibinfo {volume} {63}},\ \bibinfo {pages} {67} (\bibinfo {year} {2022})},\
  \Eprint {http://arxiv.org/abs/2202.01105} {arXiv:2202.01105 [nucl-th]}
  \BibitemShut {NoStop}%
\bibitem [{\citenamefont {Lev}\ \emph {et~al.}(1998)\citenamefont {Lev},
  \citenamefont {Pace},\ and\ \citenamefont {Salme}}]{Lev:1998qz}%
  \BibitemOpen
  \bibfield  {author} {\bibinfo {author} {\bibfnamefont {F.~M.}\ \bibnamefont
  {Lev}}, \bibinfo {author} {\bibfnamefont {E.}~\bibnamefont {Pace}}, \ and\
  \bibinfo {author} {\bibfnamefont {G.}~\bibnamefont {Salme}},\ }\href
  {\doibase 10.1016/S0375-9474(98)00469-2} {\bibfield  {journal} {\bibinfo
  {journal} {Nucl. Phys. A}\ }\textbf {\bibinfo {volume} {641}},\ \bibinfo
  {pages} {229} (\bibinfo {year} {1998})},\ \Eprint
  {http://arxiv.org/abs/hep-ph/9807255} {arXiv:hep-ph/9807255} \BibitemShut
  {NoStop}%
\bibitem [{\citenamefont {Chang}\ \emph {et~al.}(2018)\citenamefont {Chang},
  \citenamefont {Davoudi}, \citenamefont {Detmold}, \citenamefont {Gambhir},
  \citenamefont {Orginos}, \citenamefont {Savage}, \citenamefont {Shanahan},
  \citenamefont {Wagman},\ and\ \citenamefont {Winter}}]{Chang:2017eiq}%
  \BibitemOpen
  \bibfield  {author} {\bibinfo {author} {\bibfnamefont {E.}~\bibnamefont
  {Chang}}, \bibinfo {author} {\bibfnamefont {Z.}~\bibnamefont {Davoudi}},
  \bibinfo {author} {\bibfnamefont {W.}~\bibnamefont {Detmold}}, \bibinfo
  {author} {\bibfnamefont {A.~S.}\ \bibnamefont {Gambhir}}, \bibinfo {author}
  {\bibfnamefont {K.}~\bibnamefont {Orginos}}, \bibinfo {author} {\bibfnamefont
  {M.~J.}\ \bibnamefont {Savage}}, \bibinfo {author} {\bibfnamefont {P.~E.}\
  \bibnamefont {Shanahan}}, \bibinfo {author} {\bibfnamefont {M.~L.}\
  \bibnamefont {Wagman}}, \ and\ \bibinfo {author} {\bibfnamefont
  {F.}~\bibnamefont {Winter}} (\bibinfo {collaboration} {NPLQCD}),\ }\href
  {\doibase 10.1103/PhysRevLett.120.152002} {\bibfield  {journal} {\bibinfo
  {journal} {Phys. Rev. Lett.}\ }\textbf {\bibinfo {volume} {120}},\ \bibinfo
  {pages} {152002} (\bibinfo {year} {2018})},\ \Eprint
  {http://arxiv.org/abs/1712.03221} {arXiv:1712.03221 [hep-lat]} \BibitemShut
  {NoStop}%
\bibitem [{\citenamefont {Winter}\ \emph {et~al.}(2017)\citenamefont {Winter},
  \citenamefont {Detmold}, \citenamefont {Gambhir}, \citenamefont {Orginos},
  \citenamefont {Savage}, \citenamefont {Shanahan},\ and\ \citenamefont
  {Wagman}}]{Winter:2017bfs}%
  \BibitemOpen
  \bibfield  {author} {\bibinfo {author} {\bibfnamefont {F.}~\bibnamefont
  {Winter}}, \bibinfo {author} {\bibfnamefont {W.}~\bibnamefont {Detmold}},
  \bibinfo {author} {\bibfnamefont {A.~S.}\ \bibnamefont {Gambhir}}, \bibinfo
  {author} {\bibfnamefont {K.}~\bibnamefont {Orginos}}, \bibinfo {author}
  {\bibfnamefont {M.~J.}\ \bibnamefont {Savage}}, \bibinfo {author}
  {\bibfnamefont {P.~E.}\ \bibnamefont {Shanahan}}, \ and\ \bibinfo {author}
  {\bibfnamefont {M.~L.}\ \bibnamefont {Wagman}},\ }\href {\doibase
  10.1103/PhysRevD.96.094512} {\bibfield  {journal} {\bibinfo  {journal} {Phys.
  Rev. D}\ }\textbf {\bibinfo {volume} {96}},\ \bibinfo {pages} {094512}
  (\bibinfo {year} {2017})},\ \Eprint {http://arxiv.org/abs/1709.00395}
  {arXiv:1709.00395 [hep-lat]} \BibitemShut {NoStop}%
\bibitem [{\citenamefont {Detmold}\ \emph {et~al.}(2019)\citenamefont
  {Detmold}, \citenamefont {Edwards}, \citenamefont {Dudek}, \citenamefont
  {Engelhardt}, \citenamefont {Lin}, \citenamefont {Meinel}, \citenamefont
  {Orginos},\ and\ \citenamefont {Shanahan}}]{Detmold:2019ghl}%
  \BibitemOpen
  \bibfield  {author} {\bibinfo {author} {\bibfnamefont {W.}~\bibnamefont
  {Detmold}}, \bibinfo {author} {\bibfnamefont {R.~G.}\ \bibnamefont
  {Edwards}}, \bibinfo {author} {\bibfnamefont {J.~J.}\ \bibnamefont {Dudek}},
  \bibinfo {author} {\bibfnamefont {M.}~\bibnamefont {Engelhardt}}, \bibinfo
  {author} {\bibfnamefont {H.-W.}\ \bibnamefont {Lin}}, \bibinfo {author}
  {\bibfnamefont {S.}~\bibnamefont {Meinel}}, \bibinfo {author} {\bibfnamefont
  {K.}~\bibnamefont {Orginos}}, \ and\ \bibinfo {author} {\bibfnamefont
  {P.}~\bibnamefont {Shanahan}} (\bibinfo {collaboration} {USQCD}),\ }\href
  {\doibase 10.1140/epja/i2019-12902-4} {\bibfield  {journal} {\bibinfo
  {journal} {Eur. Phys. J. A}\ }\textbf {\bibinfo {volume} {55}},\ \bibinfo
  {pages} {193} (\bibinfo {year} {2019})},\ \Eprint
  {http://arxiv.org/abs/1904.09512} {arXiv:1904.09512 [hep-lat]} \BibitemShut
  {NoStop}%
\bibitem [{\citenamefont {Detmold}\ \emph {et~al.}(2021)\citenamefont
  {Detmold}, \citenamefont {Illa}, \citenamefont {Murphy}, \citenamefont
  {Oare}, \citenamefont {Orginos}, \citenamefont {Shanahan}, \citenamefont
  {Wagman},\ and\ \citenamefont {Winter}}]{Detmold:2020snb}%
  \BibitemOpen
  \bibfield  {author} {\bibinfo {author} {\bibfnamefont {W.}~\bibnamefont
  {Detmold}}, \bibinfo {author} {\bibfnamefont {M.}~\bibnamefont {Illa}},
  \bibinfo {author} {\bibfnamefont {D.~J.}\ \bibnamefont {Murphy}}, \bibinfo
  {author} {\bibfnamefont {P.}~\bibnamefont {Oare}}, \bibinfo {author}
  {\bibfnamefont {K.}~\bibnamefont {Orginos}}, \bibinfo {author} {\bibfnamefont
  {P.~E.}\ \bibnamefont {Shanahan}}, \bibinfo {author} {\bibfnamefont {M.~L.}\
  \bibnamefont {Wagman}}, \ and\ \bibinfo {author} {\bibfnamefont
  {F.}~\bibnamefont {Winter}} (\bibinfo {collaboration} {NPLQCD}),\ }\href
  {\doibase 10.1103/PhysRevLett.126.202001} {\bibfield  {journal} {\bibinfo
  {journal} {Phys. Rev. Lett.}\ }\textbf {\bibinfo {volume} {126}},\ \bibinfo
  {pages} {202001} (\bibinfo {year} {2021})},\ \Eprint
  {http://arxiv.org/abs/2009.05522} {arXiv:2009.05522 [hep-lat]} \BibitemShut
  {NoStop}%
\bibitem [{\citenamefont {Sun}\ \emph {et~al.}(2022)\citenamefont {Sun},
  \citenamefont {Detmold}, \citenamefont {Luo},\ and\ \citenamefont
  {Shanahan}}]{Sun:2022frr}%
  \BibitemOpen
  \bibfield  {author} {\bibinfo {author} {\bibfnamefont {X.}~\bibnamefont
  {Sun}}, \bibinfo {author} {\bibfnamefont {W.}~\bibnamefont {Detmold}},
  \bibinfo {author} {\bibfnamefont {D.}~\bibnamefont {Luo}}, \ and\ \bibinfo
  {author} {\bibfnamefont {P.~E.}\ \bibnamefont {Shanahan}},\ }\href {\doibase
  10.1103/PhysRevD.105.074508} {\bibfield  {journal} {\bibinfo  {journal}
  {Phys. Rev. D}\ }\textbf {\bibinfo {volume} {105}},\ \bibinfo {pages}
  {074508} (\bibinfo {year} {2022})},\ \Eprint
  {http://arxiv.org/abs/2202.03530} {arXiv:2202.03530 [nucl-th]} \BibitemShut
  {NoStop}%
\bibitem [{\citenamefont {Ivanov}\ \emph
  {et~al.}(2020{\natexlab{b}})\citenamefont {Ivanov}, \citenamefont {Kalugin},
  \citenamefont {Ogarkova},\ and\ \citenamefont
  {Ogarkov}}]{ivanov2020functional}%
  \BibitemOpen
  \bibfield  {author} {\bibinfo {author} {\bibfnamefont {M.~G.}\ \bibnamefont
  {Ivanov}}, \bibinfo {author} {\bibfnamefont {A.~E.}\ \bibnamefont {Kalugin}},
  \bibinfo {author} {\bibfnamefont {A.~A.}\ \bibnamefont {Ogarkova}}, \ and\
  \bibinfo {author} {\bibfnamefont {S.~L.}\ \bibnamefont {Ogarkov}},\
  }\href@noop {} {\bibfield  {journal} {\bibinfo  {journal} {Symmetry}\
  }\textbf {\bibinfo {volume} {12}},\ \bibinfo {pages} {1657} (\bibinfo {year}
  {2020}{\natexlab{b}})}\BibitemShut {NoStop}%
\bibitem [{\citenamefont {Kiefer}\ and\ \citenamefont
  {Wipf}(1994)}]{kiefer1994functional}%
  \BibitemOpen
  \bibfield  {author} {\bibinfo {author} {\bibfnamefont {C.}~\bibnamefont
  {Kiefer}}\ and\ \bibinfo {author} {\bibfnamefont {A.}~\bibnamefont {Wipf}},\
  }\href@noop {} {\bibfield  {journal} {\bibinfo  {journal} {Annals of
  Physics}\ }\textbf {\bibinfo {volume} {236}},\ \bibinfo {pages} {241}
  (\bibinfo {year} {1994})}\BibitemShut {NoStop}%
\bibitem [{\citenamefont {Drews}\ and\ \citenamefont
  {Weise}(2017)}]{drews2017functional}%
  \BibitemOpen
  \bibfield  {author} {\bibinfo {author} {\bibfnamefont {M.}~\bibnamefont
  {Drews}}\ and\ \bibinfo {author} {\bibfnamefont {W.}~\bibnamefont {Weise}},\
  }\href@noop {} {\bibfield  {journal} {\bibinfo  {journal} {Progress in
  Particle and Nuclear Physics}\ }\textbf {\bibinfo {volume} {93}},\ \bibinfo
  {pages} {69} (\bibinfo {year} {2017})}\BibitemShut {NoStop}%
\bibitem [{\citenamefont {Finelli}\ \emph {et~al.}(2006)\citenamefont
  {Finelli}, \citenamefont {Kaiser}, \citenamefont {Vretenar},\ and\
  \citenamefont {Weise}}]{finelli2006relativistic}%
  \BibitemOpen
  \bibfield  {author} {\bibinfo {author} {\bibfnamefont {P.}~\bibnamefont
  {Finelli}}, \bibinfo {author} {\bibfnamefont {N.}~\bibnamefont {Kaiser}},
  \bibinfo {author} {\bibfnamefont {D.}~\bibnamefont {Vretenar}}, \ and\
  \bibinfo {author} {\bibfnamefont {W.}~\bibnamefont {Weise}},\ }\href@noop {}
  {\bibfield  {journal} {\bibinfo  {journal} {Nuclear Physics A}\ }\textbf
  {\bibinfo {volume} {770}},\ \bibinfo {pages} {1} (\bibinfo {year}
  {2006})}\BibitemShut {NoStop}%
\bibitem [{\citenamefont {Meng}(2016)}]{meng2016relativistic}%
  \BibitemOpen
  \bibfield  {author} {\bibinfo {author} {\bibfnamefont {J.}~\bibnamefont
  {Meng}},\ }\href@noop {} {\emph {\bibinfo {title} {Relativistic density
  functional for nuclear structure}}},\ Vol.~\bibinfo {volume} {10}\ (\bibinfo
  {publisher} {World Scientific},\ \bibinfo {year} {2016})\BibitemShut
  {NoStop}%
\bibitem [{\citenamefont {Hotta}\ and\ \citenamefont
  {Tanisaki}(2007)}]{hotta2007d}%
  \BibitemOpen
  \bibfield  {author} {\bibinfo {author} {\bibfnamefont {R.}~\bibnamefont
  {Hotta}}\ and\ \bibinfo {author} {\bibfnamefont {T.}~\bibnamefont
  {Tanisaki}},\ }\href@noop {} {\emph {\bibinfo {title} {D-modules, perverse
  sheaves, and representation theory}}},\ Vol.\ \bibinfo {volume} {236}\
  (\bibinfo  {publisher} {Springer Science \& Business Media},\ \bibinfo {year}
  {2007})\BibitemShut {NoStop}%
\bibitem [{\citenamefont {Ferry}\ and\ \citenamefont
  {Weinberger}(2013)}]{ferry2013quantitative}%
  \BibitemOpen
  \bibfield  {author} {\bibinfo {author} {\bibfnamefont {S.}~\bibnamefont
  {Ferry}}\ and\ \bibinfo {author} {\bibfnamefont {S.}~\bibnamefont
  {Weinberger}},\ }\href@noop {} {\bibfield  {journal} {\bibinfo  {journal}
  {Proceedings of the National Academy of Sciences}\ }\textbf {\bibinfo
  {volume} {110}},\ \bibinfo {pages} {19246} (\bibinfo {year}
  {2013})}\BibitemShut {NoStop}%
\bibitem [{\citenamefont {Michor}\ and\ \citenamefont
  {Mumford}(2013)}]{michor2013zoo}%
  \BibitemOpen
  \bibfield  {author} {\bibinfo {author} {\bibfnamefont {P.~W.}\ \bibnamefont
  {Michor}}\ and\ \bibinfo {author} {\bibfnamefont {D.}~\bibnamefont
  {Mumford}},\ }\href@noop {} {\bibfield  {journal} {\bibinfo  {journal}
  {Annals of Global Analysis and Geometry}\ }\textbf {\bibinfo {volume} {44}},\
  \bibinfo {pages} {529} (\bibinfo {year} {2013})}\BibitemShut {NoStop}%
\bibitem [{\citenamefont {Hjorth}(2000)}]{hjorth2000classification}%
  \BibitemOpen
  \bibfield  {author} {\bibinfo {author} {\bibfnamefont {G.}~\bibnamefont
  {Hjorth}},\ }\href@noop {} {\emph {\bibinfo {title} {Classification and orbit
  equivalence relations}}},\ \bibinfo {number} {75}\ (\bibinfo  {publisher}
  {American Mathematical Soc.},\ \bibinfo {year} {2000})\BibitemShut {NoStop}%
\bibitem [{\citenamefont {Delabaere}\ and\ \citenamefont
  {Pham}(1999)}]{delabaere1999resurgent}%
  \BibitemOpen
  \bibfield  {author} {\bibinfo {author} {\bibfnamefont {E.}~\bibnamefont
  {Delabaere}}\ and\ \bibinfo {author} {\bibfnamefont {F.}~\bibnamefont
  {Pham}},\ }in\ \href@noop {} {\emph {\bibinfo {booktitle} {Annales de l'IHP
  Physique th{\'e}orique}}},\ Vol.~\bibinfo {volume} {71}\ (\bibinfo {year}
  {1999})\ pp.\ \bibinfo {pages} {1--94}\BibitemShut {NoStop}%
\bibitem [{\citenamefont {Filipuk}\ and\ \citenamefont
  {Halburd}(2009)}]{filipuk2009movable}%
  \BibitemOpen
  \bibfield  {author} {\bibinfo {author} {\bibfnamefont {G.}~\bibnamefont
  {Filipuk}}\ and\ \bibinfo {author} {\bibfnamefont {R.}~\bibnamefont
  {Halburd}},\ }\href@noop {} {\bibfield  {journal} {\bibinfo  {journal}
  {Journal of mathematical physics}\ }\textbf {\bibinfo {volume} {50}},\
  \bibinfo {pages} {023509} (\bibinfo {year} {2009})}\BibitemShut {NoStop}%
\bibitem [{\citenamefont {Gentile}\ and\ \citenamefont
  {Gallavotti}(2005)}]{gentile2005degenerate}%
  \BibitemOpen
  \bibfield  {author} {\bibinfo {author} {\bibfnamefont {G.}~\bibnamefont
  {Gentile}}\ and\ \bibinfo {author} {\bibfnamefont {G.}~\bibnamefont
  {Gallavotti}},\ }\href@noop {} {\bibfield  {journal} {\bibinfo  {journal}
  {Communications in Mathematical Physics}\ }\textbf {\bibinfo {volume}
  {257}},\ \bibinfo {pages} {319} (\bibinfo {year} {2005})}\BibitemShut
  {NoStop}%
\bibitem [{\citenamefont {Ashok}\ \emph {et~al.}(2020)\citenamefont {Ashok},
  \citenamefont {Jatkar}, \citenamefont {Raman} \emph
  {et~al.}}]{ashok2020aspects}%
  \BibitemOpen
  \bibfield  {author} {\bibinfo {author} {\bibfnamefont {S.~K.}\ \bibnamefont
  {Ashok}}, \bibinfo {author} {\bibfnamefont {D.~P.}\ \bibnamefont {Jatkar}},
  \bibinfo {author} {\bibfnamefont {M.}~\bibnamefont {Raman}},  \emph
  {et~al.},\ }\href@noop {} {\bibfield  {journal} {\bibinfo  {journal} {SIGMA.
  Symmetry, Integrability and Geometry: Methods and Applications}\ }\textbf
  {\bibinfo {volume} {16}},\ \bibinfo {pages} {001} (\bibinfo {year}
  {2020})}\BibitemShut {NoStop}%
\bibitem [{\citenamefont {Chakravarty}\ and\ \citenamefont
  {Ablowitz}(2010)}]{chakravarty2010parameterizations}%
  \BibitemOpen
  \bibfield  {author} {\bibinfo {author} {\bibfnamefont {S.}~\bibnamefont
  {Chakravarty}}\ and\ \bibinfo {author} {\bibfnamefont {M.~J.}\ \bibnamefont
  {Ablowitz}},\ }\href@noop {} {\bibfield  {journal} {\bibinfo  {journal}
  {Studies in Applied Mathematics}\ }\textbf {\bibinfo {volume} {124}},\
  \bibinfo {pages} {105} (\bibinfo {year} {2010})}\BibitemShut {NoStop}%
\bibitem [{\citenamefont {Pachucki}(1994)}]{pachucki1994complete}%
  \BibitemOpen
  \bibfield  {author} {\bibinfo {author} {\bibfnamefont {K.}~\bibnamefont
  {Pachucki}},\ }\href@noop {} {\bibfield  {journal} {\bibinfo  {journal}
  {Physical review letters}\ }\textbf {\bibinfo {volume} {72}},\ \bibinfo
  {pages} {3154} (\bibinfo {year} {1994})}\BibitemShut {NoStop}%
\bibitem [{\citenamefont {Eides}\ \emph {et~al.}(2001)\citenamefont {Eides},
  \citenamefont {Grotch},\ and\ \citenamefont {Shelyuto}}]{eides2001theory}%
  \BibitemOpen
  \bibfield  {author} {\bibinfo {author} {\bibfnamefont {M.~I.}\ \bibnamefont
  {Eides}}, \bibinfo {author} {\bibfnamefont {H.}~\bibnamefont {Grotch}}, \
  and\ \bibinfo {author} {\bibfnamefont {V.~A.}\ \bibnamefont {Shelyuto}},\
  }\href@noop {} {\bibfield  {journal} {\bibinfo  {journal} {Physics Reports}\
  }\textbf {\bibinfo {volume} {342}},\ \bibinfo {pages} {63} (\bibinfo {year}
  {2001})}\BibitemShut {NoStop}%
\bibitem [{\citenamefont {Constanda}(2020)}]{constanda2020direct}%
  \BibitemOpen
  \bibfield  {author} {\bibinfo {author} {\bibfnamefont {C.}~\bibnamefont
  {Constanda}},\ }\href@noop {} {\emph {\bibinfo {title} {Direct and indirect
  boundary integral equation methods}}}\ (\bibinfo  {publisher} {Chapman and
  Hall/CRC},\ \bibinfo {year} {2020})\BibitemShut {NoStop}%
\bibitem [{\citenamefont {Harnad}\ and\ \citenamefont
  {Its}(2002)}]{harnad2002integrable}%
  \BibitemOpen
  \bibfield  {author} {\bibinfo {author} {\bibfnamefont {J.}~\bibnamefont
  {Harnad}}\ and\ \bibinfo {author} {\bibfnamefont {A.~R.}\ \bibnamefont
  {Its}},\ }\href@noop {} {\bibfield  {journal} {\bibinfo  {journal}
  {Communications in mathematical physics}\ }\textbf {\bibinfo {volume}
  {226}},\ \bibinfo {pages} {497} (\bibinfo {year} {2002})}\BibitemShut
  {NoStop}%
\bibitem [{\citenamefont {Toledano-Laredo}\ and\ \citenamefont
  {Xu}(2022)}]{toledano2022stokes}%
  \BibitemOpen
  \bibfield  {author} {\bibinfo {author} {\bibfnamefont {V.}~\bibnamefont
  {Toledano-Laredo}}\ and\ \bibinfo {author} {\bibfnamefont {X.}~\bibnamefont
  {Xu}},\ }\href@noop {} {\bibfield  {journal} {\bibinfo  {journal} {arXiv
  preprint arXiv:2202.10298}\ } (\bibinfo {year} {2022})}\BibitemShut {NoStop}%
\bibitem [{\citenamefont {Bridgeland}\ and\ \citenamefont
  {Laredo}(2013)}]{bridgeland2013stokes}%
  \BibitemOpen
  \bibfield  {author} {\bibinfo {author} {\bibfnamefont {T.}~\bibnamefont
  {Bridgeland}}\ and\ \bibinfo {author} {\bibfnamefont {V.~T.}\ \bibnamefont
  {Laredo}},\ }\href@noop {} {\bibfield  {journal} {\bibinfo  {journal}
  {Journal f{\"u}r die reine und angewandte Mathematik (Crelles Journal)}\
  }\textbf {\bibinfo {volume} {2013}},\ \bibinfo {pages} {89} (\bibinfo {year}
  {2013})}\BibitemShut {NoStop}%
\bibitem [{\citenamefont {Barkatou}\ \emph {et~al.}(2017)\citenamefont
  {Barkatou}, \citenamefont {Jaroschek},\ and\ \citenamefont
  {Maddah}}]{barkatou2017formal}%
  \BibitemOpen
  \bibfield  {author} {\bibinfo {author} {\bibfnamefont {M.~A.}\ \bibnamefont
  {Barkatou}}, \bibinfo {author} {\bibfnamefont {M.}~\bibnamefont {Jaroschek}},
  \ and\ \bibinfo {author} {\bibfnamefont {S.~S.}\ \bibnamefont {Maddah}},\
  }\href@noop {} {\bibfield  {journal} {\bibinfo  {journal} {Journal of
  Symbolic Computation}\ }\textbf {\bibinfo {volume} {81}},\ \bibinfo {pages}
  {41} (\bibinfo {year} {2017})}\BibitemShut {NoStop}%
\bibitem [{\citenamefont {Katzarkov}\ \emph {et~al.}(2015)\citenamefont
  {Katzarkov}, \citenamefont {Noll}, \citenamefont {Pandit},\ and\
  \citenamefont {Simpson}}]{katzarkov2015harmonic}%
  \BibitemOpen
  \bibfield  {author} {\bibinfo {author} {\bibfnamefont {L.}~\bibnamefont
  {Katzarkov}}, \bibinfo {author} {\bibfnamefont {A.}~\bibnamefont {Noll}},
  \bibinfo {author} {\bibfnamefont {P.}~\bibnamefont {Pandit}}, \ and\ \bibinfo
  {author} {\bibfnamefont {C.}~\bibnamefont {Simpson}},\ }\href@noop {}
  {\bibfield  {journal} {\bibinfo  {journal} {Communications in Mathematical
  Physics}\ }\textbf {\bibinfo {volume} {336}},\ \bibinfo {pages} {853}
  (\bibinfo {year} {2015})}\BibitemShut {NoStop}%
\bibitem [{\citenamefont {Sargsian}\ and\ \citenamefont
  {Strikman}(2006)}]{Sargsian:2005rm}%
  \BibitemOpen
  \bibfield  {author} {\bibinfo {author} {\bibfnamefont {M.}~\bibnamefont
  {Sargsian}}\ and\ \bibinfo {author} {\bibfnamefont {M.}~\bibnamefont
  {Strikman}},\ }\href {\doibase 10.1016/j.physletb.2006.05.091} {\bibfield
  {journal} {\bibinfo  {journal} {Phys. Lett. B}\ }\textbf {\bibinfo {volume}
  {639}},\ \bibinfo {pages} {223} (\bibinfo {year} {2006})},\ \Eprint
  {http://arxiv.org/abs/hep-ph/0511054} {arXiv:hep-ph/0511054} \BibitemShut
  {NoStop}%
\bibitem [{\citenamefont {Cosyn}\ and\ \citenamefont
  {Weiss}(2020)}]{Cosyn:2020kwu}%
  \BibitemOpen
  \bibfield  {author} {\bibinfo {author} {\bibfnamefont {W.}~\bibnamefont
  {Cosyn}}\ and\ \bibinfo {author} {\bibfnamefont {C.}~\bibnamefont {Weiss}},\
  }\href {\doibase 10.1103/PhysRevC.102.065204} {\bibfield  {journal} {\bibinfo
   {journal} {Phys. Rev. C}\ }\textbf {\bibinfo {volume} {102}},\ \bibinfo
  {pages} {065204} (\bibinfo {year} {2020})},\ \Eprint
  {http://arxiv.org/abs/2006.03033} {arXiv:2006.03033 [hep-ph]} \BibitemShut
  {NoStop}%
\bibitem [{\citenamefont {Jentsch}\ \emph {et~al.}(2021)\citenamefont
  {Jentsch}, \citenamefont {Tu},\ and\ \citenamefont
  {Weiss}}]{Jentsch:2021qdp}%
  \BibitemOpen
  \bibfield  {author} {\bibinfo {author} {\bibfnamefont {A.}~\bibnamefont
  {Jentsch}}, \bibinfo {author} {\bibfnamefont {Z.}~\bibnamefont {Tu}}, \ and\
  \bibinfo {author} {\bibfnamefont {C.}~\bibnamefont {Weiss}},\ }\href
  {\doibase 10.1103/PhysRevC.104.065205} {\bibfield  {journal} {\bibinfo
  {journal} {Phys. Rev. C}\ }\textbf {\bibinfo {volume} {104}},\ \bibinfo
  {pages} {065205} (\bibinfo {year} {2021})},\ \Eprint
  {http://arxiv.org/abs/2108.08314} {arXiv:2108.08314 [hep-ph]} \BibitemShut
  {NoStop}%
\bibitem [{\citenamefont {Palli}\ \emph {et~al.}(2009)\citenamefont {Palli},
  \citenamefont {Ciofi~degli Atti}, \citenamefont {Kaptari}, \citenamefont
  {Mezzetti},\ and\ \citenamefont {Alvioli}}]{Palli:2009it}%
  \BibitemOpen
  \bibfield  {author} {\bibinfo {author} {\bibfnamefont {V.}~\bibnamefont
  {Palli}}, \bibinfo {author} {\bibfnamefont {C.}~\bibnamefont {Ciofi~degli
  Atti}}, \bibinfo {author} {\bibfnamefont {L.~P.}\ \bibnamefont {Kaptari}},
  \bibinfo {author} {\bibfnamefont {C.~B.}\ \bibnamefont {Mezzetti}}, \ and\
  \bibinfo {author} {\bibfnamefont {M.}~\bibnamefont {Alvioli}},\ }\href
  {\doibase 10.1103/PhysRevC.80.054610} {\bibfield  {journal} {\bibinfo
  {journal} {Phys. Rev. C}\ }\textbf {\bibinfo {volume} {80}},\ \bibinfo
  {pages} {054610} (\bibinfo {year} {2009})},\ \Eprint
  {http://arxiv.org/abs/0911.1377} {arXiv:0911.1377 [nucl-th]} \BibitemShut
  {NoStop}%
\bibitem [{\citenamefont {Strikman}\ and\ \citenamefont
  {Weiss}(2018)}]{Strikman:2017koc}%
  \BibitemOpen
  \bibfield  {author} {\bibinfo {author} {\bibfnamefont {M.}~\bibnamefont
  {Strikman}}\ and\ \bibinfo {author} {\bibfnamefont {C.}~\bibnamefont
  {Weiss}},\ }\href {\doibase 10.1103/PhysRevC.97.035209} {\bibfield  {journal}
  {\bibinfo  {journal} {Phys. Rev. C}\ }\textbf {\bibinfo {volume} {97}},\
  \bibinfo {pages} {035209} (\bibinfo {year} {2018})},\ \Eprint
  {http://arxiv.org/abs/1706.02244} {arXiv:1706.02244 [hep-ph]} \BibitemShut
  {NoStop}%
\bibitem [{\citenamefont {Goncalves}\ \emph {et~al.}(2016)\citenamefont
  {Goncalves}, \citenamefont {Navarra},\ and\ \citenamefont
  {Spiering}}]{Goncalves:2015mbf}%
  \BibitemOpen
  \bibfield  {author} {\bibinfo {author} {\bibfnamefont {V.~P.}\ \bibnamefont
  {Goncalves}}, \bibinfo {author} {\bibfnamefont {F.~S.}\ \bibnamefont
  {Navarra}}, \ and\ \bibinfo {author} {\bibfnamefont {D.}~\bibnamefont
  {Spiering}},\ }\href {\doibase 10.1103/PhysRevD.93.054025} {\bibfield
  {journal} {\bibinfo  {journal} {Phys. Rev. D}\ }\textbf {\bibinfo {volume}
  {93}},\ \bibinfo {pages} {054025} (\bibinfo {year} {2016})},\ \Eprint
  {http://arxiv.org/abs/1512.06594} {arXiv:1512.06594 [hep-ph]} \BibitemShut
  {NoStop}%
\bibitem [{\citenamefont {Freese}\ \emph {et~al.}(2015)\citenamefont {Freese},
  \citenamefont {Sargsian},\ and\ \citenamefont {Strikman}}]{Freese:2014zda}%
  \BibitemOpen
  \bibfield  {author} {\bibinfo {author} {\bibfnamefont {A.~J.}\ \bibnamefont
  {Freese}}, \bibinfo {author} {\bibfnamefont {M.~M.}\ \bibnamefont
  {Sargsian}}, \ and\ \bibinfo {author} {\bibfnamefont {M.~I.}\ \bibnamefont
  {Strikman}},\ }\href {\doibase 10.1140/epjc/s10052-015-3755-4} {\bibfield
  {journal} {\bibinfo  {journal} {Eur. Phys. J. C}\ }\textbf {\bibinfo {volume}
  {75}},\ \bibinfo {pages} {534} (\bibinfo {year} {2015})},\ \Eprint
  {http://arxiv.org/abs/1411.6605} {arXiv:1411.6605 [hep-ph]} \BibitemShut
  {NoStop}%
\bibitem [{\citenamefont {Cosyn}\ and\ \citenamefont
  {Sargsian}(2011)}]{Cosyn:2011jnm}%
  \BibitemOpen
  \bibfield  {author} {\bibinfo {author} {\bibfnamefont {W.}~\bibnamefont
  {Cosyn}}\ and\ \bibinfo {author} {\bibfnamefont {M.}~\bibnamefont
  {Sargsian}},\ }\href {\doibase 10.1103/PhysRevC.84.014601} {\bibfield
  {journal} {\bibinfo  {journal} {Phys. Rev. C}\ }\textbf {\bibinfo {volume}
  {84}},\ \bibinfo {pages} {014601} (\bibinfo {year} {2011})},\ \Eprint
  {http://arxiv.org/abs/1012.0293} {arXiv:1012.0293 [nucl-th]} \BibitemShut
  {NoStop}%
\bibitem [{\citenamefont {Barbieri}\ \emph {et~al.}(1995)\citenamefont
  {Barbieri}, \citenamefont {Hall},\ and\ \citenamefont
  {Strumia}}]{Barbieri:1995tw}%
  \BibitemOpen
  \bibfield  {author} {\bibinfo {author} {\bibfnamefont {R.}~\bibnamefont
  {Barbieri}}, \bibinfo {author} {\bibfnamefont {L.~J.}\ \bibnamefont {Hall}},
  \ and\ \bibinfo {author} {\bibfnamefont {A.}~\bibnamefont {Strumia}},\ }\href
  {\doibase 10.1016/0550-3213(95)00208-A} {\bibfield  {journal} {\bibinfo
  {journal} {Nucl. Phys. B}\ }\textbf {\bibinfo {volume} {445}},\ \bibinfo
  {pages} {219} (\bibinfo {year} {1995})},\ \Eprint
  {http://arxiv.org/abs/hep-ph/9501334} {arXiv:hep-ph/9501334} \BibitemShut
  {NoStop}%
\bibitem [{\citenamefont {Abada}\ \emph {et~al.}(2007)\citenamefont {Abada},
  \citenamefont {Biggio}, \citenamefont {Bonnet}, \citenamefont {Gavela},\ and\
  \citenamefont {Hambye}}]{Abada:2007ux}%
  \BibitemOpen
  \bibfield  {author} {\bibinfo {author} {\bibfnamefont {A.}~\bibnamefont
  {Abada}}, \bibinfo {author} {\bibfnamefont {C.}~\bibnamefont {Biggio}},
  \bibinfo {author} {\bibfnamefont {F.}~\bibnamefont {Bonnet}}, \bibinfo
  {author} {\bibfnamefont {M.~B.}\ \bibnamefont {Gavela}}, \ and\ \bibinfo
  {author} {\bibfnamefont {T.}~\bibnamefont {Hambye}},\ }\href {\doibase
  10.1088/1126-6708/2007/12/061} {\bibfield  {journal} {\bibinfo  {journal}
  {JHEP}\ }\textbf {\bibinfo {volume} {12}},\ \bibinfo {pages} {061} (\bibinfo
  {year} {2007})},\ \Eprint {http://arxiv.org/abs/0707.4058} {arXiv:0707.4058
  [hep-ph]} \BibitemShut {NoStop}%
\bibitem [{\citenamefont {Alonso}\ \emph {et~al.}(2013)\citenamefont {Alonso},
  \citenamefont {Dhen}, \citenamefont {Gavela},\ and\ \citenamefont
  {Hambye}}]{Alonso:2012ji}%
  \BibitemOpen
  \bibfield  {author} {\bibinfo {author} {\bibfnamefont {R.}~\bibnamefont
  {Alonso}}, \bibinfo {author} {\bibfnamefont {M.}~\bibnamefont {Dhen}},
  \bibinfo {author} {\bibfnamefont {M.~B.}\ \bibnamefont {Gavela}}, \ and\
  \bibinfo {author} {\bibfnamefont {T.}~\bibnamefont {Hambye}},\ }\href
  {\doibase 10.1007/JHEP01(2013)118} {\bibfield  {journal} {\bibinfo  {journal}
  {JHEP}\ }\textbf {\bibinfo {volume} {01}},\ \bibinfo {pages} {118} (\bibinfo
  {year} {2013})},\ \Eprint {http://arxiv.org/abs/1209.2679} {arXiv:1209.2679
  [hep-ph]} \BibitemShut {NoStop}%
\bibitem [{\citenamefont {Cirigliano}\ \emph {et~al.}(2005)\citenamefont
  {Cirigliano}, \citenamefont {Grinstein}, \citenamefont {Isidori},\ and\
  \citenamefont {Wise}}]{Cirigliano:2005ck}%
  \BibitemOpen
  \bibfield  {author} {\bibinfo {author} {\bibfnamefont {V.}~\bibnamefont
  {Cirigliano}}, \bibinfo {author} {\bibfnamefont {B.}~\bibnamefont
  {Grinstein}}, \bibinfo {author} {\bibfnamefont {G.}~\bibnamefont {Isidori}},
  \ and\ \bibinfo {author} {\bibfnamefont {M.~B.}\ \bibnamefont {Wise}},\
  }\href {\doibase 10.1016/j.nuclphysb.2005.08.037} {\bibfield  {journal}
  {\bibinfo  {journal} {Nucl. Phys. B}\ }\textbf {\bibinfo {volume} {728}},\
  \bibinfo {pages} {121} (\bibinfo {year} {2005})},\ \Eprint
  {http://arxiv.org/abs/hep-ph/0507001} {arXiv:hep-ph/0507001} \BibitemShut
  {NoStop}%
\bibitem [{\citenamefont {Raidal}\ \emph {et~al.}(2008)\citenamefont {Raidal}
  \emph {et~al.}}]{Raidal:2008jk}%
  \BibitemOpen
  \bibfield  {author} {\bibinfo {author} {\bibfnamefont {M.}~\bibnamefont
  {Raidal}} \emph {et~al.},\ }\href {\doibase 10.1140/epjc/s10052-008-0715-2}
  {\bibfield  {journal} {\bibinfo  {journal} {Eur. Phys. J. C}\ }\textbf
  {\bibinfo {volume} {57}},\ \bibinfo {pages} {13} (\bibinfo {year} {2008})},\
  \Eprint {http://arxiv.org/abs/0801.1826} {arXiv:0801.1826 [hep-ph]}
  \BibitemShut {NoStop}%
\bibitem [{\citenamefont {de~Gouvea}\ and\ \citenamefont
  {Vogel}(2013)}]{deGouvea:2013zba}%
  \BibitemOpen
  \bibfield  {author} {\bibinfo {author} {\bibfnamefont {A.}~\bibnamefont
  {de~Gouvea}}\ and\ \bibinfo {author} {\bibfnamefont {P.}~\bibnamefont
  {Vogel}},\ }\href {\doibase 10.1016/j.ppnp.2013.03.006} {\bibfield  {journal}
  {\bibinfo  {journal} {Prog. Part. Nucl. Phys.}\ }\textbf {\bibinfo {volume}
  {71}},\ \bibinfo {pages} {75} (\bibinfo {year} {2013})},\ \Eprint
  {http://arxiv.org/abs/1303.4097} {arXiv:1303.4097 [hep-ph]} \BibitemShut
  {NoStop}%
\bibitem [{\citenamefont {Calibbi}\ and\ \citenamefont
  {Signorelli}(2018)}]{Calibbi:2017uvl}%
  \BibitemOpen
  \bibfield  {author} {\bibinfo {author} {\bibfnamefont {L.}~\bibnamefont
  {Calibbi}}\ and\ \bibinfo {author} {\bibfnamefont {G.}~\bibnamefont
  {Signorelli}},\ }\href {\doibase 10.1393/ncr/i2018-10144-0} {\bibfield
  {journal} {\bibinfo  {journal} {Riv. Nuovo Cim.}\ }\textbf {\bibinfo {volume}
  {41}},\ \bibinfo {pages} {71} (\bibinfo {year} {2018})},\ \Eprint
  {http://arxiv.org/abs/1709.00294} {arXiv:1709.00294 [hep-ph]} \BibitemShut
  {NoStop}%
\bibitem [{\citenamefont {Gonderinger}\ and\ \citenamefont
  {Ramsey-Musolf}(2010)}]{Gonderinger:2010yn}%
  \BibitemOpen
  \bibfield  {author} {\bibinfo {author} {\bibfnamefont {M.}~\bibnamefont
  {Gonderinger}}\ and\ \bibinfo {author} {\bibfnamefont {M.~J.}\ \bibnamefont
  {Ramsey-Musolf}},\ }\href {\doibase 10.1007/JHEP11(2010)045} {\bibfield
  {journal} {\bibinfo  {journal} {JHEP}\ }\textbf {\bibinfo {volume} {11}},\
  \bibinfo {pages} {045} (\bibinfo {year} {2010})},\ \bibinfo {note} {[Erratum:
  JHEP 05, 047 (2012)]},\ \Eprint {http://arxiv.org/abs/1006.5063}
  {arXiv:1006.5063 [hep-ph]} \BibitemShut {NoStop}%
\bibitem [{\citenamefont {Cirigliano}\ \emph {et~al.}(2021)\citenamefont
  {Cirigliano}, \citenamefont {Fuyuto}, \citenamefont {Lee}, \citenamefont
  {Mereghetti},\ and\ \citenamefont {Yan}}]{Cirigliano:2021img}%
  \BibitemOpen
  \bibfield  {author} {\bibinfo {author} {\bibfnamefont {V.}~\bibnamefont
  {Cirigliano}}, \bibinfo {author} {\bibfnamefont {K.}~\bibnamefont {Fuyuto}},
  \bibinfo {author} {\bibfnamefont {C.}~\bibnamefont {Lee}}, \bibinfo {author}
  {\bibfnamefont {E.}~\bibnamefont {Mereghetti}}, \ and\ \bibinfo {author}
  {\bibfnamefont {B.}~\bibnamefont {Yan}},\ }\href {\doibase
  10.1007/JHEP03(2021)256} {\bibfield  {journal} {\bibinfo  {journal} {JHEP}\
  }\textbf {\bibinfo {volume} {03}},\ \bibinfo {pages} {256} (\bibinfo {year}
  {2021})},\ \Eprint {http://arxiv.org/abs/2102.06176} {arXiv:2102.06176
  [hep-ph]} \BibitemShut {NoStop}%
\bibitem [{\citenamefont {Zhang}\ \emph
  {et~al.}(2022{\natexlab{b}})\citenamefont {Zhang} \emph
  {et~al.}}]{Zhang:2022zuz}%
  \BibitemOpen
  \bibfield  {author} {\bibinfo {author} {\bibfnamefont {J.~L.}\ \bibnamefont
  {Zhang}} \emph {et~al.},\ }\href@noop {} {\  (\bibinfo {year}
  {2022}{\natexlab{b}})},\ \Eprint {http://arxiv.org/abs/2207.10261}
  {arXiv:2207.10261 [hep-ph]} \BibitemShut {NoStop}%
\bibitem [{\citenamefont {Banerjee}\ \emph {et~al.}(2022)\citenamefont
  {Banerjee} \emph {et~al.}}]{Banerjee:2022xuw}%
  \BibitemOpen
  \bibfield  {author} {\bibinfo {author} {\bibfnamefont {S.}~\bibnamefont
  {Banerjee}} \emph {et~al.},\ }\href@noop {} {\  (\bibinfo {year} {2022})},\
  \Eprint {http://arxiv.org/abs/2203.14919} {arXiv:2203.14919 [hep-ph]}
  \BibitemShut {NoStop}%
\bibitem [{\citenamefont {Ellis}\ \emph {et~al.}(2021)\citenamefont {Ellis},
  \citenamefont {Madigan}, \citenamefont {Mimasu}, \citenamefont {Sanz},\ and\
  \citenamefont {You}}]{Ellis:2020unq}%
  \BibitemOpen
  \bibfield  {author} {\bibinfo {author} {\bibfnamefont {J.}~\bibnamefont
  {Ellis}}, \bibinfo {author} {\bibfnamefont {M.}~\bibnamefont {Madigan}},
  \bibinfo {author} {\bibfnamefont {K.}~\bibnamefont {Mimasu}}, \bibinfo
  {author} {\bibfnamefont {V.}~\bibnamefont {Sanz}}, \ and\ \bibinfo {author}
  {\bibfnamefont {T.}~\bibnamefont {You}},\ }\href {\doibase
  10.1007/JHEP04(2021)279} {\bibfield  {journal} {\bibinfo  {journal} {JHEP}\
  }\textbf {\bibinfo {volume} {04}},\ \bibinfo {pages} {279} (\bibinfo {year}
  {2021})},\ \Eprint {http://arxiv.org/abs/2012.02779} {arXiv:2012.02779
  [hep-ph]} \BibitemShut {NoStop}%
\bibitem [{\citenamefont {Ethier}\ \emph {et~al.}(2021)\citenamefont {Ethier},
  \citenamefont {Magni}, \citenamefont {Maltoni}, \citenamefont {Mantani},
  \citenamefont {Nocera}, \citenamefont {Rojo}, \citenamefont {Slade},
  \citenamefont {Vryonidou},\ and\ \citenamefont {Zhang}}]{Ethier:2021bye}%
  \BibitemOpen
  \bibfield  {author} {\bibinfo {author} {\bibfnamefont {J.~J.}\ \bibnamefont
  {Ethier}}, \bibinfo {author} {\bibfnamefont {G.}~\bibnamefont {Magni}},
  \bibinfo {author} {\bibfnamefont {F.}~\bibnamefont {Maltoni}}, \bibinfo
  {author} {\bibfnamefont {L.}~\bibnamefont {Mantani}}, \bibinfo {author}
  {\bibfnamefont {E.~R.}\ \bibnamefont {Nocera}}, \bibinfo {author}
  {\bibfnamefont {J.}~\bibnamefont {Rojo}}, \bibinfo {author} {\bibfnamefont
  {E.}~\bibnamefont {Slade}}, \bibinfo {author} {\bibfnamefont
  {E.}~\bibnamefont {Vryonidou}}, \ and\ \bibinfo {author} {\bibfnamefont
  {C.}~\bibnamefont {Zhang}} (\bibinfo {collaboration} {SMEFiT}),\ }\href
  {\doibase 10.1007/JHEP11(2021)089} {\bibfield  {journal} {\bibinfo  {journal}
  {JHEP}\ }\textbf {\bibinfo {volume} {11}},\ \bibinfo {pages} {089} (\bibinfo
  {year} {2021})},\ \Eprint {http://arxiv.org/abs/2105.00006} {arXiv:2105.00006
  [hep-ph]} \BibitemShut {NoStop}%
\bibitem [{\citenamefont {Falkowski}\ \emph {et~al.}(2017)\citenamefont
  {Falkowski}, \citenamefont {Gonz\'alez-Alonso},\ and\ \citenamefont
  {Mimouni}}]{Falkowski:2017pss}%
  \BibitemOpen
  \bibfield  {author} {\bibinfo {author} {\bibfnamefont {A.}~\bibnamefont
  {Falkowski}}, \bibinfo {author} {\bibfnamefont {M.}~\bibnamefont
  {Gonz\'alez-Alonso}}, \ and\ \bibinfo {author} {\bibfnamefont
  {K.}~\bibnamefont {Mimouni}},\ }\href {\doibase 10.1007/JHEP08(2017)123}
  {\bibfield  {journal} {\bibinfo  {journal} {JHEP}\ }\textbf {\bibinfo
  {volume} {08}},\ \bibinfo {pages} {123} (\bibinfo {year} {2017})},\ \Eprint
  {http://arxiv.org/abs/1706.03783} {arXiv:1706.03783 [hep-ph]} \BibitemShut
  {NoStop}%
\bibitem [{\citenamefont {Alte}\ \emph {et~al.}(2019)\citenamefont {Alte},
  \citenamefont {K\"onig},\ and\ \citenamefont {Shepherd}}]{Alte:2018xgc}%
  \BibitemOpen
  \bibfield  {author} {\bibinfo {author} {\bibfnamefont {S.}~\bibnamefont
  {Alte}}, \bibinfo {author} {\bibfnamefont {M.}~\bibnamefont {K\"onig}}, \
  and\ \bibinfo {author} {\bibfnamefont {W.}~\bibnamefont {Shepherd}},\ }\href
  {\doibase 10.1007/JHEP07(2019)144} {\bibfield  {journal} {\bibinfo  {journal}
  {JHEP}\ }\textbf {\bibinfo {volume} {07}},\ \bibinfo {pages} {144} (\bibinfo
  {year} {2019})},\ \Eprint {http://arxiv.org/abs/1812.07575} {arXiv:1812.07575
  [hep-ph]} \BibitemShut {NoStop}%
\bibitem [{\citenamefont {Boughezal}\ \emph {et~al.}(2020)\citenamefont
  {Boughezal}, \citenamefont {Petriello},\ and\ \citenamefont
  {Wiegand}}]{Boughezal:2020uwq}%
  \BibitemOpen
  \bibfield  {author} {\bibinfo {author} {\bibfnamefont {R.}~\bibnamefont
  {Boughezal}}, \bibinfo {author} {\bibfnamefont {F.}~\bibnamefont
  {Petriello}}, \ and\ \bibinfo {author} {\bibfnamefont {D.}~\bibnamefont
  {Wiegand}},\ }\href {\doibase 10.1103/PhysRevD.101.116002} {\bibfield
  {journal} {\bibinfo  {journal} {Phys. Rev. D}\ }\textbf {\bibinfo {volume}
  {101}},\ \bibinfo {pages} {116002} (\bibinfo {year} {2020})},\ \Eprint
  {http://arxiv.org/abs/2004.00748} {arXiv:2004.00748 [hep-ph]} \BibitemShut
  {NoStop}%
\bibitem [{\citenamefont {Boughezal}\ \emph {et~al.}(2022)\citenamefont
  {Boughezal}, \citenamefont {Emmert}, \citenamefont {Kutz}, \citenamefont
  {Mantry}, \citenamefont {Nycz}, \citenamefont {Petriello}, \citenamefont
  {\c{S}im\c{s}ek}, \citenamefont {Wiegand},\ and\ \citenamefont
  {Zheng}}]{Boughezal:2022pmb}%
  \BibitemOpen
  \bibfield  {author} {\bibinfo {author} {\bibfnamefont {R.}~\bibnamefont
  {Boughezal}}, \bibinfo {author} {\bibfnamefont {A.}~\bibnamefont {Emmert}},
  \bibinfo {author} {\bibfnamefont {T.}~\bibnamefont {Kutz}}, \bibinfo {author}
  {\bibfnamefont {S.}~\bibnamefont {Mantry}}, \bibinfo {author} {\bibfnamefont
  {M.}~\bibnamefont {Nycz}}, \bibinfo {author} {\bibfnamefont {F.}~\bibnamefont
  {Petriello}}, \bibinfo {author} {\bibfnamefont {K.}~\bibnamefont
  {\c{S}im\c{s}ek}}, \bibinfo {author} {\bibfnamefont {D.}~\bibnamefont
  {Wiegand}}, \ and\ \bibinfo {author} {\bibfnamefont {X.}~\bibnamefont
  {Zheng}},\ }\href {\doibase 10.1103/PhysRevD.106.016006} {\bibfield
  {journal} {\bibinfo  {journal} {Phys. Rev. D}\ }\textbf {\bibinfo {volume}
  {106}},\ \bibinfo {pages} {016006} (\bibinfo {year} {2022})},\ \Eprint
  {http://arxiv.org/abs/2204.07557} {arXiv:2204.07557 [hep-ph]} \BibitemShut
  {NoStop}%
\bibitem [{\citenamefont {Weinberg}(1989)}]{PhysRevLett.63.2333}%
  \BibitemOpen
  \bibfield  {author} {\bibinfo {author} {\bibfnamefont {S.}~\bibnamefont
  {Weinberg}},\ }\href {\doibase 10.1103/PhysRevLett.63.2333} {\bibfield
  {journal} {\bibinfo  {journal} {Phys. Rev. Lett.}\ }\textbf {\bibinfo
  {volume} {63}},\ \bibinfo {pages} {2333} (\bibinfo {year}
  {1989})}\BibitemShut {NoStop}%
\bibitem [{\citenamefont {Cirigliano}\ \emph {et~al.}(2020)\citenamefont
  {Cirigliano}, \citenamefont {Mereghetti},\ and\ \citenamefont
  {Stoffer}}]{Cirigliano:2020msr}%
  \BibitemOpen
  \bibfield  {author} {\bibinfo {author} {\bibfnamefont {V.}~\bibnamefont
  {Cirigliano}}, \bibinfo {author} {\bibfnamefont {E.}~\bibnamefont
  {Mereghetti}}, \ and\ \bibinfo {author} {\bibfnamefont {P.}~\bibnamefont
  {Stoffer}},\ }\href {\doibase 10.1007/JHEP09(2020)094} {\bibfield  {journal}
  {\bibinfo  {journal} {JHEP}\ }\textbf {\bibinfo {volume} {09}},\ \bibinfo
  {pages} {094} (\bibinfo {year} {2020})},\ \Eprint
  {http://arxiv.org/abs/2004.03576} {arXiv:2004.03576 [hep-ph]} \BibitemShut
  {NoStop}%
\bibitem [{\citenamefont {Hatta}(2021)}]{Hatta:2020riw}%
  \BibitemOpen
  \bibfield  {author} {\bibinfo {author} {\bibfnamefont {Y.}~\bibnamefont
  {Hatta}},\ }\href {\doibase 10.1016/j.physletb.2021.136126} {\bibfield
  {journal} {\bibinfo  {journal} {Phys. Lett. B}\ }\textbf {\bibinfo {volume}
  {814}},\ \bibinfo {pages} {136126} (\bibinfo {year} {2021})},\ \Eprint
  {http://arxiv.org/abs/2012.01865} {arXiv:2012.01865 [hep-ph]} \BibitemShut
  {NoStop}%
\bibitem [{\citenamefont {Liu}\ \emph {et~al.}(2018)\citenamefont {Liu},
  \citenamefont {Zhao},\ and\ \citenamefont {Gao}}]{Liu:2017olr}%
  \BibitemOpen
  \bibfield  {author} {\bibinfo {author} {\bibfnamefont {T.}~\bibnamefont
  {Liu}}, \bibinfo {author} {\bibfnamefont {Z.}~\bibnamefont {Zhao}}, \ and\
  \bibinfo {author} {\bibfnamefont {H.}~\bibnamefont {Gao}},\ }\href {\doibase
  10.1103/PhysRevD.97.074018} {\bibfield  {journal} {\bibinfo  {journal} {Phys.
  Rev. D}\ }\textbf {\bibinfo {volume} {97}},\ \bibinfo {pages} {074018}
  (\bibinfo {year} {2018})},\ \Eprint {http://arxiv.org/abs/1704.00113}
  {arXiv:1704.00113 [hep-ph]} \BibitemShut {NoStop}%
\bibitem [{\citenamefont {Radici}\ and\ \citenamefont
  {Bacchetta}(2018)}]{Radici:2018iag}%
  \BibitemOpen
  \bibfield  {author} {\bibinfo {author} {\bibfnamefont {M.}~\bibnamefont
  {Radici}}\ and\ \bibinfo {author} {\bibfnamefont {A.}~\bibnamefont
  {Bacchetta}},\ }\href {\doibase 10.1103/PhysRevLett.120.192001} {\bibfield
  {journal} {\bibinfo  {journal} {Phys. Rev. Lett.}\ }\textbf {\bibinfo
  {volume} {120}},\ \bibinfo {pages} {192001} (\bibinfo {year} {2018})},\
  \Eprint {http://arxiv.org/abs/1802.05212} {arXiv:1802.05212 [hep-ph]}
  \BibitemShut {NoStop}%
\bibitem [{\citenamefont {Cammarota}\ \emph {et~al.}(2020)\citenamefont
  {Cammarota}, \citenamefont {Gamberg}, \citenamefont {Kang}, \citenamefont
  {Miller}, \citenamefont {Pitonyak}, \citenamefont {Prokudin}, \citenamefont
  {Rogers},\ and\ \citenamefont {Sato}}]{Cammarota:2020qcw}%
  \BibitemOpen
  \bibfield  {author} {\bibinfo {author} {\bibfnamefont {J.}~\bibnamefont
  {Cammarota}}, \bibinfo {author} {\bibfnamefont {L.}~\bibnamefont {Gamberg}},
  \bibinfo {author} {\bibfnamefont {Z.-B.}\ \bibnamefont {Kang}}, \bibinfo
  {author} {\bibfnamefont {J.~A.}\ \bibnamefont {Miller}}, \bibinfo {author}
  {\bibfnamefont {D.}~\bibnamefont {Pitonyak}}, \bibinfo {author}
  {\bibfnamefont {A.}~\bibnamefont {Prokudin}}, \bibinfo {author}
  {\bibfnamefont {T.~C.}\ \bibnamefont {Rogers}}, \ and\ \bibinfo {author}
  {\bibfnamefont {N.}~\bibnamefont {Sato}} (\bibinfo {collaboration} {Jefferson
  Lab Angular Momentum}),\ }\href {\doibase 10.1103/PhysRevD.102.054002}
  {\bibfield  {journal} {\bibinfo  {journal} {Phys. Rev. D}\ }\textbf {\bibinfo
  {volume} {102}},\ \bibinfo {pages} {054002} (\bibinfo {year} {2020})},\
  \Eprint {http://arxiv.org/abs/2002.08384} {arXiv:2002.08384 [hep-ph]}
  \BibitemShut {NoStop}%
\bibitem [{\citenamefont {Boughezal}\ \emph {et~al.}(2023)\citenamefont
  {Boughezal}, \citenamefont {de~Florian}, \citenamefont {Petriello},\ and\
  \citenamefont {Vogelsang}}]{Boughezal:2023ooo}%
  \BibitemOpen
  \bibfield  {author} {\bibinfo {author} {\bibfnamefont {R.}~\bibnamefont
  {Boughezal}}, \bibinfo {author} {\bibfnamefont {D.}~\bibnamefont
  {de~Florian}}, \bibinfo {author} {\bibfnamefont {F.}~\bibnamefont
  {Petriello}}, \ and\ \bibinfo {author} {\bibfnamefont {W.}~\bibnamefont
  {Vogelsang}},\ }\href@noop {} {\  (\bibinfo {year} {2023})},\ \Eprint
  {http://arxiv.org/abs/2301.02304} {arXiv:2301.02304 [hep-ph]} \BibitemShut
  {NoStop}%
\bibitem [{\citenamefont {Brady}\ \emph {et~al.}(2012)\citenamefont {Brady},
  \citenamefont {Accardi}, \citenamefont {Melnitchouk},\ and\ \citenamefont
  {Owens}}]{Brady_2012}%
  \BibitemOpen
  \bibfield  {author} {\bibinfo {author} {\bibfnamefont {L.~T.}\ \bibnamefont
  {Brady}}, \bibinfo {author} {\bibfnamefont {A.}~\bibnamefont {Accardi}},
  \bibinfo {author} {\bibfnamefont {W.}~\bibnamefont {Melnitchouk}}, \ and\
  \bibinfo {author} {\bibfnamefont {J.~F.}\ \bibnamefont {Owens}},\ }\href
  {\doibase 10.1007/jhep06(2012)019} {\bibfield  {journal} {\bibinfo  {journal}
  {J. High Energy Phys.}\ }\textbf {\bibinfo {volume} {2012}} (\bibinfo {year}
  {2012}),\ 10.1007/jhep06(2012)019}\BibitemShut {NoStop}%
\bibitem [{\citenamefont {Xie}\ \emph {et~al.}(2022)\citenamefont {Xie},
  \citenamefont {Yang}, \citenamefont {Fu}, \citenamefont {Liu}, \citenamefont
  {Han}, \citenamefont {Hou}, \citenamefont {Dulat},\ and\ \citenamefont
  {Yuan}}]{Xie:2022tzo}%
  \BibitemOpen
  \bibfield  {author} {\bibinfo {author} {\bibfnamefont {M.}~\bibnamefont
  {Xie}}, \bibinfo {author} {\bibfnamefont {S.}~\bibnamefont {Yang}}, \bibinfo
  {author} {\bibfnamefont {Y.}~\bibnamefont {Fu}}, \bibinfo {author}
  {\bibfnamefont {M.}~\bibnamefont {Liu}}, \bibinfo {author} {\bibfnamefont
  {L.}~\bibnamefont {Han}}, \bibinfo {author} {\bibfnamefont {T.-J.}\
  \bibnamefont {Hou}}, \bibinfo {author} {\bibfnamefont {S.}~\bibnamefont
  {Dulat}}, \ and\ \bibinfo {author} {\bibfnamefont {C.~P.}\ \bibnamefont
  {Yuan}},\ }\href@noop {} {\  (\bibinfo {year} {2022})},\ \Eprint
  {http://arxiv.org/abs/2209.13143} {arXiv:2209.13143 [hep-ex]} \BibitemShut
  {NoStop}%
\bibitem [{\citenamefont {Gao}\ \emph {et~al.}(2022{\natexlab{c}})\citenamefont
  {Gao}, \citenamefont {Liu},\ and\ \citenamefont {Xie}}]{Gao:2022wxk}%
  \BibitemOpen
  \bibfield  {author} {\bibinfo {author} {\bibfnamefont {J.}~\bibnamefont
  {Gao}}, \bibinfo {author} {\bibfnamefont {D.}~\bibnamefont {Liu}}, \ and\
  \bibinfo {author} {\bibfnamefont {K.}~\bibnamefont {Xie}},\ }\href {\doibase
  10.1088/1674-1137/ac930b} {\bibfield  {journal} {\bibinfo  {journal} {Chin.
  Phys. C}\ }\textbf {\bibinfo {volume} {46}},\ \bibinfo {pages} {123110}
  (\bibinfo {year} {2022}{\natexlab{c}})},\ \Eprint
  {http://arxiv.org/abs/2205.03942} {arXiv:2205.03942 [hep-ph]} \BibitemShut
  {NoStop}%
\bibitem [{\citenamefont {Gao}\ \emph {et~al.}(2013)\citenamefont {Gao},
  \citenamefont {Guzzi},\ and\ \citenamefont {Nadolsky}}]{Gao:2013wwa}%
  \BibitemOpen
  \bibfield  {author} {\bibinfo {author} {\bibfnamefont {J.}~\bibnamefont
  {Gao}}, \bibinfo {author} {\bibfnamefont {M.}~\bibnamefont {Guzzi}}, \ and\
  \bibinfo {author} {\bibfnamefont {P.~M.}\ \bibnamefont {Nadolsky}},\ }\href
  {\doibase 10.1140/epjc/s10052-013-2541-4} {\bibfield  {journal} {\bibinfo
  {journal} {Eur. Phys. J. C}\ }\textbf {\bibinfo {volume} {73}},\ \bibinfo
  {pages} {2541} (\bibinfo {year} {2013})},\ \Eprint
  {http://arxiv.org/abs/1304.3494} {arXiv:1304.3494 [hep-ph]} \BibitemShut
  {NoStop}%
\bibitem [{\citenamefont {Hobbs}\ and\ \citenamefont
  {Melnitchouk}(2008)}]{Hobbs:2008mm}%
  \BibitemOpen
  \bibfield  {author} {\bibinfo {author} {\bibfnamefont {T.}~\bibnamefont
  {Hobbs}}\ and\ \bibinfo {author} {\bibfnamefont {W.}~\bibnamefont
  {Melnitchouk}},\ }\href {\doibase 10.1103/PhysRevD.77.114023} {\bibfield
  {journal} {\bibinfo  {journal} {Phys. Rev. D}\ }\textbf {\bibinfo {volume}
  {77}},\ \bibinfo {pages} {114023} (\bibinfo {year} {2008})},\ \Eprint
  {http://arxiv.org/abs/0801.4791} {arXiv:0801.4791 [hep-ph]} \BibitemShut
  {NoStop}%
\bibitem [{\citenamefont {Liu}\ \emph {et~al.}(2021{\natexlab{b}})\citenamefont
  {Liu}, \citenamefont {Melnitchouk}, \citenamefont {Qiu},\ and\ \citenamefont
  {Sato}}]{Liu_2021}%
  \BibitemOpen
  \bibfield  {author} {\bibinfo {author} {\bibfnamefont {T.}~\bibnamefont
  {Liu}}, \bibinfo {author} {\bibfnamefont {W.}~\bibnamefont {Melnitchouk}},
  \bibinfo {author} {\bibfnamefont {J.-W.}\ \bibnamefont {Qiu}}, \ and\
  \bibinfo {author} {\bibfnamefont {N.}~\bibnamefont {Sato}},\ }\href {\doibase
  10.1103/physrevd.104.094033} {\bibfield  {journal} {\bibinfo  {journal}
  {Phys. Rev. D}\ }\textbf {\bibinfo {volume} {104}} (\bibinfo {year}
  {2021}{\natexlab{b}}),\ 10.1103/physrevd.104.094033}\BibitemShut {NoStop}%
\bibitem [{\citenamefont {Rizik}\ \emph {et~al.}(2020)\citenamefont {Rizik},
  \citenamefont {Monahan},\ and\ \citenamefont {Shindler}}]{Rizik:2020naq}%
  \BibitemOpen
  \bibfield  {author} {\bibinfo {author} {\bibfnamefont {M.~D.}\ \bibnamefont
  {Rizik}}, \bibinfo {author} {\bibfnamefont {C.~J.}\ \bibnamefont {Monahan}},
  \ and\ \bibinfo {author} {\bibfnamefont {A.}~\bibnamefont {Shindler}}
  (\bibinfo {collaboration} {SymLat}),\ }\href {\doibase
  10.1103/PhysRevD.102.034509} {\bibfield  {journal} {\bibinfo  {journal}
  {Phys. Rev. D}\ }\textbf {\bibinfo {volume} {102}},\ \bibinfo {pages}
  {034509} (\bibinfo {year} {2020})},\ \Eprint
  {http://arxiv.org/abs/2005.04199} {arXiv:2005.04199 [hep-lat]} \BibitemShut
  {NoStop}%
\bibitem [{\citenamefont {Syritsyn}\ \emph {et~al.}(2019)\citenamefont
  {Syritsyn}, \citenamefont {Izubuchi},\ and\ \citenamefont
  {Ohki}}]{Syritsyn:2019vvt}%
  \BibitemOpen
  \bibfield  {author} {\bibinfo {author} {\bibfnamefont {S.}~\bibnamefont
  {Syritsyn}}, \bibinfo {author} {\bibfnamefont {T.}~\bibnamefont {Izubuchi}},
  \ and\ \bibinfo {author} {\bibfnamefont {H.}~\bibnamefont {Ohki}},\ }\href
  {\doibase 10.22323/1.336.0194} {\bibfield  {journal} {\bibinfo  {journal}
  {PoS}\ }\textbf {\bibinfo {volume} {Confinement2018}},\ \bibinfo {pages}
  {194} (\bibinfo {year} {2019})},\ \Eprint {http://arxiv.org/abs/1901.05455}
  {arXiv:1901.05455 [hep-lat]} \BibitemShut {NoStop}%
\bibitem [{\citenamefont {Bhattacharya}\ \emph {et~al.}(2021)\citenamefont
  {Bhattacharya}, \citenamefont {Cirigliano}, \citenamefont {Gupta},
  \citenamefont {Mereghetti},\ and\ \citenamefont
  {Yoon}}]{Bhattacharya:2021lol}%
  \BibitemOpen
  \bibfield  {author} {\bibinfo {author} {\bibfnamefont {T.}~\bibnamefont
  {Bhattacharya}}, \bibinfo {author} {\bibfnamefont {V.}~\bibnamefont
  {Cirigliano}}, \bibinfo {author} {\bibfnamefont {R.}~\bibnamefont {Gupta}},
  \bibinfo {author} {\bibfnamefont {E.}~\bibnamefont {Mereghetti}}, \ and\
  \bibinfo {author} {\bibfnamefont {B.}~\bibnamefont {Yoon}},\ }\href {\doibase
  10.1103/PhysRevD.103.114507} {\bibfield  {journal} {\bibinfo  {journal}
  {Phys. Rev. D}\ }\textbf {\bibinfo {volume} {103}},\ \bibinfo {pages}
  {114507} (\bibinfo {year} {2021})},\ \Eprint
  {http://arxiv.org/abs/2101.07230} {arXiv:2101.07230 [hep-lat]} \BibitemShut
  {NoStop}%
\bibitem [{\citenamefont {Courtoy}\ \emph {et~al.}(2015)\citenamefont
  {Courtoy}, \citenamefont {Bae\ss{}ler}, \citenamefont {Gonz\'alez-Alonso},\
  and\ \citenamefont {Liuti}}]{Courtoy:2015haa}%
  \BibitemOpen
  \bibfield  {author} {\bibinfo {author} {\bibfnamefont {A.}~\bibnamefont
  {Courtoy}}, \bibinfo {author} {\bibfnamefont {S.}~\bibnamefont
  {Bae\ss{}ler}}, \bibinfo {author} {\bibfnamefont {M.}~\bibnamefont
  {Gonz\'alez-Alonso}}, \ and\ \bibinfo {author} {\bibfnamefont
  {S.}~\bibnamefont {Liuti}},\ }\href {\doibase 10.1103/PhysRevLett.115.162001}
  {\bibfield  {journal} {\bibinfo  {journal} {Phys. Rev. Lett.}\ }\textbf
  {\bibinfo {volume} {115}},\ \bibinfo {pages} {162001} (\bibinfo {year}
  {2015})},\ \Eprint {http://arxiv.org/abs/1503.06814} {arXiv:1503.06814
  [hep-ph]} \BibitemShut {NoStop}%
\bibitem [{\citenamefont {Bhattacharya}\ \emph {et~al.}(2015)\citenamefont
  {Bhattacharya}, \citenamefont {Cirigliano}, \citenamefont {Cohen},
  \citenamefont {Gupta}, \citenamefont {Joseph}, \citenamefont {Lin},\ and\
  \citenamefont {Yoon}}]{Bhattacharya:2015wna}%
  \BibitemOpen
  \bibfield  {author} {\bibinfo {author} {\bibfnamefont {T.}~\bibnamefont
  {Bhattacharya}}, \bibinfo {author} {\bibfnamefont {V.}~\bibnamefont
  {Cirigliano}}, \bibinfo {author} {\bibfnamefont {S.}~\bibnamefont {Cohen}},
  \bibinfo {author} {\bibfnamefont {R.}~\bibnamefont {Gupta}}, \bibinfo
  {author} {\bibfnamefont {A.}~\bibnamefont {Joseph}}, \bibinfo {author}
  {\bibfnamefont {H.-W.}\ \bibnamefont {Lin}}, \ and\ \bibinfo {author}
  {\bibfnamefont {B.}~\bibnamefont {Yoon}} (\bibinfo {collaboration} {PNDME}),\
  }\href {\doibase 10.1103/PhysRevD.92.094511} {\bibfield  {journal} {\bibinfo
  {journal} {Phys. Rev. D}\ }\textbf {\bibinfo {volume} {92}},\ \bibinfo
  {pages} {094511} (\bibinfo {year} {2015})},\ \Eprint
  {http://arxiv.org/abs/1506.06411} {arXiv:1506.06411 [hep-lat]} \BibitemShut
  {NoStop}%
\bibitem [{\citenamefont {Gupta}\ \emph {et~al.}(2018)\citenamefont {Gupta},
  \citenamefont {Jang}, \citenamefont {Yoon}, \citenamefont {Lin},
  \citenamefont {Cirigliano},\ and\ \citenamefont
  {Bhattacharya}}]{Gupta:2018qil}%
  \BibitemOpen
  \bibfield  {author} {\bibinfo {author} {\bibfnamefont {R.}~\bibnamefont
  {Gupta}}, \bibinfo {author} {\bibfnamefont {Y.-C.}\ \bibnamefont {Jang}},
  \bibinfo {author} {\bibfnamefont {B.}~\bibnamefont {Yoon}}, \bibinfo {author}
  {\bibfnamefont {H.-W.}\ \bibnamefont {Lin}}, \bibinfo {author} {\bibfnamefont
  {V.}~\bibnamefont {Cirigliano}}, \ and\ \bibinfo {author} {\bibfnamefont
  {T.}~\bibnamefont {Bhattacharya}},\ }\href {\doibase
  10.1103/PhysRevD.98.034503} {\bibfield  {journal} {\bibinfo  {journal} {Phys.
  Rev. D}\ }\textbf {\bibinfo {volume} {98}},\ \bibinfo {pages} {034503}
  (\bibinfo {year} {2018})},\ \Eprint {http://arxiv.org/abs/1806.09006}
  {arXiv:1806.09006 [hep-lat]} \BibitemShut {NoStop}%
\bibitem [{\citenamefont {Aoki}\ \emph {et~al.}(2022)\citenamefont {Aoki} \emph
  {et~al.}}]{FlavourLatticeAveragingGroupFLAG:2021npn}%
  \BibitemOpen
  \bibfield  {author} {\bibinfo {author} {\bibfnamefont {Y.}~\bibnamefont
  {Aoki}} \emph {et~al.} (\bibinfo {collaboration} {Flavour Lattice Averaging
  Group (FLAG)}),\ }\href {\doibase 10.1140/epjc/s10052-022-10536-1} {\bibfield
   {journal} {\bibinfo  {journal} {Eur. Phys. J. C}\ }\textbf {\bibinfo
  {volume} {82}},\ \bibinfo {pages} {869} (\bibinfo {year} {2022})},\ \Eprint
  {http://arxiv.org/abs/2111.09849} {arXiv:2111.09849 [hep-lat]} \BibitemShut
  {NoStop}%
\bibitem [{\citenamefont {Courtoy}\ \emph {et~al.}(2022)\citenamefont
  {Courtoy}, \citenamefont {Miramontes}, \citenamefont {Avakian}, \citenamefont
  {Mirazita},\ and\ \citenamefont {Pisano}}]{Courtoy:2022kca}%
  \BibitemOpen
  \bibfield  {author} {\bibinfo {author} {\bibfnamefont {A.}~\bibnamefont
  {Courtoy}}, \bibinfo {author} {\bibfnamefont {A.~S.}\ \bibnamefont
  {Miramontes}}, \bibinfo {author} {\bibfnamefont {H.}~\bibnamefont {Avakian}},
  \bibinfo {author} {\bibfnamefont {M.}~\bibnamefont {Mirazita}}, \ and\
  \bibinfo {author} {\bibfnamefont {S.}~\bibnamefont {Pisano}},\ }\href
  {\doibase 10.1103/PhysRevD.106.014027} {\bibfield  {journal} {\bibinfo
  {journal} {Phys. Rev. D}\ }\textbf {\bibinfo {volume} {106}},\ \bibinfo
  {pages} {014027} (\bibinfo {year} {2022})},\ \Eprint
  {http://arxiv.org/abs/2203.14975} {arXiv:2203.14975 [hep-ph]} \BibitemShut
  {NoStop}%
\bibitem [{\citenamefont {Boehnlein}\ \emph {et~al.}(2022)\citenamefont
  {Boehnlein} \emph {et~al.}}]{Boehnlein:2021eym}%
  \BibitemOpen
  \bibfield  {author} {\bibinfo {author} {\bibfnamefont {A.}~\bibnamefont
  {Boehnlein}} \emph {et~al.},\ }\href {\doibase 10.1103/RevModPhys.94.031003}
  {\bibfield  {journal} {\bibinfo  {journal} {Rev. Mod. Phys.}\ }\textbf
  {\bibinfo {volume} {94}},\ \bibinfo {pages} {031003} (\bibinfo {year}
  {2022})},\ \Eprint {http://arxiv.org/abs/2112.02309} {arXiv:2112.02309
  [nucl-th]} \BibitemShut {NoStop}%
\bibitem [{\citenamefont {Abdul~Khalek}\ \emph {et~al.}(2019)\citenamefont
  {Abdul~Khalek}, \citenamefont {Ethier},\ and\ \citenamefont
  {Rojo}}]{AbdulKhalek:2019mzd}%
  \BibitemOpen
  \bibfield  {author} {\bibinfo {author} {\bibfnamefont {R.}~\bibnamefont
  {Abdul~Khalek}}, \bibinfo {author} {\bibfnamefont {J.~J.}\ \bibnamefont
  {Ethier}}, \ and\ \bibinfo {author} {\bibfnamefont {J.}~\bibnamefont {Rojo}}
  (\bibinfo {collaboration} {NNPDF}),\ }\href {\doibase
  10.1140/epjc/s10052-019-6983-1} {\bibfield  {journal} {\bibinfo  {journal}
  {Eur. Phys. J. C}\ }\textbf {\bibinfo {volume} {79}},\ \bibinfo {pages} {471}
  (\bibinfo {year} {2019})},\ \Eprint {http://arxiv.org/abs/1904.00018}
  {arXiv:1904.00018 [hep-ph]} \BibitemShut {NoStop}%
\bibitem [{\citenamefont {Lai}\ \emph {et~al.}(2022{\natexlab{c}})\citenamefont
  {Lai}, \citenamefont {Neill}, \citenamefont {P\l{}osko\'n},\ and\
  \citenamefont {Ringer}}]{Lai:2020byl}%
  \BibitemOpen
  \bibfield  {author} {\bibinfo {author} {\bibfnamefont {Y.~S.}\ \bibnamefont
  {Lai}}, \bibinfo {author} {\bibfnamefont {D.}~\bibnamefont {Neill}}, \bibinfo
  {author} {\bibfnamefont {M.}~\bibnamefont {P\l{}osko\'n}}, \ and\ \bibinfo
  {author} {\bibfnamefont {F.}~\bibnamefont {Ringer}},\ }\href {\doibase
  10.1016/j.physletb.2022.137055} {\bibfield  {journal} {\bibinfo  {journal}
  {Phys. Lett. B}\ }\textbf {\bibinfo {volume} {829}},\ \bibinfo {pages}
  {137055} (\bibinfo {year} {2022}{\natexlab{c}})},\ \Eprint
  {http://arxiv.org/abs/2012.06582} {arXiv:2012.06582 [hep-ph]} \BibitemShut
  {NoStop}%
\bibitem [{\citenamefont {Pang}\ \emph {et~al.}(2018)\citenamefont {Pang},
  \citenamefont {Zhou}, \citenamefont {Su}, \citenamefont {Petersen},
  \citenamefont {St\"ocker},\ and\ \citenamefont {Wang}}]{Pang:2016vdc}%
  \BibitemOpen
  \bibfield  {author} {\bibinfo {author} {\bibfnamefont {L.-G.}\ \bibnamefont
  {Pang}}, \bibinfo {author} {\bibfnamefont {K.}~\bibnamefont {Zhou}}, \bibinfo
  {author} {\bibfnamefont {N.}~\bibnamefont {Su}}, \bibinfo {author}
  {\bibfnamefont {H.}~\bibnamefont {Petersen}}, \bibinfo {author}
  {\bibfnamefont {H.}~\bibnamefont {St\"ocker}}, \ and\ \bibinfo {author}
  {\bibfnamefont {X.-N.}\ \bibnamefont {Wang}},\ }\href {\doibase
  10.1038/s41467-017-02726-3} {\bibfield  {journal} {\bibinfo  {journal}
  {Nature Commun.}\ }\textbf {\bibinfo {volume} {9}},\ \bibinfo {pages} {210}
  (\bibinfo {year} {2018})},\ \Eprint {http://arxiv.org/abs/1612.04262}
  {arXiv:1612.04262 [hep-ph]} \BibitemShut {NoStop}%
\bibitem [{\citenamefont {Kanwar}\ \emph {et~al.}(2020)\citenamefont {Kanwar},
  \citenamefont {Albergo}, \citenamefont {Boyda}, \citenamefont {Cranmer},
  \citenamefont {Hackett}, \citenamefont {Racani\`ere}, \citenamefont
  {Rezende},\ and\ \citenamefont {Shanahan}}]{Kanwar:2020xzo}%
  \BibitemOpen
  \bibfield  {author} {\bibinfo {author} {\bibfnamefont {G.}~\bibnamefont
  {Kanwar}}, \bibinfo {author} {\bibfnamefont {M.~S.}\ \bibnamefont {Albergo}},
  \bibinfo {author} {\bibfnamefont {D.}~\bibnamefont {Boyda}}, \bibinfo
  {author} {\bibfnamefont {K.}~\bibnamefont {Cranmer}}, \bibinfo {author}
  {\bibfnamefont {D.~C.}\ \bibnamefont {Hackett}}, \bibinfo {author}
  {\bibfnamefont {S.}~\bibnamefont {Racani\`ere}}, \bibinfo {author}
  {\bibfnamefont {D.~J.}\ \bibnamefont {Rezende}}, \ and\ \bibinfo {author}
  {\bibfnamefont {P.~E.}\ \bibnamefont {Shanahan}},\ }\href {\doibase
  10.1103/PhysRevLett.125.121601} {\bibfield  {journal} {\bibinfo  {journal}
  {Phys. Rev. Lett.}\ }\textbf {\bibinfo {volume} {125}},\ \bibinfo {pages}
  {121601} (\bibinfo {year} {2020})},\ \Eprint
  {http://arxiv.org/abs/2003.06413} {arXiv:2003.06413 [hep-lat]} \BibitemShut
  {NoStop}%
\bibitem [{\citenamefont {Winterhalder}\ \emph {et~al.}(2022)\citenamefont
  {Winterhalder}, \citenamefont {Magerya}, \citenamefont {Villa}, \citenamefont
  {Jones}, \citenamefont {Kerner}, \citenamefont {Butter}, \citenamefont
  {Heinrich},\ and\ \citenamefont {Plehn}}]{Winterhalder:2021ngy}%
  \BibitemOpen
  \bibfield  {author} {\bibinfo {author} {\bibfnamefont {R.}~\bibnamefont
  {Winterhalder}}, \bibinfo {author} {\bibfnamefont {V.}~\bibnamefont
  {Magerya}}, \bibinfo {author} {\bibfnamefont {E.}~\bibnamefont {Villa}},
  \bibinfo {author} {\bibfnamefont {S.~P.}\ \bibnamefont {Jones}}, \bibinfo
  {author} {\bibfnamefont {M.}~\bibnamefont {Kerner}}, \bibinfo {author}
  {\bibfnamefont {A.}~\bibnamefont {Butter}}, \bibinfo {author} {\bibfnamefont
  {G.}~\bibnamefont {Heinrich}}, \ and\ \bibinfo {author} {\bibfnamefont
  {T.}~\bibnamefont {Plehn}},\ }\href {\doibase 10.21468/SciPostPhys.12.4.129}
  {\bibfield  {journal} {\bibinfo  {journal} {SciPost Phys.}\ }\textbf
  {\bibinfo {volume} {12}},\ \bibinfo {pages} {129} (\bibinfo {year} {2022})},\
  \Eprint {http://arxiv.org/abs/2112.09145} {arXiv:2112.09145 [hep-ph]}
  \BibitemShut {NoStop}%
\bibitem [{\citenamefont {Grigsby}\ \emph {et~al.}(2021)\citenamefont
  {Grigsby}, \citenamefont {Kriesten}, \citenamefont {Hoskins}, \citenamefont
  {Liuti}, \citenamefont {Alonzi},\ and\ \citenamefont
  {Burkardt}}]{Grigsby:2020auv}%
  \BibitemOpen
  \bibfield  {author} {\bibinfo {author} {\bibfnamefont {J.}~\bibnamefont
  {Grigsby}}, \bibinfo {author} {\bibfnamefont {B.}~\bibnamefont {Kriesten}},
  \bibinfo {author} {\bibfnamefont {J.}~\bibnamefont {Hoskins}}, \bibinfo
  {author} {\bibfnamefont {S.}~\bibnamefont {Liuti}}, \bibinfo {author}
  {\bibfnamefont {P.}~\bibnamefont {Alonzi}}, \ and\ \bibinfo {author}
  {\bibfnamefont {M.}~\bibnamefont {Burkardt}},\ }\href {\doibase
  10.1103/PhysRevD.104.016001} {\bibfield  {journal} {\bibinfo  {journal}
  {Phys. Rev. D}\ }\textbf {\bibinfo {volume} {104}},\ \bibinfo {pages}
  {016001} (\bibinfo {year} {2021})},\ \Eprint
  {http://arxiv.org/abs/2012.04801} {arXiv:2012.04801 [hep-ph]} \BibitemShut
  {NoStop}%
\bibitem [{\citenamefont {Almaeen}\ \emph {et~al.}(2022)\citenamefont
  {Almaeen}, \citenamefont {Grigsby}, \citenamefont {Hoskins}, \citenamefont
  {Kriesten}, \citenamefont {Li}, \citenamefont {Lin},\ and\ \citenamefont
  {Liuti}}]{Almaeen:2022imx}%
  \BibitemOpen
  \bibfield  {author} {\bibinfo {author} {\bibfnamefont {M.}~\bibnamefont
  {Almaeen}}, \bibinfo {author} {\bibfnamefont {J.}~\bibnamefont {Grigsby}},
  \bibinfo {author} {\bibfnamefont {J.}~\bibnamefont {Hoskins}}, \bibinfo
  {author} {\bibfnamefont {B.}~\bibnamefont {Kriesten}}, \bibinfo {author}
  {\bibfnamefont {Y.}~\bibnamefont {Li}}, \bibinfo {author} {\bibfnamefont
  {H.-W.}\ \bibnamefont {Lin}}, \ and\ \bibinfo {author} {\bibfnamefont
  {S.}~\bibnamefont {Liuti}},\ }\href@noop {} {\  (\bibinfo {year} {2022})},\
  \Eprint {http://arxiv.org/abs/2207.10766} {arXiv:2207.10766 [hep-ph]}
  \BibitemShut {NoStop}%
\bibitem [{\citenamefont {Del~Debbio}\ \emph {et~al.}(2021)\citenamefont
  {Del~Debbio}, \citenamefont {Giani}, \citenamefont {Karpie}, \citenamefont
  {Orginos}, \citenamefont {Radyushkin},\ and\ \citenamefont
  {Zafeiropoulos}}]{DelDebbio:2020rgv}%
  \BibitemOpen
  \bibfield  {author} {\bibinfo {author} {\bibfnamefont {L.}~\bibnamefont
  {Del~Debbio}}, \bibinfo {author} {\bibfnamefont {T.}~\bibnamefont {Giani}},
  \bibinfo {author} {\bibfnamefont {J.}~\bibnamefont {Karpie}}, \bibinfo
  {author} {\bibfnamefont {K.}~\bibnamefont {Orginos}}, \bibinfo {author}
  {\bibfnamefont {A.}~\bibnamefont {Radyushkin}}, \ and\ \bibinfo {author}
  {\bibfnamefont {S.}~\bibnamefont {Zafeiropoulos}},\ }\href {\doibase
  10.1007/JHEP02(2021)138} {\bibfield  {journal} {\bibinfo  {journal} {JHEP}\
  }\textbf {\bibinfo {volume} {02}},\ \bibinfo {pages} {138} (\bibinfo {year}
  {2021})},\ \Eprint {http://arxiv.org/abs/2010.03996} {arXiv:2010.03996
  [hep-ph]} \BibitemShut {NoStop}%
\bibitem [{\citenamefont {Cichy}\ \emph {et~al.}(2019)\citenamefont {Cichy},
  \citenamefont {Del~Debbio},\ and\ \citenamefont {Giani}}]{Cichy:2019ebf}%
  \BibitemOpen
  \bibfield  {author} {\bibinfo {author} {\bibfnamefont {K.}~\bibnamefont
  {Cichy}}, \bibinfo {author} {\bibfnamefont {L.}~\bibnamefont {Del~Debbio}}, \
  and\ \bibinfo {author} {\bibfnamefont {T.}~\bibnamefont {Giani}},\ }\href
  {\doibase 10.1007/JHEP10(2019)137} {\bibfield  {journal} {\bibinfo  {journal}
  {JHEP}\ }\textbf {\bibinfo {volume} {10}},\ \bibinfo {pages} {137} (\bibinfo
  {year} {2019})},\ \Eprint {http://arxiv.org/abs/1907.06037} {arXiv:1907.06037
  [hep-ph]} \BibitemShut {NoStop}%
\bibitem [{\citenamefont {Gao}\ \emph {et~al.}(2022{\natexlab{d}})\citenamefont
  {Gao}, \citenamefont {Hanlon}, \citenamefont {Holligan}, \citenamefont
  {Karthik}, \citenamefont {Mukherjee}, \citenamefont {Petreczky},
  \citenamefont {Syritsyn},\ and\ \citenamefont {Zhao}}]{Gao:2022uhg}%
  \BibitemOpen
  \bibfield  {author} {\bibinfo {author} {\bibfnamefont {X.}~\bibnamefont
  {Gao}}, \bibinfo {author} {\bibfnamefont {A.~D.}\ \bibnamefont {Hanlon}},
  \bibinfo {author} {\bibfnamefont {J.}~\bibnamefont {Holligan}}, \bibinfo
  {author} {\bibfnamefont {N.}~\bibnamefont {Karthik}}, \bibinfo {author}
  {\bibfnamefont {S.}~\bibnamefont {Mukherjee}}, \bibinfo {author}
  {\bibfnamefont {P.}~\bibnamefont {Petreczky}}, \bibinfo {author}
  {\bibfnamefont {S.}~\bibnamefont {Syritsyn}}, \ and\ \bibinfo {author}
  {\bibfnamefont {Y.}~\bibnamefont {Zhao}},\ }\href@noop {} {\  (\bibinfo
  {year} {2022}{\natexlab{d}})},\ \Eprint {http://arxiv.org/abs/2212.12569}
  {arXiv:2212.12569 [hep-lat]} \BibitemShut {NoStop}%
\bibitem [{\citenamefont {Khan}\ \emph {et~al.}(2022)\citenamefont {Khan},
  \citenamefont {Liu},\ and\ \citenamefont {Sufian}}]{Khan:2022vot}%
  \BibitemOpen
  \bibfield  {author} {\bibinfo {author} {\bibfnamefont {T.}~\bibnamefont
  {Khan}}, \bibinfo {author} {\bibfnamefont {T.}~\bibnamefont {Liu}}, \ and\
  \bibinfo {author} {\bibfnamefont {R.~S.}\ \bibnamefont {Sufian}},\
  }\href@noop {} {\  (\bibinfo {year} {2022})},\ \Eprint
  {http://arxiv.org/abs/2211.15587} {arXiv:2211.15587 [hep-lat]} \BibitemShut
  {NoStop}%
\bibitem [{\citenamefont {Bauer}\ \emph {et~al.}(2022)\citenamefont {Bauer}
  \emph {et~al.}}]{Bauer:2022hpo}%
  \BibitemOpen
  \bibfield  {author} {\bibinfo {author} {\bibfnamefont {C.~W.}\ \bibnamefont
  {Bauer}} \emph {et~al.},\ }\href@noop {} {\  (\bibinfo {year} {2022})},\
  \Eprint {http://arxiv.org/abs/2204.03381} {arXiv:2204.03381 [quant-ph]}
  \BibitemShut {NoStop}%
\bibitem [{\citenamefont {Klco}\ \emph {et~al.}(2018)\citenamefont {Klco},
  \citenamefont {Dumitrescu}, \citenamefont {McCaskey}, \citenamefont {Morris},
  \citenamefont {Pooser}, \citenamefont {Sanz}, \citenamefont {Solano},
  \citenamefont {Lougovski},\ and\ \citenamefont {Savage}}]{Klco:2018kyo}%
  \BibitemOpen
  \bibfield  {author} {\bibinfo {author} {\bibfnamefont {N.}~\bibnamefont
  {Klco}}, \bibinfo {author} {\bibfnamefont {E.~F.}\ \bibnamefont
  {Dumitrescu}}, \bibinfo {author} {\bibfnamefont {A.~J.}\ \bibnamefont
  {McCaskey}}, \bibinfo {author} {\bibfnamefont {T.~D.}\ \bibnamefont
  {Morris}}, \bibinfo {author} {\bibfnamefont {R.~C.}\ \bibnamefont {Pooser}},
  \bibinfo {author} {\bibfnamefont {M.}~\bibnamefont {Sanz}}, \bibinfo {author}
  {\bibfnamefont {E.}~\bibnamefont {Solano}}, \bibinfo {author} {\bibfnamefont
  {P.}~\bibnamefont {Lougovski}}, \ and\ \bibinfo {author} {\bibfnamefont
  {M.~J.}\ \bibnamefont {Savage}},\ }\href {\doibase
  10.1103/PhysRevA.98.032331} {\bibfield  {journal} {\bibinfo  {journal} {Phys.
  Rev. A}\ }\textbf {\bibinfo {volume} {98}},\ \bibinfo {pages} {032331}
  (\bibinfo {year} {2018})},\ \Eprint {http://arxiv.org/abs/1803.03326}
  {arXiv:1803.03326 [quant-ph]} \BibitemShut {NoStop}%
\bibitem [{\citenamefont {Klco}\ \emph {et~al.}(2020)\citenamefont {Klco},
  \citenamefont {Stryker},\ and\ \citenamefont {Savage}}]{Klco:2019evd}%
  \BibitemOpen
  \bibfield  {author} {\bibinfo {author} {\bibfnamefont {N.}~\bibnamefont
  {Klco}}, \bibinfo {author} {\bibfnamefont {J.~R.}\ \bibnamefont {Stryker}}, \
  and\ \bibinfo {author} {\bibfnamefont {M.~J.}\ \bibnamefont {Savage}},\
  }\href {\doibase 10.1103/PhysRevD.101.074512} {\bibfield  {journal} {\bibinfo
   {journal} {Phys. Rev. D}\ }\textbf {\bibinfo {volume} {101}},\ \bibinfo
  {pages} {074512} (\bibinfo {year} {2020})},\ \Eprint
  {http://arxiv.org/abs/1908.06935} {arXiv:1908.06935 [quant-ph]} \BibitemShut
  {NoStop}%
\bibitem [{\citenamefont {Kharzeev}\ and\ \citenamefont
  {Kikuchi}(2020)}]{Kharzeev:2020kgc}%
  \BibitemOpen
  \bibfield  {author} {\bibinfo {author} {\bibfnamefont {D.~E.}\ \bibnamefont
  {Kharzeev}}\ and\ \bibinfo {author} {\bibfnamefont {Y.}~\bibnamefont
  {Kikuchi}},\ }\href {\doibase 10.1103/PhysRevResearch.2.023342} {\bibfield
  {journal} {\bibinfo  {journal} {Phys. Rev. Res.}\ }\textbf {\bibinfo {volume}
  {2}},\ \bibinfo {pages} {023342} (\bibinfo {year} {2020})},\ \Eprint
  {http://arxiv.org/abs/2001.00698} {arXiv:2001.00698 [hep-ph]} \BibitemShut
  {NoStop}%
\bibitem [{\citenamefont {Ikeda}\ \emph {et~al.}(2021)\citenamefont {Ikeda},
  \citenamefont {Kharzeev},\ and\ \citenamefont {Kikuchi}}]{Ikeda:2020agk}%
  \BibitemOpen
  \bibfield  {author} {\bibinfo {author} {\bibfnamefont {K.}~\bibnamefont
  {Ikeda}}, \bibinfo {author} {\bibfnamefont {D.~E.}\ \bibnamefont {Kharzeev}},
  \ and\ \bibinfo {author} {\bibfnamefont {Y.}~\bibnamefont {Kikuchi}},\ }\href
  {\doibase 10.1103/PhysRevD.103.L071502} {\bibfield  {journal} {\bibinfo
  {journal} {Phys. Rev. D}\ }\textbf {\bibinfo {volume} {103}},\ \bibinfo
  {pages} {L071502} (\bibinfo {year} {2021})},\ \Eprint
  {http://arxiv.org/abs/2012.02926} {arXiv:2012.02926 [hep-ph]} \BibitemShut
  {NoStop}%
\bibitem [{\citenamefont {De~Jong}\ \emph {et~al.}(2021)\citenamefont
  {De~Jong}, \citenamefont {Metcalf}, \citenamefont {Mulligan}, \citenamefont
  {P\l{}osko\'n}, \citenamefont {Ringer},\ and\ \citenamefont
  {Yao}}]{DeJong:2020riy}%
  \BibitemOpen
  \bibfield  {author} {\bibinfo {author} {\bibfnamefont {W.~A.}\ \bibnamefont
  {De~Jong}}, \bibinfo {author} {\bibfnamefont {M.}~\bibnamefont {Metcalf}},
  \bibinfo {author} {\bibfnamefont {J.}~\bibnamefont {Mulligan}}, \bibinfo
  {author} {\bibfnamefont {M.}~\bibnamefont {P\l{}osko\'n}}, \bibinfo {author}
  {\bibfnamefont {F.}~\bibnamefont {Ringer}}, \ and\ \bibinfo {author}
  {\bibfnamefont {X.}~\bibnamefont {Yao}},\ }\href {\doibase
  10.1103/PhysRevD.104.L051501} {\bibfield  {journal} {\bibinfo  {journal}
  {Phys. Rev. D}\ }\textbf {\bibinfo {volume} {104}},\ \bibinfo {pages}
  {051501} (\bibinfo {year} {2021})},\ \Eprint
  {http://arxiv.org/abs/2010.03571} {arXiv:2010.03571 [hep-ph]} \BibitemShut
  {NoStop}%
\bibitem [{\citenamefont {de~Jong}\ \emph {et~al.}(2022)\citenamefont
  {de~Jong}, \citenamefont {Lee}, \citenamefont {Mulligan}, \citenamefont
  {Plosko\'n}, \citenamefont {Ringer},\ and\ \citenamefont
  {Yao}}]{deJong:2021wsd}%
  \BibitemOpen
  \bibfield  {author} {\bibinfo {author} {\bibfnamefont {W.~A.}\ \bibnamefont
  {de~Jong}}, \bibinfo {author} {\bibfnamefont {K.}~\bibnamefont {Lee}},
  \bibinfo {author} {\bibfnamefont {J.}~\bibnamefont {Mulligan}}, \bibinfo
  {author} {\bibfnamefont {M.}~\bibnamefont {Plosko\'n}}, \bibinfo {author}
  {\bibfnamefont {F.}~\bibnamefont {Ringer}}, \ and\ \bibinfo {author}
  {\bibfnamefont {X.}~\bibnamefont {Yao}},\ }\href {\doibase
  10.1103/PhysRevD.106.054508} {\bibfield  {journal} {\bibinfo  {journal}
  {Phys. Rev. D}\ }\textbf {\bibinfo {volume} {106}},\ \bibinfo {pages}
  {054508} (\bibinfo {year} {2022})},\ \Eprint
  {http://arxiv.org/abs/2106.08394} {arXiv:2106.08394 [quant-ph]} \BibitemShut
  {NoStop}%
\bibitem [{\citenamefont {Farrell}\ \emph
  {et~al.}(2022{\natexlab{a}})\citenamefont {Farrell}, \citenamefont
  {Chernyshev}, \citenamefont {Powell}, \citenamefont {Zemlevskiy},
  \citenamefont {Illa},\ and\ \citenamefont {Savage}}]{Farrell:2022wyt}%
  \BibitemOpen
  \bibfield  {author} {\bibinfo {author} {\bibfnamefont {R.~C.}\ \bibnamefont
  {Farrell}}, \bibinfo {author} {\bibfnamefont {I.~A.}\ \bibnamefont
  {Chernyshev}}, \bibinfo {author} {\bibfnamefont {S.~J.~M.}\ \bibnamefont
  {Powell}}, \bibinfo {author} {\bibfnamefont {N.~A.}\ \bibnamefont
  {Zemlevskiy}}, \bibinfo {author} {\bibfnamefont {M.}~\bibnamefont {Illa}}, \
  and\ \bibinfo {author} {\bibfnamefont {M.~J.}\ \bibnamefont {Savage}},\
  }\href@noop {} {\  (\bibinfo {year} {2022}{\natexlab{a}})},\ \Eprint
  {http://arxiv.org/abs/2207.01731} {arXiv:2207.01731 [quant-ph]} \BibitemShut
  {NoStop}%
\bibitem [{\citenamefont {Farrell}\ \emph
  {et~al.}(2022{\natexlab{b}})\citenamefont {Farrell}, \citenamefont
  {Chernyshev}, \citenamefont {Powell}, \citenamefont {Zemlevskiy},
  \citenamefont {Illa},\ and\ \citenamefont {Savage}}]{Farrell:2022vyh}%
  \BibitemOpen
  \bibfield  {author} {\bibinfo {author} {\bibfnamefont {R.~C.}\ \bibnamefont
  {Farrell}}, \bibinfo {author} {\bibfnamefont {I.~A.}\ \bibnamefont
  {Chernyshev}}, \bibinfo {author} {\bibfnamefont {S.~J.~M.}\ \bibnamefont
  {Powell}}, \bibinfo {author} {\bibfnamefont {N.~A.}\ \bibnamefont
  {Zemlevskiy}}, \bibinfo {author} {\bibfnamefont {M.}~\bibnamefont {Illa}}, \
  and\ \bibinfo {author} {\bibfnamefont {M.~J.}\ \bibnamefont {Savage}},\
  }\href@noop {} {\  (\bibinfo {year} {2022}{\natexlab{b}})},\ \Eprint
  {http://arxiv.org/abs/2209.10781} {arXiv:2209.10781 [quant-ph]} \BibitemShut
  {NoStop}%
\bibitem [{\citenamefont {Kharzeev}(2021{\natexlab{b}})}]{Kharzeev:2021nzh}%
  \BibitemOpen
  \bibfield  {author} {\bibinfo {author} {\bibfnamefont {D.~E.}\ \bibnamefont
  {Kharzeev}},\ }\href {\doibase 10.1098/rsta.2021.0063} {\bibfield  {journal}
  {\bibinfo  {journal} {Phil. Trans. A. Math. Phys. Eng. Sci.}\ }\textbf
  {\bibinfo {volume} {380}},\ \bibinfo {pages} {20210063} (\bibinfo {year}
  {2021}{\natexlab{b}})},\ \Eprint {http://arxiv.org/abs/2108.08792}
  {arXiv:2108.08792 [hep-ph]} \BibitemShut {NoStop}%
\bibitem [{\citenamefont {Kovner}\ \emph {et~al.}(2019)\citenamefont {Kovner},
  \citenamefont {Lublinsky},\ and\ \citenamefont {Serino}}]{Kovner:2018rbf}%
  \BibitemOpen
  \bibfield  {author} {\bibinfo {author} {\bibfnamefont {A.}~\bibnamefont
  {Kovner}}, \bibinfo {author} {\bibfnamefont {M.}~\bibnamefont {Lublinsky}}, \
  and\ \bibinfo {author} {\bibfnamefont {M.}~\bibnamefont {Serino}},\ }\href
  {\doibase 10.1016/j.physletb.2018.10.043} {\bibfield  {journal} {\bibinfo
  {journal} {Phys. Lett. B}\ }\textbf {\bibinfo {volume} {792}},\ \bibinfo
  {pages} {4} (\bibinfo {year} {2019})},\ \Eprint
  {http://arxiv.org/abs/1806.01089} {arXiv:1806.01089 [hep-ph]} \BibitemShut
  {NoStop}%
\bibitem [{\citenamefont {Armesto}\ \emph {et~al.}(2019)\citenamefont
  {Armesto}, \citenamefont {Dominguez}, \citenamefont {Kovner}, \citenamefont
  {Lublinsky},\ and\ \citenamefont {Skokov}}]{Armesto:2019mna}%
  \BibitemOpen
  \bibfield  {author} {\bibinfo {author} {\bibfnamefont {N.}~\bibnamefont
  {Armesto}}, \bibinfo {author} {\bibfnamefont {F.}~\bibnamefont {Dominguez}},
  \bibinfo {author} {\bibfnamefont {A.}~\bibnamefont {Kovner}}, \bibinfo
  {author} {\bibfnamefont {M.}~\bibnamefont {Lublinsky}}, \ and\ \bibinfo
  {author} {\bibfnamefont {V.}~\bibnamefont {Skokov}},\ }\href {\doibase
  10.1007/JHEP05(2019)025} {\bibfield  {journal} {\bibinfo  {journal} {JHEP}\
  }\textbf {\bibinfo {volume} {05}},\ \bibinfo {pages} {025} (\bibinfo {year}
  {2019})},\ \Eprint {http://arxiv.org/abs/1901.08080} {arXiv:1901.08080
  [hep-ph]} \BibitemShut {NoStop}%
\bibitem [{\citenamefont {Lipatov}(1995)}]{Lipatov:1995pn}%
  \BibitemOpen
  \bibfield  {author} {\bibinfo {author} {\bibfnamefont {L.~N.}\ \bibnamefont
  {Lipatov}},\ }\href {\doibase 10.1016/0550-3213(95)00390-E} {\bibfield
  {journal} {\bibinfo  {journal} {Nucl. Phys. B}\ }\textbf {\bibinfo {volume}
  {452}},\ \bibinfo {pages} {369} (\bibinfo {year} {1995})},\ \Eprint
  {http://arxiv.org/abs/hep-ph/9502308} {arXiv:hep-ph/9502308} \BibitemShut
  {NoStop}%
\bibitem [{\citenamefont {Andreev}\ \emph {et~al.}(2021)\citenamefont {Andreev}
  \emph {et~al.}}]{H1:2020zpd}%
  \BibitemOpen
  \bibfield  {author} {\bibinfo {author} {\bibfnamefont {V.}~\bibnamefont
  {Andreev}} \emph {et~al.} (\bibinfo {collaboration} {H1}),\ }\href {\doibase
  10.1140/epjc/s10052-021-08896-1} {\bibfield  {journal} {\bibinfo  {journal}
  {Eur. Phys. J. C}\ }\textbf {\bibinfo {volume} {81}},\ \bibinfo {pages} {212}
  (\bibinfo {year} {2021})},\ \Eprint {http://arxiv.org/abs/2011.01812}
  {arXiv:2011.01812 [hep-ex]} \BibitemShut {NoStop}%
\bibitem [{\citenamefont {Kharzeev}\ and\ \citenamefont
  {Levin}(2021)}]{Kharzeev:2021yyf}%
  \BibitemOpen
  \bibfield  {author} {\bibinfo {author} {\bibfnamefont {D.~E.}\ \bibnamefont
  {Kharzeev}}\ and\ \bibinfo {author} {\bibfnamefont {E.}~\bibnamefont
  {Levin}},\ }\href {\doibase 10.1103/PhysRevD.104.L031503} {\bibfield
  {journal} {\bibinfo  {journal} {Phys. Rev. D}\ }\textbf {\bibinfo {volume}
  {104}},\ \bibinfo {pages} {L031503} (\bibinfo {year} {2021})},\ \Eprint
  {http://arxiv.org/abs/2102.09773} {arXiv:2102.09773 [hep-ph]} \BibitemShut
  {NoStop}%
\bibitem [{\citenamefont {Hentschinski}\ and\ \citenamefont
  {Kutak}(2022)}]{Hentschinski:2021aux}%
  \BibitemOpen
  \bibfield  {author} {\bibinfo {author} {\bibfnamefont {M.}~\bibnamefont
  {Hentschinski}}\ and\ \bibinfo {author} {\bibfnamefont {K.}~\bibnamefont
  {Kutak}},\ }\href {\doibase 10.1140/epjc/s10052-022-10056-y} {\bibfield
  {journal} {\bibinfo  {journal} {Eur. Phys. J. C}\ }\textbf {\bibinfo {volume}
  {82}},\ \bibinfo {pages} {111} (\bibinfo {year} {2022})},\ \Eprint
  {http://arxiv.org/abs/2110.06156} {arXiv:2110.06156 [hep-ph]} \BibitemShut
  {NoStop}%
\bibitem [{\citenamefont {Mueller}\ \emph {et~al.}(2020)\citenamefont
  {Mueller}, \citenamefont {Tarasov},\ and\ \citenamefont
  {Venugopalan}}]{Mueller:2019qqj}%
  \BibitemOpen
  \bibfield  {author} {\bibinfo {author} {\bibfnamefont {N.}~\bibnamefont
  {Mueller}}, \bibinfo {author} {\bibfnamefont {A.}~\bibnamefont {Tarasov}}, \
  and\ \bibinfo {author} {\bibfnamefont {R.}~\bibnamefont {Venugopalan}},\
  }\href {\doibase 10.1103/PhysRevD.102.016007} {\bibfield  {journal} {\bibinfo
   {journal} {Phys. Rev. D}\ }\textbf {\bibinfo {volume} {102}},\ \bibinfo
  {pages} {016007} (\bibinfo {year} {2020})},\ \Eprint
  {http://arxiv.org/abs/1908.07051} {arXiv:1908.07051 [hep-th]} \BibitemShut
  {NoStop}%
\bibitem [{\citenamefont {Lamm}\ \emph {et~al.}(2020)\citenamefont {Lamm},
  \citenamefont {Lawrence},\ and\ \citenamefont {Yamauchi}}]{Lamm:2019uyc}%
  \BibitemOpen
  \bibfield  {author} {\bibinfo {author} {\bibfnamefont {H.}~\bibnamefont
  {Lamm}}, \bibinfo {author} {\bibfnamefont {S.}~\bibnamefont {Lawrence}}, \
  and\ \bibinfo {author} {\bibfnamefont {Y.}~\bibnamefont {Yamauchi}} (\bibinfo
  {collaboration} {NuQS}),\ }\href {\doibase 10.1103/PhysRevResearch.2.013272}
  {\bibfield  {journal} {\bibinfo  {journal} {Phys. Rev. Res.}\ }\textbf
  {\bibinfo {volume} {2}},\ \bibinfo {pages} {013272} (\bibinfo {year}
  {2020})},\ \Eprint {http://arxiv.org/abs/1908.10439} {arXiv:1908.10439
  [hep-lat]} \BibitemShut {NoStop}%
\bibitem [{\citenamefont {Barata}\ \emph {et~al.}(2021)\citenamefont {Barata},
  \citenamefont {Mueller}, \citenamefont {Tarasov},\ and\ \citenamefont
  {Venugopalan}}]{Barata:2020jtq}%
  \BibitemOpen
  \bibfield  {author} {\bibinfo {author} {\bibfnamefont {J.~a.}\ \bibnamefont
  {Barata}}, \bibinfo {author} {\bibfnamefont {N.}~\bibnamefont {Mueller}},
  \bibinfo {author} {\bibfnamefont {A.}~\bibnamefont {Tarasov}}, \ and\
  \bibinfo {author} {\bibfnamefont {R.}~\bibnamefont {Venugopalan}},\ }\href
  {\doibase 10.1103/PhysRevA.103.042410} {\bibfield  {journal} {\bibinfo
  {journal} {Phys. Rev. A}\ }\textbf {\bibinfo {volume} {103}},\ \bibinfo
  {pages} {042410} (\bibinfo {year} {2021})},\ \Eprint
  {http://arxiv.org/abs/2012.00020} {arXiv:2012.00020 [hep-th]} \BibitemShut
  {NoStop}%
\bibitem [{\citenamefont {Li}\ \emph {et~al.}(2020{\natexlab{c}})\citenamefont
  {Li}, \citenamefont {Zhao}, \citenamefont {Maris}, \citenamefont {Chen},
  \citenamefont {Li}, \citenamefont {Tuchin},\ and\ \citenamefont
  {Vary}}]{Li:2020uhl}%
  \BibitemOpen
  \bibfield  {author} {\bibinfo {author} {\bibfnamefont {M.}~\bibnamefont
  {Li}}, \bibinfo {author} {\bibfnamefont {X.}~\bibnamefont {Zhao}}, \bibinfo
  {author} {\bibfnamefont {P.}~\bibnamefont {Maris}}, \bibinfo {author}
  {\bibfnamefont {G.}~\bibnamefont {Chen}}, \bibinfo {author} {\bibfnamefont
  {Y.}~\bibnamefont {Li}}, \bibinfo {author} {\bibfnamefont {K.}~\bibnamefont
  {Tuchin}}, \ and\ \bibinfo {author} {\bibfnamefont {J.~P.}\ \bibnamefont
  {Vary}},\ }\href {\doibase 10.1103/PhysRevD.101.076016} {\bibfield  {journal}
  {\bibinfo  {journal} {Phys. Rev. D}\ }\textbf {\bibinfo {volume} {101}},\
  \bibinfo {pages} {076016} (\bibinfo {year} {2020}{\natexlab{c}})},\ \Eprint
  {http://arxiv.org/abs/2002.09757} {arXiv:2002.09757 [nucl-th]} \BibitemShut
  {NoStop}%
\bibitem [{\citenamefont {Yao}(2022)}]{Yao:2022eqm}%
  \BibitemOpen
  \bibfield  {author} {\bibinfo {author} {\bibfnamefont {X.}~\bibnamefont
  {Yao}},\ }\href@noop {} {\  (\bibinfo {year} {2022})},\ \Eprint
  {http://arxiv.org/abs/2205.07902} {arXiv:2205.07902 [hep-ph]} \BibitemShut
  {NoStop}%
\bibitem [{\citenamefont {Barata}\ \emph {et~al.}(2022)\citenamefont {Barata},
  \citenamefont {Du}, \citenamefont {Li}, \citenamefont {Qian},\ and\
  \citenamefont {Salgado}}]{Barata:2022wim}%
  \BibitemOpen
  \bibfield  {author} {\bibinfo {author} {\bibfnamefont {J.~a.}\ \bibnamefont
  {Barata}}, \bibinfo {author} {\bibfnamefont {X.}~\bibnamefont {Du}}, \bibinfo
  {author} {\bibfnamefont {M.}~\bibnamefont {Li}}, \bibinfo {author}
  {\bibfnamefont {W.}~\bibnamefont {Qian}}, \ and\ \bibinfo {author}
  {\bibfnamefont {C.~A.}\ \bibnamefont {Salgado}},\ }\href {\doibase
  10.1103/PhysRevD.106.074013} {\bibfield  {journal} {\bibinfo  {journal}
  {Phys. Rev. D}\ }\textbf {\bibinfo {volume} {106}},\ \bibinfo {pages}
  {074013} (\bibinfo {year} {2022})},\ \Eprint
  {http://arxiv.org/abs/2208.06750} {arXiv:2208.06750 [hep-ph]} \BibitemShut
  {NoStop}%
\bibitem [{\citenamefont {Barata}\ and\ \citenamefont
  {Salgado}(2021)}]{Barata:2021yri}%
  \BibitemOpen
  \bibfield  {author} {\bibinfo {author} {\bibfnamefont {J.~a.}\ \bibnamefont
  {Barata}}\ and\ \bibinfo {author} {\bibfnamefont {C.~A.}\ \bibnamefont
  {Salgado}},\ }\href {\doibase 10.1140/epjc/s10052-021-09674-9} {\bibfield
  {journal} {\bibinfo  {journal} {Eur. Phys. J. C}\ }\textbf {\bibinfo {volume}
  {81}},\ \bibinfo {pages} {862} (\bibinfo {year} {2021})},\ \Eprint
  {http://arxiv.org/abs/2104.04661} {arXiv:2104.04661 [hep-ph]} \BibitemShut
  {NoStop}%
\bibitem [{\citenamefont {Florio}\ \emph {et~al.}(2023)\citenamefont {Florio},
  \citenamefont {Frenklakh}, \citenamefont {Ikeda}, \citenamefont {Kharzeev},
  \citenamefont {Korepin}, \citenamefont {Shi},\ and\ \citenamefont
  {Yu}}]{Florio:2023dke}%
  \BibitemOpen
  \bibfield  {author} {\bibinfo {author} {\bibfnamefont {A.}~\bibnamefont
  {Florio}}, \bibinfo {author} {\bibfnamefont {D.}~\bibnamefont {Frenklakh}},
  \bibinfo {author} {\bibfnamefont {K.}~\bibnamefont {Ikeda}}, \bibinfo
  {author} {\bibfnamefont {D.}~\bibnamefont {Kharzeev}}, \bibinfo {author}
  {\bibfnamefont {V.}~\bibnamefont {Korepin}}, \bibinfo {author} {\bibfnamefont
  {S.}~\bibnamefont {Shi}}, \ and\ \bibinfo {author} {\bibfnamefont
  {K.}~\bibnamefont {Yu}},\ }\href@noop {} {\  (\bibinfo {year} {2023})},\
  \Eprint {http://arxiv.org/abs/2301.11991} {arXiv:2301.11991 [hep-ph]}
  \BibitemShut {NoStop}%
\bibitem [{\citenamefont {Alexeev}\ \emph {et~al.}(2021)\citenamefont {Alexeev}
  \emph {et~al.}}]{Alexeev:2020xrq}%
  \BibitemOpen
  \bibfield  {author} {\bibinfo {author} {\bibfnamefont {Y.}~\bibnamefont
  {Alexeev}} \emph {et~al.},\ }\href {\doibase 10.1103/PRXQuantum.2.017001}
  {\bibfield  {journal} {\bibinfo  {journal} {PRX Quantum}\ }\textbf {\bibinfo
  {volume} {2}},\ \bibinfo {pages} {017001} (\bibinfo {year} {2021})},\ \Eprint
  {http://arxiv.org/abs/1912.07577} {arXiv:1912.07577 [quant-ph]} \BibitemShut
  {NoStop}%
\end{thebibliography}%

\end{document}